\pdfoutput=1
\newcommand\lbnedoc{PWG-004}

\documentclass[aps,onecolumn,superscriptaddress,showpacs,preprintnumbers]{revtex4-1}

\usepackage{graphicx,color}
\usepackage{amssymb,amsmath}
\usepackage{enumitem}
\usepackage{float}
\usepackage{color}
\usepackage{colortbl}
\usepackage{appendix}

\newcommand{\mev}{\,\mathrm{MeV}}

\newcommand{\gev}{\,\mathrm{GeV}}

\newcommand{\nubar}{\bar{\nu}}
\newcommand{\numu}{\nu_{\mu}}
\newcommand{\numubar}{\bar{\nu}_{\mu}}
\newcommand{\nue}{\nu_e}
\newcommand{\nuebar}{\bar{\nu}_e}
\newcommand{\nutau}{\nu_{\tau}}

\newcommand{\mutoe}{\numu\rightarrow\nue}

\newcommand{\evsq}{{eV}^{2}}

\newcommand{\gtwid}{\mathrel{\raise.3ex\hbox{$>$\kern-.75em\lower1ex\hbox{$\sim$}}}}
\newcommand{\ltwid}{\mathrel{\raise.3ex\hbox{$<$\kern-.75em\lower1ex\hbox{$\sim$}}}}

\def\be{\begin{equation}}
\def\ee{\end{equation}}
\def\ba{\begin{eqnarray}}
\def\ea{\end{eqnarray}}
\def\lsi{\raise0.3ex\hbox{$<$\kern-0.75em\raise-1.1ex\hbox{$\sim$}}}
\def\gsi{\raise0.3ex\hbox{$>$\kern-0.75em\raise-1.1ex\hbox{$\sim$}}}

\newcommand{\eps}{\varepsilon}
\renewcommand{\vec}[1]{{\mathbf{#1}}}

\begin{document}

\preprint{LBNE-\lbnedoc}

\pagenumbering{roman}

\title{\vspace{1cm} The 2010 Interim Report of the Long-Baseline Neutrino Experiment Collaboration Physics Working Groups}



\newcommand{\Alabama}{Univ. of Alabama, Tuscaloosa, AL 35487-0324, USA}
\newcommand{\Argonne}{Argonne National Laboratory, Argonne, IL 60437, USA}
\newcommand{\Boston}{Boston Univ., Boston, MA 02215, USA}
\newcommand{\Brookhaven}{Brookhaven National Laboratory, Upton, NY 11973-5000,
USA}
\newcommand{\Davis}{Univ. of California at Davis, Davis, CA 95616, USA}
\newcommand{\Irvine}{Univ. of California at Irvine, Irvine, CA 92697-4575,
USA}
\newcommand{\UCLA}{Univ. of California at Los Angeles, Los Angeles, CA
90095-1547, USA}
\newcommand{\Caltech}{California Inst. of Tech., Pasadena, CA 91109,
USA}
\newcommand{\Cambridge}{Univ. of Cambridge, Madingley Road, Cambridge CB3
0HE, United Kingdom}
\newcommand{\Catania}{Univ. of Catania and INFN, I-95129 Catania, Italy}
\newcommand{\UChicago}{Univ. of Chicago, Chicago, IL 60637-1434, USA}
\newcommand{\CSU}{Colorado State Univ., Fort Collins, CO 80521, USA}
\newcommand{\CUBoulder}{Univ. of Colorado, Boulder, CO 80309 USA}
\newcommand{\Columbia}{Columbia Univ., New York, NY 10027 USA}
\newcommand{\Dakota}{Dakota State University, Brookings, SD 57007, USA}
\newcommand{\Drexel}{Drexel Univ., Philadelphia, PA 19104, USA}
\newcommand{\Duke}{Duke Univ., Durham, NC 27708, USA}
\newcommand{\Fermilab}{Fermilab, Batavia, IL 60510-500, USA}
\newcommand{\Hawaii}{Univ. of Hawai'i, Honolulu, HI 96822-2216, USA}

\newcommand{\Houston}{Univ. of Houston, Houston, Texas, USA}

\newcommand{\VARANASI}{Banaras Hindu Univ., Varanasi UP 221005, India}
\newcommand{\Delhi}{Univ. of Delhi, Delhi 110007, India}
\newcommand{\GUWAHATI}{Indian Institute of Technology, North Guwahata,
Guwahata 781039, Assam, India}
\newcommand{\CHANDIGARH}{Panjab Univ., Chandigarh 160014, U.T., India}
\newcommand{\Indiana}{Indiana Univ., Bloomington, Indiana 47405, USA}
\newcommand{\Tokyo}{Institute for the Physics and Mathematics of the Universe
University of Tokyo, Chiba 277-8568, Japan}
\newcommand{\ISU}{Iowa State Univ., Ames, IA 50011, USA}
\newcommand{\KSU}{Kansas State Univ., Manhattan, KS 66506, USA}
\newcommand{\LBL}{Lawrence Berkeley National Lab., Berkeley, CA 94720-8153, USA}
\newcommand{\LLNL}{Lawrence Livermore National Lab., Livermore, CA
94551, USA}
\newcommand{\UCL}{University College London, London, WIC1E 6BT, England, UK}
\newcommand{\LANL}{Los Alamos National Lab., Los Alamos, NM 87545, USA}
\newcommand{\LSU}{Louisiana State Univ., Baton Rouge, LA 70803-4001, USA}
\newcommand{\UMD}{Univ. of Maryland, College Park, MD 20742-4111, USA}
\newcommand{\MSU}{Michigan State Univ., East Lansing, MI 48824, USA}
\newcommand{\UMN}{Univ. of Minnesota, Minneapolis, MN 55455, USA}
\newcommand{\Crookston}{Univ. of Minnesota, Crookston, Crookston, MN
56716-5001, USA}
\newcommand{\Duluth}{Univ. of Minnesota, Duluth, Dululth, MN 55812, USA}
\newcommand{\MIT}{MIT Massachusetts Inst. of Technology, Cambridge, MA 02139-
4307, USA}
\newcommand{\NGA}{National Geospatial-Intelligence Agency, Reston, VA 20191,
USA}
\newcommand{\NewMexico}{New Mexico State Univ., Albuquerque, NM 87131, USA}
\newcommand{\NotreDame}{Univ. of Notre Dame, Notre Dame, IN 46556-5670,
USA}
\newcommand{\Oxford}{Univ. of Oxford, Oxford OX1 3RH England, UK}
\newcommand{\Pennsylvania}{Univ. of Pennsylvania, Philadelphia, PA 19104-
6396, USA}
\newcommand{\Pittsburgh}{University of Pittsburgh, Pittsburgh, PA 15260, USA}
\newcommand{\Princeton}{Princeton University, Princeton, NJ 08544-0708, USA}
\newcommand{\Rensselaer}{Rensselaer Polytechnic Inst., Troy, NY 12180-
3590, USA}
\newcommand{\Rochester}{Univ. of Rochester, Rochester, NY 14627-0171, USA}

\newcommand{\Sheffield}{Univ. of Sheffield, Sheffield, S3 7RH, England, UK
}

\newcommand{\SouthCarolina}{Univ. of South Carolina, Orangeburg, SC 29117,
USA}
\newcommand{\SDSMT}{South Dakota School of Mines and Technology, Rapid City,
SD 57701, USA}
\newcommand{\SDState}{South Dakota State Univ., Brookings, SD 57007, USA}
\newcommand{\SMU}{Southern Methodist Univ., Dallas, TX 75275, USA}
\newcommand{\Syracuse}{Syracuse Univ., Syracuse, NY 13244-1130, USA}
\newcommand{\UTexas}{Univ. of Texas, Austin, Texas 78712, USA}
\newcommand{\Tuffs}{Tuffs Univ., Medford, Massachusetts 02155, USA}
\newcommand{\Virgina}{Virginia Tech., Blacksburg, VA 24061-0435, USA}
\newcommand{\Washington}{Univ. of Washington, Seattle, WA 98195-1560,
USA}
\newcommand{\Wisconsin}{Univ. of Wisconsin, Madison, WI 53706, USA}
\newcommand{\Yale}{Yale Univ., New Haven, CT 06520, USA}

\newcommand{\UWO}{University of Western Ontario, London, Canada}
\newcommand{\Rome}{``Sapienza'' Universit\`a di Roma, I-00185 Roma, Italy}
\newcommand{\Tata}{Tata Institute of Fundamental Research, Homi Bhabha Road, Colaba, Mumbai 400005, India}
\newcommand{\INT}{Institute for Nuclear Theory, University of Washington, Seattle, WA 98195, USA}
\newcommand{\INR}{Institute for Nuclear Research of the Russian Academy of Sciences, Moscow 117312, Russia}
\newcommand{\NCS}{North Carolina State University, Raleigh, North Carolina 27695, USA}
\newcommand{\ASU}{Arizona State University, Tempe, AZ 85287-1504}
\newcommand{\JLab}{Jefferson Lab, Newport News, Virginia 23606, USA}
\newcommand{\EPFL}{Ecole Polytechnique F\'ed\'erale de Lausanne, CH-1015 Lausanne, Switzerland}


\affiliation{\Alabama}
\affiliation{\Argonne}
\affiliation{\ASU} 
\affiliation{\Boston}
\affiliation{\Brookhaven}
\affiliation{\Davis}
\affiliation{\Irvine}
\affiliation{\JLab} 
\affiliation{\UCLA}
\affiliation{\Caltech}
\affiliation{\Cambridge}
\affiliation{\Catania}
\affiliation{\UChicago}
\affiliation{\CSU}
\affiliation{\CUBoulder}
\affiliation{\Columbia}
\affiliation{\Dakota}
\affiliation{\Drexel}
\affiliation{\Duke}
\affiliation{\Fermilab}
\affiliation{\Hawaii}
\affiliation{\Houston}
\affiliation{\VARANASI}
\affiliation{\Delhi}
\affiliation{\GUWAHATI}
\affiliation{\CHANDIGARH}
\affiliation{\Indiana}
\affiliation{\Tokyo}
\affiliation{\ISU}
\affiliation{\KSU}
\affiliation{\EPFL} 
\affiliation{\LBL}
\affiliation{\LLNL}
\affiliation{\UCL}
\affiliation{\LANL}
\affiliation{\LSU}
\affiliation{\UMD}
\affiliation{\MSU}
\affiliation{\UMN}
\affiliation{\Crookston}
\affiliation{\Duluth}
\affiliation{\MIT}
\affiliation{\INR} 
\affiliation{\NGA}
\affiliation{\NewMexico}
\affiliation{\NotreDame}
\affiliation{\NCS} 
\affiliation{\Oxford}
\affiliation{\Pennsylvania}
\affiliation{\Pittsburgh}
\affiliation{\Princeton}
\affiliation{\Rensselaer}
\affiliation{\Rochester}
\affiliation{\Rome} 
\affiliation{\Sheffield}
\affiliation{\SouthCarolina}
\affiliation{\SDSMT}
\affiliation{\SDState}
\affiliation{\SMU}
\affiliation{\Syracuse}
\affiliation{\Tata}  
\affiliation{\UTexas}
\affiliation{\Tuffs}
\affiliation{\Virgina}
\affiliation{\Washington}
\affiliation{\INT} 
\affiliation{\UWO} 
\affiliation{\Wisconsin}
\affiliation{\Yale}

\author{T.~Akiri}
\affiliation{\Duke}
\author{D.~Allspach}
\affiliation{\Fermilab}
\author{M.~Andrews}
\affiliation{\Fermilab}
\author{K.~Arisaka}
\affiliation{\UCLA}
\author{E.~Arrieta-Diaz}
\affiliation{\MSU}
\author{M.~Artuso}
\affiliation{\Syracuse}
\author{X.~Bai}
\affiliation{\SDSMT}
\author{B.~Balantekin}
\affiliation{\Wisconsin}
\author{B.~Baller}
\affiliation{\Fermilab}
\author{W.~Barletta}
\affiliation{\MIT}
% ...

\collaboration{The Long-Baseline Neutrino Experiment Science Collaboration (LBNE)}
\noaffiliation

\author{A.~Beck}\affiliation{\Duke} 
\author{O.~Benhar}\affiliation{\Rome} 
\author{F.~Beroz}\affiliation{\Duke} 
\author{A.~Dighe}\affiliation{\Tata}  
\author{H.~Duan}\affiliation{\NewMexico} 
\author{D.~Gorbunov}\affiliation{\INR} 
\author{P.~Huber}\affiliation{\Virgina} 
\author{J.~Kneller}\affiliation{\NCS} 
\author{J.~Kopp}\affiliation{\Fermilab} 
\author{C.~Lunardini}\affiliation{\ASU} 
\author{W.~Melnitchouk}\affiliation{\JLab} 
\author{A.~Moss}\affiliation{\Duke} 
\author{M.~Shaposhnikov}\affiliation{\EPFL} 
\collaboration{Additional Contributors}
\noaffiliation

\smallskip
\date{\today}

\newpage
\eject
%
\begin{abstract}
\vfill\eject
\centerline{\bf Abstract}
In early 2010, the Long-Baseline Neutrino Experiment (LBNE) science collaboration initiated a
study to investigate the physics potential of the experiment with a broad set of different beam, near- and far-detector configurations.  Nine initial topics were identified as scientific areas that motivate construction of a long-baseline neutrino experiment with a very large far detector.  We summarize the scientific justification for each topic and the estimated performance for a set of far detector reference configurations. We report also on a study of optimized beam parameters and the physics capability of proposed Near Detector configurations. This document was presented to the collaboration in fall 2010 and updated with minor modifications in early 2011.

\end{abstract}

\pacs{14.60.Lm, 14.60.Pq, 95.85.Ry, 13.15+g, 13.30.Ce, 11.30.Fs, 14.20.Dh, 26.65.+t, 25.30.Pt, 29.40.Ka, 29.40.Gx} %


\maketitle
\tableofcontents

\vfill\eject

\pagenumbering{arabic}
%
\section{Introduction}\label{intro}

This report is the first of an anticipated series of documents from the Long-Baseline Neutrino Experiment (LBNE) Science Collaboration Physics Working Group (PWG) that are intended to assist the collaboration and LBNE Project~\cite{LBNE_fnal} with establishing the best possible science case. This first document in the series focuses on the relative performance of a set of Far Detector configurations with large water Cerenkov and liquid argon detectors.


Nine initial topics (Table~\ref{TG:list}) were identified as scientific areas that motivate construction of a long-baseline neutrino experiment with a very large far detector.
In each section of this report we summarize the scientific justification for each topic, the expected state of knowledge in each area from current and planned experiments, and the estimated performance in these areas for each of a set of reference configurations described in Section~\ref{fdref}.
In each section the performance parameters most relevant to that topic are presented--it must be emphasized that these parameters are in various stages of development and will evolve as the detector groups develop more sophisticated simulations.

Although the primary focus of this report is on the Far Detector configurations, we have included a substantial chapter on the physics requirements for the Near Detector complex.  Though the emphasis is on topics that most impact the long-baseline mixing parameter measurements, we summarize also some of the broad range of additional neutrino interaction studies that would be enabled with an enhanced complex and higher neutrino flux than assumed for the long-baseline studies.

\begin{table} [h!]
\begin{center}
\begin{tabular}{|l|l|} \hline
&\\
~Long-Baseline Physics: & ~Proton decay   \\
~~~(a)~Mass Hierarchy and CP violation &  ~UHE neutrinos   \\
~~~(b)~Theta13 measurement& ~Atmospheric neutrinos \\
~~~(c)~Oscillation parameters precision measurement& ~Solar neutrinos   \\
~~~(d)~New Phenomena       &  ~Geo- \& Reactor neutrinos   \\
~Supernova Burst Neutrinos & ~Supernova Relic Neutrinos\\
&\\
\hline
\end{tabular}
\caption{\label{TG:list} List of topics investigated.}
\end{center}
\end{table}

%
\section{Far Detector Reference Configurations}\label{fdref}

In order to explore the sensitivity of LBNE in the physics topic areas, a set of ``reference'' detectors are defined. A reference detector is not a specific detector design, and likely not an optimal one, but rather is a set of performance parameters based on preliminary designs for LBNE, simulations in some cases and from previous experience (Super-Kamiokande water Cerenkov and ICARUS liquid argon detectors, for example). These parameterized detectors can be used to study the physics case for various configurations, where a configuration is defined as a set of reference detectors at specified locations at the near and far sites. The depth requirements for a massive detector at Homestake are discussed in detail in Ref.~\cite{depth}.

Previous studies have indicated that the expected sensitivity to neutrino oscillation parameter measurements for a 17-kt fiducial mass liquid argon detector is roughly equivalent to a 100-kt water Cerenkov detector. For the purposes of this comparative study, we have considered the
fourteen far detector configurations, listed below, that total 300-kt Water Cerenkov Equivalent (WCE) fiducial mass. Although the enumerated configurations are 300-kt WCE, most sections first calculate the performance for 100-kt WCE modules and these results are combined for larger configurations; this means that the relative performance of many other combinations of lower and higher mass can be deduced relatively straightforwardly. In Section~\ref{lbl}, Long-Baseline Physics, the focus is on a 200-kt WCE configuration to match the LBNE Project reference designs. Table~\ref{tab:refconfigs} shows the configuration list in a more compact form that will be referred to throughout the document.

\begin{enumerate}
\item
Three 100-kt fiducial water Cherenkov detectors
at a depth of 4850~feet at DUSEL. Photosensitive area coverage is 15\% of total surface area with 10~inch High-QE PMT's.

1a.  Same as 1, but with 30\% coverage.

1b.  Same as 1a, but with gadolinium loading also.

\item
Three 17-kt fiducial liquid argon detectors
at a depth of 4850~feet at DUSEL. Assume a scintillation photon trigger is available for proton decay and supernova neutrinos.

2a.  Same as 2, but with depth 300 feet and no photon trigger.

2b.  Same as 2, but with depth 800 feet and no photon trigger.

\item
Two 100-kt fiducial water Cherenkov detectors at 4850 foot depth as specified in 1, plus one 17-kt fiducial liquid argon detector at 300 foot depth as specified in 2a.

3a.   Same as 3, but with the water Cherenkov modules as in 1a.

3b.   Same as 3, but with one water Cherenkov modules as in 1b.

\item
Two 100-kt fiducial water Cherenkov detectors at 4850 foot depth as specified in 1, plus one 17-kt fiducial liquid argon detector at 800 foot depth as specified in 2b.

4a.   Same as 4, but with the water Cherenkov modules as in 1a.

4b.   Same as 4, but with one water Cherenkov modules as in 1b.

\item
One 100-kt fiducial gadolinium loaded water Cherenkov detector at 4850 depth as specified in 1b. Two 17-kt liquid argon modules at 300 feet as
specified as in 2a.

\item
One 100-kt fiducial gadolinium loaded water Cherenkov detector at 4850 depth as specified in 1b. Two 17-kt liquid argon modules at 800 feet as
specified as in 2b.

\end{enumerate}

\begin{table}[h!]
\begin{tabular}{|p{1.2cm}|l|} \hline
Config. Number & Detector Configuration Description  \\
\hline
  1   &Three 100~kt WC, 15\% \\
  1a  &Three 100~kt WC, 30\% \\
  1b  &Three 100~kt WC, 30\% with Gd \\
  2   &Three 17~kt LAr, 4850~ft, $\gamma$ trig \\
  2a  &Three 17~kt LAr, 300~ft, no $\gamma$ trig  \\
  2b  &Three 17~kt, LAr, 800~ft, $\gamma$ trig  \\
  3   &Two 100~kt WC, 15\% + One 17~kt LAr, 300~ft, no $\gamma$ trig \\
  3a  &Two 100~kt WC, 30\%  + One 17~kt LAr, 300~ft, no $\gamma$ trig \\
  3b  &One 100~kt WC, 15\% + One  100~kt WC, 30\% \& Gd  + One 17~kt LAr, 300~ft, no $\gamma$ trig \\
  4   &Two 100~kt WC, 15\% + One 17~kt LAr, 800~ft, $\gamma$ trig \\
  4a  &Two 100~kt WC, 30\%  + One 17~kt LAr, 800~ft, $\gamma$ trig \\
  4b  &One 100~kt WC, 15\% + One  100~kt WC, 30\% \& Gd  + One 17~kt LAr, 800~ft, $\gamma$ trig \\
  5   &One 100~kt WC, 30\% \& Gd + Two 17~kt LAr, 300~ft, no $\gamma$ trig \\
  6   &One 100~kt WC, 30\% \& Gd + Two 17~kt LAr, 800~ft, $\gamma$ trig  \\
  \hline
\end{tabular}
\caption{\label{tab:refconfigs} Summary of the far detector reference configurations.}
\end{table}

\vfill\eject 
%

\section{Long-Baseline Physics}\label{lbl}

\subsection{Motivation and Scientific Impact}\label{lbp_motivation}
Long-baseline neutrino oscillation physics is the primary focus for the
Long-Baseline Neutrino Experiment (LBNE); the motivation and scientific impact
has been well-discussed in numerous documents~\cite{USLBNEStudy:2007} so it
will not be repeated here. In each of the following sections,
we summarize the motivation for the specific measurement and discuss the
precision expected from current and planned experiments worldwide.

\subsection{Optimization of the LBNE Beam Design}\label{lbl_beam_configs}

The neutrino beamline is the central component of the Long-Baseline Neutrino
Experiment. For several years, beamline designs have been investigated in an
effort to optimize the physics reach of the experiment. In this section, we
report on the most recent work showing the direct impact of different beam
design on the sensitivity to neutrino oscillation parameters.

The LBNE beamline will be a new neutrino beamline that uses the Main Injector
(MI) 120~GeV proton accelerator. The longest baseline neutrino oscillation
experiment currently in operation is the Main Injector Neutrino Oscillation
Search (MINOS) experiment based at Fermilab. It uses the NuMI (Neutrinos at
the Main Injector)~\cite{NUMITDR} beamline from the MI. The NuMI beamline has
been operational since Jan 21, 2005 and delivered in excess of
$1 \times 10^{21}$ protons-on-target (POT) to the MINOS experiment through
2010~\cite{MINOS1}. The GEANT~\cite{GEANT} based simulation of the NuMI
beamline has been validated using data from the MINOS
experiment. The NuMI simulation software has proven to be a remarkable
success at predicting the measured neutrino charged-current (CC)
interaction rates observed in the MINOS near detector with the level of
agreement between the data and simulation CC interaction rates within
$10\%$ in the region of interest to the MINOS experiment. The current
LBNE beamline design is based on the NuMI design and uses the same
simulation framework.

\begin{figure}[!h]
\centerline{
\includegraphics[width=0.5\textwidth]{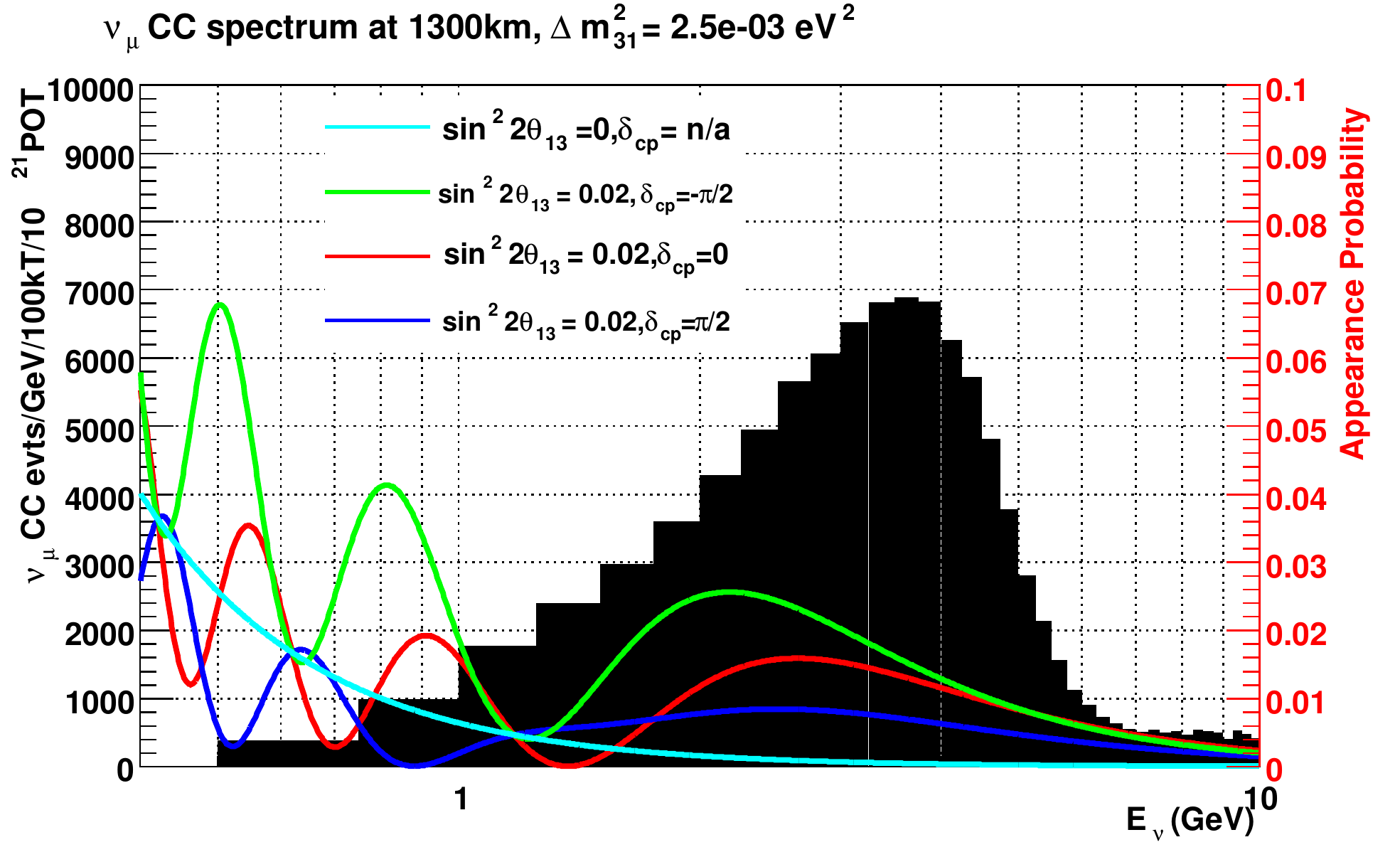}
\includegraphics[width=0.5\textwidth]{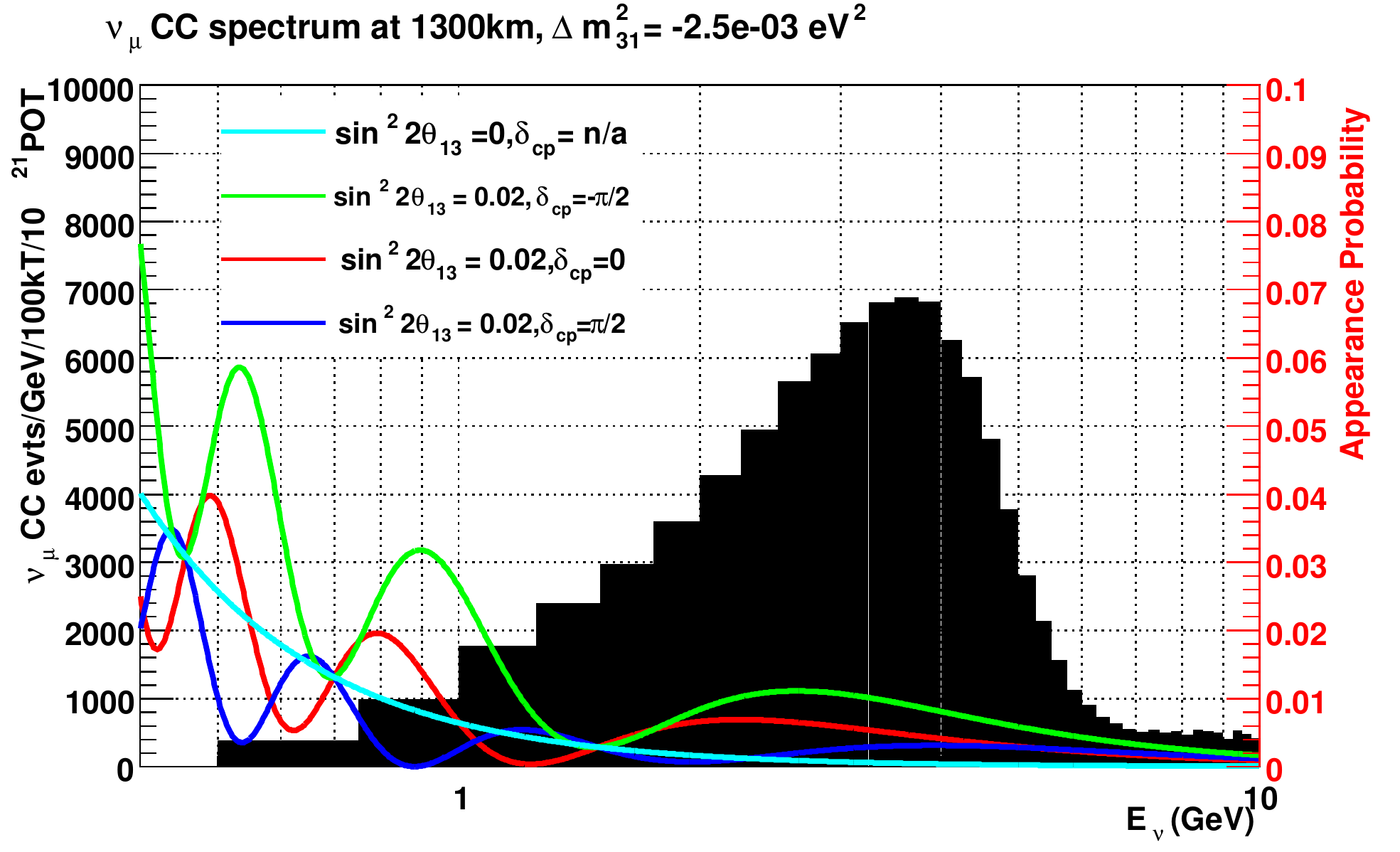}
}
\caption{The $\numu \rightarrow \nue$ oscillation probability for the LBNE
to DUSEL baseline of 1300~km for different mixing parameter with normal
hierarchy (left) and inverted hierarchy (right), is shown as colored curves.
The unoscillated CC $\numu$ spectrum from an LBNE candidate beam is shown as
the solid black histogram.}
\label{fig:lbl_beam_prob1}
\end{figure}

The design specifications of the LBNE neutrino beamline is driven by
the physics of $\nu_{\mu}\rightarrow \nu_{\mu}$, and
$\nu_{\mu}\rightarrow \nu_e/\nu_{\tau}$ oscillations.
In Fig.~\ref{fig:lbl_beam_prob1},
the $\nu_{\mu}\rightarrow \nu_e$ oscillation probability for the LBNE
to DUSEL baseline of 1300~km for different mixing parameters is
shown as colored curves. The total CC $\nu_{\mu}$ spectrum from an
LBNE candidate beam is shown as the black
solid histogram. In principle, the ideal LBNE
neutrino beam would be one that has a wide energy band that covers the
energy region from low energies to the energy of the first ($\pi/2$)
oscillation maximum and minimal flux beyond the region of
interest. Low flux at high neutrino energies is desired to eliminate
neutral-current backgrounds from high energy neutrinos that are not
sensitive to oscillations but still produce significant background at
low observed energies in the neutrino detectors.

In 2008/2009 we specified the following broad requirements for the
LBNE beam based on examination
of the oscillation nodes in Fig.~\ref{fig:lbl_beam_prob1}:
\begin{enumerate}
\item We require the highest possible neutrino fluxes to encompass at least
  the 1st and 2nd oscillation nodes, the maxima of which occur at 2.4 and
  0.8~GeV respectively.
\item Since neutrino cross sections scale with energy, larger fluxes
  at lower energies are desirable to achieve the physics
  sensitivities using effects at the 2nd oscillation node and beyond.
\item To detect $\nu_{\mu} \rightarrow \nu_e$ events at the far
  detector, it is critical to minimize the neutral-current
  contamination at lower energy, therefore it is highly desirable to
  minimize the flux of neutrinos with energies greater than 5~GeV
  where there is little sensitivity to the oscillation parameters
  (including the CP phase and the mass hierarchy).
\item The irreducible background to $\nu_{\mu} \rightarrow \nu_e$
  appearance signal comes from beam generated $\nu_e$ events,
  therefore, a high purity $\nu_{\mu}$ beam (lowest
  possible $\nu_e$ contamination) is required.
\end{enumerate}

We studied the physics performance of several conventional horn
focused neutrino beam designs in 2008-2010. We considered different
decay pipe geometries and different aluminum horn designs (AGS, NuMI, T2K)
and different beam tunes (horn/target placement and horn currents). A
solid cylindrical water cooled graphite target, two
nuclear-interaction-lengths long was chosen as the initial target material
and geometry for use with the 700~kW LBNE proton beam pending the result of
radiation damage studies with different target materials. An
optimization of the target material and geometry is underway.

We found that a two horn focusing design using parabolic horns
similar to the NuMI horns gave the best performance of the
conventional horn focused beams. We chose the radius of the decay pipe
to be 2~m to maximize the yield of low energy neutrinos in the
oscillation region ($<$~5~GeV). Ideally the length of the decay pipe should
be chosen such that the pions generating neutrinos in the
oscillation energy range to decay will decay.
The 1st oscillation maxima at 2.5~GeV is generated by 6~GeV pions,
for which the decay length is
333~m. The length of the decay pipe was chosen to be only 250~m to
reduce the excavation volume required. The LBNE decay pipe in the ``March 2010''
design was chosen to be air cooled to mitigate the risks associated
with a water cooled decay pipe - such as NuMI's. However, the air in the decay
pipe absorbs some of the pions in the beam before they decay so for this
study we used an evacuated or helium-filled decay pipe. A technical design
for a helium-filled decay pipe is currently being assessed.

The current (2010) best candidate LBNE beam spectrum obtained using a 120~GeV
proton beam is shown as the solid black histogram in
Fig.~\ref{fig:lbl_beam_prob1}. As shown in the figure, the rate of
$\nu_{\mu}$ CC events in the region of the 2nd oscillation maxima ($<~1.5$~GeV) is much lower than at the 1st oscillation maxima. Currently,
we do not have a conventional horn focused beamline design that can
provide sufficient neutrino flux at the 2nd maxima ($<$~1.5~GeV) or
below where the impact of the CP violating phase is maximal. Therefore
for this study, we have focused on optimizing the neutrino flux
coverage in the oscillation region of the 1st maximum (1.5--6~GeV).

\begin{figure}[!h]
\centerline{\includegraphics[width=0.6\textwidth]{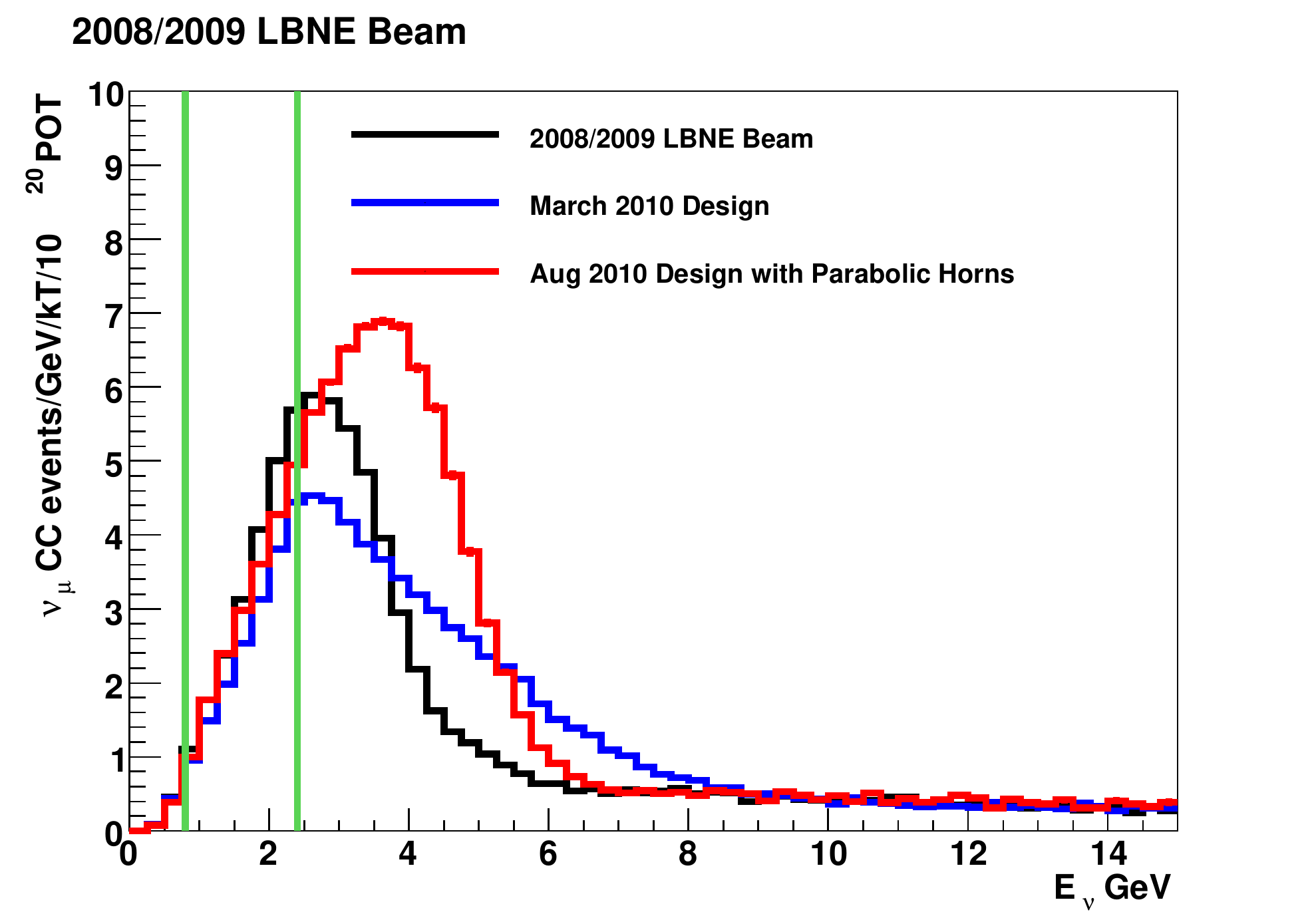}}
\caption{Various $\numu$ CC spectra obtained using variations of the two horn
LBNE design. In black is the 2008/2009 LBNE design using an embedded high
density carbon target 0.6~cm in radius and 80~cm in length and the two
NuMI horns with an evacuated decay pipe 2~m in radius and 280~m in
length. In blue is an LBNE design from ``March 2010'', which has an
embedded graphite target with 0.77~cm radius and 96~cm in length, a
modified NuMI horn~1 with a cylindrical front end and NuMI horn~2
operating at 300~kA. The decay pipe in the ``March 2010'' design is air
filled, 2~m in radius and 250~m in length. In red is the ``August 2010'' LBNE
candidate beam design, which has two parabolic NuMI horns operating at 250~kA
with the target pulled back 30~cm from the face of horn~1.}
\label{fig:fig_beam_spectra1}
\end{figure}

The unoscillated $\numu$ CC spectra
optimized for $\nu_{\mu} \rightarrow \nu_e$ appearance obtained
using variations of the conventional two horn LBNE beamline design are
shown in Fig.~\ref{fig:fig_beam_spectra1}. The beam spectrum used for the $\nu_e$ appearance studies and
$\nu_{\mu}$ disappearance studies reported in this document is shown
 in red.

\begin{figure}[!h]
\centerline{
\includegraphics[width=0.5\textwidth]{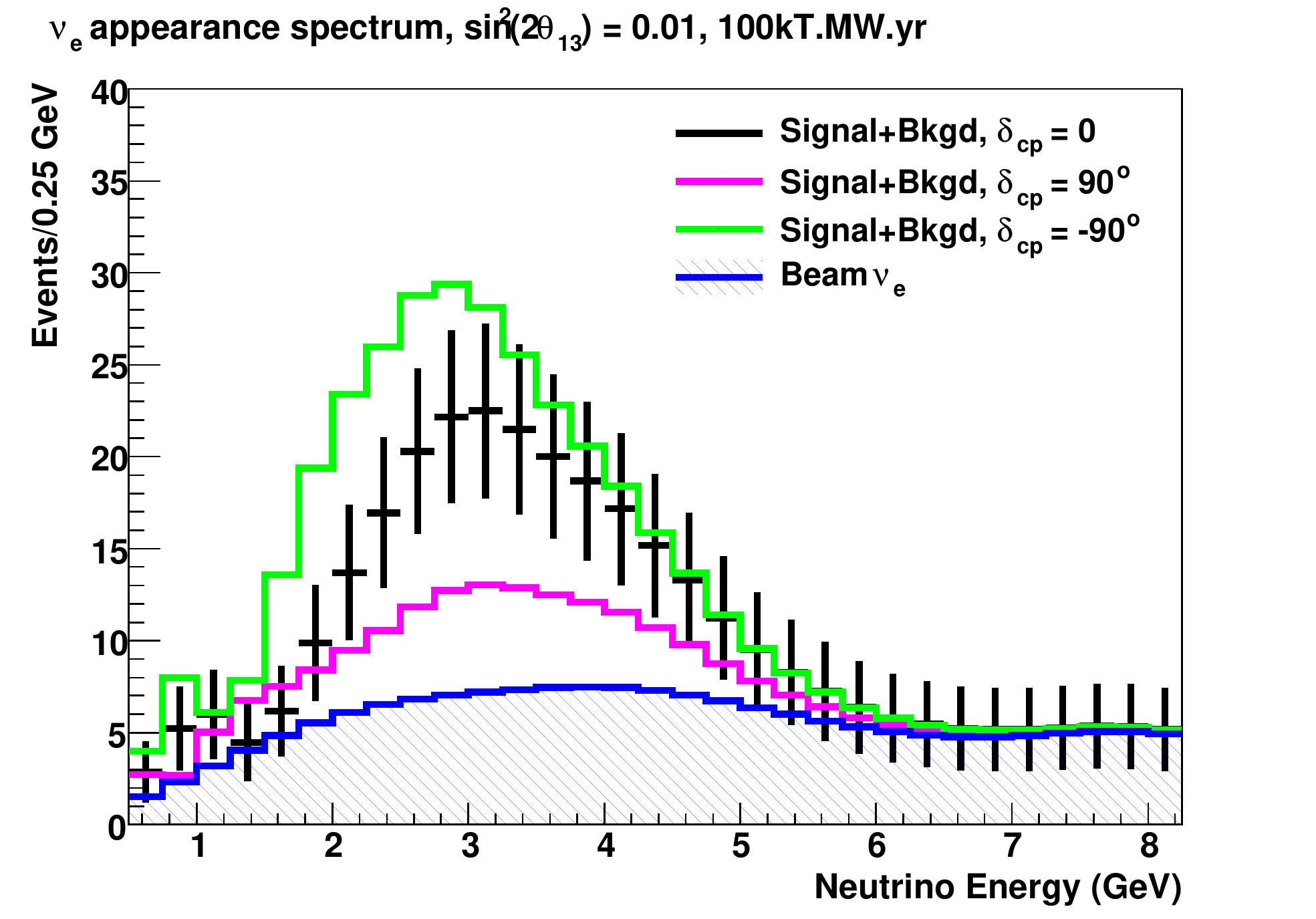}
\includegraphics[width=0.5\textwidth]{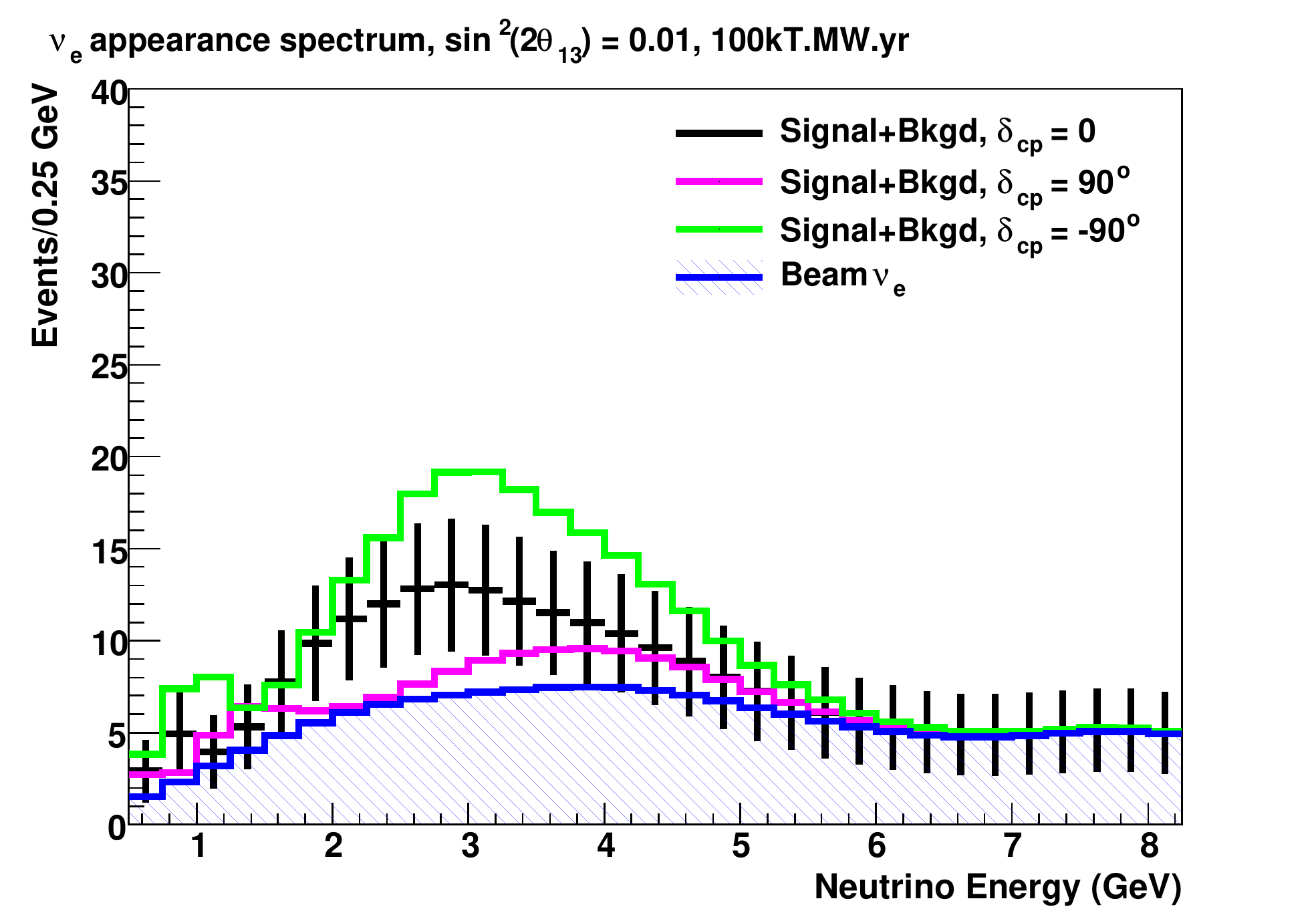}}
\caption{The $\nu_e$ appearance spectrum from the 2008/2009 reference beam
for an exposure of 100~kT.MW.yr,  $\sin^2(2\theta_{13}) = 0.01$,
normal hierarchy (left), and inverted hierarchy (right). No detector
effects are included.}
\label{fig:beam_nueappear}
\end{figure}

\begin{figure}[ht]
\centering\includegraphics[height=0.45\textwidth,angle=90]{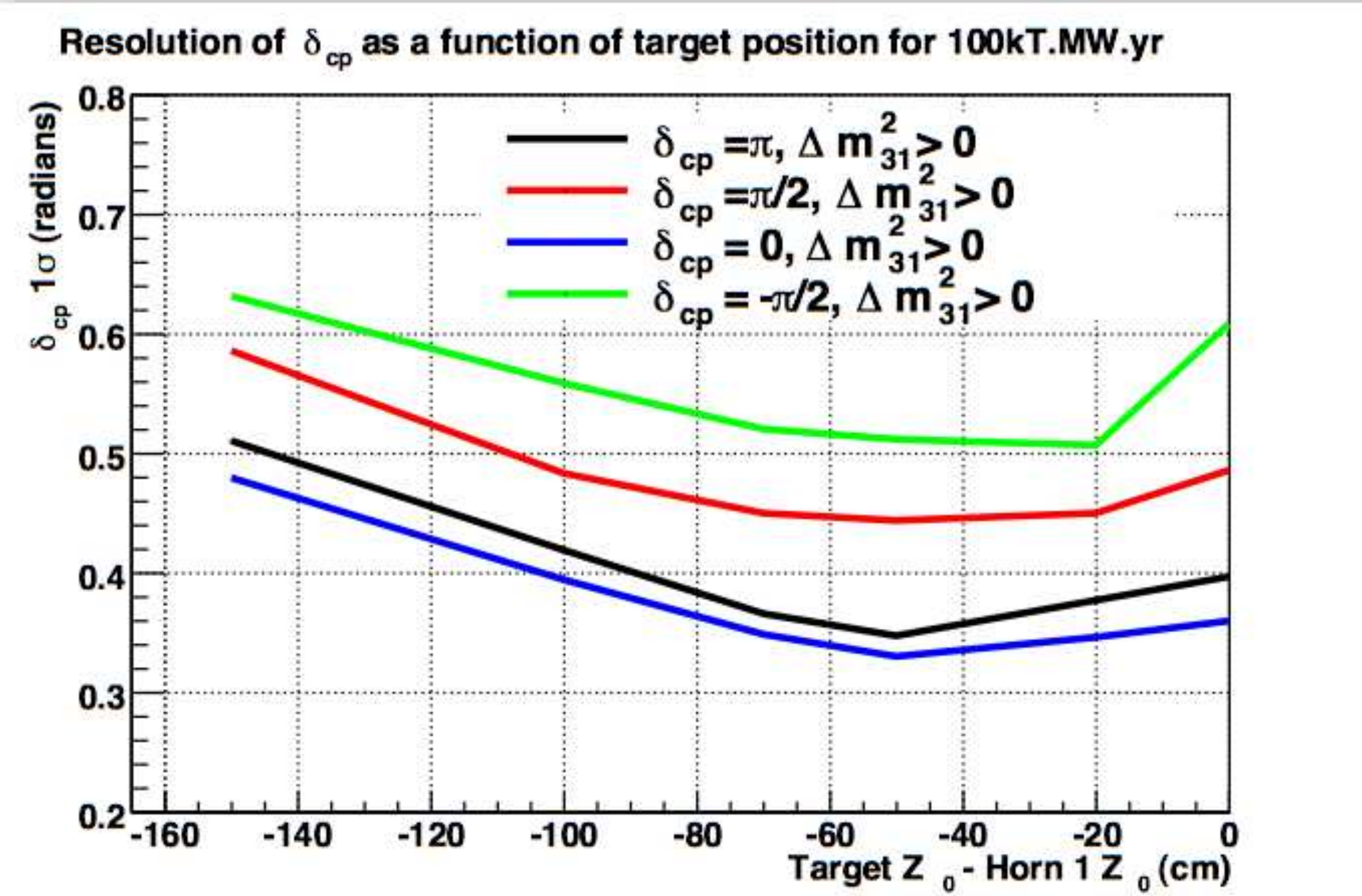}
\centering\includegraphics[height=0.45\textwidth,angle=90]{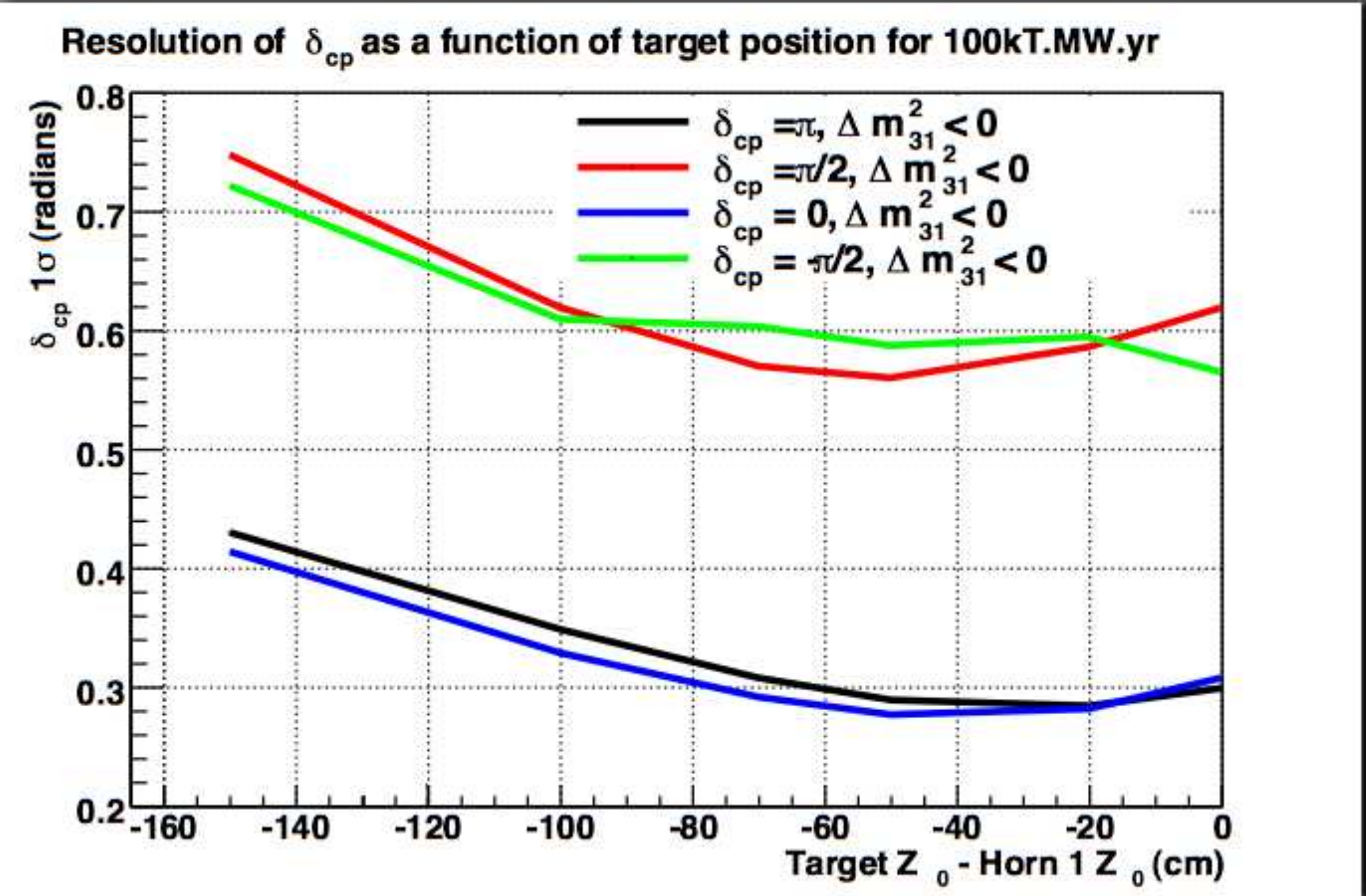}
\caption{The $1 \sigma$ resolution of  $\delta_{CP}$ as the target position is
changed in the 2008/2009 LBNE beam design. Normal hierarchy (left) and
inverted hierarchy (right). An exposure of 100~kt.MW.yr is assumed
with $\nu: \bar{\nu}$ running in the ratio 1:1. No detector effects included.}
\label{fig:dcpres_vs_horni}
\end{figure}

\begin{figure}[ht]
\centering\includegraphics[height=0.45\textwidth,angle=90]{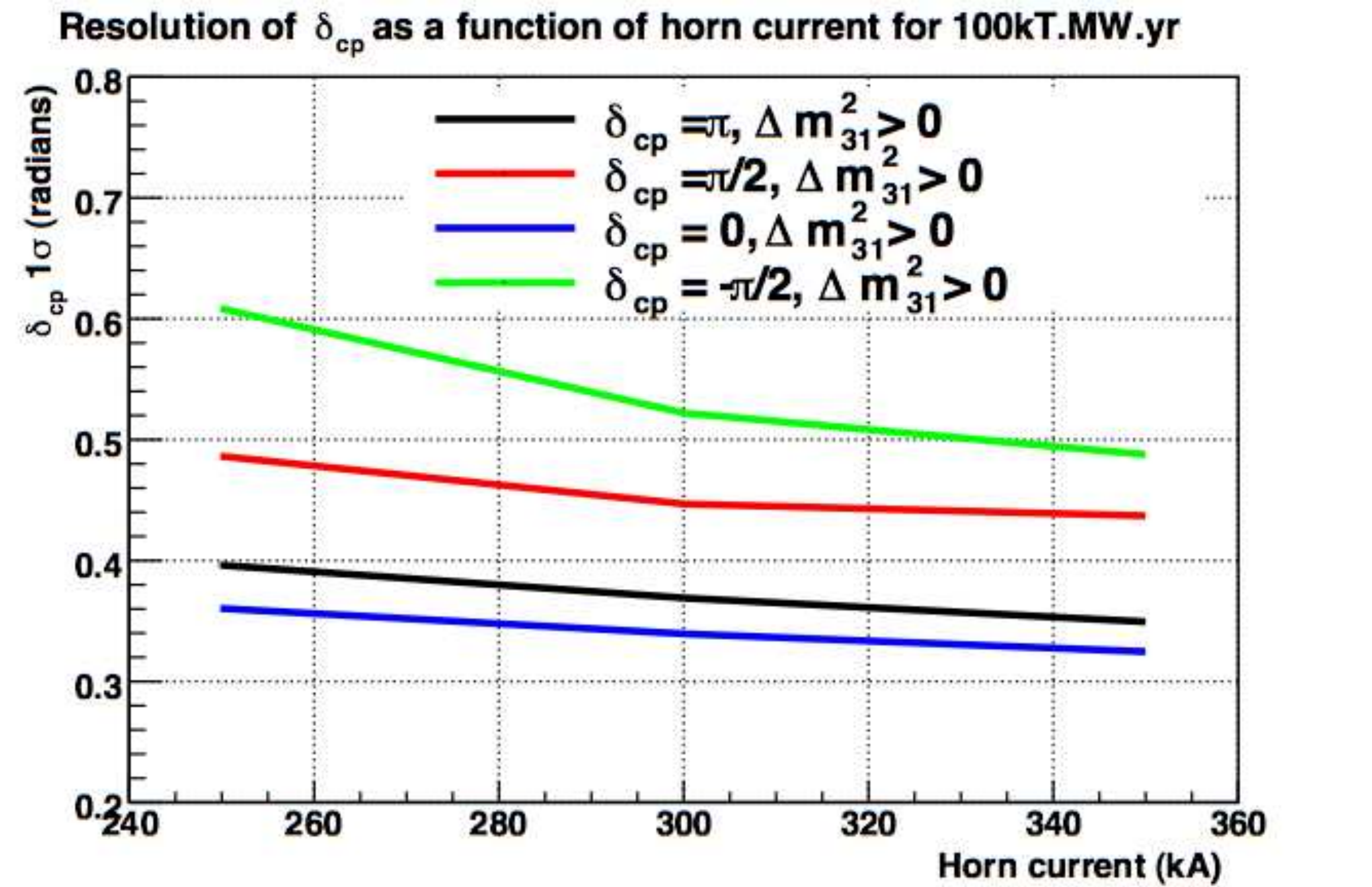}
\centering\includegraphics[height=0.45\textwidth,angle=90]{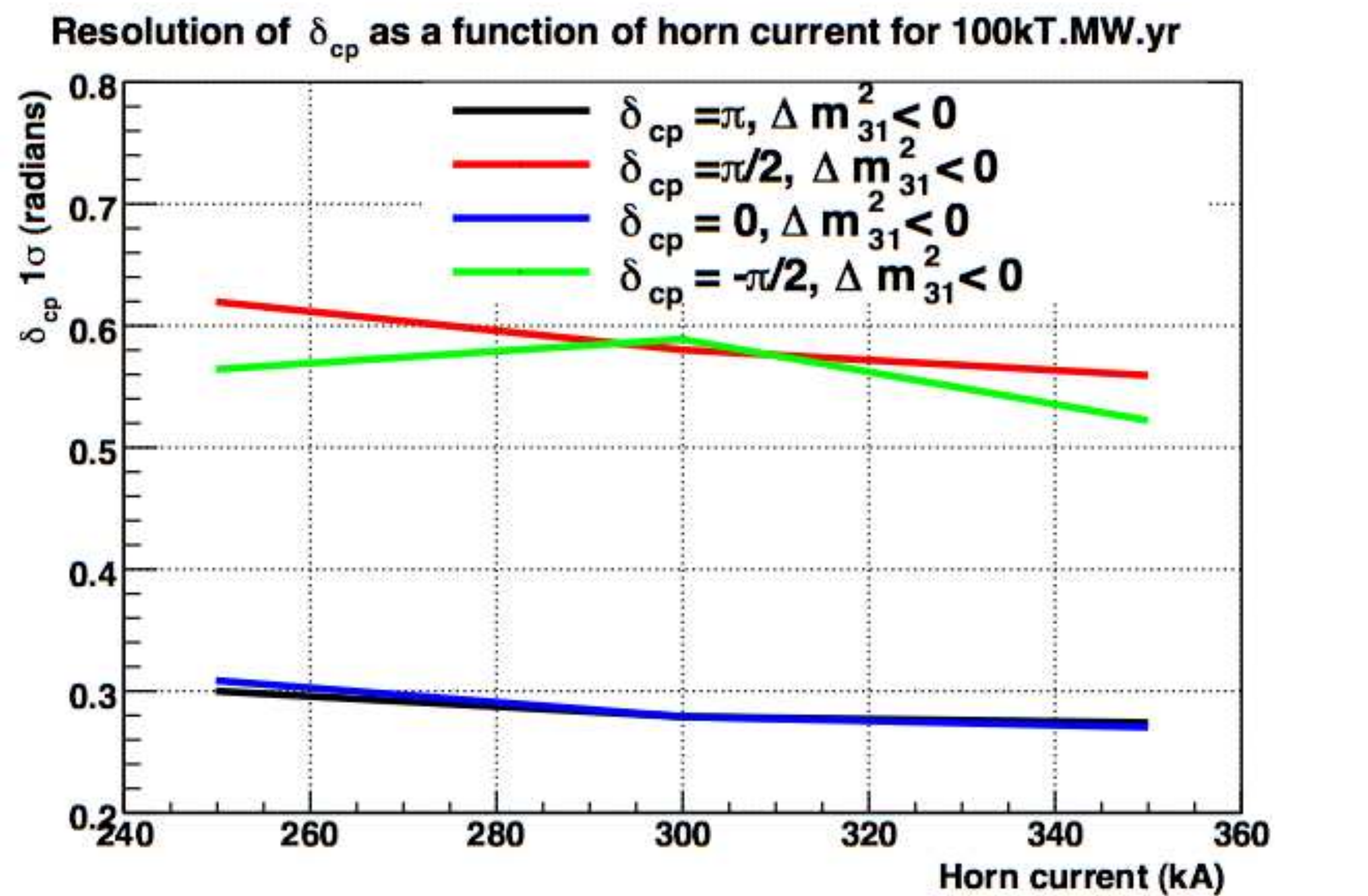}
\caption{The $1 \sigma$ resolution of  $\delta_{CP}$ as the horn current is
changed in the 2008/2009 LBNE beam design. Normal hierarchy (left) and
inverted hierarchy (right). An exposure of 100~kt.MW.yr is assumed
with $\nu: \bar{\nu}$ running in the ratio 1:1. No detector effects included.}
\label{fig:dcpres_vs_tgtpos}
\end{figure}

As a first step, we studied the impact of possible beam design
variations on the {\it resolution} of the value of $\sin^2 2\theta_{13}$ and
$\delta_{CP}$. For this initial study, a simultaneous $\chi^2$ fit to the
binned neutrino and anti-neutrino appearance spectrum
(Fig.~\ref{fig:beam_nueappear}) was performed for each candidate beam -
including the irreducible $\nu_e$ background from the beam.
In the fit, the parameters
$\sin^2 2\theta_{13}$ and $\delta_{CP}$ were allowed to vary, while the mass
hierarchy and the other oscillation parameters were fixed to their correct
values.  A $2\%$ systematic uncertainty on the $\nu_e$ background was
included. All other uncertainties included statistical errors only. The
1$\sigma$ $\delta_{CP}$ uncertainty returned from the fit was then examined
for different beam configurations. We considered the impact of horn currents
and the graphite target position w.r.t to horn~1 on the resolution of
$\delta_{CP}$ as shown in Figs.~\ref{fig:dcpres_vs_horni} and
\ref{fig:dcpres_vs_tgtpos}, respectively. The $\delta_{CP}$ measurement
is discussed in more detail in Section~\ref{lbl_cp}. From this, we see that
the ``March 2010'' design and the 2008/2009 reference design (shown in
Fig.~\ref{fig:fig_beam_spectra1}) have similar resolution for $\delta_{CP}$
and $\sin^22\theta_{13}$ in the absence of NC backgrounds and detector
effects. These same studies indicate that the ``August 2010'' reference beam design
with either a 350~kA current and embedded target OR a 250~kA with the target approximately
-0.5~m from the face of horn~1 produces significantly improved
resolution for $\delta_{CP}$ and $\sin^22 \theta_{13}$ as compared to the
``March 2010'' or 2008/2009 reference designs.

To further pursue these apparent gains, full oscillation sensitivity fits
were performed using a GLoBES-based~\cite{globes} LBNE analysis, which includes
detector effects and background sources as descrined in Appendix~\ref{lbl_appendix}. Fig.~\ref{fig:beamcomp-wc}
shows the resultant comparison of LBNE oscillation sensitivities in WC for
the 2008/2009 reference design beam and the newer ``August 2010'' beam.
Fig.~\ref{fig:beamcomp-lar} shows the same for LAr. For both WC and LAr
detectors, the sensitivities are very similar for the two beam designs.
Adding detector effects and NC backgrounds may have lessened the gains
we expected to see based on the resolution studies
(Figs.~\ref{fig:dcpres_vs_horni} and \ref{fig:dcpres_vs_tgtpos}); however,
the fact that the sensitivities are also so similar for LAr suggests that
the apparent gains were more likely lost once parameter degeneracies were
taken into account in the full oscillation fits. Recall that in the resolution
studies, only a single oscillation parameter was studied at a time. This
highlights the importance of performing full oscillation fits before making
final conclusions about a given beam design.

\begin{figure}[ht]
\centering\includegraphics[height=0.45\textwidth]{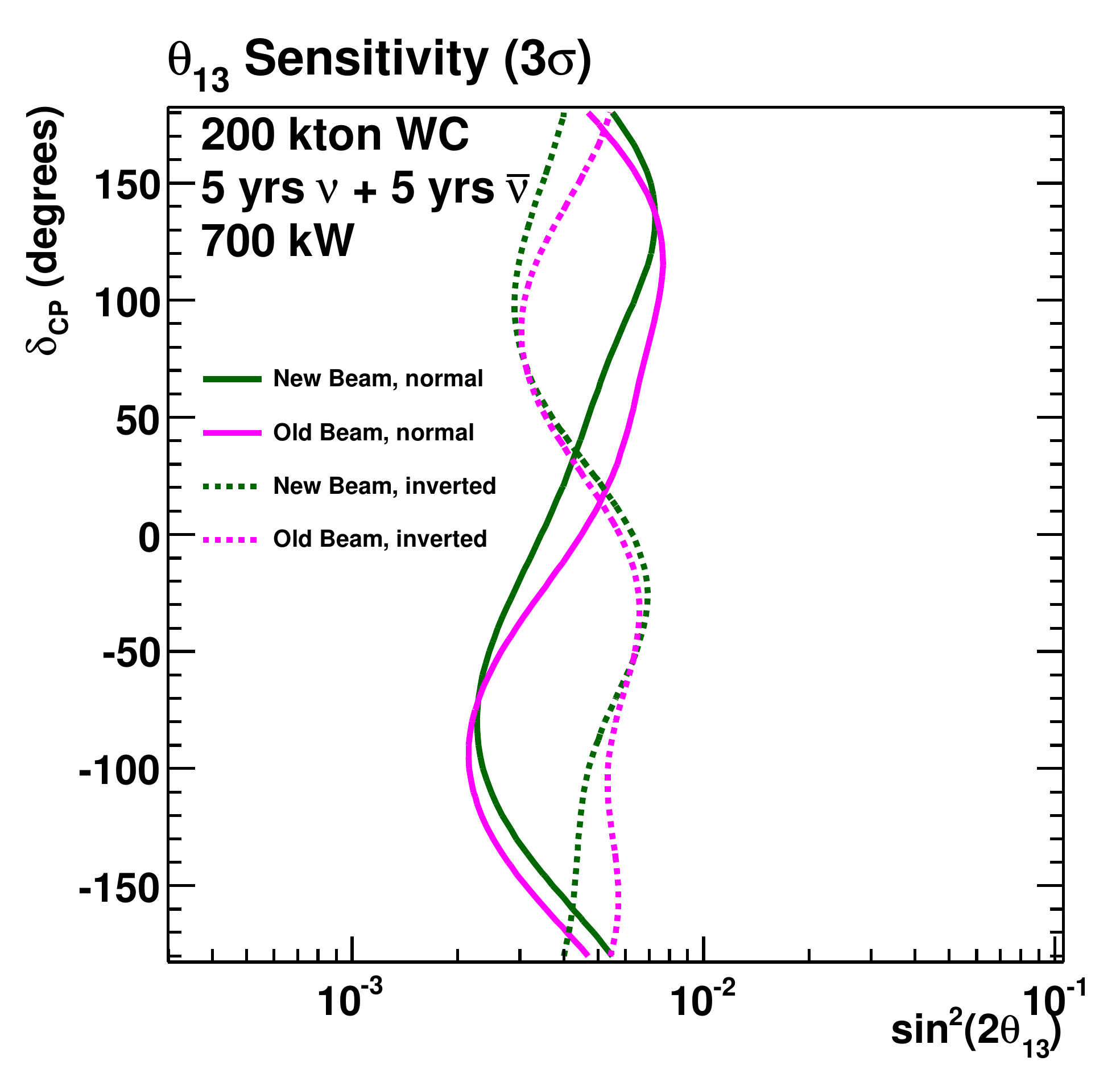}
\centering\includegraphics[height=0.45\textwidth]{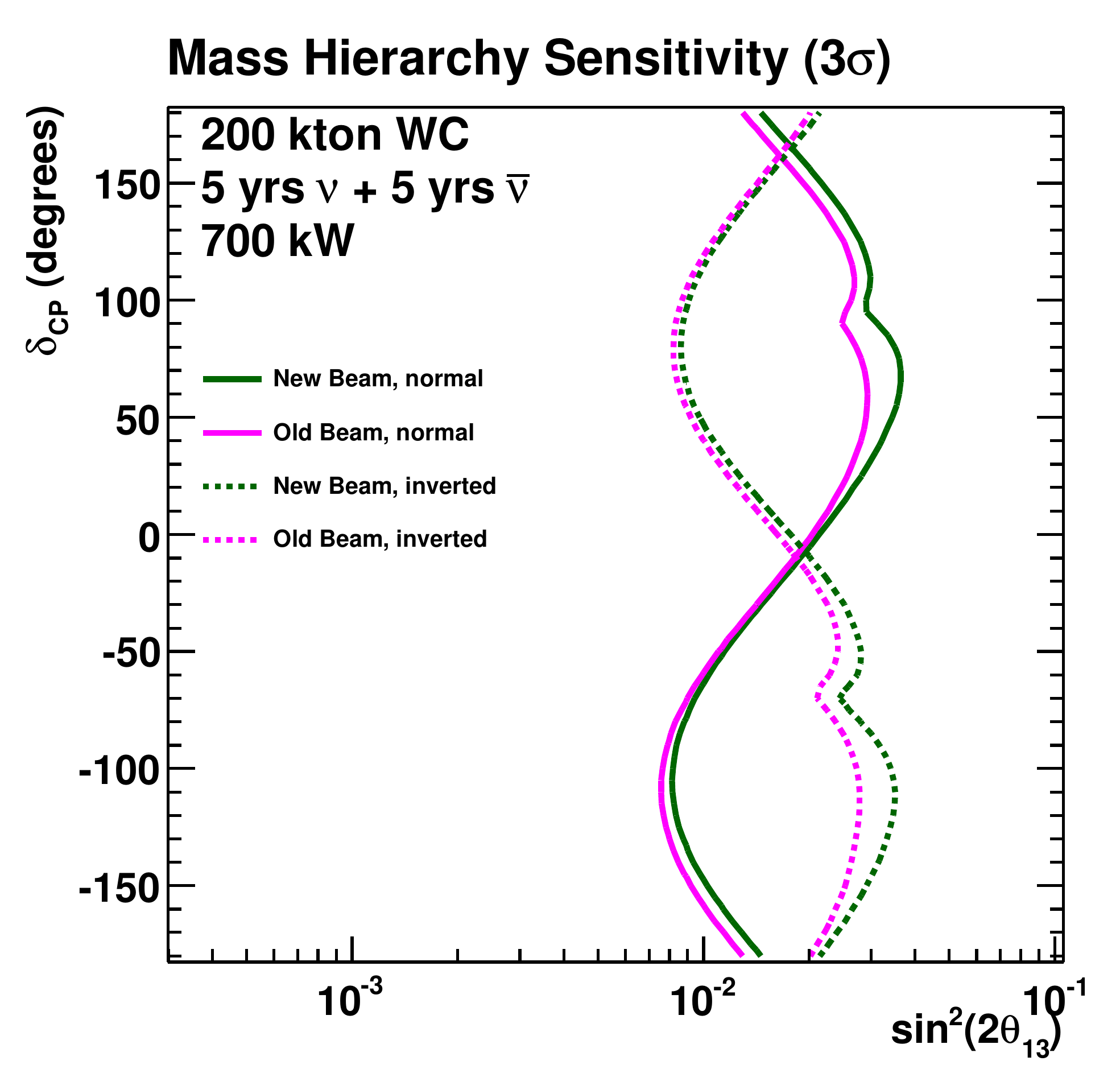}
\centering\includegraphics[height=0.45\textwidth]{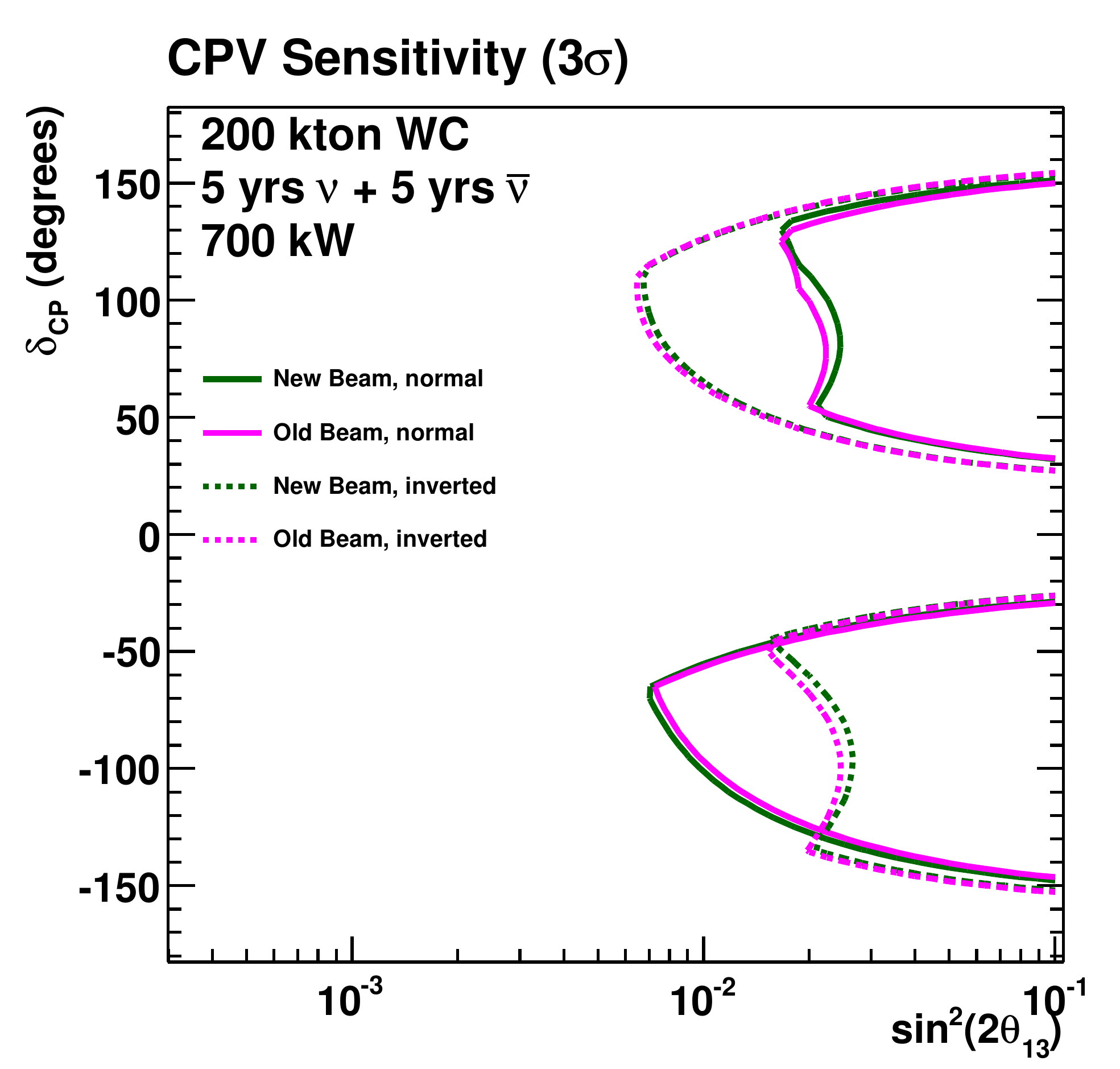}
\caption{Comparison of the sensitivity of LBNE to discovering
$\sin^22\theta_{13}\neq0$, the mass hierarchy, and CP violation at $3\sigma$
for two different LBNE beam designs. Sensitivities are shown for 200 kt WC
in 5+5 years of $\nu+\nubar$ running in a 700 kW beam for both normal
(solid) and inverted (dashed) mass hierarchies. The projections include
detector effects, all background sources and their uncertainties
(see Appendix~\ref{lbl_appendix}). ``Old beam'' (magenta) refers to the
2008/2009 LBNE reference design and corresponds to the flux plotted in black
in Fig.~\ref{fig:fig_beam_spectra1}. ``New beam'' (black) refers to the
``August 2010'' beam and corresponds to the flux plotted in red in
Fig.~\ref{fig:fig_beam_spectra1}.}
\label{fig:beamcomp-wc}
\end{figure}

\begin{figure}[ht]
\centering\includegraphics[height=0.45\textwidth]{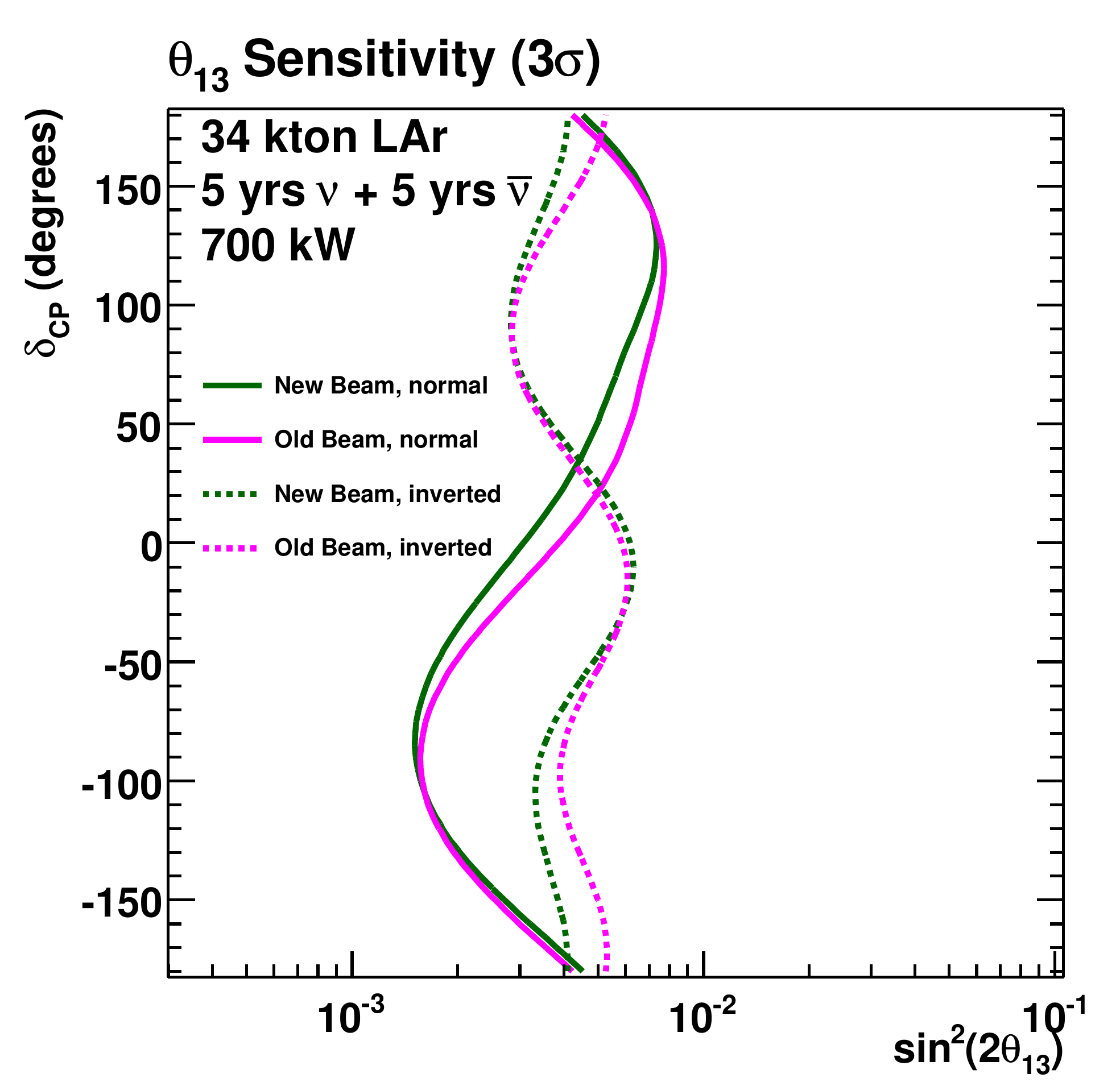}
\centering\includegraphics[height=0.45\textwidth]{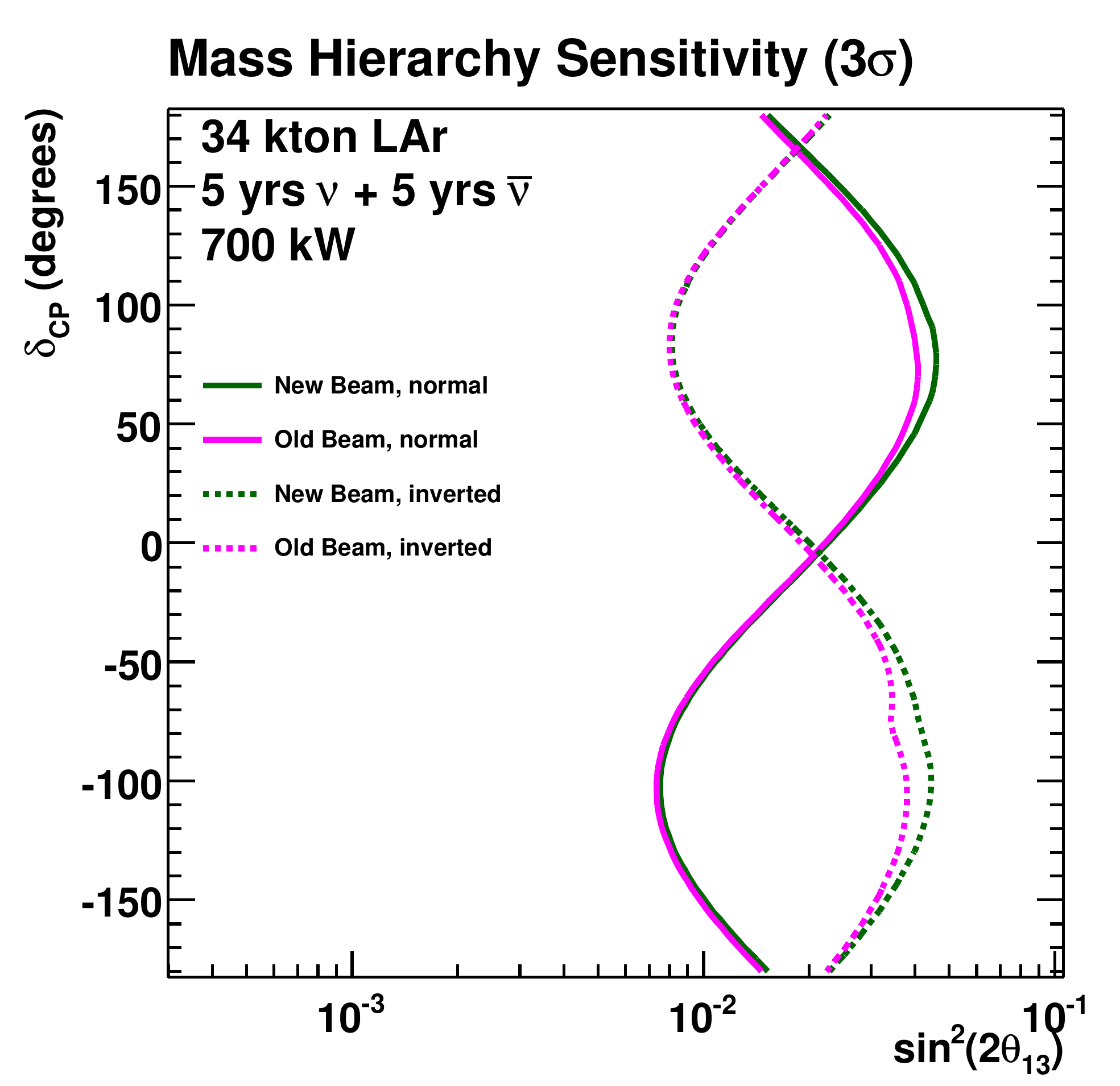}
\centering\includegraphics[height=0.45\textwidth]{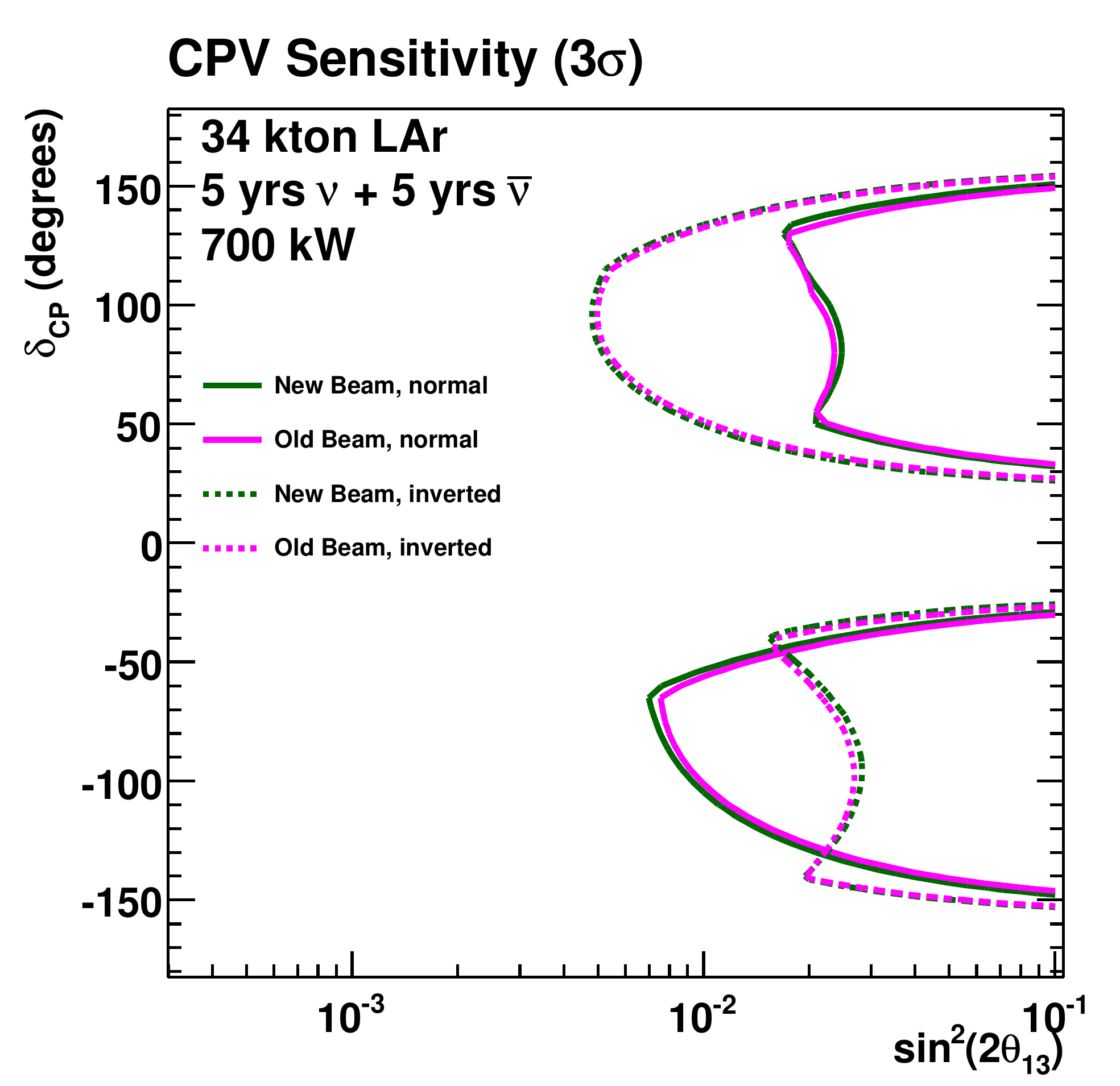}
\caption{Comparison of the sensitivity of LBNE to discovering
$\sin^22\theta_{13}\neq0$, the mass hierarchy, and CP violation at $3\sigma$
for two different LBNE beam designs. Sensitivities are shown for 34 kt LAr
in 5+5 years of $\nu+\nubar$ running in a 700 kW beam for both normal
(solid) and inverted (dashed) mass hierarchies. The projections include
detector effects, all background sources and their uncertainties
(see Appendix~\ref{lbl_appendix}). Same labelling convention
as Fig.~\ref{fig:beamcomp-wc}.}
\label{fig:beamcomp-lar}
\end{figure}

\subsubsection{Conclusion}

Despite the fact that the ``August 2010'' beam design has about a $30\%$
higher flux, it appears to yield similar sensitivity to
$\sin^22\theta_{13}\neq0$, CP violation, and the mass hierarchy as the
original 2008/2009 reference beam. This holds for both WC and LAr detectors.
This suggests that LBNE's long-baseline oscillation sensitivity may not
be very sensitive to the exact shape of the flux between 3 and 6~GeV.

For the physics studies reported here, we have chosen the ``August 2010'' beam
design with two NuMI parabolic horns $\sim$6~m apart, 250~kA current and
with the target upstream end -30~cm from the upstream face of horn~1 (red
spectrum in Fig.~\ref{fig:fig_beam_spectra1}). [The decay pipe is
assumed to be evacuated or helium filled to allow the maximum number of pions
to decay.]  The advantage of this beam is that it may be technically easier
to build and maintain the target/focusing system.

We note that none of the beams studied so far have sufficient flux at the
second maxima to impact the measurement of $\delta_{CP}$ and
$\sin^22\theta_{13}$. Preliminary estimates indicate that we would require
at least 5X more flux at the 2nd maxima to significantly improve the
measurement of $\delta_{CP}$ and $\sin^22\theta_{13}$~\cite{deb}.

\subsection{$\nue$ Appearance}\label{lbl_nue_appearance}
Motivated by an exciting history of discoveries in neutrino oscillations
over the course of the past decade, a major experimental effort is underway
to probe the last unknown mixing angle, $\theta_{13}$. To date,
only an upper limit on $\theta_{13}$ exists
(Table~\ref{table:lbl_current_knowledge}).
Information from reactor (Daya Bay, Double CHOOZ, RENO) and
accelerator-based (NOvA, T2K) experiments will be able to test whether
$\sin^22\theta_{13}$ is non-zero down to the $\sim0.01$ level in the coming
years. The next generation of experiments will need to be able to provide:

\begin{table} [htb]
\begin{center}
\begin{tabular}{cc} 
Parameter  & Value  \\ \hline
$\sin^22\theta_{12}$  & $0.87 \pm 0.03$ \\ 
$\sin^22\theta_{23}$  & $>0.91$ (at $90\%$ CL) \\ 
$\sin^22\theta_{13}$  & $<0.15$ (at $90\%$ CL) \\ 
$\Delta m^2_{21}$ & $(7.59 \pm 0.20) \times10^{-5}$ eV$^2$ \\ 
$\Delta m^2_{32}$ & $(2.35^{+0.11}_{-0.08})\times10^{-3}$ eV$^2$  \\ 
$\delta_{\mathrm CP}$     & no measurement \\ 
\end{tabular}
\caption{\label{table:lbl_current_knowledge} Current knowledge of neutrino
oscillation parameters. Values are from the Particle Data Group~\cite{rpp2010}. $\sin^2\theta_{23}$
and $\Delta m^2_{32}$ have been updated to reflect the new measurements
from MINOS~\cite{MINOS:Nu2010}.}
\end{center}
\end{table}

\begin{itemize}
  \item a more precise measurement of $\theta_{13}$ (or extension of the limit in the case of non-observation);
  \item a determination of the neutrino mass hierarchy, assuming
        non-zero $\theta_{13}$ (i.e. to determine whether the mass ordering is normal $\Delta m^2_{31}>0$ or
        inverted $\Delta m^2_{31}<0$);
  \item a measurement of leptonic CP violation (through measurement of the CP violating phase, $\delta_{\mathrm CP}$, assuming non-zero $\theta_{13}$).
\end{itemize}

The following sections detail measurement capabilities of LBNE in each of these three areas for water Cherenkov and liquid argon TPC detector technologies as a function of beam exposure. For context, we discuss the state of knowledge on the target parameters expected by the time LBNE would be operational.

\subsubsection{Current and Planned Experiments $\theta_{13}$ Reach}\label{project_lbl_theta13}

Before discussing the $\theta_{13}$ reach of LBNE, we provide an overview of the precisions that might be achieved by current and planned experiments. A valuable resource for this is the 2009 EURONU annual report~\cite{euronu}, which included their updated projections.
The EURONU calculations were based, to the extent possible, on
official statements from the collaborations and assume that the data are continuously
analyzed and the results made available immediately.

\begin{figure}[htb]
 \centering\includegraphics[width=.45\textwidth]{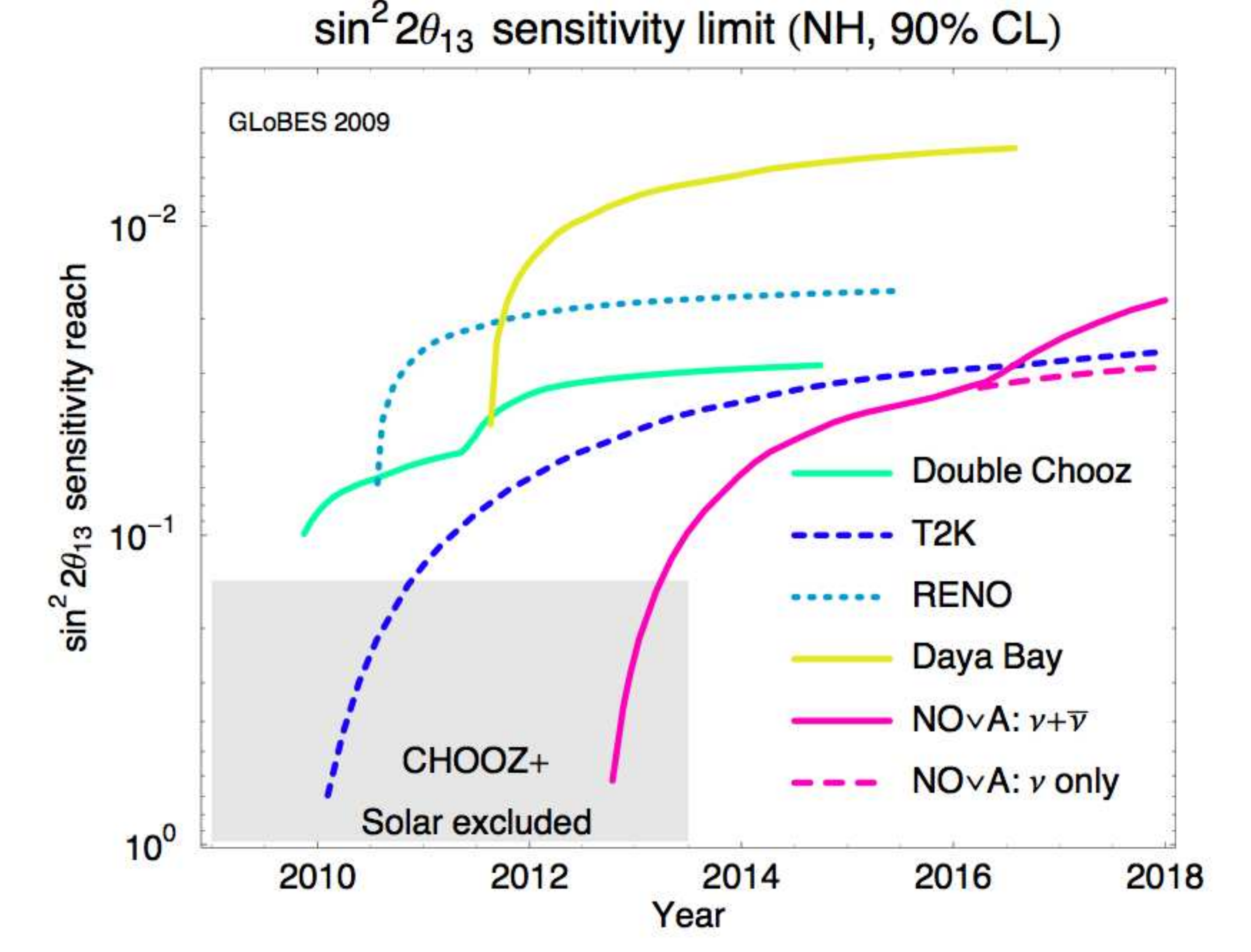}
 \centering\includegraphics[width=.45\textwidth]{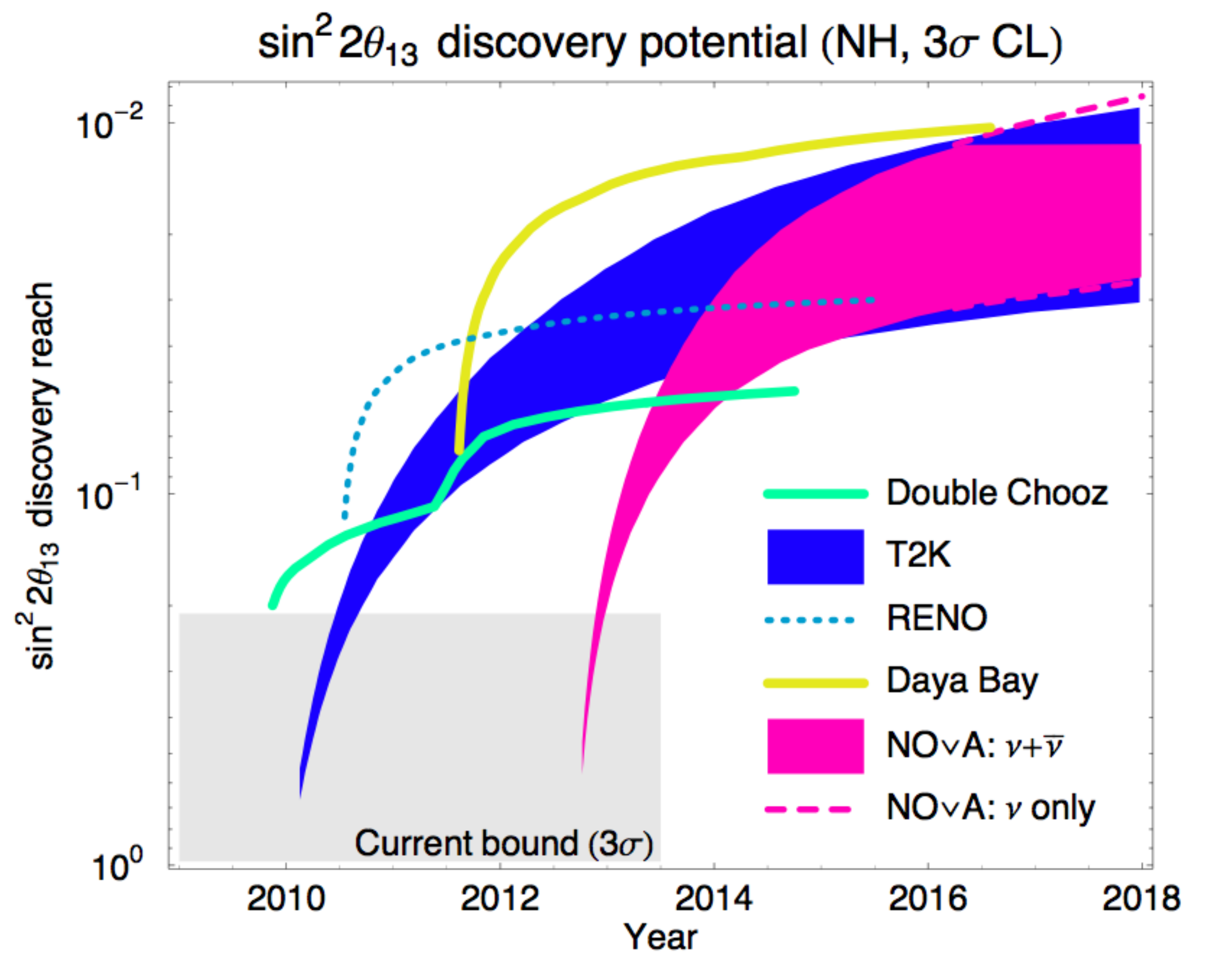}
  \caption{Projected $\theta_{13}$ sensitivity (left) and discovery potential (right) as a function
           of time for experiments that will run before LBNE.
           Sensitivity is defined as the $90\%$ CL limit that will be obtained if the true
           $\theta_{13}$ is zero.
           Discovery potential is defined as the smallest value of $\theta_{13}$ that
           can be distinguished from zero at 3$\sigma$. A normal mass hierarchy
           is assumed and the bands reflect the variation for different values
           of $\delta_{\mathrm CP}$ for the accelerator-based experiments.
           Plots are reproduced from the 2009 EURONU annual report~\cite{euronu};
           experiment schedules shown at Neutrino 2010~\cite{nu2010} imply the horizontal axis should
           be delayed by about one year from those shown for most of the experiments.}
  \label{fig:lbl_world_theta13_sensitivity}
\end{figure}

Fig.~\ref{fig:lbl_world_theta13_sensitivity}(left) shows the evolution of
the $\theta_{13}$ sensitivity limit ($90\%$ CL) as a function time
from upcoming reactor and accelerator-based experiments (we note that at Neutrino 2010~\cite{nu2010}, the schedule for most of the experiments had slipped by about one year).
As can be seen from the figure, the global sensitivity will be dominated
by the reactor experiments, with Daya Bay possibly reaching
$\sin^22\theta_{13}$ sensitivity down to $\sim0.006$ (at $90\%$ CL) before LBNE operation.

Fig.~\ref{fig:lbl_world_theta13_sensitivity}(right) shows the projected
discovery potential from these same experiments, also plotted as a function
of time. For the accelerator-based experiments, there is some ambiguity
resulting from the mass hierarchy and $\delta_{\mathrm CP}$, so in their case,
a normal mass hierarchy is assumed and the dependence on $\delta_{\mathrm CP}$
is indicated by colored bands. Note that, for some values of $\delta_{\mathrm CP}$,
the discovery reach of T2K and NOvA can approach that of Daya Bay. From these
projections, upcoming experiments should be able to distinguish $\theta_{13}$
from zero at the $3\sigma$ level for $\sin^22\theta_{13}$ values down to 0.01
by the year 2016-2018.

\subsubsection{LBNE $\theta_{13}$ Reach}\label{lbl_theta13}

\begin{figure}[htb]
 \centering\includegraphics[width=.45\textwidth]{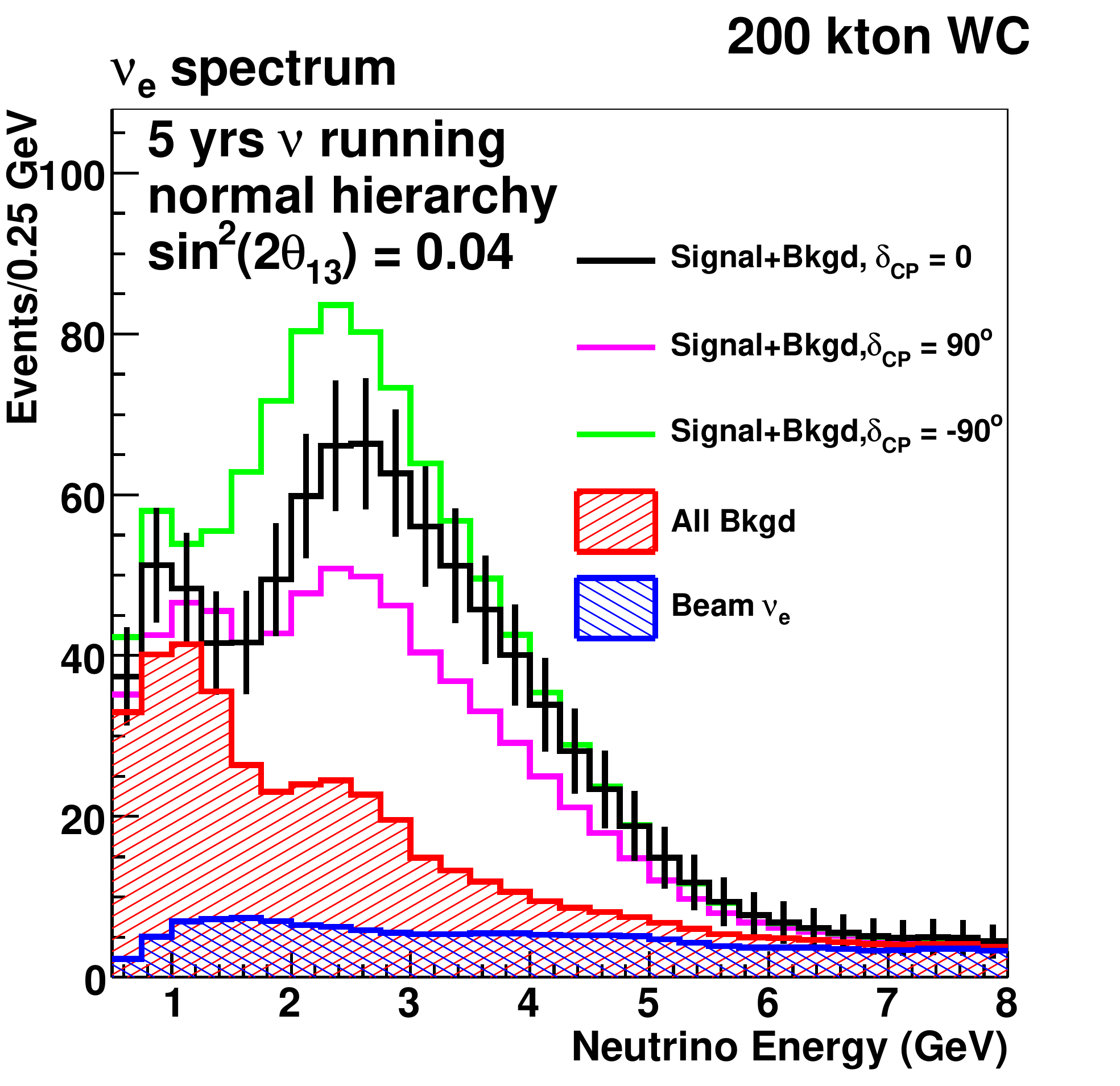}
 \centering\includegraphics[width=.45\textwidth]{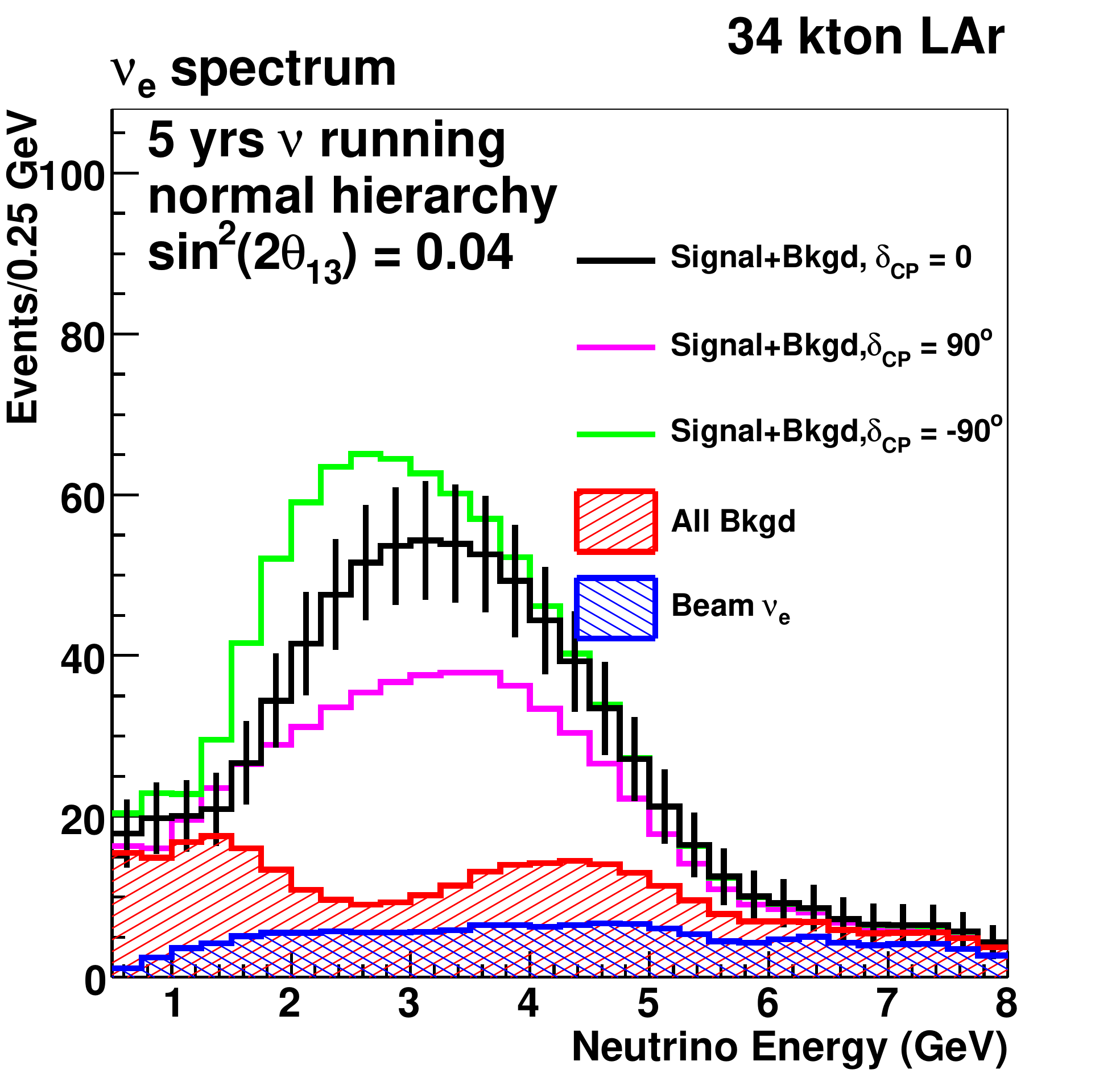}
  \caption{The expected $\nue$ appearance spectrum for a 200~kt WC (left)
  and 34~kt LAr (right) detector for $\sin^22\theta_{13}=0.04$ and 5 years
  of neutrino running in a 700~kW beam assuming a {\em normal} mass hierarchy.
  The black points assume $\delta_{CP}=0$ while the green and pink lines
  are for $\delta_{CP}=\pm 90^0$. The different background contributions are
  indicated by the hatched histograms.}
  \label{fig:lbl_spectrum_nu_normal}
\end{figure}

\begin{figure}[htb]
 \centering\includegraphics[width=.45\textwidth]{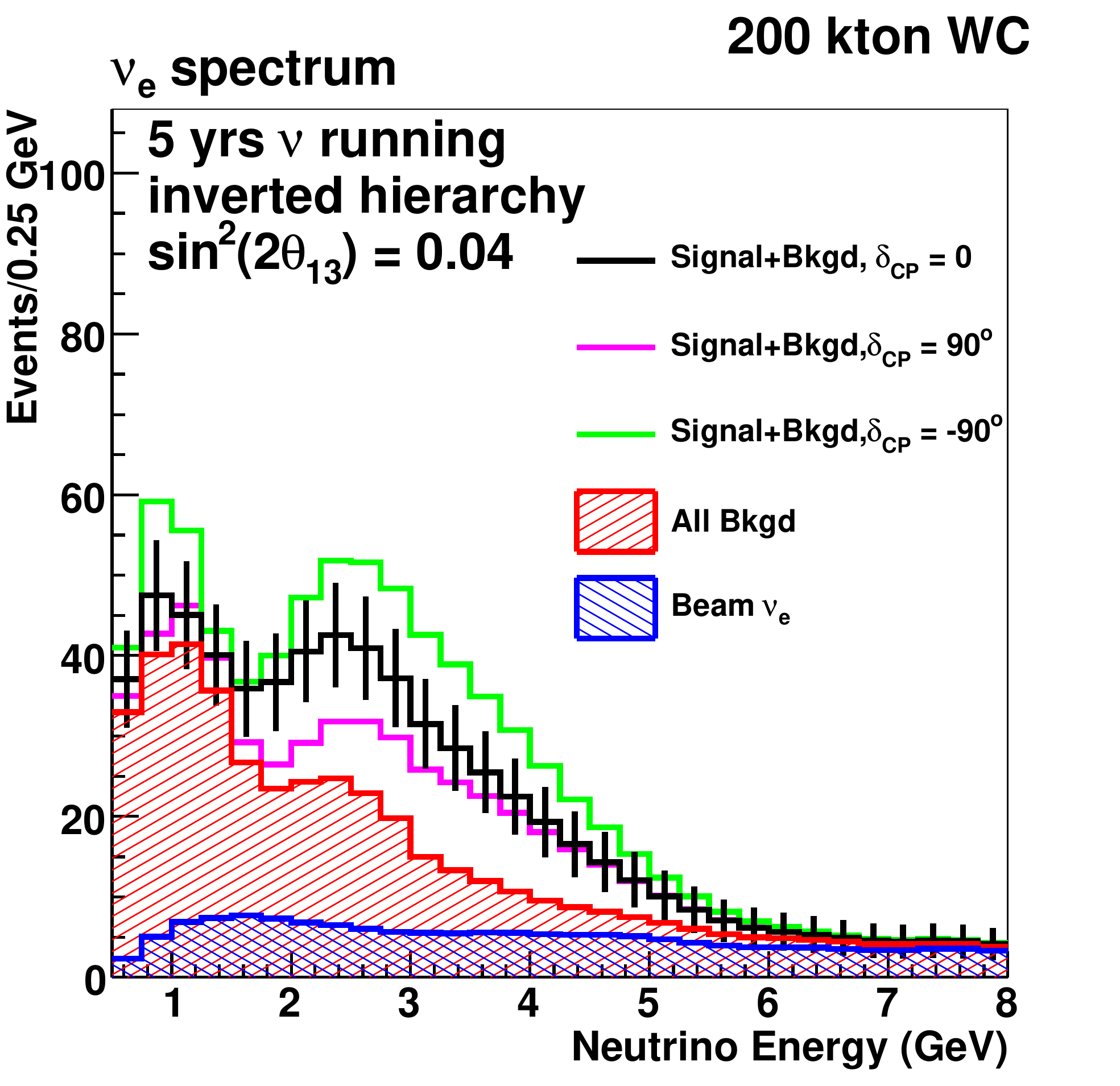}
 \centering\includegraphics[width=.45\textwidth]{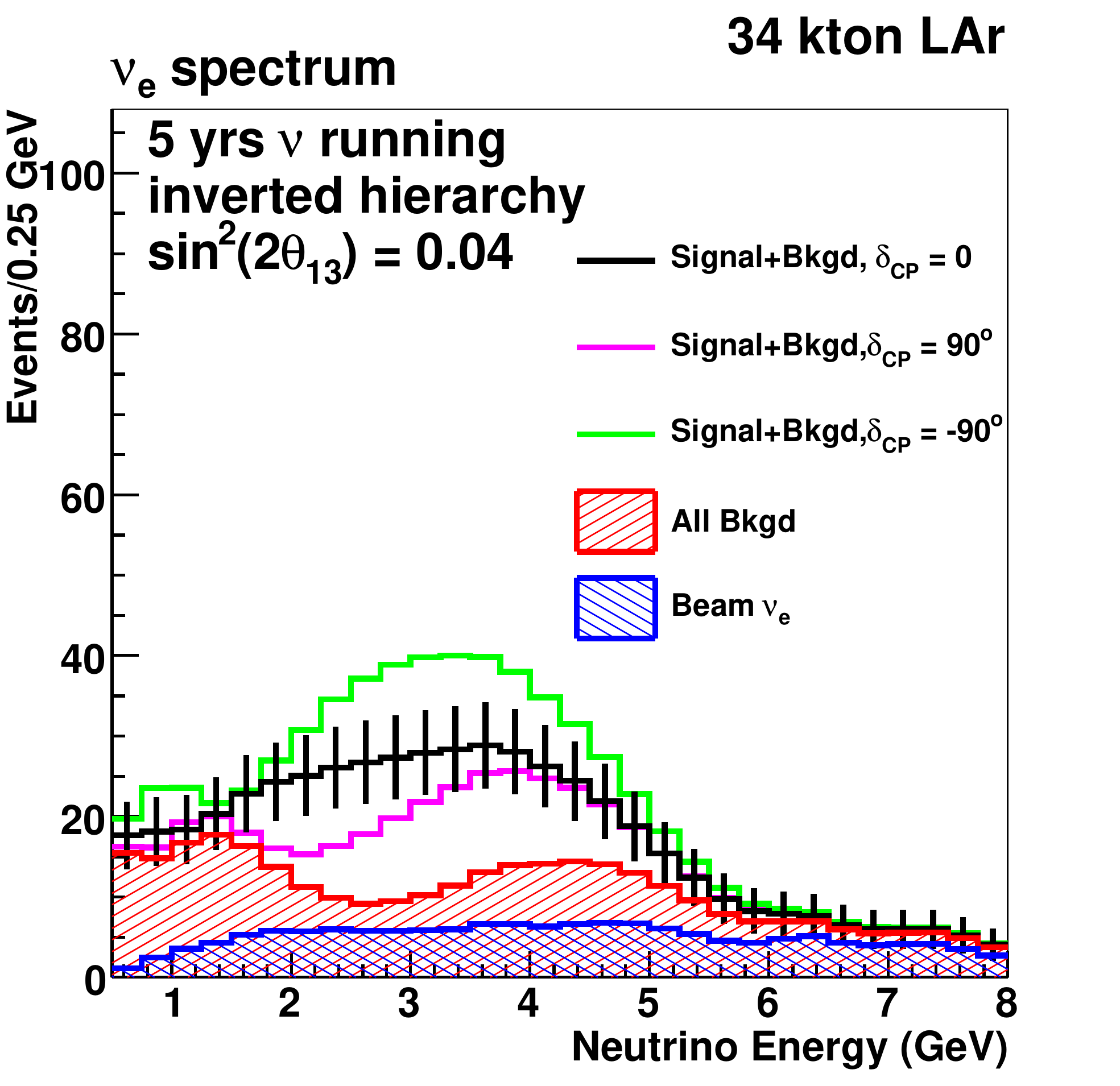}
 \caption{The $\nue$ appearance spectrum as described for Fig.~\ref{fig:lbl_spectrum_nu_normal}
 except for an {\em inverted} mass hierarchy.}
  \label{fig:lbl_spectrum_nu_inverted}
\end{figure}

\begin{figure}[htb]
 \centering\includegraphics[width=.45\textwidth]{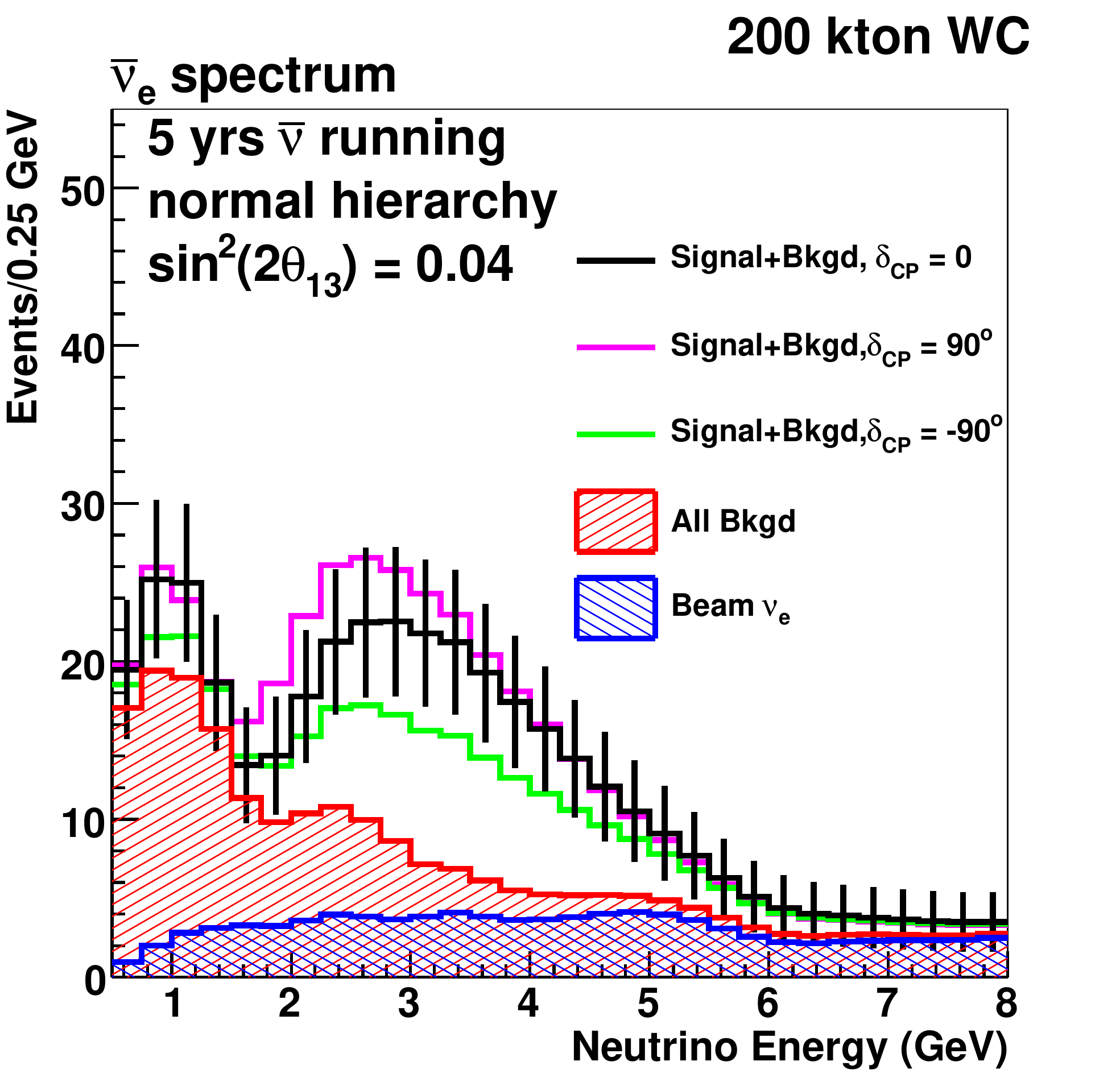}
 \centering\includegraphics[width=.45\textwidth]{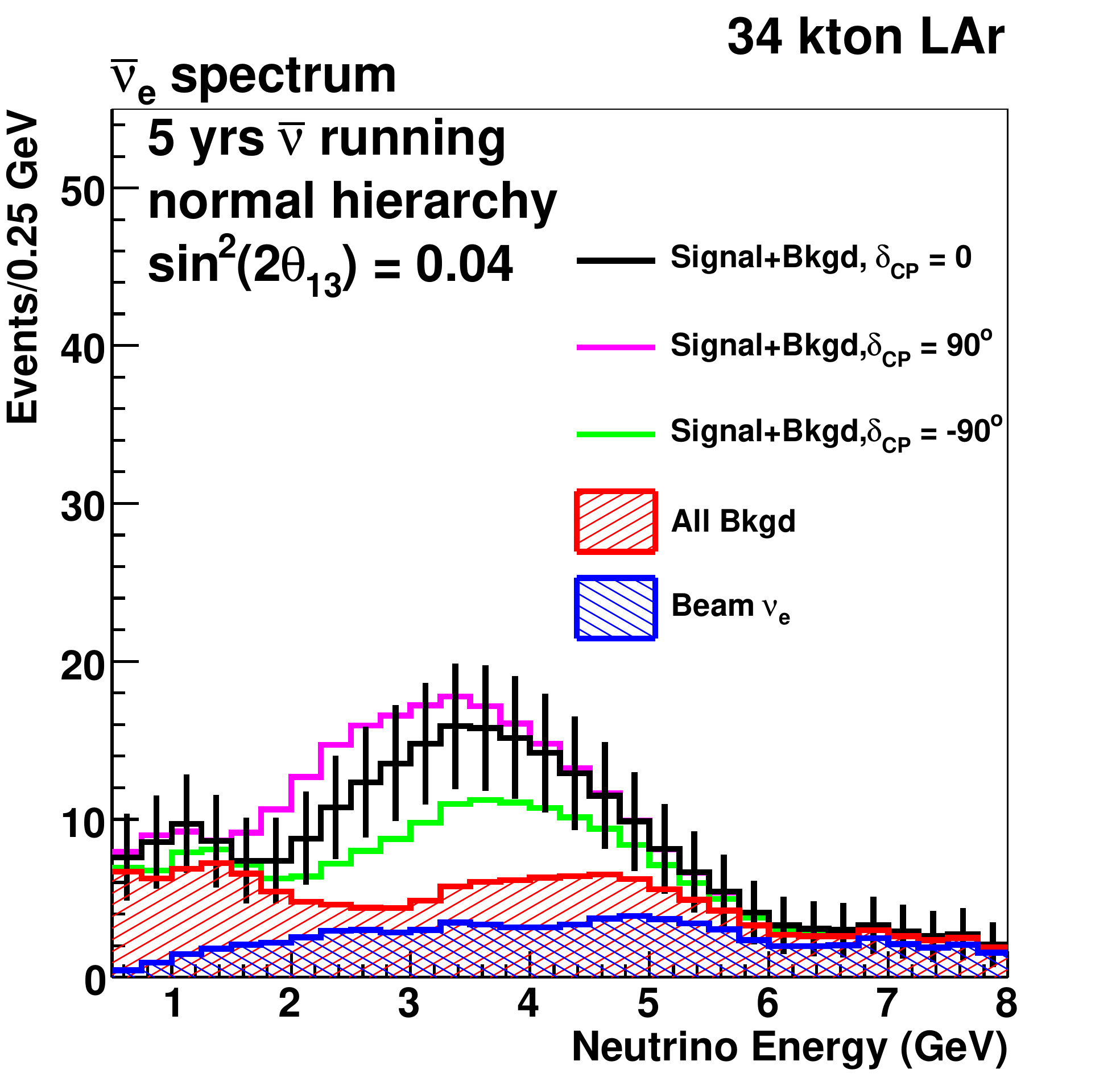}
 \caption{The expected $\nuebar$ appearance spectrum for a 200~kt WC (left)
  and 34~kt LAr (right) detector for $\sin^22\theta_{13}=0.04$ and 5 years
  of antineutrino running in a 700~kW beam assuming a {\em normal} mass hierarchy.
  The black points assume $\delta_{CP}=0$ while the green and pink lines
  are for $\delta_{CP}=\pm 90^0$. The different background contributions are
  indicated by the hatched histograms. [Note the vertical axis scale change compared to neutrino running.]}
  \label{fig:lbl_spectrum_nubar_normal}
\end{figure}

\begin{figure}[htb]
 \centering\includegraphics[width=.45\textwidth]{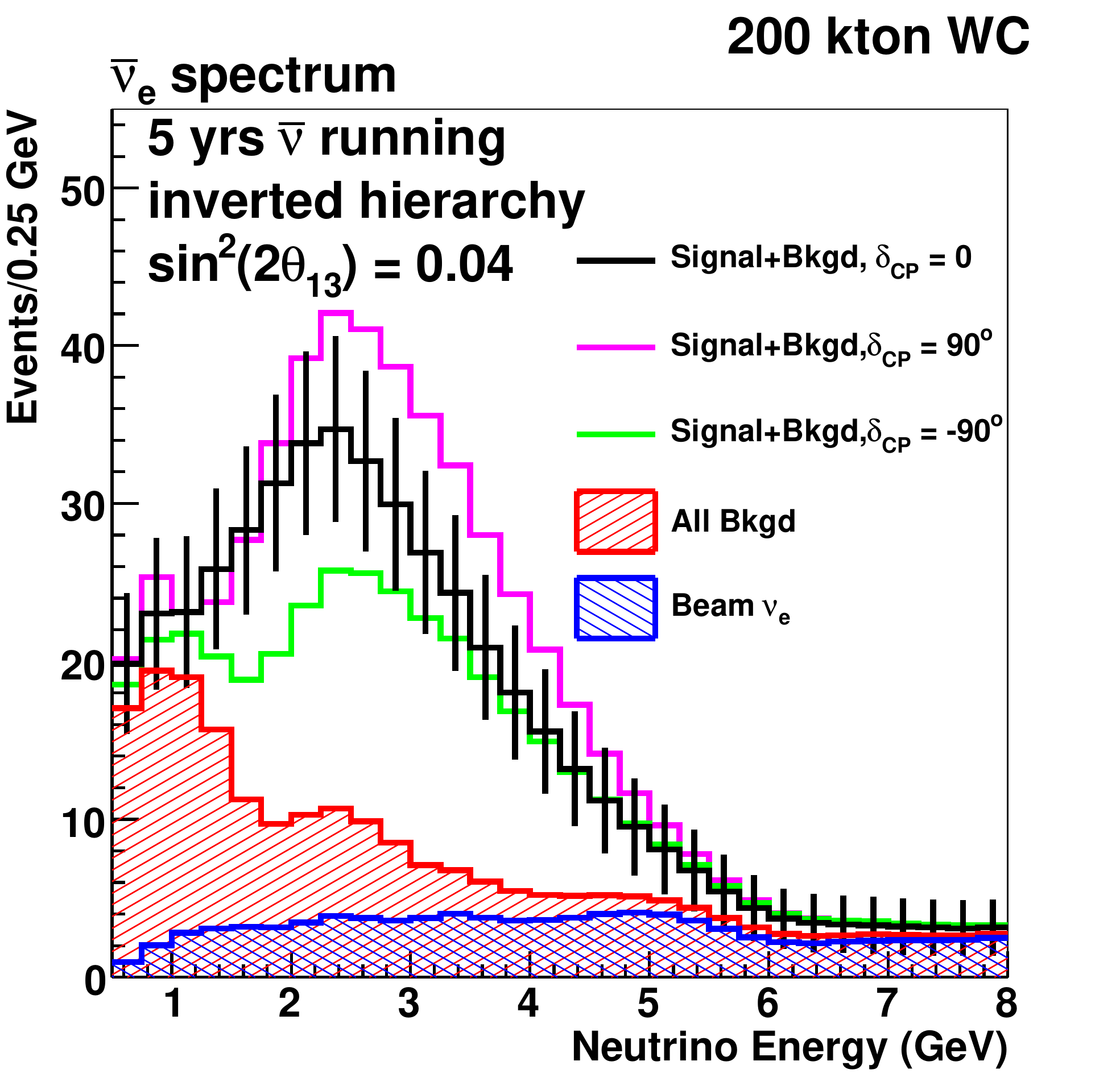}
 \centering\includegraphics[width=.45\textwidth]{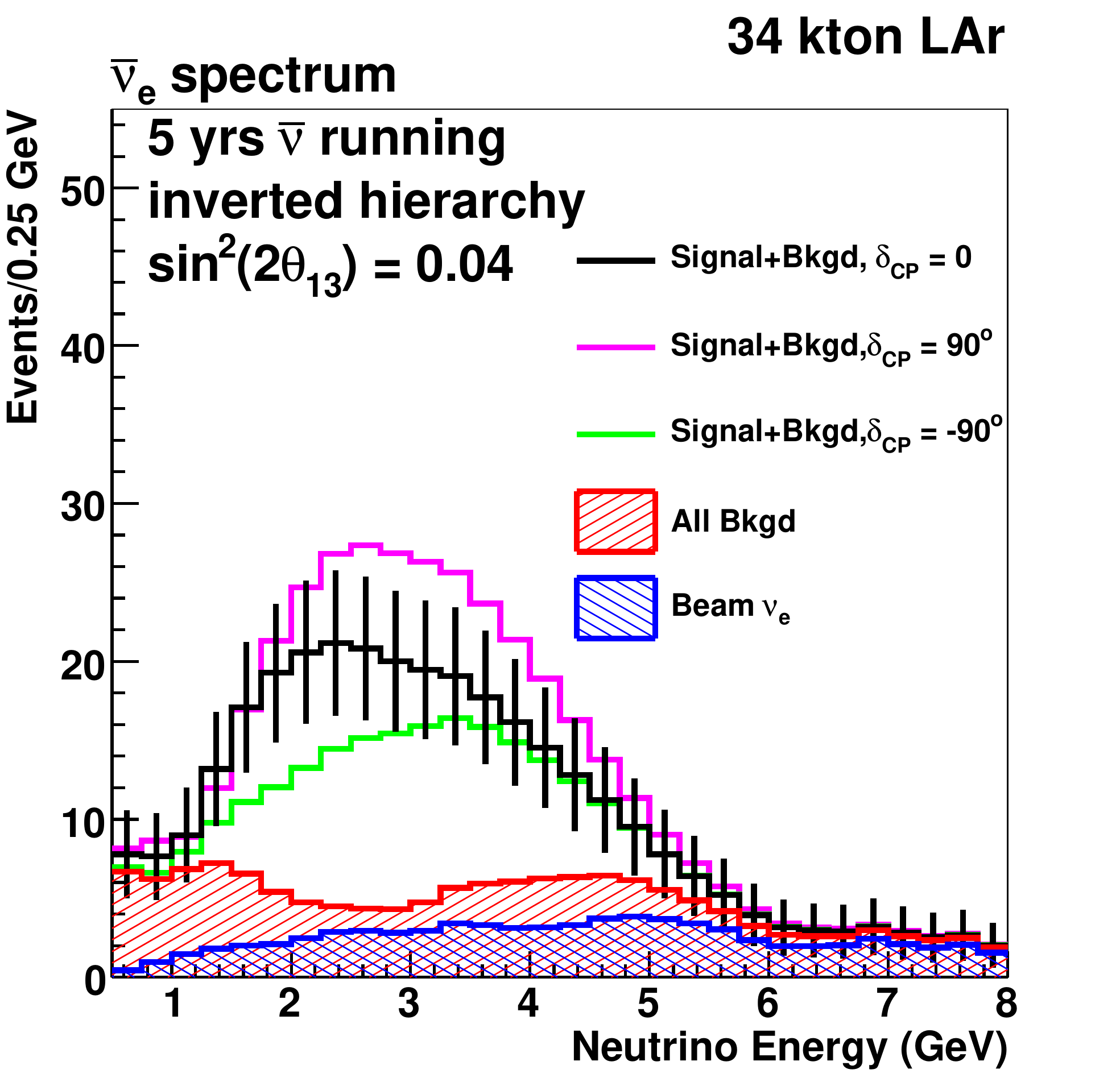}
 \caption{The $\nuebar$ appearance spectrum as described for Fig.~\ref{fig:lbl_spectrum_nubar_normal}
 except for an {\em inverted} mass hierarchy.}
  \label{fig:lbl_spectrum_nubar_inverted}
\end{figure}

The LBNE detectors will have excellent sensitivity to $\theta_{13}$, further
extending the reach of upcoming reactor and accelerator-based experiments
by roughly an order of magnitude. Observation of $\numu \rightarrow \nue$
oscillations in LBNE will be the key to measuring $\theta_{13}$ at this level.
Figs.~\ref{fig:lbl_spectrum_nu_normal} and
\ref{fig:lbl_spectrum_nu_inverted} show the expected event rates for $\nue$
appearance measurements in LBNE in both a WC and LAr far detector for normal
and inverted mass hierarchies, respectively.
Figs.~\ref{fig:lbl_spectrum_nubar_normal} and
\ref{fig:lbl_spectrum_nubar_inverted} show the same for antineutrino running.
As expected, the rates are higher for the normal mass hierarchy in the
case of neutrinos and for the inverted hierarchy in the case of antineutrinos.
We also see that the effect of positive and negative $\delta_{\mathrm CP}$ phase is opposite for neutrinos that has more events for negative phase (both normal and inverted hierarchy), compared to antineutrinos that has more events for positive phase. The detector performance parameters (such as background levels and efficiency) used to produce these plots are detailed in Appendix~\ref{lbl_appendix}.

\begin{figure}[htb]
 \centering\includegraphics[width=.4\textwidth]{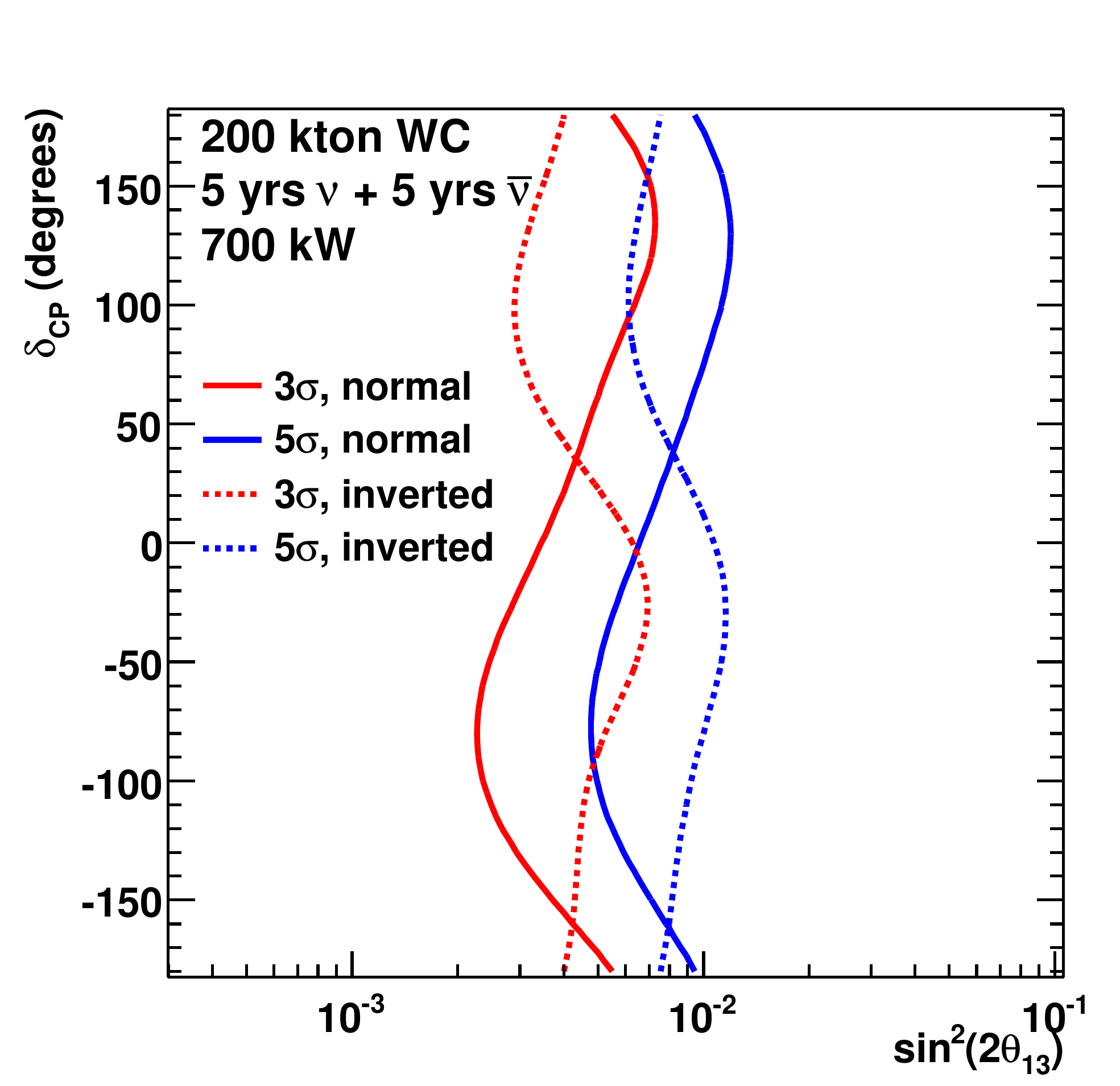}
 \centering\includegraphics[width=.4\textwidth]{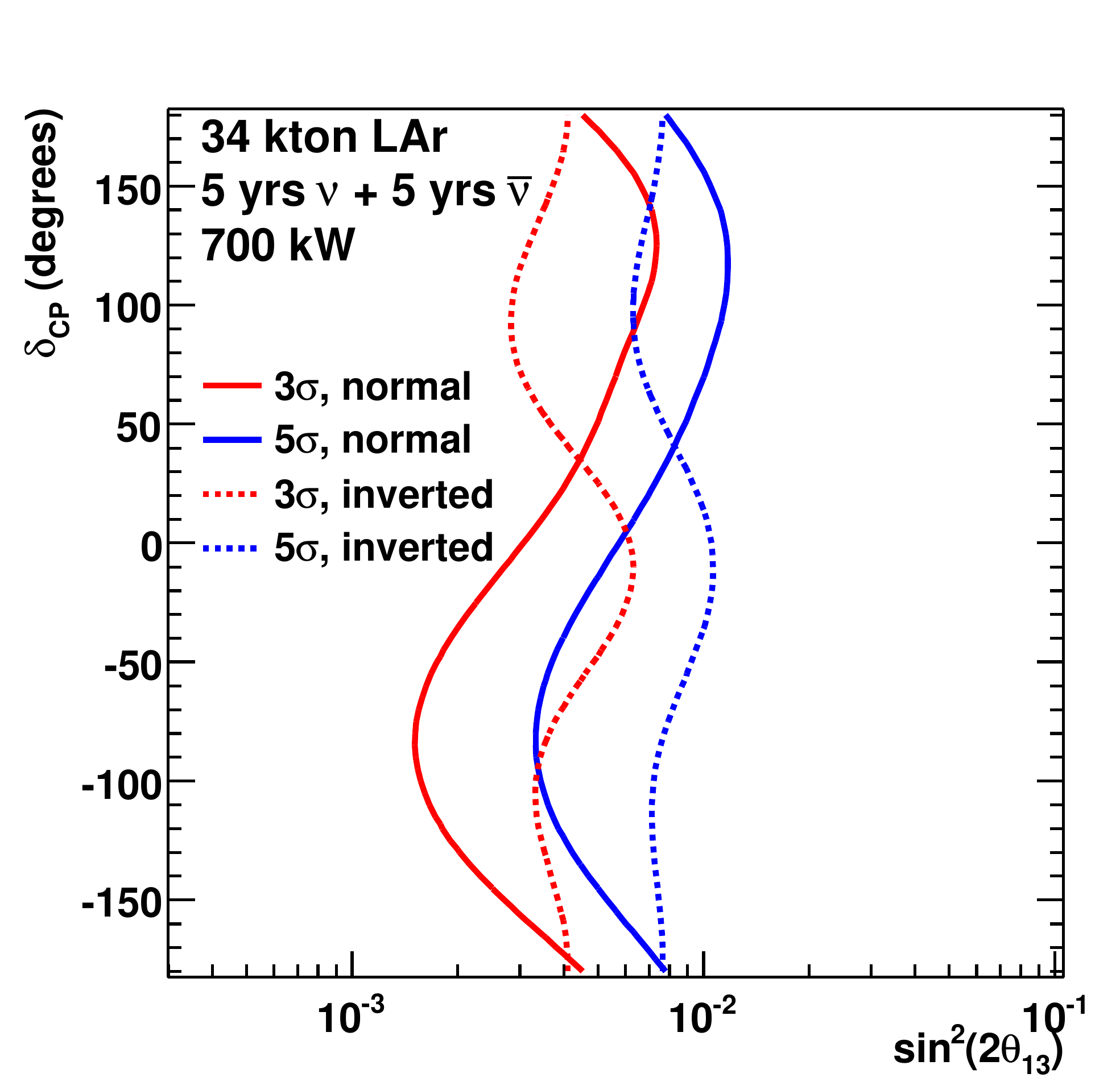}
  \caption{$3\sigma$ (red) and $5\sigma$ (blue) sensitivity of LBNE
           to $\sin\theta_{13}\neq0$ as a function of $\delta_{\mathrm CP}$ for 200~kt
           of WC (top) and 34~kt of LAr (bottom). This assumes 5+5 years of
           $\nu$ and $\nubar$ running in a 700~kW beam. Curves are shown for
           both normal (solid) and inverted (dashed) mass hierarchies.}
  \label{fig:lbl_theta13_sensitivity}
\end{figure}

Fig.~\ref{fig:lbl_theta13_sensitivity} shows the sensitivity of LBNE to
$\theta_{13} \neq 0$ as a function of $\delta_{\mathrm CP}$ for each of the mass
orderings. Here, the discovery reach for $\sin^22\theta_{13}$ is defined
as the minimum value of $\sin^22\theta_{13}$ for which LBNE can rule
out $\sin^22\theta_{13}=0$ at the 3$\sigma$ and 5$\sigma$ levels.
The results are dependent on the value of $\delta_{\mathrm CP}$ and
the mass hierarchy. As can be seen, the sensitivity is better if the mass
hierarchy is normal rather than inverted unless $\delta_{\mathrm CP}$=$45-180^o$,
in which case the reverse is true.

\begin{figure}[htb]
 \centering\includegraphics[width=.4\textwidth]{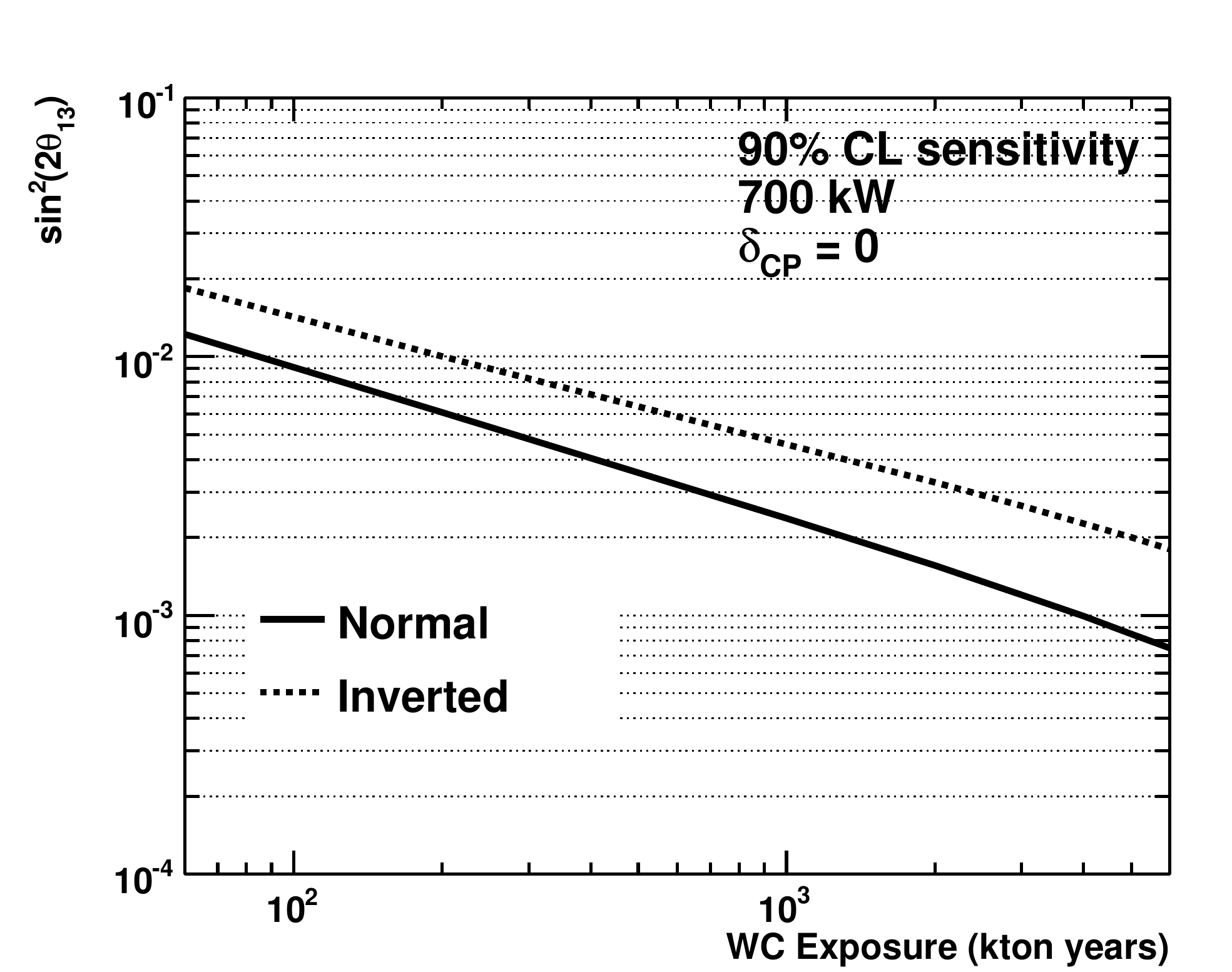}
 \centering\includegraphics[width=.4\textwidth]{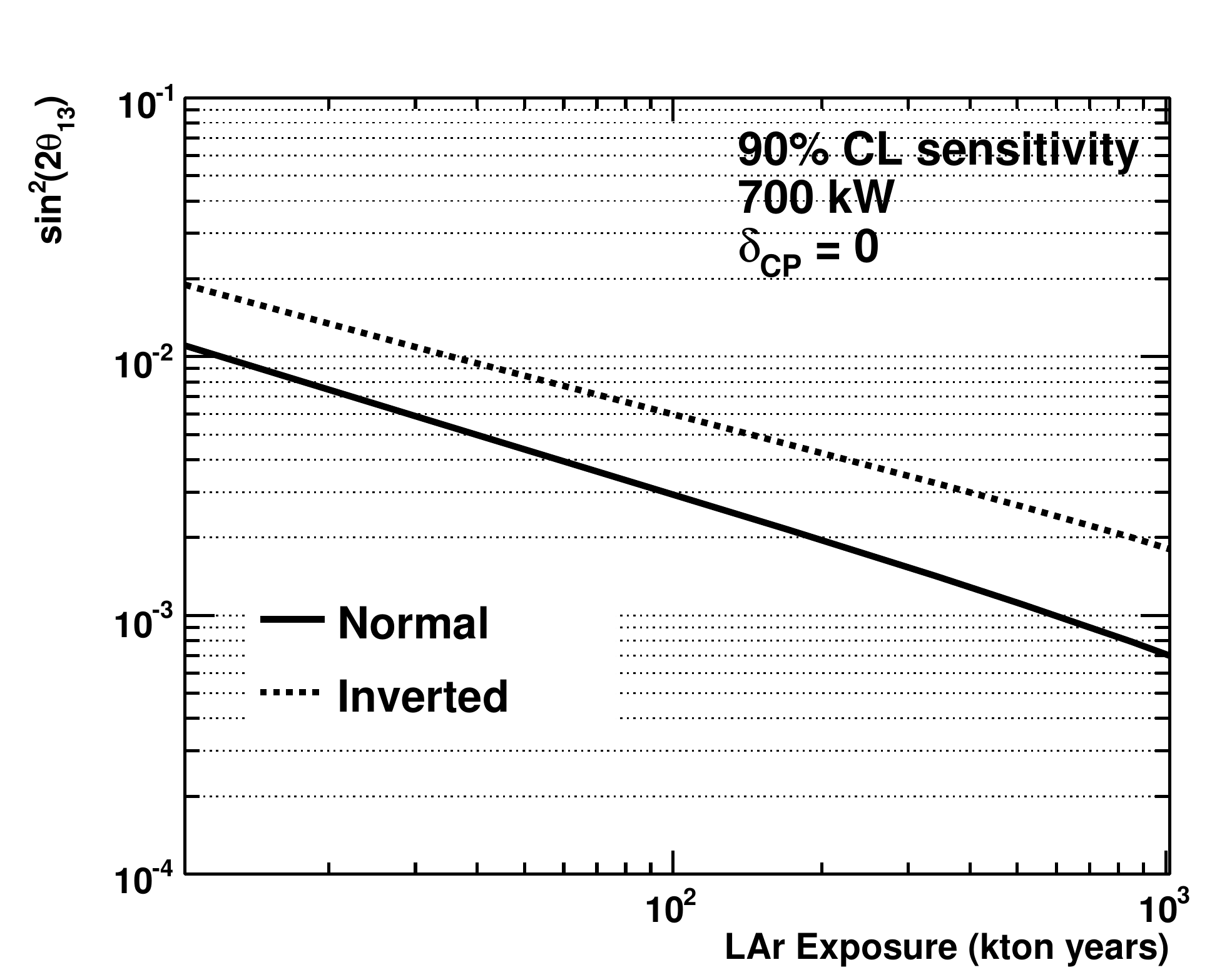}
  \caption{Sensitivity of LBNE to determining non-zero $\theta_{13}$ at the
  $90\%$ CL as a function of exposure for both WC (left) and LAr (right).
  The plots show the projections for $\delta_{CP}=0$ for normal (solid)
  and inverted (dashed) mass hierarchies.}
  \label{fig:lbl_theta13_exposure_90per}
\end{figure}

\begin{figure}[htb]
 \centering\includegraphics[width=.4\textwidth]{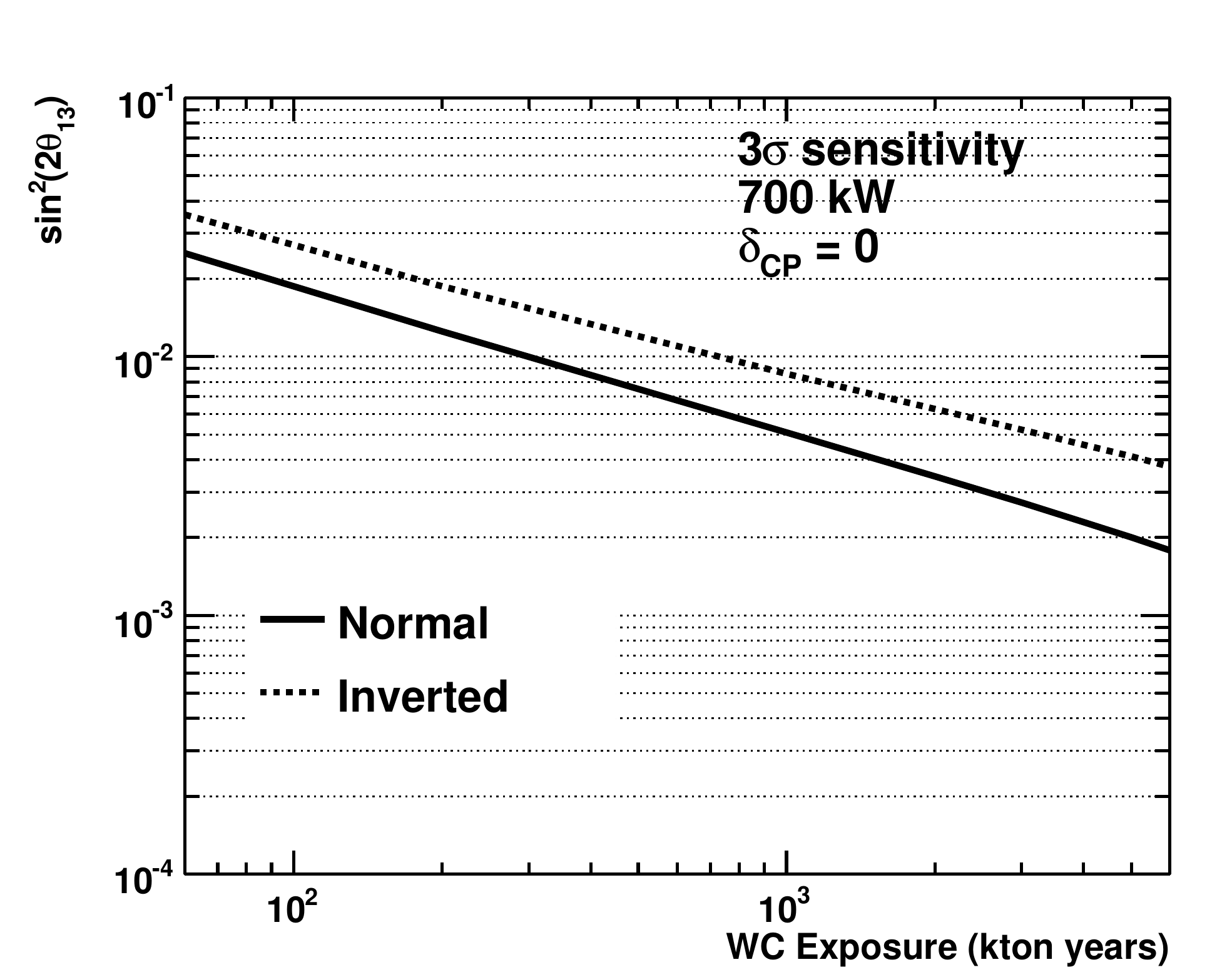}
 \centering\includegraphics[width=.4\textwidth]{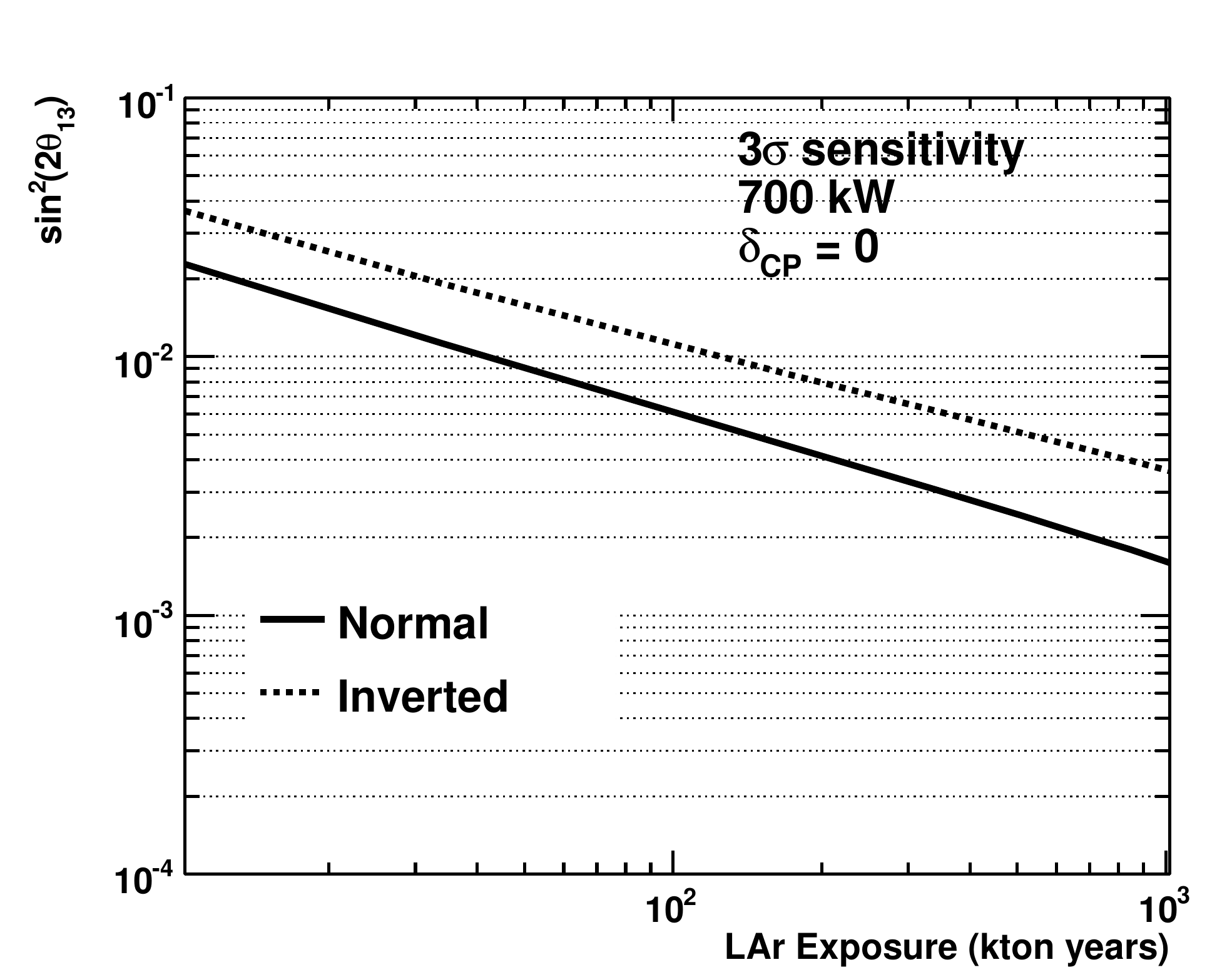}
  \caption{Sensitivity of LBNE to determining non-zero $\theta_{13}$ at
  $3\sigma$ as a function of exposure for both WC (left) and LAr (right).
  The plots show the projections for $\delta_{CP}=0$ for normal (solid)
  and inverted (dashed) mass hierarchies.}
  \label{fig:lbl_theta13_exposure_3sig}
\end{figure}

Of course, the sensitivity to determining a non-zero value of $\theta_{13}$
increases with exposure. Figs.~\ref{fig:lbl_theta13_exposure_90per}
and \ref{fig:lbl_theta13_exposure_3sig} show the sensitivity as a function of
exposure at $90\%$ and $3\sigma$ CL, respectively. Both detectors
can probe $\sin^22\theta_{13}$ down to the $10^{-3}$ level with reasonable
exposures. A WC detector is sensitive to $\sin^22\theta_{13}\neq0$ at
$3\sigma$ down to a $\sin^22\theta_{13}$ value of 0.008 for $100\%$ of all
possible $\delta_{\mathrm CP}$ values assuming an exposure of 2000~kt-yrs. The same
is true for LAr assuming an exposure of 340~kt-yrs. Hence, LAr appears
to have similar $\theta_{13}$ reach as WC with about 1/6 the exposure.
[The main parameter that determines this factor is the ratio of the detector signal efficiencies
near the 1st maximum.]

\clearpage

\subsubsection{Neutrino Mass Hierarchy}\label{lbl_mh}

\begin{figure}[htb]
 \centering\includegraphics[width=.7\textwidth]{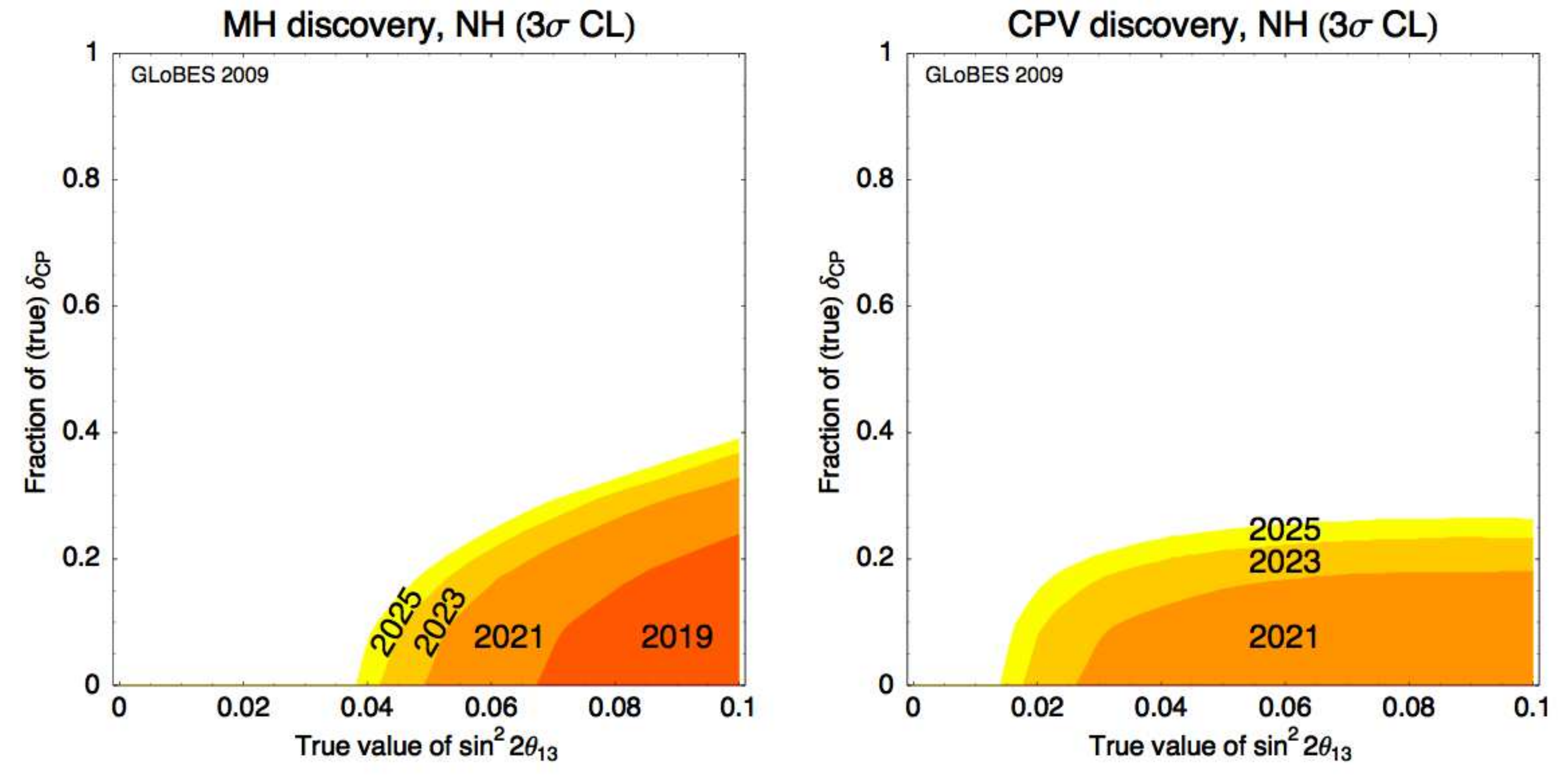}
  \caption{Mass hierarchy (left) and CP violation (right) discovery potentials
  at $3\sigma$ as a function of true $\sin^22\theta_{13}$ for T2K+NOvA+reactor
  including major beam upgrades to both T2K and NOvA with a global $\nu$/$\nubar$
  running optimization (plot reproduced from~\cite{euronu}).
  Different colors correspond to different projections in time after the upgrades.}
  \label{fig:lbl_world_cp_mh}
\end{figure}

While the primary goal of upcoming neutrino oscillation experiments is
the discovery of the yet unknown mixing angle, $\theta_{13}$, they may also provide information on
the mass hierarchy and CP violation if $\theta_{13}$ is relatively
large. In the 2009 EURONU report~\cite{euronu}, referred to in the previous section, ``modest upgrades'' to the T2K and NOvA experiments were considered; specifically, increase of the T2K beam power from 0.75~MW to 1.66~MW starting in 2015 and a linear beam power increase for NOvA from 0.7~MW to 2.3~MW (Project~X) starting in 2018.
Fig.~\ref{fig:lbl_world_cp_mh} illustrates that the projected $3\sigma$ discovery regions for
mass hierarchy and CP violation for upgraded T2K+NOvA are quite limited, despite the assumption of rather aggressive beam power start times, a global optimization of neutrino and antineutrino running in both beams, and combined limits with information from reactor data.
The EURONU report does indicate that these experiments might see hints of the mass hierarchy and CP violation at $90\%$ CL for $\sin^22\theta_{13}>0.05$ and most values of $\delta_{\mathrm CP}$ but concludes:
``Although `minor upgrades' of existing facilities
may provide a non-negligible sensitivity to the mass hierarchy and CP
violation, there is high risk associated with this strategy, since for
$\sim75\%$ of all possible values of $\delta_{\mathrm CP}$, no discovery would
be possible at the $3\sigma$ level. Therefore, we conclude that the upcoming
generation of oscillation experiments may lead to interesting indications
for the mass hierarchy and CP violation, but it is very likely that an
experiment beyond the upcoming superbeams (including reasonable upgrades)
will be required to confirm these hints.''

\begin{figure}[htb]
 \centering\includegraphics[width=.45\textwidth]{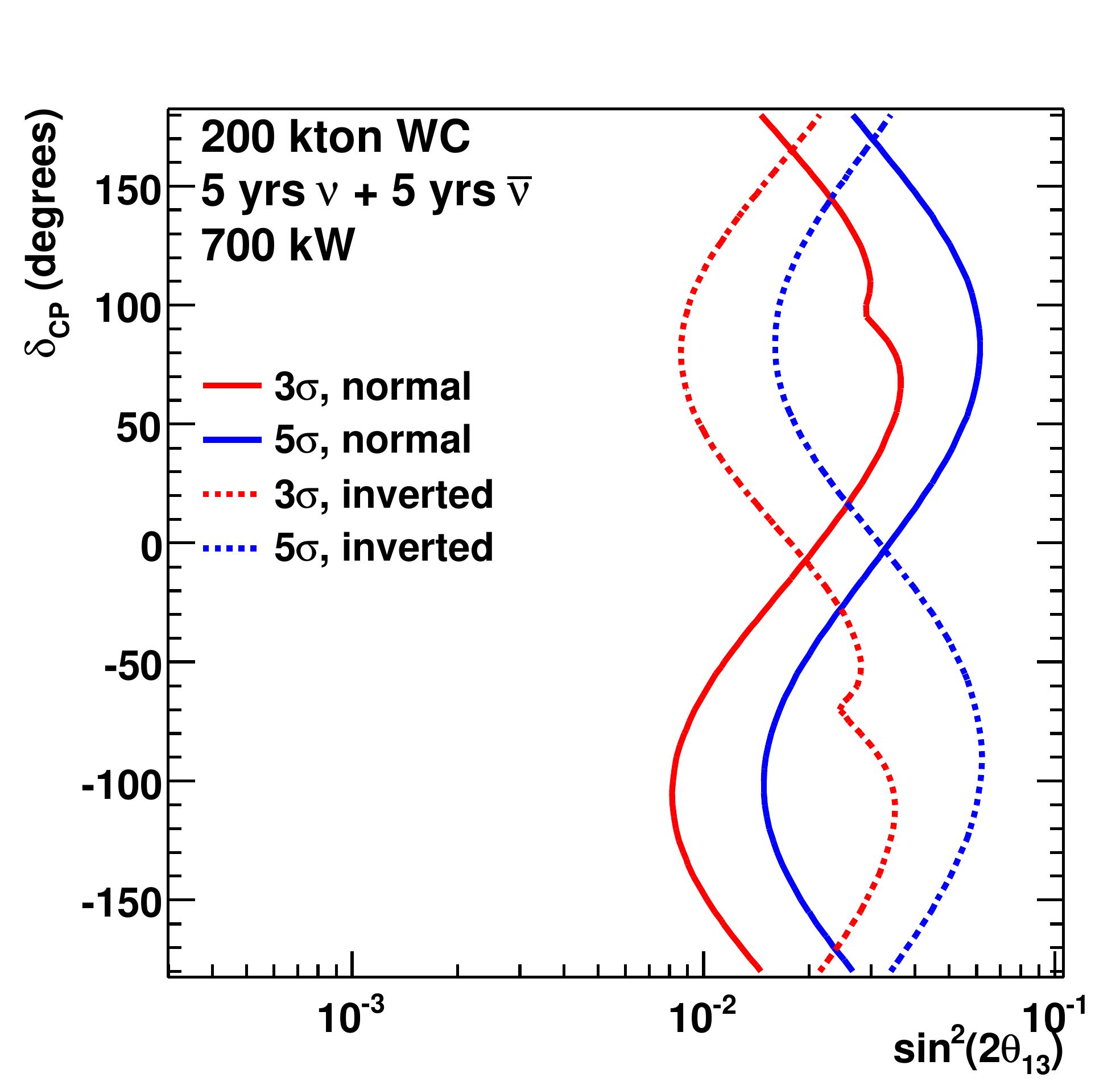}
 \centering\includegraphics[width=.45\textwidth]{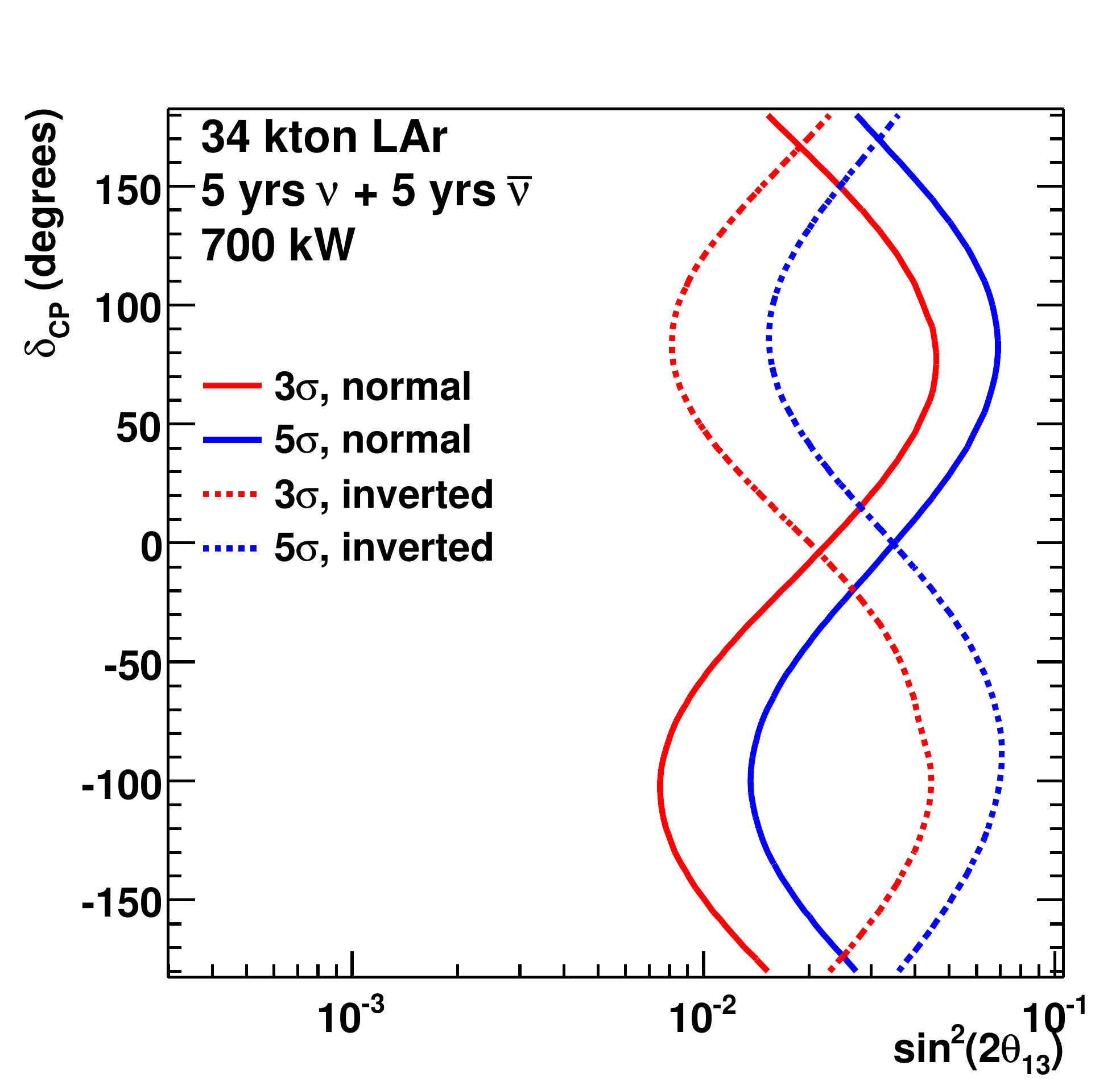}
  \caption{Resolution of the mass hierarchy for 200~kt of WC
           (top) and 34~kt of LAr (bottom). This assumes 5+5 years of
           $\nu$ and $\nubar$ running in a 700~kW beam.
           To the right of the curves, the normal (solid) or inverted (dashed)
           mass hierarchy can be excluded at the $3\sigma$ (red) or
           $5\sigma$ (blue) level for the indicated
           values of true $\sin^22\theta_{13}$ and $\delta_{\mathrm CP}$.}
  \label{fig:lbl_mh_sensitivity}
\end{figure}

With poor coverage of the mass hierarchy and CP violation even with factors of 2-3 increase in the beam power delivered to experiments like T2K and NOvA, it is clear that an experiment like LBNE is needed to take
the next step in physics reach. Fig.~\ref{fig:lbl_mh_sensitivity} shows LBNE's projected sensitivity to the mass ordering as a function of $\theta_{13}$ and the CP-violating phase for both WC and LAr. Here, the mass hierarchy discovery reach is defined as the minimum value of $\sin^22\theta_{13}$ for which the wrong hierarchy can be excluded for a given value of $\delta_{\mathrm CP}$. \\

\begin{figure}
 \centering\includegraphics[width=.45\textwidth]{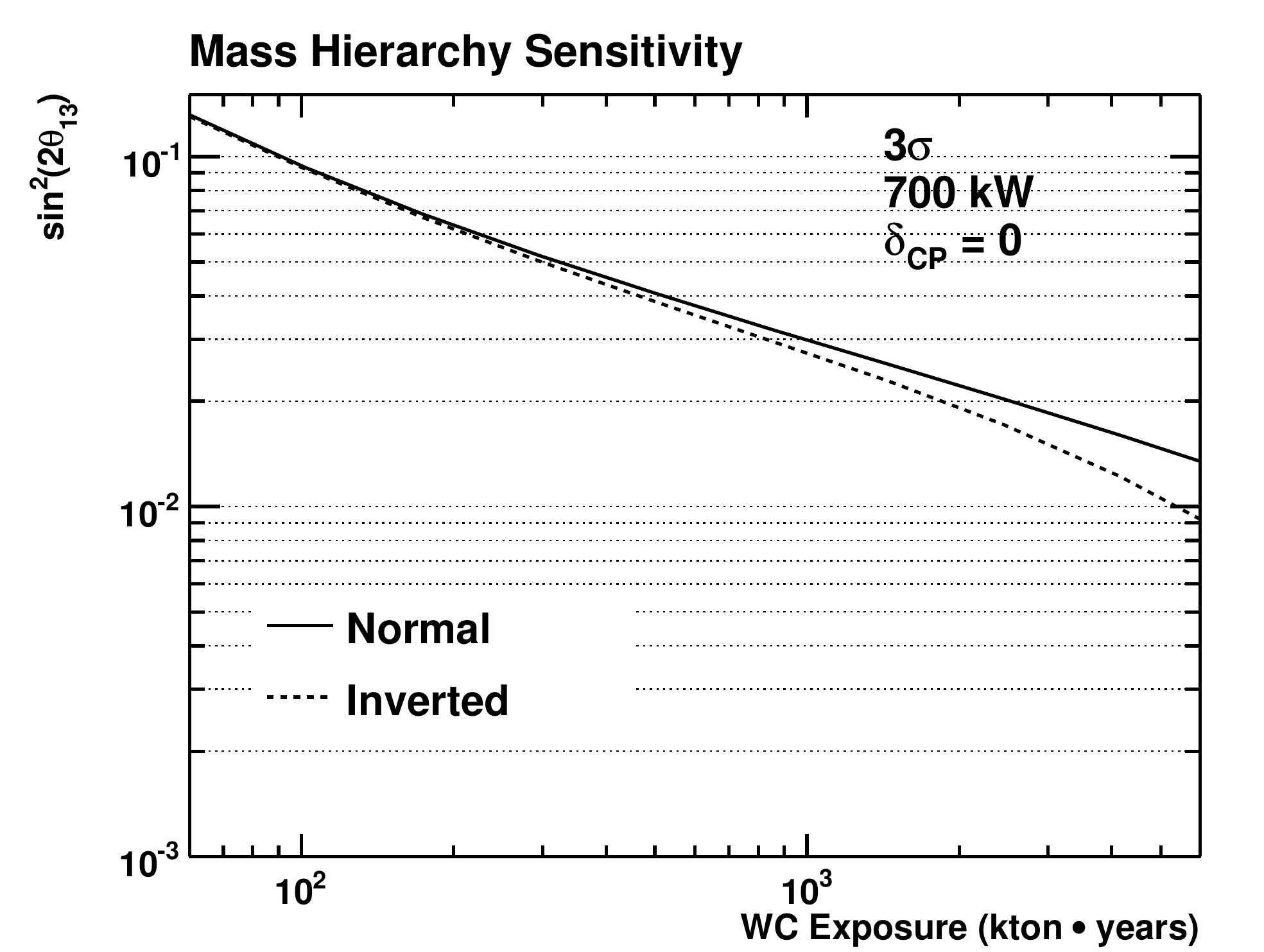}
 \centering\includegraphics[width=.45\textwidth]{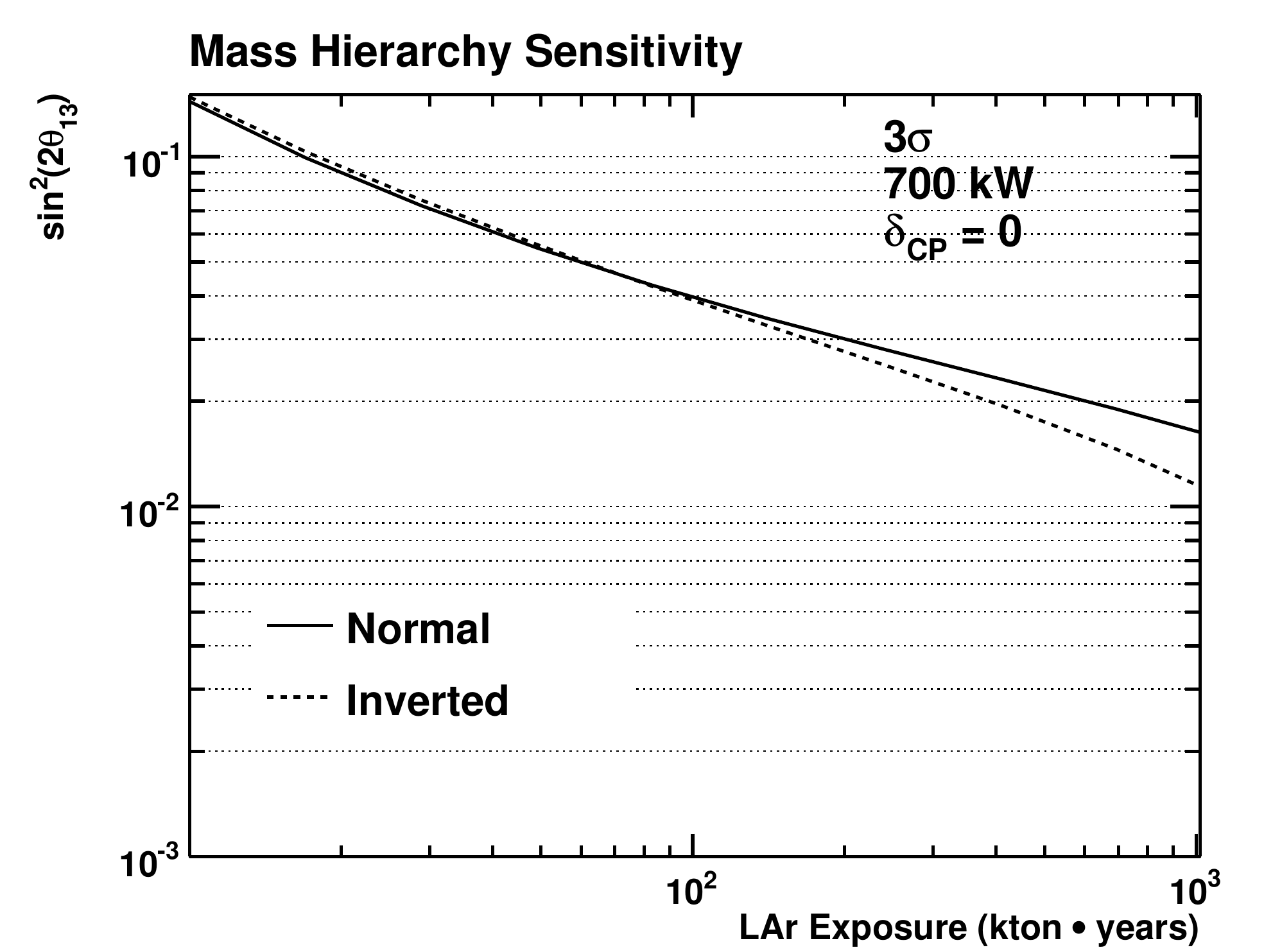}
  \caption{Sensitivity of LBNE to resolving the mass hierarchy at
  $3\sigma$ as a function of exposure for both WC (left) and LAr (right).
  The plots show the projections for $\delta_{CP}=0$ for normal (solid)
  and inverted (dashed) and inverted (dashed) mass hierarchies.}
  \label{fig:lbl_mh_exposure_3sig}
\end{figure}

Fig.~\ref{fig:lbl_mh_exposure_3sig} shows the sensitivity of LBNE for resolving
the mass hierarchy at $3\sigma$ as a function of exposure so one can see how
the reach improves with time and/or detector size. In this case,
WC can resolve the mass hierarchy at $3\sigma$ for $100\%$ of all
$\delta_{\mathrm CP}$ values for a $\sin^22\theta_{13}$ value down to 0.04 in
an exposure of 2000~kt-yrs. The same can be achieved in LAr for
$\sin^22\theta_{13}$ down to 0.05 in an exposure of 340~kt-yrs.

\clearpage

\subsubsection{CP Violation}\label{lbl_cp}

\begin{figure}[hb]
 \centering\includegraphics[width=.45\textwidth]{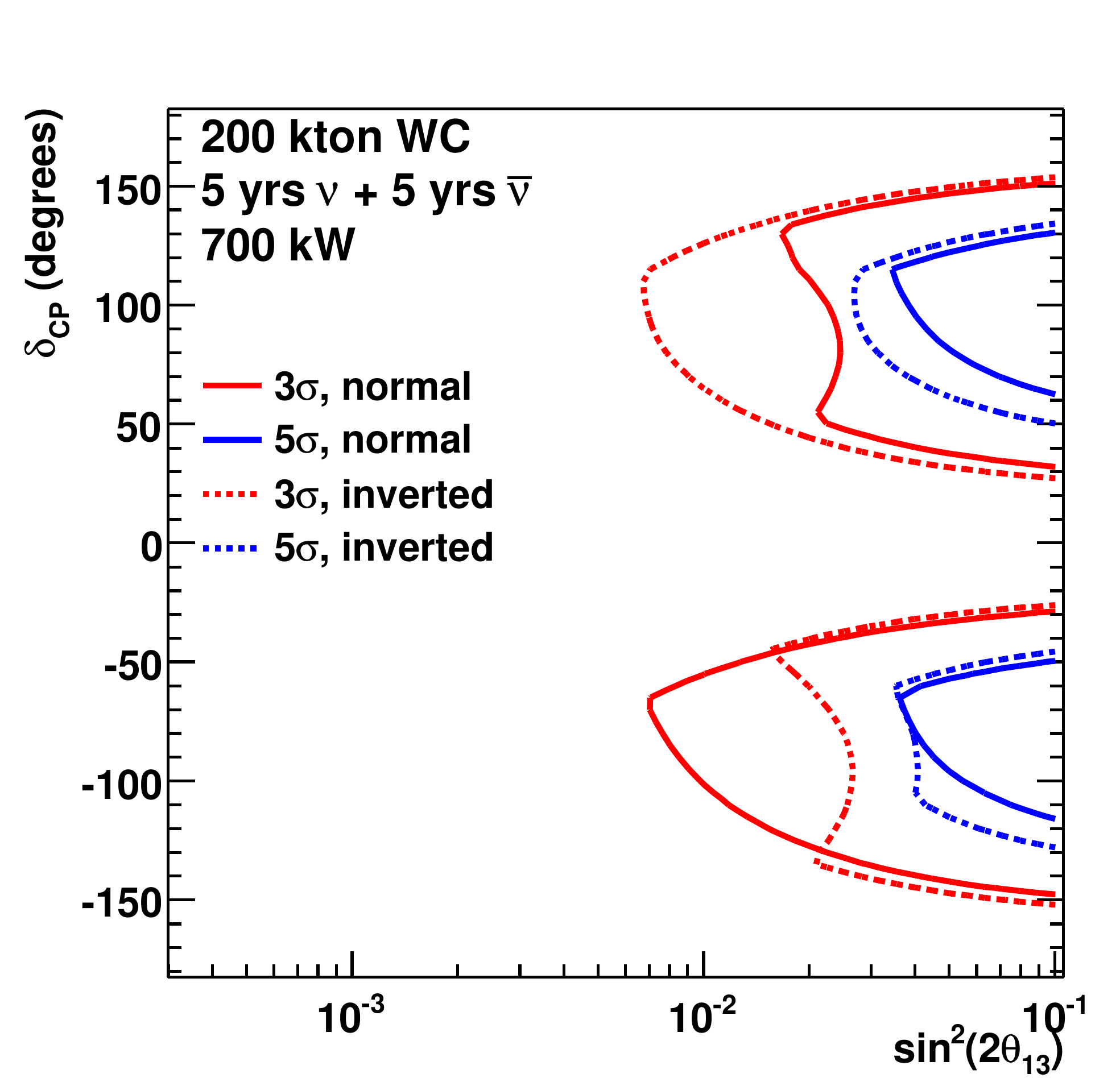}
 \centering\includegraphics[width=.45\textwidth]{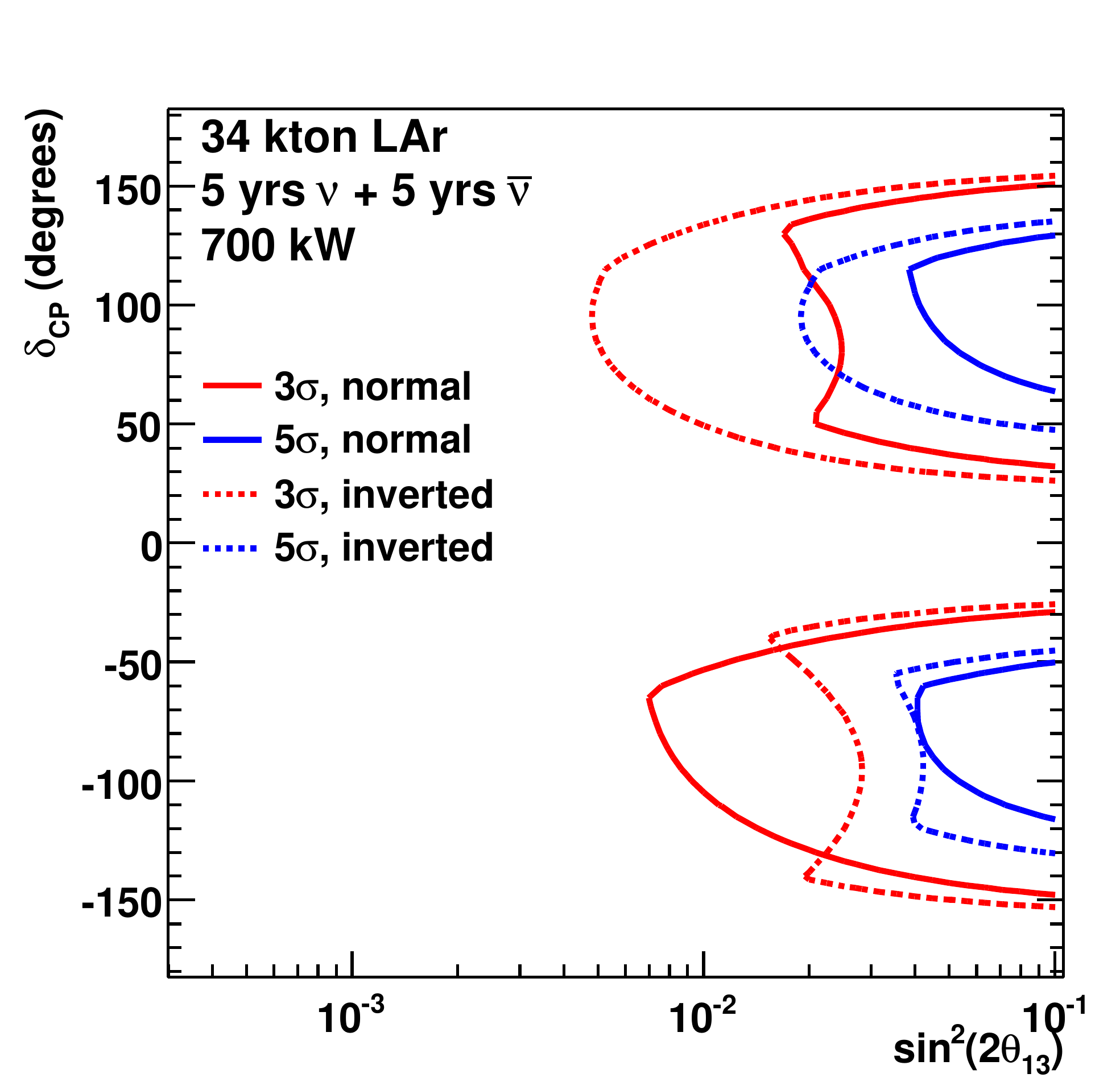}
  \caption{$3\sigma$ (red) and $5\sigma$ (blue) sensitivity of LBNE
           to CP violation for 200~kt of WC (top) and
           34~kt of LAr (bottom). This assumes 5+5 years of $\nu$ and $\nubar$
           running in a 700~kW beam. Curves are shown for both normal (solid)
           and inverted (dashed) mass hierarchies.}
  \label{fig:lbl_cp_sensitivity}
\end{figure}

Section~\ref{lbl_mh} summarizes the CP violation discovery
potential of upcoming oscillation experiments, T2K+NOvA.
Fig.~\ref{fig:lbl_cp_sensitivity} shows the  CP violation reach. Here,
we define CP violation discovery potential as the range of $\delta_{\mathrm CP}$
values as a function of $\sin^22\theta_{13}$ for which one can exclude the
CP conserving solutions for $\delta_{CP}=0^o$ and $\delta_{CP}=180^o$.
In the case of LBNE, a WC detector can make a $3\sigma$ discovery of CP
violation for $50\%$ of all $\delta_{\mathrm CP}$ values  for $\sin^22\theta_{13}$
values down to 0.03 assuming an exposure of 2000~kt-yrs. The same holds
for LAr assuming a 340~kt-yr exposure.

\begin{figure}[htb]
 \centering\includegraphics[width=.45\textwidth]{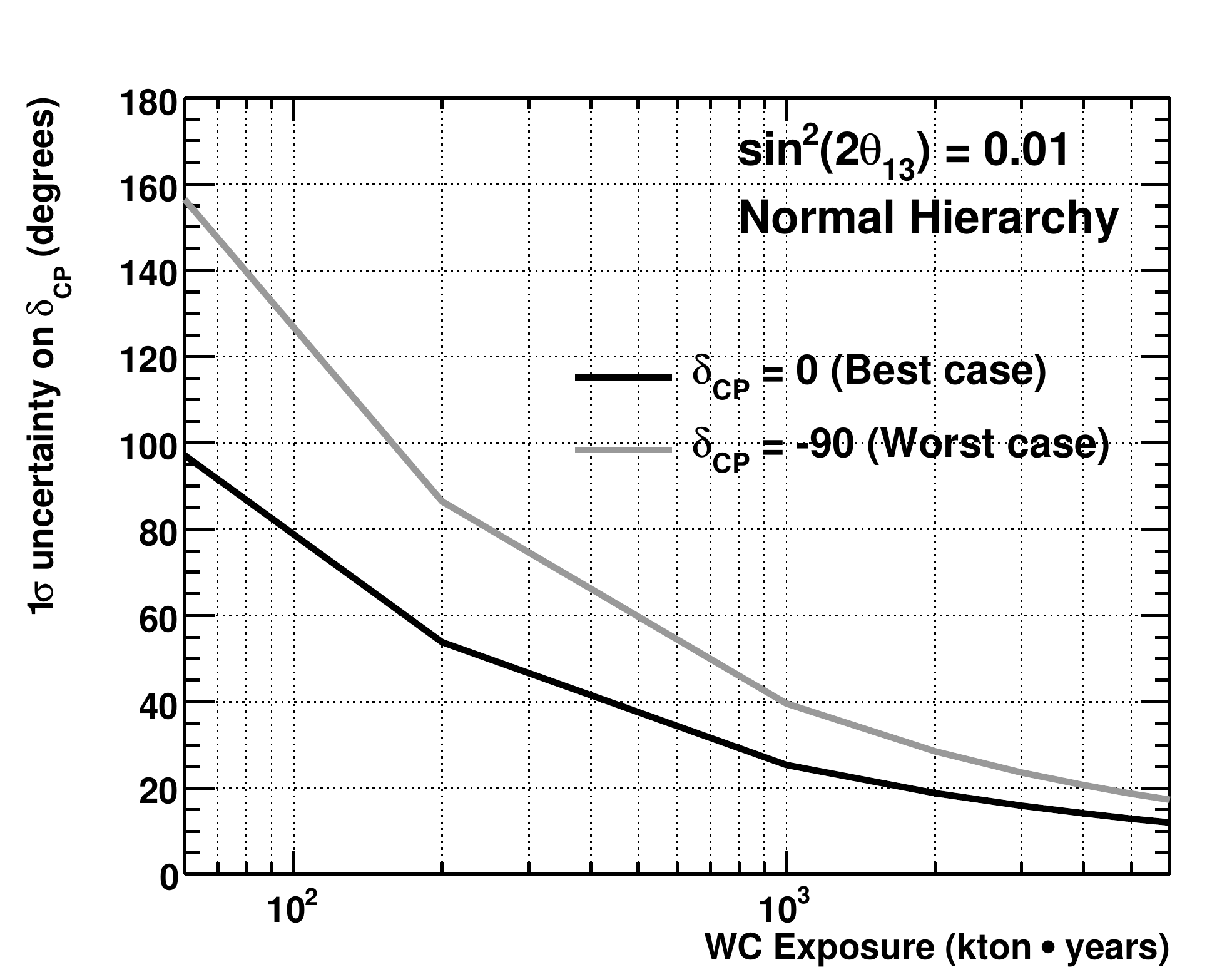}
 \centering\includegraphics[width=.45\textwidth]{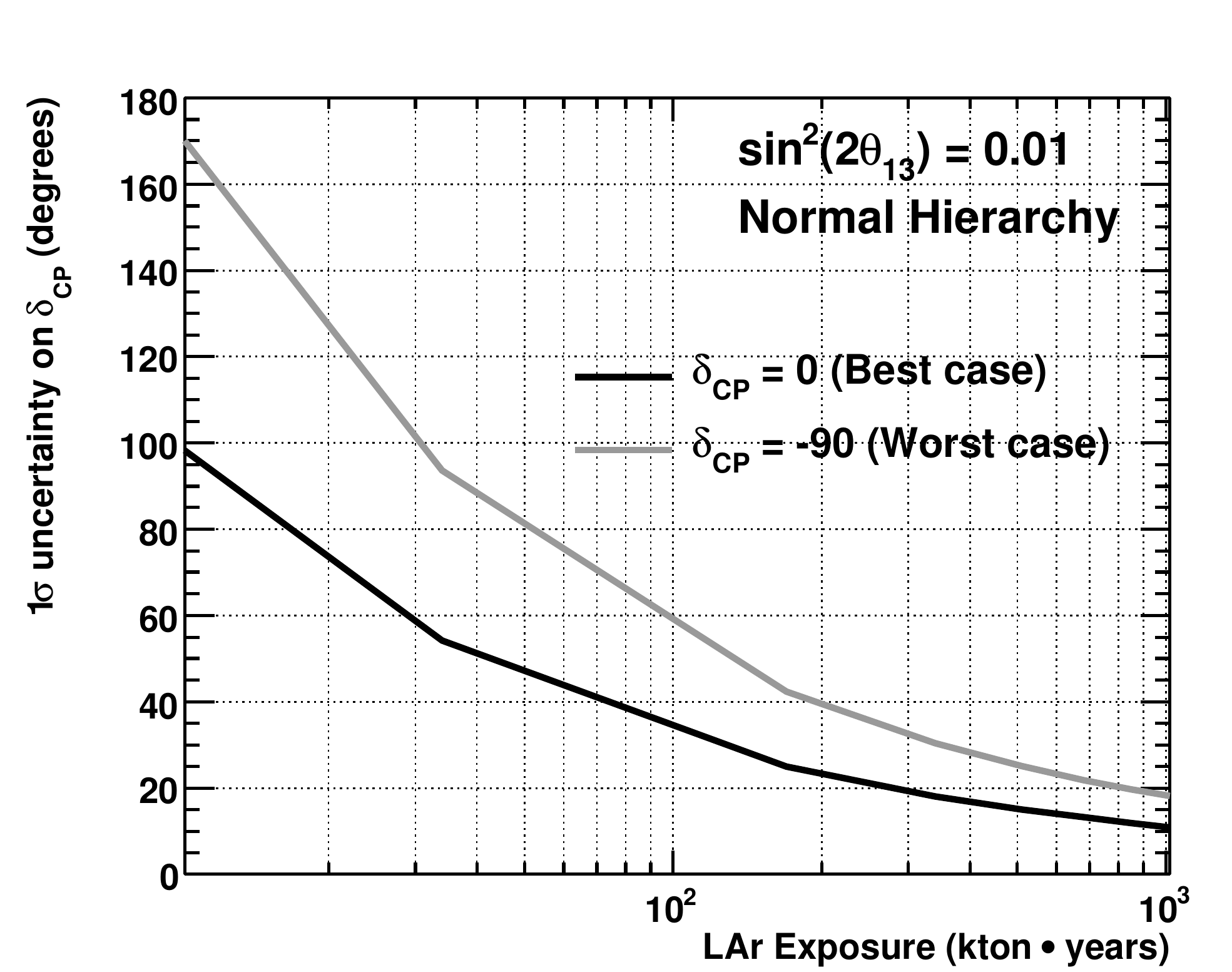}
  \caption{$1\sigma$ resolution on the measurement of $\delta_{\mathrm CP}$ in LBNE
  for both WC (left) and LAr (right) assuming $\sin^22\theta_{13}=0.01$ and
  normal mass hierarchy. Projections for both $\delta_{CP}=0$ (black) and
  $\delta_{CP}=-90^0$ (gray) are separately shown.}
  \label{fig:lbl_delta_resolution_exposure}
\end{figure}

Fig.~\ref{fig:lbl_delta_resolution_exposure} shows the resolution on
LBNE's ability to measure $\delta_{\mathrm CP}$ as a function of exposure for
both WC and LAr. Assuming a normal mass hierarchy, $\sin^22\theta_{13}=0.01$,
and $\delta_{CP}=0$, a WC detector can measure $\delta_{\mathrm CP}$ to within
$\pm19^\circ$ (at $1\sigma$) assuming a 2000~kt-yr exposure. Similarly, LAr
can measure $\delta_{\mathrm CP}$ to the same precision with about 1/6 the exposure
of WC.

\begin{figure}[htb]
 \centering\includegraphics[width=.45\textwidth]{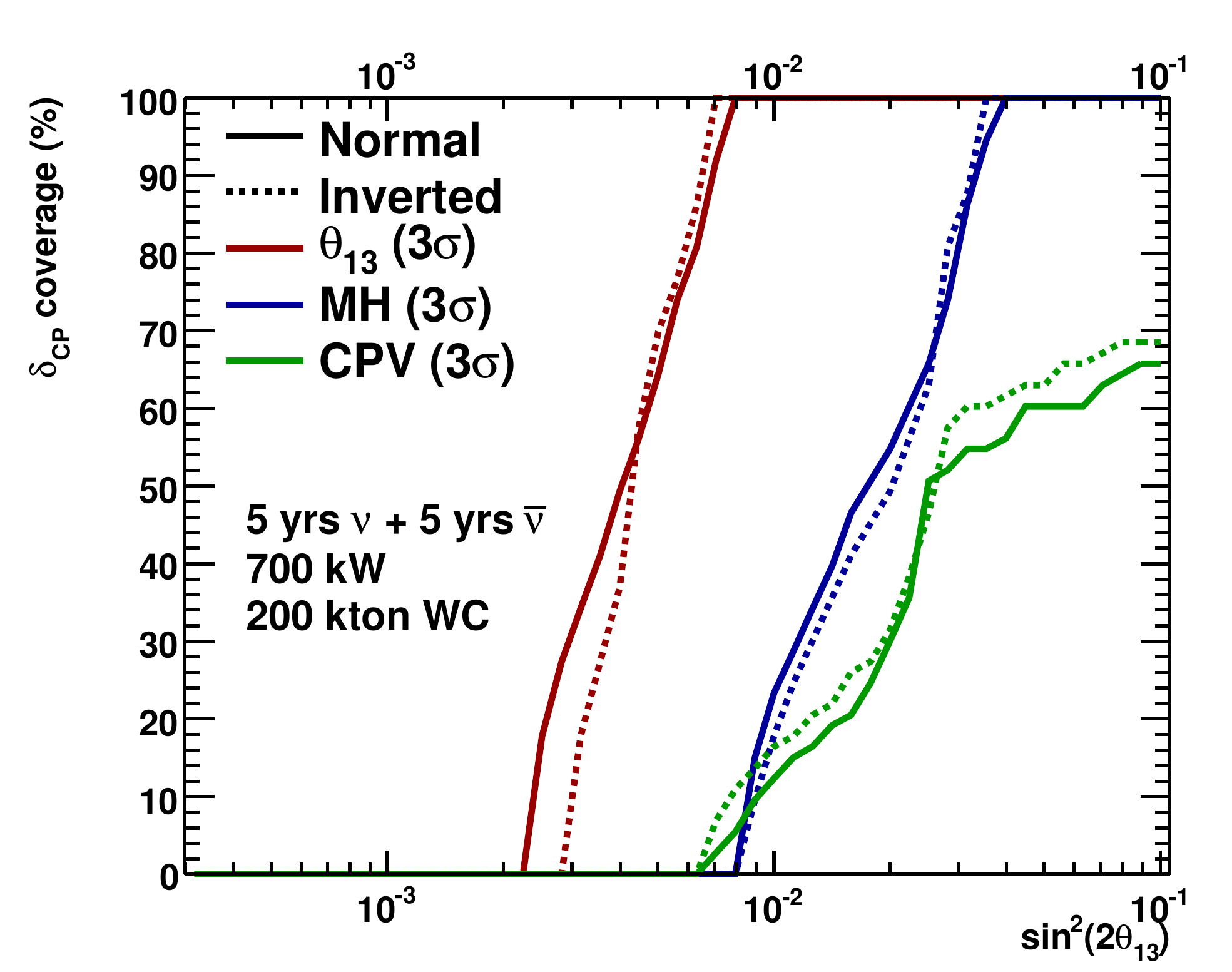}
 \centering\includegraphics[width=.45\textwidth]{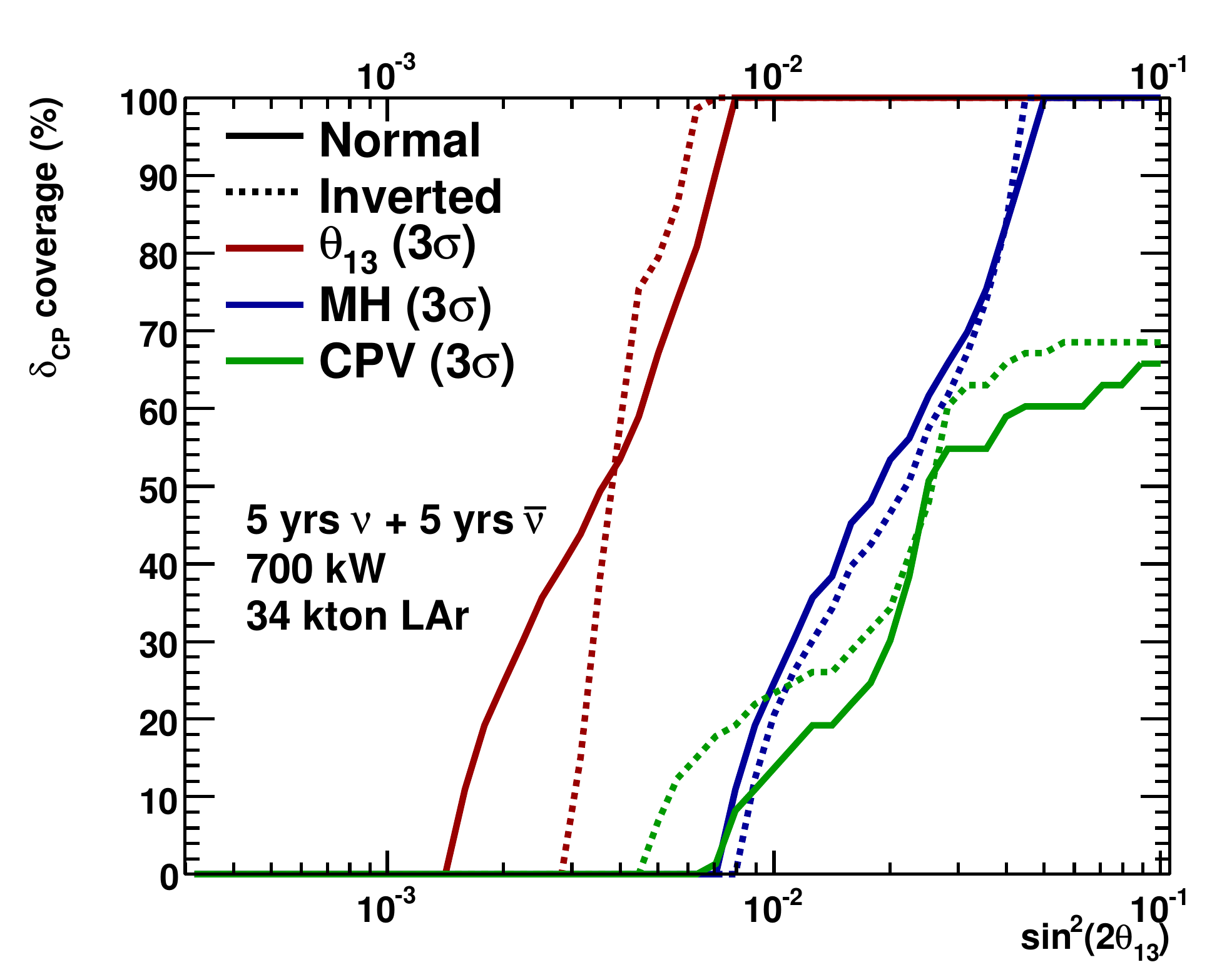}
  \caption{$3\sigma$ discovery potential of LBNE for determining
  $\sin^22\theta_{13}\neq0$ (red), the mass hierarchy (blue), and CP
  violation (green) as function of $\sin^22\theta_{13}$ and the fraction
  of $\delta_{\mathrm CP}$ coverage. Here the fraction of $\delta_{\mathrm CP}$ reflects
  the fraction of all true values of $\delta_{\mathrm CP}$ for which the corresponding
  quantity can be measured. Sensitivities are shown for both
  the normal (solid) and inverted (dashed) mass hierarchies for
  5+5 years of $\nu$+$\nubar$ running in a 700~kW beam. Results for both
  200~kt of WC (top) and 34~kt LAr (bottom) are displayed.}
  \label{fig:lbl_cp_coverage}
\end{figure}

Combining these, Fig.~\ref{fig:lbl_cp_coverage} summarizes the
overall discovery reach of LBNE to determine $\theta_{13}\neq0$, the
mass hierarchy, and CP violation as a fraction of $\delta_{\mathrm CP}$ coverage.
This side-by-side comparison shows what can be achieved with 200~kt of WC
and 34~kt of LAr under the same set of beam exposure assumptions. Similar
sensitivities can be achieved in the two cases.

\clearpage

\subsection{$\numu$ Disappearance}\label{lbl_numu_disappearance}

\begin{figure}[htb]
 \centering\includegraphics[width=.45\textwidth]{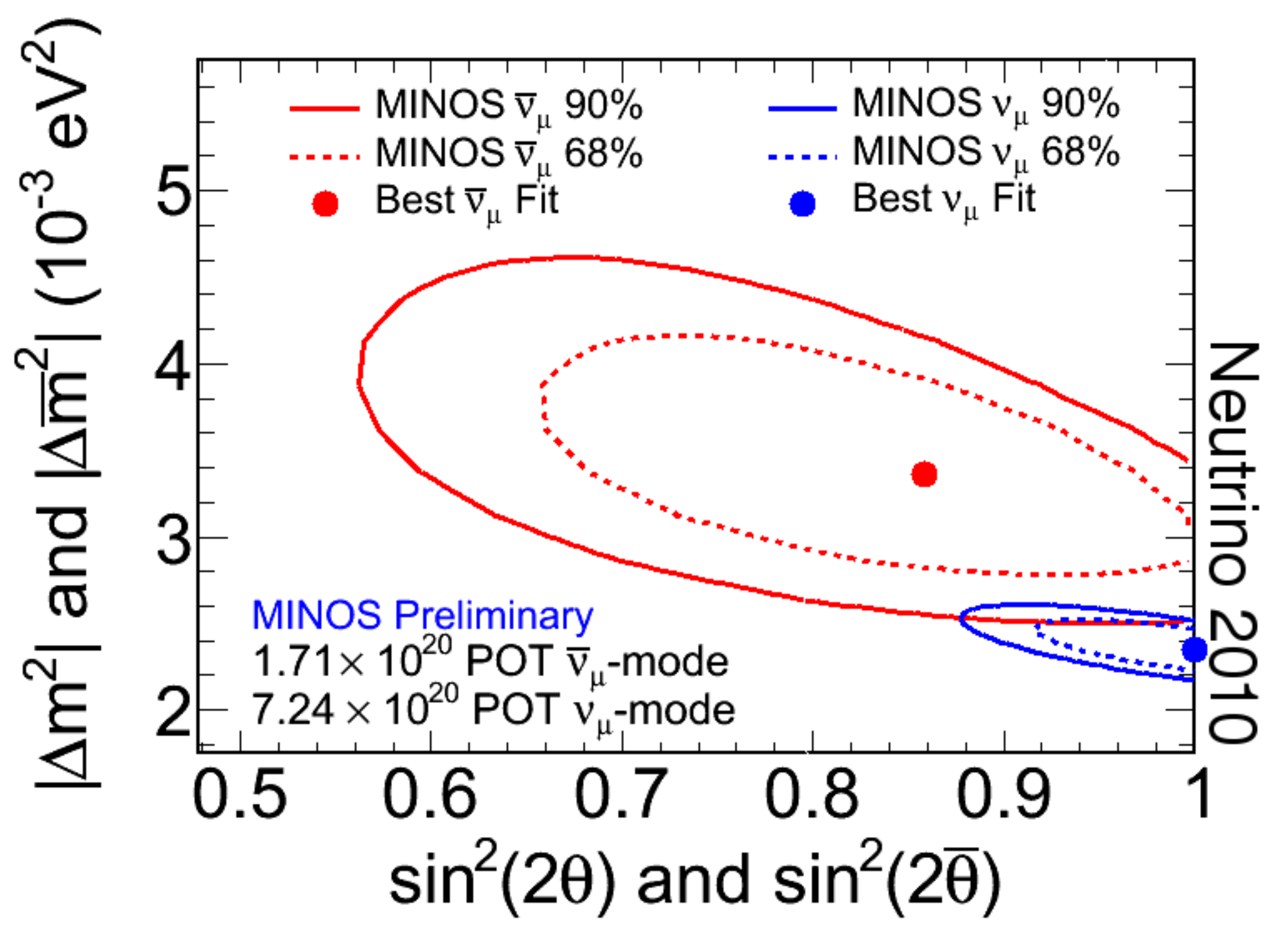}
  \caption{Confidence interval contours from fits to the MINOS far
          detector neutrino (blue) and antineutrino (red) data to a
          two-flavor neutrino oscillation hypothesis (Neutrino 2010). The solid (dashed)
          curve gives the $90\%$ ($68\%$) contours.
          }
  \label{fig:lbl_minos_disapp}
\end{figure}

In addition to the $\nue$ appearance measurements, LBNE will also be able to
provide precise measurements of the atmospheric oscillation parameters
through measurement of both $\numu$ and $\numubar$ disappearance. The most
precise constraints on the atmospheric oscillation parameters are currently
set by the MINOS experiment operating at a baseline distance of 735~km.
Figure~\ref{fig:lbl_minos_disapp} shows the present constraints on
$\sin^22\theta_{23}$ and $\Delta m^2_{32}$ from MINOS as shown at Neutrino
2010~\cite{MINOS:Nu2010}. Currently, a $5\%$ measurement of $\Delta m^2_{32}$
has been obtained and the results suggest a value of $\theta_{23}$ that is
very close to maximal. The MINOS $\numu$ disappearance results
constrain the oscillation parameters to be:
$\Delta m^2_{32}=(2.35^{+0.11}_{-0.08})\times10^{-3}$ eV$^2$, $\sin^22\theta_{23}=1.0$ ($68\%$ CL) and
$\sin^22\theta_{23}>0.91$ at $90\%$ CL.
They have also performed fits to $\numubar$ disappearance. The results in
this case are less precise and are currently $\sim2\sigma$ from the
parameters obtained in neutrino mode:
$\overline{\Delta m^2_{32}}=(3.36^{+0.45}_{-0.40}$ (stat)$\pm 0.06$ (syst))$\times10^{-3}$ eV$^2$ ($68\%$ CL) and
$\sin^22\overline{\theta_{23}}=0.86 \pm 0.11$ (stat)$\pm 0.01$ (syst) ($68\%$ CL).

\begin{figure}[htb]
 \centering\includegraphics[width=.45\textwidth]{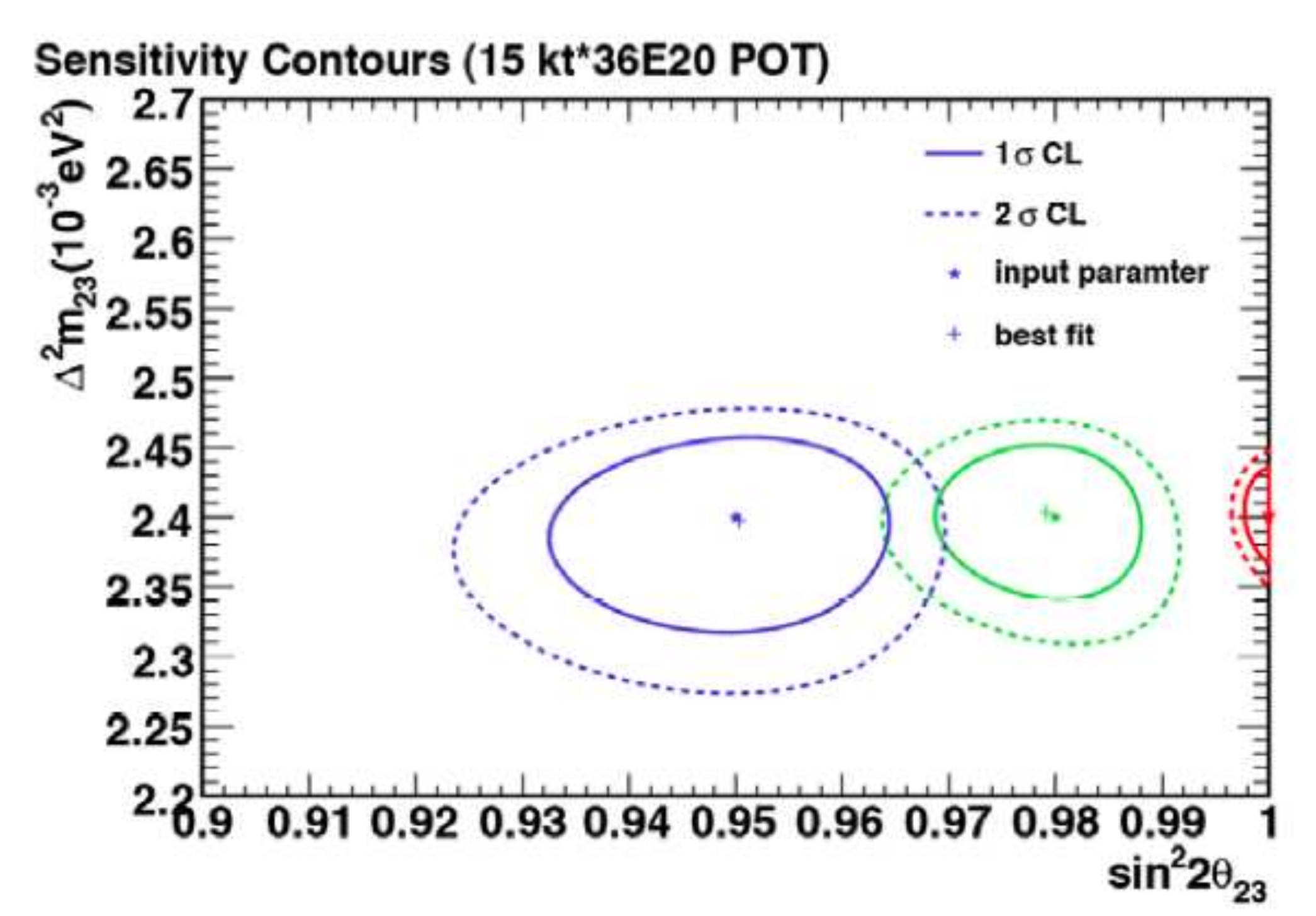}
  \caption{$1\sigma$ and $2\sigma$ contours for a simultaneous measurement
  of $\Delta m^2_{32}$ and $\sin^22\theta_{23}$ in NOvA for a 6-year run
  at 700~kW equally divided between neutrinos and antineutrinos. Plot is
  from~\cite{nova}.}
  \label{fig:lbl_nova_disapp}
\end{figure}

In the coming years, next generation experiments, such as T2K and NOvA,
will be able to push beyond the current values and obtain even more precise
measurements of these parameters. For example, with an exposure of
3.75~MW $\times 10^7$ sec, T2K hopes to make a $1\%$ measurement of
$\sin^22\theta_{23}$ and a measurement of $\Delta m^2_{32}$ with an error
$<4\%$~\cite{t2k:nu2010}. NOvA also aims for more precise measurement of these
parameters. Figure~\ref{fig:lbl_nova_disapp} shows the projected sensitivity
contours for 6 years of NOvA running in the 700~kW NuMI beam. For maximal
mixing and after 6 years of $\nu$+$\nubar$ running, NOvA plans to measure
$\sin^22\theta_{23}$ to $\sim0.3\%$ and $\Delta m^2_{32}$ to
$\sim1\%$~\cite{nova}.

\clearpage

\begin{figure}[htb]
 \centering\includegraphics[width=.45\textwidth]{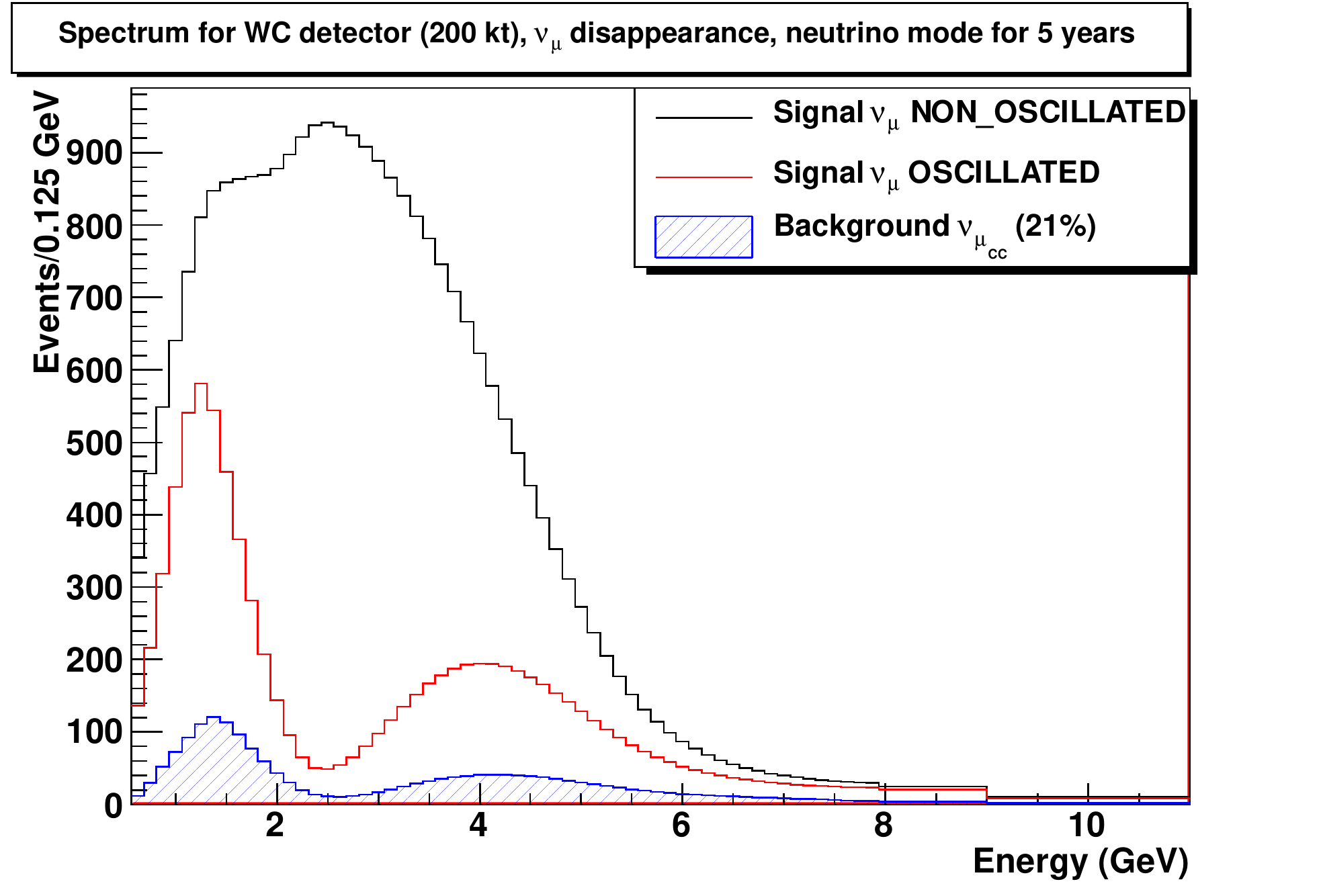}
 \centering\includegraphics[width=.45\textwidth]{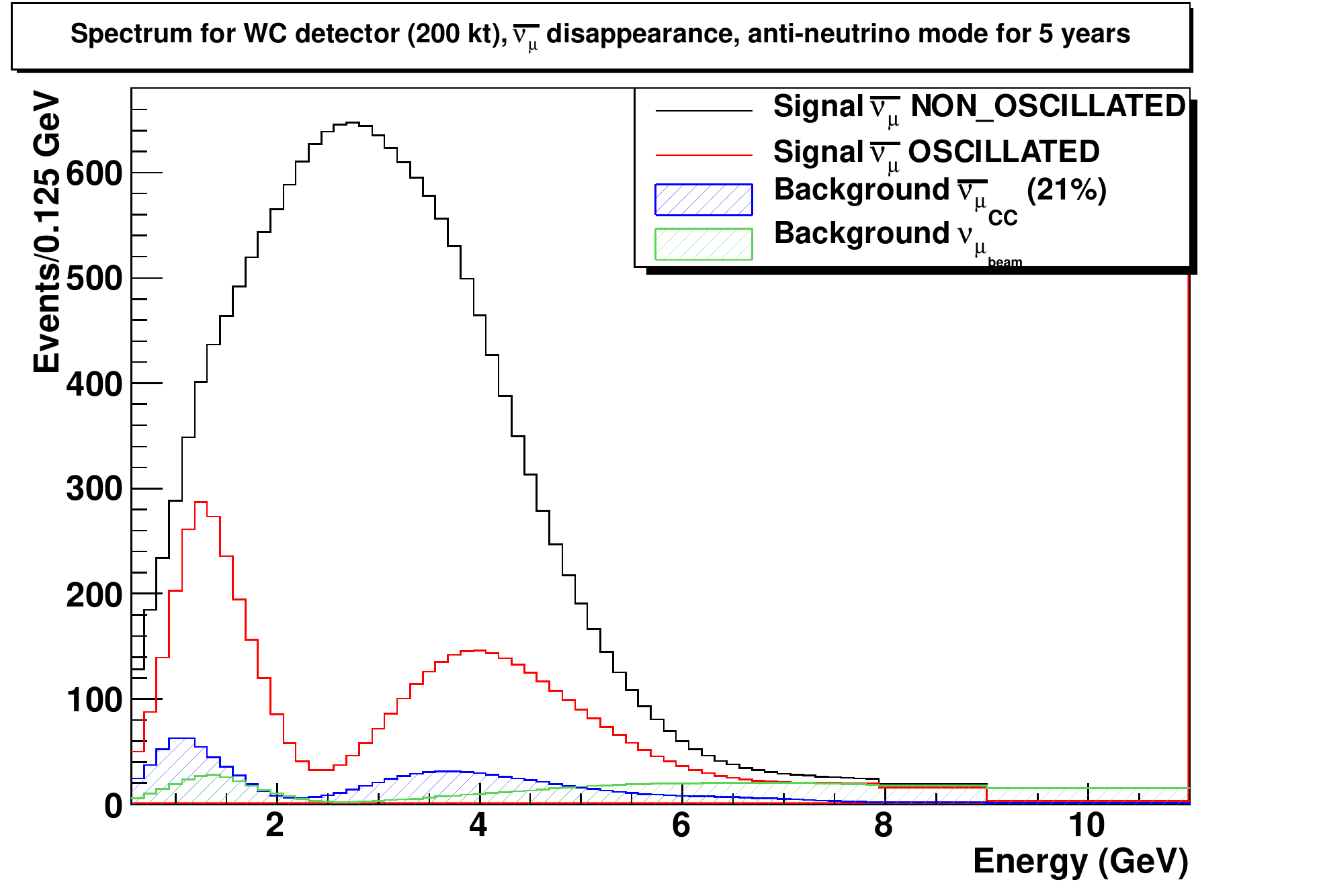}
  \caption{Number of events expected with (red) and
          without (black) oscillations as observed by a 200~kt
          water Cerenkov detector in 5 years of neutrino (left) or 5 years of
          antineutrino (right) running in a 120~GeV 700~kW beam.
          In the current set of assumptions, a $\numu$ QE sample is
          used for the signal channel. The backgrounds, are assumed
          to come from CC $\pi^+$ events, are shown in blue. In
          the case of antineutrino running, there is also an additional
          background contribution from $\numu$ events shown in green.}
  \label{fig:lbl_disapp_events_wc}
\end{figure}

\begin{figure}[htb]
 \centering\includegraphics[width=.45\textwidth]{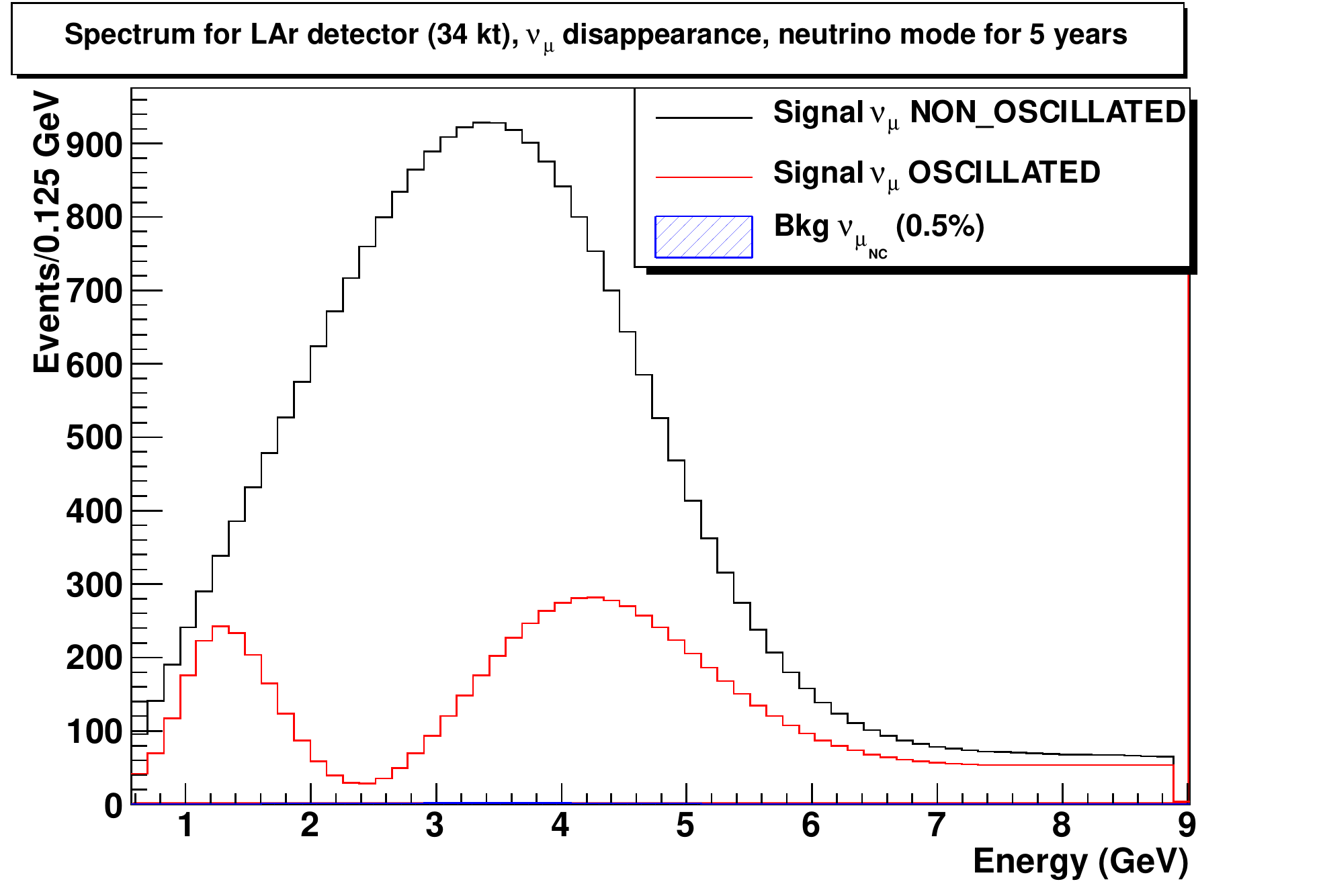}
 \centering\includegraphics[width=.45\textwidth]{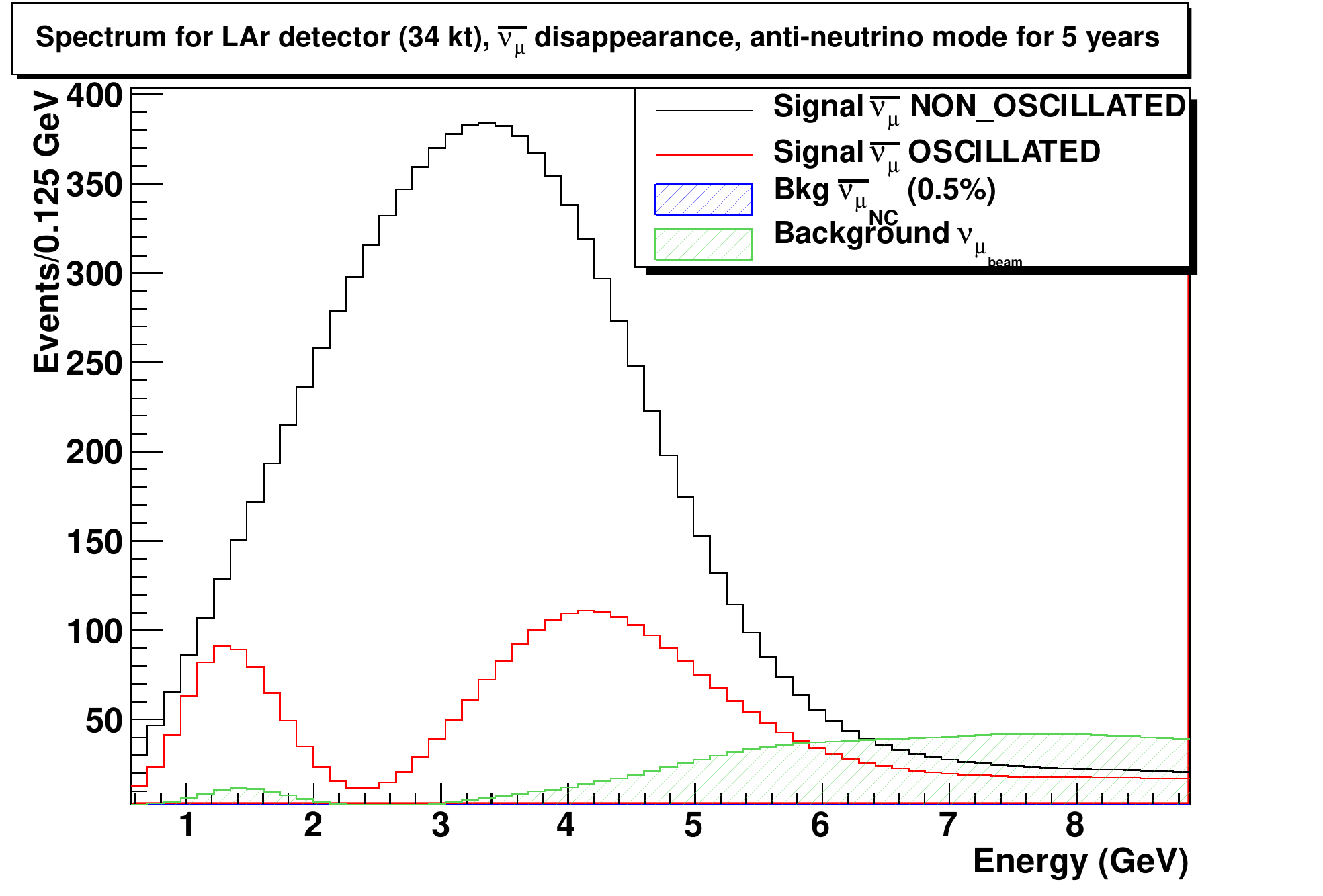}
  \caption{The same plots and beam conditions as in Fig.~\ref{fig:lbl_disapp_events_wc} except for a 34~kt liquid argon detector (but note the different energy axis scales).
          In the current set of assumptions, a $\numu$ CC sample is
          used for the signal channel. The assumed NC backgrounds are also
          plotted but are too small to be visible. In the case of antineutrino
          running, there is an additional contribution from $\numu$ events
          which is taken into account and shown in green.}
  \label{fig:lbl_disapp_events_lar}
\end{figure}

LBNE will provide an even more sensitive test of the atmospheric oscillation
parameters through its measurement of $\numu$ and $\numubar$ disappearance.
One advantage of the long-baseline in LBNE is that the multiple oscillation
pattern in the spectrum will be clearly detectable. This should offer some
advantage when it comes to systematics. As such, LBNE should clearly
show the bi-modal structure in $\Delta m^2_{32}$
(note that KamLAND has already observed this for $\Delta m^2_{21}$).
Figures~\ref{fig:lbl_disapp_events_wc} and \ref{fig:lbl_disapp_events_lar}
show the expected $\numu$ and $\numubar$ event distributions at the LBNE
far detector site for both WC and LAr detectors. In both cases, the statistics
and the size of the expected signal (i.e. distortion in the spectrum) are
large.

The bimodal structure is slightly different for WC and LAr due to the assumed
signal selection. For WC, the signal is chosen to be a $\numu$ quasi-elastic
(QE) sample. This channel is selected because it provides a determination of
the incoming neutrino energy based solely on the reconstructed muon
kinematics. Background events are then predominately CC non-QE interactions
where additional particles are produced but not detected. These are assumed
to be largely CC $\pi^+$ interactions. They account for $21\%$ of the
$\numu$ QE  sample (based on estimates of background contamination in K2K\cite{k2k:ahn}
and MiniBooNE\cite{miniboone:disapp}). For WC, a $97\%$ $\numu$ QE signal efficiency is assumed.

The signal is assigned a
$5\%$ normalization error and a $3\%$ energy scale uncertainty. The CC $\pi^+$
backgrounds are assigned a $10\%$ normalization error (based on the current
accuracy of existing CC $\pi^+$/QE measurements) and a $3\%$ energy scale
uncertainty. For LAr, inclusive $\numu$ CC events are selected as the signal
sample. An $85\%$ signal efficiency is currently assumed with a $5\%$
normalization error and a $2\%$ energy scale uncertainty. In this case,
the backgrounds are NC events, $0.5\%$ of which are assumed to be
misidentified as CC events and assigned a $10\%$ ($2\%$) normalization
(energy scale) uncertainty. In the case of antineutrino running,
these same assumptions are applied with the addition of an extra level of $\numu$
contamination. The actual parameter assumptions and GLoBES inputs used for
the LBNE $\numu$ and $\numubar$ disappearance projections are provided for
both detectors in Appendix~\ref{lbl_appendix}.

\begin{figure}[htb]
 \centering\includegraphics[width=.45\textwidth,angle=0]{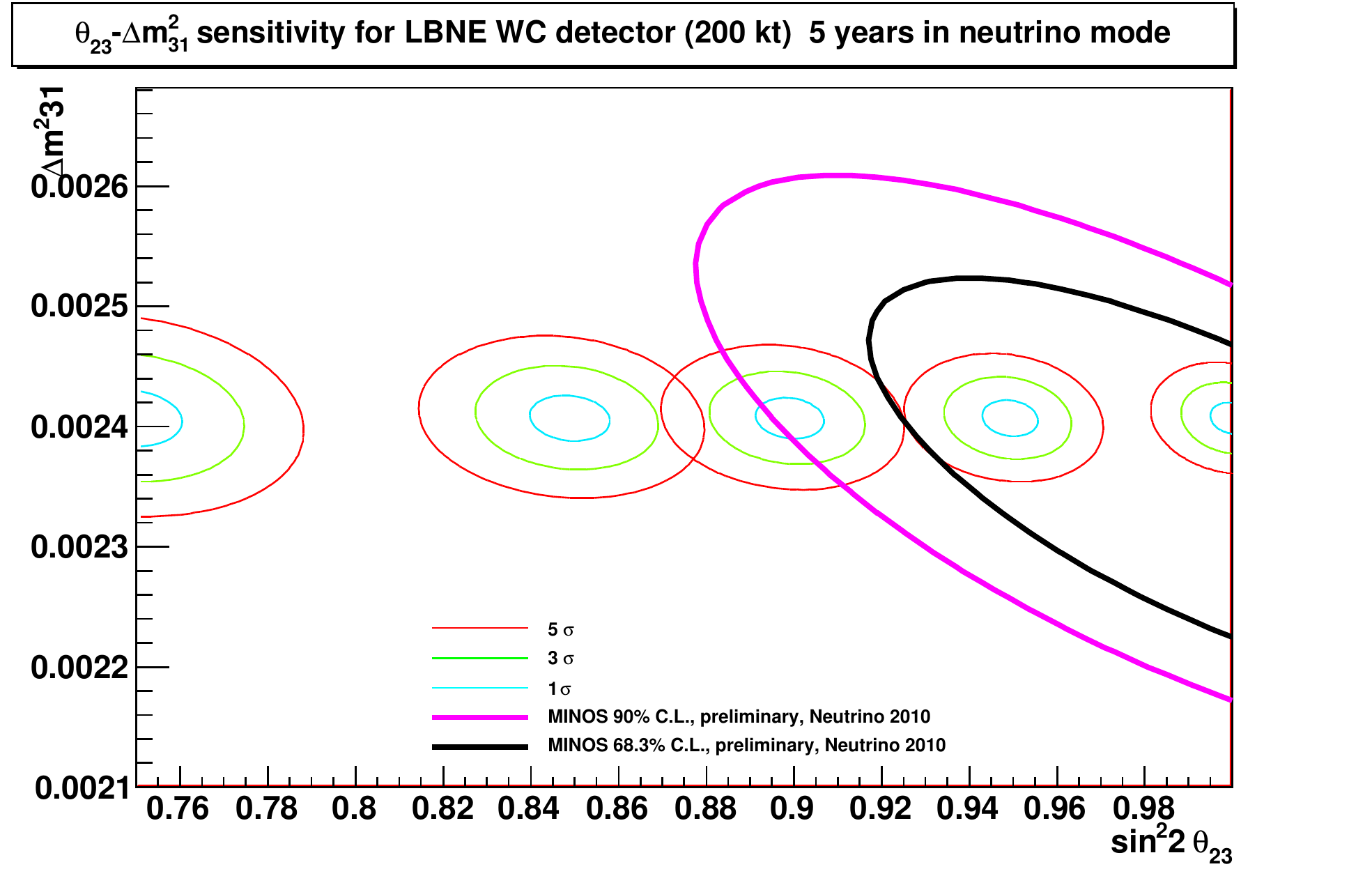}
 \centering\includegraphics[width=.45\textwidth,angle=0]{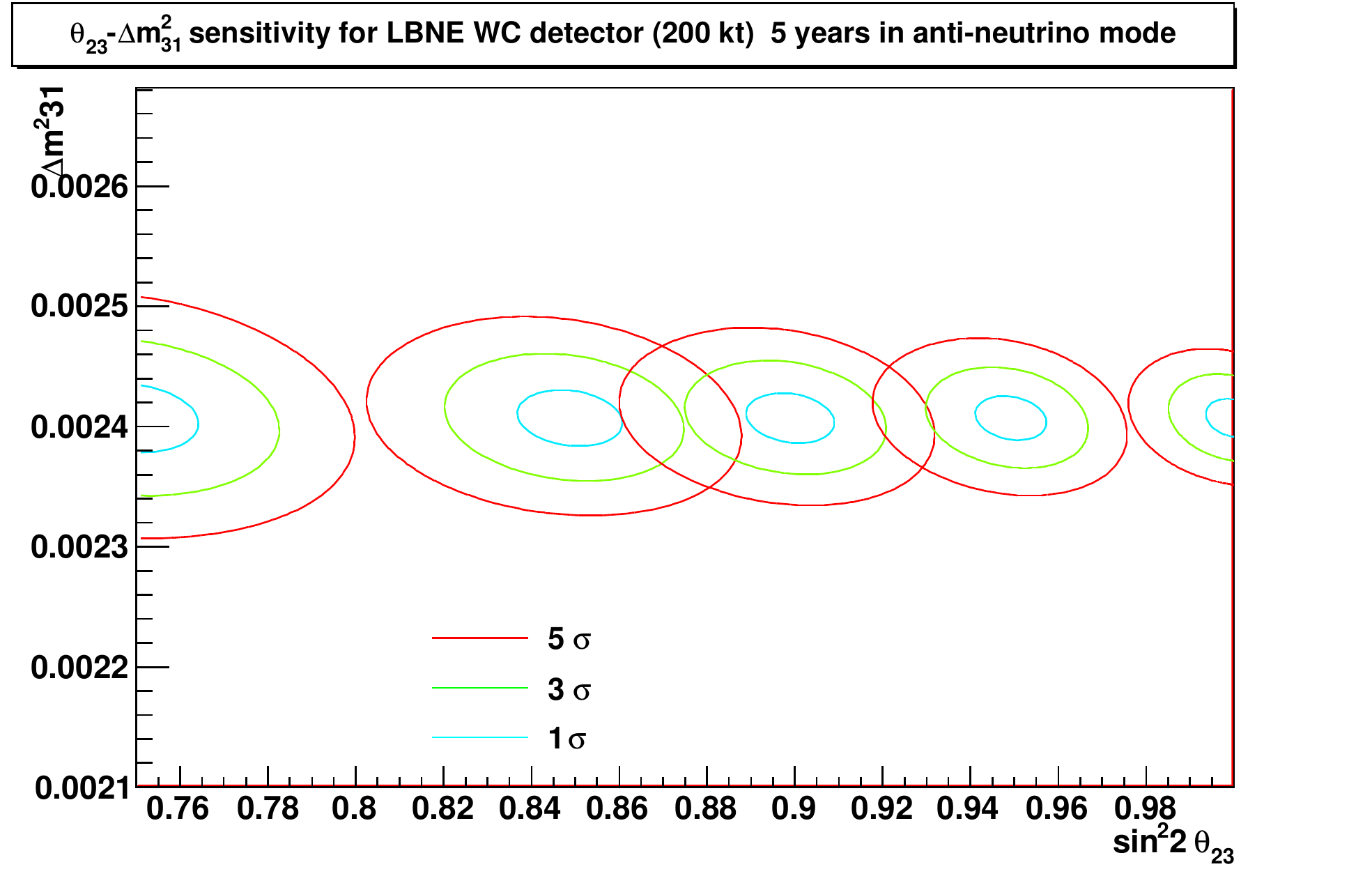}
  \caption{Sensitivity of LBNE to $\sin^22\theta_{23}$ and $\Delta m^2_{32}$
          for select values of $\sin^22\theta_{23}$ for 200~kt of
          WC. This assumes 5 years of neutrino (left) or 5 years of
          antineutrino (right) running in a 120~GeV 700~kW beam. Recent
          results from MINOS are also overlaid in the neutrino
          case~\cite{rebel}.}
  \label{fig:lbl_disapp_sensitivity_wc}
\end{figure}

\begin{figure}[htb]
 \centering\includegraphics[width=.45\textwidth,angle=0]{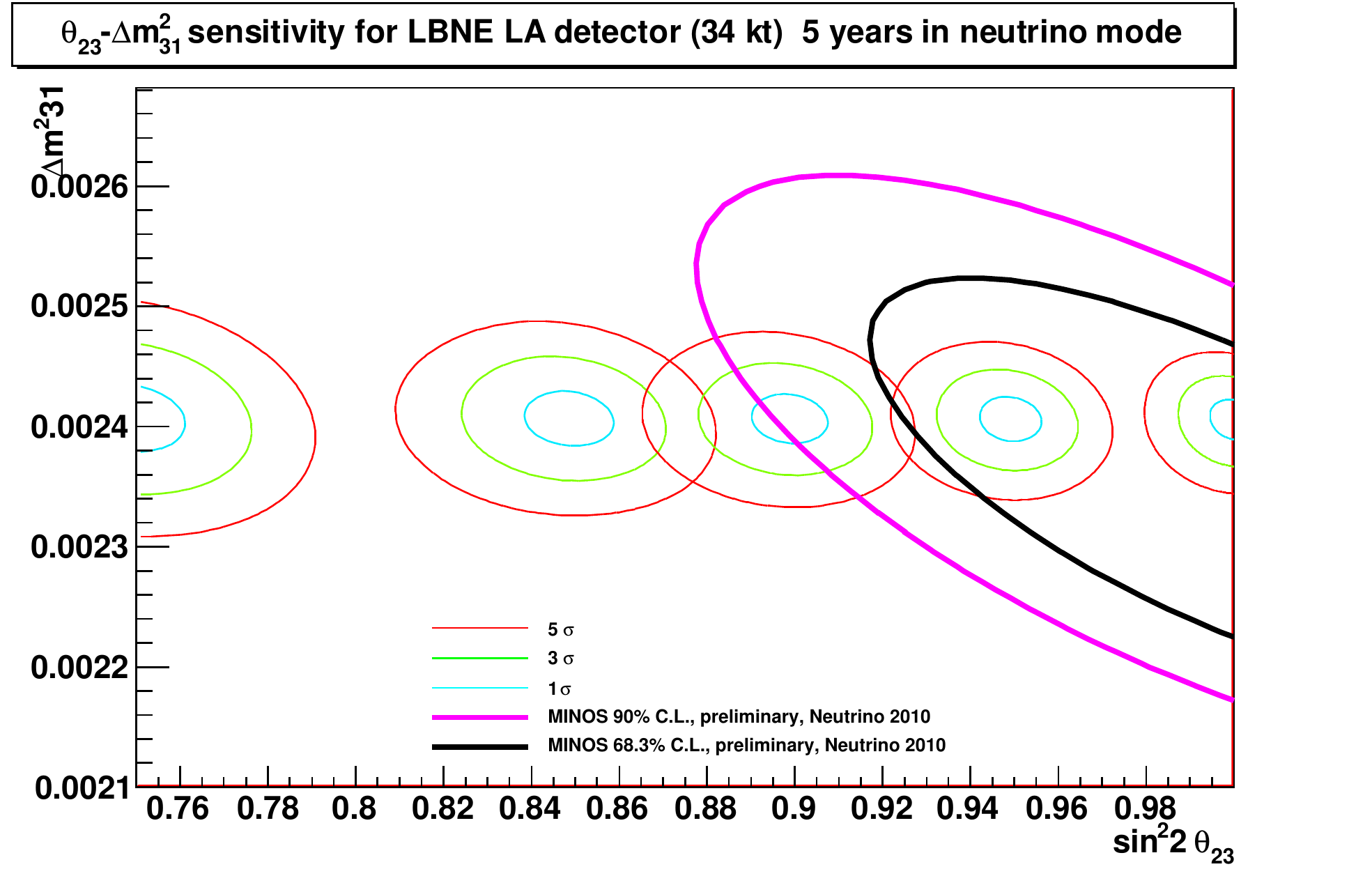}
 \centering\includegraphics[width=.45\textwidth,angle=0]{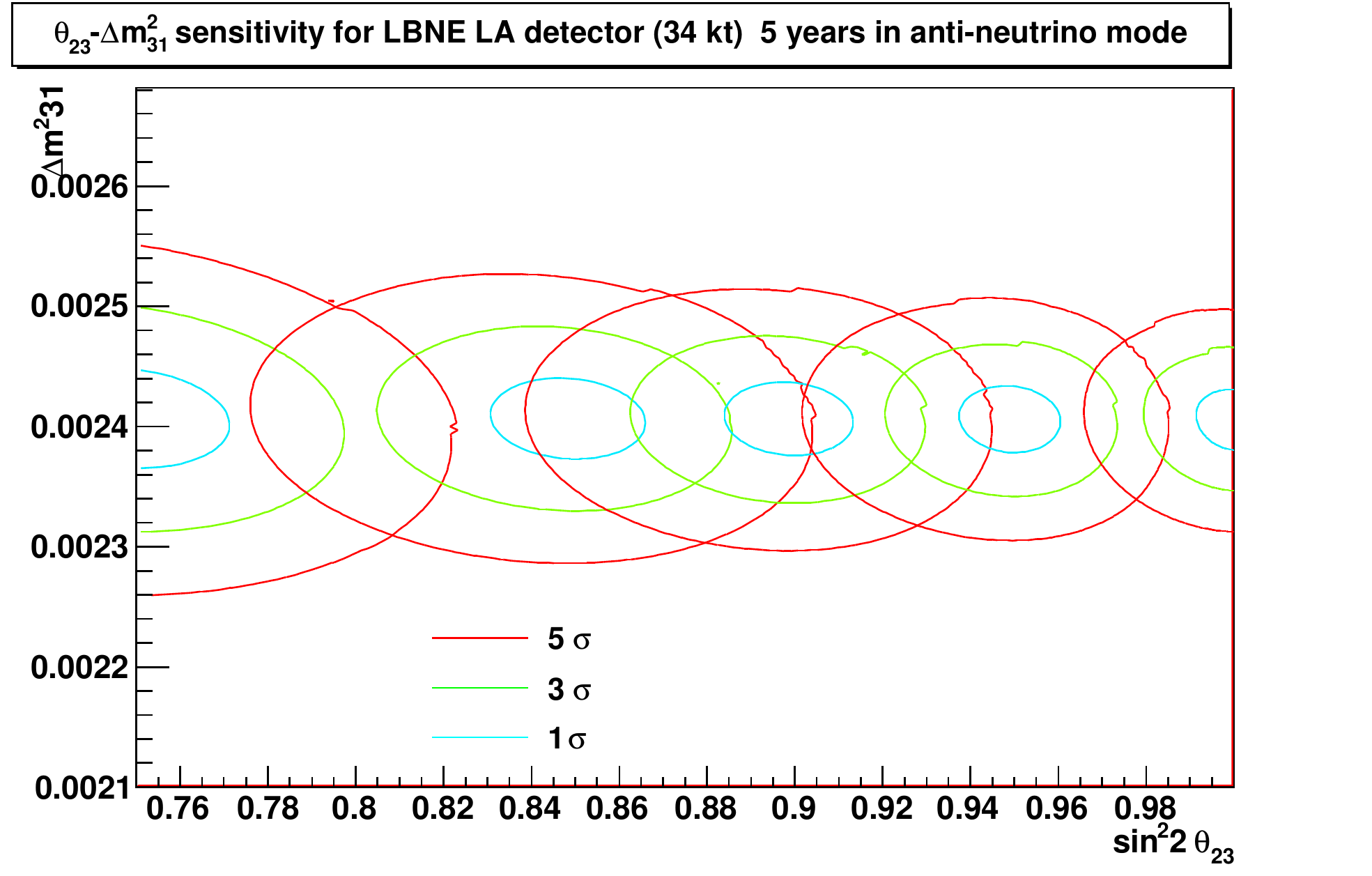}
  \caption{Same plots as Fig.~\ref{fig:lbl_disapp_sensitivity_wc} except for 34~kt of LAr.}
  \label{fig:lbl_disapp_sensitivity_lar}
\end{figure}

Under these assumptions, the sensitivity of LBNE to $\sin^22\theta_{23}$
and $\Delta m^2_{32}$ in both neutrino and antineutrino modes is shown in
Figures~\ref{fig:lbl_disapp_sensitivity_wc} and
\ref{fig:lbl_disapp_sensitivity_lar} for both WC and LAr, respectively.
As one might expect, the sensitivity improves for large values of
$\sin^22\theta_{23}$.

\begin{figure}[htb]
 \centering\includegraphics[width=0.38\textwidth]{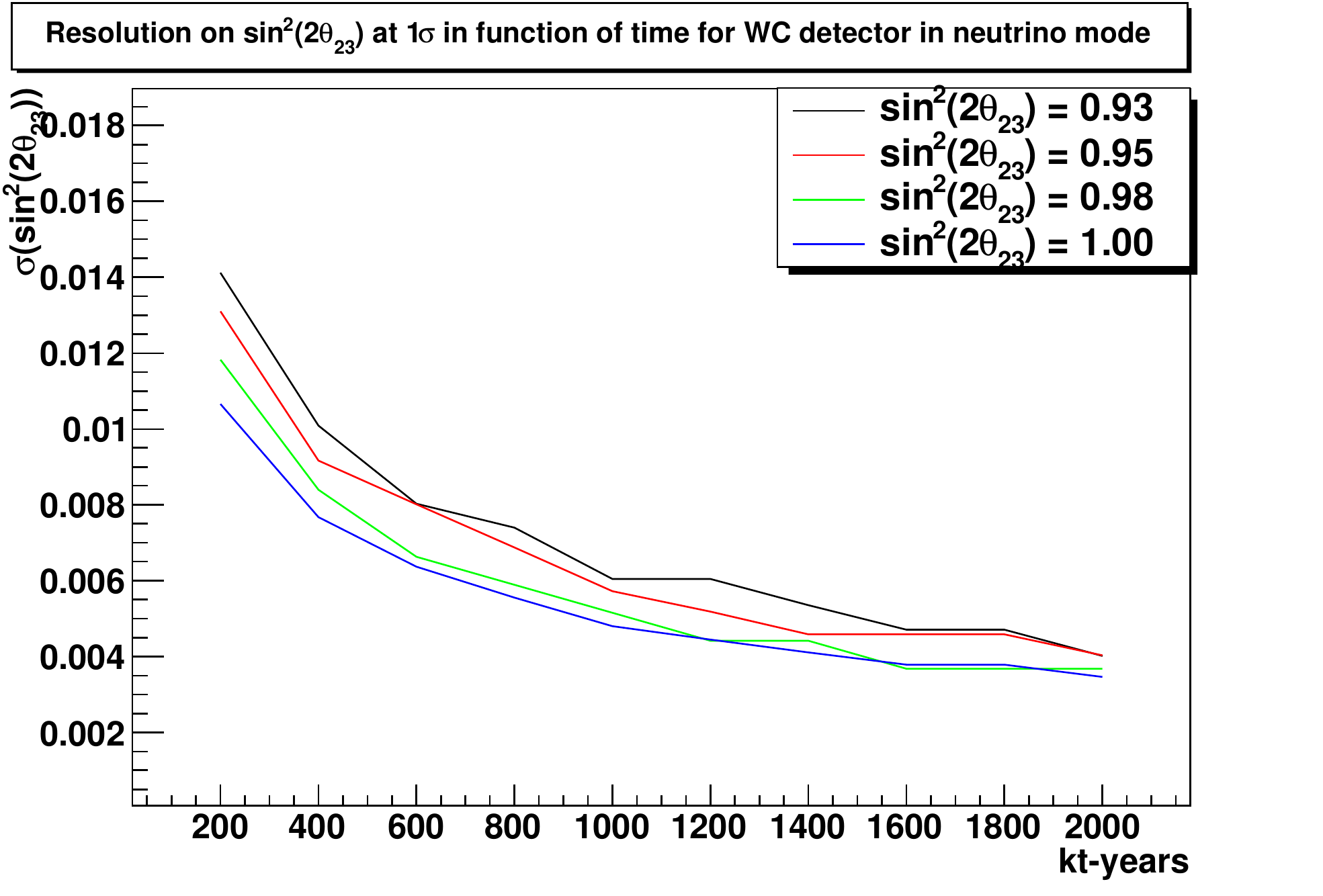}
 \centering\includegraphics[width=0.38\textwidth]{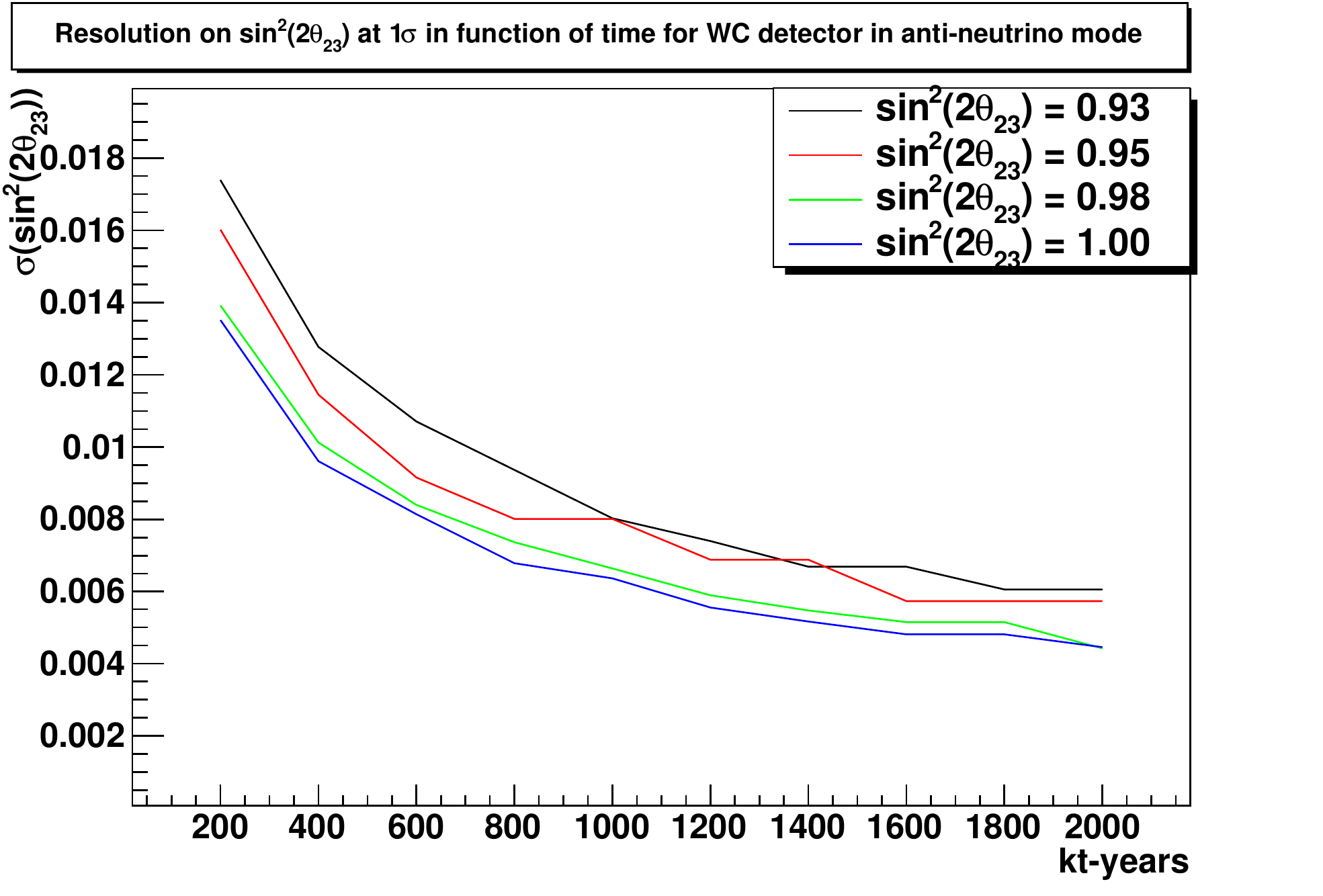}\\
 \centering\includegraphics[width=0.38\textwidth]{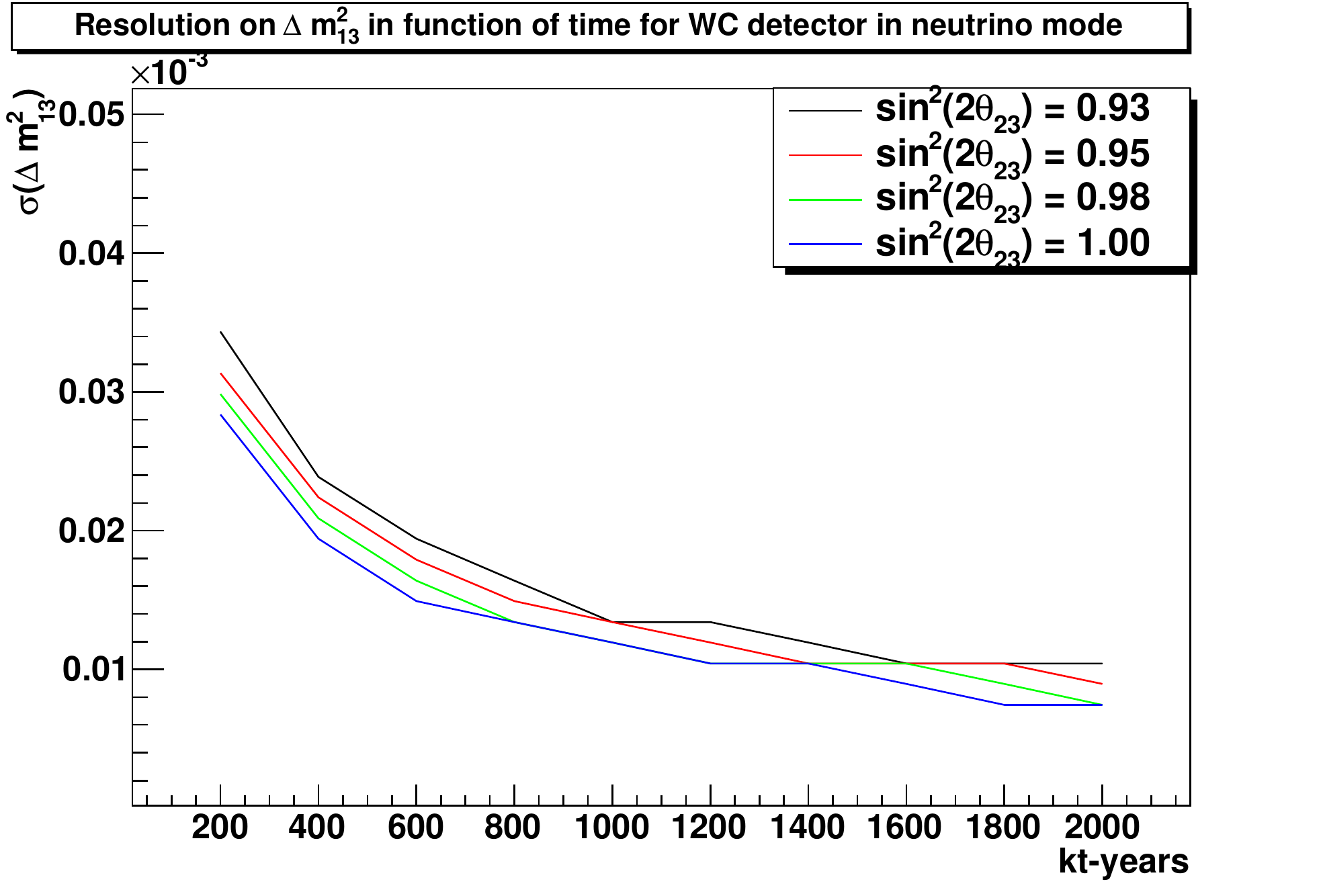}
 \centering\includegraphics[width=0.38\textwidth]{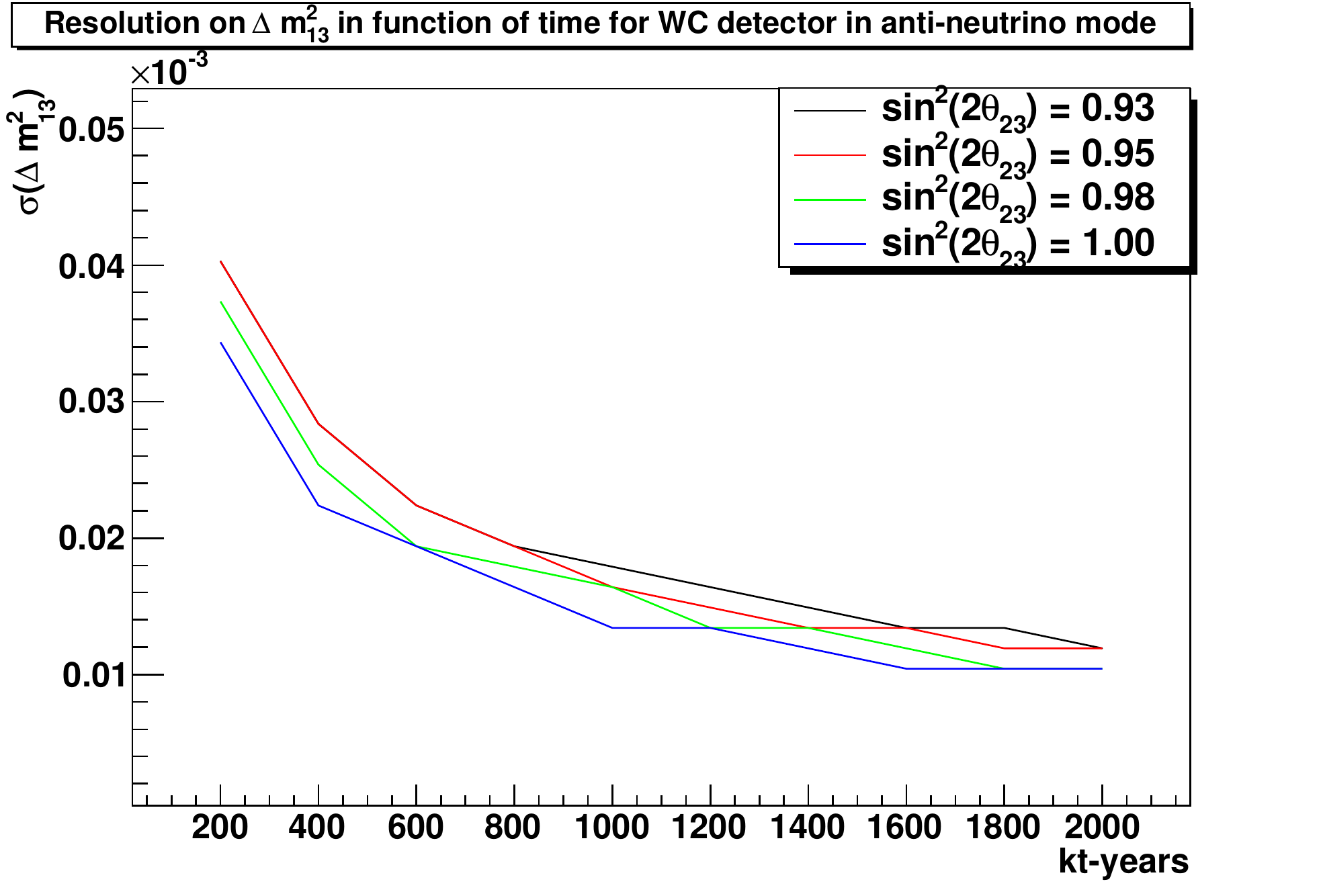}
  \caption{Resolution on $\sin^2\theta_{23}$ (top) and $\Delta m^2_{32}$
          (bottom) as a function of kt-years that could be achieved in LBNE
          at the $1\sigma$ level for a WC detector running in neutrino (left)
          and antineutrino (right) mode assuming 700~kW beams.}
  \label{fig:lbl_disapp_resolution_wc}
\end{figure}

\begin{figure}[htb]
 \centering\includegraphics[width=0.38\textwidth]{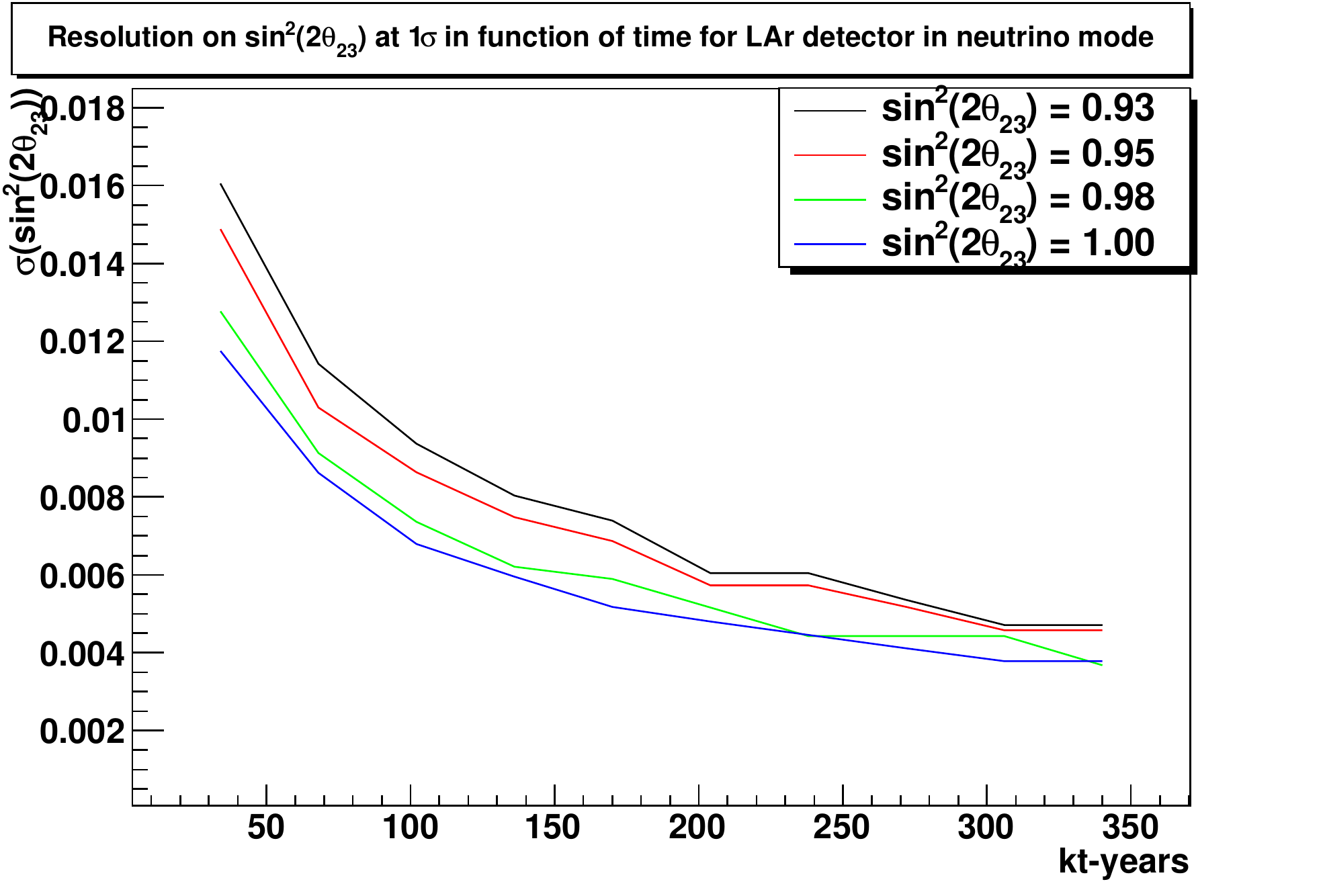}
 \centering\includegraphics[width=0.38\textwidth]{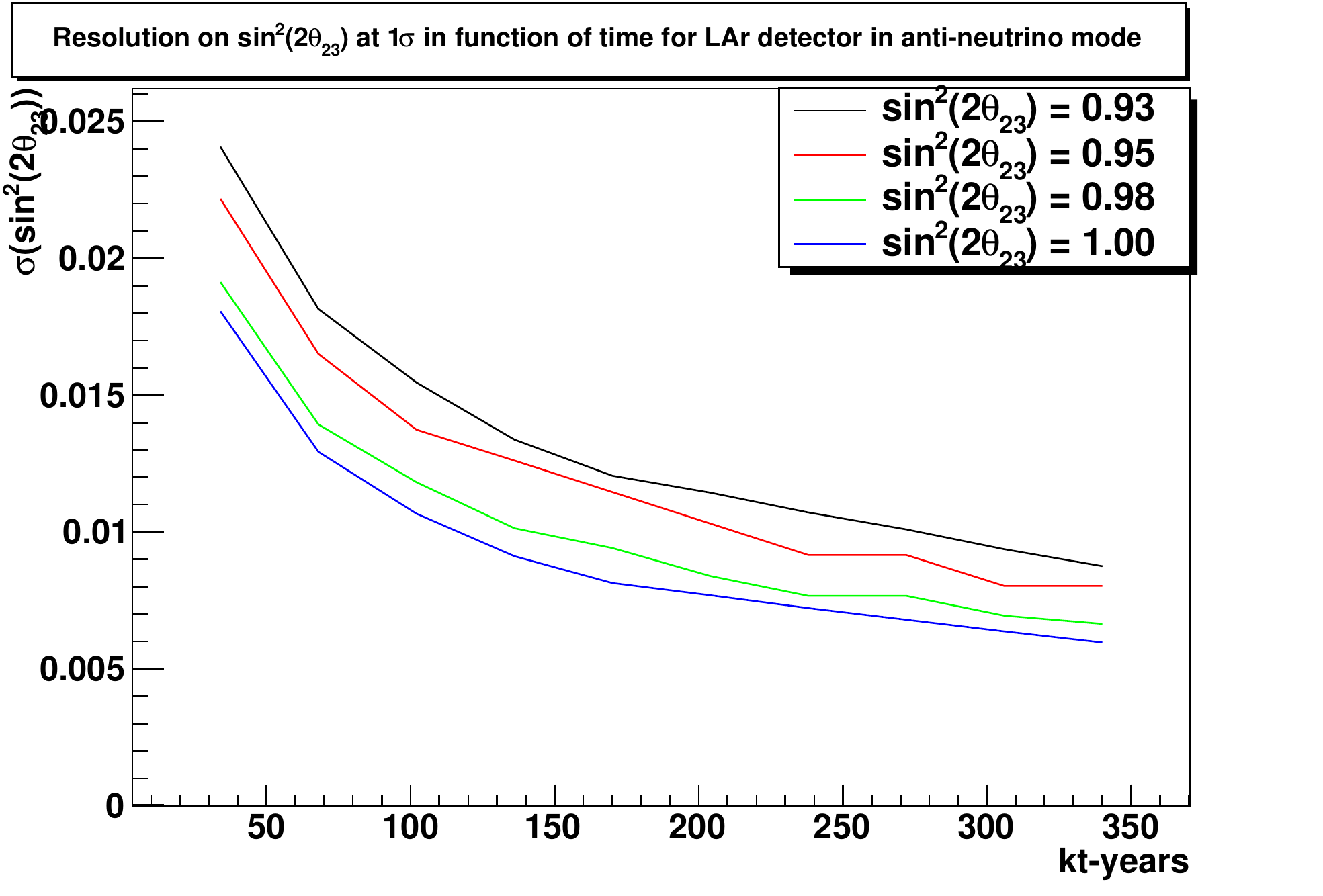}\\
 \centering\includegraphics[width=0.38\textwidth]{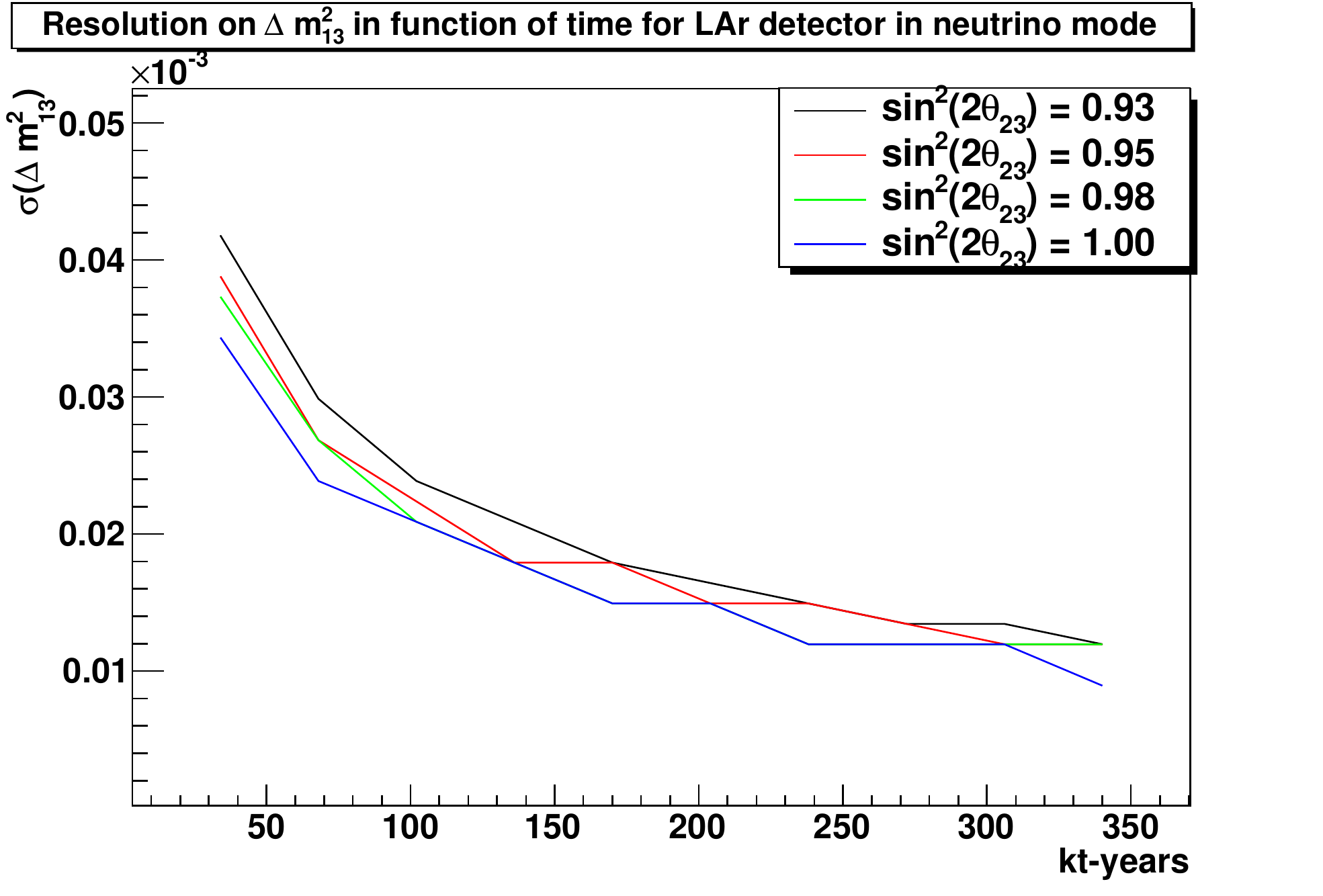}
 \centering\includegraphics[width=0.38\textwidth]{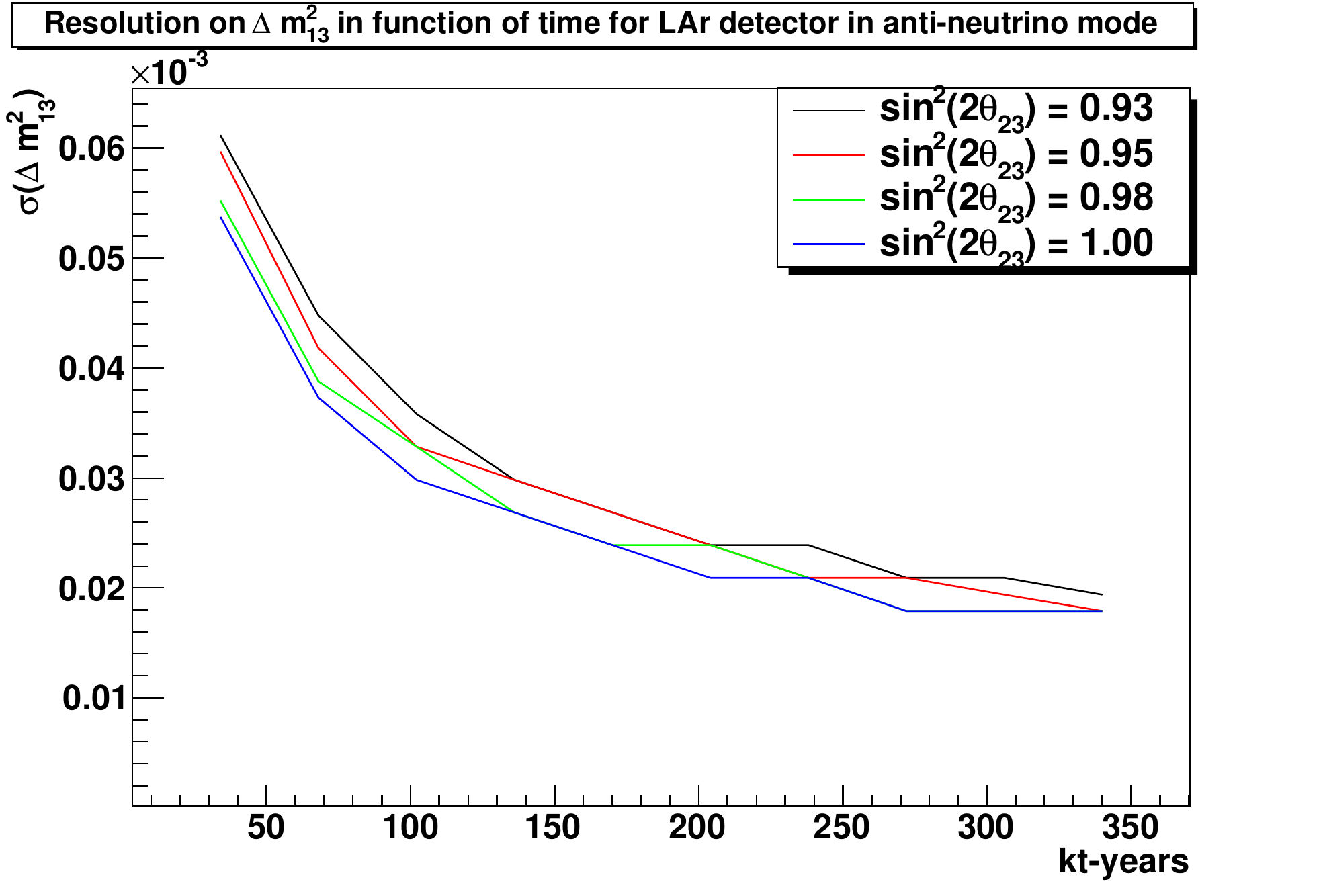}
  \caption{Resolution on $\sin^2\theta_{23}$ (top) and $\Delta m^2_{32}$
          (bottom) as a function of kt-years that could be achieved in LBNE
          at the $1\sigma$ level for a LAr detector running in neutrino (left)
          and antineutrino (right) mode assuming 700~kW beams.}
  \label{fig:lbl_disapp_resolution_lar}
\end{figure}

To further illustrate how these projections might scale with exposure,
Figures~\ref{fig:lbl_disapp_resolution_wc} and
\ref{fig:lbl_disapp_resolution_lar} plot the resolutions on
$\sin^22\theta_{23}$ and $\Delta m^2_{32}$ achievable in LBNE as a function
of kt-years. Here, the $1\sigma$ contours have been used for the resolution
calculation. In 5 years of neutrino running only and for maximal mixing,
$<1\%$ measurements of $\Delta m^2_{32}$ and $\sin^22\theta_{23}$ are
possible (at $1\sigma$) with either a 200~kt WC or 34~kt LAr detector.
Measurements of these parameters in the antineutrino disappearance channel are
possible at the $1\%$ level assuming a similar exposure.

\clearpage

\subsection{$\theta_{23}$ Octant Degeneracy}\label{lbl_octant}

\begin{figure}[htb]
 \centering\includegraphics[width=.5\textwidth]{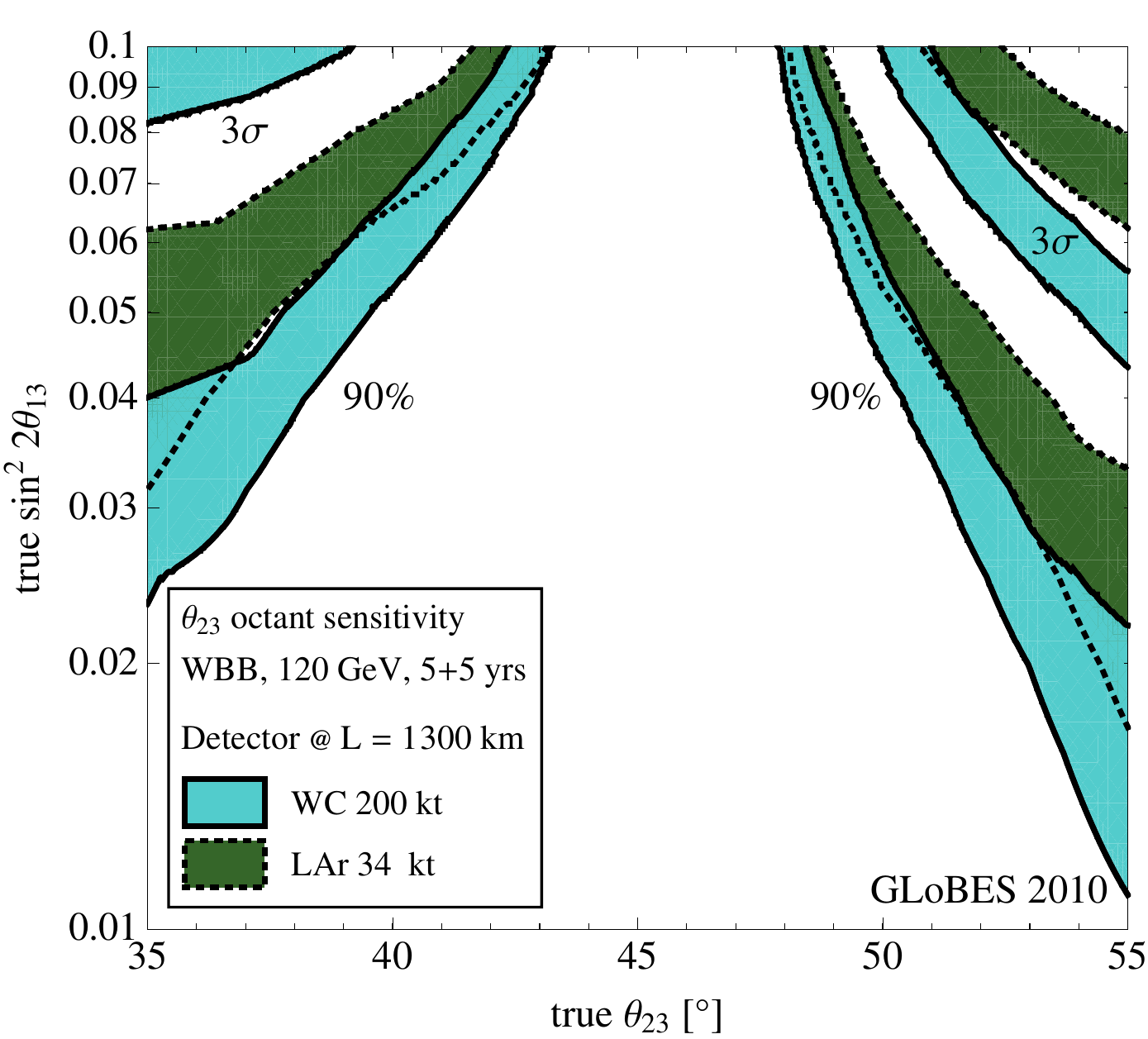}
  \caption{Sensitivity of LBNE to resolve the $\theta_{23}$ octant degeneracy
   for 5+5 years of $\nu$+$\nubar$ running at 700~kW assuming the ``August 2010''
   beam design (red curve in Fig.~\ref{fig:fig_beam_spectra1}) and normal
   mass hierarchy. The blue band shows the results for 200~kt WC and the
   green for 34~kt LAr. The width of the bands corresponds to the impact
   of different true values for $\delta_{\mathrm CP}$, ranging from a $10\%$ to
   $90\%$ fraction of $\delta_{\mathrm CP}$. In the region above the bands, the
   determination of the $\theta_{23}$ octant is possible at $90\%$ CL
   (lower bands) and $3\sigma$ (upper bands). Resolution of the octant
   degenerary is determined by using $\pi/2-\theta_{23}$ as a starting value
   and the user-defined priors in GLoBES to force the minimizer to remain in
   the wrong octant (i.e., this includes marginalization over $\theta_{23}$).}
  \label{fig:lbl_theta23_octant}
\end{figure}

Current experimental results tell us that $\sin^22\theta_{23}$ is near
maximal ($\sin^2\theta_{23}>0.91$ at $90\%$ CL~\cite{MINOS:Nu2010}), however
there exist two solutions of $\theta_{23}$ for a given set of measured
oscillation parameters, known as the $\theta_{23}$ octant ambiguity.
If the oscillation associated with $\numu$ disappearance is not maximal,
then it will be important to determine whether $\theta_{23}$ is greater
than or less than $\pi/4$. This in turn will help tell us whether the third
neutrino mass eigenstate couples more strongly to $\numu$ or $\nutau$.
Fig.~\ref{fig:lbl_theta23_octant} displays the capability of LBNE to resolve
the $\theta_{23}$ octant for both WC and LAr. Running in a 700~kW beam,
LBNE is able to resolve the $\theta_{23}$ octant degeneracy for $\theta_{23}$
values less than $40^\circ$ at $90\$$ CL and $90\%$ of $\delta_{\mathrm CP}$
values if $\sin^22\theta_{13}$ is greater than 0.070 for 200~kt WC and greater
than 0.075 for 34~kt LAr.

\subsection{$\nu_{\tau}$ Appearance}\label{lbl_nutau_appearance}

\begin{figure}[htb]
\centering\includegraphics[width=0.49\textwidth]{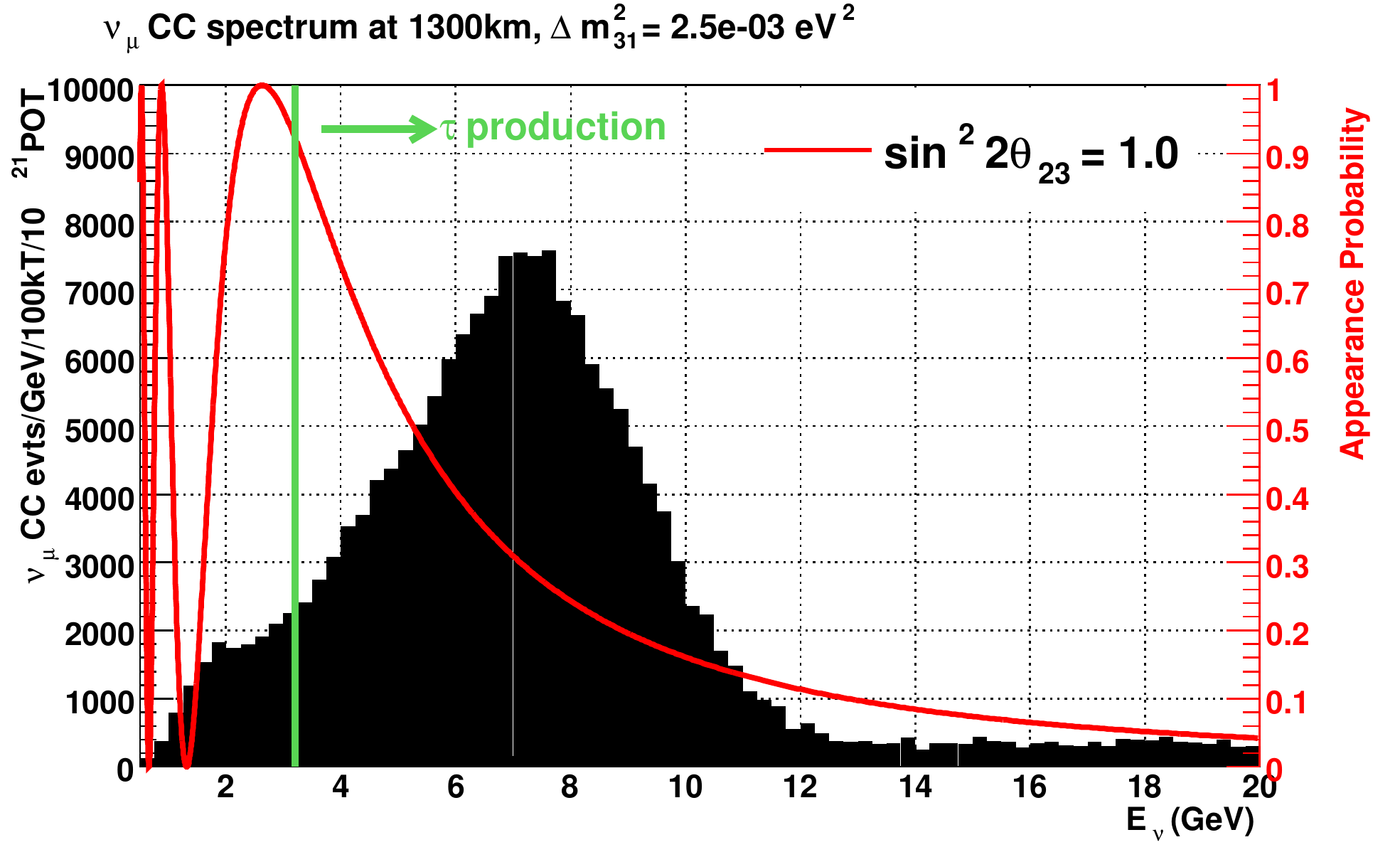}
\centering\includegraphics[width=0.49\textwidth]{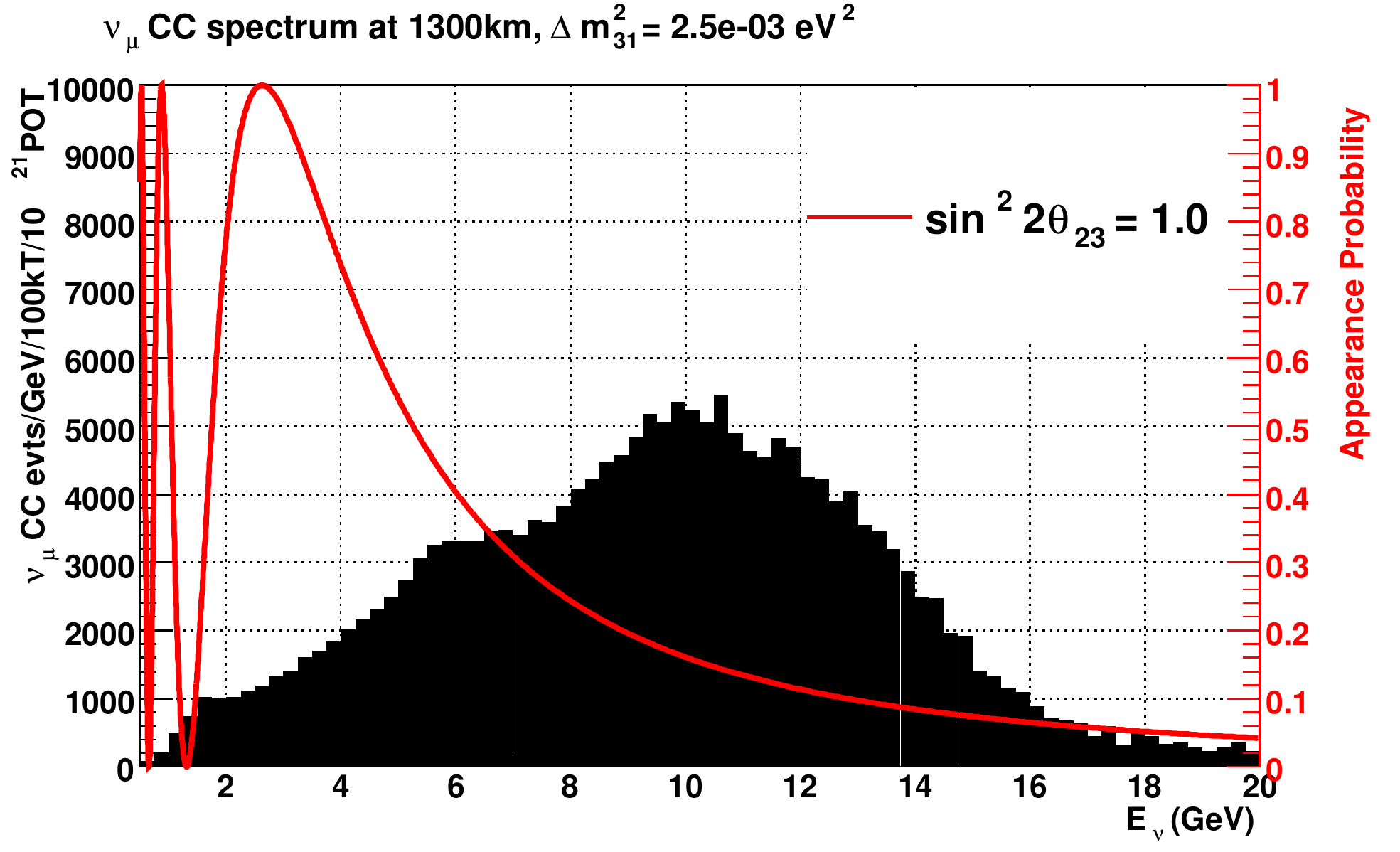}
\caption{The unoscillated $\numu$ CC spectra at DUSEL obtained by moving the target 1.5 m
  upstream of horn~1 - HE1 beam (left) and moving
  the target 2.5~m upstream of horn~2 (right) is shown as the solid
  black histogram. The $\nu_{\mu} \rightarrow \nu_{\tau}$
oscillation probability is overlaid as the red curve.}
\label{fig:lbl_beam_spectra_he}
\end{figure}

The LBNE baseline at 1300~km will be longer than any long baseline
experiment currently in operation. As a result, the oscillation
probability occurs at higher energy and in particular the energy range
is favorable to $\nu_{\mu} \rightarrow \nu_{\tau}$ appearance since
there is a large appearance probability above the $\tau$ CC production
threshold of 3.2~GeV. In this respect LBNE has a unique ability compared to
current long baseline experiments, since
oscillation between all three flavors of neutrinos can be observed in a
single experiment. To increase the $\nu_\tau$ CC appearance signal, we
are considering several high energy beam tunes produced by moving the
target further upstream of horn~1. An example of a high energy beam
spectrum produced by pulling the target back by 1.5~m is
shown in Fig.~\ref{fig:lbl_beam_spectra_he}.

\begin{figure}[htb]
\centering\includegraphics[width=0.32\textwidth]{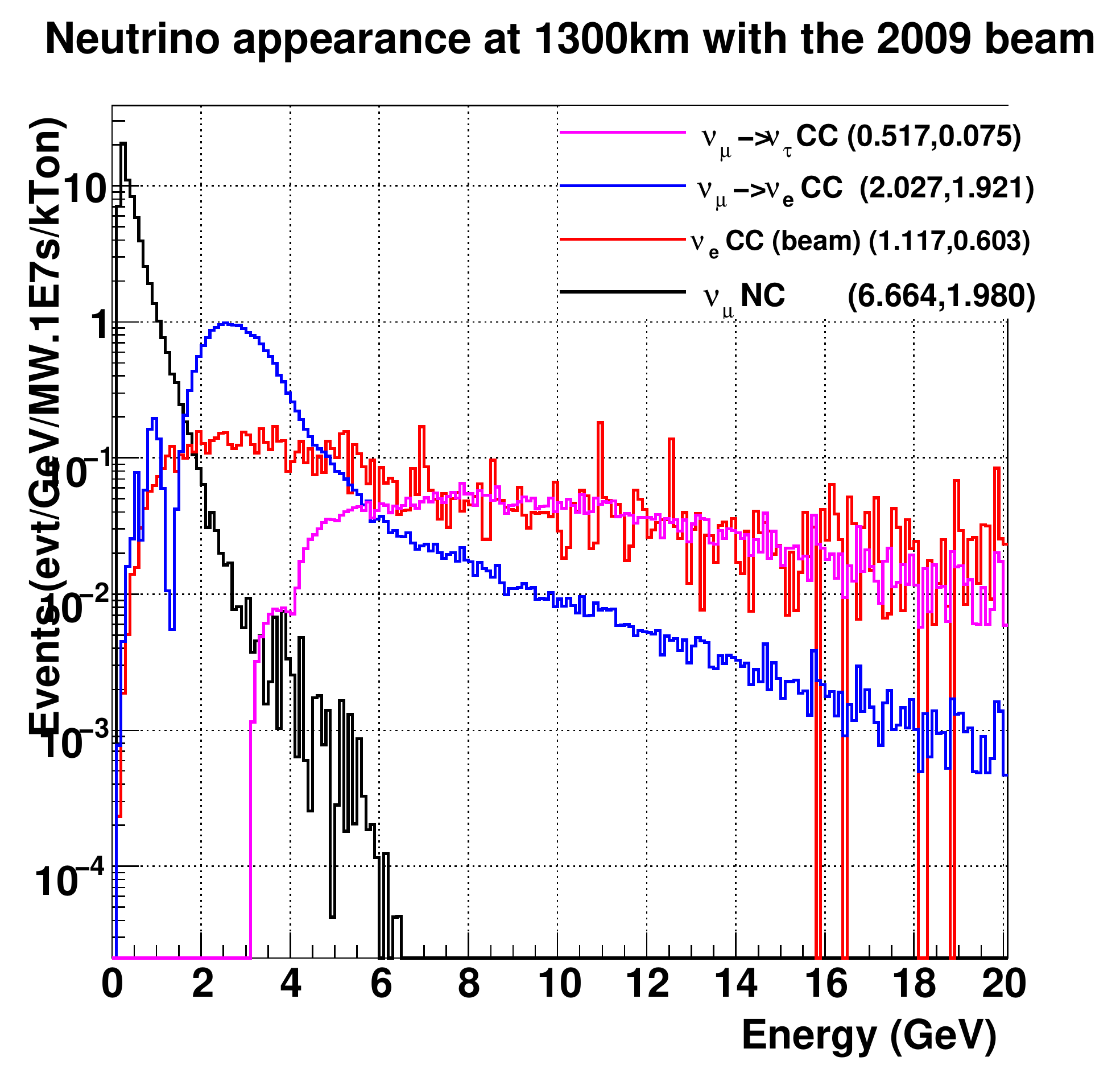}
\centering\includegraphics[width=0.32\textwidth]{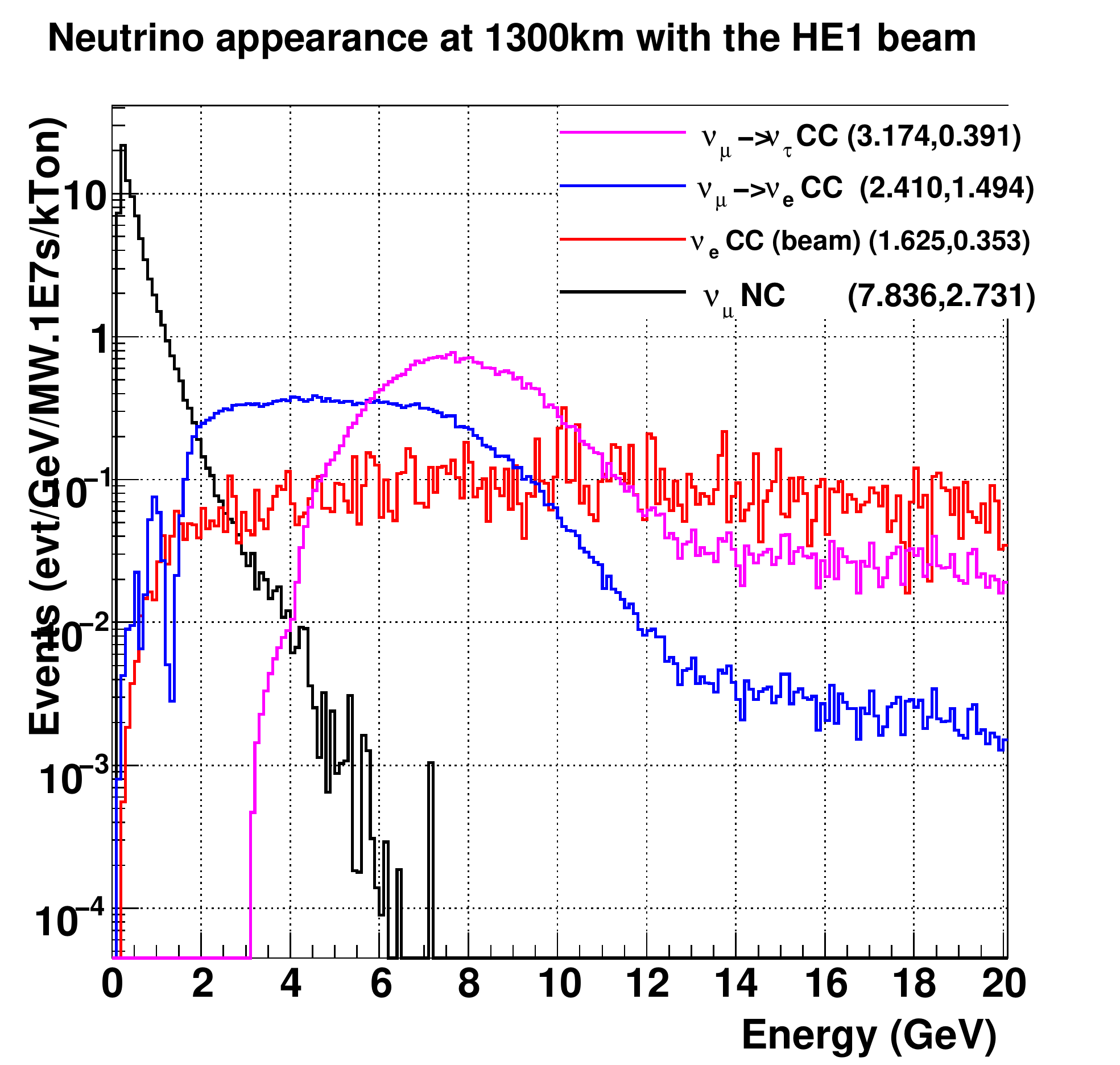}
\centering\includegraphics[width=0.32\textwidth]{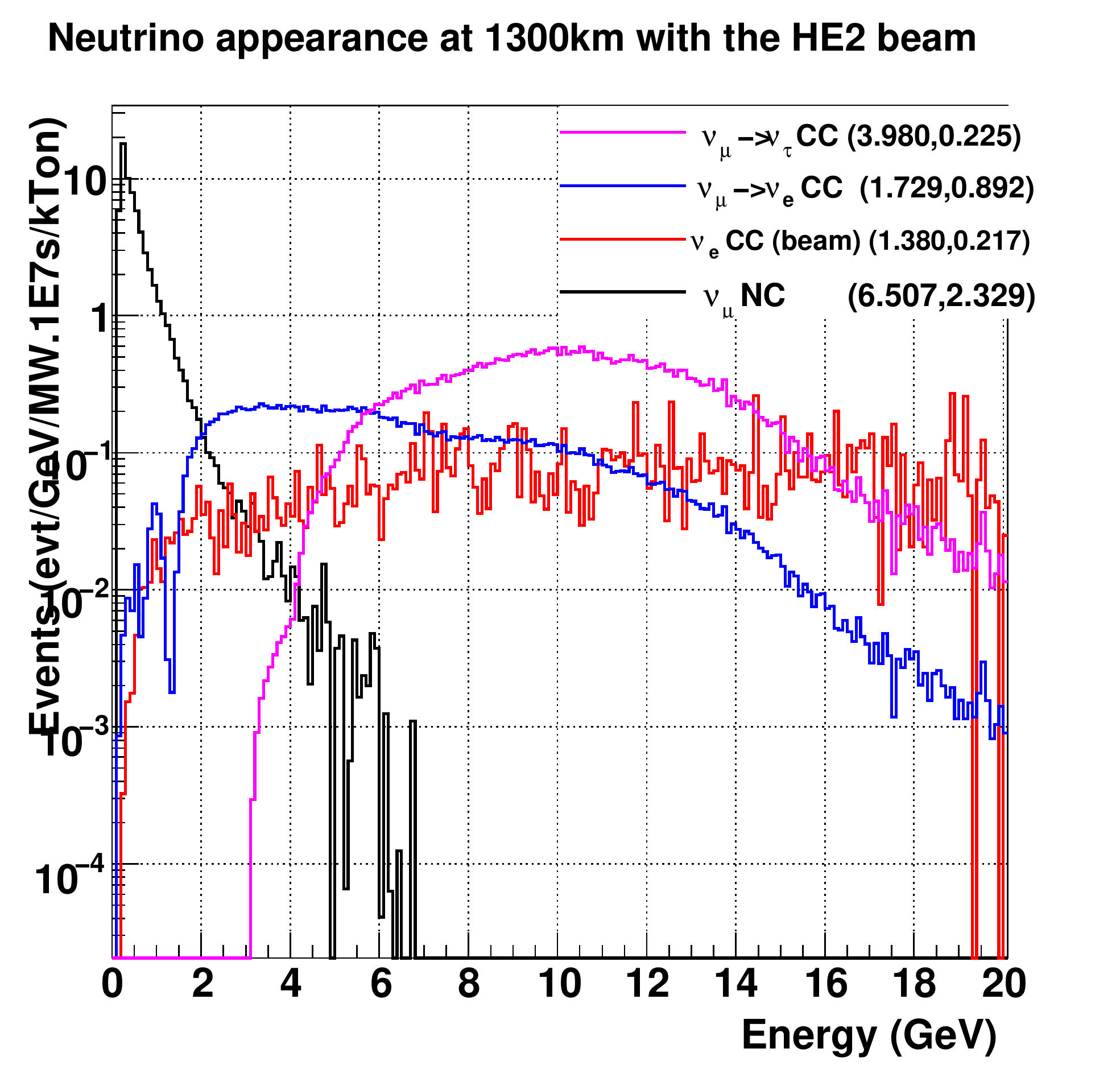}
\caption{Total $\nu_e$ and $\nu_{\tau}$ CC appearance rates at 1300~km  with normal
  hierarchy, $\Delta  m^2_{31} = 2.5 \times 10^{-3} {\rm eV}^2$, $\delta_{cp} = 0$ and $\sin^2 2\theta_{13} = 0.04$ with the
  LBNE 2009 beam with the target inserted fully into the 1st horn
  (left), the HE1 beam tune with the target -1.5~m from 1st horn (center) and
  the HE2 beam tune with the target -2.5~m from 1st horn (right).}
\label{fig:lbl_lbne_he1_appear}
\end{figure}

In Table~\ref{tab:lbl_nutau_rates}, the $\nue$ and $\nutau$ CC
appearance rates for several LBNE beam tunes are shown.  The first two
rows in Table~\ref{tab:lbl_nutau_rates} correspond to the 2009 and
``August 2010'' reference beams. The last two rows correspond to two
proposed high energy beam tunes produced by pulling the target back by
1.5~m and 2.5~m from horn~1. For LBNE, we will label these tunes as HE1 (-1.5~m) and
HE2 (-2.5~m); these tunes are very similar to the
NuMI/MINOS medium energy (ME) and high energy (HE) beam tunes respectively.

The spectrum of all $\nutau$ and $\nue$ CC events
appearing at the LBNE far detector as a function of neutrino energy
for the 2009, HE1 and HE2 beams are shown in Fig.~\ref{fig:lbl_lbne_he1_appear}. No detector effects are included. The
spectrum of NC events containing a single $\pi^0$ obtained from a
Nuance~\cite{nuance} simulation as a function of $\pi^0$ energy is also shown. In
addition, the spectrum of $\nue$ CC events from the beam contamination
is overlaid. The NC single $\pi^0$ integrated rate as well as the beam
$\nue$ CC and QE rates are shown in Table
~\ref{tab:lbl_nutau_rates}.

\begin{table}[!htb]
\begin{tabular}{l|ccccccc}
Target Position &
$\nu_{\mu}$ CC &
$\nu_{\mu}$ CC osc &
$\nu_e$ CC beam &
$\nu_e$ QE beam &
NC-$1\pi^0$ &
$\nu_{\mu} \rightarrow \nu_e$  CC &
$\nu_{\mu} \rightarrow \nu_\tau$ CC \\ \hline
    0 (2009 tune)& 206 & 78  & 2.2 & 0.26 & 4.0 & 4.0 & 1.0 \\
-0.3~m (Aug 2010 tune)& 290 & 108 & 2.6 & 0.28 & 5.2 & 5.6 & 1.4 \\
-1.5~m (HE1 tune) & 444 & 282 & 3.2 & 0.22 & 5.4 & 4.8 & 6.4 \\
-2.5~m (HE2 tune) & 466 & 350 & 2.8 & 0.16 & 4.6 & 3.4 & 8.0 \\
\end{tabular}
\caption{$\nu_{\mu}, \nu_{\tau}, \nu_e$ appearance rates per kT.MW.yr at the far
  detector in LBNE for
  different beam tunes obtained by moving the target w.r.t. horn~1. Normal hierarchy, $\sin^2 2 \theta_{13} = 0.04$. The rates are
  integrated in the region 0-20~GeV. The NC single $\pi^0$ rates
  are given for visible energies $>$0.5~GeV.}
\label{tab:lbl_nutau_rates}
\end{table}

The $\nutau$ signal in both water Cerenkov
and LAr can be observed as an excess of $e$-like events from leptonic
decays of the $\tau$ in $\nutau$ CC events where
$\tau \rightarrow e \nuebar \nutau (\gamma)$ with a total branching
fraction of 19.6\%~\cite{rpp2010}, the dominant
background will be $\nue$ CC events from $\numu \rightarrow \nue$
oscillations if $\sin^2 2 \theta_{13}$ is large, the beam
contamination and NC $\pi^0$ events. From Fig.~\ref{fig:lbl_lbne_he1_appear} and Table~\ref{tab:lbl_nutau_rates},
we can see that the total appearance rate of $\nutau$ CC events for both the HE1 and
HE2 beams is substantially larger than the irreducible beam $\nue$ CC
background. For the HE2 beam, the appearance signal is well separated from the
$\numu \rightarrow \nue$ CC signal as a function of true neutrino
energy. In addition, there is still a substantial $\nue$ appearance
signal in the HE1 beam, with a total CC appearance rate larger than the 2009 reference
beam. The actual signal and spectrum of $\nutau$ CC events in both
detector technologies is yet to evaluated and compared, but with a
higher energy beam tune the study outlined in this section
demonstrates that it is possible to obtain a large $\nutau$ appearance
signal rate, while also maintaining a significant portion of the
$\nue$ appearance signal.

\subsection{New Physics Searches in LBNE}\label{lbl_nsi}

In addition to precision measurements of the standard three-flavor neutrino
oscillation parameters, LBNE is also well-suited for new physics searches
in the neutrino sector. For example, the experiment is sensitive to
non-standard neutrino interactions and active-sterile neutrino mixing,
provided that these effects are not too weak. To illustrate the potential
of new physics searches, we will here focus on non-standard interactions
(NSI).

Theories beyond the Standard Model can induce Lagrangian operators that couple
neutrinos to normal matter in non-standard ways. In the low-energy effective
theory relevant to neutrino oscillation experiments, these non-standard
interactions manifest themselves as 4-fermion operators, either of the
charged-current (CC) type (e.g.\ $[\bar{\nu}_\alpha \gamma^\rho \ell_\beta] \,
[\bar{q} \gamma_\rho q]$, where $\ell_\beta$ is a charged lepton) or of the NC
type (e.g.\ $[\nu_\alpha \gamma^\rho \nu_\beta] \, [\bar{q} q]$). Most
(but not all) CC NSI are most easily seen in a near detector, while NC NSI
can be understood as non-standard matter effects that are visible only in a
far detector at a sufficiently long baseline. This is where LBNE has a unique
advantage compared to other long-baseline experiments (except atmospheric
neutrino experiments, which are, however, limited by systematic effects).
Therefore, and because there is no near detector definition for LBNE yet, we
will here focus on NC NSI.  They can be parameterized as new contributions to
the MSW matrix in the neutrino propagation Hamiltonian:
\begin{align}
  H &= U \begin{pmatrix}
           0 &                    & \\
             & \Delta m_{21}^2/2E & \\
             &                    & \Delta m_{31}^2/2E
         \end{pmatrix} U^\dag + \tilde{V}_{\rm MSW} \,,
\end{align}
with
\begin{align}
  \tilde{V}_{\rm MSW} = \sqrt{2} G_F N_e
  \begin{pmatrix}
    1 + \eps^m_{ee}       & \eps^m_{e\mu}       & \eps^m_{e\tau}  \\
        \eps^{m*}_{e\mu}  & \eps^m_{\mu\mu}     & \eps^m_{\mu\tau} \\
        \eps^{m*}_{e\tau} & \eps^{m*}_{\mu\tau} & \eps^m_{\tau\tau}
  \end{pmatrix} \,.
\end{align}
Here, $U$ is the leptonic mixing matrix, and the $\eps$-parameters give the
magnitude of the NSI relative to standard weak interactions.  For new physics
scales of $\text{few} \times 100$~GeV, we expect $|\eps| \lesssim 0.01$.
Model-independent bounds on NSI are typically of order
$0.01$--$1$~\cite{Davidson:2003ha, GonzalezGarcia:2007ib, Biggio:2009nt}.
However, in many concrete models, neutrino NSI are related to non-standard
effects in the charged lepton sector, which are much more strongly
constrained~\cite{Antusch:2008tz, Gavela:2008ra}. Thus, one could take the
point of view that NSI large enough to be experimentally accessible in the
foreseeable future are disfavored by theoretical arguments. On the other hand,
since theoretical prejudices about the properties of neutrinos have been wrong
in the past, once can also adopt the standpoint that our inability to
construct a simple model featuring large NSI does not necessarily mean they
cannot exist. Also, NSI provide one
explanation~\cite{Mann:2010jz,Akhmedov:2010vy,Kopp:2010} for the interesting
recent observations by the MINOS~\cite{MINOS:Nu2010} and
MiniBooNE~\cite{AguilarArevalo:2010wv} collaborations that could point to
a new source of CP violation, but are not significant enough yet to draw any
firm conclusions.

\begin{figure}
  \begin{center}
    \includegraphics[width=0.4\textwidth]{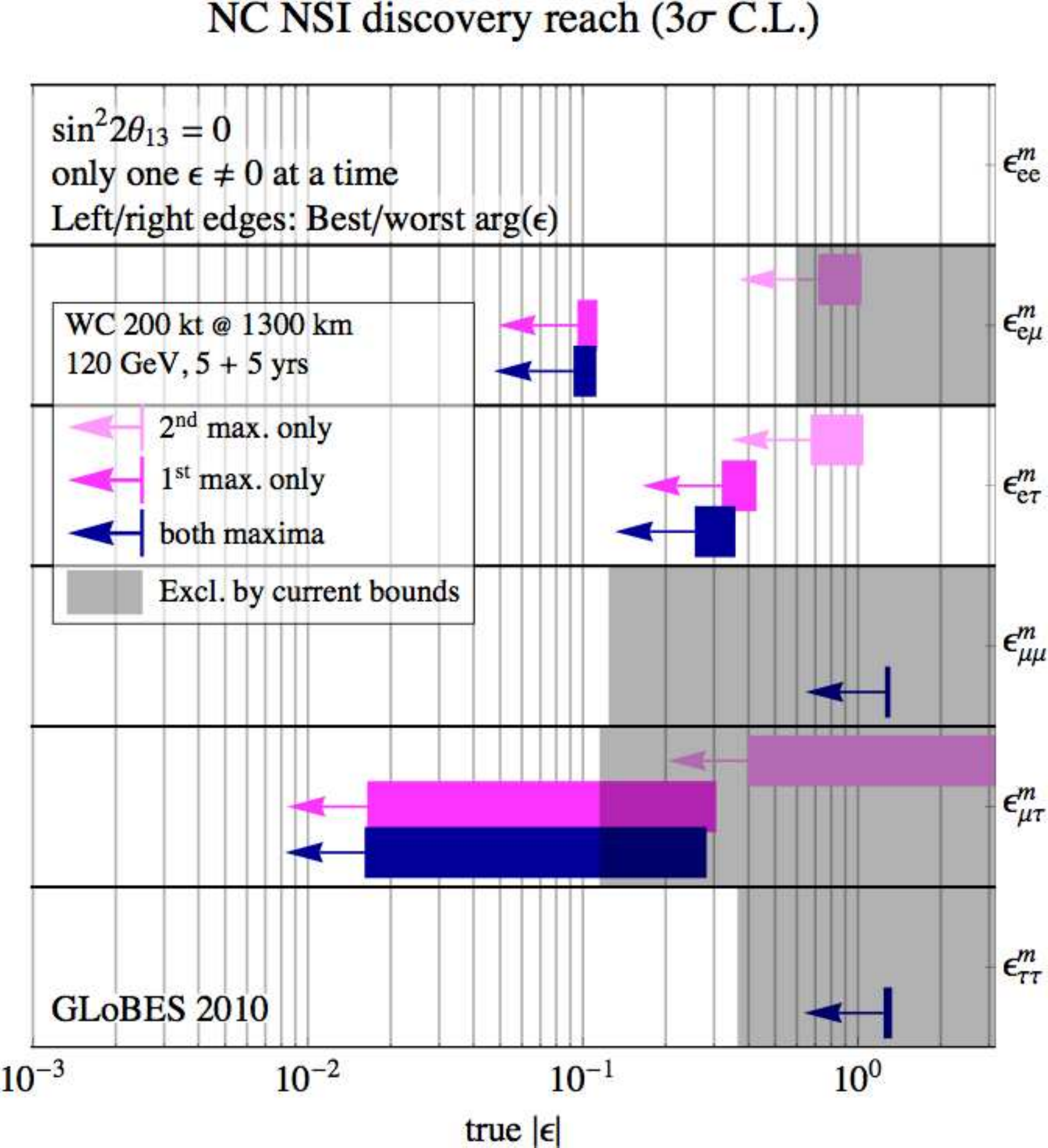} \hspace{0.5cm}
    \includegraphics[width=0.4\textwidth]{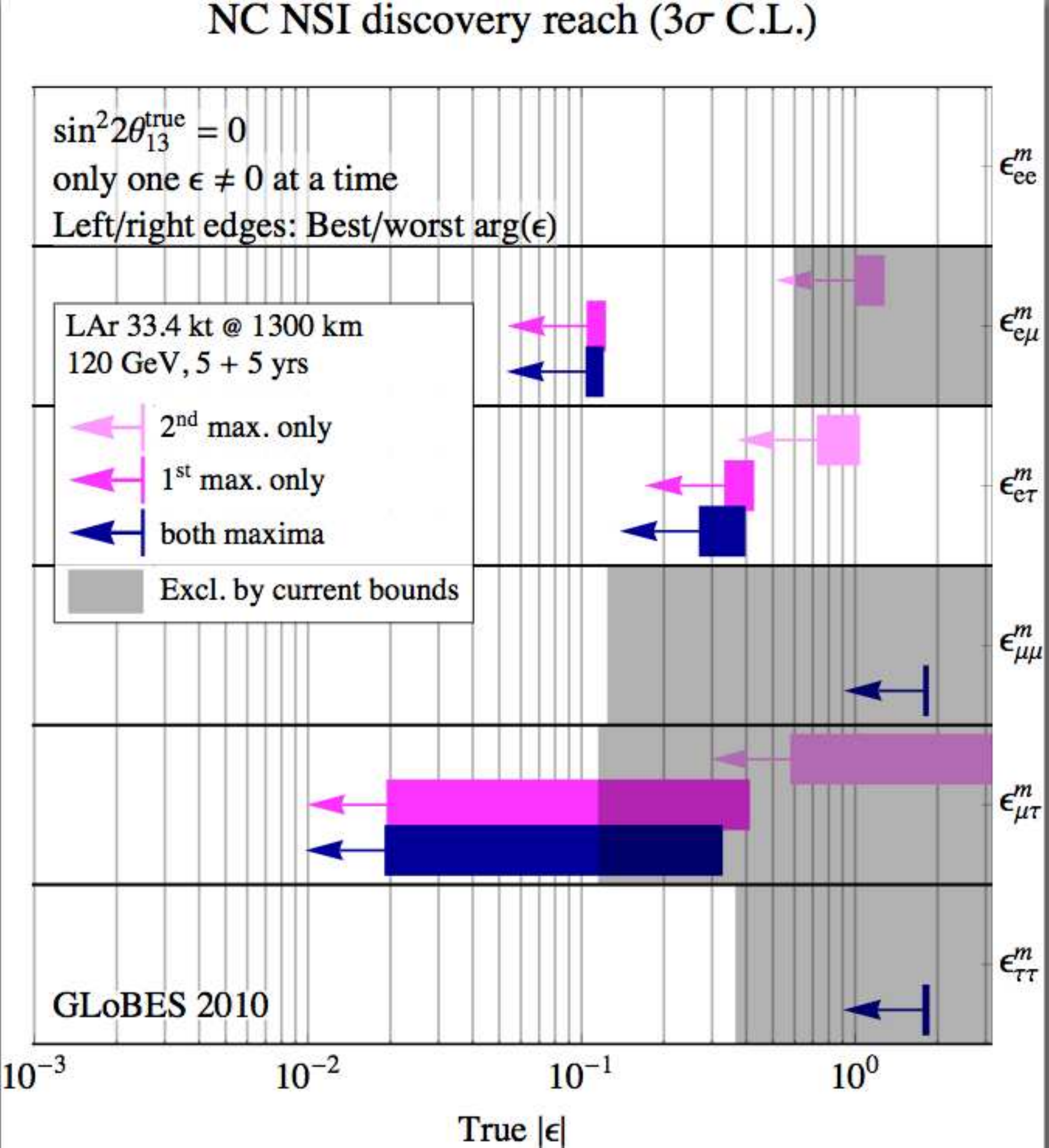}
  \end{center}
  \caption{NSI discovery reach in a WC detector (left) and a LAr detector
    (right). The left and right edges of the error bars correspond to the
    most favorable and the most unfavorable values for the complex phase
    of the respective NSI parameters.  The gray shaded regions indicate
    the current model-independent limits on the different
    parameters at 3 $\sigma$~\cite{Davidson:2003ha, GonzalezGarcia:2007ib,
    Biggio:2009nt}.  For some of them, limits are extremely weak.  Note that
    model-dependent limits can be several orders of magnitude stronger. In
    this plot, we have assumed only one NSI parameter to be non-negligible
    at a time.  Comparing the dark and light pink bars, we see that the NSI
    sensitivity comes mainly from the data in the first oscillation maximum;
    the second one is too polluted by backgrounds to contribute appreciably.
    The performance of the WC and LAr detectors is very similar.}
  \label{fig:NSI-WC}
\end{figure}

To assess the sensitivity of LBNE to NC NSI, we define the NSI discovery reach
in the following way: We simulate the expected event spectra, assuming given
``true'' values for the NSI parameters, and then attempt a fit assuming no
NSI. If the fit is incompatible with the simulated data at a given confidence
level, we say that the chosen ``true'' values of the NSI parameters are
within the experimental discovery reach.  In Fig.~\ref{fig:NSI-WC}, we show
the NSI discovery reach of LBNE for the case where only one of the
$\eps^m_{\alpha\beta}$ parameters is non-negligible at a time.

We conclude from the figure that LBNE will be able to improve
model-independent bounds on NSI in the $e$--$\mu$ sector by a factor of two,
and in the $e$--$\tau$ sectors by an order of magnitude. Bounds on
non-standard effects in the $\mu$--$\tau$ sector are already quite strong
because of the sensitivity of atmospheric neutrino experiments, but LBNE may
be able to improve also some of the bounds in this sector, and in any case,
LBNE bounds will be more robust than the ones derived from atmospheric
neutrino oscillations.  In particular, it has been shown
in~\cite{Friedland:2005vy} that atmospheric neutrino bounds
can become significantly weaker if the possibility of several NSI being
non-zero is taken into account. Since LBNE provides a precise measurement of
both the $\nue$ appearance and $\numu$ disappearance channels and can be
operated in neutrino and antineutrino mode, we expect it to be more robust
with respect to these correlations.

The sensitivity of LBNE to non-unitarity effects in the leptonic mixing
matrix (which can be recast into a special case of non-standard interactions)
(see e.g.~\cite{Antusch:2008tz}) is not expected to be competitive with existing
limits, which are already of order $10^{-3}$. Searches for short-baseline
oscillations into sterile neutrinos, on the other hand, would certainly be
a worthwhile effort at a near detector. While near detector searches
for oscillations between $\nu_e$ and $\nu_\mu$ will be limited by systematic
uncertainties in our knowledge of the beam spectrum and flavor composition,
oscillations into $\nu_\tau$ would provide a very clean signature. However,
this would require the construction of a dedicated near detector that is
able to identify $\nu_\tau$. It remains to be seen by how much such a detector
could improve existing bounds from the NOMAD experiment~\cite{Astier:2001yj}.

\subsection{Next Steps}
While there has been much progress in the past year, we list here some of
the known issues requiring further study.

\begin{itemize} [parsep=-1pt]
  \item For the $\nue$ appearance estimates:
        \begin{itemize}[parsep=-1pt]
          \item What is the impact of improved WC reconstruction
                (similar to what has been accomplished for T2K and MiniBooNE)
                on our assumed selection efficiencies when applied at
                LBNE energies?
          \item What is the impact of improved LAr reconstruction
                on our assumed selection efficiencies when applied at
                LBNE energies? In particular, we need to evaulate energy
                dependent signal efficiencies and background rejection levels
                along with improved estimates of energy resolutions
                in LAr.
          \item We need to modify GLoBES to include a more realistic
                estimate of background uncertainties that are both process
                and energy dependent.
        \end{itemize}
  \item For the $\numu$ disappearance estimates:
        \begin{itemize}[parsep=-1pt]
           \item We need to improve the energy smearing for CC $\pi^+$
                 background events for WC (the current
                 smearing seems too symmetrical).
           \item We need to examine whether a $97\%$ $\numu$ QE signal
                 efficiency is the correct number for WC. Our guess is that
                 this number should be lower.
           \item We need to assess a more realistic estimate of the neutrino
                 energy resolution for $\numu$ CC events in LAr. We are
                 presently using an estimate from ICARUS evaluated for
                 neutrino energies $<1.25$~GeV.
           \item We need to study whether using a QE or CC sample optimally
                 gives better sensitivity to $\sin^22\theta_{23}$ and
                 $\Delta m^2_{32}$ for each detector type, WC and LAr.
        \end{itemize}
  \item Can additional gains be made on our long-baseline oscillation
        sensitivities if we are able to separate $\nu$ from $\nubar$ events
        in a LAr far detector? For now, we have not accounted for
        potential $\nu$/$\nubar$ separation capabilities in LAr.
  \item What is the impact of photocathode coverage on our long-baseline
        oscillation sensitivities in the case of WC? For this, we would need
        revised efficiencies and energy resolutions for different PMT
        coverages.
  \item What is the impact of different likelihood cut placements? Is there
        a more optimized selection possible for the LBNE beam configuration?
\end{itemize}

\subsection{Conclusions}\label{lbl_conclusions}

The LBNE project, as currently defined, comprises two 100~kt water Cerenkov-equivalent detectors.
Since the $\mutoe$ signal is dominated by statistical uncertainties, we chose the mass of the LAr detector to give equivalent signal statistics in both.
As described in Appendix~\ref{lbl_appendix}, the energy dependent WC $\nue$ signal efficiency used in this study is based on Super-K analysis and for LAr we have used a constant signal efficiency of 85\%.
Based on these assumptions, a fiducial mass of $\sim17$~kt LAr TPC produces similar sensitivity for oscillation parameter measurements as a 100~kt WC fiducial mass. The mass ratio of 1/6 is largely a reflection of the ratio of the WC and LAr signal efficiencies (15\% and 85\%, respectively) at the energy of the first $\numu$ oscillation maxima for a 1300~km baseline and $\Delta m_{31}^2 = 2.5 \times 10^{-3} {\rm eV}^2$ ($\sim$2.4~GeV).

Published studies for the T2KK long baseline experiment~\cite{dufour:2010}, carried out using Super-K simulation and data, indicate that a factor of two increase in the WC $\nue$ signal efficiency may be
possible, albeit with a significant increase in NC backgrounds.
For both the LBNE and T2KK studies, the  WC detector $\nue$ reconstruction
efficiencies are based on searching for signal events with a single
electron like ring. The LBNE $\nue$ appearance signal is at neutrino
energies with significant contributions from deep inelastic interactions
with multiple particles in the final state. So future increases in
the WC $\nue$ signal efficiency may be possible when multi-ring signal
events with pions are included.
The net effect of increased efficiency and backgrounds on oscillation
parameter sensitivities is being investigated will be presented in a future document.

We conclude this section with a series of tables to summarize the results of the chapter.
Table~\ref{table:lbl_conclusions_theta13_sensitivity} shows the LBNE sensitivity to non-zero
$\sin^22\theta_{13}$ for different exposures, including a 10 year operation of the reference configuration (three `100-kt equivalent' LBNE modules), the CDR `200~kt,' and a single `100~kt' module. We see that the reference configurations reach a sensitivity almost a factor of two higher than a single module; higher beam power would reduce the elapsed time to reach the same sensitivity. Table~\ref{table:lbl_conclusions_cp_1sigma} provides examples of the $1\sigma$ resolution on the CP phase for these configurations; in Fig.~\ref{fig:lbl_delta_resolution_exposure} we see that higher mass (or beam power) provides a rapid improvement in resolution in the early years of running.

Tables ~\ref{table:lbl_conclusions_sensitivity},
~\ref{table:lbl_conclusions_resolution_app},
~\ref{table:lbl_conclusions_resolution_disapp} and
~\ref{table:lbl_conclusions_octant} summarize the sensitivity of the 200~kT
WC and 34~kT LAr detectors to $\sin^22\theta_{13}\neq0$, the mass
hierarchy, CP violation, resolution of $\sin^22\theta_{23}$ ($\nu$ and
$\nubar$), resolution of $\Delta m^2_{32}$ ($\nu$ and
$\nubar$), and the resolution of the $\theta_{23}$ octant. We assumed
a running time of 5 years in $\nu$ mode and 5 years in $\nubar$
mode at 700~kW. We find that given the current assumptions on detector
performance, both the water Cerenkov and liquid argon TPC technologies
have similar physics sensitivities to the $\numu$ oscillation
parameters.
In the case of mixed technology Far Detector configurations, we assume that the results from each of the detector modules can be combined with no loss of precision compared to single technology configurations.

\clearpage

\begin{table}
\begin{center}
\begin{tabular}{cc|cc}
           WC  & $\sin^22\theta_{13}\neq0$ & LAr & $\sin^22\theta_{13}\neq0$\\
\hline
1000~kt-yrs& 0.0050  & 170~kt-yrs& 0.0044  \\
2000~kt-yrs& 0.0033  & 340~kt-yrs& 0.0030  \\
3000~kt-yrs& 0.0026  & 510~kt-yrs& 0.0023  \\
\end{tabular}
\caption{Sensitivity of LBNE to non-zero $\sin^22\theta_{13}$ to $3\sigma$ significance after 5+5 years $\nu+\nubar$ running at 700~kW for one, two, three ``100-kt equivalent'' LBNE modules if $\delta_{CP}=0$ and normal mass hierarchy (see Fig.~\ref{fig:lbl_theta13_exposure_3sig}).}
\label{table:lbl_conclusions_theta13_sensitivity}
\end{center}
\end{table}

\begin{table}
\begin{center}
\begin{tabular}{cc|cc}
           WC  & $\delta_{\mathrm CP}$ & LAr & $\delta_{\mathrm CP}$\\
\hline
1000~kt-yrs& 25$^\circ$  & 170~kt-yrs& 25$^\circ$  \\
2000~kt-yrs& 19$^\circ$  & 340~kt-yrs& 18$^\circ$  \\
3000~kt-yrs& 16$^\circ$  & 510~kt-yrs& 15$^\circ$  \\
\end{tabular}
\caption{$1\sigma$ resolution on the measurement of $\delta_{\mathrm CP}$ for one, two, three ``100-kt equivalent'' LBNE modules assuming $\sin^22\theta_{13}=0.01$, $\delta_{CP}=0$ and normal mass hierarchy (see Fig.~\ref{fig:lbl_delta_resolution_exposure}).}
\label{table:lbl_conclusions_cp_1sigma}
\end{center}
\end{table}

\begin{table}
\begin{center}
\begin{tabular}{c|ccc}
        & $\sin^22\theta_{13}\neq0$  & Mass Hierarchy & CP violation \\
\hline
2000~kt-yrs WC & 0.008 & 0.04 & 0.03 \\
340~kt-yrs LAr & 0.008 & 0.05 & 0.03 \\
\end{tabular}
\caption{Sensitivity comparisons for an exposure of 2000~kt-yrs of WC (e.g.,
200~kt WC, 5+5 years $\nu+\nubar$ running at 700~kW) and 340~kt-yrs of LAr
(e.g., 34~kt LAr, 5+5 years $\nu+\nubar$ running at 700~kW).  These numbers
represent the value of $\sin^22\theta_{13}$ where a $3\sigma$ determination
of $\sin^22\theta_{13}\neq0$, the sign of $\Delta m^2_{31}$, and CP violation
can be made for $100\%$ of the possible values of $\delta_{\mathrm CP}$. For CP
violation, the value is quoted for $50\%$ of possible $\delta_{\mathrm CP}$ values.
These numbers were calculated assuming a normal mass hierarchy.}
\label{table:lbl_conclusions_sensitivity}
\end{center}
\end{table}

\begin{table}
\begin{center}
\begin{tabular}{c|cc}
        & $\sin^22\theta_{13}$ & $\delta_{\mathrm CP}$
\\ \hline
2000~kt-yrs WC & 0.002 & $19^\circ$ \\
340~kt-yrs LAr & 0.002 & $18^\circ$ \\
\end{tabular}
\caption{$1\sigma$ resolution on the measurement of $\nue$ appearance
parameters in LBNE assuming an exposure of 2000~kt-yrs
for WC (e.g., 200~kt WC, 5+5 years $\nu+\nubar$ running at 700~kW) and 340~kt-yrs for LAr
(e.g., 34~kt LAr, 5+5 years $\nu+\nubar$ running at 700~kW).
Values are quoted assuming a normal mass hierarchy, $\sin^22\theta_{13}=0.01$,
and $\delta_{CP}=0$.}
\label{table:lbl_conclusions_resolution_app}
\end{center}
\end{table}

\begin{table}
\begin{center}
\begin{tabular}{c|cc|cc}
        & $\delta (\sin^22\theta_{23})$ ($\nu$)
        & $\delta (\Delta m^2_{32})$ ($\nu$)
        & $\delta (\sin^22\theta_{23})$ ($\nubar$)
        & $\delta (\Delta m^2_{32})$ ($\nubar$)
\\ \hline
1000~kt-yrs WC & 0.007 & 0.015 & 0.008 & 0.020  \\
170~kt-yrs LAr & 0.007 & 0.017 & 0.011 & 0.025 \\
\end{tabular}
\caption{$1\sigma$ resolution on the measurement of $\numu$ (and $\numubar$)
disappearance parameters in LBNE assuming an exposure of 1000~kt-yrs
for WC (e.g., 200~kt WC, 5 years $\nu$ (or $\nubar$) running at 700~kW)
and 170~kt-yrs for LAr (e.g., 34~kt LAr, 5 years $\nu$ (or $\nubar$) running
at 700~kW). Values are quoted for $\sin^22\theta_{23}=1.0$.}
\label{table:lbl_conclusions_resolution_disapp}
\end{center}
\end{table}

\begin{table}
\begin{center}
\begin{tabular}{c|cc}
 & $\theta_{23}$  & $\sin^22\theta_{13}$ \\ \hline
 2000~kt-yrs WC & $<40^\circ$ & $>0.070$ \\
 340~kt-yrs LAr & $<40^\circ$ & $>0.075$ \\
\end{tabular}
\caption{LBNE can resolve the $\theta_{23}$ octant at $90\%$ CL for values
of $\theta_{23}<40^\circ$ given $\sin^22\theta_{13}$ greater than the indicated
values assuming an exposure of 2000~kt-yrs for WC (e.g., 200~kt WC,
5+5 years $\nu+\nubar$ running at 700~kW) and 340~kt-yrs for LAr (e.g.,
34~kt LAr, 5+5 years $\nu+\nubar$ running at 700~kW). Numbers are quoted
for a normal mass hierarchy and $90\%$ of $\delta_{\mathrm CP}$ values.}
\label{table:lbl_conclusions_octant}
\end{center}
\end{table}

\clearpage
\vfill\eject

\section{Proton Decay}\label{PDK}

Proton decay, bound neutron decay, and similar processes such as
dinucleon decay and neutron-antineutron oscillation test the apparent
but unexplained conservation law of baryon number. These decays are
already known to be rare based on decades of prior searches, all of
which have been negative. If measurable event rates or even single
candidate events are found, one immediately concludes that they must
have proceeded via unknown virtual processes based on physics beyond
the standard model. The impact of demonstrating the existence of a
baryon number violating process would be profound.

The class of theories known as Grand Unified Theories (GUTs) make
predictions about baryon number violation and the life of the proton
that may be within reach of the LBNE detectors. Early GUTs were the
original motivation for putting kiloton-scale detectors
underground. The 22.5 kiloton Super-Kamiokande experiment extended the
search for proton decay by more than an order of magnitude. Although
there has been no sign of proton decay, the strict limits from these
experiments constrain the construction of contemporary GUTs and
indeed, a tension between experiment and theory is now commonly
discussed. It is very natural to continue the search with
100-kiloton-scale detectors.

\subsection{Motivation and Scientific Impact of Future Measurements}

The grand unified theoretical motivation for the study of proton decay
has a long and distinguished
history~\cite{Pati:1973rp,Georgi:1974sy,Dimopoulos:1981dw}, and has
been reviewed many
times~\cite{Langacker:1980js,deBoer:1994dg,Nath:2006ut}. Contemporary
reviews~\cite{Raby:2008pd,Senjanovic:2009kr,Li:2010dp} discuss the
strict limits already set by Super-Kamiokande and the context of
proposed multi-100-kiloton scale experiments such as Hyper-Kamiokande and
LBNE. Here are some of the key points related to scientific impact:

\begin{itemize}
\item Conservation of baryon number is unexplained, corresponding to
no known long-range force.
\item Baryon number non-conservation has cosmological consequences,
  such as a role in inflation and the baryon asymmetry of the
  universe.
\item Proton decay is predicted by a wide range of GUTs.
\item Grand unified theories are also often able to accommodate
  massive neutrinos with characteristics as discovered over the last
  decade.
\item GUTs incorporate other unexplained features of
  the standard model such as the relationship of quark and lepton
  electric charges.
\item The unification scale is suggested experimentally and
  theoretically by the apparent convergence of the running coupling
  constants of the Standard Model. It is in excess of $10^{15}$~GeV.
\item The unification scale is not accessible by any accelerator
  experiment, and can only be probed by virtual processes such a
  proton decay.
\item GUTs usually predict the relative branching
  fractions of different nucleon decay modes, requiring of course
  a sizeable sample of proton decay events to test.
\item The dominant proton decay mode is often sufficient to roughly
  identify the likely characteristics of the GUT,
  such as gauge mediation or the involvement of supersymmetry.
\end{itemize}

\noindent In summary, the observation of even a single unambiguous
proton decay event would signal that the ideas of grand unification
are correct and would give guidance as to which models most likely
describe the universe. One or two events would also give guidance to
what even larger size detector would be needed to explore the physics
in more detail.

From the body of literature, two decay modes emerge that dominate our
experimental design. First, there is the decay mode of $p \rightarrow
e^+ \pi^0$ that arises from gauge mediation. This is the most famous
proton decay mode, often predicted to have the highest branching
fraction, and also demonstrably the most straightforward experimental
signature for a water Cherenkov detector. The total mass of the proton
is converted into the electromagnetic shower energy of the positron
and the two photons from $\pi^0$ decay, with a net momentum vector
near zero.

The second key mode is $p \rightarrow K^+ \nu$. This mode is dominant
in most supersymmetric-GUTs, which also often favor several other
modes involving kaons in the final state. The decay mode with a
charged kaon is notable because it presents the unique opportunity for
a liquid argon TPC to detect it with extremely high efficiency. This
is because the momentum of the kaon will result in high ionization
density which can be compared to the range of the kaon, not to mention
the unique final states of $K^+$ decay that should be fully
reconstructed.

There are a number of other proton decay channels to consider, but
they will not influence the design of a next-generation experiment
beyond the above decay modes.  There are 27 allowed modes of proton or
bound neutron into anti-lepton plus meson (conserving $B-L$). The most
stringent limits besides $p \rightarrow e^+ \pi^0$ include $p
\rightarrow \mu^+ \pi^0$ and $p \rightarrow e^+ \eta$, both of which
must have partial lifetimes greater than $4 \times 10^{33}$ years. Any
experiment that will do well for $e^+ \pi^0$ will do well for these
decay modes. The decay $p \rightarrow \nu \pi^+$ or $n \rightarrow \nu
\pi^0$ may have large theoretically predicted branching fractions but
are experimentally difficult due to sizeable backgrounds from
atmospheric neutrino interactions. The decay $p \rightarrow \mu^+ K^0$
is detected relatively efficiently by either water Cherenkov or LAr
TPC detectors. There are a number of other possibilities such as modes
that conserve $B+L$, or violate only baryon number, or that decay into
only leptons. These possibilities are less well-motivated
theoretically, as they do not appear as frequently in the
literature. In any case, they can be accommodated with equal ease or
difficulty by the large detectors considered here.

\begin{figure}[t]
\centering
\includegraphics[width=0.9\textwidth]
{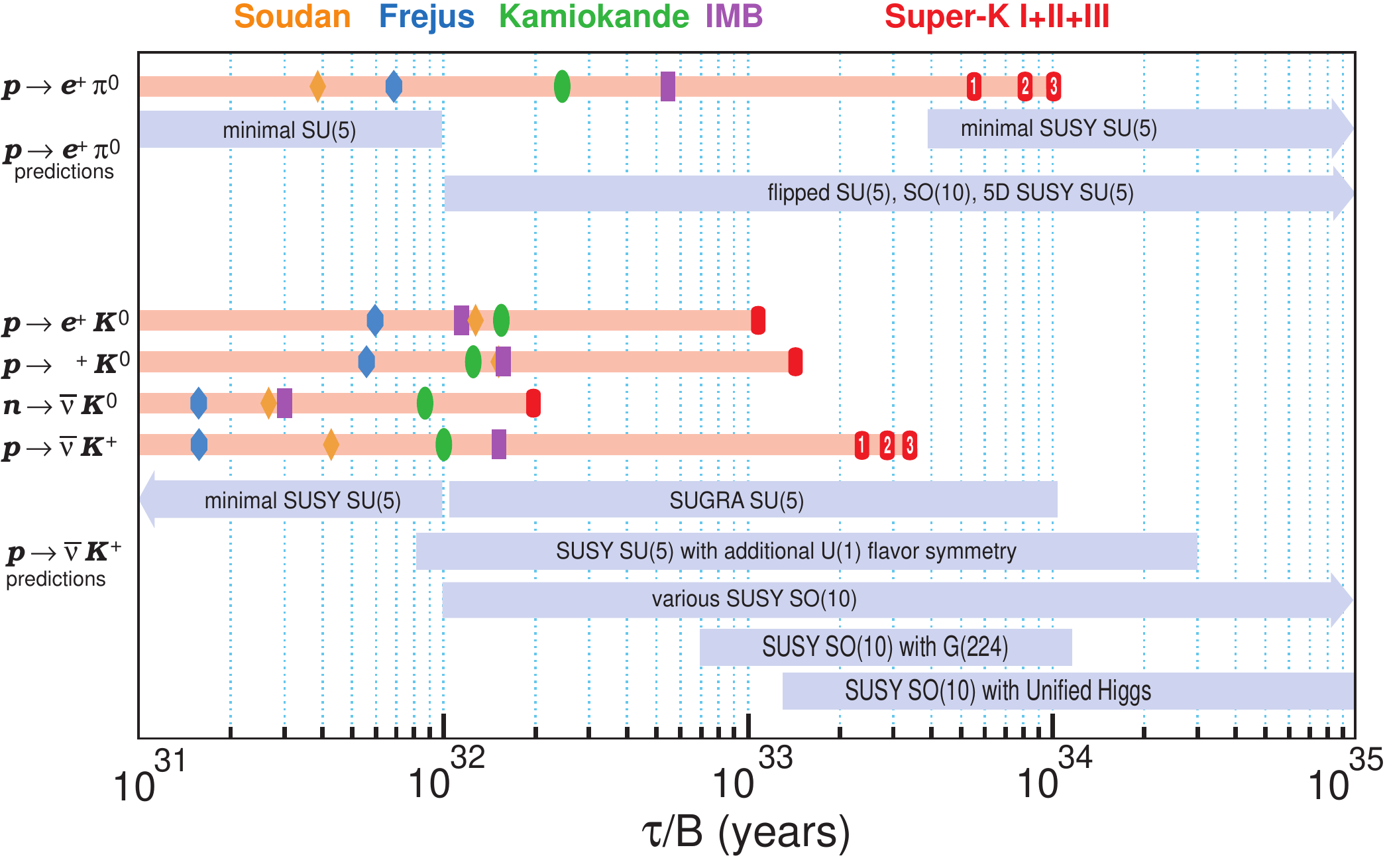}
\caption{Proton decay lifetime limits compared to lifetime ranges
  predicted by Grand Unified Theories. The upper section is for
  $p \rightarrow e^+ \pi^0$, most commonly caused by gauge mediation.
  The lower section is for SUSY motivated models, which commonly
  predict decay modes with kaons in the final state. The
  marker symbols indicate published limits by experiments, as indicated
  by the sequence and colors on top of the figure.}
\label{PDK-limits-theory}
\end{figure}

Fig.~\ref{PDK-limits-theory} shows experimental limits, dominated by
recent results from Super-Kamiokande, compared to the ranges of
lifetimes predicted by an assortment of GUTs. At this time, the theory
literature does not attempt to precisely predict lifetimes,
concentrating instead on suggesting the dominant decay modes and
relative branching fractions. The uncertainty in the lifetime
predictions come from details of the theory, such as unknown heavy
particles masses and coupling constants, as well as poorly known
details of matrix elements for quarks within the nucleon.

It is apparent from this figure that a continued search for proton
decay is by no means assured of success. In addition to the lifetime
ranges shown, there are models that predict essentially no proton
decay or lifetimes out of reach of likely experiments. With that
caveat, an experiment with sensitivity between $10^{33}$ and $10^{35}$
years is searching in the right territory over a wide range of
GUTs and even if no proton decay is detected, the stringent lifetime
limits will restrict efforts to build such theories.  Minimal
SU(5) was ruled out by the early work of IMB and Kamiokande; minimal
SUSY~SU(5) is considered to be ruled out by Super-Kamiokande.  In most
cases, another order of magnitude in limit will not rule out specific
theories, but will constrain their allowed parameters, perhaps leading
to the conclusion that some are fine-tuned.

\subsection{Sensitivity of Reference Configurations}

The experimental requirements of the search for proton decay can be
found in the basic formula for the partial lifetime $\tau$ for
branching fraction $B$:
\begin{equation}
\frac{\tau}{B} = \frac{N_0~\Delta t~\epsilon}{n_{obs}-n_{bg}},
\label{PDK:lifetime}
\end{equation}
\noindent where $N_0$ is the number of nucleons exposed, $\Delta t$ is
the exposure time, $\epsilon$ is the detection efficiency, $n_{obs}$
is the observed number of events, and $n_{bg}$ is the estimated number
of background events. To measure $\tau / B$, one would like the
numerator to be as large as possible, which calls for the largest
possible exposure of nucleons as well as the highest possible
efficiency.

The sensitivity for a detector configuration is determined by the
detector mass, efficiency, expected background, and running time,
following Eq.~\ref{PDK:lifetime}. For the purpose of generating
sensitivity curves, we calculate the 90\% C.L. lifetime limit one
would publish after a given exposure under the assumption that the
number of detected events exactly equals the number of background
events, and the background is subtracted. The efficiency and
background estimates are drawn from Table~\ref{PDK-effic-bg}.

The lifetime limit is calculated for the 90\% confidence level based
on the {\it Poisson processes with background} method from the 1996
Review of Particle Properties~\cite{Barnett:1996hr}. This method does
not take into account systematic uncertainty; doing so typically
weakens these limits by 20\%.

\begin{table} [ht]
\begin{center}
\begin{tabular}{|c|c|c|c|c|}
  \hline
  & \multicolumn{2}{c|}{Water Cherenkov} & \multicolumn{2}{c|}{Liquid Argon} \\
  Mode & Efficiency & Background Rate (evts/100~kt-y)
       & Efficiency & Background Rate (evts/100~kt-y)\\ \hline
  $p \rightarrow e^+ \pi^0$ & 45\% $\pm$ 19\%
                            & 0.2($\pm$40\%)
                            & 45\% & 0.1 \\
  $p \rightarrow \nu K^+$ & 13.4\% $\pm$ 22\%
                          & 0.67($\pm$30\%) (SK1)
                          & 97\% & 0.1 \\
  $p \rightarrow \nu K^+$ & 10.6\% $\pm$ 22\%
                          & 0.83($\pm$30\%) (SK2) &  &  \\
  \hline
\end{tabular}
\caption{\label{PDK-effic-bg}Efficiency and background numbers used for
  sensitivity calculations. The water Cherenkov numbers are based on
  published or preliminary Super-Kamiokande studies. The systematic
  uncertainties are included for reference but play no role in the
  sensitivity calculation. The liquid argon numbers come from the
  paper by Bueno~{\it et~al.}~\cite{Bueno:2007um}.}
\end{center}

\end{table}

\subsubsection{Proton decay to $e^+ \pi^0$}

In the decay $p \rightarrow e^+ \pi^0$, the total mass of the proton
is converted into the electromagnetic shower energy of the positron
and the two photons from $\pi^0$ decay, with a net momentum vector near
zero. For the decay of a free proton, the momentum should be zero
within the limits of detector resolution; for the decay of a bound
proton, the momentum is smeared up to the fermi level (225~MeV/$c$ in
$^{16}$O). No muon-decay electron should be present, a requirement that
eliminates a great deal of atmospheric neutrino background. Compared
to other possible nucleon decay modes, this is a very clean signature.

For $e^+ \pi^0$, the detection efficiency is dominated by nuclear
absorption when the proton decays in $^{16}$O, with $37\%$ of the
pions being absorbed or undergoing charge exchange; in either case the
signature is lost. Decay of the free proton is detected efficiently,
however, with an experimental efficiency of $87$\%. Overall, a proton
decay to $e^+ \pi^0$ in water event will pass the standard set of
Super-Kamiokande cuts with an efficiency of $45$\%.

\begin{figure}[t]
\centering
\includegraphics[width=0.7\textwidth]
{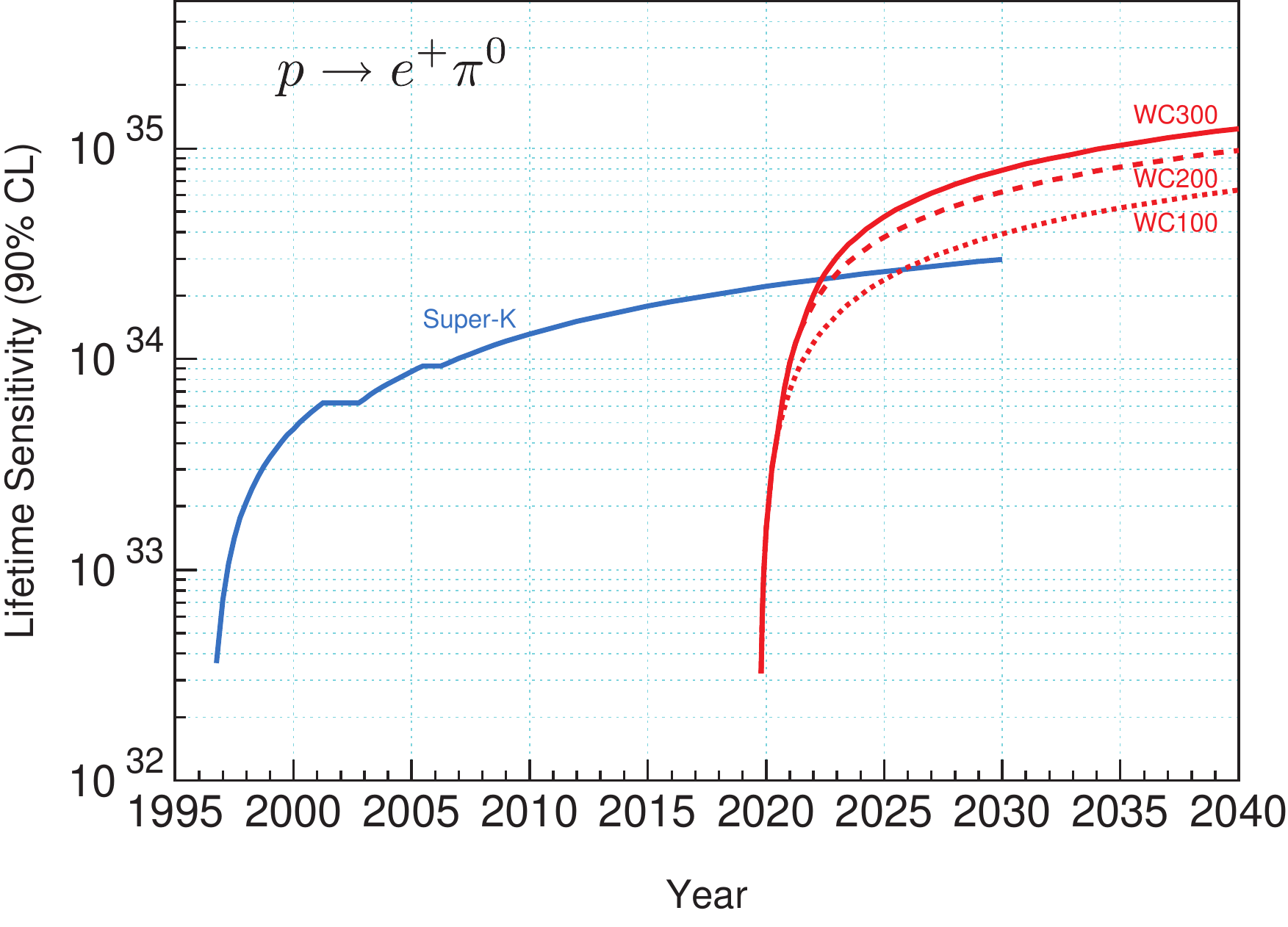}
\caption{Proton decay lifetime limit for $p \rightarrow e^+ \pi^0$ as
  a function of time for Super-Kamiokande compared to 300~kt of
  water Cherenkov detector starting in 2019. The water Cherenkov
  detector is assumed to commission 100~kt each year for the first
  three years; the limits from the partial detector masses of 100~kt
  or 200~kt is indicated with dashed lines. All limits use the same
  detection efficiency of 45\% and background rate of 0.2 events/100
 ~kt-years; systematic uncertainties are not included. The limits
  are at the 90\% C.L., calculated for a poisson process including
  background assuming the detected events equals the expected
  background.}
\label{PDK-peppi0-wc}
\end{figure}

Figure~\ref{PDK-peppi0-wc} shows the 90\% sensitivity curve for
$p \rightarrow e^+ \pi^0$ plotted as a function of calendar year.  The
leftmost curve is that for Super-K. The first smooth section
reflects the initial running period known as Super-K--I (SK1) that
started in May 1996. Then there is a flat period reflecting a planned
small shutdown in 2001 that was lengthened due to the PMT chain
reaction accident. The subsequent smooth curve is SK2, followed by a
brief shutdown, and then SK3 changing smoothly into SK4 in 2008. The
Super-K official limit for these three running periods
(SK1+2+3) is $1 \times 10^{34}$ years~\cite{Nishino:2009gd,skwww}. Future
running with SK4 may have slightly higher efficiency or lower
backgrounds due to new electronics, but the improvement levels are not
finally estimated and assumed to be small.

The curves for LBNE assume a start date for the first 100 kilotons
starting in mid-2019 with subsequent detector masses turning on one
year and two years later. The efficiencies and background rates were
taken to be identical to those for Super-K. Based on the SK2 studies
of $p \rightarrow e^+ \pi^0$~\cite{Nishino:2009gd}, the efficiency and
background rates for 20\% photocoverage were indistinguishable from
40\% photocoverage (SK1 and SK3). Therefore, we take the LBNE curves
to represent configurations with either 15\%-HQE or 30\%-HQE.

Fig.~\ref{PDK-peppi0-wc} does not include sensitivity curves
for LAr for $p \rightarrow e^+ \pi^0$. This is because the detector
mass, even in the optimistic case of 51 kilotons, cannot compete with the
long exposure of Super-Kamiokande. Based on
Bueno~{\it~et~al.}~\cite{Bueno:2007um}, an efficiency of 45\% with a
background rate of 0.1 events per 100~kt-years results in a
background subtracted limit at the 90\%~CL of $2.6 \times 10^{34}$
years for a 500~kt-year exposure. So a ten year run starting in 2020
of a 51~kt LAr detector does not exceed the projected Super-K limit
of $3 \times 10^{34}$ in 2030. Although there is no reason not to
search for this mode if such a detector is built, we do not consider
$e^+ \pi^0$ to be a critical physics topic for a 51~kt or smaller
liquid argon detector.

Hybrid configurations can be roughly appreciated from the 100, 200,
and 300~kt water Cherenkov curves. For example, configurations 5 or
6, 100~kt WC plus 34~kt LAr will lie between the WC100 and WC200
lines in Fig.~\ref{PDK-peppi0-wc}.

After 10 years, a 300~kt water Cherenkov detector would have an
expected background of 6 events given our assumed background rate.
This has a significant impact on the 90\% C.L. limit we would set, or
conversely, the ability to identify one or two candidate events in
such an exposure.  It is possible that atmospheric neutrino
backgrounds could be reduced in a detector with gadolinium such that
it can detect coincident neutron capture. This assumes that (a) proton
decay does not eject neutrons from a $^{16}O$ nucleus and (b)
atmospheric neutrino interactions are frequently accompanied by
primary or secondary neutrons. If the background rate could be
convincingly reduced by a factor of two, from 0.2 events to 0.1 events
per 100~kt-years, then a 10-year exposure would set a limit of
$1.0\times10^{35}$ years instead of $0.8\times10^{35}$ years.

\subsubsection{Proton decay to $\nu K^+$}

\begin{figure}[t]
\centering
\includegraphics[width=0.7\textwidth]
{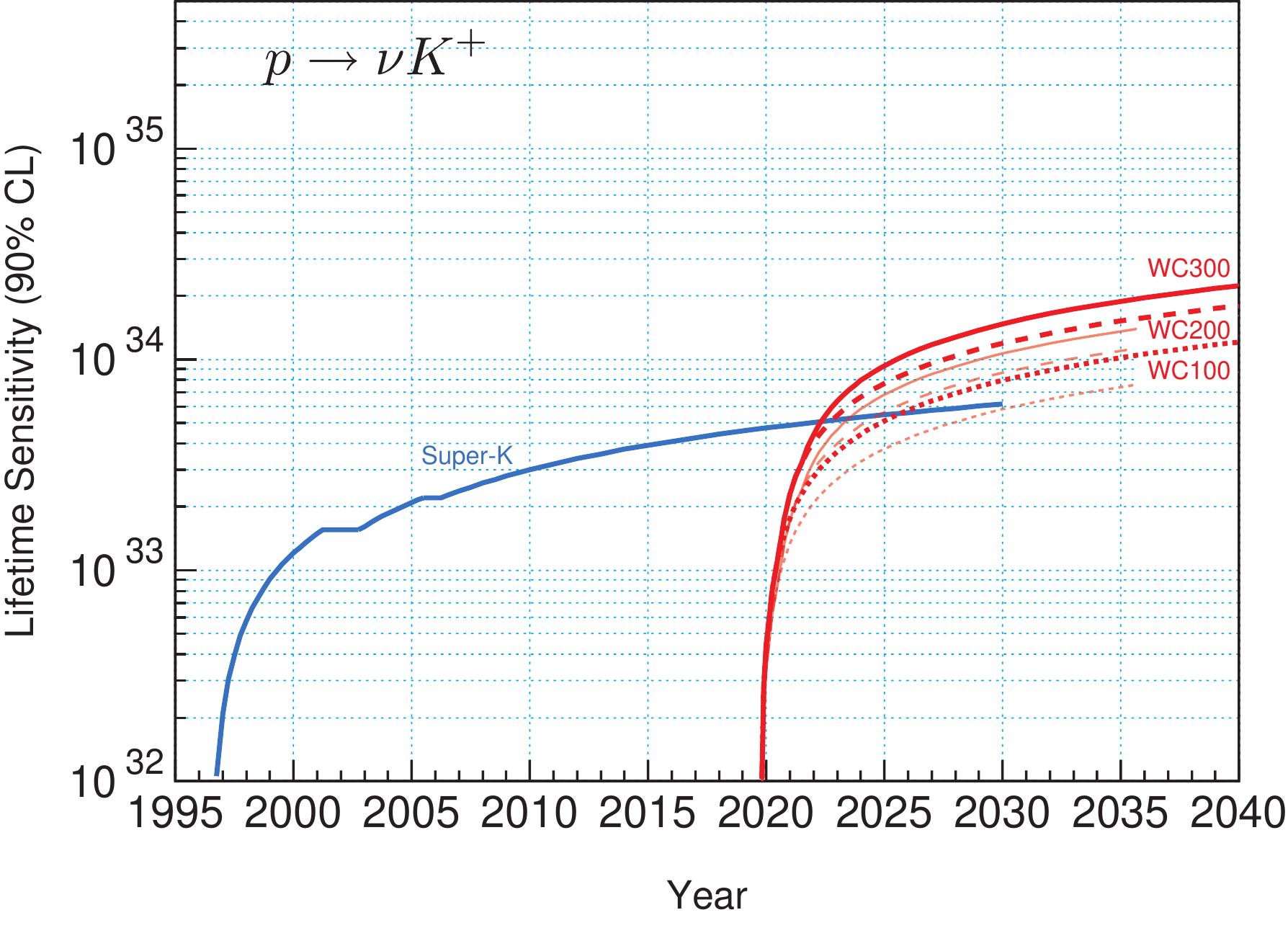}
\caption{Proton decay lifetime limit for $p \rightarrow \nu K^+$ as a
  function of time for Super-Kamiokande compared to 300~kt of water
  Cherenkov detector starting in 2019. The water Cherenkov detector is
  assumed to commission 100~kt each year for the first three years;
  the limits from the partial detector masses of 100~kt or 200~kt
  is indicated with dashed lines. The bold lines use the efficiency
  and background numbers for the SK1 analysis; the thin lines use the
  efficiency and background numbers for the SK2 analysis. The limits
  are at the 90\% C.L., calculated for a poisson process including
  background assuming the detected events equals the expected
  background.}
\label{PDK-nukp-wc}
\end{figure}

Fig.~\ref{PDK-nukp-wc} shows the 90\% sensitivity curve for $p
\rightarrow \nu K^+$ plotted as a function of calendar year.  The
leftmost curve is that for Super-K as described above. The
Super-K analysis is described in several
publications~\cite{Hayato:1999az,Kobayashi:2005pe}.

The Super-K analysis uses three methods: (i) gamma tag with $K^+
\rightarrow \mu+ \nu$, (ii) $K^+ \rightarrow \pi^+ \pi^0$, and (iii) a
background limited search for a monoenergetic muon. For the purpose of
the Super-K and LBNE water Cherenkov curves, only (i) and (ii) are
used, because the the background limited search contributes very
little for large exposures. The SK1 analysis using methods (i) and
(ii) has a relatively high background rate of 0.67 events per
100~kt-year. It is likely that some re-optimization would benefit
large exposures, but for the sake of argument, we assume the final
sensitivity would not change much, as the most likely alteration would
be some loss in proton decay signal efficiency in exchange for lower
atmospheric neutrino background. It is also possible that the LBNE
detector with smaller PMTs and better timing could result in a sharper
set of cuts to find the gamma ray tag. In other words, there is some
hope that a well-instrumented LBNE water cherenkov detector would
perform slightly better than Super-K for $p \rightarrow \nu K^+$,
which is not likely to be true for $e^+ \pi^0$. But we have no
estimates of this yet, so we conservatively use the SK efficiencies
and backgrounds as our benchmark.

As seen in Table~\ref{PDK-effic-bg}, the performance of the Super-K
analysis is markedly worse for SK2 (20\% photocoverage) than SK1 (40\%
photocoverage). The efficiency is lower and the background rate is
slightly higher. To study this difference, two sets of LBNE water
Cherenkov curves are provided in Fig.~\ref{PDK-nukp-wc}, one set for
each case.

\begin{figure}[t]
\centering
\includegraphics[width=0.7\textwidth]
{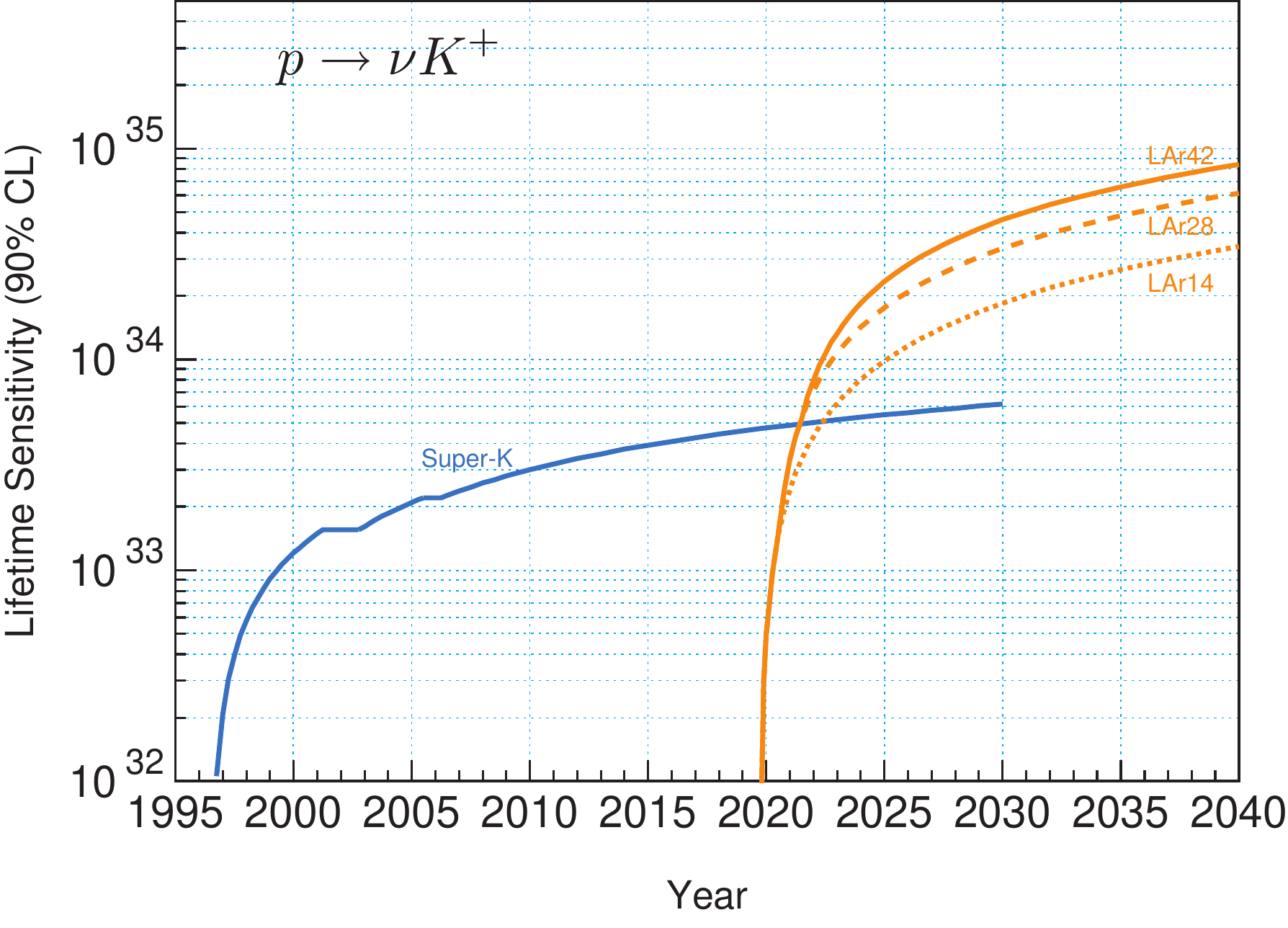}
\caption{Proton decay lifetime limit for $p \rightarrow \nu K^+$ as a
  function of time for Super-Kamiokande compared to 60 kilotons total, 42~kt
  fiducial, of LAr TPC starting in 2019. The LAr detector modules
  are assumed to commission 20~kt each year for the first three
  years; the limits from the partial detector fiducial masses of 14
  and 28~kt are indicated with dashed lines. The limits are at the
  90\% C.L., calculated for a poisson process including background
  assuming the detected events equals the expected background.}
\label{PDK-nukp-lars}
\end{figure}

Shown on Fig.~\ref{PDK-nukp-lars} is the projected sensitivity for a
liquid argon TPC of various masses. The efficiency and background
rates are taken from Bueno {\it et al.}~\cite{Bueno:2007um}. The high
efficiency for $p \rightarrow \nu K^+$ is a classic strong point in
favor of liquid argon. Such a proton decay event signature would be
highly described by such a detector, with a unique range and
ionization level for the charged kaon. The charged kaon will decay at
rest to fully reconstructible final states, for example, a subsequent
muon with no other particle, indicating $K^+ \rightarrow \mu^+ \nu$
(predominant, with 65\% branching fraction) should have a muon with
momentum reconstructed at 236~MeV/$c$. Therefore a high efficiency in
excess of 90\%, with very low background, is quite plausible.

The most serious background is cosmogenic neutral kaons undergoing
charge exchange in the sensitive volume. These could result in the
appearance of a charged kaon, mimicking proton decay if the $K^+$ had
just the right momentum (339~MeV/$c$). This background process could
be effectively studied by measuring the rate of such events in
momentum sidebands and close to the detector walls. In any case, a
fiducial mass reduction is anticipated to eliminate such events, by
cutting out candidates near the side walls. In the Bueno {\it et al.}
paper~\cite{Bueno:2007um}, several different overburden and active
veto scenarios were considered, with fiducial cuts as much as 6 meters
from the wall, resulting in fiducial mass reductions ranging from 66\%
to 90\%. For the sensitivity curves plotted in
Fig.~\ref{PDK-nukp-lars} we will take a generic value of 70\%. This
corresponds to a 2 meter cut from the sidewalls of the planned LBNE
detector, reducing the volume from 14$\times$15$\times$71 meters$^3$
to 14$\times$11$\times$67 meters$^3$. At shallow depths, a key
detector element required to allow even this fiducial restriction is
some sort of active muon tracking veto that extends wider than the LAr
detector itself. This is used to track nearby cosmic ray muons that
could produce neutral kaons or neutrons. Whether such a veto could be
constructed with ease and reasonable cost at either the 300-ft level
or 800-ft level is unknown, and not the topic of this report, although
it can be safely assumed that no external tracking veto is needed at
the 4850-ft level. Therefore, one 20-kiloton LAr module, assumed to
have a 17~kt fiducial mass for long-baseline neutrinos, is taken to
have a 14~kt fiducial mass for this proton decay mode if at shallow
depth, but the full 17~kt at the 4850-ft level.

\subsection{Conclusions}

The case for proton decay rests on either discovering a small number
of events or extending existing lifetime limits after an extensive
search by Super-Kamiokande. We assume a Super-K limit of $3 \times
10^{34}$ years for $p \rightarrow e^+\pi^0$ and a limit of $6 \times
10^{33}$ years for $p \rightarrow e^+\pi^0$ in 2030. To characterize
the power of a search using LBNE detectors, we look at the improvement
factor of a lifetime limit by LBNE in 2020 over these hypothetical
Srper-K limits.

Even the most massive configurations being considered, 300 kilotons of
water Cherenkov detector, are projected to improve the lifetime limit
on $p \rightarrow e^+ \pi^0$ by only a factor of 2.7 after a
decade-long run starting around 2020. Such a detector will cross the
symbolic milestone of $10^{35}$ years in 2035, after running for 15
years. Because a 10-year exposure of 300 kilotons would have an
expected background of six events, background rejection using neutron
tagging via gadolinium is desirable. Assuming a background reduction
factor of two, the improvement factor is raised to 3.4.

For $p \rightarrow e^+ \pi^0$, a liquid argon TPC of fiducial mass 42
kilotons makes no improvement over the projected Super-K limit by itself.
As supplementary detector mass to water Cherenkov, liquid argon will
contribute slightly.

For $p \rightarrow \nu K^+$ and 300 kilotons of water Cherenkov, the
improvement on the lifetime limit ranges from 1.8 to 3.3 depending on
photocathode coverage and the addition of neutron tagging using
gadolinium (again assuming a factor of two reduction in background
rate with Gd).

The improvement factor in the lifetime limit is largest for a 50~kt
scale liquid argon detector. Assuming that the high performance
characteristics projected for a large LAr TPC are realized, such a
detector could achieve an improvement factor of 8 to 9 on the lifetime
limit for $p \rightarrow K^+ \nu$ after 10 years and could continue to
do well with further exposure, having accumulated a modest background
estimate of roughly one-half event after that time. However, a LAr TPC
would have an impact on fewer decay modes than a water Cherenkov
detector of a few times greater mass. The LAr detector is effective
chiefly for the SUSY mode of $p \rightarrow K^+ \nu$, due to the high
detection efficiency.

The experimental prognosis for a large LAr TPC is less certain than
that for a large water Cherenkov detector, particularly if operated at
shallow depth. A highly effective cosmic ray veto shield is essential
at shallow depths. For the purpose of the tables below, no attempt is
made to distinguish the 300-ft and 800-ft depths, although option 2 is
given the benefit of a full 51 kilotons at 4850 feet. With respect to
the LAr detector configurations, the naive assumption would be that an
array of photomultiplier tubes that trigger based on scintillation
light in the liquid argon is required for non-accelerator contained
events such as proton decay. However, work is underway to develop a
pattern recognition trigger that uses the TPC data, dubbed {\em
  K-select}, that would efficiently trigger on kaon tracks in the
region of interest for proton decay. Since the prospect of such a
trigger is hopeful, albeit speculative, no attempt is made to compare
relative efficiencies with and without the photon trigger.
Tables~\ref{tab:PDK-configs-eppi0} and \ref{tab:PDK-configs-nuK+} summarize the results of this section.

\medskip
\medskip
\medskip

\begin{table}[h]
\begin{tabular}{|l|c|c|c|c|c|c|c|r|} \hline
Configuration  & WC Mass & PMT & Gd & LAr Mass & WC BG evts & LAr BG evts & 10-yr Limit ($\times 10^{34} yr$) & Factor \\ \hline
  1   & 300~kt & 15\% & N &       &   6 &      &  8.2 & 2.7 \\
  1a  & 300~kt & 30\% & N &       &   6 &      &  8.2 & 2.7 \\
  1b  & 300~kt & 30\% & Y &       &   3 &      & 10.3 & 3.4 \\
  2   &        &      &   & 51~kt &     & 0.51 &  2.7 & 0.9 \\
  2a  &        &      &   & 51~kt &     & 0.51 &  2.7 & 0.9 \\
  2b  &        &      &   & 51~kt &     & 0.51 &  2.7 & 0.9 \\
  3   & 200~kt & 15\% & N & 14~kt &   4 & 0.14 &  7.1 & 2.4 \\
  3a  & 200~kt & 30\% & N & 14~kt &   4 & 0.14 &  7.1 & 2.4 \\
  3b  & 200~kt & 15\%/30\% & N/Y & 14~kt &     3 & 0.14 &  9.3 & 3.1 \\
  4   & 200~kt & 15\% & N & 14~kt &   4 & 0.14 &  7.1 & 2.4 \\
  4a  & 200~kt & 30\% & N & 14~kt &   4 & 0.14 &  7.1 & 2.4 \\
  4b  & 200~kt & 15\%/30\% & N/Y & 14~kt &     3 & 0.14 &  9.3 & 3.1 \\
  5   & 100~kt & 30\% & Y & 28~kt &   1 & 0.28 & 6.2 & 2.1 \\
  6   & 100~kt & 30\% & Y & 28~kt &   1 & 0.28 & 6.2 & 2.1 \\ \hline

\end{tabular}
\caption{ Summary of $p \rightarrow e^+ \pi^0$ proton decay results of the reference configurations
  (see Table~\ref{tab:refconfigs} for more details). The background number of events and 90\% C.L. limit
  is calculated assuming a 10-year exposure of the tabulated mass. Efficiencies and background rates are
  documented in the narrative. For hybrid configurations, the limits from WC and LAr are combined.
  The factor is compared to the projected Super-K limit in 2030 of $3 \times 10^{34}$ years.}
\label{tab:PDK-configs-eppi0}
\end{table}

\begin{table}[h]
\begin{tabular}{|l|c|c|c|c|c|c|c|r|} \hline
Configuration  & WC Mass & PMT & Gd & LAr Mass & WC BG evts & LAr BG evts & 10-yr Limit ($\times 10^{34} yr$) & Factor \\ \hline
  1   & 300~kt & 15\% & N &       &  25 &      &  1.1 & 1.8 \\
  1a  & 300~kt & 30\% & N &       &  20 &      &  1.5 & 2.6 \\
  1b  & 300~kt & 30\% & Y &       &  10 &      &  2.0 & 3.4 \\
  2   &        &      &   & 51~kt &     & 0.51 &  5.7 & 9.5 \\
  2a  &        &      &   & 42~kt &     & 0.42 &  4.8 & 8.1 \\
  2b  &        &      &   & 42~kt &     & 0.42 &  4.8 & 8.1 \\
  3   & 200~kt & 15\% & N & 14~kt &  17 & 0.14 &  2.7 & 4.5 \\
  3a  & 200~kt & 30\% & N & 14~kt &  13 & 0.14 &  3.0 & 5.0 \\
  3b  & 200~kt & 15\%/30\% & N/Y & 14~kt &  12 & 0.14 &  3.4 & 5.6 \\
  4   & 200~kt & 15\% & N & 14~kt &  17 & 0.14 &  2.7 & 4.5 \\
  4a  & 200~kt & 30\% & N & 14~kt &  13 & 0.14 &  3.0 & 5.0 \\
  4b  & 200~kt & 15\%/30\% & N/Y & 14~kt &  12 & 0.14 &  3.4 & 5.6 \\
  5   & 100~kt & 30\% & Y & 28~kt & 3.4 & 0.28 & 4.4 & 7.3 \\
  6   & 100~kt & 30\% & Y & 28~kt & 3.4 & 0.28 & 4.4 & 7.3 \\ \hline

\end{tabular}
\caption{ Summary of $p \rightarrow \nu K^+$ proton decay results of the reference configurations
  (see Table~\ref{tab:refconfigs} for more details). The background number of events and 90\% C.L. limit
  is calculated assuming a 10-year exposure of the tabulated mass. Efficiencies and background rates are
  documented in the narrative. For hybrid configurations, the limits from WC and LAr are combined.
  The factor is compared to the projected Super-K limit in 2030 of $0.6 \times 10^{34}$ years.}
\label{tab:PDK-configs-nuK+}
\end{table}

\vfill\eject
%

\section{Supernova Burst Physics}\label{snbintro}

\subsection{Motivation and Scientific Impact of Future Measurements}

A nearby core collapse supernova will provide a wealth of information
via its neutrino signal (see~\cite{Scholberg:2007nu,Dighe:2008dq} for
reviews).  The neutrinos are emitted in a burst of a few tens of
seconds duration, with about half in the first second. Energies are in
the few tens of MeV range, and luminosity is divided roughly equally
between flavors.  The baseline model of core collapse was confirmed by
the observation of 19 neutrino events in two water Cherenkov detectors
for SN1987A in the Large Magellanic Cloud, 55~kpc
away~\cite{Bionta:1987qt,Hirata:1987hu}. An observed high-statistics
core collapse neutrino signal will shed light on a variety of physics and astrophysics topics.

Core collapses are rare events: the expected rate is 2-3 per century in the Milky Way.  As for the Homestake and Super-Kamiokande detectors, the large LBNE detector(s), once constructed, may operate for decades.  On this time scale, there is a significant likelihood of a supernova exploding in our galaxy.  In a 20-year run of an experiment, the probability of observing a collapse event is about 40\%.  The detection of the neutrino burst from such an event would dramatically expand the science reach of these detectors: from measuring the neutrino mass hierarchy and $\theta_{13}$ mixing angle, to observing the development of the explosion in the core of the star, to probing the equation of state of matter at nuclear densities, to constraining physics beyond the Standard Model.  Each of these questions represents an important outstanding problem in modern physics, worthy of a separate, dedicated experiment. The possibility to target them all at once is very attractive, especially since it may come only at incremental cost to the project.
 The expected harvest of physics is rich
enough that is essential to prepare to collect as much information as
possible when a burst happens.

In contrast to the SN1987A, for which a few dozen neutrinos were observed,
the detectors currently on the drawing board would register thousands or tens of
thousands of interactions from the burst. The exact type of
interactions depends on the detector technology: a water-Cherenkov
detector would be primarily sensitive to the electron antineutrinos,
while a liquid argon detector has an excellent sensitivity to electron
neutrinos. In each case, the high event rates imply that it should be
possible to measure not only the time-integrated spectra, but also
their second-by-second evolution. This is the key reason behind the impressive
physics potential of the planned detectors.

The interest in observationally establishing the supernova explosion
mechanism comes from the key role supernova explosions play in the
history of the universe. In fact, it would not be an exaggeration to
say that the ancient supernovae have in a very large measure shaped
our world. For example, modern simulations of galaxy formation cannot
reproduce the structure of our galactic disk without taking the
supernova feedback into account. Shock waves from ancient supernovae
triggered further rounds of star formation. The iron in our blood was
once synthesized inside a massive star and ejected in a supernova
explosion.

For over half a century, researchers have been grappling to understand
the physics of the explosion. The challenge of reconstructing the
explosion mechanism from the light curves and the structure of the
remnants is akin to reconstructing the cause of a plane crash from a
debris field, without a black box. In fact, the supernova neutrinos
are just like a black box: they record the information about the
physical processes in the center of the explosion during the first
several seconds, as it happens.

The explosion mechanism is thought to have three distinct stages: the collapse of the iron core, with the formation of the shock and its breakout through the neutrinosphere; the accretion phase, in which the shock temporarily stalls at the radius of about 200~km, while the material keeps raining in; and the cooling stage, in which the hot proto-neutron star loses its energy and trapped lepton number, while the reenergized shock expands to push out the rest of the star. All these stages are predicted to have distinct signatures in the neutrino signal. Thus, it should be possible to directly observe, for example, how long the shock is stalled.  More exotic features of the collapse may be observable in the neutrino flux as well, such as possible transitions to quark matter or to a black hole.
(An observation in conjunction with a gravitational wave detection would be especially interesting.)

To correctly interpret the neutrino signal, one needs to take into
account neutrino flavor oscillations. Over the last decades, the
oscillations have been firmly established in solar neutrinos and a
variety of terrestrial sources, which means that including them in the
supernova case is no longer optional. As it turns out, however, the
physics of the oscillations in the supernova environment is much
richer than in any of the cases measured to date. Neutrinos travel
through the changing profile of the explosion, with stochastic density
fluctuations behind the expanding shock. Their flavor states are also
coupled due to their coherent scattering off each other: ``collective'' neutrino
effects may be dramatic.  The net
result is a problem that requires supercomputers, as well as
state-of-the-art analytical models, to understand.

The effort to understand this complicated evolution has its reward:
the oscillation patterns come out very differently for the normal and
inverted mass hierarchies. There are also several smoking gun
signatures one can look for: for example, the expanding shock and
turbulence leave a unique imprint in the neutrino signal. The
supernova signal also has a very high sensitivity to values of
$\theta_{13}$, down to the levels inaccessible in any
laboratory experiment.
Additional information on oscillation parameters, free of supernova model-dependence, will be available if Earth matter effects can be observed in detectors at different locations around the Earth~\cite{Mirizzi:2006xx,Choubey:2010up}.
The observation of this potentially copious source of neutrinos will also allow limits on coupling to axions, large extra dimensions, and other exotic physics (\textit{e.g.}~\cite{Raffelt:1997ac,Hannestad:2001jv}).


Two comments need to be made at this point. First, it would be
extremely valuable to detect both the neutrinos and antineutrinos with
high statistics, as the oscillations occur very different in the two
channels. In the neutrino channel the oscillation features
are in general more pronounced, since the initial spectra of $\nu_e$ and
$\nu_\mu$ ($\nu_\tau$) are always significantly different. Second, the problem
is truly multidisciplinary and the neutrino physics and astrophysics
go hand-in-hand. One needs to model both, and the payout one gets is
simultaneous for both fields. For instance, one learns the sign of
the neutrino hierarchy, the value of $\theta_{13}$, the speed at which the
shock expands, and the density profile of the star, ``all in one package''.
The better one understands the astrophysics, the better the quality of
information about neutrino physics, and vice versa.  Hence it is
essential to gather as much high-quality information as possible, and to optimize ability to disentangle the flavor components of the flux.

As a final note, because the neutrinos emerge promptly after core
 collapse, in contrast to the electromagnetic radiation which must
 beat its way out of the stellar envelope, an observed neutrino
 signal can provide a prompt supernova alert~\cite{Antonioli:2004zb,
   Scholberg:2008fa}.  This will allow astronomers to find the
  supernova in early light turn-on stages, which may yield information
  about the progenitor (in turn important for understanding
  oscillations).  The LBNE detector(s) should be designed to allow prompt alert
capability.

Several other experiments sensitive to supernova neutrinos will be online over the next few decades~\cite{Scholberg:2007nu,Scholberg:2010zz}.  However one should not consider these to be ``competition'' for a supernova detection by LBNE: more experiments online during a supernova burst will only enhance the science yield from a supernova, and the ability to measure fluxes at different locations around the Earth will make the whole more than the sum of the parts~\cite{Mirizzi:2006xx}.


\subsection{Sensitivity of Reference Configurations}

The predicted event rate from a supernova burst may be calculated by folding expected neutrino differential spectra with cross sections for the relevant channels, and with detector response.   Although WCsim~\cite{wcsim}, the LBNE water Cherenkov simulation package, is nearly mature and can be used for some studies, LAr simulation packages are not yet ready for detailed studies of low energy response.  Furthermore, neutrino interaction generators which properly handle products from interactions with nuclei in the tens-of-MeV range are currently lacking.  For this reason, for this study we have chosen to do the event rate computation by using parameterized detector responses with a software package called SNOwGLoBES~\cite{snowglobes}, written for this purpose, which makes use of GLoBES software~\cite{globes}.  This package employs only the front-end rate engine part of GLoBES, and not the oscillation sensitivity part.  SNOwGLoBES takes as input fluxes, cross sections, ``smearing matrices'' and post-smearing efficiencies.  The smearing matrices incorporate both interaction product spectra and detector response.

\subsubsection{Supernova Neutrino Flux Models}

We have examined several flux models.
We assume fluxes at 10~kpc, which is just beyond the center of the Galaxy: event rates just scale as $1/D^2$, where $D$ is the distance to the supernova.


We consider here the ``Livermore'' model~\cite{Totani:1997vj},  and the ``GKVM'' model~\cite{Gava:2009pj}.   The Livermore model was digitized using Fig.~1 of reference~\cite{Totani:1997vj}, assuming spectra given by Eqn.~10 of that reference.
The model is somewhat out of date; however it is one of the few fluxes available for the full burst time interval, and it appears frequently in the literature, so it is considered for comparison purposes.
The GKVM flux includes shock and collective effects.  We
consider also ``Duan'' fluxes~\cite{Duan:2010bf,Keil:2002in} for which different
oscillation hypotheses have been applied: see Section~\ref{osc_compare}.  The Duan flux represents only a single late time slice of the supernova burst and not the full flux.

\begin{figure}[htb]
  \centering\includegraphics[width=.49\textwidth]{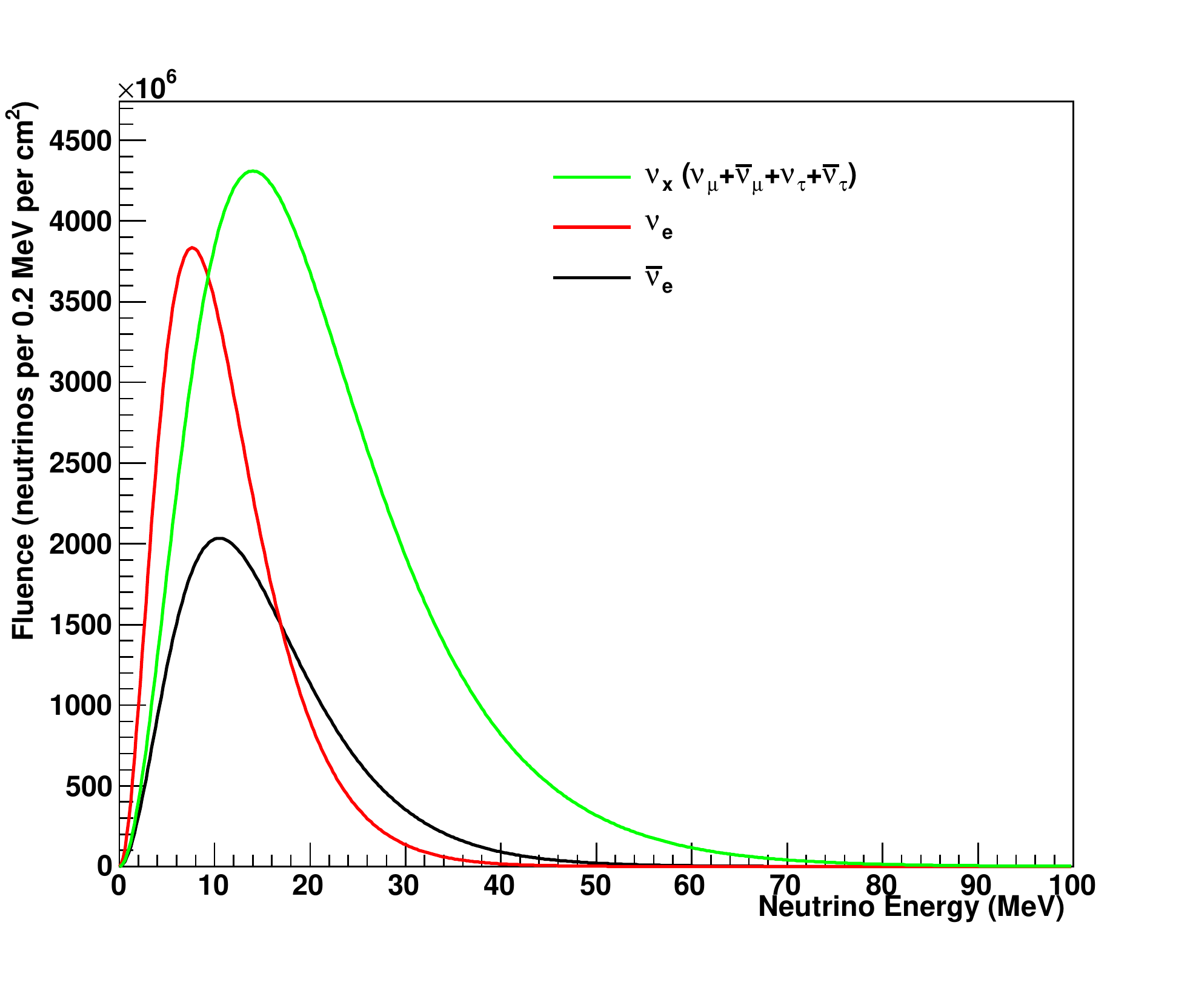}
  \centering\includegraphics[width=.49\textwidth]{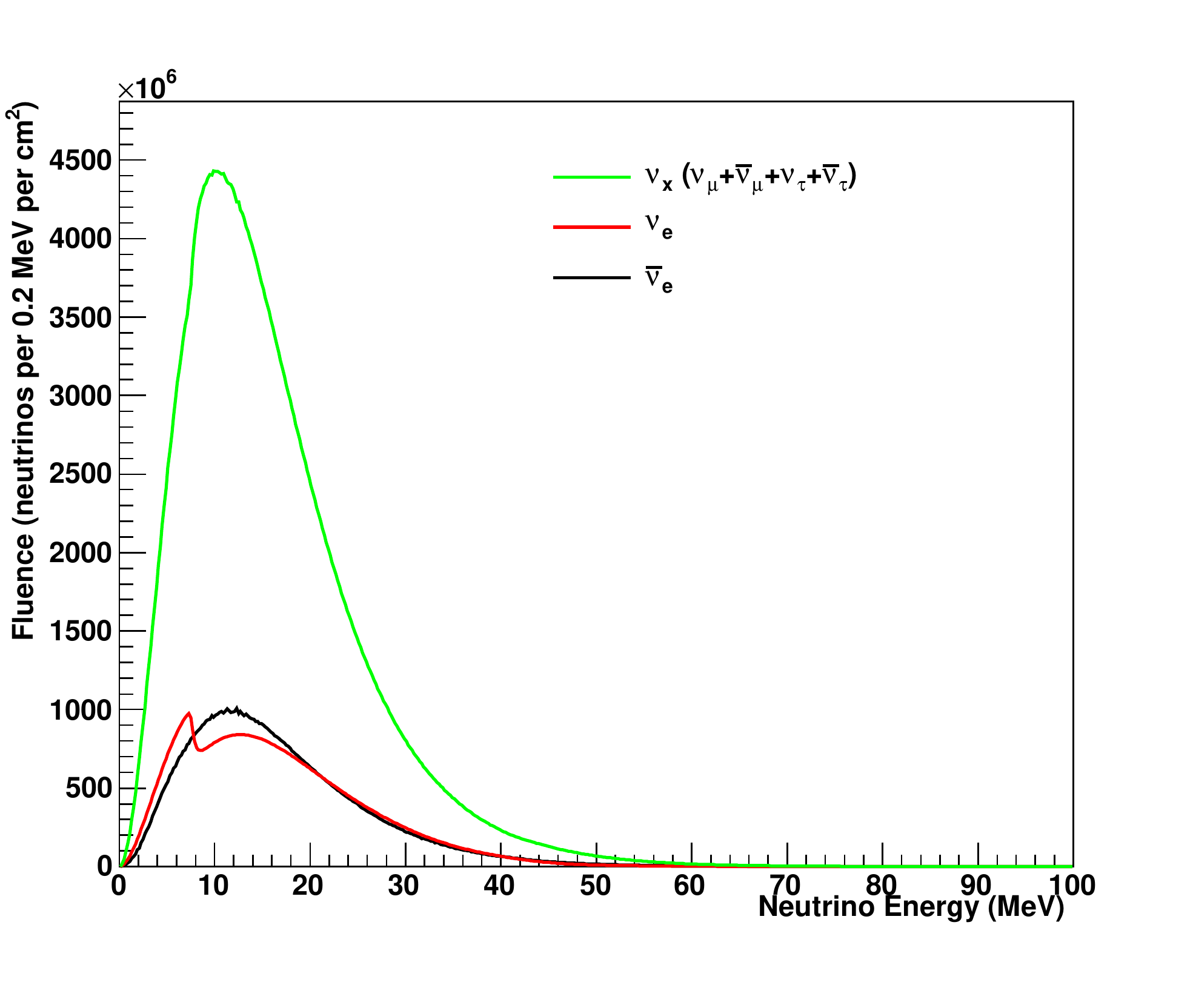}
 \caption{Flavor components of the fluxes used for this study: red is $\nu_e$, black is $\bar{\nu}_e$ and green is the sum of all other flavors.  The left plot shows the Livermore model,
integrated from $t=0$ to $t=14$~seconds.  The right plot shows the GKVM model,
integrated over 10~seconds.}
  \label{fig:fluxes}
\end{figure}

\subsubsection{Event Rates in Water}

Detector response assumptions for water are described in Appendix~\ref{snb_assumptions}.

\begin{figure}[htb]
   \centering\includegraphics[width=.5\textwidth]{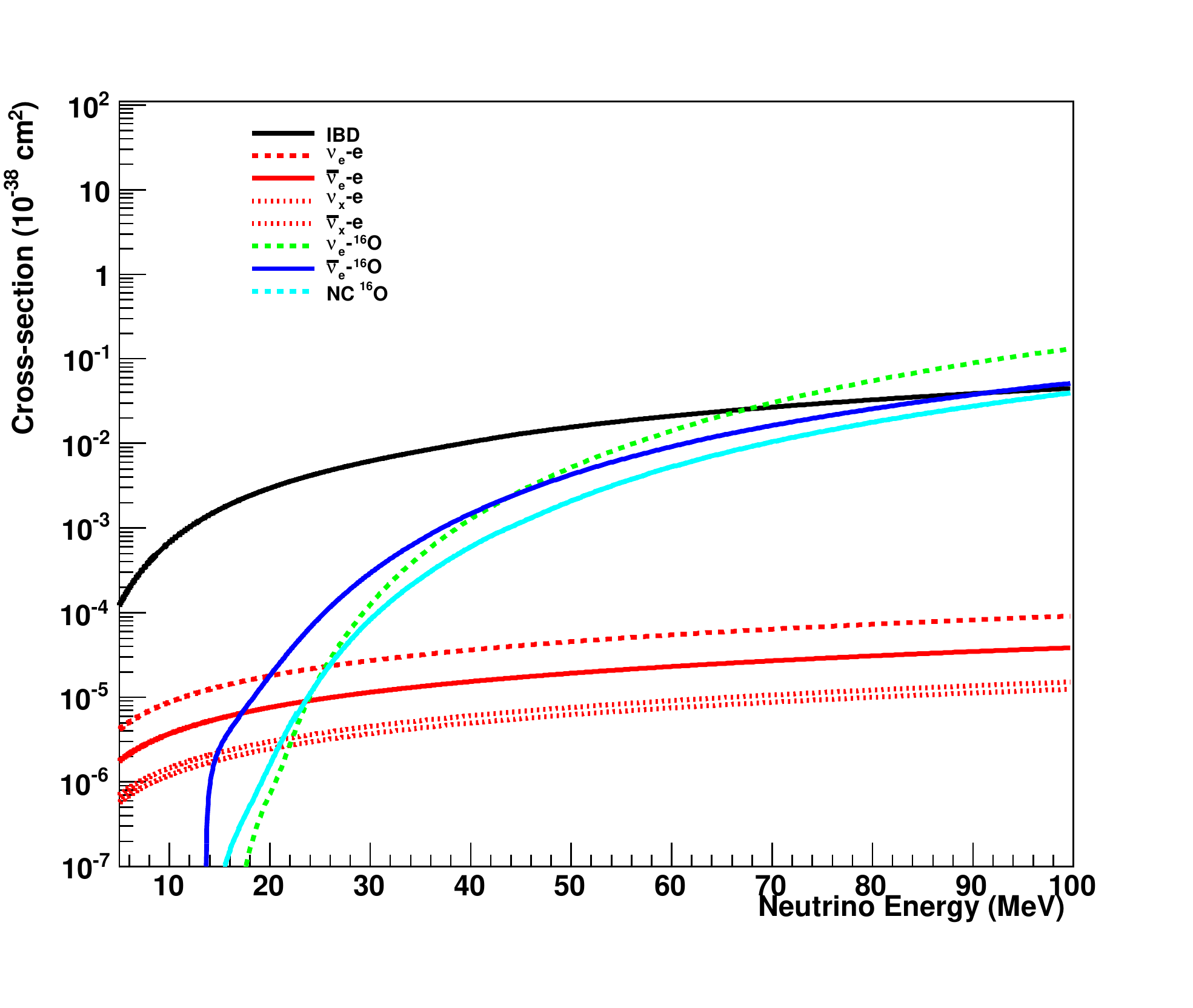}
   \caption{Cross sections for relevant processes in water.}
   \label{fig:water_xscns}
\end{figure}

The cross sections for relevant interactions in water are shown in
Fig.~\ref{fig:water_xscns}.  Some of these cross sections-- in
particular, inverse beta decay $\bar{\nu}_e+ p \rightarrow e^+ + n$
 (IBD) and elastic scattering (ES) of
neutrinos on electrons $\nu_{e,x} + e^- \rightarrow \nu_{e,x} + e^-$
 (both NC and CC) are known to few percent or better level.
In contrast, others have relatively large uncertainties, and have
never been measured in the few tens-of-MeV energy range.

In particular, interactions on oxygen,
$\nu_e + ^{16}{\rm O} \rightarrow e^- + ^{16}{\rm F}$, $\bar{\nu}_e + ^{16}{\rm O} \rightarrow e^+ + ^{16}{\rm N}$,
 have diverse final states, including ejected nucleons and deexcitation gammas.   For this study, we are considering only
the lepton in the final state response for the CC interactions, taking into account the energy threshold.  For the NC interaction with $^{16}$O,
$\nu_x + ^{16}{\rm O} \rightarrow \nu_x+ ^{16}{\rm O}^*$, we are using a simplified model
of the resulting deexcitation gammas by assuming relative final energy levels
according to reference~\cite{Kolbe:2002gk}.  Because this reference does not provide differential final state information,
we assume the distribution of these levels is independent of
neutrino energy (which is an incorrect assumption, but probably not a terrible approximation).  The resulting gamma cascade was simulated using relative probabilities of the transitions for a given excited state;  the resulting gamma spectrum was then run through WCsim detector simulation.  We found rather poor efficiency for detecting these gammas, in contrast to the results in reference~\cite{Langanke:1995he}, due to the fact that gammas frequently scatter electrons below Compton threshold.

\begin{figure}[htb]
 \centering
\includegraphics[width=.49\textwidth]{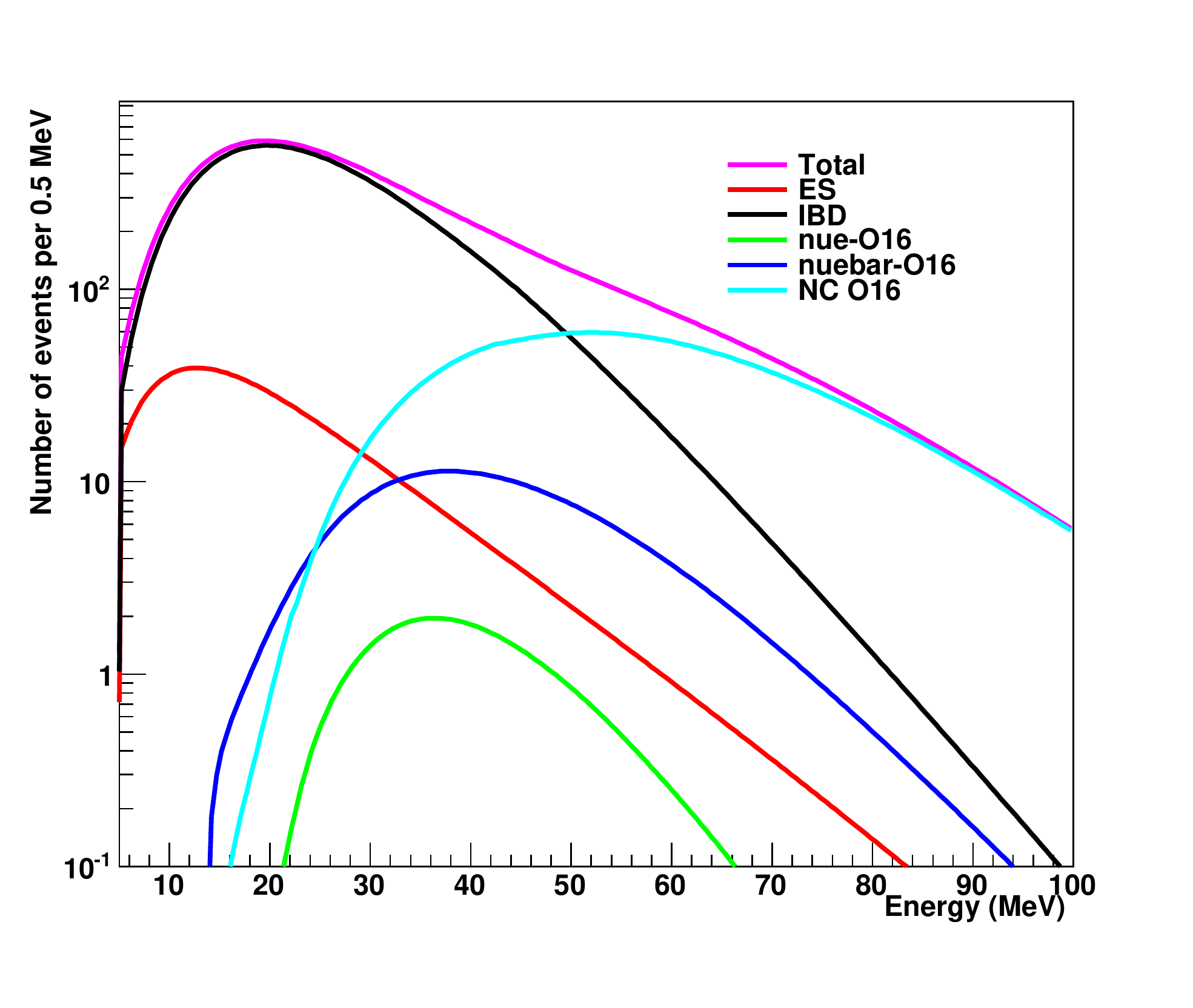}
\includegraphics[width=.49\textwidth]{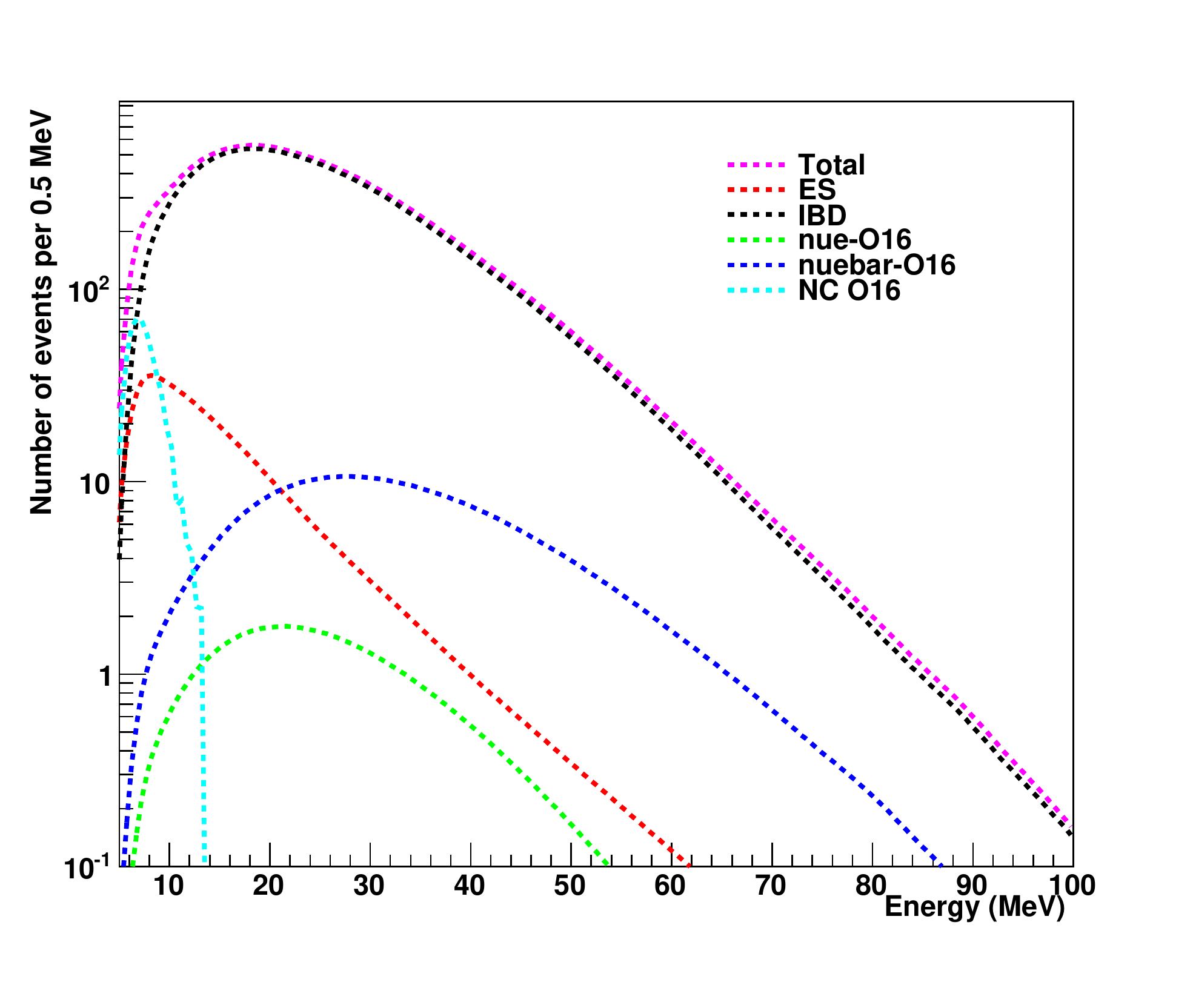}
\caption{Event rates in water, for the Livermore model and 30\% coverage (events per 0.5~MeV).}
\label{fig:water_events}
\end{figure}

Figure~\ref{fig:water_events} shows the resulting differential energy spectra for the different channels.  The plot on the left shows the interaction rates as a function of neutrino energy.  The plot on the right shows the distribution of observed event energies in the detector.   Table~\ref{tab:water_events} shows the breakdown of detected event channels, for two different specific supernova models.

\begin{table}[h]
\centering
\begin{tabular}{|c|c|c|}
\hline
Channel & Events, Livermore model & Events, GKVM model  \\
\hline
   $\bar{\nu}_e+ p \rightarrow e^+ + n$                  &  27116 &   16210\\
$\nu_x + e^- \rightarrow \nu_x + e^-$                           & 868 &   534\\
$\nu_e + ^{16}{\rm O} \rightarrow e^- + ^{16}{\rm F}$                         & 88  &  378  \\
$\bar{\nu}_e + ^{16}{\rm O} \rightarrow e^+ + ^{16}{\rm N}$  & 700 &  490 \\

$\nu_x + ^{16}{\rm O} \rightarrow \nu_x+ ^{16}{\rm O}^*$
                         &  513 &  124 \\ \hline
Total &  29284 & 17738 \\ \hline
\end{tabular}
\caption{Event rates for different models in 100~kt of water, for the 30\% coverage reference configuration. }
\label{tab:water_events}
\end{table}

These results show that IBD is overwhelmingly dominant: water Cherenkov is primarily sensitive to the $\bar{\nu}_e$ component of the flux.
However there are non-negligible contributions from other channels.   IBD positrons are emitted nearly isotropically; however, because ES and CC interactions on oxygen have anisotropic angular distributions, one may be able to use the reconstructed Cherenkov angular information to help disentangle
the flavor components (see Section~\ref{flavor_tagging}. (Or, if the direction of the supernova is unknown, which is likely at early times, the angular information can be used to point to it~\cite{Beacom:1998fj,Tomas:2003xn}.)

We also note that different flux models can give substantially different event rates.  In particular, because of the thresholds of the $^{16}$O interactions, the rates of the CC interactions on oxygen are quite sensitive to the $\nu_e$ and $\bar{\nu}_e$ spectra.

\begin{figure}[htb]
  \centering
\includegraphics[width=.5\textwidth]{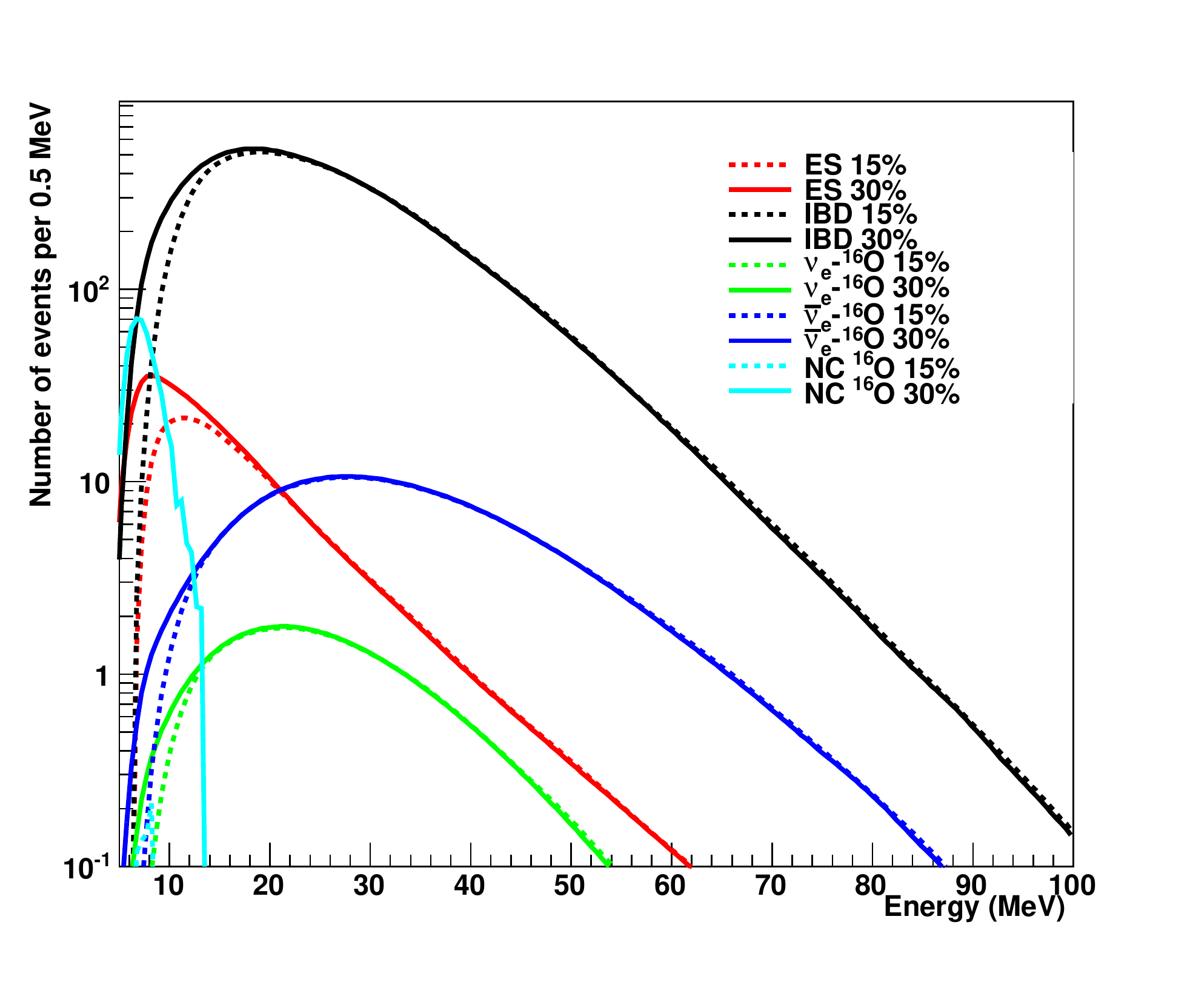}
\caption{Comparison of event rates for 15\% and 30\% PMT coverage configurations in 100~kt of water.}
\label{fig:water_comparison}
\end{figure}

Figure~\ref{fig:water_comparison} shows the difference in observed event rates between the 15\% and 30\% PMT coverage reference configurations.  For the 15\% configuration, one loses about 9\% of self-triggered events below $\sim$10~MeV.  The loss includes most of the NC excitation events.  (We note that clever triggering may mitigate this loss.)

The addition of Gd to a water detector will not substantially change event rates, but will enhance ability to determine the flavor composition of an observed signal
by allowing tagging of IBD events: see Section~\ref{flavor_tagging} (although note that interactions on $^{16}$O may produce ejected neutrons as well).

Because all of the supernova burst events arrive in a time window of a few tens of seconds, background is a much less serious issue than for relic supernova neutrino searches.  For water detectors, it should be nearly negligible for Galactic bursts.  To estimate it, we scale from Super-K~\cite{ikeda_thesis,Ikeda:2007sa}: the rate in 22.5~kt with loose selection cuts is about $3\times 10^{-2}$~Hz at a 7~MeV threshold.  Scaling by mass, this gives only about 4 background events in a 30~second burst.  For a distant supernova search~\cite{Ando:2005ka}, background becomes more important and limits the distance sensitivity.

\subsubsection{Event Rates in Argon}

Detector response assumptions for LAr are described in Appendix~\ref{snb_assumptions}.

\begin{figure}[htb]
  \centering\includegraphics[width=.5\textwidth]{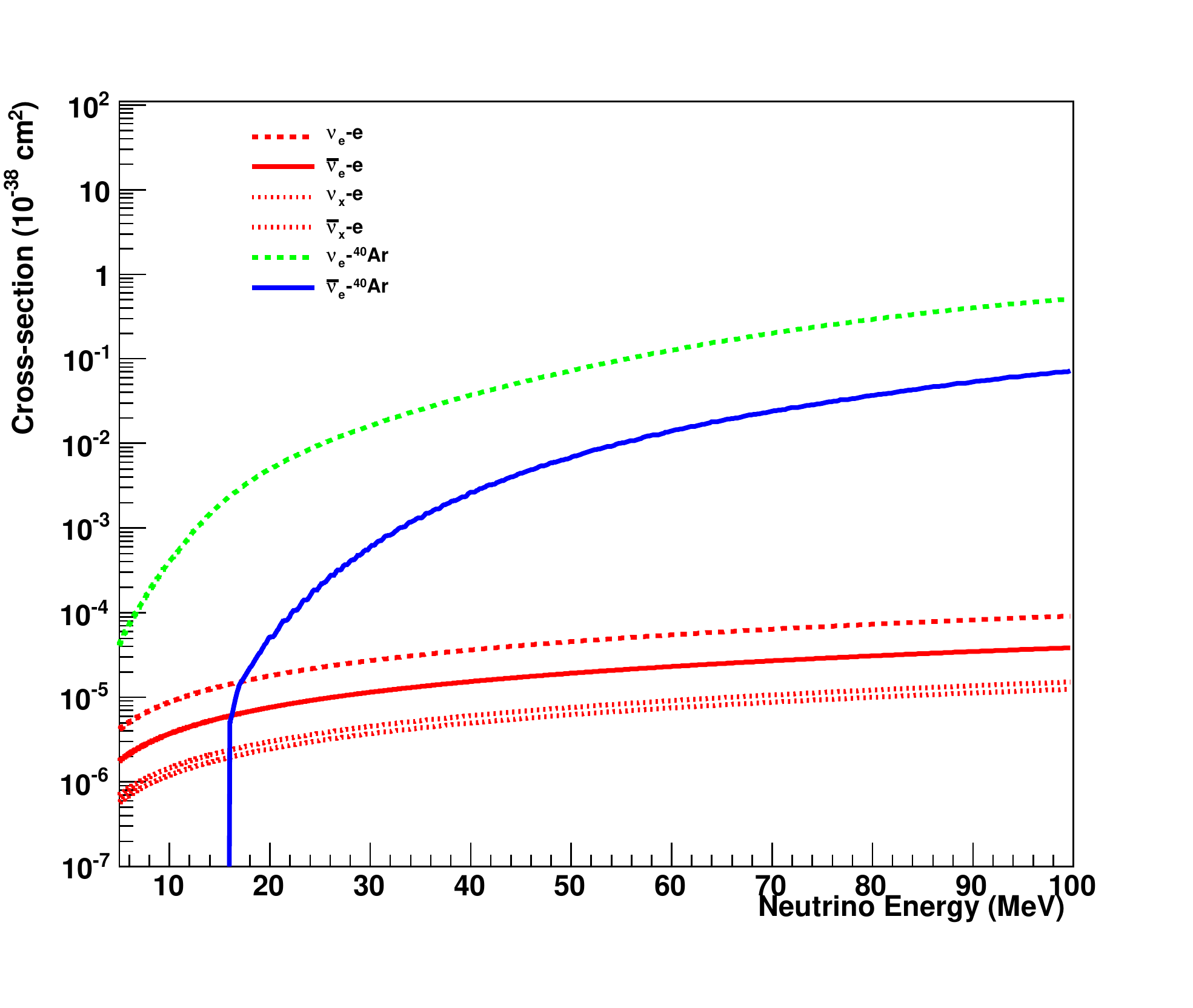}
  \caption{Cross sections for relevant processes in argon.}
  \label{fig:argon_xscns}
\end{figure}

The cross sections for interactions in argon~\cite{GilBotella:2004bv,Kolbe:2003ys},  are shown in Fig.~\ref{fig:argon_xscns}.   The uncertainties for the recent calculations are at around the 10-20\% level.  For the CC channels we have included energy deposition of the leading lepton;  in the detector response, we also incorporate additional visible energy from deexcitation gammas (these gammas may also possibly help to tag the $\nu_e$ or $\bar{\nu}_e$ channels, although we made no assumptions about such tagging).     We found no information in the literature about resulting excited levels for the NC interactions, so for the moment this channel is
not included in the study, even though event rates may be fairly large.


\begin{figure}[htb]
\includegraphics[width=.49\textwidth]{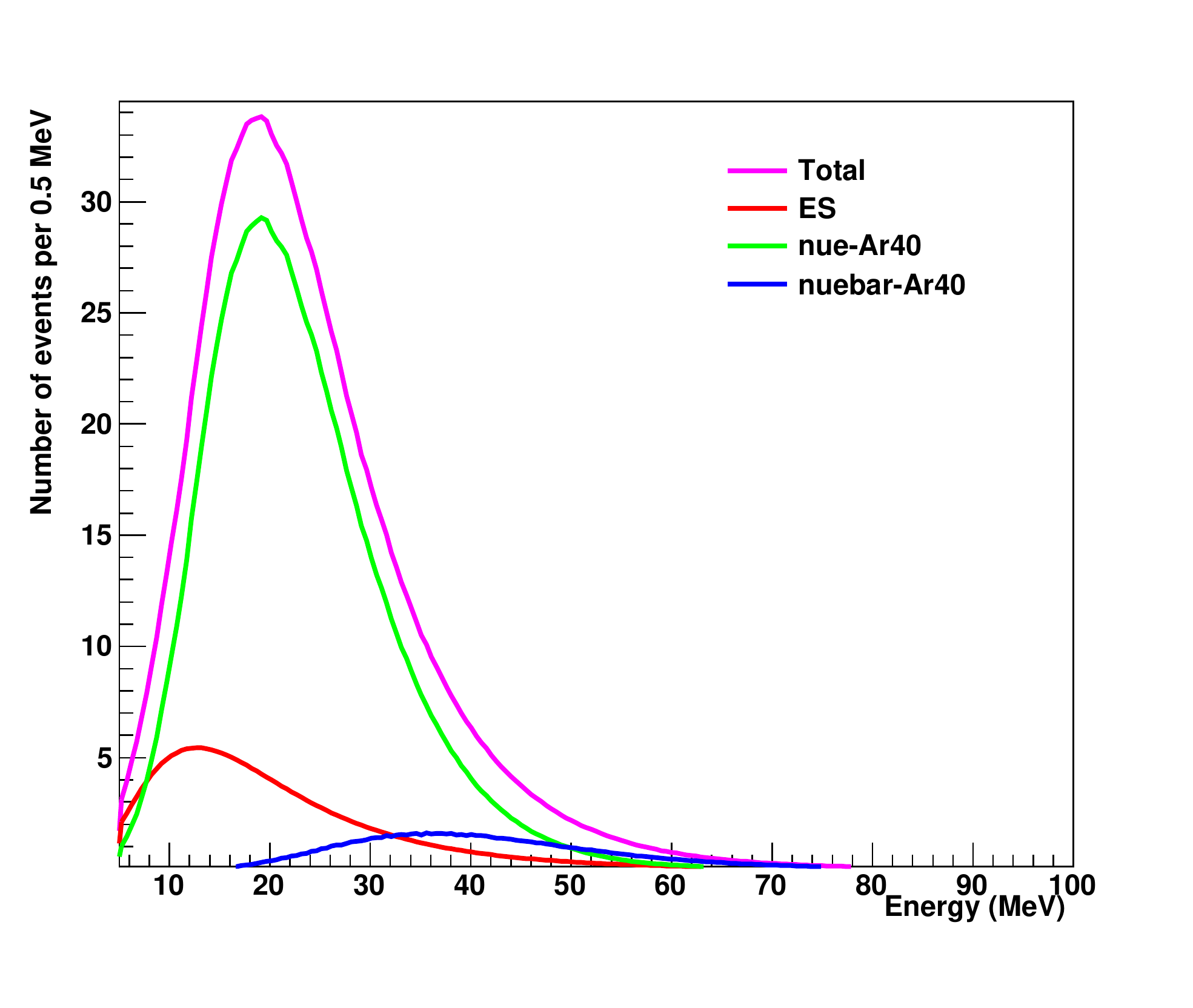}
\includegraphics[width=.49\textwidth]{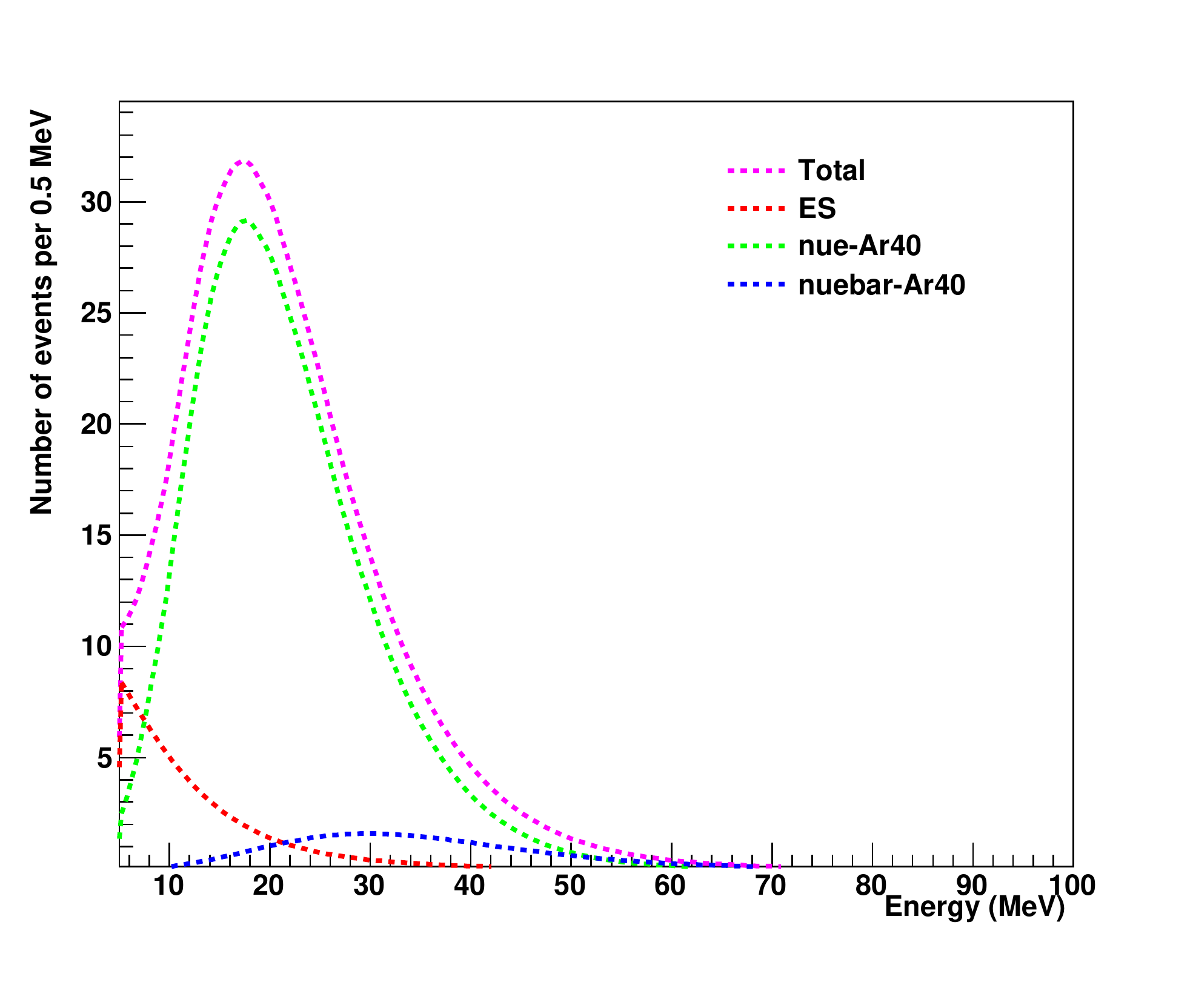}

\caption{Event rates in 17~kt of argon (events per 0.5~MeV).}
\label{fig:argon_rates}
\end{figure}

Figure~\ref{fig:argon_rates} shows resulting interaction rates as a function of
neutrino energy (left) and distribution of observed energies (right) in argon, for the Livermore model.  Table~\ref{tab:argon_events} gives a table of event rates for two models.  Note here that primary sensitivity is to the $\nu_e$ component for argon.

\begin{table}[h]
\centering
\begin{tabular}{|c|c|c|} \hline
Channel & Events, Livermore model & Events, GKVM model  \\
\hline

$\nu_e + ^{40}{\rm Ar} \rightarrow e^- + ^{40}{\rm K^*}$ & 1154  & 1424 \\

$\bar{\nu}_e + ^{40}{\rm Ar} \rightarrow e^+ + ^{40}{\rm Cl^*}$ & 97& 67\\

$\nu_x + e^- \rightarrow \nu_x + e^-$                           & 148 &   89\\

\hline

Total &  1397 & 1580 \\ \hline
\end{tabular}
\caption{Event rates for different models in 17~kt of LAr. }
\label{tab:argon_events}
\end{table}


For liquid argon, we have little information about backgrounds at the time of this writing, although again we can assume that they will be less of an issue for burst than for relic supernova neutrinos; furthermore backgrounds will be well known and can be statistically subtracted from a burst signal.
We can assume that cosmic ray muons will be easily identifiable as long tracks, and Michel electrons can be tagged in association with muons.
Backgrounds for supernova neutrinos in the range from 5-100~MeV will include events from radioactive products associated with muon spallation (some of which can be substantially delayed with respect to their parent muon).
 The distribution of spallation products in argon is currently unknown; however, most radioactive decays will deposit less than about 10~MeV.
At the 300~ft level, the muon rate in one 17~kt module will be
about 800~Hz;  at 800~ft it is about 200~Hz.
Assuming that the fraction of muons producing radioisotopes which decay in the supernova neutrino range of interest (and which cannot be vetoed using space and time correlation information) is less than $\sim0.01$, we can assume that this background will not be overwhelming during a nearby supernova burst, even at 300 ft.


\subsubsection{Comparing Oscillation Scenarios}\label{osc_compare}

As described in the introduction to this section, there will likely be significant and observable imprints of oscillation parameters on the observed spectrum of burst supernova neutrino events.  For oscillation sensitivity, ability to measure and tag the different flavor components of the spectrum is essential.

\begin{figure}[htb]
 \centering\includegraphics[width=.32\textwidth]{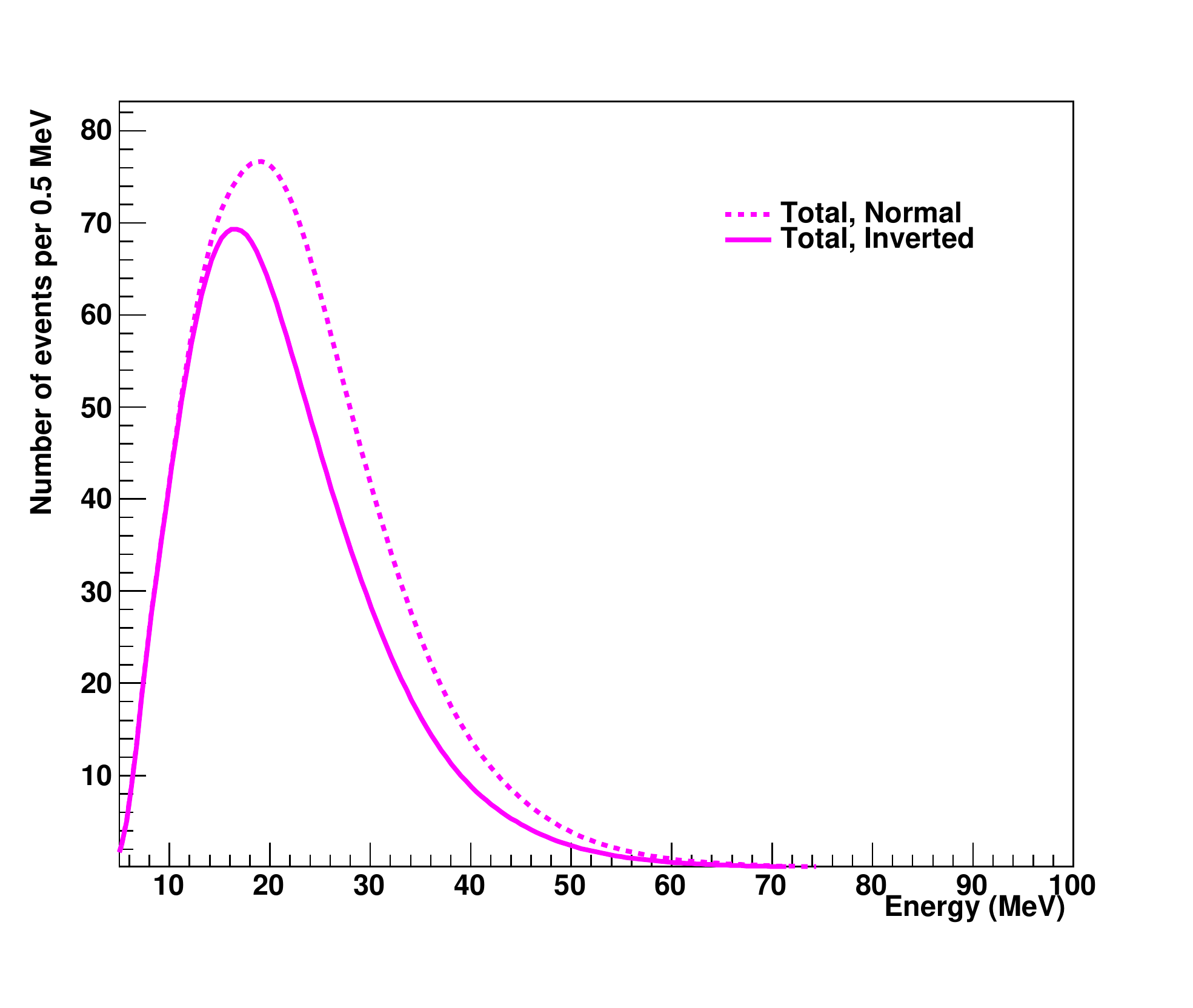}
 \centering\includegraphics[width=.32\textwidth]{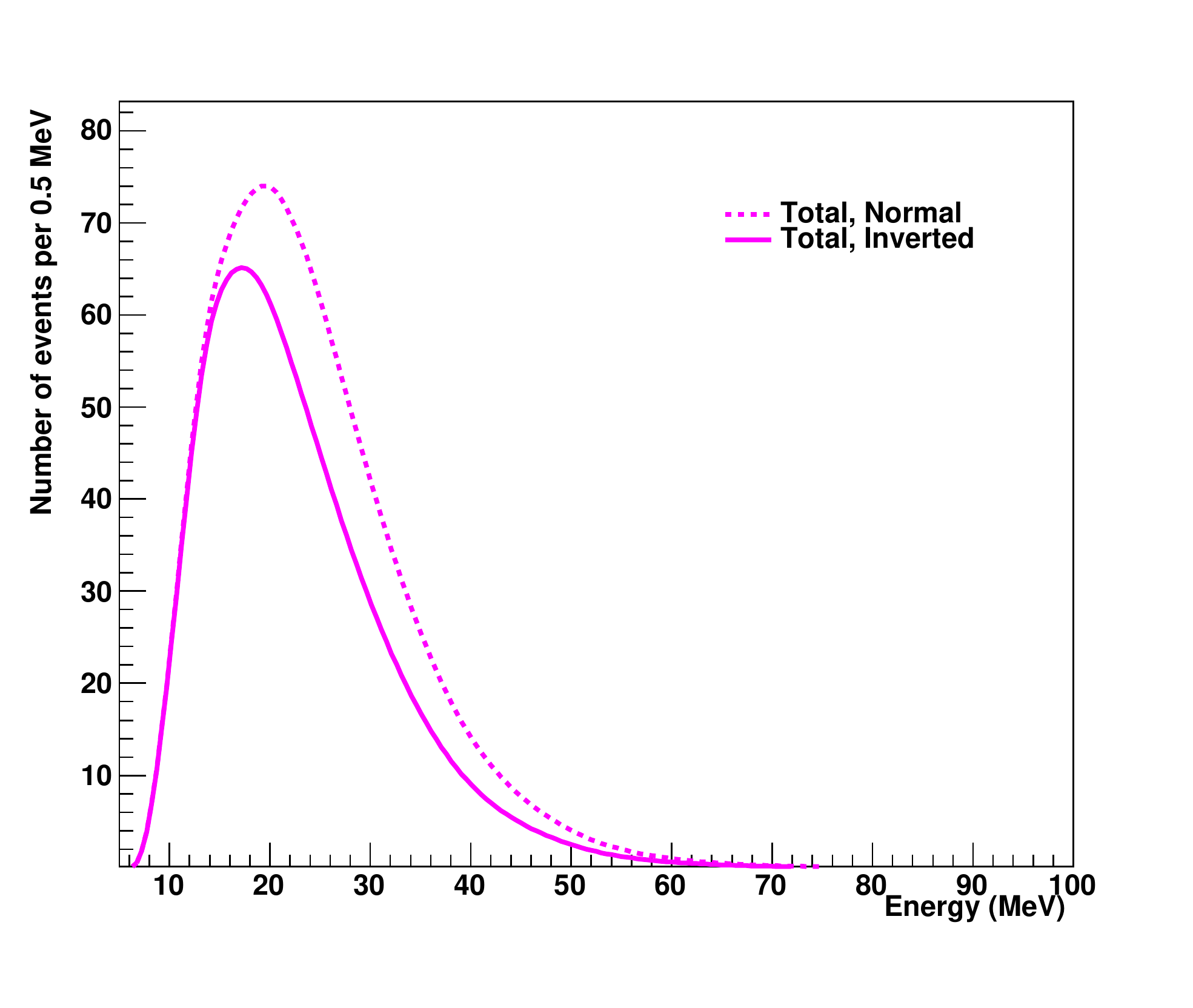}
 \centering\includegraphics[width=.32\textwidth]{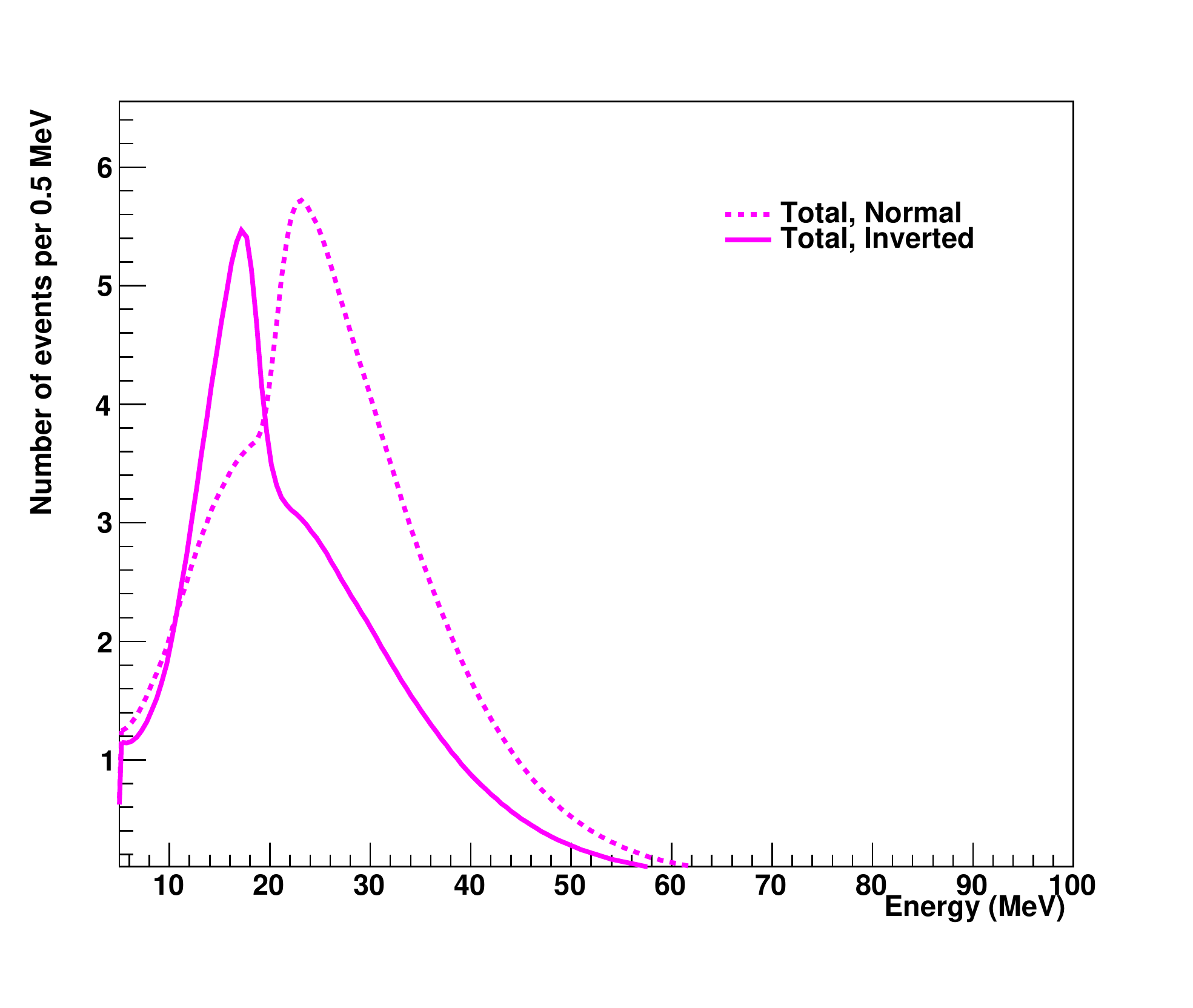}
\caption{Comparison of total event rates for normal and inverted hierarchy, for the Duan flux (a late time slice, not the full flux) , for WC 30\% (left), WC 15\% (center) and 17~kt LAr (right) configurations, in events per 0.5~MeV.}
\label{fig:hierarchy_comparison}
\end{figure}

Figure~\ref{fig:hierarchy_comparison} compares event rates for normal and inverted hierarchy, for a particular spectrum (a late time slice, not the full flux) provided by Huaiyu Duan~\cite{Duan:2010bf}.
While information about the hierarchy is clearly present in the water spectrum, which is mostly $\bar{\nu}_e$, the difference between the hierarchies is most dramatic in the observed mostly-$\nu_e$ argon spectrum.

\begin{figure}[htb]
\includegraphics[width=.4\textwidth]{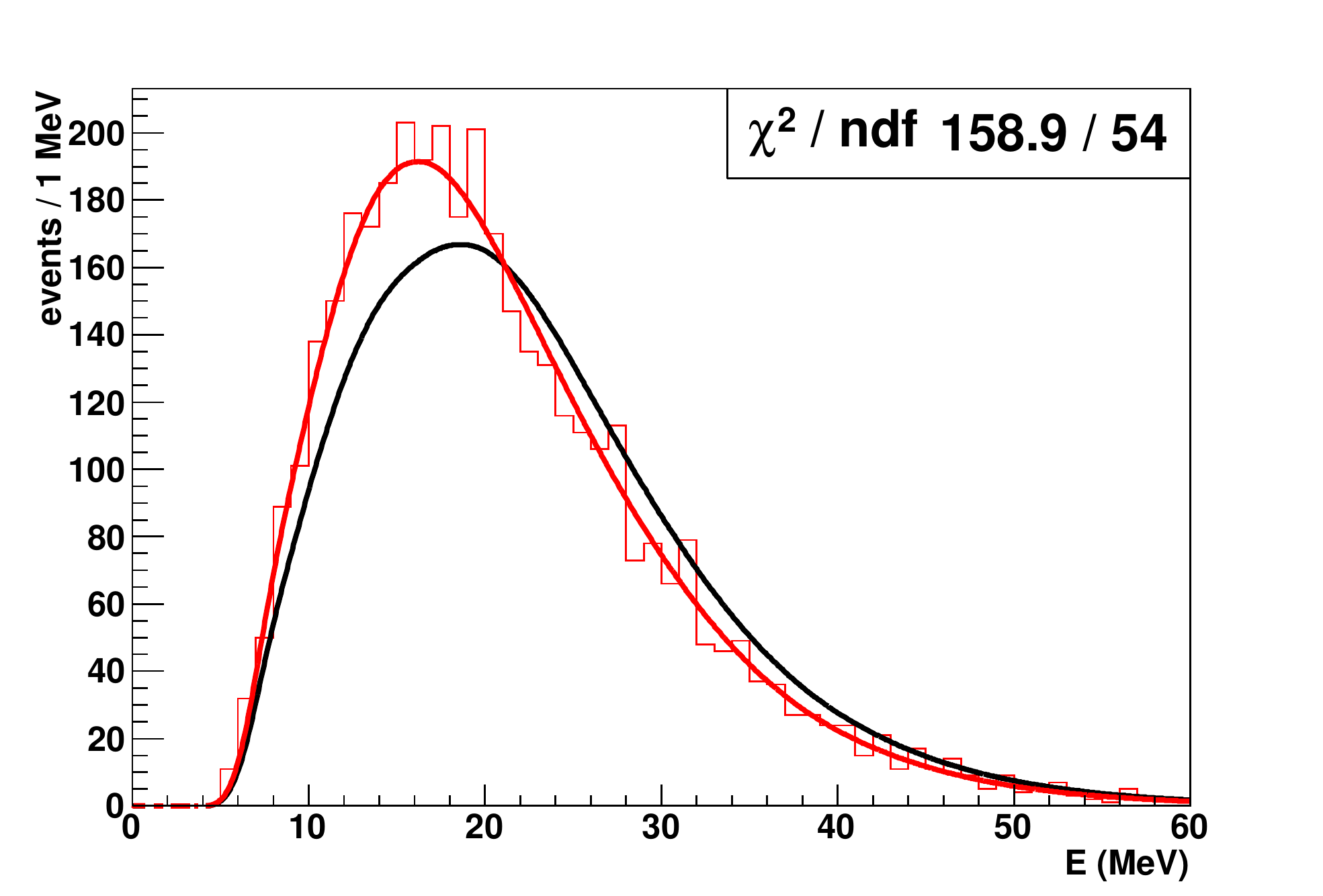}
\includegraphics[width=.4\textwidth]{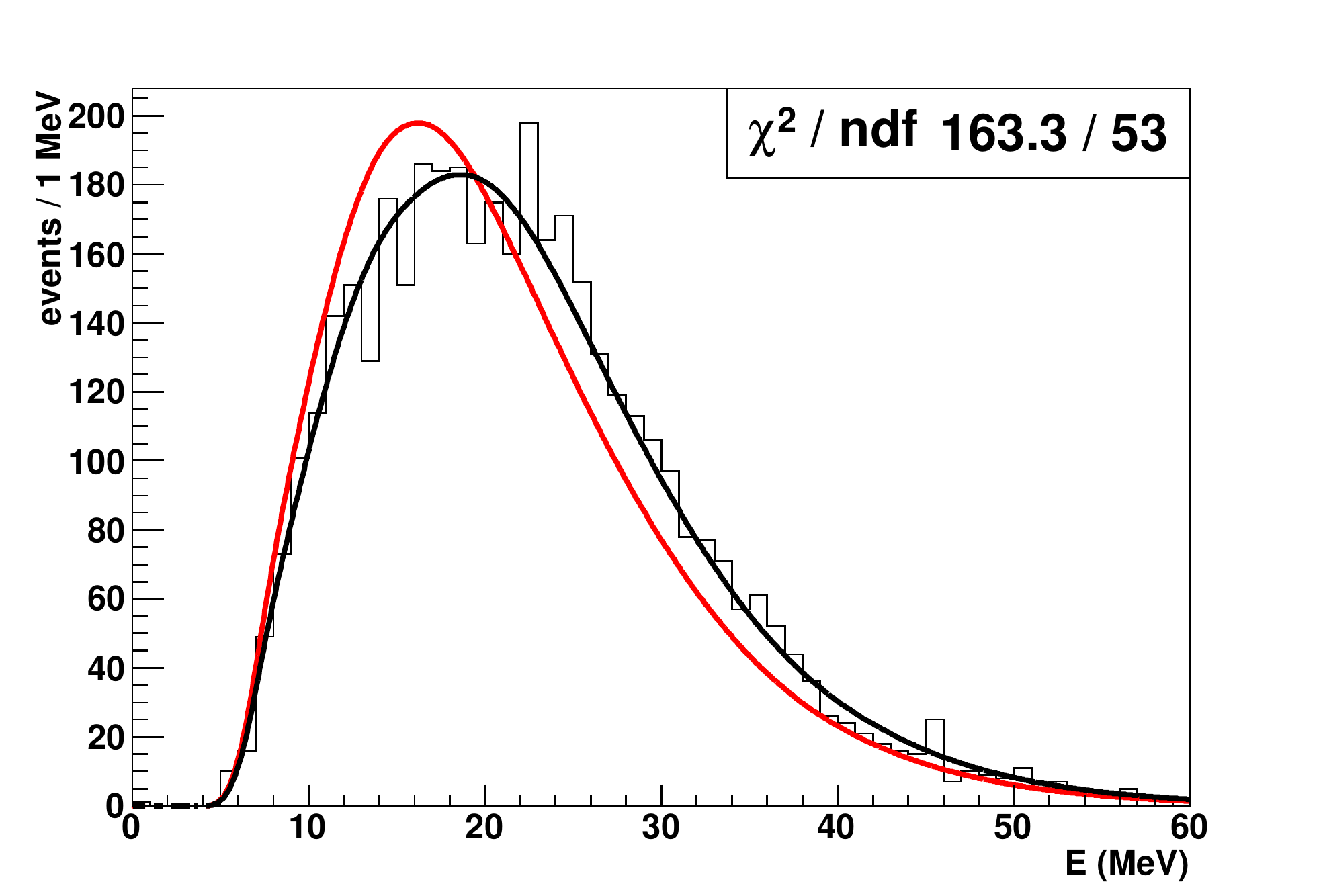}
\includegraphics[width=.4\textwidth]{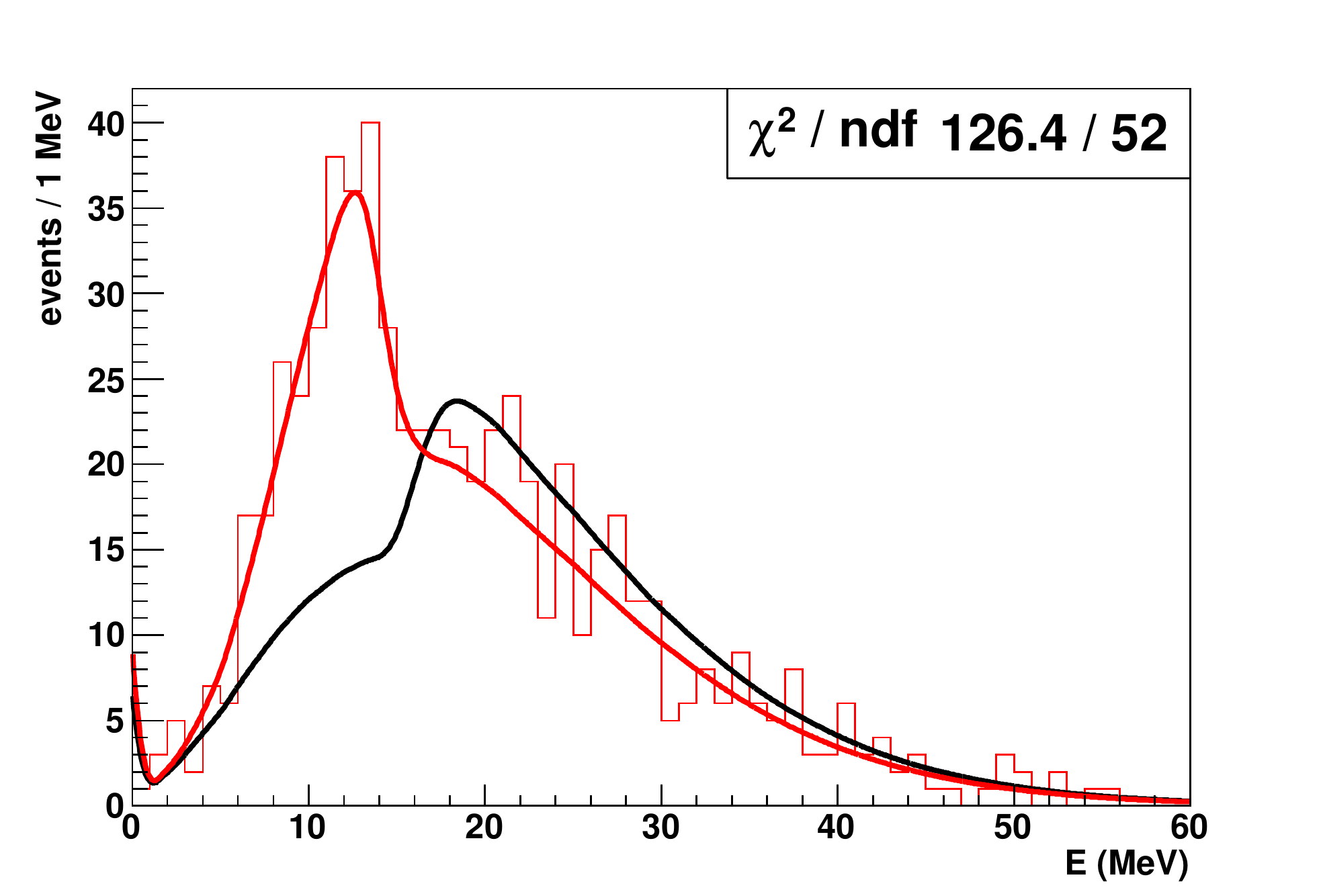}
\includegraphics[width=.4\textwidth]{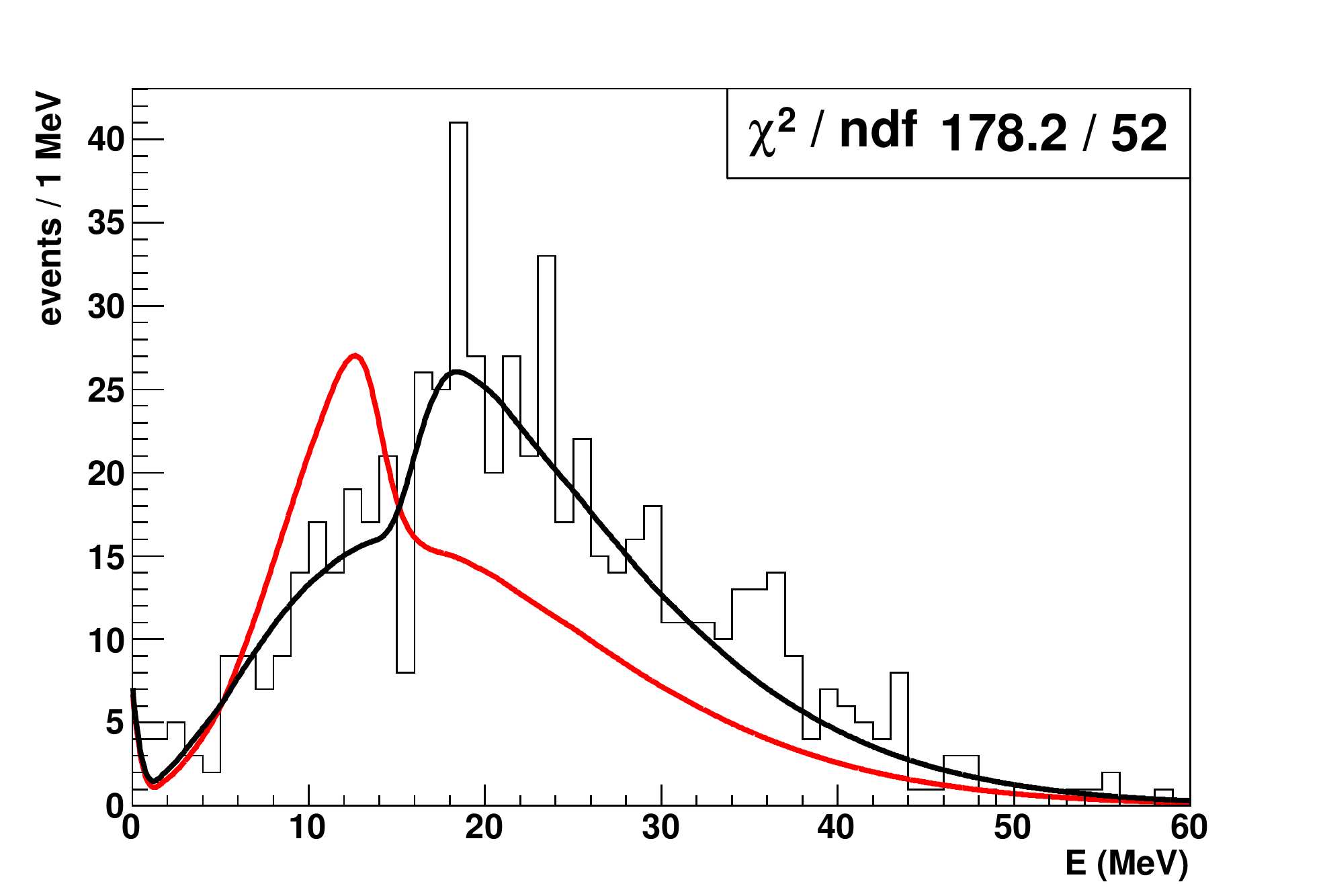}
\caption{Example fits to the expected spectral shapes for normal and inverted hierarchies for the Duan model. Top plots: 100~kt water, 30\% PMT coverage, assuming 4000 events.  Bottom plots: 17~kt argon, assuming 630 events observed.  Left plots: true hierarchy is inverted.  Right plots: true hierarchy is normal.  The $\chi^2$/dof is given for the fit to the ``wrong'' hierarchy. }
\label{fig:stat_compare}
\end{figure}

We have attempted a simple quantification of the relative sensitivity of the different single detector configurations to mass hierarchy.
Because fluxes with oscillation signatures are at this time only available representing a fraction of the total flux, we cannot evaluate the full statistical sensitivity.
However we have done the following: we have determined the minimum statistics for which normal hierarchy is distinguishable from inverted
hierarchy, for the Duan multi-angle spectrum~\cite{Duan:2010bf}.
For water Cherenkov (either 15\% or 30\% coverage), approximately 3500 events are required to distinguish the hierarchy at 3$\sigma$; 15\% and 30\% PMT coverage configurations are equally sensitive, because the differences occur at relatively high energy.  For LAr, about 550 events are required.

Figure~\ref{fig:stat_compare} shows examples of observed spectra
for the different configurations and hierarchies, for statistics near distinguishability.
Since number of events scales
as inverse square of the distance to the supernova, we convert this to a \textit{relative} figure of merit based on the distance at which hierarchy is distinguishable:
assigning a value of 1 to the distance sensitivity of one 17~kt module of LAr, we assign relative distance sensitivity of other single-detector configurations according to $D_{\rm max}({\rm detector~configuration})/D_{\rm max}({\rm 17~kt~LAr})$. The results are reported in the last column of Table~\ref{tab:designs}.


Although this study was done for one specific model, the relative evaluation of the configurations based on statistical reach can be considered relatively robust.

\subsubsection{Flavor Tagging}\label{flavor_tagging}
In order to
extract interesting physics from a supernova burst signal, it is
desirable to be able to determine the flavor content as a function of
energy and time.
In practice, given a real supernova burst signal, one would perform a fit to energy, angle and time distributions, making use of all available information, in order to determine flavor and interaction channel content.  At this point we have not yet developed the tools for such a comprehensive analysis, so we focus on a few possible signatures which could be exploited to help disentangle flavor content of a supernova burst signal in a WC detector.
The statistics considered in these studies are for a 100~kt WC detector with 30\% PMT coverage.

\textit{Neutron Tagging:} In a water detector, the bulk of detected events will be IBD,
$\bar{\nu}_e+ p \rightarrow e^+ + n$.
If gadolinium (Gd) is added to a water Cherenkov detector, we expect some
fraction of  IBD events to be tagged using neutron capture
on Gd, which produces about 4.3~MeV of energy in the detector, on a
$\sim$20~$\mu$s timescale~\cite{Beacom:2003nk}.
The efficiency of tagging is expected to be about 67\%~\cite{Watanabe:2008ru}.    The tagged sample will
represent a highly enriched $\bar{\nu}_e$ sample.
Events with no neutron tag will be enriched in flavors other than $\bar{\nu}_e$, and in
principle, the IBD contribution can be subtracted from the overall
signal using the tagged rates.
[Note that with sufficient PMT coverage it may be possible to use 2.2~MeV gammas from neutron capture on protons even in the absence of Gd doping, but the efficiency would be lower.]

To get a general idea of the value of neutron tagging of $\bar{\nu}_e$ we
have done a simple study: we looked at flavor
composition for tagged and untagged events.   We assume that 67\% of
the true IBD events will be tagged; we also assume that no events without a neutron
will be falsely tagged as having a neutron (the false tagging rate
should be $\sim 10^{-4}$ according to reference~\cite{Watanabe:2008ru}).   We also
take into account CC and NC reactions of neutrinos on $^{16}$O, for which some
final states have neutrons.  To estimate this contribution we use
tables II, III and IV from reference~\cite{Kolbe:2002gk} and (in absence of differential
final state information) we assume that the relative
fractions of final states with neutrons are independent of neutrino energy.

\begin{figure}[!ht]
\begin{centering}
\includegraphics[height=2.7in]{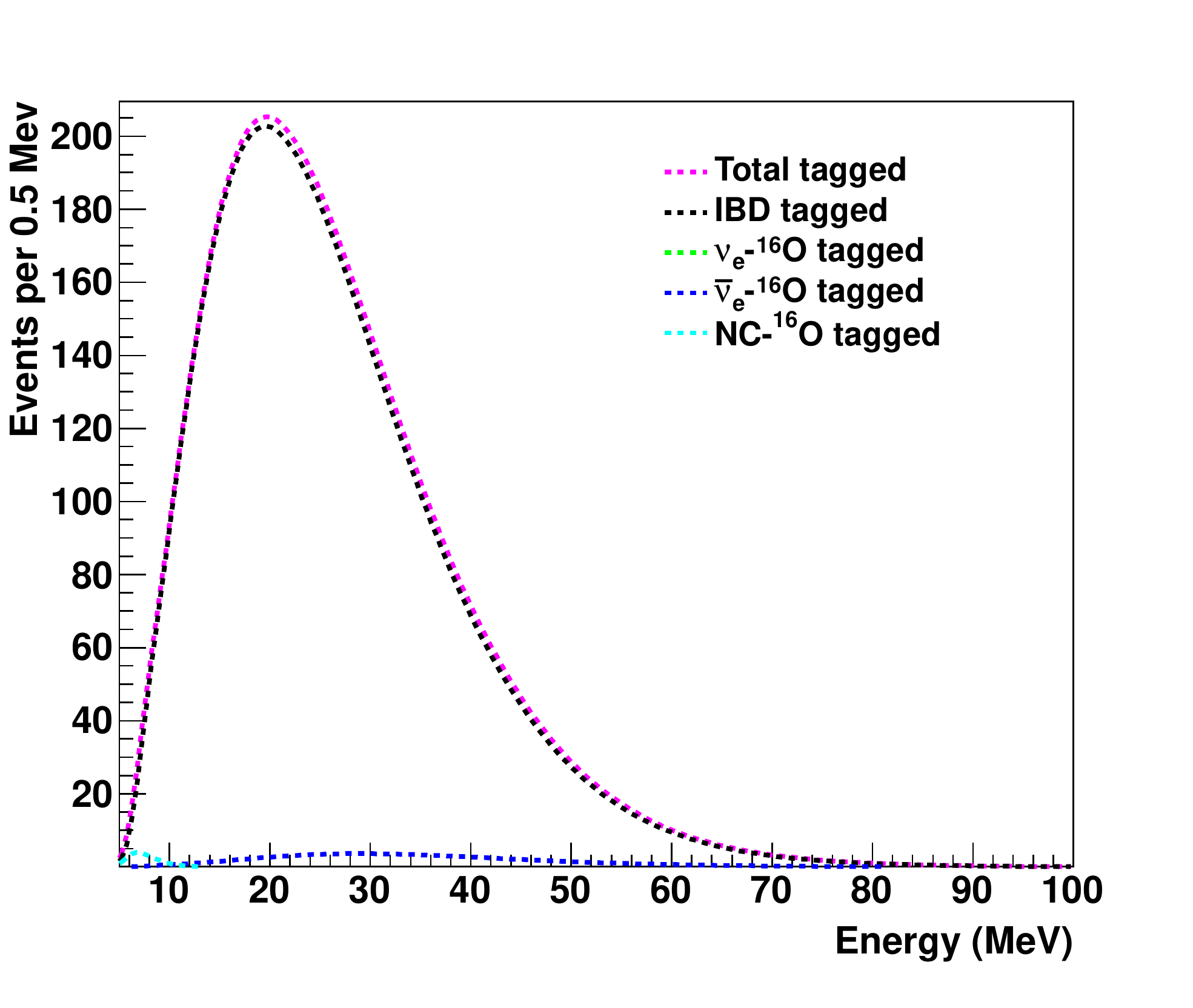}
\includegraphics[height=2.7in]{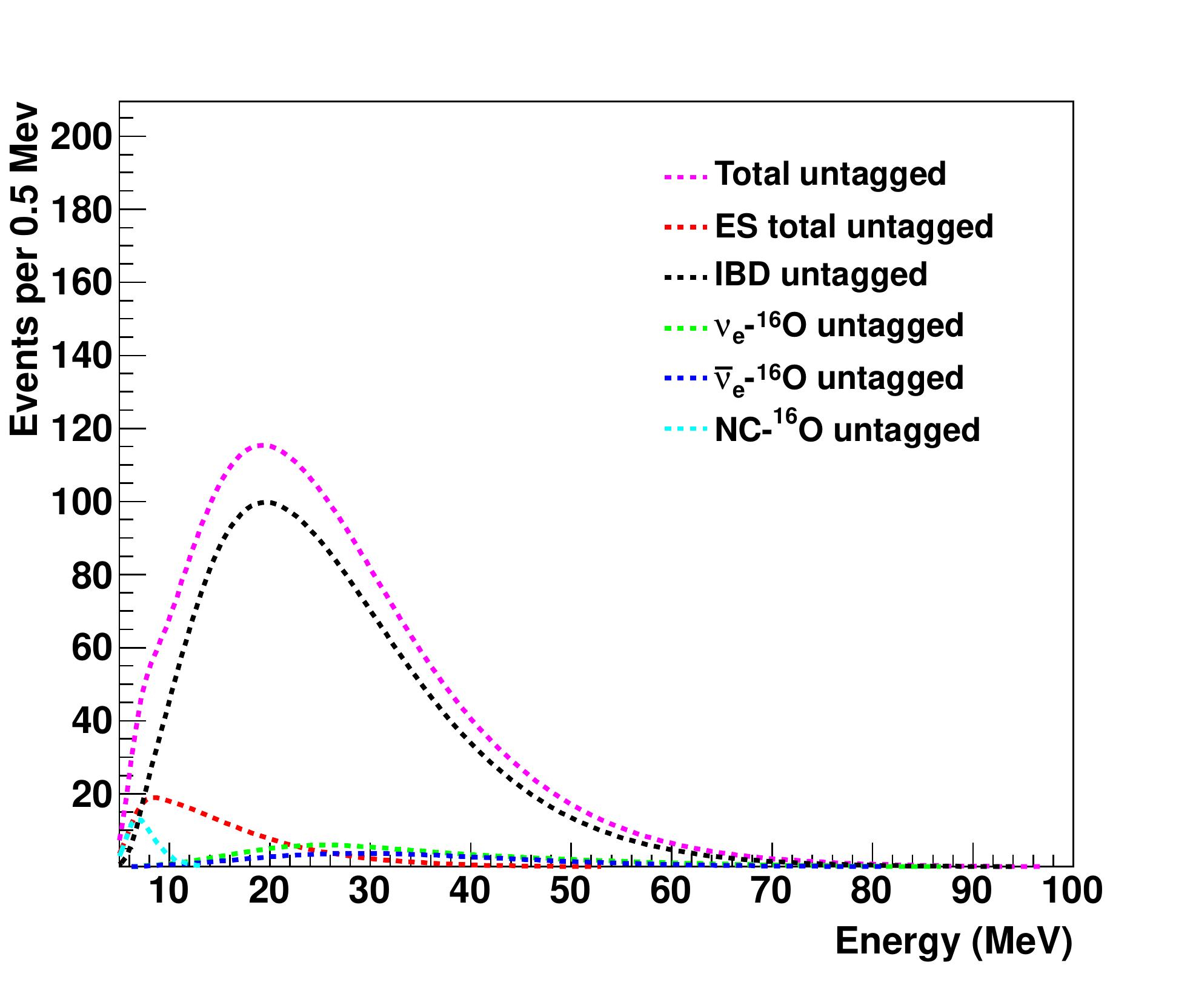}
\caption{Total events in WC showing contribution from the different
 interaction channels, for neutron-tagged (left) and untagged (right)
events.}
\label{fig:interaction_content}
\end{centering}
\end{figure}

Figure~\ref{fig:interaction_content} shows the contributions of the
different interaction channels for tagged and untagged events, for the GKVM flux.
The neutron-tagged event rate is a nearly-pure IBD sample.
The untagged event rate has contributions from elastic scattering (ES), and from
CC and NC interactions on $^{16}$O, but is dominated by untagged IBD.

\begin{figure}[!ht]
\begin{centering}
\includegraphics[height=2.7in]{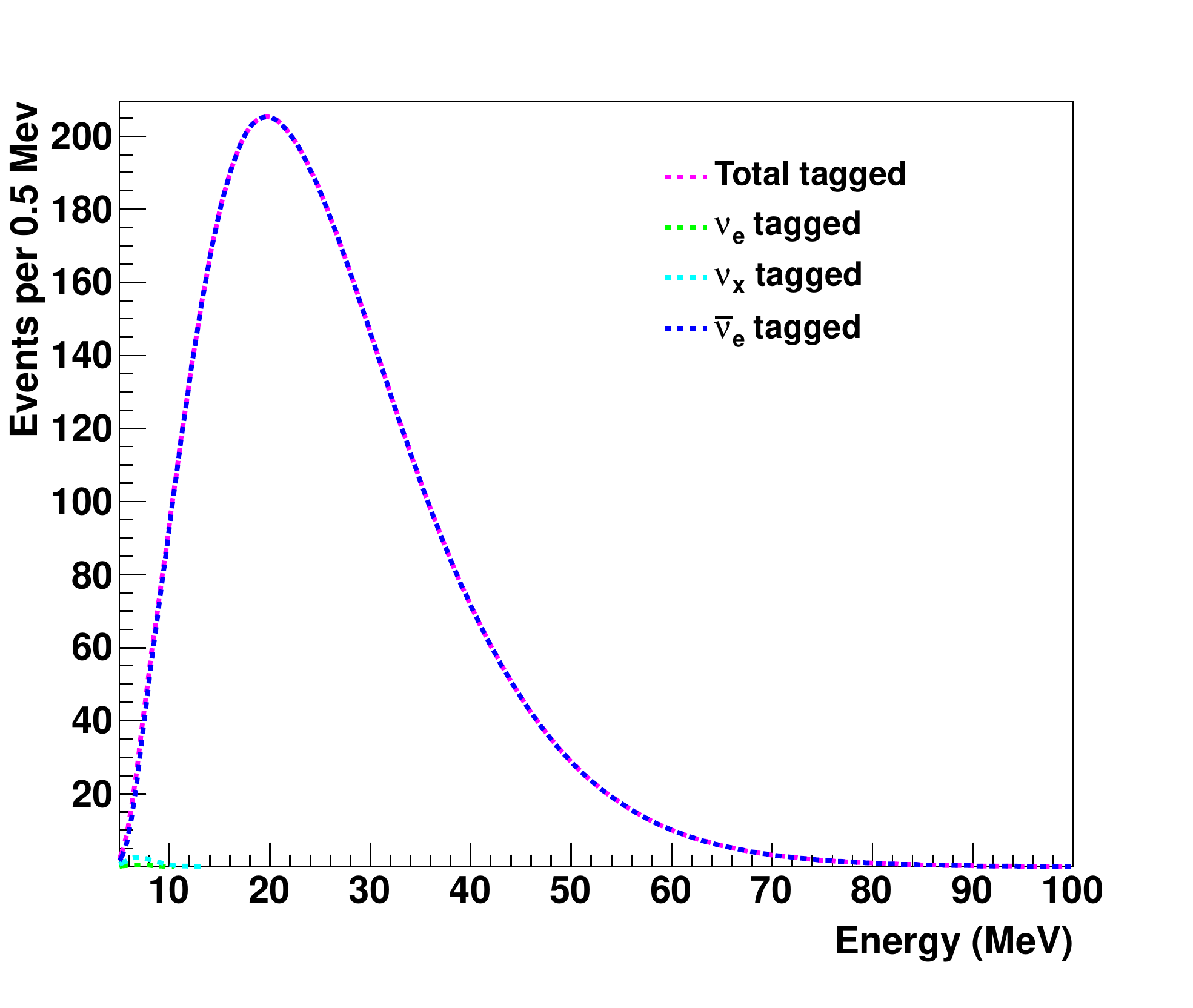}
\includegraphics[height=2.7in]{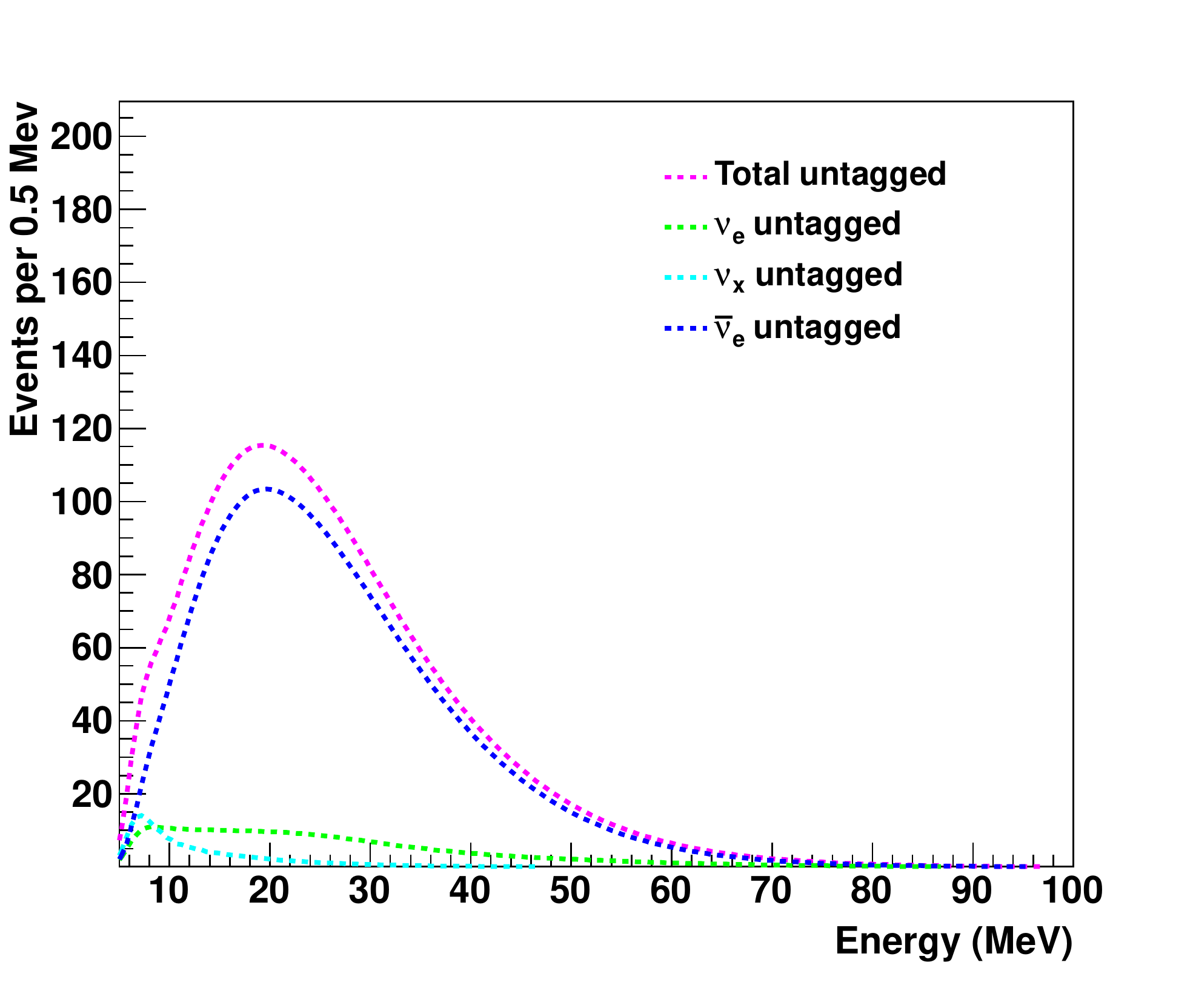}
\caption{Total events in WC showing contribution from the different
flavors, for neutron-tagged (left) and untagged (right).}
\label{fig:flavor_content}
\end{centering}
\end{figure}

Figure~\ref{fig:flavor_content} shows the contributions of the
different neutrino flavors for tagged and untagged events.   The
tagged sample is nearly pure $\bar{\nu}_e$.  The untagged sample has
contributions from other flavors, and large contamination from
untagged IBD $\bar{\nu}_e$.

\begin{figure}[!ht]
\begin{centering}
\includegraphics[height=2.7in]{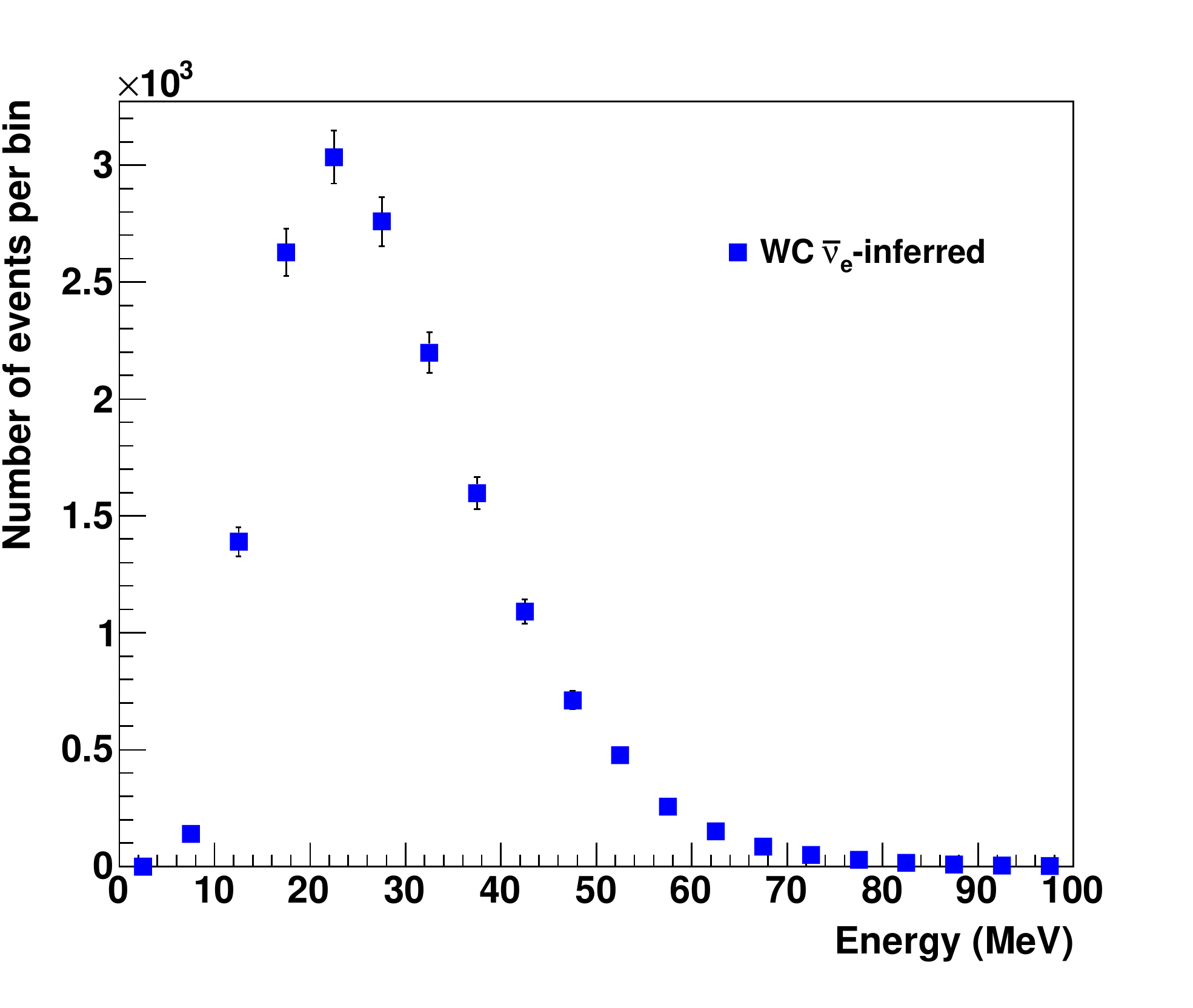}
\includegraphics[height=2.7in]{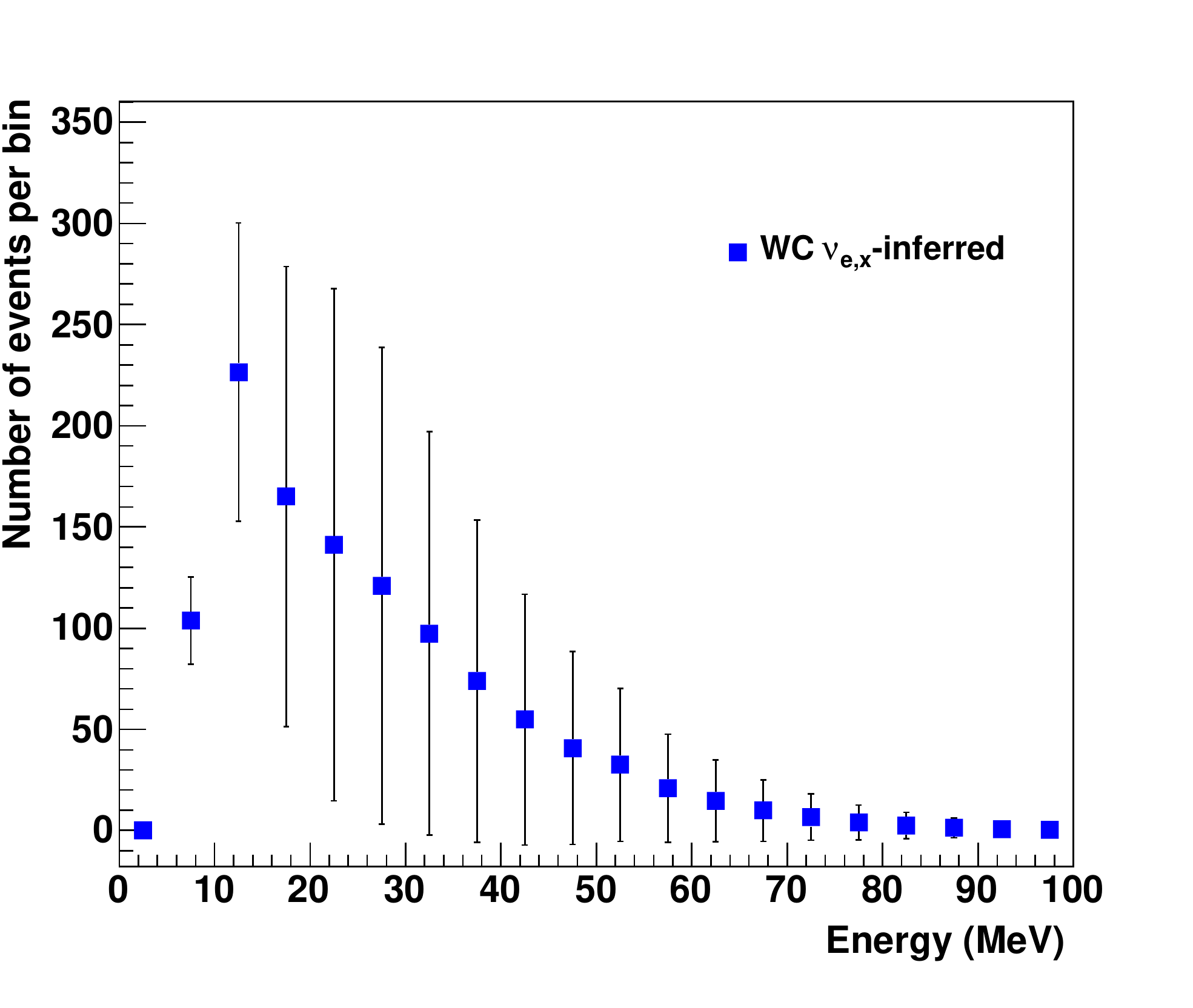}
\caption{ Inferred $\bar{\nu}_e$ (left) and $\nu_{e,x}$ (right) signals using neutron-tagging information, with uncertainties, for the GKVM flux. }
\label{fig:subtracted}
\end{centering}
\end{figure}

Given a burst neutrino signal, one would estimate flavor composition using the known tagging fraction.  The tagged sample represents a $\bar{\nu}_e$-enriched sample (in fact, it is a nearly pure one).  To estimate non-$\bar{\nu}_e$ (\textit{i.e.} $\nu_{e,x}$; $x$ refers to any non-electron flavor) content in the untagged sample (which has more $\nu_e$ than any other flavor) one can subtract the IBD contribution measured using the tagged rate.  Figure~\ref{fig:subtracted} shows plots with error bars of estimated $\bar{\nu}_e$ flux and $\nu_{e,x}$ flux.

Tagging here clearly allows selection of a clean $\bar{\nu}_e$ sample; however the $\nu_{e,x}$-enriched sample suffers significant contamination, and quality of determination of
non-$\bar{\nu}_e$ component of the flux will be dependent on tagging fraction uncertainty (as well as on relative size of that component).    The mean number of $\nu_{e,x}$ events must be greater than the uncertainty on the
number of untagged $\bar{\nu}_e$ events
for the
non-$\bar{\nu}_e$ component to be measurable.  Figure~\ref{fig:subtracted} clearly shows that given our assumptions, it will be possible to determine the size of the $\nu_{e,x}$ component for the GKVM flux at 10~kpc.  However the untagged sample represents a mix of flavors, and the spectral information for any given flavor is not cleanly measured.

\textit{Selection of Elastic Scattering Events:}\label{es}
Another potential method for selecting a flavor-enhanced sample in WC is to use the directionality of the neutrino-electron scattering signal:   electrons are scattered away from the supernova.  (This anisotropy is likely the best method for pointing to the supernova~\cite{Beacom:1998fj,Tomas:2003xn}.)  The ES sample is enriched in $\nu_e$ and $\nu_x$ relative
to $\bar{\nu}_e$.
The relative amounts of the different flavors are sensitive to the
neutrino spectrum, and the fractions vary substantially from model to model.

\begin{figure}[!ht]
\begin{centering}
\includegraphics[height=2.7in]{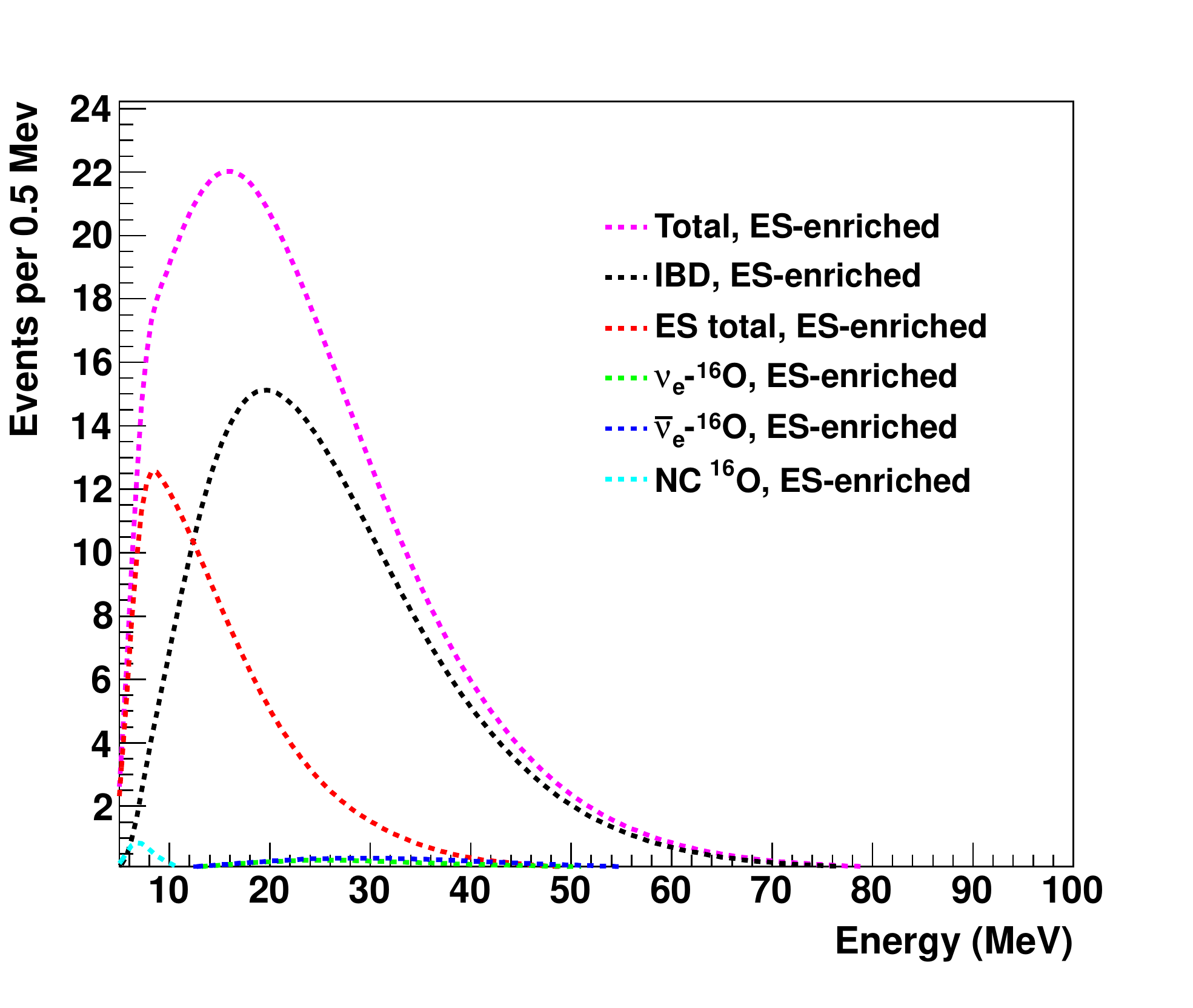}
\includegraphics[height=2.7in]{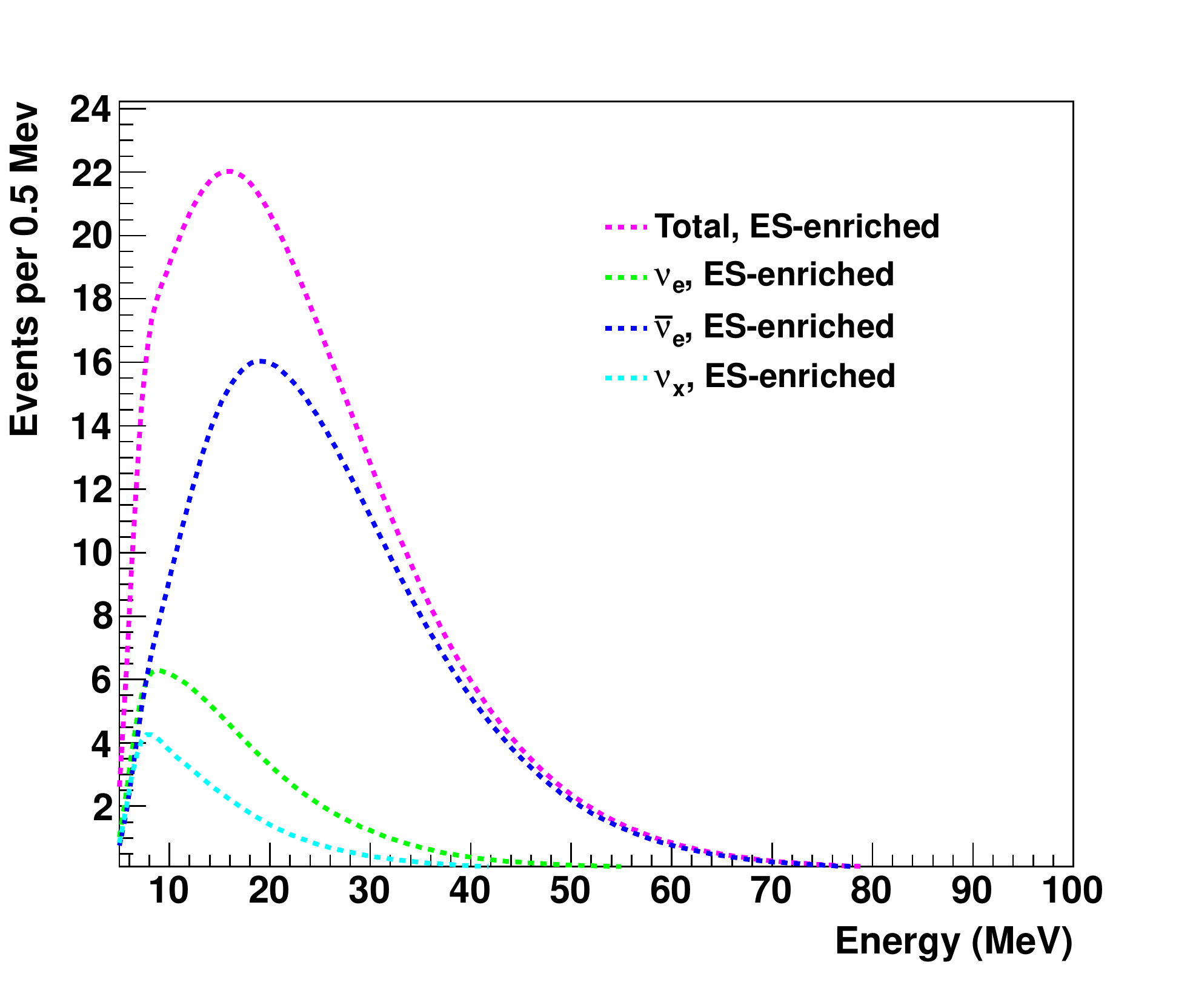}
\caption{ For the GKVM model, interaction (left) and flavor (right) composition of the ES-enriched sample. }
\label{fig:es_enriched}
\end{centering}
\end{figure}

\begin{figure}[!ht]
\begin{centering}
\includegraphics[height=2.7in]{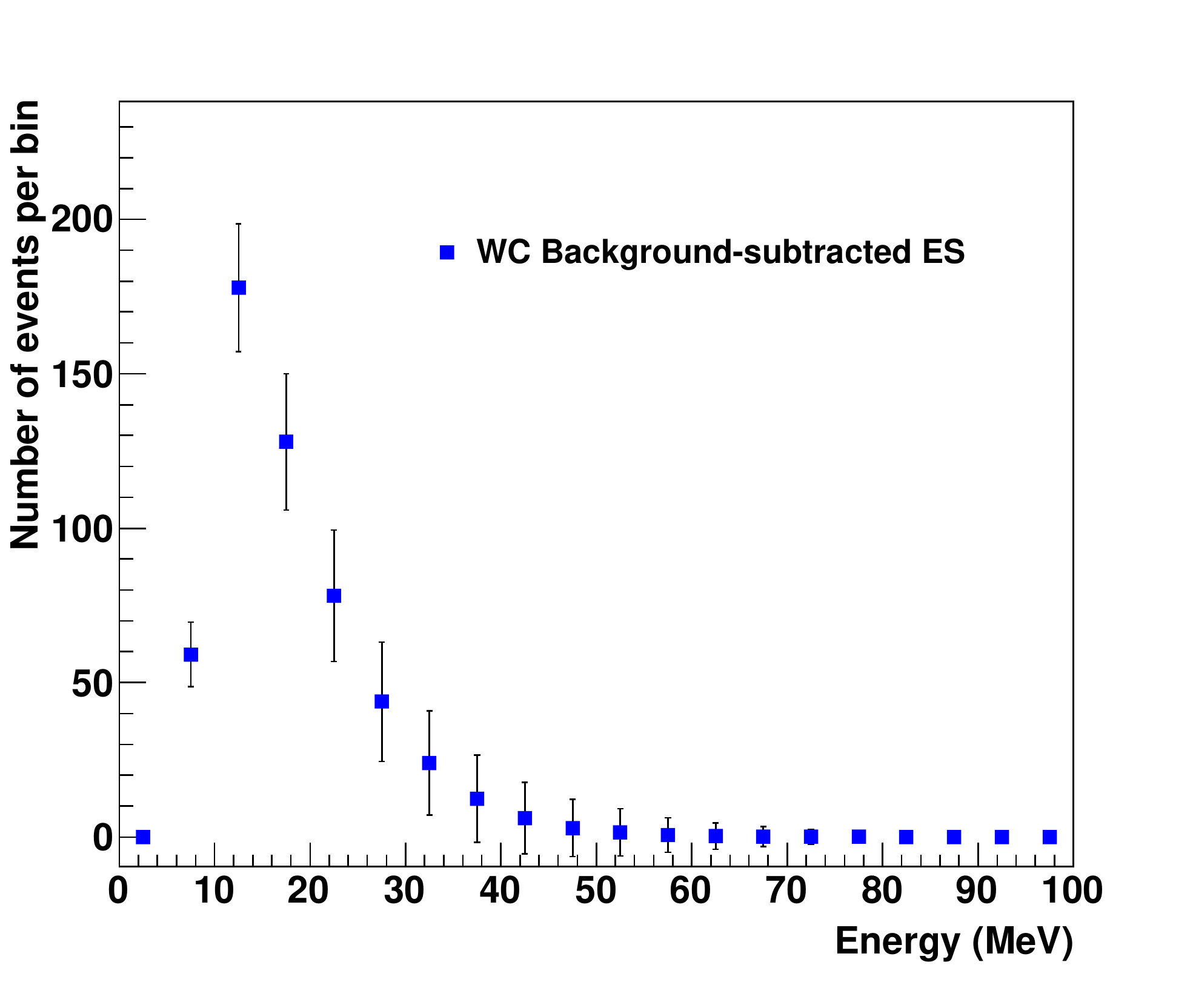}
\caption{ Background-subtracted ES signal in 5~MeV bins, for the GKVM flux.  Error bars are statistical.}
\label{fig:es_enriched_bg_subtracted}
\end{centering}
\end{figure}

We estimate the quality of a flavor-enriched ES sample by assuming that a fraction $\epsilon_s= 0.66$ of ES events will have $\cos\theta>0.9$, where $\theta$ is the reconstructed angle of the scattered event~\cite{Beacom:1998fj}.  Such a cut will reduce isotropic background by 95\%.
Figure~\ref{fig:es_enriched} shows the interaction and flavor compositions of the ``ES-enriched'' sample selected by an angular cut, for the GKVM model.
The non-ES  background can be determined by counting events outside of the angular cut window
(where we assume for the purpose of determining statistical uncertainty on the ES background subtraction that non-ES events have isotropic background, although that is not completely true-- IBD events have a weak asymmetry, and interactions on $^{16}$O have a backwards asymmetry).
Figure~\ref{fig:es_enriched_bg_subtracted} shows the background-subtracted ES-enriched signal, with statistical error bars, for the GKVM flux.

\textit{Combining Neutron Tagging and Angular Information:}\label{combining}
It should also be possible to combine angular selection with
neutron-tagging to further enhance flavor information in the ES signal.  First, because ES events do not have neutrons, the isotropic background in this sample can be reduced by selecting only untagged events.

\begin{figure}[!ht]
\begin{centering}
\includegraphics[height=2.7in]{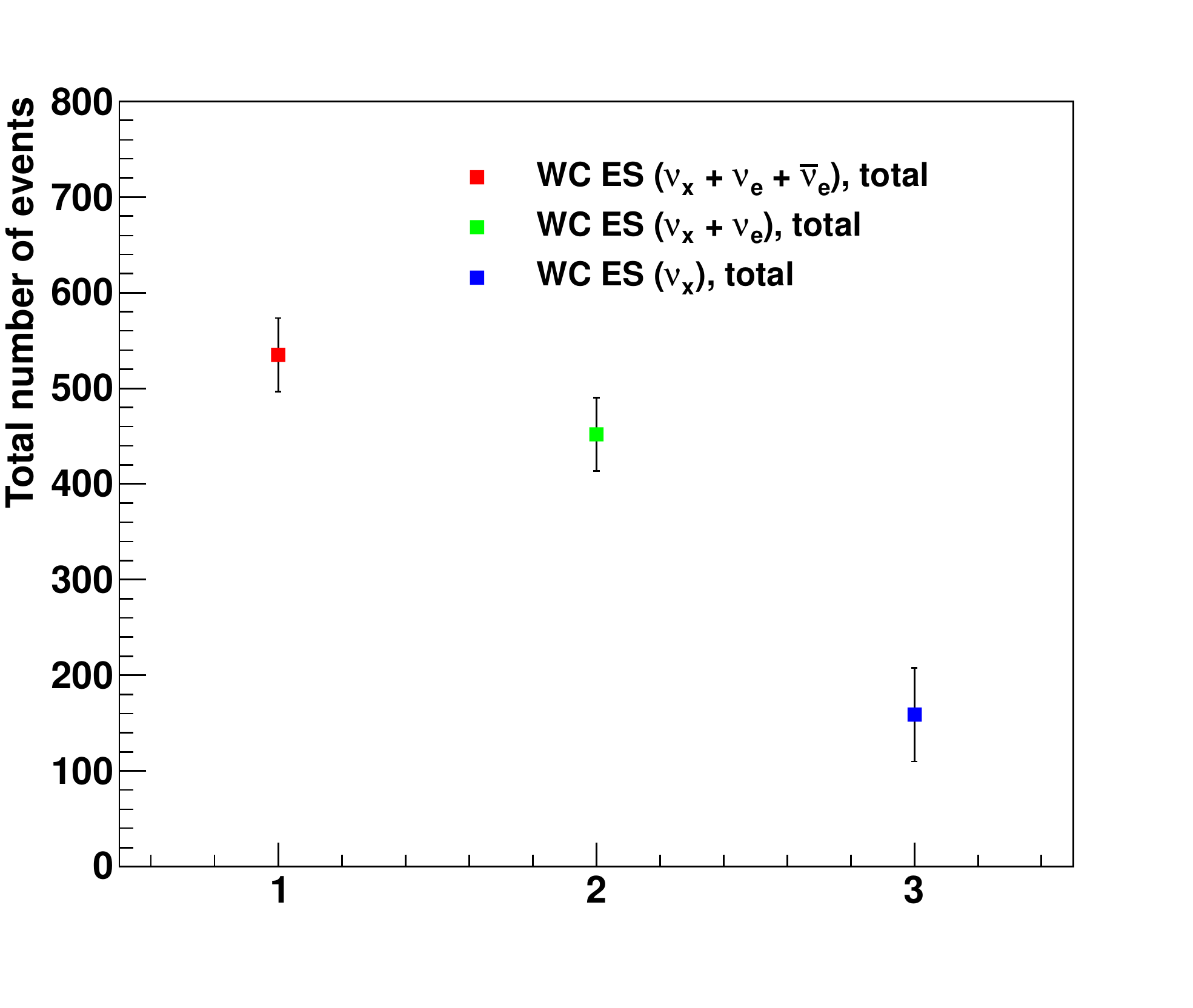}

\caption{Inferred flavor components of a WC ES signal, assuming neutron-tagging, angular selection, and a $\nu_e$ measurement from LAr. 1:  total ES signal from all flavors;  2: non-$\bar{\nu}_e$ flavors; 3: $\nu_x$ flavor.}
\label{fig:compare_es_enriched_bg_subtracted}
\end{centering}
\end{figure}

Second, one can use the tagged sample to determine the
$\bar{\nu}_e$ flux, and then subtract the $\bar{\nu}_e$ component
from the ES sample to determine the $\nu_{e,x}$ content.
Going one step farther:
if there were an independent measurement of the $\nu_e$
flux from LAr (or some other detector), one could in principle determine
the $\nu_x$ flux, of considerable interest in itself.
Figure~\ref{fig:compare_es_enriched_bg_subtracted} summarizes the  total ES, $\nu_{e,xES}$, and $\nu_{xES}$ scattering rates for the GKVM flux, assuming neutron-tagging and angular selection.

\subsection{Next Steps}

We have made preliminary estimates of event rates and simple ``anecdotal'' evidence of observability of oscillation features for a supernova burst signal, as well as simple studies of information available from flavor tagging.  There are several improvements and further studies in progress.

\begin{itemize}

\item We will refine the detector response parametrization as simulations improve.

\item The information on products (especially deexcitation gammas and ejected nucleons) of interactions on $^{16}$O and $^{40}$Ar is quite sparse in the literature.    We will work on improving our modeling of these interactions.  Other isotopes of oxygen and argon, although making a small contribution to the total signal, should also be considered.
 We will also study the effect of systematic uncertainties (\textit{e.g.} on the cross sections) on the physics sensitivity.

\item Angular distributions of products will provide valuable information.  ES and CC interactions on oxygen have pronounced anisotropy (IBD is weakly anisotropic).  These anisotropies can be exploited for pointing to the core collapse~\cite{Beacom:1998fj,Tomas:2003xn} (for an early alert, or to aid in finding the remnant in the case of weak supernova signal), for disentangling flavor components, and for making more precise measurements of
the neutrino energy spectrum.  We will evaluate the angular distribution of the expected signal.

\item In practice, given a burst signal, one would perform a multiparameter fit to all available energy, angle and flavor information in order to extract supernova and oscillation physics.  We will develop such an analysis and explore the physics sensitivity given different models.

\item Sensitivity to numerous other physics features will be explored.  For example, matter oscillations in the Earth may provide additional information about neutrino oscillation parameters.  We will explore the potential to observe physics signatures, possibly in conjunction with other experiments likely to be running over the next decades.  Other examples include: observability of the breakout peak, the accretion-to-cooling transition, and transition to a black hole.

\item At the moment we have very little information on the nature of spallation background in argon,  on the potential quality of signal tagging  in LAr using gammas, and on background reduction in an LAr detector.  These issues are critical
for evaluation of relic supernova neutrino sensitivity, and they may also be relevant for burst supernova neutrinos, especially for low-statistics bursts for core collapses beyond 10~kpc.  We will continue to investigate these issues.

\end{itemize}

\subsection{Conclusions}

Table~\ref{tab:designs} shows overall evaluation of the different reference configurations.

\begin{table}[h]
\begin{tabular}{|l|c|c|c|} \hline
Configuration  & Events in water & Events in argon & Relative \\
Number         &  at 10~kpc       &   at 10 kpc     &   hierarchy\\
               &                  &                 &    sensitivity \\  \hline
  1   & 60,000 & N/A & 2.6 \\
  1a  & 66,000  & N/A  &  2.6\\
  1b  & 66,000; enhanced flavor ID&  N/A& 2.6+ \\
  2   & N/A & 4500 &  1.7 \\
  2a  & N/A & 4500 & 1.7  \\
  2b  & N/A & 4500 & 1.7 \\
  3   & 40,000 & 1500 & 2.1+ \\
  3a  & 44,000  & 1500 &  2.1+ \\
  3b  & 44,000; enhanced flavor ID & 1500  & 2.1++  \\
  4   & 40,000 & 1500 &  2.1+ \\
  4a  & 44,000 & 1500 &  2.1+ \\
  4b  & 44,000; enhanced flavor ID & 1500 & 2.1++  \\
  5   & 22,000; enhanced flavor ID & 3000 & 1.5++ \\
  6   & 22,000; enhanced flavor ID & 3000 &  1.5++ \\ \hline

\end{tabular}
\caption{ Summary of supernova burst capabilities of the reference configurations (see Table~\ref{tab:refconfigs} for more details).
See text for explanation of the last column, which refers to a specific model. A ``+'' is included if Gd is present to enhance flavor tagging, and ``++'' indicates a configuration in which both LAr and WC are running simultaneously.}
\label{tab:designs}
\end{table}

Sensitivity to physics in a supernova burst is good for any of the configurations; it improves with larger active mass.  The 15\% coverage gives somewhat degraded performance with respect to the 30\% coverage: about 10\% of the total number of events are lost.  All of the loss is below 10~MeV and includes a very large fraction of NC $^{16}$O excitation events.  The addition of Gd will improve flavor tagging.
A combination of different detector types offers the best physics sensitivity, because of ability to distinguish different flavor components of the supernova burst flux.

\vfill\eject
%
\clearpage

\section{Supernova Relic Neutrinos}\label{SRNsect}

All of the neutrinos which have ever been emitted by every supernova since the onset of stellar formation suffuse the universe.  These supernova relic neutrinos [SRN], also known as the diffuse supernova neutrino background [DSNB], have not yet been observed but they may be just of reach of the current generation of operating detectors. In this section we will see that with the appropriate technology the relic neutrinos can play a unique and powerful role in the physics output that can be expected from LBNE.

\subsection{Motivation and Scientific Impact of Future Measurements}

\subsubsection{Motivation for the Measurement}

Supernova neutrinos carry unique information about one of the most dramatic processes in the stellar life-cycle, a process responsible for the production and dispersal of all the heavy elements (i.e., just about everything above helium) in the universe, and therefore a process absolutely essential not only to the look and feel of the universe as we know it, but also to life itself.

As a gauge of the level of interest in these particular particles, it is worth noting that, based upon the world sample of twenty or so neutrinos detected from SN1987A, there has on average been a paper published once every ten days... for the last twenty-three years.  After a quarter of a century, this handful of events remain the only recorded neutrinos known to have originated from a more distant source than our own Sun (by an easily-remembered factor of 10$^{10}$).

\begin{figure}[htb]
     \centering\includegraphics[width=.6\textwidth]{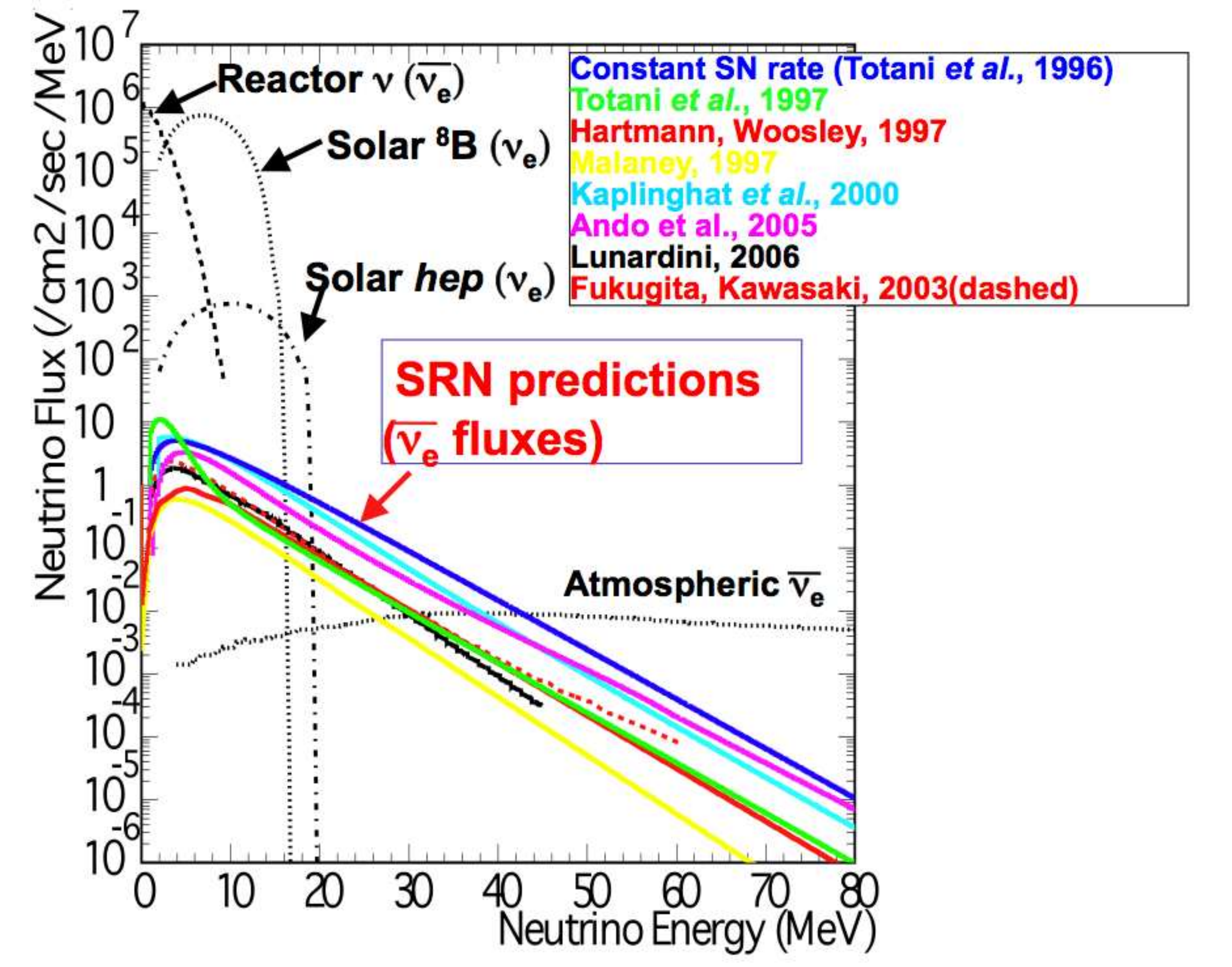}
     \caption{A variety of predicted SRN spectra and some key neutrino backgrounds.}
     \label{fig:pred_spec}
\end{figure}

Unfortunately, galactic supernovas are relatively rare, occurring somewhere between once and four times a century (Section \ref{snbintro}). However, while nearby supernovas are rare, supernovas themselves are not -- there are thousands of neutrino-producing explosions every hour in the universe as a whole.
There is a small but dedicated industry devoted to trying to predict the flux of these relic supernova neutrinos here on Earth; a representative selection of modeled spectra~\cite{Totani:1995dw,Sato:1997sc,Hartmann:1997qe,Malaney:1996ar,Kaplinghat:1999xi,Ando:2005ka,Lunardini:2006pd,Fukugita:2002qw} are shown in Fig.~\ref{fig:pred_spec}, along with some of the key physics backgrounds one might face.

\begin{figure}[htb]
     \centering\includegraphics[width=.5\textwidth]{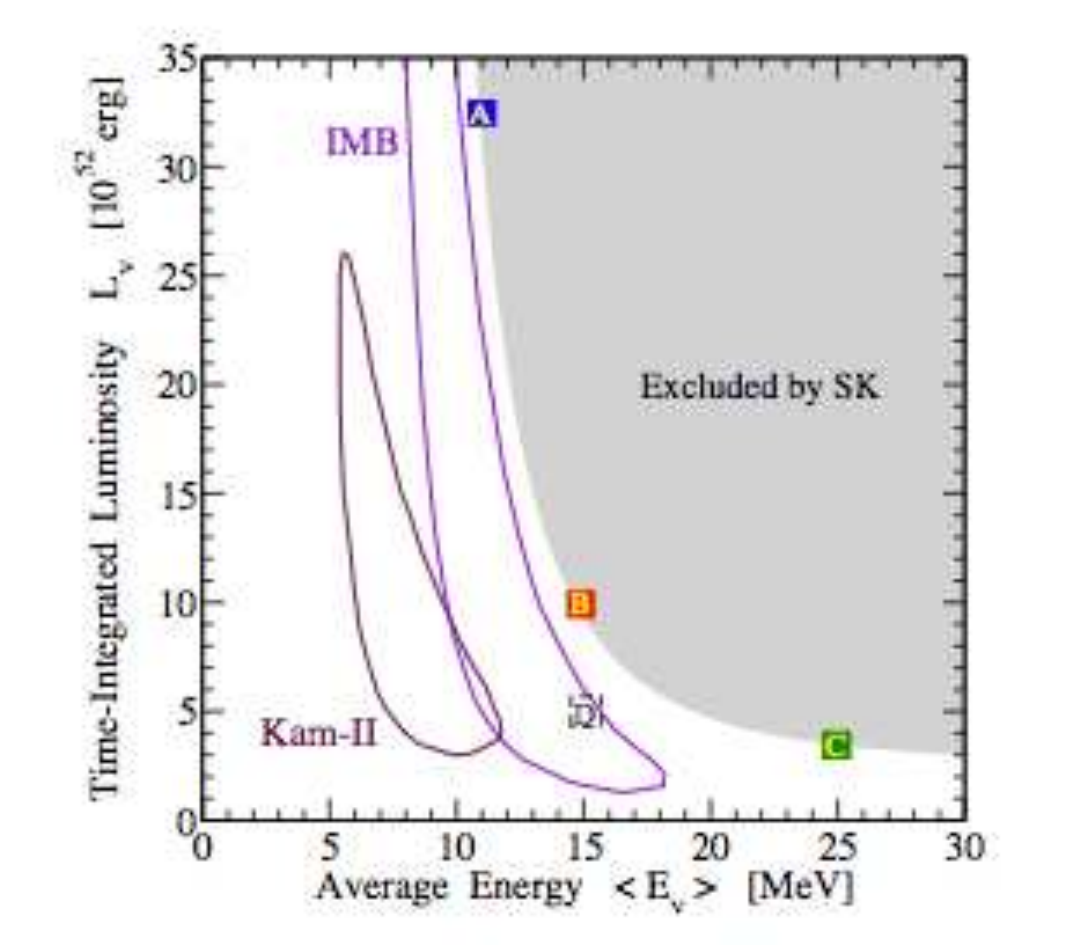}
     \caption{The parameter space of total supernova neutrino luminosity vs. average energy showing the approximate allowed regions from the SN1987A data, along with the region excluded by the 2003 Super--K SRN limit.}
     \label{fig:87_relic}
\end{figure}

\begin{figure}[htb]
     \centering\includegraphics[width=.5\textwidth]{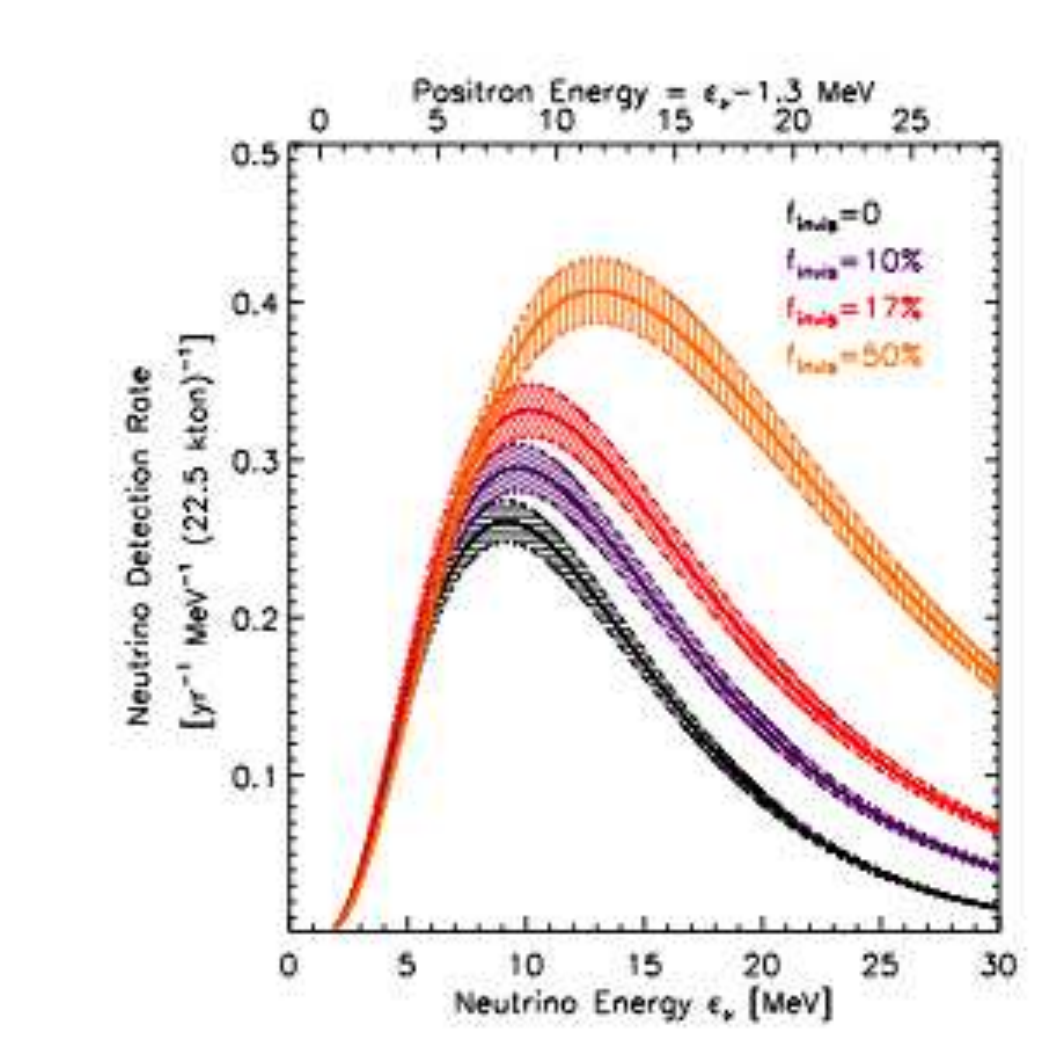}
     \caption{Expected SRN spectra in 22.5~kt of water Cherenkov volume per year as a function of different invisible supernova fractions within the range currently allowed.  The width of the bands represents the expected state of knowledge (from upcoming supernova survey missions) of the stellar core collapse rate by the time LBNE turns on, about a 5\% uncertainty.}
     \label{fig:invisible}
\end{figure}

So, in addition to being the first to measure this new flux of neutrinos, what scientific benefits would such a measurement bring?  Aside from any potential surprises, which we have often encountered when peering into a new corner of the neutrino sector, measuring the relics can shed light on the following topics:

\begin{itemize}
\item Understanding supernovas, central to understanding many aspects of the present physical universe, requires the detection of their neutrino emissions. More supernova neutrino data is strongly needed, but galactic supernova explosions are rare; the SRN will provide a continuous stream of input to theoretical and computational models.
\item The shape of the SRN spectrum will provide a test of the uniformity of neutrino emissions in core-collapse supernovas, determining both the total and average neutrino energy emitted.
\item Was SN1987A a ``normal'' explosion or not?  The sparse, 23-year-old data concerning a single neutrino burst cannot say, but the SRN data can (see Fig.~\ref{fig:87_relic}).
\item How common are optically dark explosions?  No one knows. Comparing the SRN rate with optical data of distant supernovas can tell us, and is probably the only way to get this information (see Fig.~\ref{fig:invisible}).
\end{itemize}

Contrary to what is often stated, measuring the SRN flux will not uniquely determine the cosmic core-collapse (and hence star formation) rate,  a key factor in cosmology, stellar evolution, and nucleosynthesis.  Rather, the SRN measurement will be a new and independent probe of this rate, which will be well constrained by current and near-future astronomical observations~\cite{Lien:2010yb}.

\subsubsection{Predicting the Relic Flux}

Predicting what the flux of neutrinos generated by these distant explosions will look like is a complex and subtle business, and the details of the calculations are well beyond the scope of this brief section.  However, there are two extensive reviews of the subject~\cite{Beacom:2010kk,Lunardini:2010ab} from which many of this SRN section's theoretical plots and event rate predictions are drawn.

That being said, here are the basic components that necessarily must go into any such calculation:

\begin{enumerate}
\item The initial supernova collapse model, and its neutrino emission.  Computational models have famously resisted producing robust, realistic explosions, so there is no one ``standard'' explosion model as of yet. These models encode both the total and average neutrino energies, and are  responsible for  much of the variation seen in predicted relic rates.
\item The oscillations and self-interactions of the neutrinos as they exit the dying star.  This determines the flavor mix which will eventually reach Earth, a key component in predicting various detectors' responses.  There is still considerable uncertainty in the precise blend of flavors which will arrive at Earth, which is also reflected in the width of the predicted rates for each detector configuration.
\item The stellar core collapse rate, which is expected to be proportional to the star formation rate.  This is now determined from astronomical observations to $\pm40$\%, and the next generation of synoptic sky surveys are expected to narrow this uncertainty to the 5\% level within the next few years.  This contributes an overall (though steadily shrinking) normalization uncertainty to the predictions.
\end{enumerate}

Once the relic flux and spectrum at the Earth is predicted, then the rate of neutrino interactions in each detector must be calculated using the best available knowledge of the relevant cross sections.  Finally, performance characteristics of the individual detectors and their locations on Earth must be taken into account, such that backgrounds can be estimated and appropriate energy windows for relic detection defined.

What all this boils down to is an uncertainty on the predicted relic rates of about a factor of 12 for a conventional (SK-style) water Cherenkov detector, and about a factor of six for a gadolinium-loaded water Cherenkov detector (described in more detail later) due to its widened energy window covering more of the competing models' spectral ranges.

While its energy window is quite similar to that of a regular water Cherenkov detector, a liquid argon detector has an uncertainty in the predicted relic rate of only about a factor of seven.  This is due to more solid predictions for the survival probability at Earth of the ${\nu}_{e}$ seen in argon as opposed to the $ \overline{\nu}_{e}$ seen in water.

\subsubsection{Current and Future Experimental Status}

The only significant competition for  relic neutrino discovery comes from the Super--Kamiokande experiment [generally referred to as Super--K or just SK], which is located at a depth of 3300 feet in an old zinc mine near Kamioka, Japan.  As a 50 kiloton (22.5~kt fiducial) water Cherenkov detector, Super--K's sensitivity to the SRN is strictly through the inverse beta reaction, $\overline{\nu}_{e} + p \rightarrow e^+ + n$.  The outgoing positron makes Cherenkov light which can be detected (note that WC detectors generally cannot differentiate between matter and antimatter), while the neutron is eventually invisibly absorbed by a hydrogen nucleus, forming a deuteron.

\begin{figure}[htb]
     \centering\includegraphics[width=.45\textwidth]{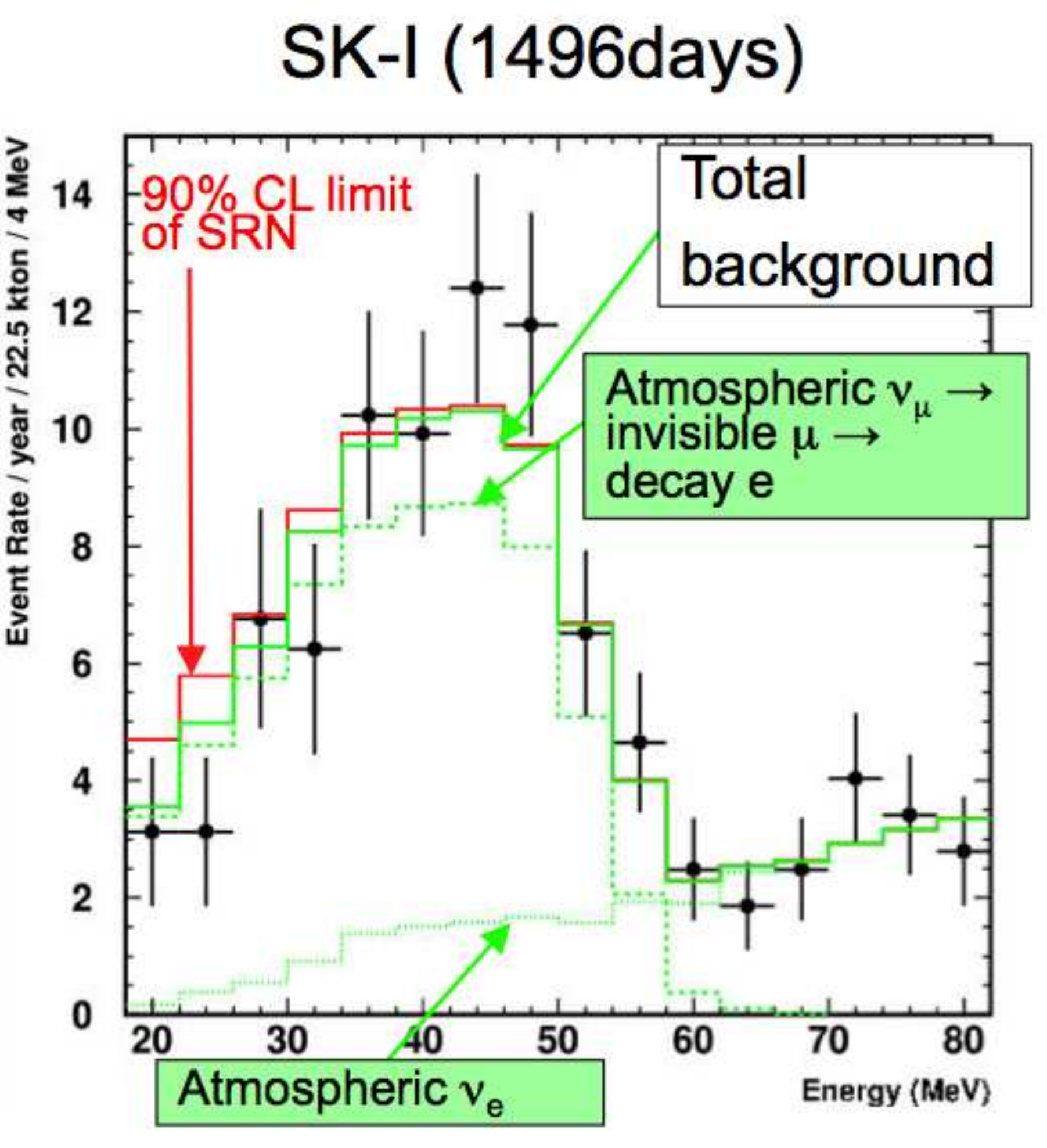}
     \caption{The data (points) from the 2003 Super--Kamiokande relic search.  The dashed histogram peak centered around 42~MeV is that expected from decays of sub-Cherenkov muons into electrons, while the dotted line slanting down from right to left is that expected to be produced by atmospheric electron neutrinos and antineutrinos.  The (green) lower solid line is the sum of these two backgrounds, while the (red) upper solid line represents the 90\% upper limit on potential excess caused by the relic neutrinos.}
     \label{fig:SKI_relic}
\end{figure}

In 2003, Super--Kamiokande published the results of a search for the supernova relic neutrinos~\cite{Malek:2002ns}.  However, as seen in Fig.~\ref{fig:SKI_relic} this study was strongly background limited, especially  by Michel decay electrons from sub-Cherenkov threshold muons produced by atmospheric neutrino interactions in the detector.  The low energy cutoff of the analysis was imposed not due to elastically scattered electrons from solar neutrinos, which SK can identify by their angular correlation with the direction back to the Sun, but due to by the many cosmic ray muon-induced spallation events below 19~MeV which unfortunately swamped any possible DSNB signal in that most likely energy range.  This plot is the result of 1496 days of data, or about 92~kt-years.  It took just over five years of continuous data-taking (April 1996 through July 2001) to collect 4.1 years of usable data for the relic analysis, a rather typical real-world duty cycle of 80\%.

\begin{figure}[htb]
     \centering
     \includegraphics[width=0.8\textwidth]{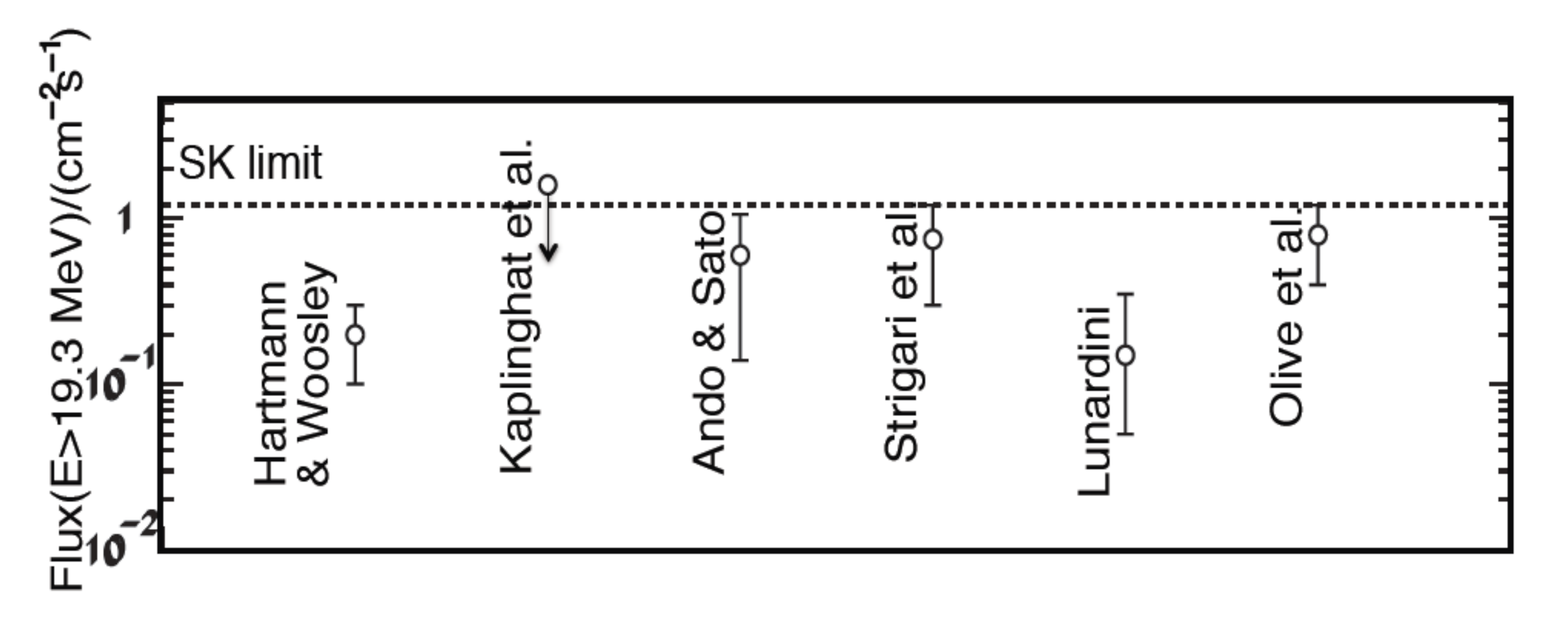}
     \caption{Super--K relic flux limit from 2003 compared with some theoretically predicted ranges.}
     \label{fig:predictions}
\end{figure}

\begin{figure}[htb]
      \includegraphics[width=0.8\textwidth]{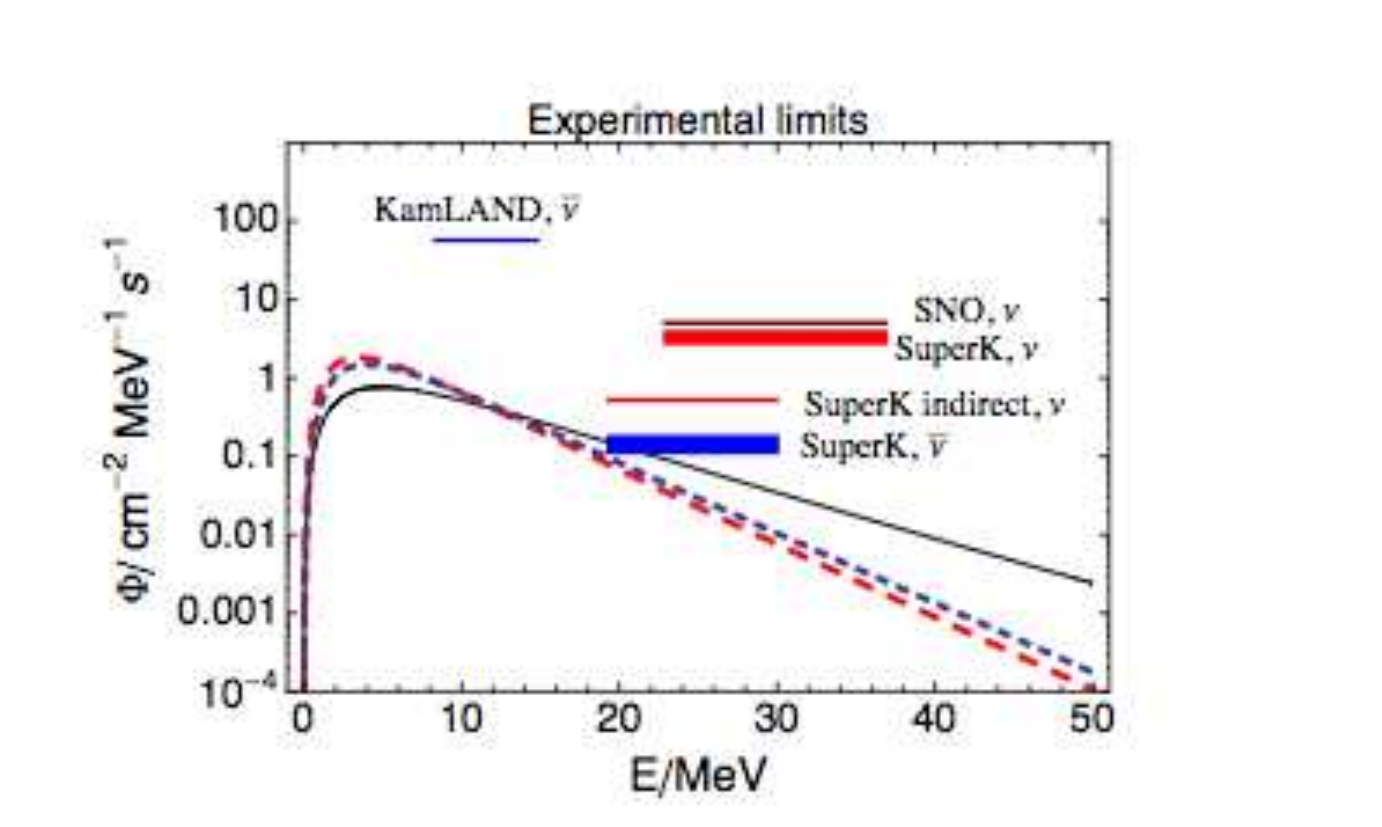}
     \caption{Experimental SRN limits are approaching theoretical curves.  The Super--K ${\nu}_{e}$ lines have been theoretically extracted based on the measured $\overline{\nu}_{e}$ results.}
     \label{fig:exp_limits}
\end{figure}

Consequently, this Super--K study could see no statistically significant excess of events and therefore was only able to set the world's most stringent upper limits on the relic flux:  $<1.2$~ $\overline{\nu}_{e}$~cm$^{-2}$~s$^{-1}$ for $E_{\nu} > 19.3$~MeV.  Fig.~\ref{fig:predictions} indicates how close this Super--K limit is to some of the predictions including their theoretical uncertainties, while Fig.~\ref{fig:exp_limits} compares the range of predicted spectra to several experimental limits. Note in both of these figures how tantalizingly close to theory the experimental limits from Super--K are getting: perhaps we are on the brink of discovery.

However, after seven more years of Super--K data-taking and intensive efforts to improve the relic analysis, it now appears unlikely that any existing, unmodified detector will be able to make this discovery.

Fifteen years from now, Super--K will have only roughly doubled its present (as yet largely unpublished) statistics.  In the useful energy range shown in Fig.~\ref{fig:SKI_relic} one expects between 0.25 and 2.8 signal events per year depending on the model~\cite{Ando:2005ka,Lunardini:2006pd,Beacom:2010kk,Strigari:2003ig}, as compared to a measured 14 background events. Therefore, in the best possible case -- which means the flux lies just below the current published SK limit -- fifteen years hence,  Super--K would have  recorded 54 relic events and 269 background events in the energy window between 19 and 30~MeV.  This would be a 3.3-$\sigma$ effect, and they could have beaten LBNE to the discovery.  However, if the relic flux is lower, say 1.4 events a year (which is still the {\em maximum} value for Lunardini's predicted range in Super--K), then after 29 years of operations Super--Kamiokande would have just 27 events and a mere 1.6-$\sigma$ effect; certainly no discovery. These numbers are based on 80\% typical detector livetime over 24 years, with the five year period between mid-2001 and mid-2006 having been removed from consideration for reasons apparent in Fig.~\ref{fig:SKII_relic}.

\subsection{Sensitivity of Reference Configurations}

In this section we consider the performance and sensitivity of the various reference configurations in R\&D Document 643v2 (Table~\ref{tab:sensitivity_summary} provides a summary). In the case of the relics this largely equates to discovery potential.

{\em Reference Configuration 1: Three 100~kt fiducial water Cherenkov detectors. Depth 4850 foot level of DUSEL. Coverage 15\% 10 inch high QE PMT's.}

Since 15\% coverage with high quantum efficiency photomultiplier tubes is nearly equivalent to 19\% coverage with normal efficiency tubes, it is instructive to examine the case of Super--Kamiokande--II.

\begin{figure}[htb]
     \centering\includegraphics[width=.45\textwidth]{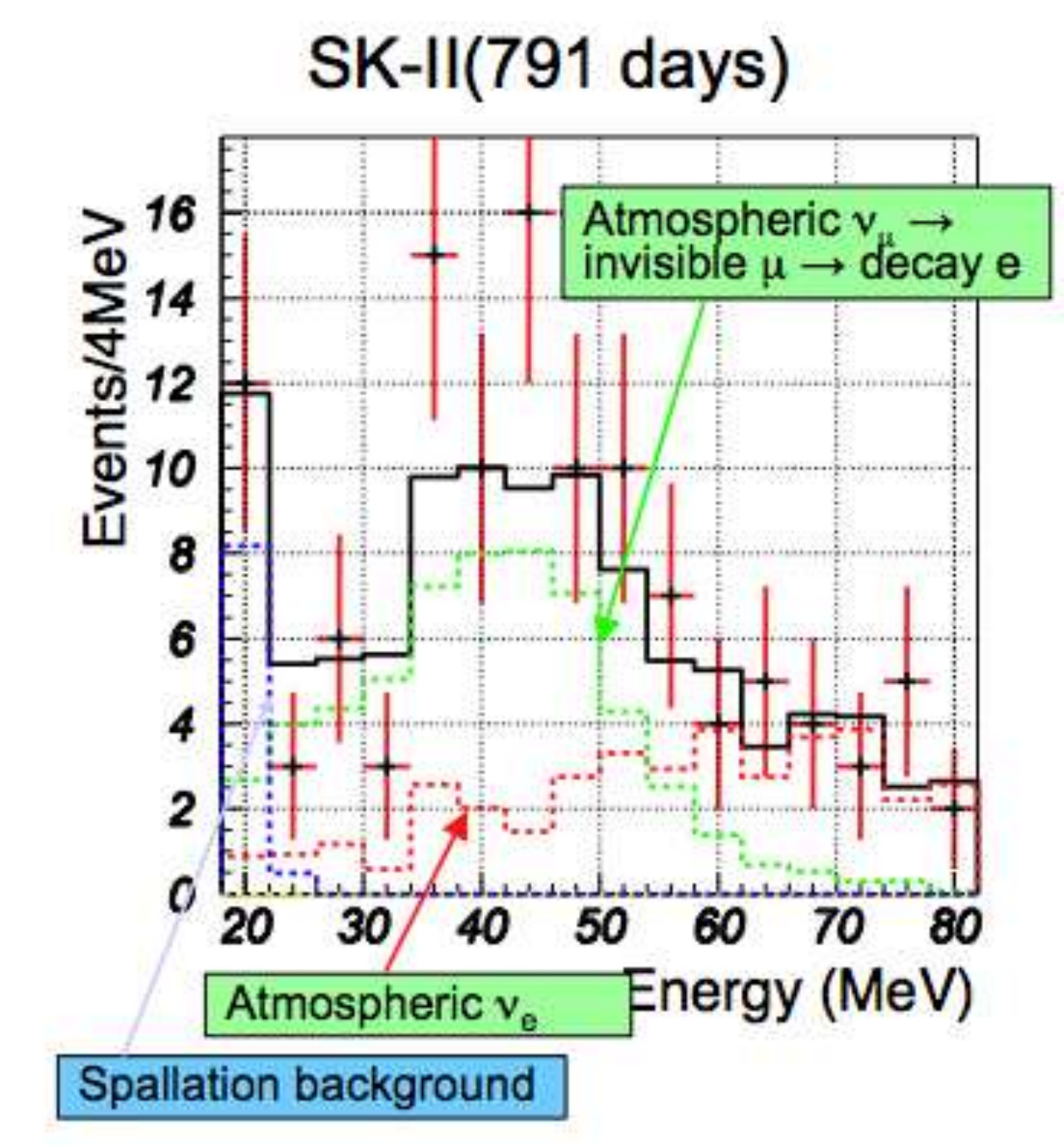}
     \caption{The Super--Kamiokande--II relic search.  The reduced photocathode coverage of SK--II (19\% vs. SK--I's 40\%) and shorter running time makes this a much less useful data set than that seen in Figure ~\ref{fig:SKI_relic}.}
     \label{fig:SKII_relic}
\end{figure}

After the Super--K accident in 2001, the detector was initially rebuilt with about half ($\sim$19\%) of the original photocathode coverage of 40\%.  This phase of operation, known as Super--K--II, yielded 49~kt-years of data before the full density of phototubes could be recovered in 2006.  Fig.~\ref{fig:SKII_relic} graphically demonstrates how much lower quality this data set was for the SK--II phase relic search as compared to the SK--I data shown in Fig.~ \ref{fig:SKI_relic}.  Due to impaired energy resolution, spallation events with true energies below 19~MeV can be seen leaking into the signal region.  This was reflected in this period's much degraded upper limit on the SRN flux:  $<3.7$  $\overline{\nu}_{e}$~cm$^{-2}$~s$^{-1}$ for $E_{\nu} > 19.3$~MeV.

Increasing the overburden from that at Kamioka to Homestake means there will be about a factor of 15 less spallation (see the DUSEL depth document), but a factor of 1.5 more atmospheric neutrinos (see Wurm et al.).  This will serve to make reference configuration 1's performance very similar to Super--Kamiokande--I's. The spallation leakage in Fig.~\ref{fig:SKII_relic}  will be nearly eliminated, but the backgrounds per unit volume seen in Fig.~\ref{fig:SKI_relic} are increased by 50\%.

Therefore, expect 280 background events per year in 300 kilotons, and somewhere between 3 and 38 relic events per year. This range in the relic flux reflects the range of the various models' predictions, with the upper bound somewhat below the current SK limit. It is evident that this configuration will probably not be much of a discovery machine in terms of relic detection. In the best possible case, reaching 3.0-$\sigma$ would take a minimum of 2.2 years of operation.  If the true relic rate is half of the maximum prediction (but still six times the minimum!) then getting to 3.0-$\sigma$ would take a painful but possible 9 years.  However, if the true relic rate falls at the low end of the predicted range, then this particular reference configuration would take 350 years to reach 3.0-$\sigma$.\\

{\em Reference Configuration 1a: Three 100~kt fiducial water Cherenkov detectors. Depth 4850 foot level of DUSEL. Coverage 30\% 10 inch high QE PMT's.}

Since 30\% coverage with high quantum efficiency photomultiplier tubes is nearly equivalent to 40\% coverage with normal efficiency tubes, it is instructive to examine the case of Super--Kamiokande--I.

The lowered spallation rates mean that the energy window can be extended 2.5~MeV below that used in SK--I and shown in Fig.~\ref{fig:SKI_relic} , but the higher atmospheric backgrounds are again troublesome. The enlarged energy window means that we should expect  320 background events per year in 300~kt, while at the same time increasing the expected relic flux by a factor of 1.4, making the predicted range of relic events per year fall between 5 and 52.

Therefore, repeating the same calculations as above, in the best case the 3.0-$\sigma$ level would take just 1.3 years of running at 80\% livetime.
If the true relic rate is half of the maximum prediction (but still five times the minimum) then getting to 3.0-$\sigma$ would take a tolerable 5.3 years.  Unfortunately, at the bottom of the predicted SRN range, getting to  3.0-$\sigma$  would take 144 years.

If the relic signal turns out to lie not far from the current experimental limits, giving us a strong hint after just a few years, this detector configuration could -- in principle -- be improved by adding gadolinium in order to quickly achieve a conclusive discovery.  Which brings us naturally to...\\

{\em Reference Configuration 1b: Three 100~kt fiducial water Cherenkov detectors. Depth 4850 foot level of DUSEL. Coverage 30\% 10 inch high QE PMTs. Gadolinium dissolved in water.}

\begin{figure}[htb]
     \centering
     \includegraphics[width=0.5\textwidth]{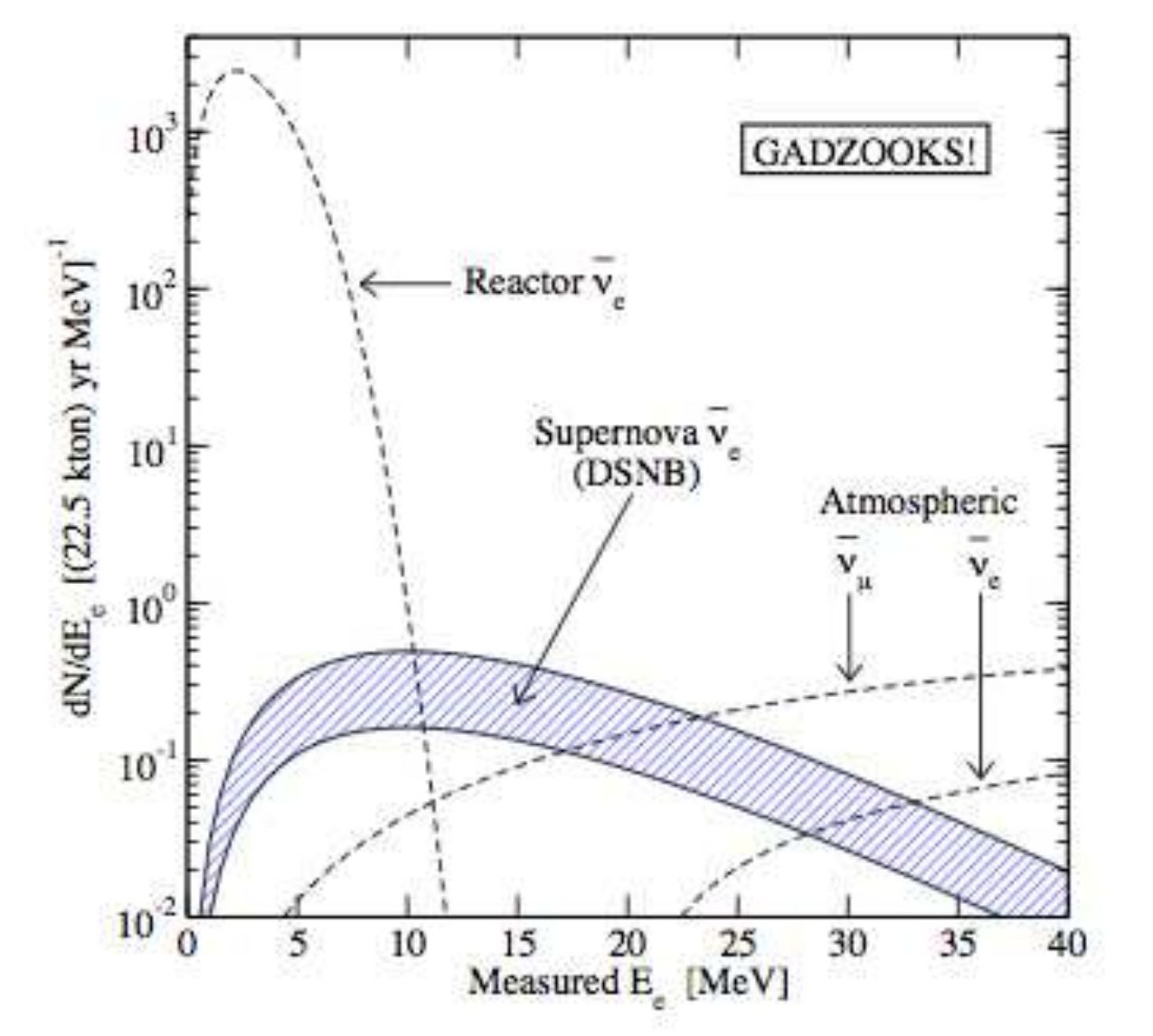}
     \caption{Expected spectrum of positrons seen in coincidence with neutron capture signals in a gadolinium-loaded Super--Kamiokande detector.  The band shows the theoretical range of predictions for the diffuse supernova neutrino background (relic) flux.  GADZOOKS! stands for
\underline{G}adolinium \underline{A}ntineutrino \underline{D}etector
\underline{Z}ealously \underline{O}utperforming \underline{O}ld
\underline{K}amiokande, \underline{S}uper\underline{!}. }
     \label{fig:gad_sk}
\end{figure}

Inspired by the 2003 Super--K SRN limit, adding 0.2\% by mass of a soluble gadolinium compound like GdCl$_3$ or Gd$_2$(SO$_4$)$_3$ to water Cherenkov detectors  has been suggested~\cite{Beacom:2003nk}.   The neutrons produced by inverse beta reactions would thermalize in the water and then be captured on gadolinium, emitting an 8~MeV gamma cascade in the process.   In coincidence with the prompt position signal, this delayed neutron capture signal would serve to dramatically lower backgrounds from spallation and atmospheric neutrinos, at the same time allowing an enlarged energy window for detection of the relic neutrinos.  Fig.~\ref{fig:gad_sk} shows the expected signals in a gadolinium-loaded Super--K.  Such a 22.5~kt detector would expect to see between 1 and 5.6 relic events a year, with about 4 background events.

While this proposal has drawn both considerable interest and funding support, after more than half a decade of study~\cite{Watanabe:2008ru,Kibayashi:2009ih} it has not yet received approval from the Super--K leadership to move beyond the R\&D stage.   Consequently, its immediate future in Japan is uncertain, especially with regards to potentially perturbing the newly operational T2K experiment by attempting to modify Super--K in the early stages of the  long-baseline  experimental program.  For this reason and others (such as a desire to first fix a slow leak in the detector's main  tank) it {\em is} relatively certain that no gadolinium will be added to Super--K during the next five years.  Beyond that it is unknown what will happen, but if Super--K does add gadolinium in 2016 it will be able to extract meaningful spectral information before LBNE comes on-line only if the true relic flux is in the upper half  of the currently allowed range.  Regardless what decisions are made by the Super--K Collaboration in the coming years, higher statistics SRN data, which only a  next-generation detector can provide, will be urgently desired by the middle of the next decade.

Transferring these results to LBNE, the coincident detection made possible by gadolinium greatly reduces spallation, which is rarely accompanied by a neutron within the expected timing window~\cite{Koshio:1998yd}. This allows the SRN window to be opened all the way down to the irreducible wall formed by the antineutrinos from nuclear power reactors around the United States.  Note that, primarily due to the remoteness of its South Dakota location, this reactor flux is a factor of 24 times lower in Homestake than it is in Kamioka~\cite{Dye:liqscintwp}.  However, a 300~kt detector has a fiducial volume 13.3 times bigger than Super--K, so LBNE would observe about half the reactor rate that Super--K does.  As one can see from Fig.~\ref{fig:gad_sk}, a factor of two reduction in the reactor curve has very little effect on the available energy window for relic supernova neutrinos, perhaps gaining us another MeV or so.  At any rate, opening up the energy window down to 11.3~MeV means that we would expect between 13 and 74 SRN events per year in 300 kilotons of Gd-loaded detector. The coincident technique reduces the atmospheric neutrino background by about a factor of five, and so by comparison with configuration 1a we would expect 64 background events a year across the entire energy spectrum, with 87\% of this background in the bins above 19~MeV.

Repeating the same calculations as above over the entire energy window, in the best case the 3.0-$\sigma$ would take only 0.13 years (48 days) of running at  80\% livetime. If the true relic rate is half of the maximum prediction (and in this case just three times bigger than the minimum due to the more complete spectral range of the models contained in the window) then getting to 3.0-$\sigma$ would take an entirely painless 0.53 years.  Even in the most pessimistic case it would still only take us 4.3 years to reach the 3.0-$\sigma$ level.

The signal to noise ratio improves as one goes down in energy, however, so considering the entire spectral window is really a worst case analysis in terms of time to discovery.  As an example, if one restricts the range to between 11.3~MeV and 19.3~MeV, we would expect between 10 and 36 relic events per year in 300 kilotons, and 8 background events.  This means we would reach the 3.0-$\sigma$ level in less than one year, even for the worst case.

Clearly this is an SRN discovery configuration.  What's more, it will be able to extract spectral information in addition to a simple flux measurement.  The spectrum is a necessary input for the supernova modelers as it encodes the total neutrino energy and the average neutrino energy of bursts.  In concert with astronomical observations it also provides a way to determine the rate of invisible explosions (see Fig.~\ref{fig:invisible}~\cite{Lien:2010yb}).

Conversely, and perhaps even more interesting, if this configuration sees {\em no} relic signal after a year's time, then we would have a smoking gun for new physics, strongly implying the existence of neutrino decay, new particles, or something similar. This ``Diffuse Supernova Neutrino Problem''
would arise from a non-detection because we know from SN1987A that supernovas do emit neutrinos, and we know from the sky surveys about how many explosions there are, what types, and at what distances.  But thus far we have never seen neutrinos older than those from 170,000 light-years away. The relic neutrinos have had to travel four orders of magnitude further, and are therefore a more sensitive probe of unusual processes.\\

{\em Reference Configuration 2: Three 17~kt fiducial liquid argon detectors. Depth 4850 foot level of DUSEL. Assume a scintillation photon trigger for proton decay and supernova neutrinos.}

There are different issues to consider in the case of liquid argon, which detects relic supernova neutrinos primarily via the charged-current process
${\nu}_{e} + ^{40}$Ar  $\rightarrow$  $^{40}$K$^{*} + e^{-}$.   The electron track should be accompanied by evidence -- shorter tracks sharing a common vertex -- of ionization from the de-excitation of the potassium; this is expected to be helpful in reducing backgrounds.  Unlike inverse beta decay, whose cross section is known to the several-percent level in the energy range of interest~\cite{Vogel:1999zy,Strumia:2003zx}, the cross section for neutrino interactions on argon is uncertain at the 20\% level~\cite{Ormand:1994js,Kolbe:2003ys,SajjadAthar:2004yf}. Another wrinkle is that the solar {\em hep} neutrinos, which have an endpoint at 18.8~MeV, will determine the lower bound of the SRN search window.  The upper bound is determined by the atmospheric ${\nu}_{e}$ flux.

In the case of liquid argon no multi-kiloton detectors have yet been built.  In addition to having to rely on estimates of efficiencies and downtime in a huge new detector, this necessarily forces us to rely heavily on some important assumptions concerning potential background processes which might fall inside the 18 to 30~MeV window defined by the solar {\em hep}  and atmospheric ${\nu}_{e}$ fluxes.

For our analysis, we will adopt the following strong assumptions~\cite{Cocco:2004ac}.  They have been listed from least speculative (most certain) most speculative (least certain):

\begin{itemize}
\item No nuclear recoils from fast neutrons will be able to produce a signal-like event in the energy range of interest.
\item Unlike in water Cherenkov detectors, liquid argon detectors do not suffer from sub-Cherenkov muons decaying into electrons and faking the SRN signal, as no muons (or evidence of their decays) should escape detection in the detector.
\item No spallation products will be produced that generate electrons in the energy range of interest without clear evidence of their parent muon allowing the event to be removed from consideration.  (The full family of spallation daughters of argon does not seem to be known, but it must include all possible oxygen spallation products.)
\item No radioactive background or impurity, electronic effect in the detector, track-finding inefficiency, particle misidentification, or failed event reconstruction will ever be able to lead to a signal in the energy range of interest.
\end{itemize}

Before the first long-term, high-statistics, real-world neutrino data in liquid argon becomes available, this last requirement in particular might strike some as a rather optimistic.  Also it is worth noting that, contrary to what some references have stated~\cite{Cocco:2004ac}, there {\em are} spallation decays in this energy range, i.e. $^{11}$Li, which has a $Q$ value of 20.6~MeV and is a beta emitter~\cite{11Li}, so efficient identification of the parent muon is vital.

\begin{figure}[htb]
     \centering
     \includegraphics[width=0.55\textwidth]{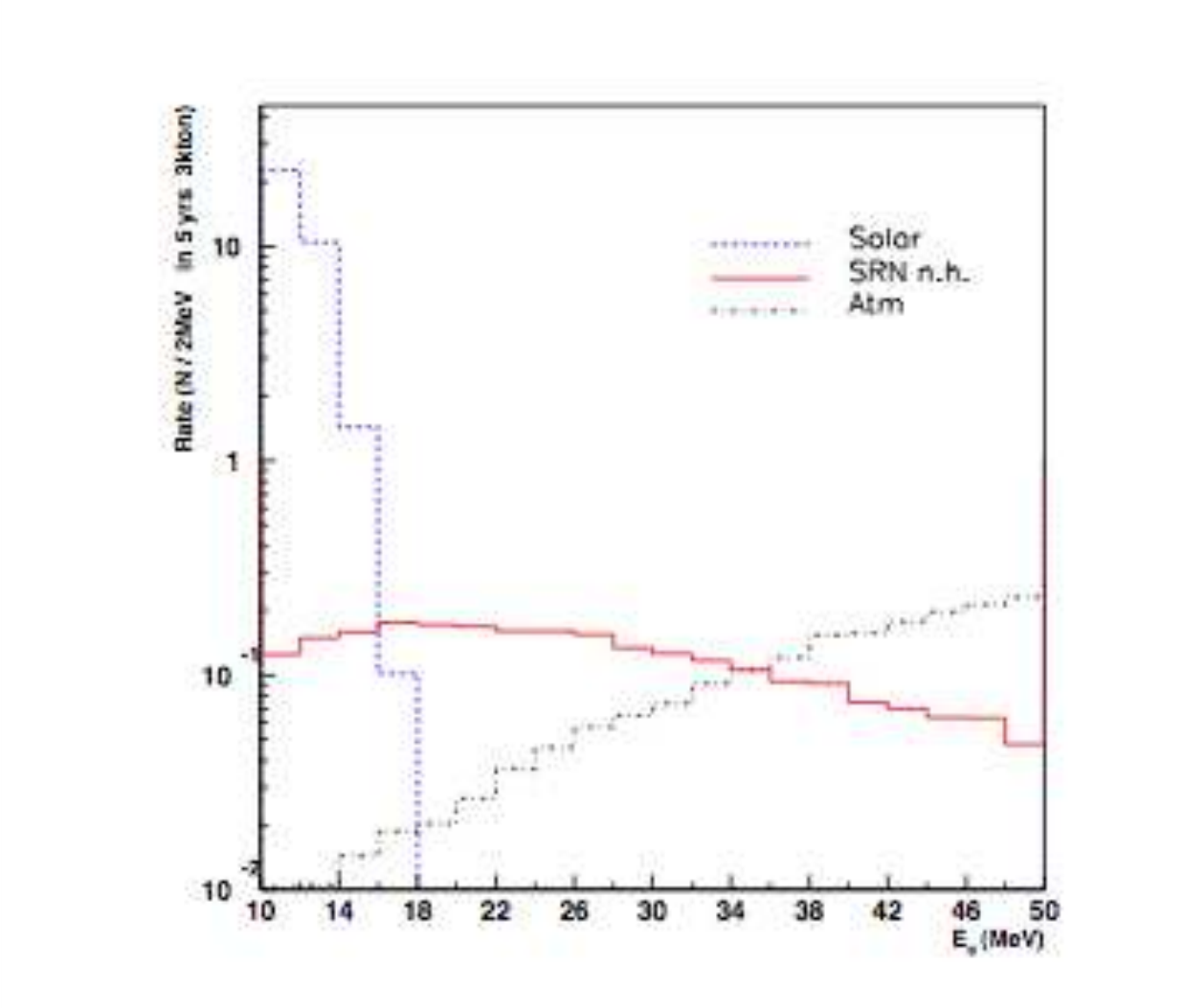}
     \caption{Expected relic supernova ${\nu}_{e}$ spectrum for a 5 year exposure of a 3 kiloton liquid argon detector in Gran Sasso, along with the limiting backgrounds of solar and atmospheric neutrinos.}
     \label{fig:relic_argon}
\end{figure}

Based on these potentially ``best case'' assumptions we reach the following performance estimates.
In Fig.~\ref{fig:relic_argon} we see the expected relic supernova ${\nu}_{e}$
spectrum for a 5 year exposure of a 3 kiloton liquid argon detector in Gran Sasso, along with the limiting backgrounds of solar and atmospheric neutrinos.  For the relic flux,  normal hierarchy and a large value for sin$^2$$\theta_{13}$ ($\geq 10^{-3}$) has been assumed; inverted hierarchy or very small sin$^2$$\theta_{13}$  ($\leq 10^{-6}$) would result in a relic flux  some 25\% lower.
For this 15~kt-year exposure, 1.7 $\pm$ 1.6 relic events are expected in the energy window between 16~MeV and 40~MeV.

Transferring this to to LBNE, and using a more beneficial (in terms of signal to background) energy window of 18 to 30~MeV, we would expect between 0.5 and 3.3 relic events per year in 51 kilotons of liquid argon, and 0.3 background events from atmospheric neutrinos.  These numbers are sufficiently small that quoting expectations in terms of sigma has little meaning; they are solidly in the Poissonian domain. Suffice it to say that any events in the signal region would have to be taken very seriously, since in all cases the signal is expected to exceed the background rate, a very attractive feature.

On the other hand, this serves to underline just how vital the assumptions listed above regarding backgrounds are for liquid argon, since even a single unexpected false event a year combined with the expected background could easily equal or exceed the true signal's rate in all but the more optimistic cases. This equates to an event-purity requirement of less than one false event per kiloton per century.  Until kiloton-centuries of LAr data have been taken and are shown to be background-free, this uncertainty is the biggest risk inherent in configuration 2.

Given these caveats, this configuration has discovery potential for a non-zero SRN flux. Unfortunately, in even the most optimistic high-rate cases useful spectral information would only become available after a decade of running due to both the low number of events observed and the flatness of the predicted spectrum in the sensitive region.

{\em Reference Configuration 2a: Three 17~kt fiducial liquid argon detectors. Depth 300 foot level of DUSEL.  No photon trigger.}

For the previous configuration we considered the best-case scenario for 51 kilotons of LAr at the 4850 level of DUSEL.  All assumptions about zero background are predicated on control of the environment and understanding of the detector.  As that configuration assumed both a photon trigger and a high degree of depth shielding from cosmic ray muons, it is clearly the most favorable design to meet the stringent requirements of no more than one false event per kiloton per century.

However, at 300 feet the cosmic ray muon rate will be 32,600 times the rate at the 4850 level (see the official DUSEL Depth Document). As a result, a detector at this depth will suffer an effective loss of fiducial volume due to the intense muon flux.  For a liquid argon detector, it has been estimated~\cite{Rubbia:2009md} that at 300 feet approximately 20\% of the effective volume will be lost due to cutting out a 10 centimeter slice in each 2D view around all muon tracks.  This will serve to reduce the numbers in the previous section (both the relic flux and the atmospheric background) by 20\%: between 0.4 and 2.6 relic events and 0.2 background events per year.

Additionally and more critically, because the drift time in a large liquid argon detector is on the order of a few milliseconds, the several kilohertz of through-going cosmic ray muon tracks at this depth means that there will always be several lines of charge drifting through the fiducial volume of the detector at any given moment. Multiple muons from a single cosmic ray interaction will likely be especially troublesome.  Without a photon trigger to establish the exact $t_0$ of each track,
it may not be possible to guarantee that there will be no failed track reconstructions, very low energy muons misidentified as electrons, or untagged spallation debris capable of mimicking a relic signal in this configuration,
at the extremely stringent levels required.
In fact, since the expected relic signal will be reduced by 20\% due to spallation cuts, protecting against background signals is that much more vital for this configuration.

The uncertainty from lack of high-exposure operation and exquisite sensitivity to this issue makes it difficult to consider this configuration capable of making a convincing SRN discovery at this time.\\

{\em Reference Configuration 2b: Three 17~kt fiducial liquid argon detectors. Depth 800 foot level of DUSEL. Assume a scintillation photon trigger for proton decay and supernova neutrinos.}

At 800 feet the cosmic ray muon flux is reduced by a factor of about four as compared to the 300 foot level. The good news is that there is almost no loss of fiducial volume required due to cutting around muon tracks~\cite{Rubbia:2009md}, so the 20\% of events lost in configuration 2a is recovered here. We would therefore once again expect between 0.5 and 3.3 relic events per year, and 0.3 background events from atmospheric neutrinos. Also good is that in this configuration the photon trigger is restored, which should help considerably with low energy triggering.

Though significantly lower than at 300 feet, the muon flux at 800 feet is still some 8000 times more intense than at the 4850 foot level.  Even with the benefit of photon triggering,
it is challenging to guarantee that there will be no false events at the level of less than one a year arising from  failed track reconstructions, very low energy muons misidentified as electrons, or spallation debris capable of mimicking a relic signal in this configuration.  In terms of cosmic ray-related background rejection (but, sadly, not signal collection), every successful year at 800 feet is equivalent to false-background-free exposure of 408 Mt-years on the 4850 level.

Again here, some high-exposure proof-of-principle data, in this case with a photon trigger, is needed to properly assess the viability and hence the potential of this configuration for SRN discovery.

{\em Reference Configuration 3: Two 100~kt fiducial water Cherenkov detectors at 4850 feet as specified in configuration 1, plus one 17~kt fiducial liquid argon detector at 300 feet with no photon trigger as specified in configuration 2a.}

This begins a long series of combinatoric configurations where the various detectors and depths of the the previous configurations are mixed and matched.  In all cases one or two components are dominant in terms of  SRN discovery potential.  The expected rates and known backgrounds will be listed; details can be found in the discussions of configurations 1, 1a, 1b, 2, 2a, and 2b.

By referring to configuration 1, the two 100~kt WC modules of this design would expect to record between 2 and 26 relic events a year with 187 background events.  By comparison with configuration 2a, a single liquid argon detector at 300 feet would be expected to see between 0.1 and 0.9 relic events a year, adding little to the configuration even if detector-related backgrounds could be well controlled.

WC: SRN signal = 2 -- 26 events/year, background = 187 events/year

LAr: SRN signal = 0.1 -- 0.9 events/year, atmospheric $\nu_e$ background = 0.1 events/year\\

{\em Reference Configuration 3a: Two 100~kt fiducial water Cherenkov detectors at 4850 feet as specified in configuration 1a, plus one 17~kt fiducial liquid argon detector at 300 feet with no photon trigger as specified in configuration 2a.}

WC: SRN signal = 3 -- 34 events/year, background = 214 events/year

LAr: SRN signal = 0.1 -- 0.9 events/year, atmospheric $\nu_e$ background = 0.1 events/year\\

{\em Reference Configuration 3b: One 100~kt fiducial water Cherenkov detector at 4850 feet as specified in configuration 1, plus one 100~kt fiducial water Cherenkov detector at 4850 feet with gadolinium as specified in configuration 1b, plus one 17~kt fiducial liquid argon detector at 300 feet with no photon trigger as specified in configuration 2a.}

WC: SRN signal = 1 -- 13 events/year, background = 93 events/year

WC+Gd: SRN signal = 4 -- 25 events/year, background = 21 events/year

LAr: SRN signal = 0.1 -- 0.9 events/year, atmospheric $\nu_e$ background = 0.1 events/year\\

{\em Reference Configuration 4: Two 100~kt fiducial water Cherenkov detectors at 4850 feet as specified in configuration 1, plus one 17~kt fiducial liquid argon detector at 800 feet with a photon trigger as specified in configuration 2b.}

WC: SRN signal = 2 -- 26 events/year, background = 187 events/year

LAr: SRN signal = 0.2 -- 1.1 events/year, atmospheric $\nu_e$ background = 0.1 events/year\\

{\em Reference Configuration 4a: Two 100~kt fiducial water Cherenkov detectors at 4850 feet as specified in configuration 1a, plus one 17~kt fiducial liquid argon detector at 800 feet with a photon trigger as specified in configuration 2b.}

WC: SRN signal = 3 -- 34 events/year, background = 214 events/year

LAr: SRN signal = 0.2 -- 1.1 events/year, atmospheric $\nu_e$ background = 0.1 events/year\\

{\em Reference Configuration 4b: One 100~kt fiducial water Cherenkov detector at 4850 feet as specified in configuration 1, plus one 100~kt fiducial water Cherenkov detector at 4850 feet with gadolinium as specified in configuration 1b, plus one 17~kt fiducial liquid argon detector at 800 feet with a photon trigger as specified in configuration 2b.}

WC: SRN signal = 1 -- 13 events/year, background = 93 events/year

WC+Gd: SRN signal = 4 -- 25 events/year, background = 21 events/year

LAr: SRN signal = 0.2 -- 1.1 events/year, atmospheric $\nu_e$ background = 0.1 events/year\\

{\em Reference Configuration 5: One 100~kt fiducial water Cherenkov detector at 4850 feet with gadolinium as specified in configuration 1b, plus two 17~kt fiducial liquid argon detectors at 300 feet with no photon trigger as specified in configuration 2a.}

WC+Gd: SRN signal = 4 -- 25 events/year, background = 21 events/year

LAr: SRN signal = 0.2 -- 1.8 events/year, atmospheric $\nu_e$ background = 0.2 events/year\\

{\em Reference Configuration 6: One 100~kt fiducial water Cherenkov detector at 4850 feet with gadolinium as specified in configuration 1b, plus two 17~kt fiducial liquid argon detectors at 800 feet with photon trigger as specified in configuration 2b.}

WC+Gd: SRN signal = 4 -- 25 events/year, background = 21 events/year

LAr: SRN signal = 0.3 -- 2.2 events/year, atmospheric $\nu_e$ background = 0.2 events/year\\

\begin{table} []  
\begin{center}
\begin{tabular}{|c|c|c|c|c|} \hline
Reference & Expected  & Expected  & Years of LBNE Data  & Years of LBNE Data  \\
Configuration & Annual & Annual  & Needed for a 3.0-$\sigma$   & Needed for a 3.0-$\sigma$  \\
Number  & SRN Signal & Background & Signal Assuming & Signal Assuming \\
  & (events/year)  & (events/year) & Maximum SRN Flux & Minimum SRN Flux \\ \hline
    1     & 3 -- 38 & 280 & 2.2 & 350 \\
    1a   & 5 -- 52 & 320 & 1.3 & 144 \\
    1b  & 13 -- 74  & 64 &  0.13 & 0.9 \\
    2  &  0.5 -- 3.3 & 0.3 & $\sim$1 & unknown \\
    2a  & 0.4 -- 2.6 & 0.2  & $\sim$2 & unknown \\
    2b  & 0.5 -- 3.3 & 0.3 &  $\sim$1 & unknown \\
    3  & 2 -- 27 & 187 & 2.9 & 526 \\
    3a  & 3 -- 35 & 214 & 2.0 & 268 \\
    3b  & 5 -- 39 &  114 & 0.35 & 3 \\
    4  & 2 -- 27 & 187 & 2.9 & 526 \\
    4a  & 3 -- 35 & 214 & 2.0 & 268 \\
    4b  & 5 -- 39 & 114 & 0.35 & 3 \\
    5  & 4 -- 27 & 21 & 0.32  & 3 \\
     6  & 4 -- 27 & 21 & 0.32 & 3 \\ \hline
\end{tabular}
\caption{\label{tab:sensitivity_summary} Summary of the reference configurations' sensitivity to detecting the supernova relic neutrino flux.}
\end{center}
\end{table}

\subsection{Conclusions}

To observe the supernova relic neutrinos we will clearly need large detector mass, low irreducible backgrounds, and well understood detector systematics.

From Table~\ref{tab:sensitivity_summary} one can see that if
the actual relic flux in the universe is at the high end of predictions and the detectors function with possibly optimistic assumptions, we could observe the relic flux with almost any configuration.
However, it should be noted that if the flux is very high then a long-running Super--K could make the discovery before LBNE is fully operational.

The best option for timely supernova relic neutrino flux discovery requires one of the configurations containing at least one 100~kt gadolinium-loaded water Cherenkov detector (i.e., configurations 1b, 3b, 4b, 5, and 6 in R\&D Document 643v2).  No matter where in the range of model predictions the true SRN flux lies, these configurations would need no more than three years, and in some cases much less, to make a 3.0-$\sigma$ discovery.  Furthermore, these configurations will allow us to extract useful physics data in the form of spectral information; in the worst cases this could require ten years or so of running, but it should be possible nevertheless.

\vfill\eject

%

\section{Atmospheric Neutrinos}\label{AtNu_intro}

In this section we summarize the potential contributions that atmospheric neutrinos studies could contribute to the overall LBNE physics program.

\subsection{Motivation and Scientific Impact}

Atmospheric neutrinos have played a crucial role in the discovery of neutrino oscillations and the measurement of neutrino masses and mixing parameters. Atmospheric neutrinos are sensitive, at least in principle, to
all of the physics remaining to be discovered in the PMNS matrix; the flux consists of neutrinos and anti-neutrinos of all flavors, and passes through significant densities of material, introducing modifications due to matter effects. The size of the earth is nearly optimal for the study of the large neutrino mass splitting, with large oscillation probabilities in the dominant channel. Three-flavor matter-enhanced atmospheric neutrino mixing is described by a rich phenomenology~\cite{Peres:PRD79,Bernabeu:NPB669,GonzalezGarcia:PRD70,Peres:PLB456,Akhmedov:JHEP0806,Petcov:NPB740,PalomaresRuiz:NPB712} and offers significant opportunities for discovery. In addition to the precision measurements of the dominant oscillation parameters~\cite{Ashie:PRL93,Ashie:PRD71},
the super-Kamiokande experiment has already carried out studies searching for
tau appearance~\cite{Abe:PRL97}, mixing at the solar mass scale~\cite{Wendell:1002}, three flavor oscillations~\cite{Hosaka:PRD74} , as well as more exotic scenarios~\cite{Abe:PRD77,Wang:thesis,Mitsuka:thesis}.  Magnetized underground detectors have the ability to distinguish neutrinos from anti-neutrinos with high accuracy, which
provides additional strength for resolution of the mass hierarchy and possible measurement of non-zero $\theta_{13}$ \cite{minos_atmos,ino_atmos}.

Similar studies can also be carried out with the high-statistics data sample of the
LBNE detector, and may provide an important complementary measurement of the oscillation parameters as determined using neutrinos from the accelerator beam.

Atmospheric neutrinos are unique among sources used to study oscillations:  the oscillated flux
contains neutrinos and antineutrinos of all flavors, matter effects may play a significant role,
and the oscillation phenomenology plays out over several decades in energy and path length.   These
characteristics of the atmospheric flux make it an ideal source for studying a wide range of oscillation
and mixing effects.

\subsubsection{Confirmatory Role}

Atmospheric neutrinos have the potential to play a vital role in the context of a comprehensive
program to study the lepton sector because all of the physics that one might hope to examine with
beam neutrinos can also be explored (albeit with reduced precision) using atmospheric neutrinos.
This includes oscillations at the large $\Delta m^2$, tau appearance, $\nu_\mu \rightarrow \nu_e$ mixing in
the presence of a non-zero $\theta_{13}$, the CP-violating phase, and the study of the mass hierarchy.
Because these phenomena play out over a wide range of energy and path lengths, atmospheric neutrinos are
very sensitive to alternative explanations or subdominant new physics effects that predict something other
than the characteristic (L/E) dependence predicted by oscillations in the presence of matter.  This power
has already been exploited by the Super-Kamiokande in fits that compare their data binned in terms of energy and
zenith angle to a host of new physics including CPT violation~\cite{Coleman:PLB405,KlinkHamer:IJ20},
Lorentz invariance violation~\cite{Colladay:PRD116002,Grossman:PRD72}, non-standard
interactions~\cite{GonzalezGarcia:NPB573}, Mass Varying Neutrinos (MaVaNs)~\cite{Kaplan:PRL93}, and sterile neutrinos~\cite{Abe:PRD77,Wang:thesis,Mitsuka:thesis}.  In numerous cases the best limits on exotic scenarios come from atmospheric neutrino analyses.

The breadth of physics that one can study with atmospheric neutrinos, the ability to study oscillation phenomena
in a complementary way to beam studies, and the sensitivity to small admixtures of `new physics' make atmospheric neutrinos an important part of the overall physics mission of the LBNE experiment.   One concrete example is the
study of $\nu_\tau$ appearance in the atmospheric neutrino flux.    While a large fraction of the muon neutrinos
are thought to oscillate to tau neutrinos, the overall rate of tau interactions is small due to the energy
threshold for tau production.  Tau events are expected at a rate of around 1 event per kiloton-year in the oscillated atmospheric flux.  These events can be identified on a statistical basis in water Cerenkov detectors~\cite{Abe:PRL97}, and can be selected in liquid argon detectors focussing on hadronic modes and in particular for up-down asymmetries.  A recent paper estimates that with a set of simple cuts on visible energy, reconstructed zenith angle, and energy of the highest energy pion in events lacking a charged lepton, a
4.3~$\sigma$ excess over background can be identified
in a 100~kt-yr exposure~\cite{Conrad:1008}.  Confirmation of the appearance of tau neutrinos at the expected level in the atmospheric flux will be an important consistency check on our overall oscillation picture.

\subsubsection{PMNS Matrix Measurements}

The key observable for atmospheric neutrinos will be the data binned in (energy,zenith angle) for events
separated by flavor, and ideally by neutrino/anti-neutrino.  For upgoing neutrinos ($\cos(\theta)<0$), oscillations at the
atmospheric mass splitting introduce large effects and matter effects introduce significant distortions of the spectrum,
particularly for neutrinos which pass through the Earth's core.  Mixing involving electron
neutrinos is enhanced for non-zero
$\theta_{13}$ for neutrinos (anti-neutrinos) due to matter effects if the hierarchy is normal (inverted).

The two key measurements, and the data samples that would be used to study them, are:

\begin{itemize}
\item
Octant of $\theta_{23}$:  Upward-going, sub-GeV electron neutrinos are affected by sub-dominant oscillations at the solar mass scale.  This may allow the ability to determine whether $\theta_{23}$
is less than or greater than 45 degrees, even if $\theta_{13}$ is zero.  Recent work suggests that
the effects for $\sim$~sub-GeV (E$<$100~MeV) neutrinos may be as large as 10-15\%~\cite{Peres:PRD79}.

\item
Mass hierarchy and $\theta_{13}$:  A non-zero $\theta_{13}$, combined with matter effects, leads
to a complicated structure of oscillation peaks for upgoing, roughly 1-10~GeV electron and
muon neutrinos.  Matter effects lead to an enhancement for electron neutrinos if the hierarchy
is normal, and anti-neutrinos if the hierarchy is inverted.
\end{itemize}

Many of the possible signatures in the atmospheric neutrino flux appear in the few hundred MeV to few GeV energy range.  Key performance characteristics for the detector include being able to
distinguish $\nu_\mu$ CC, $\nu_e$ CC, and NC events at these energies, as well as being able to accurately
determine the energy and direction of the incoming neutrino.

Atmospheric neutrinos, when combined with beam neutrinos in a global analysis, may help to resolve some
degeneracies.  In particular if $\theta_{23}$ is not $45^\circ$, and $\theta_{13}$ is small, the analysis
of atmospheric neutrinos may contribute significantly to resolving degeneracies present in the analysis
of beam data alone~\cite{Huber:0501}.

\subsection{Evaluation of Physics Sensitivities}

In this section we will evaluate atmospheric neutrino physics sensitivities for water Cerenkov and liquid argon detectors.
We will begin by describing the key detector performance characteristics, describe the tools developed for
carrying out these studies, and present results on two key measurements:
the octant of $\theta_{23}$ and resolution of the mass hierarchy.

\subsubsection{Water Cerenkov / Liquid Argon Differences}
Due to the success of the Super-Kamiokande experiment, the performance of water Cerenkov detectors for measurements
of atmospheric neutrinos, and the ways in which the data are to be analyzed, are well understood.
The smaller size of liquid argon detectors is compensated by several advantages resulting from
the higher imaging resolution:

\begin{enumerate}
\item
Improved energy and angular resolution for the initial neutrino.

\item
The ability to image sub-relativistic particles and
therefore improve the determine the incoming neutrino direction at sub-GeV neutrino energies.
 The capability of liquid argon detectors to image sub-relativistic particles significantly
 improves the pointing resolution for these relatively low energy interactions~\cite{Rubbia:NNN99},
 compared with the capabilities of water Cerenkov detectors which are capable of identifying protons at higher energies~\cite{Fechner:PRD79}.

\item
Improved flavor separation and NC/CC-like event separation.

\item
Somewhat improved ability to tag events as coming from neutrinos/anti-neutrinos.
\end{enumerate}

These differences may have significant ramifications.  As stated previously, the key detector performance
characteristics are whether the resolution is sufficient to resolve the features in the `oscillograph' of the
earth and distinguish neutrinos from anti-neutrinos.  If the angular resolution, energy resolution, and neutrino/anti-neutrino identification are
up to this task, then atmospheric neutrinos hold the potential to make confirmatory discoveries
even with limited
statistics.  For example, Reference~\cite{Petcov:0511277} calculates that for a detector with
excellent imaging characteristics the hierarchy can be identified at 2$\sigma$ with only around
200 events in the upgoing multi-GeV sample.

\subsubsection{Method and Tools}

For sensitivity studies we have developed a fast, four-vector level simulation tool that performs
event classification, measurement, binning, and statistical analysis.
This is done with a set of software based on the MINOS atmospheric analysis framework.

\begin{figure}[!h]
\centerline{
\includegraphics[width=0.35\textwidth]{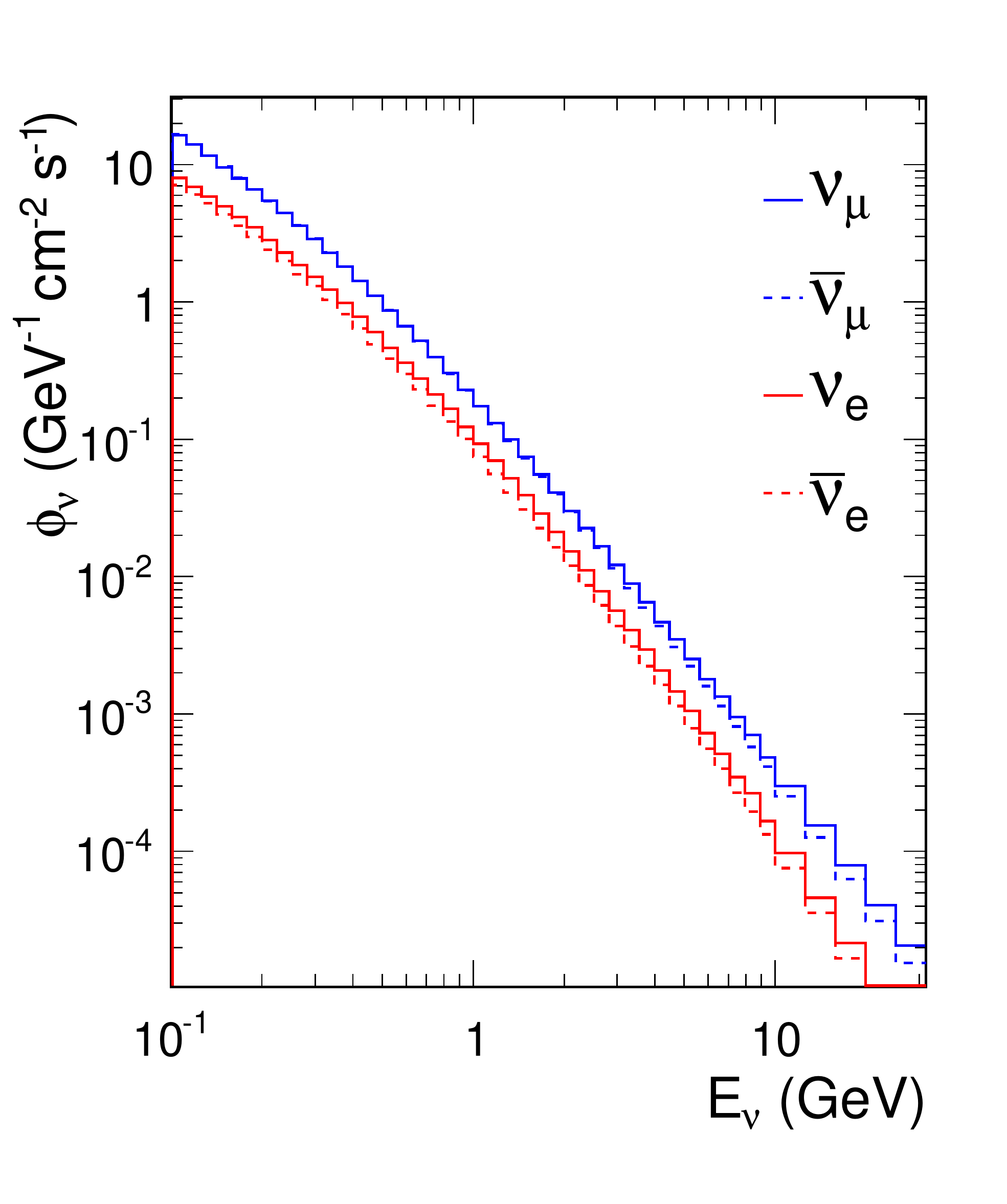}
\includegraphics[width=0.35\textwidth]{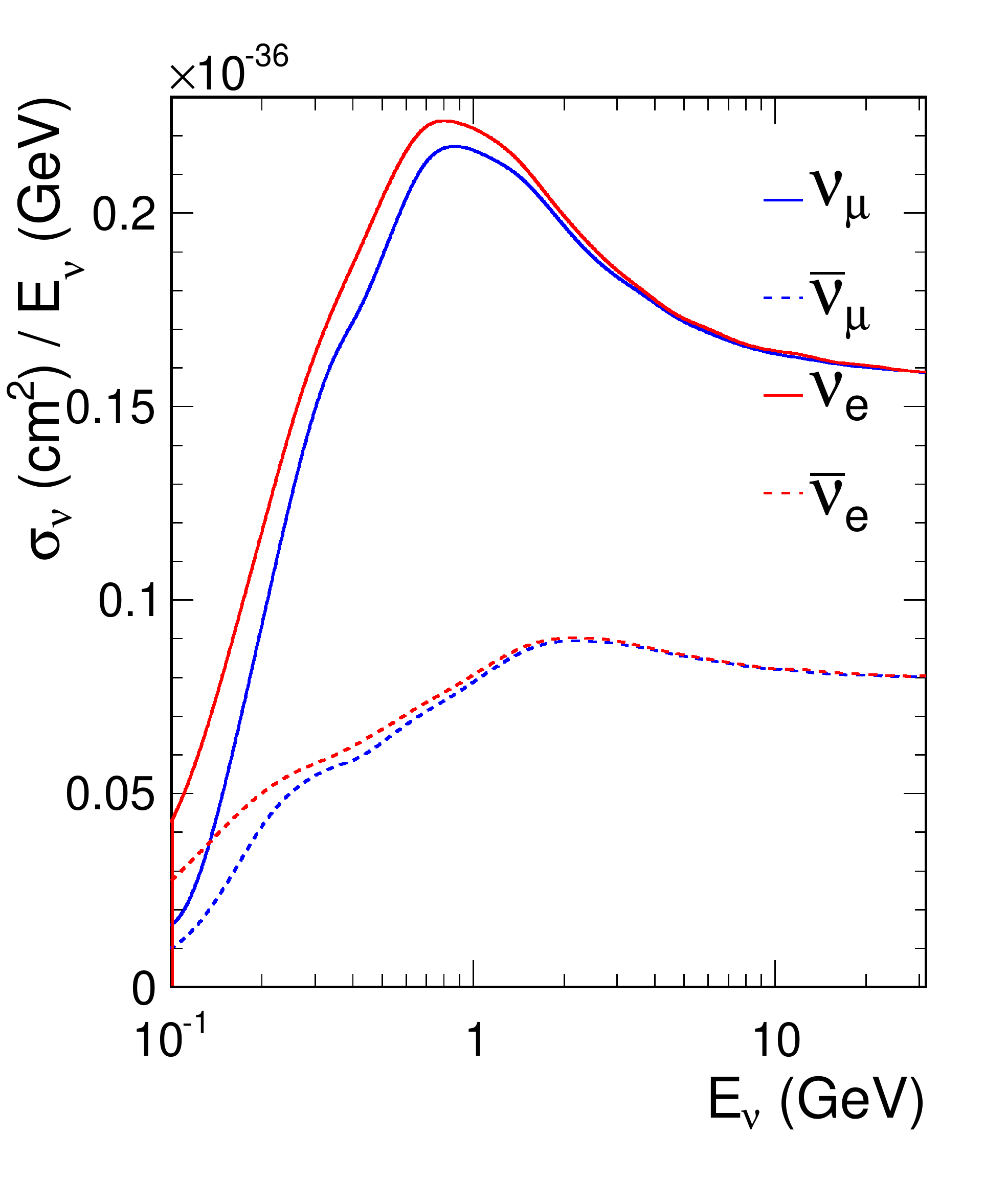}
}
\caption{The Bartol 3d atmospheric flux (left) and the GENIE neutrino-water cross sections (right).}
\label{fig:atnu_bartolgenie}
\end{figure}

The simulation proceeds in several steps:

{\bf  Event Generation:}
Four-vector level events are generated using the GENIE neutrino event generator \cite{genie}.   For this purpose new flux
drivers were developed which implement both the Bartol \cite{bartol3d} and FLUKA 3-d \cite{atmosfluka} flux calculations at the Soudan, MN site, which is a
reasonable approximation for the geomagnetic latitude of DUSEL.  Figure \ref{fig:atnu_bartolgenie} shows the two inputs
to the event rate calculation and event generation, the Bartol flux and the GENIE cross sections.    The event rate on water varies over the solar
cycle, from a minimum of 288 (275) events/kt-yr to a maximum of 331 (303) events/kt-yr as calculated using the Bartol (FLUKA) flux.
Predicted event rates in liquid argon differ from these values by less than 2\%.

\begin{figure}[!ht]
\centerline{
\includegraphics[width=0.6\textwidth]{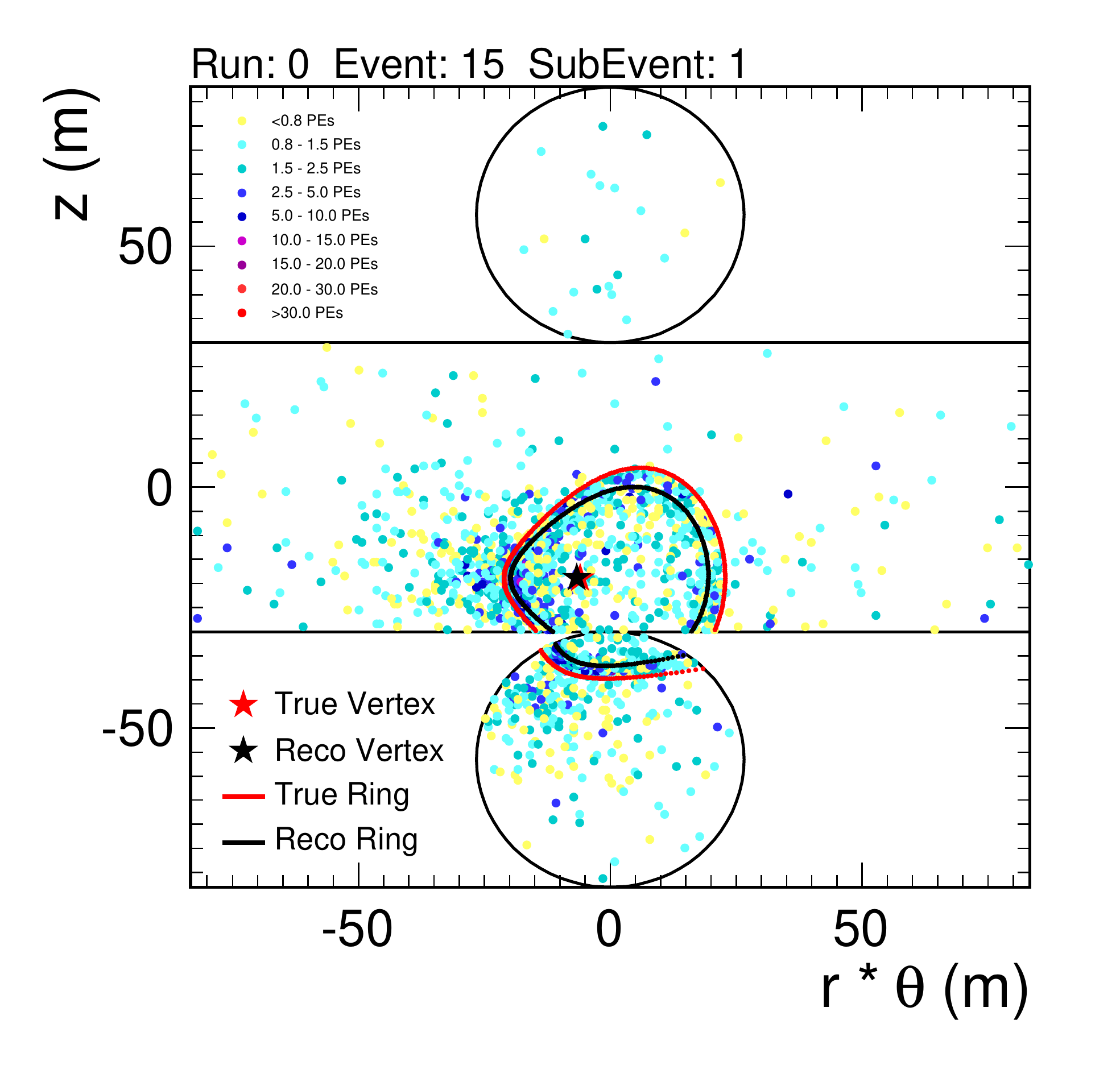}
}
\caption{A simulated 1~GeV $\nu_\mu$ CC interaction in a water Cerenkov detector.  This event was produced using the GENIE atmospheric neutrino flux driver to produce final state four-vectors,
which were converted into Nuance tracker format for input into the WCSim detector simulation and reconstruction.}
\label{fig:atnu_wcdisplay}
\end{figure}

GENIE is then used to generate large samples of interactions on argon or water.   These simulations include
all known scattering mechanisms of relevance in this energy range as well as a simulation of nuclear effects including
fermi motion, Pauli blocking, and intranuclear rescattering.   The output of these events are a set of 4-vector for particles emerging
from the struck nucleus.  These can either be input to a detector simulation or to a fast parametrized simulation of detector
response.   Figure \ref{fig:atnu_wcdisplay} shows such an event input to the WCSim detector simulation.

{\bf Pseudo-Reconstruction}
Events are passed through a fast parameterized `pseudo-reconstruction'.    In atmospheric neutrino analyses events are categorized in a number of ways:
containment (fully/partially contained), flavor (e-like/mu-like/NC-like), energy (sub/Multi-GeV),
topology (single/multi-ring, QEL/non-QEL), and neutrino/antineutrino tag.   The classification is
made based on truth level characteristics, accounting for detection thresholds and misidentification via
the following steps:

\begin{enumerate}
 \item
  Classify containment by simulating vertex and end points for
     each event using a toy detector geometry.  For the 100~kt fiducial mass WC detector the geometry is a cylinder with
    26.5~m radius and 60~m height.  For the 17~kt fiducial mass LAr detector the geometry is a box with 71~m length, 15~m width, and 14~m height.
    Vertex points are chosen at random, and energy loss formulas
      are used to obtain end points of final state particles.   A fiducial volume cut is then placed on the vertex (2~m for WC, 1~m for LAr) and track end positions (0.5~m for WC, 0.25~m for LAr)
    to determine if the event is fully or partially contained.
  \item
  Simulate trigger by selecting those events containing particles
     above threshold (``visible'' particles), here taken to be 50~MeV. In addition, charged particles in WC are required to be above Cerenkov threshold.
\item
Assign neutrino flavor to events using  true $\rightarrow$ reconstructed
classification matrices, giving probabilities for different reconstructed event types.  These matrices are adapted from versions in the
literature, for WC from Reference \cite{Ishitsuka:thesis}, for LAr from Reference \cite{Bueno:0701101}.  For the WC option the classification categories are e-like and mu-like, for
the liquid argon option they are e-like, mu-like, and NC-like.
\item
Smear energy and angle of leptons and hadronic final states, using the resolution functions
given in Table \ref{tab:atnu_resolutions}.
\begin{table}[ht]
\begin{tabular}{|c|c|c|} \hline
Resolution & WC              & LAr   \\  \hline
FC Lepton  &                     &  \\
Energy:       & $2\%+2\%/\sqrt{E}$ & $2\%+2\%/\sqrt{E}$ \\
Angle:         & $2^\circ$     &  $2^\circ$ \\ \hline
PC Lepton  &                     &  \\
Energy:       &   50\%         & 50\% \\
Angle:         & $2^\circ$     &  $2^\circ$ \\ \hline
Hadronic System &               &   \\
Energy:       & $30\%+30\%/\sqrt{E}$ & $30\%/\sqrt{E}$ \\
Angle:         & $45^\circ+15^\circ/\sqrt{E}$     &  $10^\circ$ \\ \hline
\hline
\end{tabular}
\caption{Summary of resolution functions.  Water Cerenkov resolutions were taken from, or tuned to,
SuperKamiokande resolution plots \cite{Dufour:thesis}, LAr resolutions are from \cite{ICARUS_Nu2010} and \cite{Gandhi:2008zs}.}
\label{tab:atnu_resolutions}
\end{table}
\item
Apply minimum energy cuts of 100~MeV for selected FC and 300~MeV for PC events.  The same minimum energy cuts are
used for both detector configurations.
\end{enumerate}

\begin{figure}[!h]
\centerline{
\includegraphics[width=1.0\textwidth]{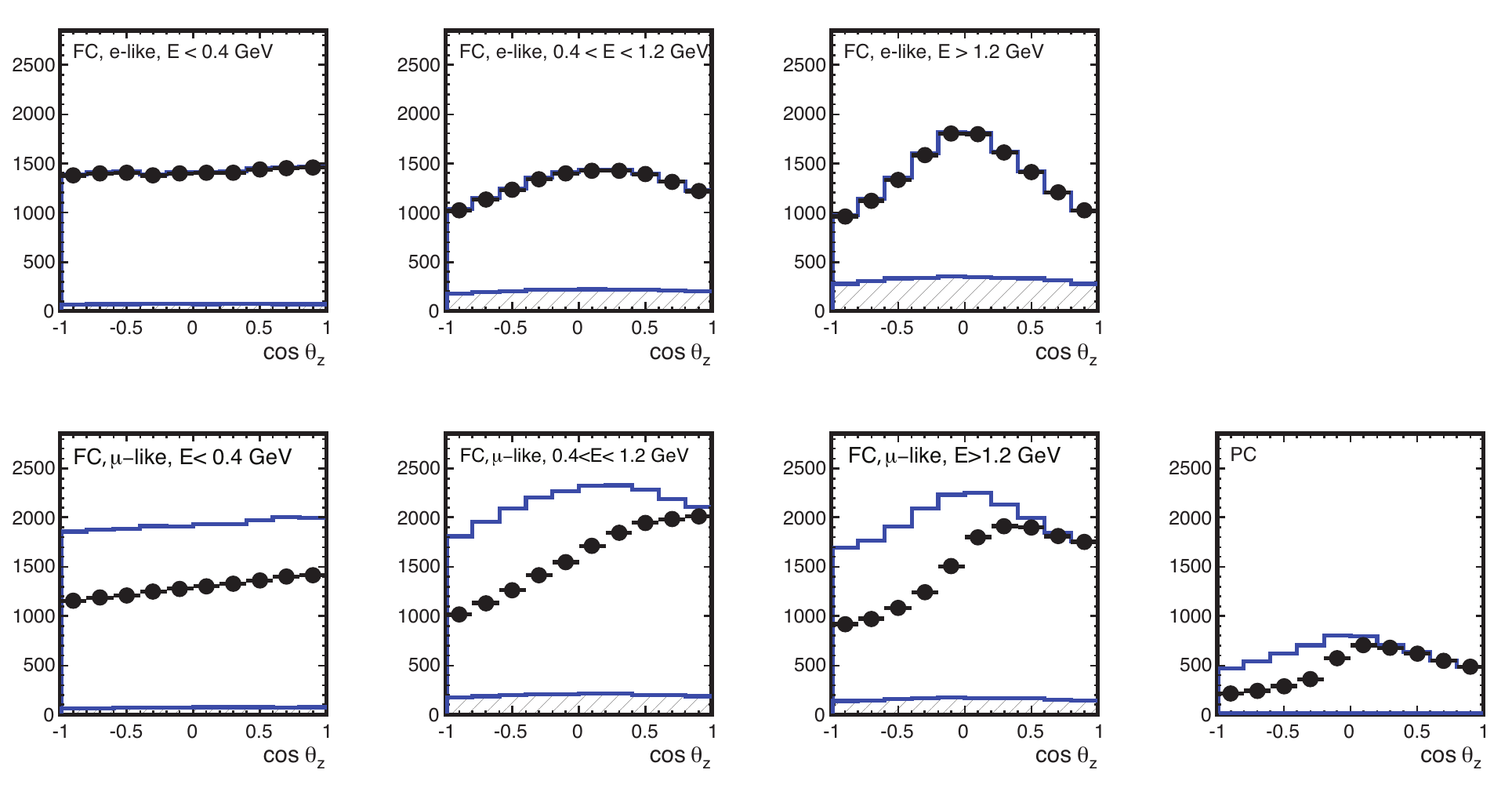}
}
\caption{Simulated zenith angle distributions for 500~kt-yrs of atmospheric neutrino data in a water Cerenkov detector.
No oscillations (solid blue), with oscillations (black), NC contribution (dashed blue).}
\label{fig:atnu_wczenith}
\end{figure}

\begin{figure}[!h]
\centerline{
\includegraphics[width=1.0\textwidth]{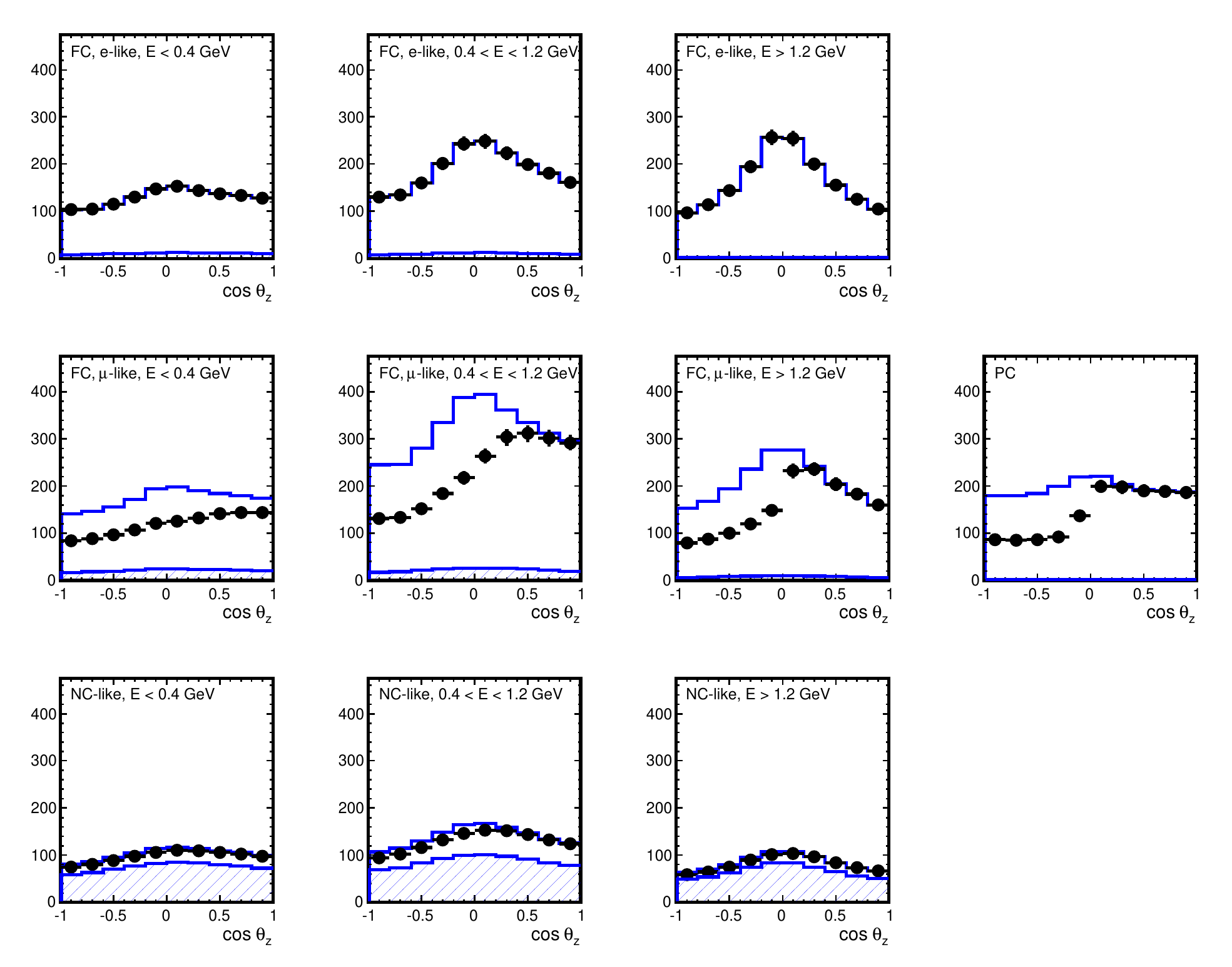}
}
\caption{Simulated zenith angle distributions for 85~kt-yrs of atmospheric neutrino data in a liquid argon detector.
No oscillations (solid blue), with oscillations (black), NC contribution (dashed blue).}
\label{fig:atnu_larzenith}
\end{figure}

Figures \ref{fig:atnu_wczenith} and \ref{fig:atnu_larzenith} show the simulated zenith angle distributions for the analysis
categories defined above, for five years of data taking in water Cerenkov and liquid argon detectors.

{\bf Sensitivity Evaluation:}  Binned pseudo-reconstructed data are then compared to oscillation hypotheses and relevant statistics are
calculated.  Three-flavor oscillation probabilities including matter effects \cite{Barger3flavor} incorporating the PREM earth model
are calculated using code provided by Mark Messier~\cite{Messier:PMNS}.   Neutrino production heights as a function of
energy and zenith angle are calculated using parametrizations and code developed by the MINOS and Soudan-2 experiments \cite{s2prodheight}.

\begin{figure}[!h]
\centerline{
\includegraphics[width=0.35\textwidth]{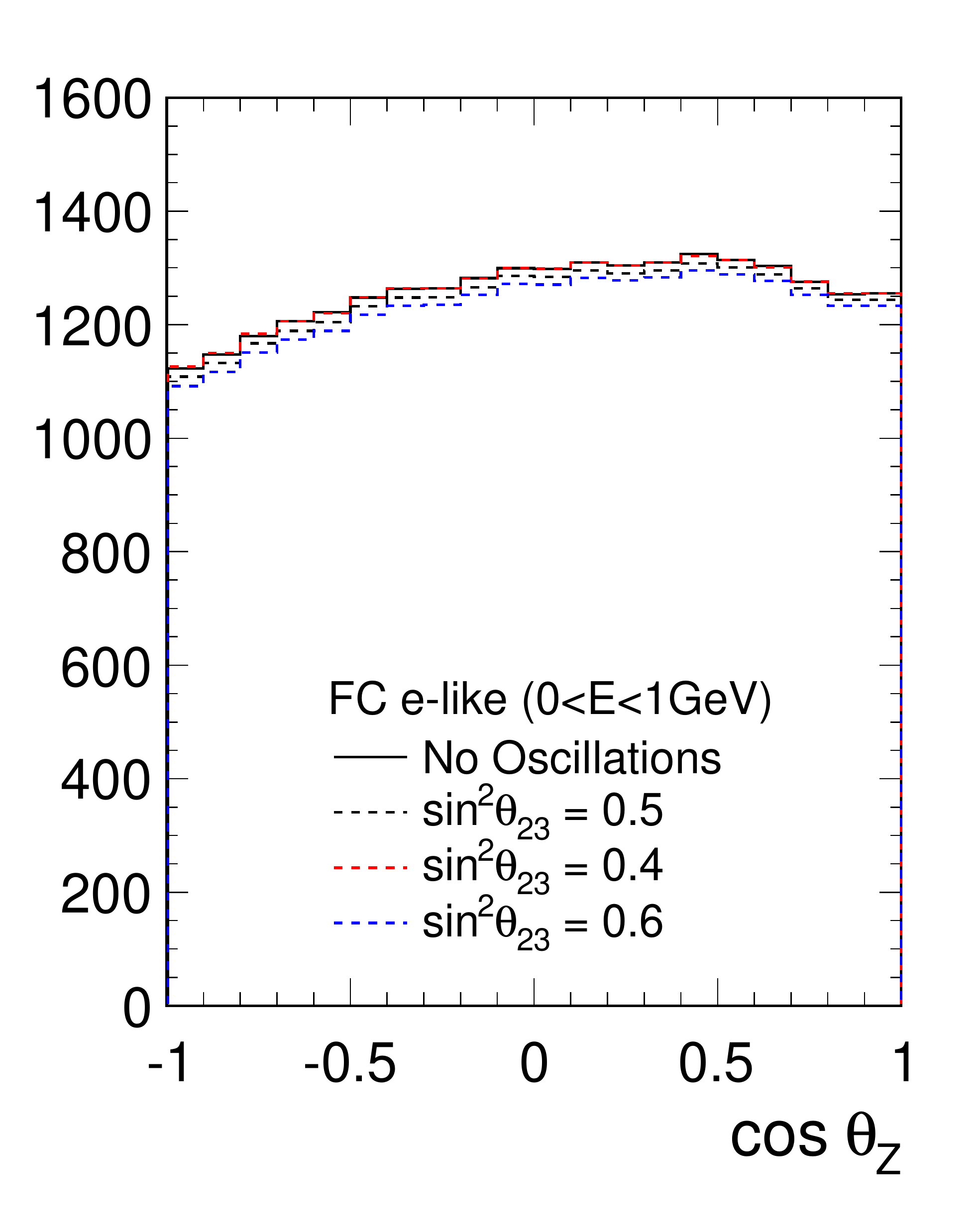}
\includegraphics[width=0.35\textwidth]{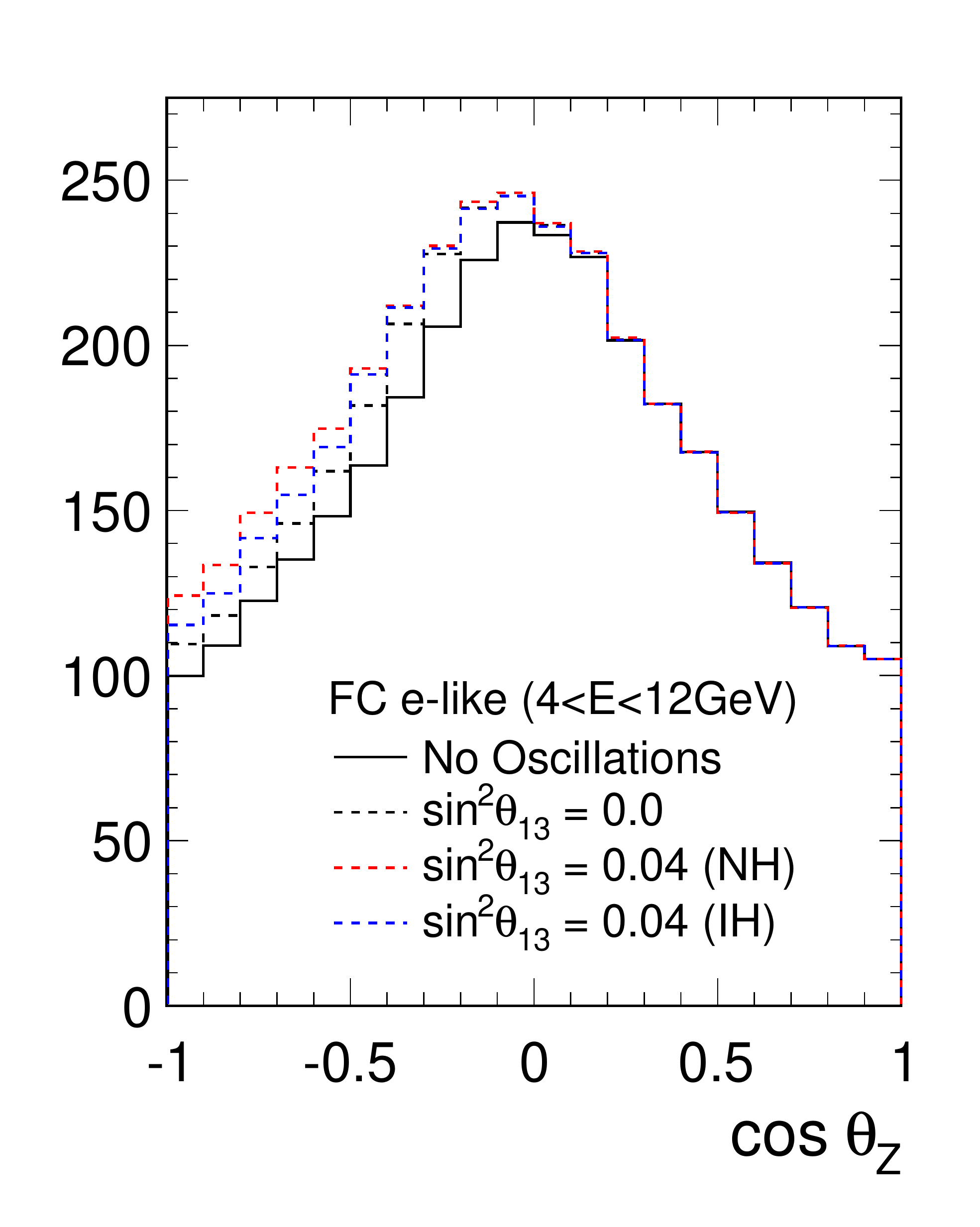}
}
\caption{Zenith angle distributions for a low energy (left) and high energy (right) WC e-like data sample.
Oscillation parameter values are given in the text. }
\label{fig:atnu_zenithwater}
\end{figure}

\begin{figure}[!h]
\centerline{
\includegraphics[width=0.35\textwidth]{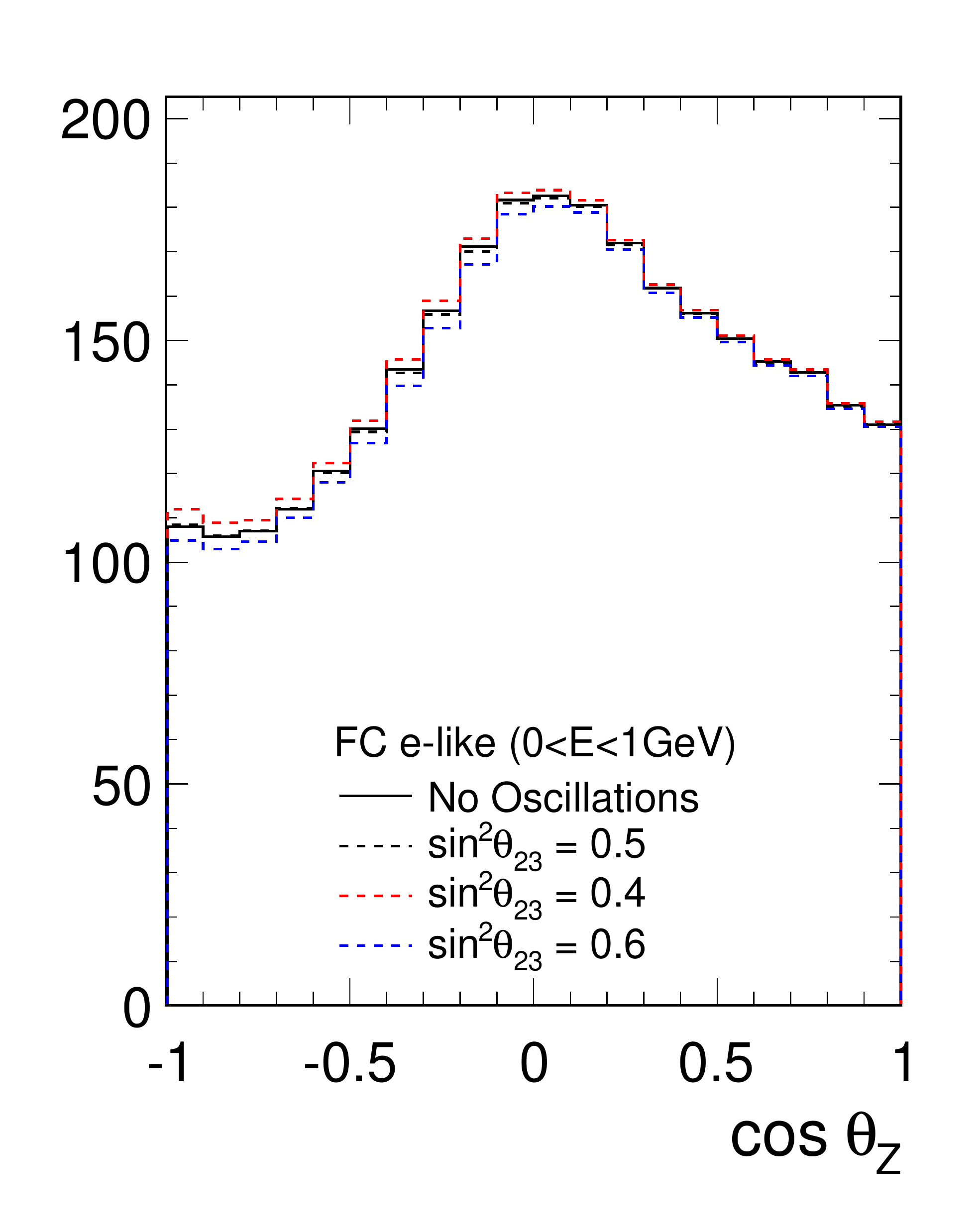}
\includegraphics[width=0.35\textwidth]{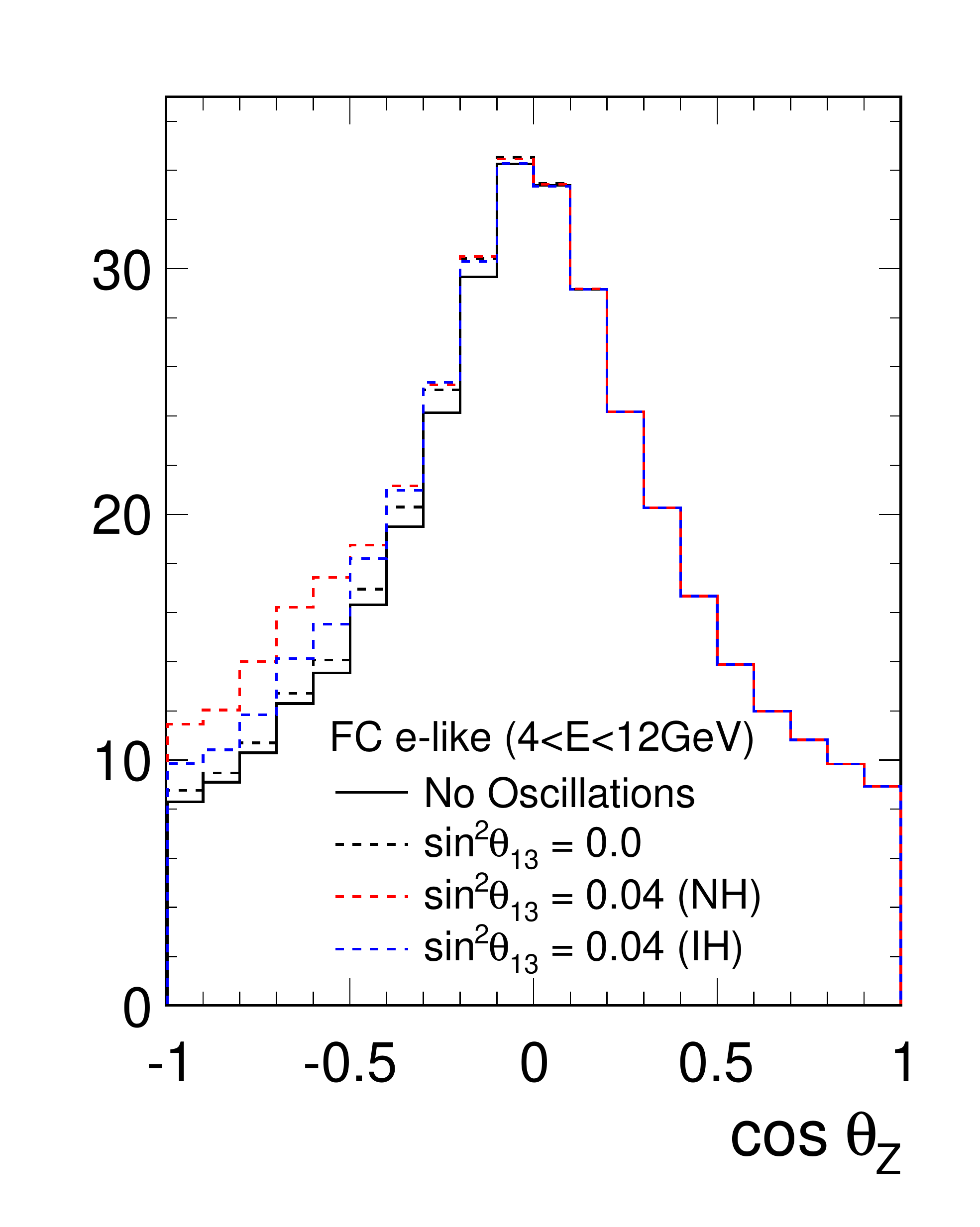}
}
\caption{Zenith angle distributions for a low energy (left) and high energy (right)  liquid argon e-like data sample.
Oscillation parameter values are given in the text. }
\label{fig:atnu_zenithlar}
\end{figure}

The effect that oscillations would have on the experimental distributions in a water Cerenkov and liquid argon detector are shown in
Figures \ref{fig:atnu_zenithwater} and
\ref{fig:atnu_zenithlar}, respectively.  The low energy ($0< E_\nu < 1$~GeV) sample (left figure in each plot) is sensitive to the octant of $\theta_{23}$
in changes in the rate of upward-going electron like events.  The high energy ($4< E_\nu < 12$ GeV) sample is sensitive to the mass hierarchy
and non-zero $\theta_{13}$ via changes in the rate of upward-going electron-like events.    In all of these plots, the default values for oscillation parameters
are $\Delta m_{32}^2= 2.3 \times 10^{-3}$  eV$^2$, $\sin^2 \theta_{23}=0.5$,
$\Delta m_{12}^2= 7.5 \times10^{-5}$  eV$^2$, $\sin^2 \theta_{12}=0.31$,
$\sin^2 \theta_{13}=0$, $\delta_{CP}=0$, and normal hierarchy.  The changes from these default values are shown on each plot.

For both detectors, data in each of the analysis categories (FC/PC, e-like/mu-like/NC-like) are binned in energy and zenith angle with
$\Delta \log_{10}(E)=0.2$ and $\Delta\cos\theta=0.1$.   For some input true value of the oscillation parameter, the log-likelihood difference
is computed between this `expected' data and data for any other set of oscillation parameters.
For the sensitivities computed here, log-likelihood curves were generated in this way for a single parameter, using statistical errors only for the chosen exposure.

To validate the method as well as the misidentification matrix and resolution functions for WC, a simulation was done of the SuperKamiokande detector
geometry for a 7.68 year exposure, in order to compare with published results \cite{Ishihara:thesis}.  Even with perfect parameterizations of detector performance one would
not expect complete agreement, in part
because we are comparing an expected sensitivity to a result derived from actual data.  This comparison does yield results in reasonable agreement, giving us
confidence in our ability to accurately calculate WC sensitivities for larger detectors and exposures.

\subsubsection{Physics Sensitivities}

For this report we have calculated sensitivities for a 100~kt fiducial mass WC and 17~kt fiducial mass liquid argon detector with five years of data.

\begin{figure}[!h]
\centerline{
\includegraphics[width=0.6\textwidth]{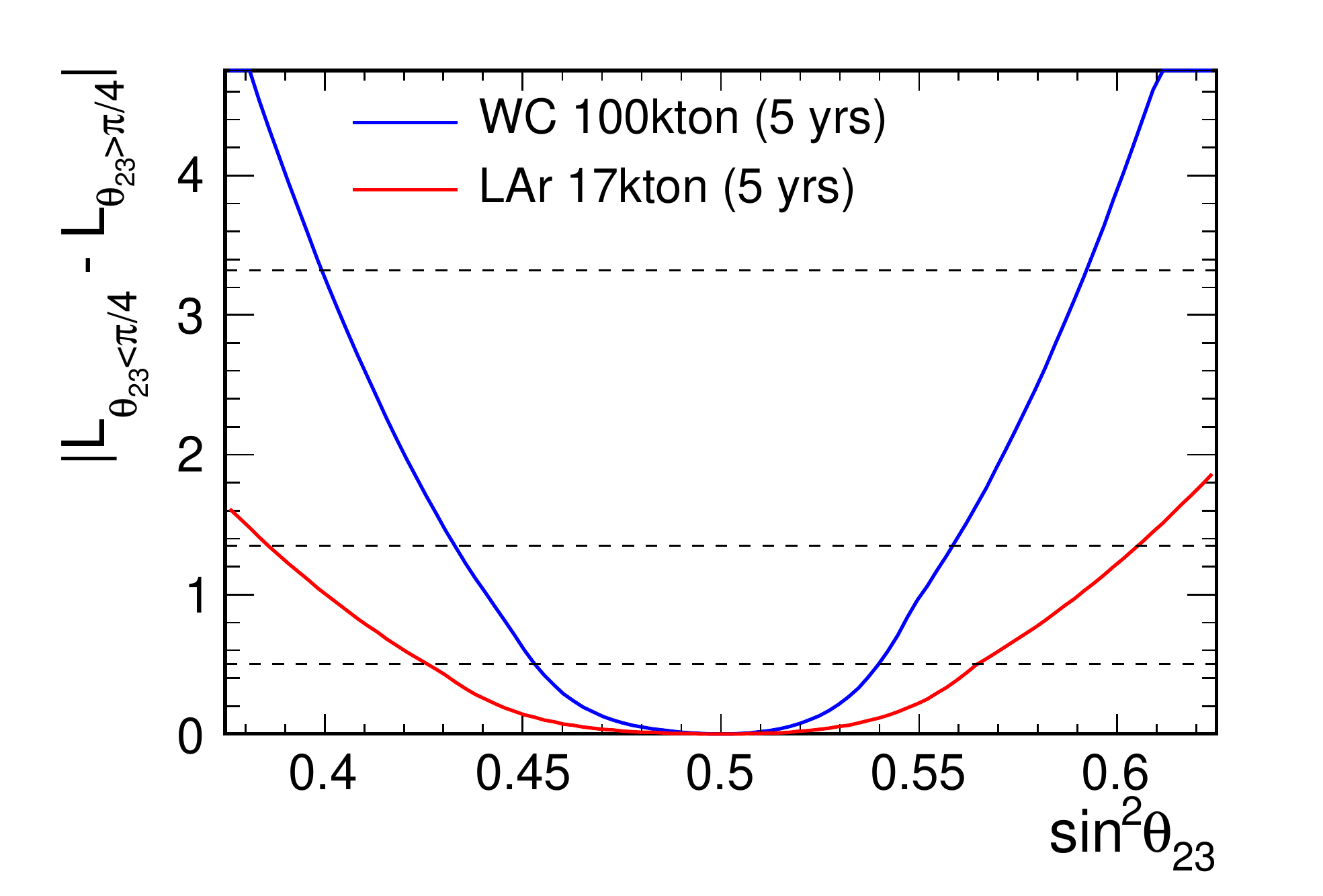}
}
\caption{Sensitivity to the octant of $\theta_{23}$ for five years of running with a 100~kt fiducial mass water Cerenkov  detector  (blue) or  a 17~kt fiducial mass liquid argon detector (red).  }
\label{fig:atnu_sensoctant}
\end{figure}

Figure \ref{fig:atnu_sensoctant} shows the sensitivity to the octant of $\theta_{23}$.   In this case the likelihood difference is calculated between the value of $\theta_{23}$ in the correct octant to that in the wrong octant.   As this plot indicates, the larger mass of the WC detector outweighs the advantages in purity and directional resolution of the liquid argon detector for this measurement.

\begin{figure}[!h]
\centerline{
\includegraphics[width=0.6\textwidth]{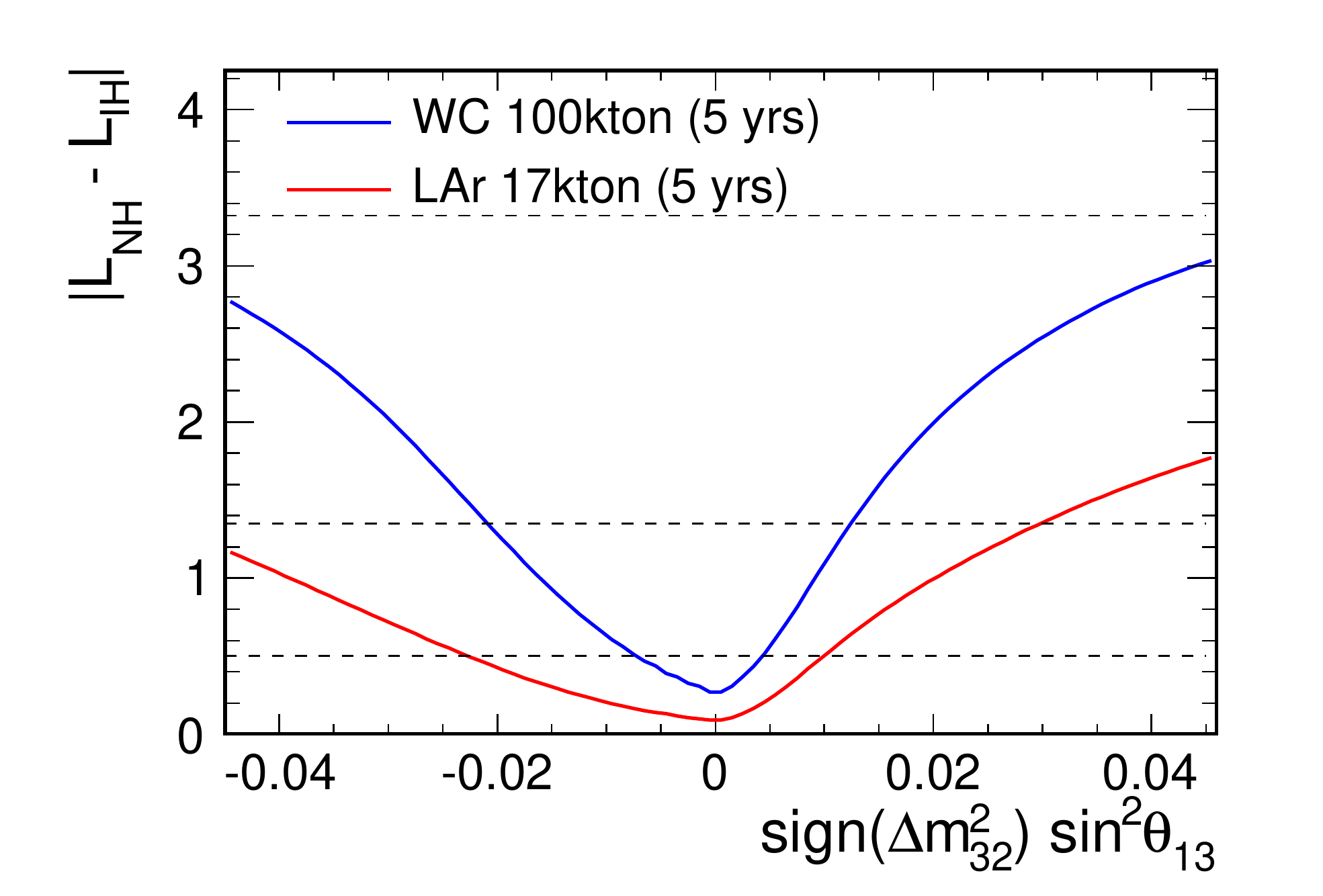}
}
\caption{Sensitivity to the determination of the mass hierarchy as a function of the true value of $\theta_{13}$ for five years of running with a 100~kt fiducial mass water Cerenkov  detector  (blue) or  a 17~kt fiducial mass liquid argon detector (red). }
\label{fig:atnu_senhierarchy}
\end{figure}

Figure \ref{fig:atnu_senhierarchy} shows
the sensitivity to the resolution of the mass hierarchy, as a function of true $\theta_{13}$.   In these plots, normal mass hierarchy corresponds to positive values, inverted hierarchy
to negative values.   One feature to note is that both technologies have a weak sensitivity to the sign of the mass hierarchy even for $\theta_{13}=0$, resulting from
interference between oscillations at the solar and atmospheric mass scales.   For either technology it is easier to resolve a normal mass hierarchy, since resonant matter effects
occur for neutrinos rather than anti-neutrinos, with larger interaction cross sections.    For the resolution of the mass hierarchy, the larger mass of the WC detector is again the dominant factor.   In future revisions
we will explore in more depth the different capabilities of the detector technologies for distinguishing neutrinos from anti-neutrinos, which may provide an additional compensating factor for liquid argon.

\subsection{Comments on Configuration Options}

Configuration options that are designed primarily to improve capabilities for lower energy neutrinos are
not expected to have a significant effect on the study of atmospheric neutrinos.
These include having gadolinium in a water Cerenkov detector, or including a photon trigger for an
atmospheric neutrino detector.

The two options for phototube coverage are also expected to have minimal impact on the atmospheric
neutrino analysis.  Since the Super-K data sets included periods where the photodetector
coverage was lower (Super-K--II), many studies were done on the differences in detection efficiency and
measurement resolution for atmospheric neutrinos when comparing the Super-K--I and II detector configurations.
These studies~\cite{Dufour:thesis} indicate that the differences are small as far as atmospheric neutrino
analyses are concerned, and for the conclusions here we are considering them to be a relatively small
perturbation.

Configuration options that involve placing a liquid argon detector at shallow depths will require
more study, but it appears at this time that the analysis and conclusions of~\cite{Bueno:0701101} which studies
cosmogenic backgrounds to proton decay are also directly for atmospheric neutrinos.  The veto shield, as proposed,
follows the conclusions of this study and should therefore provide sufficient information to work at the depth of 800~ft
with a 17~kt analysis fiducial volume, which is the fiducial volume assumed in the studies mentioned previously,
and is the fiducial volume available at 4850~ft.

\vfill\eject
%

\section{Ultra-High Energy Neutrinos}\label{UHENusect}

The field of neutrino astronomy, using high-energy neutrinos as
cosmic messengers to probe the internal mechanisms of the most energetic
astrophysical objects, offers a new window on the universe.  Where the
traditional astronomical messengers of photons and charged particles
are either absorbed by material or radiation in astrophysical environments
or are deflected into random directions by magnetic fields, neutrinos, with
no charge and a small interaction cross section, offer an unobstructed
view inside the acceleration regions where the highest energy cosmic rays are
created.  The source of these highest energy cosmic rays still remains as one
of the long-standing open questions in physics.  Any astrophysical object
capable of producing these highest energy protons will almost certainly be
a source of high energy neutrinos, as some fraction of these protons will
interact with matter or radiation fields in and near the source.  Potential
sources of high-energy neutrinos include active galactic nuclei,
supernova remnants~\cite{Learned:neutr_astr}, gamma-ray bursts
(GRBs)~\cite{Bahcall:GRBnu,Guetta:GRBNu}.  Observation
of neutrinos from these sources, in combination with photon observations at
several wavelengths, will likely be needed to identify and understand the
acceleration mechanisms at work in these sources.  Other potential sources
of high-energy neutrinos include annihilation of weakly interacting massive
particles (WIMPs), a potential dark matter candidate, after they are
gravitationally captured by objects such as the sun or earth~\cite{Drees:WIMPs}.

\subsection{Motivation and Scientific Impact of Future Measurements}
High energy neutrinos are observed by detecting the charge leptons
arising from the charged-current neutrino interaction with a nucleon:
$\nu_{l} + N \rightarrow X + l$.
In the case of $\nu_{\mu}$ interactions, the sensitive volume exceeds
the detector physical volume, as long as the resulting $\mu$ passes
through the detector volume.  This yields substantially more detectable
events at higher neutrino energies.  To quantify this effect, the
effective area of a detector to neutrinos is used as a measure of
detector sensitivity.  For a incoming neutrino flux $\Phi$, the number
of detected neutrino events is given by:
\begin{center}
$N_D = \int dt \, \int d\Omega \, \int dE \, \Phi_{\nu}(t,\Omega,E) \, A_{eff}(\Omega,E)$
\end{center}
At neutrino energies $>$ 100~GeV, the resulting muons  have a strong
angular correlation ($\sim$3 degrees or better) with the incoming neutrino direction.

\begin{figure}[htb]
     \centering\includegraphics[width=.44\textwidth, height=0.30\textwidth]
                               {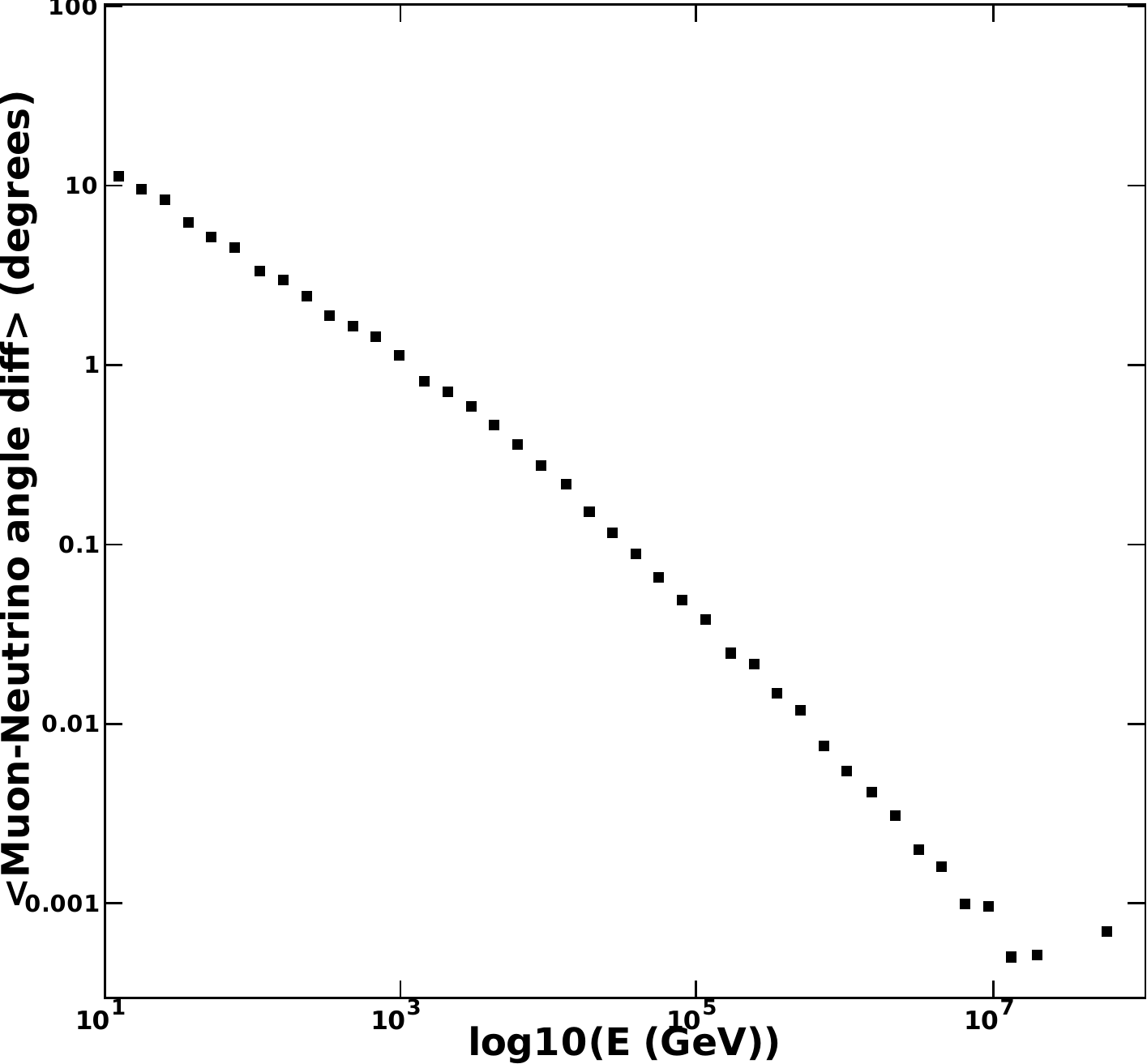}
     \vspace{0.05in}
     \caption{Mean angular separation between incoming neutrino direction and
       outgoing muon direction for charged-current interactions as a function of neutrino
       energy.  From toy Monte Carlo study, see section \ref{sect:UHESensitivity}.}
     \label{fig:uhe_neut_angle}
\end{figure}

This strong correlation between neutrino and muon directions
as neutrino energy increases is shown in Figure \ref{fig:uhe_neut_angle}.
By exploiting this angular correlation, and potentially using
time correlations with other astronomical observations of high
energy phenomena (GRB alerts, for example), the impact of the
irreducible background from atmospheric neutrino events can be
reduced, enabling searches for cosmic neutrinos and neutrino astronomy.

Previous generations of experiments have performed searches for cosmic
neutrino signals, but to date, have only set limits.  The first generation
of these detectors, AMANDA~\cite{Deyoung:AMANDA},
Baikal~\cite{Aynutdinov:BAIKAL}, and ANTARES~\cite{Antares:design},
as well as the current generation of detectors (IceCube, including the
DeepCore low energy extension~\cite{IceCube:deepcoresens}) have been specifically designed to
search for cosmic neutrino fluxes and have set the strongest limits
on cosmic neutrino fluxes from point sources~\cite{IceCube:results:ps}, diffuse
ultra-high-energy sources~\cite{IceCube:results:uhe}, coincident neutrino emission
from GRBs~\cite{IceCube:results:grb}) and WIMPs~\cite{IceCube:results:wimp}.  Super-Kamiokande has also
performed similar searches~\cite{Desai:SKSearch,Thrane:SKSearch,Swanson:SKSearch}, also with no positive
detection of a source.  Within the last year, the IceCube detector
has been augmented with the DeepCore extension (volume $\sim$15~Mt), with several
additional optical sensors added to the deepest, clearest portion
of the instrumented volume to significantly increase the sensitivity
of the detector to below 100~GeV.   In the next 5-15 years, the continued
operation of both the northern and southern hemisphere neutrino telescopes
will either result in a set of sources to follow up with more detailed
limits or stringent limits on neutrino emission from astrophysical sources
about $\sim$10~GeV.

\subsection{Sensitivity of Reference Configurations}\label{sect:UHESensitivity}

In order to evaluate the sensitivities of the proposed far detector technology
options for the LBNE experiment to a astrophysical
neutrino signal, a toy event simulation study was produced.
This toy simulation was used considering that full detector simulations are
still in early stages of readiness and there was limited manpower to perform
these studies.  These simulations made a set of assumptions that simplified
the far detectors:
\begin{itemize}[parsep=-2pt]
\item The detector fiducial volumes were modeled as spheres with radii producing the same
volumes as reference configurations.  The detector cross-sectional areas were reproduced
to within $\sim$20\%.
\item Detectors were assumed to be fully efficient at triggering and reconstructing any
muon track within the modeled fiducial volumes for muon energies $>$~100~MeV.
\item Variations in detector technology choices (depth, Gd loading, PMT
coverage) are neglected in this study, but do not greatly impact detection of muons
with E~$>$~100~MeV.
\item Backgrounds from misreconstructed atmospheric muons and atmospheric neutrinos have not been
included.
\end{itemize}
These should be included in future studies aimed to produce more accurate sensitivities.

\begin{figure}[ht]
     \centering\includegraphics[width=.6\textwidth]
                               {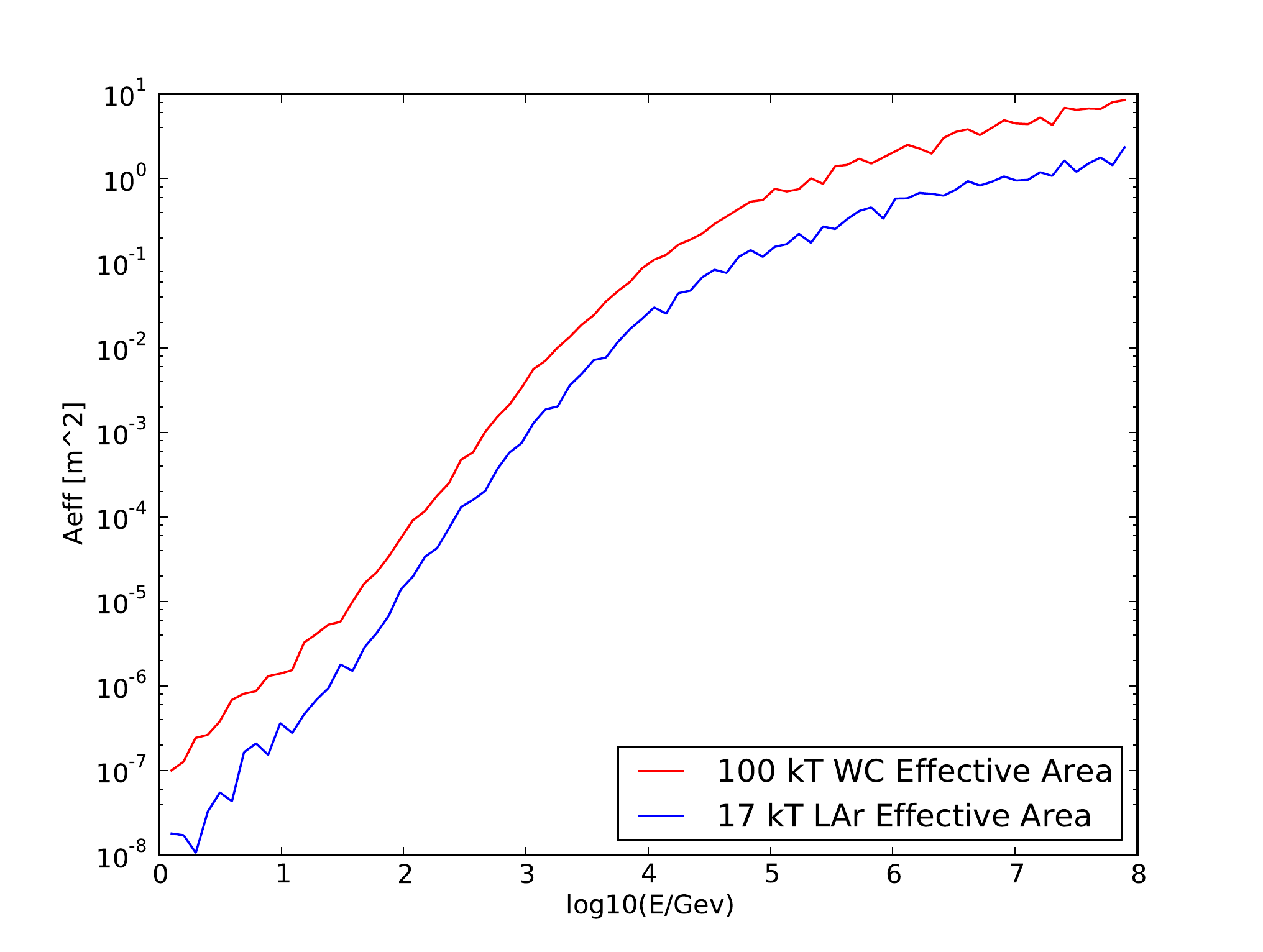}
     \caption{100~kt water cherenkov(WC) and 17~kt Liquid Argon (LAr)
       effective areas as a function of neutrino energy from the toy simulation.  For
       comparison, the IceCube effective area at 1~TeV is $\sim$1~m$^{2}$.}
     \label{fig:uhe_effective_areas}
\end{figure}

As part of this simulation, 1~GeV--100~PeV neutrinos were propagated through the earth
and weighted based on interaction probability.  Both charged-current and
neutral-current interactions are included.  Any resulting muons are propagated until they range out,
with special focus on the region surrounding the detector volume.  Events that produced
a muon of sufficient energy within the modeled detectors were counted and used to calculate
the detector neutrino effective areas as a function of energy.  The effective areas
obtained from this toy simulation are shown in Figure \ref{fig:uhe_effective_areas}
for a 100~kt water cherenkov detector and the 17~kt liquid argon detectors.
More details of these simulations are available in a UHE TG call
presentation~\cite{Blaufuss:UHECallPresentation}

This toy simulation tool was also applied to an IceCube-like detector (detector radius of
500~m and muon detection threshold of 20~GeV).  From this simulation, effective
areas and events rates that agree with current sensitivities of the IceCube detector were
obtained, verifying the approximations made and event rates obtained to an accuracy of roughly
a factor of 2-3.

\begin{figure}[ht]
     \centering\includegraphics[width=.6\textwidth]
                               {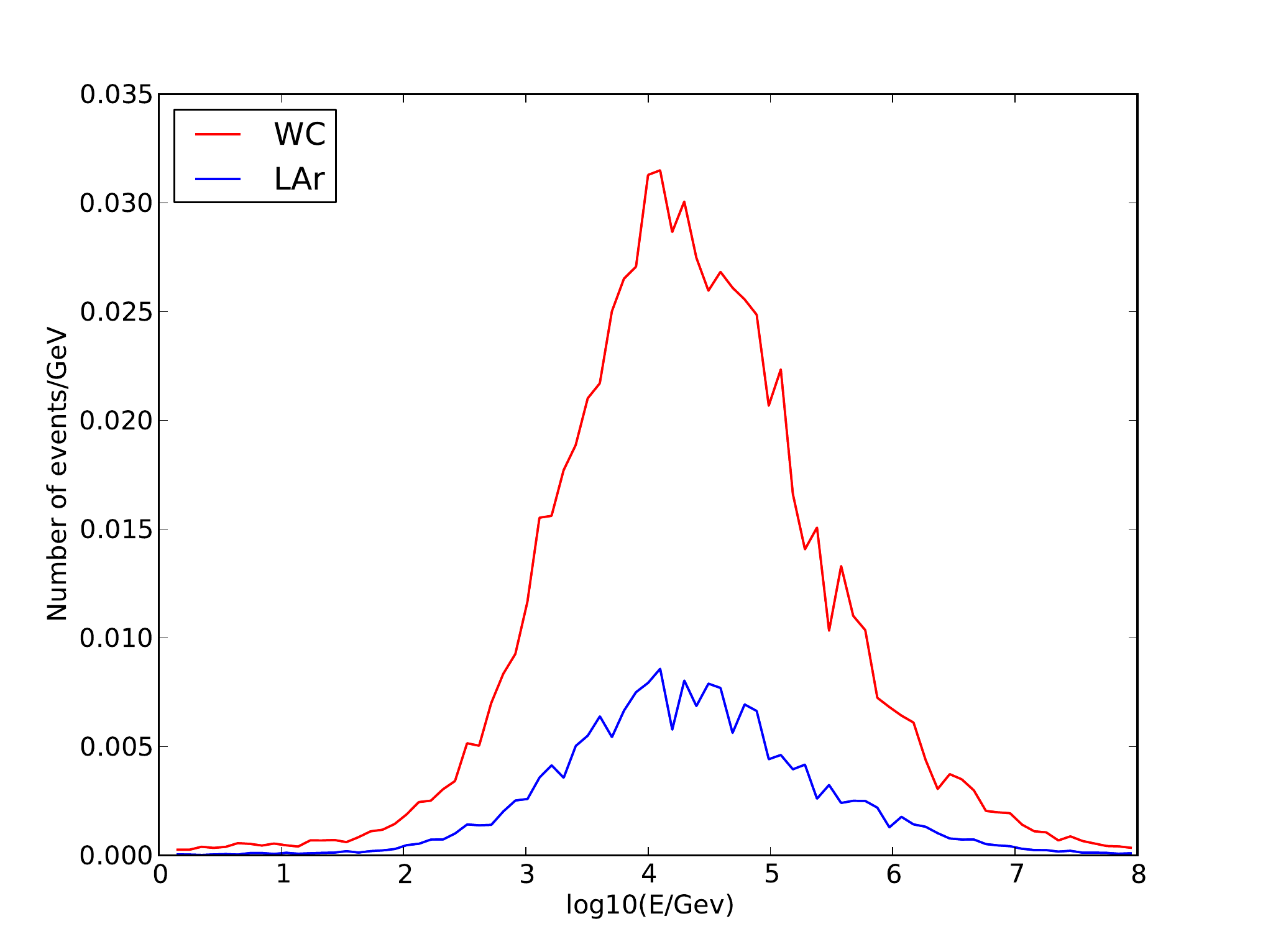}
     \caption{Energy distribution for detected events from a hypothetical
     point source with an E$^{-2}$ spectrum.  For both WC and LAr, the mean
     energy predicted is $\sim$10~TeV.}
     \label{fig:uhe_detected_events}
\end{figure}

To evaluate the sensitivity of these detector to predicted astrophysical neutrino fluxes,
three test cases where selected:

\begin{enumerate}[parsep=-1pt]
\item A generic E$^{-2}$ point source in the northern hemisphere with a flux
roughly equal to the current ANTARES point source sensitivity
($\frac{dN}{dEe} = 5\,\times\,10^{-8} E^{-2} [GeV^{-1}~cm^{-2} sec^{-1}] $)~\cite{Antares:PSResults}
\item 150 gamma-ray bursts with a neutrino fluence from the Waxman-Bahcall
fireball model~\cite{Bahcall:GRBnu}
\item 100~GeV mass WIMP annihilation at the Galactic Center (GC). This portion
of the study is still a work in progress.  Results to be updated soon. ~\cite{Beacon:WimpsGC}
\end{enumerate}
Predicted event rates are produced by convolving the neutrino effective area
for each detector with the predicted neutrino flux.  The predicted event rates
{\em per year} are
summarized in Table \ref{UHE:PrectictedCounts} and can be appropriately scaled
to the number of detectors of each type under consideration.  Even with these
smaller detector volumes, it should be noted that the most probably
neutrino energy detected is $\sim$10~TeV, as shown in
Fig. \ref{fig:uhe_detected_events}, the ``sweet spot'' when convolving a
falling neutrino spectrum with the rise in neutrino effective area.

\begin{table} [ht]  
\begin{center}
\begin{tabular}{|c|c|c|c|} \hline
             & (1) Point source   &  (2)100 WB GRB & (3) GC WIMPs \\ \hline
100~kt WC  &            0.7     &         0.07   &      TBD     \\ \hline
17~kt LAr  &            0.2     &         0.02   &      TBD     \\ \hline
Toy IceCube   &            214     &          18    &   $\sim$10s  \\ \hline
\end{tabular}
\caption{\label{UHE:PrectictedCounts} Predicted number of events per year for 3 studied
astrophysical sources.  Event rates in 1, 2 or 3 of each detector
technology can be obtained by multiplication of numbers by $N_{\rm detector}$ for each type.}
\end{center}
\end{table}

\subsection{Next Steps}

One item still needs to be completed as part of this study:
\begin{itemize}
\item Finish the calculation of expected neutrino rates from the galactic center for 100-GeV WIMPs.  This
WIMP mass is likely the best chance for an significant impact, given them threshold
of the larger neutrino observatories, but rates are still expected to be
small.
\end{itemize}

\subsection{Conclusions}
While somewhat simple in nature, this preliminary study of astrophysical neutrino
fluxes in the proposed LBNE far detector technology options indicates that neither
100-300~kt of water cherenkov detector or 17-51~kt of Liquid Argon are optimal detectors
for these neutrino signals.  Predicted event rates in the proposed detector geometries
are not competitive with the operating generation of neutrino telescopes.  There is a
slight advantage for the water-cherenkov options in terms of predicted event rates,
based on the larger detector cross-sectional area for muon detection, but
this difference is small, and should not be over-emphasized.

This study should not prevent the investigation of astrophysical neutrinos
signals in the operating detectors.  Pursuing searches that search
for signals at the edge of detector capabilities ensure that background
sensitivities (in this case, to atmospheric neutrinos) are very well
understood and provide an excellent learning opportunities for students.

\clearpage
\vfill\eject
%


\section{Solar Neutrinos}\label{sec:SolNu}

\subsection{Motivation and Scientific Impact of Future Measurements}

Even after the long standing mystery of missing solar neutrinos~\cite{davis}
was explained by data from the Super-Kamiokande and SNO~\cite{sksno}
experiments as flavor transformation of solar neutrinos, there are still
interesting open questions in solar neutrino physics. Some of these are
astrophysical (like a measurement of the fraction of energy production via
CNO cycle in the sun, or flux variations due to helio-seismological modes
which reach the solar core, or long-term stability of the solar core
temperature). But even particle physics questions remain. Can the MSW model
explain the amount of flavor transformation as a function of energy, or are
non-standard neutrino interactions required? Do solar neutrinos and reactor
anti-neutrinos oscillate with the same parameters? Some of these questions
will be answered by experimental data in the immediate future (like SNO+,
KamLAND solar phase, further Borexino data, etc.), but high statistics
measurements will be necessary to further constrain alternatives to the
standard oscillation scenario. Either a large water Cherenkov detector, or
a large liquid Argon detector, or a large liquid scintillation detector offers
unique possibilities to study these questions.

\subsection{Sensitivity of Reference Configurations}

Using event selection and reconstruction efficiency similar to Super-Kamiokande, a 300~kt water Cherenkov detector at DUSEL would collect 470,000 recoil
electrons (after selection efficiency) in ten years from $^8$B solar neutrinos
above 7~MeV. It could trace the MSW flavor transformation curve with high
precision (the uncertainties of 0.5~MeV-wide energy bins range from 0.3\% to
3\%) and should be sensitive to the change in flavor transformation due to the earth's matter
density which leads to the so-called day/night effect, the asymmetry
$A_{\mbox{\tiny DN}}=\frac{D-N}{0.5(D+N)}$ between the event rates during
the day ($=D$) and night ($=N$). This asymmetry is expected to
be only 1.5 to 2\%~\cite{skdn}; it would be measured by a 300~kt water
Cherenkov detector to 0.3\% statistical uncertainty thereby constraining
 $\Delta m^2_{\mbox{\tiny solar}}$, which is
large when measured by solar neutrinos only. This uncertainty is unlikely
to improve otherwise, since future low energy solar neutrino measurements
cannot constrain it. It is also interesting since a deviation in this
parameter between solar neutrinos and reactor anti-neutrinos cannot be
reconciled by three-flavor mixing effects (like the mixing angle).

Obviously, sparser instrumentation of this 300~kt water Cherenkov detector
immediately worsens its sensitivity to solar neutrinos. Not only will the
energy threshold increase, but also the energy resolution and therefore
background contamination worsens. Super-Kamiokande-II demonstrated solar
neutrino measurements in a large water Cherenkov detector with
$\sim3$~photo-electrons per MeV with a threshold of 7~MeV.
Radioactive background from radon decays fluctuated above this threshold
and contaminated the $\sim7,200$ solar neutrino events collected.
The statistical uncertainty of the day/night asymmetry increased to
4.2\% from the expected 2.4\%. As a conservative estimate, we can
expect the statistical accuracy of the day/night asymmetry measured
by a 300~kt water Cherenkov detector to reach 0.6\% after ten years,
if it collects only 3~photo-electrons per MeV. Super-Kamiokande (if
it runs continuously until then) will have reached $\sim0.75\%$ at
that time.

The solar neutrino physics potential of a 17~kt liquid Argon TPC largely
depends on the energy threshold and depth. Since the spallation of the
$^{40}$Ar (a rather complex nucleus compared to $^{16}$O) is likely to
produce many long-lived spallation products, only a TPC at the deepest
location has a reasonable chance of detecting solar neutrinos. Given a
10~MeV neutrino energy threshold (as reported by the ICARUS collaboration~\cite{ICARUS_Nu2010}),
it could then measure the CC/NC ratio of $^8$B solar neutrinos with high
statistical accuracy and thereby test the MSW flavor transformation curve
with high precision if the detector itself has low radioactivity levels.

The main advantage of a large, clean liquid scintillator detector is its ability to measure low
energy solar neutrinos as well. Borexino has demonstrated an energy threshold
of 3~MeV for $^8$B solar neutrinos. Since DUSEL is deep underground. A large,
clean liquid scintillator detector could precisely measure the pep flux
(constraining the solar mixing angle to ~1\%), measure the CNO fluxes and
test the MSW flavor transformation curve where it transitions from the vacuum
oscillation expectation.

\begin{table}[ht]
\begin{tabular}{|l|c|c|c|} \hline
Configuration & WC Signal Size             & WC Day/Night Asymmetry   \\
Number & p.e. per MeV & St. Acc. (10 years)       \\  \hline
   1   & $\sim3$ & 0.6\%  \\
   1a  & $\sim6$ & 0.3\%  \\
   1b  & $\sim6$ & 0.3\%  \\
   2   &   n/a\footnotemark[1]& n/a \\
   2a  &   n/a       & n/a    \\
   2b  &   n/a       & n/a    \\
   3   & $\sim3$ & 0.7\%  \\
   3a  & $\sim6$ & 0.4\%  \\
   3b  & $\sim3\&6$ & 0.5\%  \\
   4   & $\sim3$ & 0.7\%  \\
   4a  & $\sim6$ & 0.4\%  \\
   4b  & $\sim3\&6$ & 0.5\%  \\
   5   & $\sim6$ & 0.5\%  \\
   6   & $\sim6$ & 0.5\%  \\
\hline
\end{tabular}
\footnotetext[1]{Sensitivity to charged-current to neutral-current ratio depends critically on the threshold and cleanliness of the liquid argon TPC.}
\caption{Summary of solar neutrino capabilities of the reference configurations. For liquid argon TPC at 300$'$ and 800$'$, spallation backgrounds are likely to overwhelm the solar signal.
}
\label{tab:solnu_ref_configs}
\end{table}

\subsection{Conclusions}

Even after several decades of solar $^8$B neutrino experiments, the
LBNE detector at DUSEL has the potential to make unique precision
measurements. For the water Cherenkov option, the full potential is
only reached with a high density photo-sensor instrumentation yielding
at least $\sim6$~photo-electrons per MeV. A large liquid argon TPC
will only impact solar neutrino physics, if the threshold is kept low
and the detector is build with state-of-the art radiopurity requirements
(as demonstrated by the Sudbury Neutrino Observatory and the BOREXINO).
A large clean liquid scintillator detector has the best solar neutrino
physics potential: at this depth it could not only measure elastic scattering
of electrons and solar $^8$B neutrinos to very low energies
($\sim3$~MeV) but also probe CNO cycle and pep neutrinos.

\clearpage
\vfill\eject
%

\section{Geoneutrinos and Reactor Neutrinos}\label{sec:GeoRea}  

Within the earth it is believed that radioactive decays of uranium and thorium are the most significant source of heat that causes mantle convection, the fundamental geological process that regulates the thermal evolution of the earth and shapes its surface. Until recently, estimates of the total uranium and thorium content of the earth were inferred from earth formation models. However, it has been known for a long time that the uranium and thorium decays produce electron anti-neutrinos, so-called geo-neutrinos, and the detection of these geo-neutrinos near the surface of the earth can directly inform us of the deep earth uranium and thorium content. The low flux of electron anti-neutrinos from reactors, so called reactor neutrinos, at DUSEL makes it ideal for demonstrating the ability of future large detectors designed for long range nuclear reactor monitoring.

\subsection{Motivation and Scientific Impact of Future Measurements}

The geo-neutrino flux observed near the surface of the earth has a contribution from uranium and thorium in the crust and mantle. For detectors located in continental crust approximately 80\% of the signal comes from the crust, with the remainder coming from the mantle. Without the ability to determine the origin of the observed geo-neutrinos, it is not possible to determine the relative contribution from the crust and mantle with a single detector. However, with accurate knowledge of the uranium and thorium content in the crust surrounding DUSEL with approximately 200~km, it should be possible to determine the geo-neutrino contribution from the mantle although this requires higher precision due to the small fraction coming from the mantle.

KamLAND and BOREXINO have currently observed total geo-neutrinos fluxes with sensitivities limited by statistics to approximately 30\%, this sensitivity will improve only slightly with more time. The SNO+ experiment, which is currently under construction, should be able to measure the total geo-neutrino flux to a sensitivity of approximately 20\%. Because of the low reactor neutrino flux and the large size, a detector at DUSEL would be able to measure the geo-neutrino flux to approximately 5\% (limited by systematic errors) in 1 year. This should also allow for a relatively accurate measurement of the mantle geo-neutrino contribution with detailed knowledge of the local uranium and thorium crustal geo-neutrino contribution.

There are two other large geo-neutrino detectors proposed. The LENA detector would be competitive with a large detector at DUSEL, while the Hanohano detector would be complementary, as it would directly probe the mantle, so combination with a detector on the crust would determine the crustal contribution.

Aside from measuring the total uranium and thorium flux, it is predicted that the ratio of the Th/U abundance ratio in the earth is about 4/1, and measuring deviations from this value could inform us on the processing of the crustal material. Because this is a ratio, accurate measurement of this requires a larger detector than KamLAND, BOREXINO, or SNO+.

Although the flux of reactor neutrinos at DUSEL is about 24 times smaller than that at KamLAND, with a detector more than 24 times as large it should be possible to observe more reactor neutrinos. The advantage of DUSEL is that the nuclear reactors are located further away, resulting in more neutrino oscillation ``wiggles'' in the energy spectrum. With sufficient energy resolution it might be possible to measure the neutrino mixing parameter $\Delta m_{12}^2$ to approximately 1\%, an improvement upon the 3\% accuracy achieved with KamLAND.

\subsection{Sensitivity of Reference Configurations}

In a water Cherenkov detector electron anti-neutrinos can be detected by neutron inverse-beta-decay
\begin{equation}
\bar{\nu}_e + p \rightarrow n + e^+
\end{equation}
The positrons energy is equal to the antineutrinos energy minus 1.8~MeV. The maximum energy for geo-neutrinos is 3.3~MeV, although the peak in the signal is at 2.3~MeV, this results in positrons with a peak energy of 0.5~MeV and a maximum of 1.5~MeV, making it impossible to detect with a water Cherenkov detector even with the addition of gadolinium and high photocathode coverage. The energy of reactor neutrinos extends up to approximately 9~MeV making it possible to detect them with a water Cherenkov detector, however all configurations would have energy resolution below $5\%/\sqrt{E[{\rm MeV}]}$, which would wash out the oscillation spectrum, limiting the measurement to a total flux measurement. 

In a liquid Ar detector electron anti-neutrinos can be detected by Ar inverse-beta-decay
\begin{equation}
\bar{\nu}_e + ^{40}Ar \rightarrow ^{40}Cl^* + e^+
\end{equation}
The threshold for this reaction is approximately 8.5~MeV, which means that it cannot be used to detect either geo-neutrinos or reactor neutrinos. There are also elastic scattering reactions; however, these are sensitive to neutrinos as well as antineutrinos, so in order to eliminate backgrounds from solar neutrinos we need to be able to reject these by pointing at a level better than one in a thousand. At these low energies this is probably not possible with a liquid Ar detector.

\subsection{Conclusions}

The science case for detecting geo-neutrinos is very strong, it is somewhat weaker for detecting reactor neutrinos. However, it is likely not possible that these can be detected with any of the LBNE reference detector configurations, although it may be possible with a large liquid scintillator detector.

\vfill\eject

%

\section{Short Baseline Physics}
\label{SBL:sec:intro}

The design of the near detector complex is still evolving, but for the purposes of this fall 2010 report we have considered configurations with combinations of the most likely technologies based on previous studies~\cite{docdb643_ND}: a multi-ton liquid argon TPC (LAr) and two designs for Fine-Grained detectors with embedded targets of H$_2$O or D$_2$O.
In this section
we will try to evaluate the physics potential of all the proposed detectors,
with particular emphasis on the combined performance of different technologies
that could be accommodated together within the near detector complex.

One limitation of our studies is the lack of a complete GEANT simulation
for any of the reference detector options.
Therefore, we evaluate the sensitivity to different measurements based on
calculations and parameterizations of the resolutions and detector response.
Whenever possible, we try to extrapolate the detector performance from similar
existing or past experiments. This is particularly relevant for the fine-grained
tracker, for which we used results achieved by MINER$\nu$A for the scintillator
option and by NOMAD for the straw tube option (HiResM$\nu$).

\begin{figure}[h]
\centering\includegraphics[width=.8\textwidth]{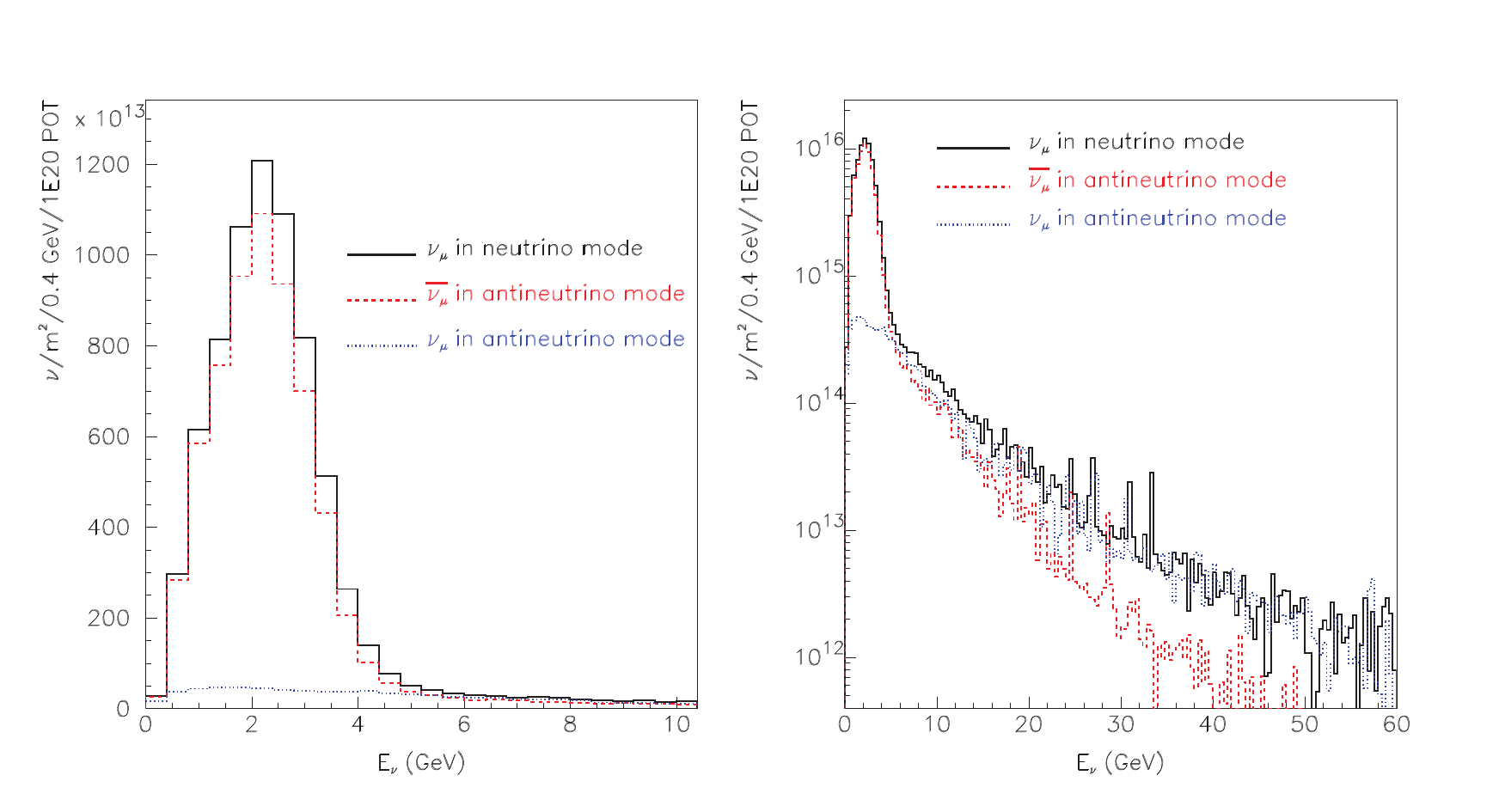}
\caption{
Comparison of predicted neutrino and antineutrino mode fluxes at the LBNE near
detector situated at 670~m on a linear (left) and log scale (right). Shown in black is the $\nu_\mu$ flux
in neutrino mode. Both the $\bar{\nu}_\mu$ (red) and $\nu_\mu$ (blue) fluxes are plotted for
antineutrino mode. All fluxes assume the 120~GeV NuMI-based design with a 250~kA horn current,
and a 2~m radius 280~m long decay region. These distributions have not been cross section weighted~\cite{docdb783}.
}
\label{fig:NDspectraLE}
\end{figure}

For the beam, we take as a reference the ``Low Energy (LE) beam'' spectrum, with the near detector
located 670~m from the target, a proton energy of 120~GeV, a horn current of
250~kA and a 2~m radius by 280~m long decay region. All the sensitivity studies
are performed with the 2009 NuMI-based beam design.
Fig.~\ref{fig:NDspectraLE} shows the resulting spectra for both the neutrino and
the anti-neutrino mode of the beam. It is worth noting that, as discussed in
Section~\ref{lbl_beam_configs}, different beam parameters resulting in different
beam spectra are being considered for the LBL oscillation analysis. The differences
with respect to our reference spectrum can be large, especially in the high energy tail,
which is particularly relevant for the short baseline measurements.
Table~\ref{tab:rates} lists the expected number of
muon neutrino interactions at the LBNE 670-m near detector site per ton of detector and for
equivalent beam exposures for both water and argon-based detectors. Neglecting
detection efficiencies and acceptance, the total raw event yields are similar
for water and argon. We evaluate the physics potential resulting from two different
assumptions for the total exposure:
\begin{itemize}
\item Scenario A: A nominal long-baseline program that delivers $7.3\times10^{20}$ protons-on-target (POT)/year for three years in neutrino mode and three years in anti-neutrino mode with a beam power of 700~kW.
\item Scenario B: A high statistics program that delivers a total $10^{22}$ POT. With a 700~kW beam this program would require 14+14 years of running (a $>$ 2~MW beam would be required to complete this scenario in a 3+3 year run).
\end{itemize}
The reference fiducial mass is assumed to be 7~t for the fine-grained tracker
(both for the scintillator and for the straw tube options), 100~t for the
LAr without a magnetic field and 25~t with a magnetic field.

For better clarity, we group the sensitivity studies in two categories:
\begin{enumerate}[parsep=-1pt]
\item Measurements to constrain systematic uncertainties in the Long-Baseline (LBL)
neutrino oscillation analyses;
\item Precision measurements of neutrino interactions.
\end{enumerate}
The design of the ND and the focus of our sensitivity studies are driven mainly by the first
category since the primary goal of the LBNE project is to perform these measurements
with the best possible sensitivity. However, the associated performance requirements for the
ND complex have not yet been established. For example, the precision required on the absolute
flux is still under investigation by the Long-Baseline and ND groups. For the purposes of
the study reported here, we have investigated the capability of several candidate ND configurations
and show that they meet or exceed the anticipated required precision based on preliminary estimates.

As will be discussed in the following sections, the two categories of measurements are highly complementary.
In general, the measurements required for the oscillation studies involve the same physics processes as the precision measurements and the accuracy needed for the latter will certainly benefit the oscillation measurements and further study may show that a similar level is actually required.
In some cases a combined analysis extracting simultaneously the underlying physics parameters is
the only way to reduce systematic uncertainties. One primary example is the
determination of the (anti)-neutrino fluxes in the presence of high $\Delta m^2$
oscillations. Similarly, a precise extraction of the (anti)-neutrino fluxes
is a necessary condition for any precision measurement of cross sections or
fundamental properties of (anti)-neutrino interactions.

%


\subsection{Measurements to Support the Far Detector Studies}
\label{SBL:sec:lblosc}

In order to allow the sensitivity of the FD to be exploited to its fullest, the Near Detector (ND)
should establish the number of expected events in the FD in the absence of oscillations with a
high precision.
It will not be sufficient to merely compare the event rates at the two detectors because the
neutrino energy distributions will be somewhat different at the two detectors. The
differences in neutrino energy between the far and near detectors are readily calculable, but
the effect of these differences on the observed event rates depends on neutrino interactions
in the material of the FD.
The rates observed at the FD depend on fluxes and cross sections, which must be known for
both $\nu$ and $\bar{\nu}$ as a function of energy, for all the processes involved in the FD
oscillation studies.
In addition, a robust computerized simulation of the response of the FD is
required.  Hence high statistics information from the ND on the energy dependence of the
yields of interactions of the various neutrino types must also be supplied.
With its greater flux and sensitivity the ND can quantify any background processes that
could interfere with the signal at the FD that is
not adequately incorporated in the simulation of the FD response.

\begin{table}[h]
\centering
\begin{tabular}{lccc}
Production mode & H$_2$O & Ar & Ar/H$_2$O ratio \\
\hline
CC QE ($\nu_\mu n \rightarrow \mu^- p$)                             & 18,977 & 23,152 & 1.22 \\
NC elastic ($\nu_\mu N \rightarrow \nu_\mu N$)                      & 7,094 & 7,165 & 1.01 \\
CC resonant $\pi^+$ ($\nu_\mu N \rightarrow \mu^- N \pi^+$)         & 25,821 & 24,014 & 0.93 \\
CC resonant $\pi^0$ ($\nu_\mu n \rightarrow \mu^- \ p \, \pi^0$)         & 6,308 & 7,696 & 1.22 \\
NC resonant $\pi^0$ ($\nu_\mu N \rightarrow \nu_\mu \, N \, \pi^0$)      & 6,261 & 6,198 & 0.99 \\
NC resonant $\pi^+$ ($\nu_\mu p \rightarrow \nu_\mu \, n \, \pi^+$)      & 2,694 & 2,182 & 0.81 \\
NC resonant $\pi^-$ ($\nu_\mu n \rightarrow \nu_\mu \, p \, \pi^-$)      & 2,325 & 2,930 & 1.26 \\
CC DIS ($\nu_\mu N \rightarrow \mu^- X$, $W>2$)                     & 29,989 & 31,788 & 1.06 \\
NC DIS ($\nu_\mu N \rightarrow \nu_\mu X$, $W>2$)                   & 10,183 & 10,285 & 1.01 \\
CC coherent $\pi^+$ ($\nu_\mu A \rightarrow \mu^- A \pi^+$)       & 1,505 & 1,505 & 1.01 \\
NC coherent $\pi^0$ ($\nu_\mu A \rightarrow \nu_\mu A \pi^0$)       & 790 & 790 & 1.01 \\
NC resonant radiative decay ($N^* \rightarrow N \gamma $)       &  41 &  &  \\
Inverse Muon Decay ($\nu_\mu e \rightarrow \mu^- \nu_e$)            & 6 & 6 & 1.00 \\
$\nu_\mu e^- \rightarrow \nu_\mu e^-$                                & 11 & 11 & 1.00 \\
Other                                                               & 17,023 & 17,193 & 1.01 \\
\hline\hline
Total CC                               & 94,948 & 100,645 & 1.06 \\
Total NC+CC                            & 129,028 & 134,189 & 1.04
\end{tabular}
\caption{Estimated $\nu_\mu$ production rates for both water and argon targets
per ton (water or argon) for $1\times10^{20}$ POT at 670~m assuming neutrino
cross sections predictions from Nuance~\cite{nuance} and a 120~GeV proton beam, 250~kA horn
current, and a 2~m radius 280~m long decay region. Processes are defined at the initial neutrino
interaction vertex and thus do not include final state effects. These estimates do not
include detector efficiencies or acceptance~\cite{docdb740,docdb783}.
}
\label{tab:rates}
\end{table}

\subsubsection{In Situ Measurement of Fluxes for the LBL Oscillation Studies}
\label{SBL:sec:fluxes}

since the ND is close to the neutrino source it sees a neutrino energy spectrum slightly, but
significantly, different from that at the FD. There are differences in the energy of the
principal components of the $\nu_\mu$ and $\bar{\nu}_\mu$ beams that must be accounted for quantitatively.
The main task of the near detector complex is the measurement of the flux as
a function of neutrino energy for each neutrino species ($\nu_\mu, \bar{\nu}_\mu, \nu_e, \bar{\nu}_e$).
Ideally this flux measurement should be an absolute measurement
using known cross sections to verify the fluxes calculated by the neutrino production
models. This verification is essential in providing confidence in the knowledge of the flux.

As discussed in the following, we will have several independent methods to determine
the (anti)-neutrino fluxes from the ND data at LBNE. The combination of the large
statistics expected at LBNE with the finely segmented detectors at the near site
allows comparable precisions from different techniques. This redundancy of measurements
is a necessary condition for the validation of the flux extraction at the level
of accuracy expected at LBNE.

The in situ determination of neutrino fluxes is not only a service measurement crucial for the
oscillation studies in the FD, but it is closely related to the precision measurements of
fundamental interactions discussed in Section~\ref{SBL:sec:nuint}. Indeed, the possibility to make
unexpected discoveries within the short baseline physics program critically depends upon the
knowledge of the incoming (anti)neutrino flux. Historically, the uncertainty on the fluxes
has always been one of the main limiting factors for all past neutrino scattering experiments.
The high intensity of the LBNE beam coupled with the excellent granularity
in the ND complex would allow a substantial reduction of the flux uncertainty. Furthermore,
as discussed in the following, the extraction of the fluxes themselves relies upon the knowledge
of specific physics processes, requiring an understanding of the theoretical and experimental
issues related to their measurements.

\begin{table}[h]
\centering
\begin{tabular}{lccccc}
Flavor & Technique & Relative  & Absolute & Relative  & Detector requirements \\
       &           & abundance & normalization & flux $\Phi(E_\nu)$ &   \\
\hline\hline
 $\nu_\mu$  & $\nu e^- \to \nu e^-$   & 1.00 & 1-3\%  & $\sim5\%$  & $e$ identification/resolution \\
   &  &  &  &  &  $e^-/e^+$ separation \\  \hline
 $\nu_\mu$  & $\nu_\mu e^- \to \mu^- \nu_e$   & 1.00 & 1-4\%  &  & $\mu^-/\mu^+$ separation  \\
   &  &  &  &  &  $\mu$ energy scale \\ \hline
 $\nu_\mu$  & $\nu_\mu n \to \mu^- p$  & 1.00 & $3-5\%$  & $3-5\%$ & $D$ target  \\
   &  $Q^2 \to 0$ &  &  &  &  $p$ angular and momentum resolution  \\  \hline
 $\nu_\mu$  & low-$\nu_0$   & 1.00 &  & 2.0\% & Magnetized detector separating $\mu^-/\mu^+$  \\ \hline
 $\nu_e$   & low-$\nu_0$ & 0.01  & 1-3\%  & 2.0\%  & $e^-/e^+$ separation ($K^0_L$) \\
\hline
\end{tabular}
\caption{Precisions achievable from in situ $\nu_\mu$ and $\nu_e$ flux measurements
in the fine-grained ND with different techniques. The reference beam configuration
with a distance from the target of 670~m, a 120~GeV proton beam, 250~kA horn
current, and a 2~m radius 280~m long decay region is assumed.
}
\label{tab:fluxes}
\end{table}

\paragraph{\bf Determination of the Absolute Flux Normalization}
\label{SBL:sec:absflux}

The experimental determination of the absolute neutrino flux relies upon the measurement of
well-known physics reactions in the ND. There are three main complementary options to be
exploited at LBNE:
\begin{enumerate}[parsep=-1pt]
\item Neutral-Current elastic scattering off electrons: $\nu e^- \to \nu e^-$;
\item Inverse Muon Decay (IMD) interactions: $\nu_\mu e \rightarrow \mu^- \nu_e$;
\item Quasi-elastic (QE) Charged-Current interactions in the limit $Q^2\to 0$:
$\nu_\mu n \rightarrow \mu^- p$.
\end{enumerate}

The total cross section for NC elastic scattering off electrons is given by~\cite{Marciano:Nuel}:
\begin{eqnarray}
\sigma (\nu_l e \to \nu_l e) & = & \frac{G_\mu^2 m_e E_\nu}{2\pi} \left[ 1 -4 \sin^2 \theta_W + \frac{16}{3} \sin^4 \theta_W \right] \\
\sigma (\bar{\nu}_l e \to \bar{\nu}_l e) & = & \frac{G_\mu^2 m_e E_\nu}{2\pi} \left[ \frac{1}{3} -\frac{4}{3} \sin^2 \theta_W + \frac{16}{3} \sin^4 \theta_W \right]
\end{eqnarray}
where $\theta_W$ is the weak mixing angle. For $\sin^2 \theta_W\simeq~0.23$ the above
cross sections are very small $\sim10^{-42} (E_\nu/\gev)~cm^2$. The NC elastic scattering
off electrons can be used to determine the absolute flux normalization since the
cross sections only depend upon the the knowledge of $\sin^2 \theta_W$.

The value of $\sin^2 \theta_W$ at the average momentum transfer expected
at LBNE $Q\sim0.07 \gev$ can be extrapolated down from the LEP/SLC measurements
with a precision of $\sin 0.2\%$ within the Standard Model (SM). However, in order to take into
account potential deviations from the SM predictions, in the flux extraction we must
initially consider a theoretical uncertainty $\leq 1\%$, obtained from direct measurements
of $\sin^2 \theta_W$ at momentum scales comparable to the LBNE one.

As discussed in Section~\ref{SBL:sec:sin2thetaW}, precision electroweak measurements
with the ND data at LBNE can determine the value of $\sin^2 \theta_W$ to better than 0.3\%.
The theoretical uncertainty on the absolute flux normalization can therefore be improved
substantially by a combined analysis with the electroweak measurements.

\begin{figure}[h]
\centering\includegraphics[width=.95\textwidth]{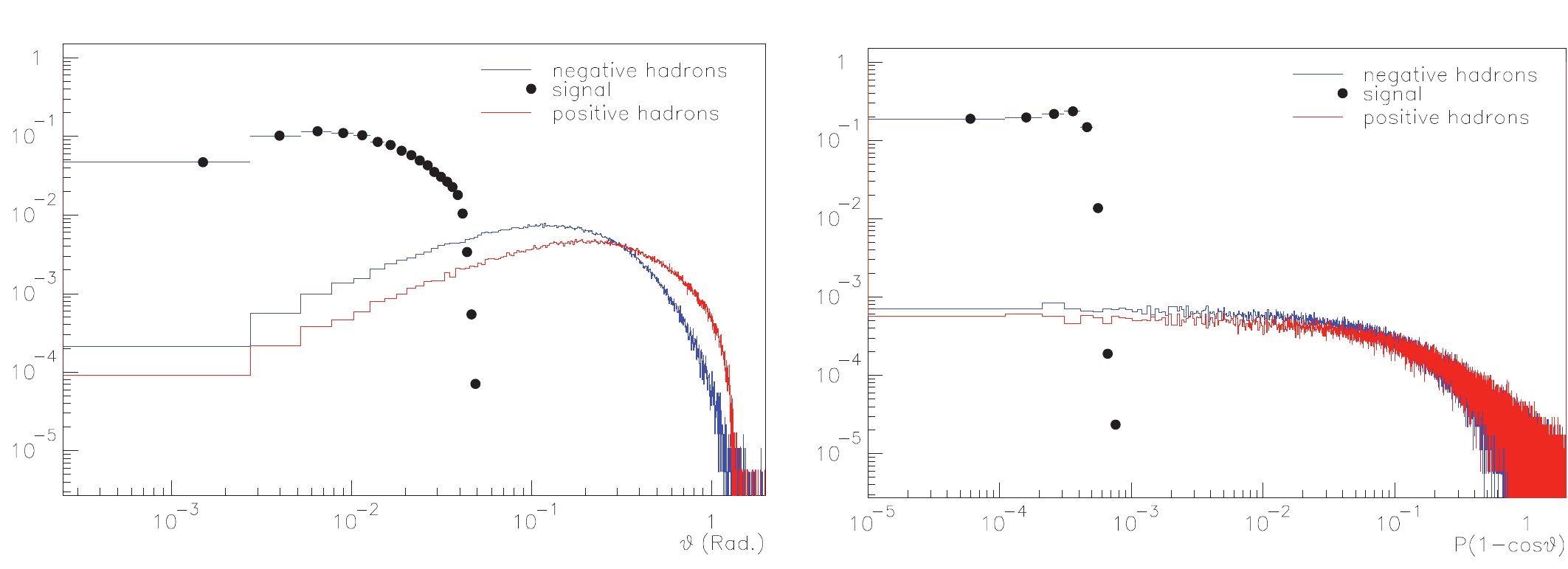}
\caption{Distributions of the angle of the electron with respect to the beam direction
(left) and of the discriminating variable $P(1 -\cos \theta)$ (right) for NC elastic
scattering off electrons and for the corresponding backgrounds in the ND.
}
\label{fig:NCnu-e}
\end{figure}

The signature of the process $\nu_l (\bar{\nu}_l) e \to \nu_l (\bar{\nu}_l) e$ is a single
electron in the final state, emitted almost collinearly with the beam direction
($\theta\sim$ mrad). The dominant backgrounds are given by NC $\pi^0$ production and
single photon production, in which one photon fakes a single electron. A smaller
background contribution is given by $\nu_e$ quasi-elastic scattering events in which the
proton is not visible. This measurement
requires a detector that can efficiently distinguish photons from electrons.

The low density magnetized tracker proposed for LBNE can identify electrons and positrons
and reconstruct the corresponding track parameters, allowing a background rejection
$\leq 10^{-6}$ with an overall efficiency for NC elastic scattering events of 64\%.
Figure~\ref{fig:NCnu-e} shows the distributions of kinematic variables for signal
and background.
It is thus possible to select a sample of NC elastic scattering events off electrons
with small background in the LBNE ND. The main limitation of such a measurement
is the statistics of the selected sample, which, for the reference beam configuration
with $22 \times 10^{20}$ pot and a fiducial mass of about 7~t, corresponds
to about 1000 events with a background of six events, giving a 3\% precision on
the flux normalization.
It must be noted that in a low density magnetized design the background originates
from asymmetric $\gamma$ conversions in which the positron is not reconstructed.
This type of background is expected to be charge-symmetric and this fact gives a
powerful tool to calibrate the $\pi^0/\gamma$ background in situ.

We can now consider the performance of a LAr TPC, which is one of the options
considered for the ND. It has good electron identification and energy resolution
for electromagnetic showers ($3\%/\sqrt{E}$) and the rejection factor of
$\pi^0$ background is at the level of $2-5 \times 10^{-3}$ with an electron
reconstruction efficiency of $80-90\%$. A measurement of NC elastic scattering off
electrons can therefore be competitive if we have a fiducial mass of the
order 100~t. This would imply about 20,000 selected events with the
reference beam configuration, giving a statistical precision on the
flux normalization of 0.7\%.

The measurement in LAr has substantially more
background than the one performed in a low density magnetized detector
and the analysis has to rely upon background subtraction as a function of the
kinematic variable $E\theta^2$ (cf. CHARM II analysis). A combined analysis
between the LAr massive detector and the low density magnetized tracker can achieve
the optimal sensitivity. The low density magnetized detector provides a precise calibration
of the background shape for the subtraction procedure, while the LAr massive detector gives the
required statistics. The difference in target nuclei (carbon vs. argon) is not critical
once the background shape is constrained as a function of $E\theta^2$, since the absolute
background scale can be determined directly in LAr with the side bands. In addition, the
NC $\pi^0/\gamma$ production can be constrained by dedicated measurements in LAr and
in the magnetized tracker. In summary, a combined analysis of the two detectors
has the potential to extract the absolute flux normalization at LBNE to $\sim1\%$ precision
with the reference beam configuration.

The measurement of NC elastic scattering off electrons can only
provide the integral of all neutrino flavors, which, for the neutrino mode, includes
about 92\% $\nu_\mu$, 7\% $\bar{\nu}_\mu$ and 1\% $\nu_e + \bar{\nu}_e$. In order to
determine the individual $\nu_\mu$ and $\bar{\nu}_\mu$ components we need to measure the
ratio $\bar{\nu}_\mu/\nu_\mu$ from CC interactions to better than 10\%. This requirement
is relatively loose since with a magnetized detector we can measure $\bar{\nu}_\mu/\nu_\mu$
to a few percent. It must be noted that the reactions $\nu_e e \to \nu_e e$ and
$\bar{\nu}_e e \to \bar{\nu}_e e$ result from combined $W$ and $Z$ boson exchange.
The corresponding cross section for $\nu_e(\bar{\nu}_e)$ is enhanced by a factor of
about six (three) with respect to $\nu_\mu (\bar{\nu}_\mu)$.
The relative $\nu_e$ contributions to the measured NC elastic scattering sample
is therefore increased to about 6\% of the total, requiring an additional measurement
of the ratio $\nu_e/\nu_\mu$ from CC interactions to $10-15\%$.

The possibility to achieve the high statistics beam exposure (Scenario B)
would have a substantial impact on the precision
achievable with the NC elastic scattering off electrons since this channel is dominated
by statistical uncertainties with the reference beam configuration. The resulting
increase in the number of protons on target by a factor of four would allow the collection of
about 80,000 events in the LAr detector, providing a statistical precision of $\sim0.3\%$.
At this level of precision the theoretical uncertainty on the knowledge of
$\sin^2 \theta_W$ will start to play a role in the determination of the
absolute flux normalization and a analysis will require a complementary
precision electroweak measurement (Section~\ref{SBL:sec:sin2thetaW}).

A second process that can be used to crosscheck the absolute flux normalization is the
Inverse Muon Decay (IMD): $\nu_\mu e^- \to \mu^- \nu_e$.
However, this reaction is characterized
by a very high energy threshold $E_\nu > (m_\mu^2 - m_e^2)/2m_e\simeq 10.9~\gev$ and
therefore we expect only about 700 events in the fine-grained tracker with the reference
beam configuration. With a fiducial mass of 100 tons we can collect about 10,000 IMD
events in the LAr detector, which would provide an independent flux normalization with
a statistical precision of $\sim1\%$. The analysis has to rely upon the subtraction
of the large background from $\nu_\mu$ CC Quasi-Elastic events with a single
reconstructed track in the final state and from the corresponding resonance and DIS events.
The identification of the charge of the muon is a crucial requirement for this measurement.
This task can be easily accomplished by the downstream magnetized tracker given that
$E_\mu \geq 10.9~\gev$. The use of IMD provides a direct measurement of the $\nu_\mu$
flux in the high energy tail of the spectrum. A relative determination of the flux as a
function of $E_\nu$ is required to fully constrain the spectrum and to extract the
corresponding total integral.

A third independent method to extract both the absolute flux normalization and the
relative flux as a function of $E_\nu$ is through the Quasi-Elastic (QE) CC scattering
$\nu_\mu n(p) \to \mu^- p(n)$. Neglecting terms in $(m_\mu/M_n)^2$, at $Q^2=0$ the
QE cross section is independent of neutrino energy for $(2E_\nu M_n)^{1/2} > m_\mu$:
\begin{equation}
\frac{d \sigma}{d Q^2} \mid Q^2 = 0 \mid = \frac{G_\mu^2 \cos^2 \theta_c}{2\pi}
\left[ F_1^2(0) + G_A^2(0) \right] = 2.08 \times 10^{-38}~cm^2\gev^{-2}
\end{equation}
which is determined by neutron $\beta$ decay and has a theoretical uncertainty $<1\%$.
The flux can be extracted experimentally by measuring low $Q^2$ QE interactions ($0-0.05~\gev$)
and extrapolating the result to the limit of $Q^2=0$. This measurement requires a deuterium
or hydrogen (for anti-neutrino) target to minimize the smearing due to Fermi motion and
other nuclear effects. This requirement can only be achieved by using both H$_2$O and
D$_2$O targets embedded into the fine-grained tracker and to extract the events produced
in deuterium by statistical subtraction of the larger oxygen component.
The experimental resolution on the muon and proton momentum and angle is crucial.
A low density tracker with $\rho\sim0.1~{\rm g/cm}^3$ would increase the
proton range by about one order of magnitude, thus increasing the proton reconstruction
efficiency in the region of $Q^2\leq 0.05~\gev^2$. The dominant
uncertainties are related to the extrapolation to $Q^2=0$, to the theoretical cross section
on deuterium, the experimental resolution, and to the statistical subtraction.
Overall, it seems feasible to achieve a precision of $3-5\%$ on the fluxes with the
current understanding of the theoretical cross sections.\\

\paragraph{\bf Determination of the Relative Flux as a Function of Energy}

A precise determination of the relative neutrino flux as a function of energy can be achieved
with the low-$\nu_0$ method.
The dynamics of neutrino-nucleon scattering implies that the number of events in a given
energy bin with hadronic energy $E_{\rm had} < \nu_0$ is proportional to the neutrino (antineutrino)
flux in that energy bin up to corrections ${\cal O}(\nu_0/E_\nu)$ and ${\cal O}(\nu_0/E_\nu)^2$.
The method follows from the general expression of the $\nu$-nucleon differential cross section:
\begin{equation}
{\cal N} (\nu < \nu_0) = C \Phi(E_\nu) \nu_0 \left[ {\cal A} +
\left( \frac{\nu_0}{E_\nu} \right) {\cal B} + \left( \frac{\nu_0}{E_\nu} \right)^2 {\cal C} +
{\cal O} \left( \frac{\nu_0}{E_\nu} \right)^3 \right]
\end{equation}
where the coefficients ${cal A} = {\cal F}_2$, ${\cal B} = ({\cal F}_2 \pm {\cal F}_3)/2$,
${\cal C} = ({\cal F}_2 \mp {\cal F}_3)/6$ and ${\cal F}_i =\int^1_0 \int^{\nu_0}_0 F_i(x) dx d\nu$
is the integral of structure function $F_i$.
The number ${\cal N}(\nu<\nu_0)$ is proportional to the flux up to correction factors of the
order ${\cal O} (\nu_0/E_\nu)$ or smaller,
which are not significant for small values of $\nu_0$ at energies $\geq \nu_0$.
It should be pointed out that the coefficients ${\cal A, B, C}$ are determined for each energy
bin and neutrino flavor within the ND data themselves. Since our primary interest is the relative
flux determination, i.e. neutrino flux in an energy bin relative to another energy bin,
variations in the coefficients do not affect the relative flux.

The prescription for the relative flux determination is simple: we count the number of $\nu$-CC
events below a certain small value of hadronic energy ($\nu_0$). The observed number of events, up
to the correction due to the finite $\nu_0$ of the order ${\cal O} (\nu_0/E_\nu)$, in each
total visible energy bin is dominated by the corresponding lepton energy, is proportional to
the relative flux. The smaller the factor $\nu_0/E_\nu$, the smaller is the correction.
It is apparent from the above discussion that this method of relative flux determination
is not very sensitive to nucleon structure, QCD corrections, and types of $\nu$-interactions such
as scaling or non-scaling. With the excellent granularity and resolution foreseen in the
low density magnetized tracker it will be possible to use a value of $\nu_0\sim0.5$~GeV or lower,
thus allowing flux predictions down to $E_\nu\sim0.5$~GeV.
For the scintillator option the value of a usable $\nu_0$ is yet to be determined
but it is expected to be higher than that for the low density tracker.
Note that the non-prompt backgrounds will be larger with the
higher density detectors, especially for the $\bar{\nu}_\mu$ component at low energy.

\begin{figure}[htb]
\centering\includegraphics[width=.55\textwidth,angle=-90]{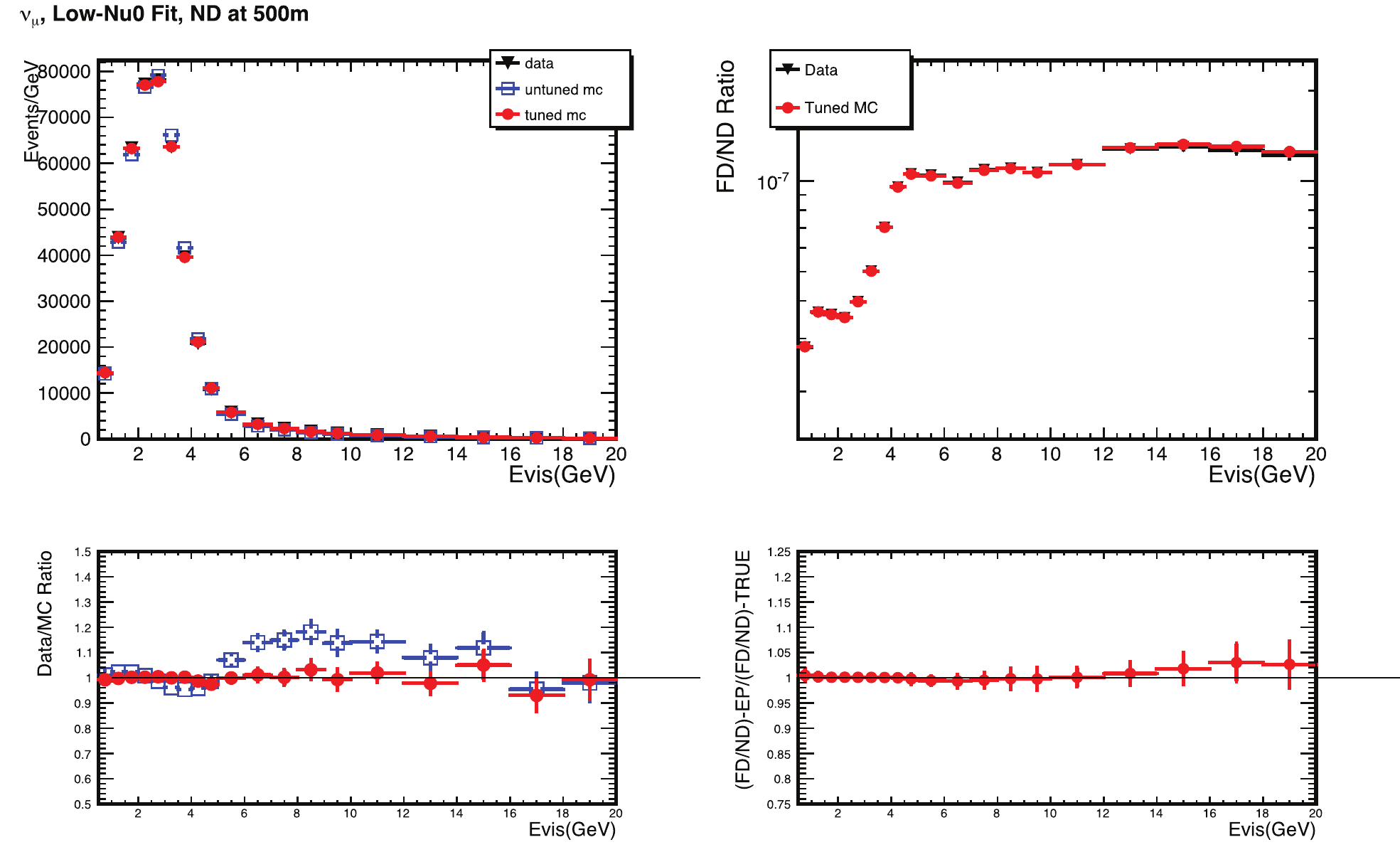}
\caption{Ratio FD/ND as determined in the ND with the low-$\nu_0$ method for the $\nu_\mu$ flux and a ND located at a distance of 500~m.
}
\label{fig:lowNu0}
\end{figure}

In this analysis, the key measurable quantities are the resolution for the low-$\nu$ events and
the systematic precision of the muon-momentum. The $\nu_\mu (\bar{\nu}_\mu)$-CC flux provides a
measure of the $\pi^+/K^+/\mu^+(\pi^-/K^-/\mu^-)$ content of the beam. We first obtain the relative
$\nu_\mu (\bar{\nu}_\mu)$ flux at the ND. We then fit the $d^2\sigma/ dx_FdP_T^2$ of the
parent mesons to the $\nu_\mu$ flux with a simple parametrization:
\begin{equation} \label{SBL:eqn:xFpT}
\frac{d^2 \sigma}{dx_FdP_T^2} = f(x_F) g(P_T) h(x_F, P_T)
\end{equation}
The ingredients to this empirical fit to the meson production
cross section (EP) are the following:
\begin{itemize} [parsep=-2pt]
\item Trace parent mesons through a simulation of the beam elements;
\item Decay the parent mesons;
\item Predict $\nu_\mu$ and $\bar{\nu}_\mu$ fluxes by folding experimental acceptance;
\item Add external constraints on $\pi/K$ from hadro-production experiments (MIPP);
\item Compare predictions to the measured spectra at the ND and minimize the
corresponding $\chi^2$.
\end{itemize}
It must be noted that the simple smoothness requirement on the functional form in
Eqn.~(\ref{SBL:eqn:xFpT}) allows to extend the flux predictions down to $E_\nu\sim\nu_0$.
In order to evaluate the sensitivity which can be achieved at LBNE with the low-$\nu_0$
method we performed the flux analysis for the neutrino beam mode (positive focusing) using
$\nu_\mu +\bar{\nu}_\mu$ CC mock-data from the low density magnetized detector
and $\nu_0<1~\gev$. The systematic error analysis included the following effects:
\begin{itemize}[parsep=-2pt]
\item Variation of the functional form in Equation~(\ref{SBL:eqn:xFpT});
\item Variation of both muon and hadron energy scales;
\item Systematic shift of the $\nu_0$ values by $\pm 20\%$;
\item Variation of quasi-elastic, resonance and DIS cross sections by $\pm 20\%$;
\item Effect of the beam-transport elements.
\end{itemize}
The beam-transport uncertainty will require additional studies after completing the
design of the beam line
since it depends on the precise beam and on the inert material that the hadrons encounter.
Depending upon our knowledge of $p/\pi/K$ nuclear collisions, this uncertainty can
become dominant for $\bar{\nu}_\mu$ production in the neutrino beam mode at low energy.
Figure~\ref{fig:lowNu0} shows the mock-data and the corresponding
fitted flux with the ND positioned at 500~m from the target. Having constrained the
$d^2\sigma/ dx_FdP_T^2$ of the pions/kaons, we can predict the flux at the ND location.
Overall we achieved a precision $\leq 2\%$ on the relative $\nu_\mu$ flux with
the low-$\nu_0$ method in the energy region $1 \leq E_\nu \leq 30~\gev$ in the fit
with $\nu_0 < 1~\gev$. Similar uncertainties are expected for the $\bar{\nu}_\mu$
component (dominant one) in the anti-neutrino beam mode (negative focusing).

The low-$\nu_0$ method was used to study the effect of different locations of the
near detector on the flux predictions at the FD site. The low-$\nu_0$ fit described
above was repeated assuming a distance of the ND from the target of 1500~m, 1000~m, 750~m,
and 500~m, respectively. It must be noted that the ratio between the spectra at the
FD and ND sites is not flat even at $L$=1000~m. However, the spectral distortions
increase substantially for $E_\nu<5~\gev$ as the distance from the target is reduced.
The low-$\nu_0$ method can correct for such spectral distortions since it
directly extracts the parent $\pi^\pm/K^\pm/K^0_L$ content from the ND data
and extrapolates the fluxes at the FD site by taking into account the beam
transport elements and by decaying the mesons. Results show that the FD/ND ratio
predicted by the low-$\nu_0$ technique can reproduce the actual FD/ND
ratio to better than $2\%$, regardless of the distance between the ND and the
target. As a consequence the ND can be located closer to the target without
increasing the systematic uncertainties for the LBL oscillation searches.

The relative flux as a function of energy could be also determined from NC elastic
scattering off electrons, which is a two body process with the initial electron being
at rest. The calculation of the incoming neutrino energy requires a resolution
of few milliradians on the angle of the outgoing electron as well as a good energy resolution.
The low density magnetized tracker can in principle fulfill such requirements. However,
the limited statistics does not allow a precision better than $\sim5\%$
on the relative flux. The angular resolution of the LAr detector for electrons $\sim8^\circ$
does not allow a precise measurement of the spectrum due to multiple scattering and
shower development.

A third independent method to determine the relative flux as a function of energy is using the
quasi-elastic interactions on a deuterium target in the limit of $Q^2\to 0$.
The precision achievable with this technique is the same as the corresponding
absolute flux measurement discussed in the previous Section.\\

\paragraph{\bf Flavor Content of the Beam: \boldmath{$\nu_\mu, \bar{\nu}_\mu, \nu_e, \bar{\nu}_e$}}
\label{SBL:sec:flavorbeam}

As discussed in the previous Section, the low-$\nu_0$ method allows the prediction of both
the relative $\nu_\mu$ and $\bar{\nu}_\mu$ at the FD location from the
measure of the $\pi^+/K^+/\mu^+(\pi^-/K^-/\mu^-)$ content of the beam at ND.
In addition, with a ND capable of identifying $\bar{\nu}_e$ CC interactions we can directly
extract the elusive $K^0_L$ content of the beam. Therefore, an accurate measurement of
$\nu_\mu, \bar{\nu}_\mu$ and $\bar{\nu}_e$ CC interactions provides an absolute
prediction of the $\nu_e$ content of the beam, which is an irreducible background for the
$\nu_e$ appearance search in the FD:
\begin{eqnarray} \label{eqn:nueparents}
\nu_e & \equiv & \mu^+(\pi^+\to \nu_\mu) \oplus K^+(\to \nu_\mu) \oplus K^0_L \\
\bar{\nu}_e & \equiv & \mu^-(\pi^-\to \bar{\nu}_\mu) \oplus K^-(\to \bar{\nu}_\mu) \oplus K^0_L
\end{eqnarray}
The $\mu$ component is well constrained from $\nu_\mu (\bar{\nu}_\mu)$ CC data at low energy,
while the $K^\pm$ component is only partially constrained by the
$\nu_\mu (\bar{\nu}_\mu)$ CC data at high energy and requires external hadro-production
measurements of $K^\pm/\pi^\pm$ ratios at low energy from MIPP.
Finally, the $K_L^0$ component can be constrained by the $\bar{\nu}_e$ CC data and by
external dedicated measurements at MIPP. The relative contributions to the $\nu_e$ spectrum
are 87\% (54\%) for the $\mu^+$, 10\% (33\%) for the $K^+$ and 3\% (15\%) for the $K_L^0$
in the energy range $1 (5) \leq E_\nu \leq 5 (15)\gev$.
Based on the NOMAD experience, we expect to achieve
a precision of $\leq 0.1\%$ on the flux ratio $\nu_e/\nu_\mu$. Taking into account the
projected precision of the $\nu_\mu$ flux discussed in the previous Section, this
translates into an absolute prediction for the $\nu_e$ flux at the level of $2\%$.
It should be pointed out that while the scintillator based ND option will be able to measure the
$\nu_\mu, \bar{\nu}_\mu, \nu_e + \bar{\nu}_e$ flavor content of the beam it will not be able to
distinguish between $\nu_e$ and $\bar{\nu}_e$. The non-prompt backgrounds in the
$\nu_e + \bar{\nu}_e$ and in the $\bar{\nu}_\mu$ would also be larger.

Finally, the fine-grained ND can directly identify $\nu_e$ CC interactions from the
LBNE beam. The relevance of this measurement is twofold: a) it provides an independent
validation for the flux predictions obtained from the low-$\nu_0$ method and b) it can
further constrain the uncertainty on the knowledge of the absolute $\nu_e$ flux.

The flux ratio $\bar{\nu}_\mu/\nu_\mu$ as a function of energy can be determined with the
low-$\nu_0$ method with an accuracy of $\leq 2\%$ in the region $1.5 < E_\nu < 30 \gev$
and of $\sim3\%$ in the region $0.5 < E_\nu < 1.5 \gev$. These numbers refer to the
neutrino beam mode (positive focusing) and are obtained from
a fit to mock-data from the low density magnetized detector to extract the parent
$\pi/K$ distributions. The beam-transport uncertainty can become dominant for
$\bar{\nu}_\mu$ production (contamination) in the neutrino beam mode.\\

\paragraph{\bf Effects of High $\Delta m^2$ Oscillations on the Flux Extraction}
\label{SBL:sec:fluxosc}

All the results described in the previous Sections were obtained under the assumption
the events observed in the ND are originated by the same (anti)-neutrino flux produced
by the decay of the parent mesons. The recent results from the MiniBooNE experiment
might suggest the possibility of relatively high $\Delta m^2$ anti-neutrino oscillations
consistent with the LSND signal. This effect, if confirmed, seem to indicate a
different behavior between neutrinos and anti-neutrinos which would imply CP or CPT
violation. The MINOS experiment also reported different oscillation parameters between
$\nu_\mu$ and $\bar{\nu}_\mu$ from the disappearance analysis.

The presence of high $\Delta m^2$ oscillations with characteristic oscillation length
comparable with the ND baseline at the LBNE energies, would imply the spectra
observed in the ND could be already distorted by neutrino oscillations.
Two main effects are expected on the flux extraction from a MiniBooNE/LSND oscillation:
\begin{itemize}[parsep=-2pt]
\item The $\nu_e$ and $\bar{\nu}_e$ CC spectra cannot be directly used to extract
the $K_L^0$ content in Equation~(\ref{eqn:nueparents}) nor to predict event rates in the FD;
\item A deficit is induced in the $\bar{\nu}_\mu$ CC spectrum from a significant disappearance rate.
\end{itemize}
Any in situ determination of the fluxes would then require the unfolding of the oscillation
effect from the measured spectra. The measurement
strategy in the ND should necessarily include a combined oscillation and flux analysis.
Since the ND cannot be easily moved, different complementary measurements are needed.
A MiniBooNE/LSND signal imposes stringent constraints on the ND design, as described
in Section~\ref{SBL:sec:highMosc}.

Several follow-up experiments have been proposed to investigate the MiniBooNE/LSND
effects: move MiniBooNE to a near detector location~\cite{BooNE}, OscSNS at the ORNL
neutron spallation source~\cite{ORNL} or a two-detector
LAr experiment at the CERN PS~\cite{CERN-PS}.
All of them are expected to cover the region in the oscillation parameters
consistent with MiniBooNE/LSND data, so that by the time LBNE will take data
we might have a confirmation or disprove of the high $\Delta m^2$
oscillation hypothesis. However, the precision that will be ultimately achieved
in the determination of the fluxes at LBNE is directly connected to the high
$\Delta m^2$ oscillation parameters.
If the oscillation is confirmed, we will need dedicated precision measurements
in the ND at LBNE.\\

\paragraph{\bf External Constraints from Hadro-production Measurements}

One of the key points for an accurate determination of the fluxes is the possibility
to exploit different complementary techniques, providing the redundancy necessary to
reach precision of the order of few percent. In the previous Sections we have
studied the measurements to be performed in situ. However, in order to fully exploit
the power of the ND data some external measurements are required to constrain the
yields of the parent mesons decaying into (anti)-neutrinos:
\begin{itemize}[parsep=-1pt]
\item $K^+/\pi^+$ as a function of $P (2 \leq P \leq 20~\gev)$ and $P_T (\leq 0.4~\gev)$
of $K^+$ and $\pi^+$;
\item $K^-/\pi^-$ as a function of $P (2 \leq P \leq 20~\gev)$ and $P_T (\leq 0.4~\gev)$
of $K^-$ and $\pi^-$;
\item $K^0/K^+$ ratio.
\end{itemize}
These can be performed with dedicated hadro-production experiments (e.g. MIPP at Fermilab).
The accuracy in the external measurements must be comparable to the precision of the ND data
and to the target precision for the flux extraction.
The program must include separate measurements of the above quantities off different targets:
\begin{itemize}[parsep=-2pt]
\item The LBNE neutrino target;
\item All the thin/thick Al, Cu, etc. targets that compose the horn and the beam elements;
\item Air (N).
\end{itemize}
The external hadro-production measurements are crucial for the LBL $\nu_e$ appearance search
since they allow a prediction of the $\nu_e(\bar{\nu}_e)$ flux independent from the
$\nu_e$ and $\bar{\nu}_e$ CC events observed in the ND, as shown
in Equation~(\ref{eqn:nueparents}).
It must be also noted that in the presence of high $\Delta m^2$ oscillations
external hadro-production measurements are the only source of information which can
be used to extract the $K_L^0$ contribution.

\subsubsection{Background Measurements for the LBL Oscillation Studies}

\paragraph{\bf Measurement of NC cross sections}

The most threatening background to the $\nu_e$ and $\bar{\nu}_e$ appearance measurements
in LBNE is NC events in which the detritus of a $\pi^0$ decay mimics an electron. Similarly, in the
$\nu_\mu$ and $\bar{\nu}_\mu$ disappearance
study the largest background comes from the muonic-decay of a pion/kaon produced in a
NC event. These backgrounds are particularly challenging in the energy region $0.5 \leq E_\nu \leq 3~{\rm GeV}$ where an additional complication arises from substantial matter induced neutrino
oscillation. It follows that the largest background to the CC events in the far detector (FD)
comes from non-prompt leptons, originating from the hadronic shower in NC and high-$y_{\rm Bj}$
CC interactions. As a first step to constrain the error on the non-prompt background to CC
events in the FD, the fine-grained ND must measure the NC cross sections for different
processes and the NC/CC ratio as function of the hadron-energy $E_{\rm had}$.\\\

\paragraph{\bf Measurement of $\pi^0$ and $\gamma$ production in NC and CC}

The near detector must determine the proportion of NC events as a function of neutrino energy, the
yield of secondary $\pi^0$ in the NC interactions, and the energy and angular distributions
of the $\pi^0$. Similarly, the ND must accurately identify single-photon yields in NC interactions
and, in particular, the radiative decays of resonances where only a
single energetic photon is evident. The prevalence of these decays are much affected
by poorly characterized final state interactions and hence will require information from the
ND. Recent theoretical work has pointed out that there are NC processes that  produce
single energetic photons at a level that could effect LBNE~\cite{Hill08,Jenkins09}. Thus the ND must provide a
detailed characterization of $\pi^0$ and $\gamma$ in every class of event: multiplicity,
energy, and transverse momentum dependence for a given $E_{\rm had}$ in NC and CC.\\

\paragraph{\bf Muonic decays of $\pi^\pm$ and $K^\pm$ from measurement of charged multiplicities}

The dominant background to $\nu_\mu$ and $\bar{\nu}_\mu$ CC
events in the FD will come from the muonic decay
of pions and kaons in NC and CC processes. Having determined the NC cross section
relative to CC, the ND must determine the charged hadron multiplicity, energy and
transverse momentum dependence, and provide an empirical determination of exclusive
charged hadron production in $\nu$-interaction.\\

\paragraph{\bf Calibration of the neutrino energy scale from reconstructed events}

The neutrino energy scale directly affects the
precision of the oscillation scale parameter. The ND measurements should provide an
in situ calibration of the $E_\nu$ scale by measurement of the final state charged lepton and the
energy in the hadronic debris.\\

\paragraph{\bf Measurement of exclusive and semi-exclusive processes and their nuclear dependence}

The ND must provide differential cross section measurements for various exclusive, semi-exclusive,
and inclusive $\nu_\mu$ and $\bar{\nu}_\mu$ CC interactions. Among these, the most important to
the FD measurement is the quasi-elastic (QE) interaction. The challenge for ND is to reduce
and quantify the error due to nuclear effects, arising from initial and final state interactions,
Fermi motion, Pauli-blocking, etc., in the QE and other CC interactions. To minimize the
systematic error, ND should also determine exclusive and inclusive processes involving
nucleon resonances. This measurement of background processes from exclusive and
semi-exclusive reactions is part of the precision studies outlined in Section~\ref{sec:xsecs}
and Section~\ref{sec:nucleff}.

\subsubsection{Measurement of Neutrino Induced Background to Proton Decay}
\label{SBL:sec:prdky}

In addition to backgrounds measurements for the long-baseline program, the ND complex can also provide an estimate of the $\nu$ induced background to the search for proton (nucleon) decay
described in Section~\ref{PDK}.
Two generic proton decay modes are under study~\cite{Shiozawa}:
the $\pi^0$-mode, $e^+ (\mu^+) + \pi^0$ and
the kaon-mode, $\nu + K^+$, and $e^+(\mu^+) + K^0$.

The atmospheric neutrino spectrum is different from than that of the LBNE beam.
However, an accurate determination of exclusive channels induced by the LBNE neutrino beam
at $E_\nu \simeq 2$ GeV will provide empirical constraints for the atmospheric-neutrino background to
proton decay at the FD, thus reducing the corresponding systematic uncertainty.\\

\paragraph*{\bf \boldmath{$\pi^0$} Decay Mode}

The K2K-ND has provided a definitive measurement of $e^+ + \pi^0$~\cite{K2Kepi}.
The ND at LBNE will have an efficient $\pi^0$ and $e^+$ identification
with purity exceeding 95\% and the identification of $\mu^+$-mode  will be
straightforward. In addition, and importantly, the ND aims to measure
the exclusive first-generation mesons such as $\eta$, $\rho^0$, etc.,
besides $\pi^0$ accompanied by an $e^+$ or a $\mu^+$.\\

\paragraph*{\bf Kaon Decay Mode}

Supersymmetric models favor the proton-to-kaon decay mode.
The main focus of the ND measurements at LBNE is the exclusive $K^0$ production, i.e.
$e^+/ \mu^+ + K^0$. There is an existence proof of precise determination of
$K^0_S$ in $\nu$ CC and NC interactions~\cite{NOMAD-CCV0, NOMAD-NCV0}.
The ND at LBNE will extend the $K^0$ measurement to lower energy. The
$\nu + K ^+$ is a challenge. However, it should be pointed out that if the
$K^0$-mode is precisely measured then the $K^+$-channel can be accurately predicted.

%


\subsection{Study of Neutrino Interactions}
\label{SBL:sec:nuint}

The unprecedented large neutrino fluxes available for the LBNE program will allow the
collection of ${\cal{O}}(10^8)$ inclusive neutrino charged-current (CC) interactions,
in the high statistics Scenario B with a goal of $10^{22}$ POT.
The reduction of systematic uncertainties for the neutrino oscillation program
requires a highly segmented near detector, thus providing excellent resolution in the
reconstruction of neutrino events. The combination of this substantial flux with a finely
segmented near detector offers a unique opportunity to produce a range of neutrino
scattering physics measurements in addition to those needed by the long base line
oscillation program. The combined statistics and precision expected in the ND will allow
precise tests of fundamental interactions and better understanding of the structure of matter.

Since the potential of the neutrino probe is largely unexplored, the substantial step forward
offered by the LBNE program also provides the opportunity for unexpected discoveries.
Given the broad energy range of the beam, a diverse range of physics measurements
is possible in the LBNE ND, complementing the physics programs using proton,
electron or ion beams from colliders to the Jefferson Laboratory.
This complementarity not only would boost the physics output of LBNE,
but it can also attract new collaborators into the LBNE project from
different physics communities.

In the following sections we list the main physics topics, grouping them into seven broad
categories. To provide a flavor for the outstanding physics potential,
we give a short description of the studies which can be performed at LBNE for
few selected topics. A more detailed and complete discussion of the short baseline
physics potential will appear in a separate physics working group paper.

\subsubsection{Structure of the Weak Current}
Main topics:
\begin{itemize}[parsep=-1pt]
\item Electroweak Physics
\item Conservation of the Vector Current (CVC)
\item PCAC and Low $Q^2$ Behavior of Cross Sections
\end{itemize}

\paragraph{\bf Electroweak Physics}
\label{SBL:sec:sin2thetaW}

\begin{figure}[htb]
\centering\includegraphics[width=.20\textwidth]{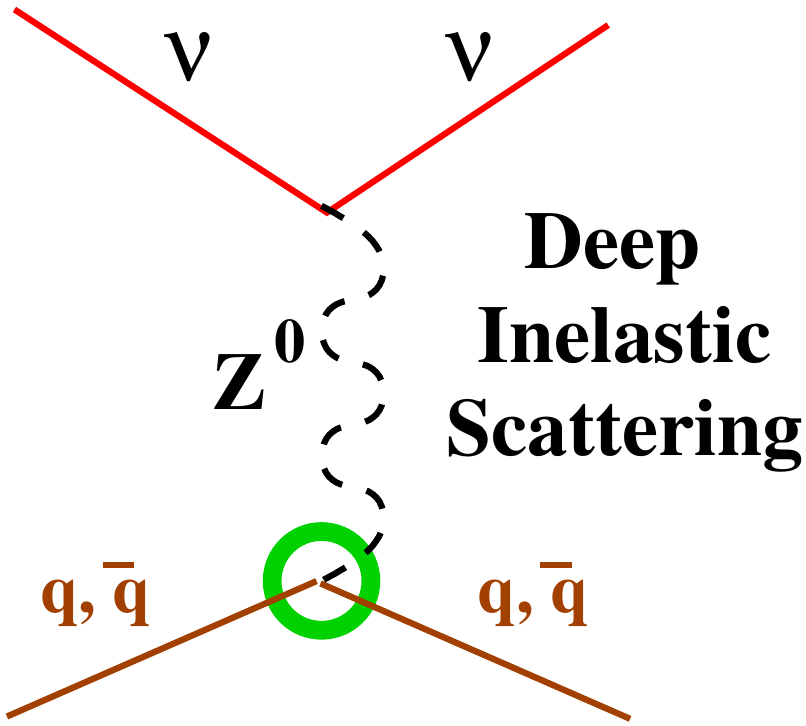}\hspace*{0.30cm}\includegraphics[width=.36\textwidth]{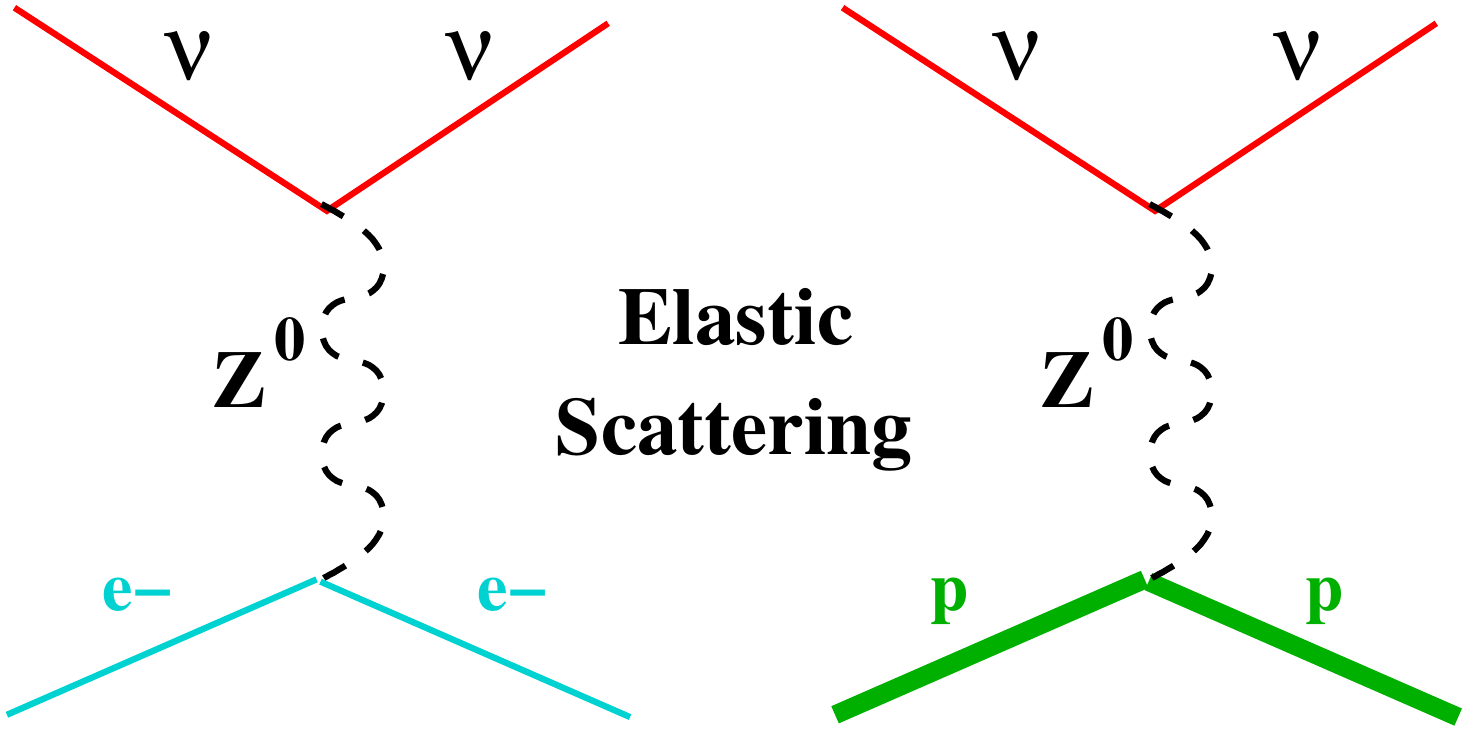}
\caption{Feynman diagrams for the three main Neutral-Current processes which can be used
to extract $\sin^2 \theta_W$ with the LBNE Near Detector complex.
}
\label{fig:graphs}
\end{figure}

Neutrinos are a natural probe for the investigation of
electroweak physics. Interest in a precise determination of the weak mixing angle ($\sin^2 \theta_W$)
at LBNE energies via neutrino scattering is twofold: a) it provides a direct measurement of
neutrino couplings to the Z boson and b) it probes a different scale of momentum transfer
than LEP by virtue of not being on the Z pole. The weak mixing angle can be extracted
experimentally from three main NC physics processes:
\begin{enumerate}[parsep=-1pt]
\item Deep Inelastic Scattering off quarks inside nucleons: $\nu N \to \nu X$;
\item Elastic Scattering off electrons: $\nu e^- \to \nu e^-$;
\item Elastic Scattering off protons: $\nu p \to \nu p$.
\end{enumerate}
Figure~\ref{fig:graphs} shows the corresponding Feynman diagrams for the three processes.

The most precise measurement of $\sin^2 \theta_W$ in
neutrino deep inelastic scattering (DIS) comes from the NuTeV experiment which reported
a value that is $3\sigma$ from the standard model~\cite{nutev-sin2w}. The LBNE ND can perform a similar
analysis in the DIS channel by measuring the ratio of NC and CC interactions induced by
neutrinos:
\begin{equation}
{\cal R}^\nu \equiv \frac{\sigma^\nu_{\rm NC}}{\sigma^\nu_{\rm CC}}
 \simeq \rho^2 \left( \frac{1}{2} - \sin^2 \theta_W +\frac{5}{9} \left(1 + r \right) \sin^4 \theta_W  \right)
\end{equation}
where $\rho$ is the relative coupling strength of the neutral to charged-current interactions
($\rho =1$ at tree level in the Standard Model) and $r$ is the ratio of anti-neutrino to
neutrino cross section ($r\sim0.5$).
The absolute sensitivity of ${\cal R}^\nu$ to $\sin^2 \theta_W$ is 0.7, which implies a measurement of
${\cal R}^\nu$ of 1\% precision would provide $\sin^2 \theta_W$ with a precision of 1.4\%.
Contrary to the NuTeV experiment the anti-neutrino interactions cannot be used for this analysis
at LBNE, due to the large number of $\nu_\mu$ DIS interactions in the $\bar{\nu}_\mu$ beam,
compared to the $\bar{\nu}_\mu$ DIS interactions. While removing some cancelation of
systematics between $\nu$ and $\bar{\nu}$, this fact reduces the statistical
uncertainty of the measurement.

The measurement of $\sin^2 \theta_W$ from DIS interactions can be only performed with the low density magnetized
tracker since an accurate reconstruction of the NC event kinematics and of the $\nu_e$ CC
interactions are crucial to keep the systematic uncertainties on the event selection
under control. The analysis selects events in the ND
after imposing a cut on the visible hadronic energy of $E_{\rm had} > 3$~GeV, as in the
NOMAD $\sin^2 \theta_W$ analysis (the CHARM analysis had $E_{\rm had} > 4$~GeV).
With the reference 700~kW beam we expect about $3.3 \times 10^6$ CC events and
$1.1 \times 10^6$ NC events, giving a statistical precision of 0.11\% on ${\cal R}^\nu$ and
0.15\% on $\sin^2 \theta_W$ (Table~\ref{tab:NuTeV-sin2tw}).

The use of a low density magnetized tracker can substantially reduce systematic uncertainties
with respect to a massive calorimeter. Table~\ref{tab:NuTeV-sin2tw} shows a comparison
of the different uncertainties on the measured ${\cal R}^\nu$ between NuTeV and LBNE.
The largest experimental systematic uncertainty in NuTeV is related to the subtraction
of the $\nu_e$ CC contamination from the NC sample. Since the low density tracker at LBNE
can efficiently reconstruct the electron tracks, the $\nu_e$ CC interactions can be
identified on an event-by-event basis, reducing the corresponding uncertainty to a
negligible level. Similarly, uncertainties related to the location of the interaction vertex,
noise, counter efficiency etc. are removed by the higher resolution and by the different
analysis selection. The experimental selection at LBNE will be dominated by two
uncertainties: the knowledge of the $\bar{\nu}_\mu$ flux and the kinematic selection
of NC interactions. The former is relevant due to the larger NC/CC ratio for
anti-neutrinos. The total experimental systematic uncertainty on $\sin^2 \theta_W$
is expected to be about 0.14\%.

\begin{table}[htb]
\begin{tabular}{r|c|c}
 & \multicolumn{2}{c}{~~~~~~~ $\delta R^{\nu}/R^{\nu}$~~~~~~~ } \\
 Source of uncertainty & NuTeV & LBNE \\ \hline\hline
 Data statistics & 0.00176 & 0.00110 \\
 Monte Carlo statistics & 0.00015   &  \\ \hline
 Total Statistics &  0.00176 &  0.00110 \\ \hline \hline
 $\nu_{e}, \bar{\nu}_{e}$ flux ($\sim1.7\%$) & 0.00064 &  0.00010 \\
 Energy measurement &  0.00038 &  0.00040 \\
 Shower length model &  0.00054 &  n.a. \\
 Counter efficiency, noise &  0.00036 &  n.a. \\
 Interaction vertex & 0.00056 &  n.a. \\
 $\bar{\nu}_\mu$ flux    &  n.a. &  0.00070 \\
 Kinematic selection    &  n.a. &  0.00060 \\ \hline
 Experimental systematics & 0.00112 &  0.00102 \\ \hline \hline
 d,s$\rightarrow$c, s-sea &  0.00227 &  0.00130  \\
 Charm sea &  0.00013  &   n.a. \\
 $r = \sigma^{\bar{\nu}}/\sigma^{\nu}$ &  0.00018 &  n.a. \\
 Radiative corrections & 0.00013 &  0.00013 \\
 Non-isoscalar target &  0.00010 &  N.A. \\
 Higher twists &  0.00031 &   0.00070  \\
 $R_{L}$ ($F_2,F_T,xF_3$) &  0.00115 &   0.00140 \\
 Nuclear correction    &        &  0.00020  \\  \hline
 Model systematics &  0.00258 &   0.00206  \\ \hline \hline
 TOTAL  &  0.00332 &    0.00255  \\ \hline \hline
\end{tabular}
\caption{Comparison of uncertainties on the ${\cal R}^\nu$ measurement between NuTeV and LBNE
with the reference beam. The corresponding relative uncertainties on $\sin^2 \theta_W$ must
be multiplied by a factor of 1.4, giving for LBNE a projected overall precision of 0.36\%.}
\label{tab:NuTeV-sin2tw}
\end{table}

The measurement of ${\cal R}^\nu$ will be dominated by model systematic
uncertainties on the structure functions of the target nucleons.
The estimate of these uncertainties for LBNE is based upon
the extensive work performed for the NOMAD analysis and includes a NNLO QCD
calculation of structure functions (NLO for charm production)~\cite{Alekhin:2007fh,Alekhin:2008ua,Alekhin:2008mb},
parton distribution functions (PDFs) extracted from dedicated low-$Q$ global fits,
high twist contributions~\cite{Alekhin:2007fh},
electroweak corrections~\cite{Arbuzov-Bardin} and nuclear corrections~\cite{Kulagin:2004ie,Kulagin:2007ju,Kulagin:2010gd}.
The charm quark production in CC, which has been
the dominant source of uncertainty in all past determinations of $\sin^2 \theta_W$
from $\nu$N DIS, is reduced to about 2.5\% of the total $\nu_\mu$ CC DIS with
$E_{\rm had}>3~\gev$ with the low energy beam spectrum at LBNE.
This number translates into a systematic uncertainty of 0.13\% on ${\cal R}^\nu$
(Table~\ref{tab:NuTeV-sin2tw}),
assuming a knowledge of the charm production cross section to 5\%.
It is worth noting the recent measurement of charm dimuon production by the
NOMAD experiment allowed a reduction of the uncertainty on the strange sea
distribution to $\sim3\%$ and on the charm quark mass $m_c$ to $\sim60~\mev$~\cite{NOMAD:DIS10}.
The lower neutrino energies available at LBNE reduce the accessible $Q^2$ values
with respect to NuTeV, increasing in turn the effect of non-perturbative
contributions (High Twists) and $R_L$. The corresponding uncertainties are
reduced by the recent studies of low-$Q$ structure functions and by
improved modeling with respect to the NuTeV analysis (NNLO vs. LO).
The total model systematic uncertainty on $\sin^2 \theta_W$ is expected to be about 0.29\%
with the reference beam configuration. The corresponding total uncertainty on the
value of $\sin^2 \theta_W$ extracted from $\nu$N DIS is 0.36\% with the 700~kW beam.

Most of the model uncertainties will be constrained by in situ dedicated measurements
using the large CC samples and employing improvements in theory that will have evolved
over the course of the experiment. In the low density tracker we shall collect
about 80,000 neutrino induced inclusive charm events with the 700~kW beam.
The precise reconstruction of charged
tracks will allow to measure exclusive decay modes of charmed hadrons (e.g. $D^{*+}$)
and to measure charm fragmentation and production parameters. The average
semileptonic branching ratio $B_\mu\sim5\%$ with the low energy LBNE beam.
The presence a 100-ton LAr TPC in front of the low density tracker will further
increase the potential of the charm analysis, allowing
the collection of about 50,000 
charm dimuon events with the reference beam.
This sample represent shall be compared with the largest existing sample of 15,400
dimuon events collected by the NOMAD experiment~\cite{NOMAD:DIS10}.
Finally, precision measurements of CC structure functions in both the
fine-grained tracker and the LAr detector would further reduce the uncertainties
on PDFs and on High Twist contributions.

The precision that can be achieved from $\nu$N DIS interactions is limited by both the
event rates and by the energy spectrum of the reference 700~kW beam configuration.
The high statistics beam exposure of Scenario B ($10^{22}$ pot) combined with a dedicated
run with the high energy beam option would increase the statistics by more than
a factor of 20. This major step forward would not only reduce the statistical
uncertainty to a negligible level, but would provide large control samples and
precision auxiliary measurements to reduce the systematic uncertainties on structure
functions. The two dominant systematic uncertainties, charm production in CC
interactions and low $Q^2$ structure functions, are essentially defined
by the available data at present.
Overall, the use of a high energy beam within the Scenario B can potentially
improve the precision achievable on $\sin^2 \theta_W$ from $\nu$N DIS to about 0.2\%.
It is worth mentioning the high energy beam is also required for the determination of the fluxes
in case high $\Delta m^2$ oscillations are present (see Section~\ref{SBL:sec:highMosc}).

A second independent measurement of $\sin^2 \theta_W$ can be obtained from NC
$\nu_\mu e$ elastic scattering. This channel has lower systematic uncertainties since it does
not depend upon the knowledge of the structure of nuclei, but has limited statistics
due to its very low cross section. The value of $\sin^2 \theta_W$ can be extracted from
the ratio of neutrino to anti-neutrino interactions~\cite{Marciano:Nuel}:
\begin{equation} \label{eqn:NCel}
{\cal R}_{\nu e} (Q^2) \equiv \frac{\sigma(\bar{\nu}_\mu e \to \bar{\nu}_\mu e)}{\sigma(\nu_\mu e \to \nu_\mu e)} (Q^2)
\simeq \frac{1 - 4 \sin^2 \theta_W + 16 \sin^4 \theta_W}{3 -12 \sin^2 \theta_W + 16 \sin^4 \theta_W}
\end{equation}
in which systematic uncertainties related to the selection and electron identification cancel out.
The absolute sensitivity of this ratio to $\sin^2 \theta_W$ is 1.79, which implies a measurement of
${\cal R}_{\nu e}$ of 1\% precision would provide $\sin^2 \theta_W$ with a precision of 0.65\%.

The event selection was described in details on Section~\ref{SBL:sec:absflux} since the NC
elastic scattering off electrons is also used for the absolute flux normalization.
This analysis can be performed only with the low density magnetized tracker and with a
large LAr detector. In the former case the total statistics available is limited
to about 1000 (600) $\nu (\bar{\nu})$ events with the minimal exposure of
Scenario A and 4500 (2800)
$\nu (\bar{\nu})$ events with the Scenario B. These numbers do not allow
a competitive determination of $\sin^2 \theta_W$ by using the magnetized tracker alone.
However, if we consider a 100~t LAr detector in the ND complex, we expect
to collect about 20,000 (12,000) $\nu (\bar{\nu})$ events with Scenario A
and 80,000 (50,000) $\nu (\bar{\nu})$ events with Scenario B.

As discussed in Section~\ref{SBL:sec:absflux} a combined analysis of both detectors
can achieve the optimal sensitivity: the fine-grained tracker is used to
reduce systematic uncertainties (measurement of backgrounds and calibration), while the LAr
ND provides the statistics required for a competitive measurement.
Overall, the use of the massive LAr detector can provide a statistical accuracy on
$\sin^2 \theta_W$ of about 0.3\% with the high statistics Scenario B.
However, the extraction of the weak
mixing angle is dominated by the systematic uncertainty on the $\bar{\nu}_\mu / \nu_\mu$
flux ratio, which is entering Equation~(\ref{eqn:NCel}). We evaluated this uncertainty
with the low-$\nu_0$ method for the flux extraction (see Section~\ref{SBL:sec:flavorbeam})
and we obtained a systematic uncertainty of about 1\% on the ratio of the
$\bar{\nu}_\mu / \nu_\mu$ flux integrals.
Therefore, the overall precision on $\sin^2 \theta_W$ achievable from
NC elastic scattering off electrons is limited to about 0.9\% in Scenario A
and 0.6\% in Scenario B.
An improvement of this result in Scenario B would require a better understanding
of the low-$\nu$ (anti)-neutrino cross sections, of the beam transport elements and
of the nuclear cross sections.

\begin{figure}[htb]
\centering\includegraphics[width=.6\textwidth]{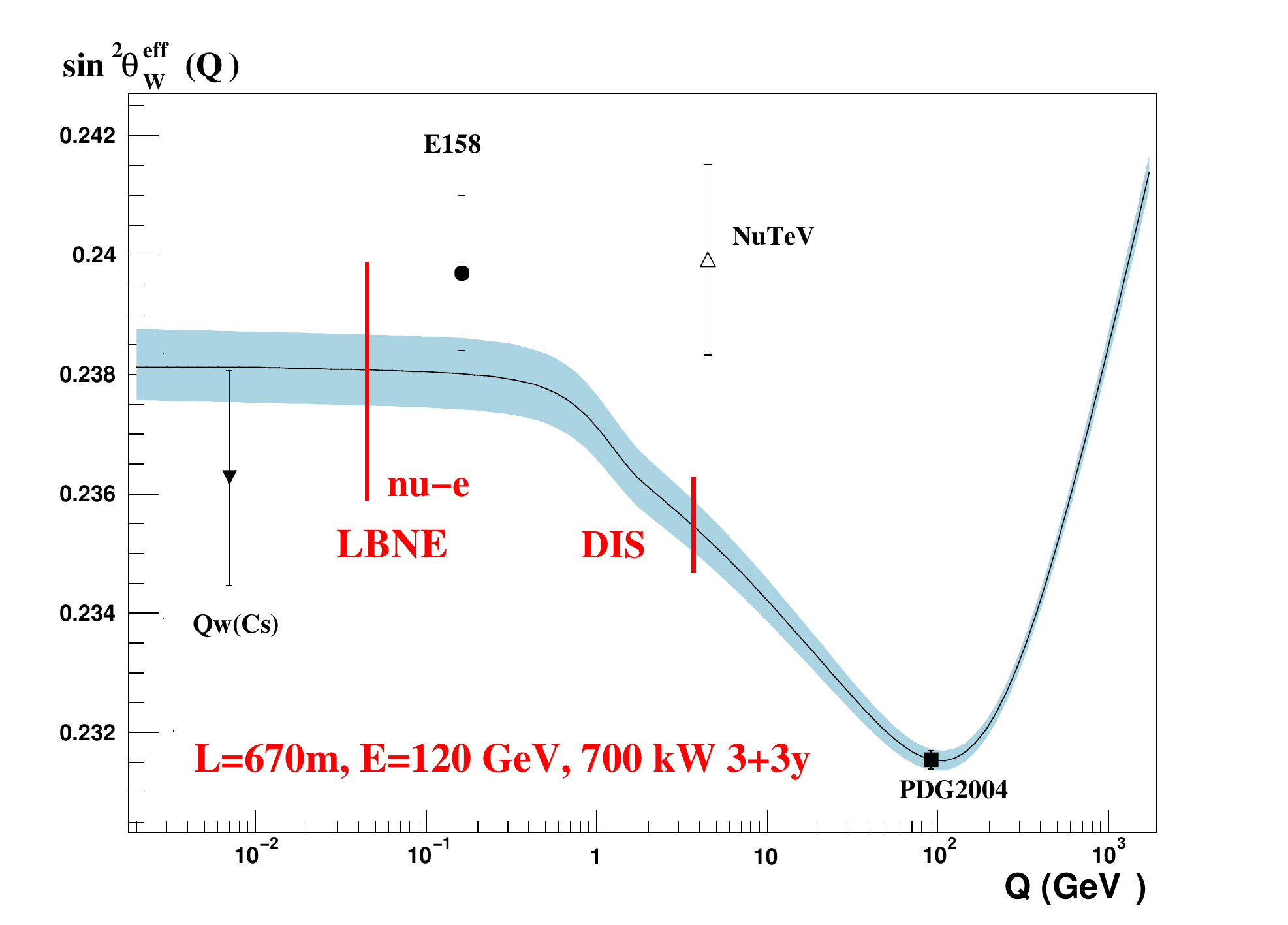}
\caption{Expected sensitivity to the measurement of $\sin^2 \theta_W$ from the LBNE ND
with the reference 700~kW beam.
The curve shows the Standard Model prediction as a function of the
momentum scale~\cite{marciano00}.
Previous measurements from Atomic Parity Violation~\cite{APV-sin2w,PDG06}, Moeller
scattering (E158~\cite{E158-sin2w}), $\nu$ DIS (NuTeV~\cite{nutev-sin2w})
and the combined $Z$ pole  measurements (LEP/SLC)~\cite{PDG06} are also shown for comparisons.
The use of a high energy beam with the maximal exposure of $10^{22}$ pot
can reduce the LBNE uncertainties by almost a factor of two.
}
\label{fig:sin2thetaw}
\end{figure}

Together, the DIS and the NC elastic scattering channels
involve substantially different scales of momentum transfer, providing a tool to test the
running of $\sin^2 \theta_W$ in a single experiment. To this end, the study of NC elastic scattering
off protons can provide additional information since it occurs at a momentum scale which is
intermediate between the two other processes.
Figure~\ref{fig:sin2thetaw} summarizes the target
sensitivity from the LBNE ND, compared with existing measurements as a function of the
momentum scale.

\subsubsection{Strange Content of the Nucleon}
\label{sec:deltas}
Main topics:
\begin{itemize}[parsep=-2pt]
\item NC Elastic Scattering and Measurement of $\Delta s$
\item Strange Form Factors
\item Charm Production and (anti)strange Parton Distribution Function
\item Strange Particle Production in NC and CC
\end{itemize}

The role of the strange quark in the
proton remains a central investigation in hadronic physics. The interesting question is
whether the strange quarks contribute substantially to the vector and axial-vector
currents of the nucleon. A large observed value of the strange quark contribution to the nucleon spin
(axial current), $\Delta s$, would require further theoretical speculations with respect to present
assumptions. The spin structure of the nucleon also affects the couplings of axions and
supersymmetric particles to dark matter.

The strange {\it vector} elastic form factors of the nucleon have been
measured to high precision in parity-violating electron scattering
(PVES) at Jefferson Lab, Mainz and elsewhere.
A recent global analysis \cite{YOUNG} of PVES data finds a strange
magnetic moment $\mu_s = 0.37 \pm 0.79$ (in units of the nucleon
magneton), so that the strange quark contribution to proton magnetic
moment is less than 10\%.
For the strange electric charge radius parameter $\rho_s$, defined in
terms of the Sachs electric form factor at low $Q^2$ as
$G_E^s = \rho_s Q^2 + \rho_s' Q^4 + {\cal O}(Q^6)$, one finds a very
small value, $\rho_s \ -0.03 \pm 0.63$~GeV$^{-2}$, consistent with zero.

Both of these results are consistent with theoretical expectations
based on lattice QCD and phenomenology \cite{LEINWEBER}.
In contrast, the strange {\it axial vector} form factors are not nearly as
well determined.  A similar global study of PVES data~\cite{YOUNG} finds
$\widetilde{G}_A^N(Q^2)
= \widetilde{g}_A^N \left( 1 + {Q^2 / M_A^2} \right)^2$,
with the effective proton and neutron axial charges
$\widetilde{g}_A^p = -0.80 \pm 1.68$ and
$\widetilde{g}_A^n =  1.65 \pm 2.62$.

The strange axial form factor at $Q^2=0$ is related to the
{\it spin} carried by strange quarks, $\Delta s$.
Currently the world data on the spin-dependent $g_1$ structure function
constrain $\Delta s$ to be $\approx -0.055$ at a scale $Q^2=1$~GeV$^2$,
with a significant fraction of this coming from the region $x < 0.001$.
In addition, the HERMES collaboration \cite{HERMES} extracted the
strange quark spin from semi-inclusive DIS data over the range
$0.02 \leq x \leq 0.6$, yielding a {\it negative} central value,
$\Delta s  = 0.037 \pm 0.019 \pm 0.027$, although still consistent
with the above global average.

\begin{table}[htb]
\centering
\begin{tabular}{c|c|c}
  A   &  B & C \\ \hline\hline
~~~~~$\frac{1}{4} \left[ G_1^2 \left( 1 +\tau \right) - \left( F_1^2 - \tau F_2^2 \right)
\left( 1 - \tau \right) + 4 \tau F_1 F_2 \right]$~~~~~   &
~~~~~$- \frac{1}{4} G_1 \left( F_1 + F_2 \right)$~~~~~   &
~~~~~$\frac{1}{16} \frac{M_p^2}{Q^2} \left( G_1^2 + F_1^2 + \tau F_2^2 \right)$~~~~~   \\ \hline
\end{tabular}
\caption{Coefficients entering Eqn.~\ref{eqn:QE} for NC elastic scattering
and CC QE interactions, with $\tau=Q^2/4M_p$.}
\label{tab:FF}
\end{table}

An independent extraction of $\Delta s$, which does not rely on the difficult
measurements of the $g_1$ structure function at very small $x$ values,
can be obtained from (anti)-neutrino NC elastic scattering off proton (see Fig.~\ref{fig:graphs}).
This process provides indeed the most direct measurement of $\Delta s$.
The differential cross section for NC elastic and CC~QE scattering of
(anti)-neutrinos from protons can be written as:
\begin{equation} \label{eqn:QE}
\frac{d \sigma}{d Q^2} = \frac{G_\mu^2}{2\pi} \frac{Q^2}{E_\nu^2} \left( A \pm BW + C W^2 \right); \;\;\;\;  W=4E_\nu/M_p - Q^2/M_p^2
\end{equation}
where the positive (negative) sign is for (anti)-neutrino scattering and the coefficients
$A, B,$ and $C$ contain the vector and axial form factors as listed in Table~\ref{tab:FF}.

The axial-vector form factor for NC scattering can be written as the sum of the known axial
form factor $G_A$ plus a strange form factor $G_A^s$:
\begin{equation}
G_1 = \left[ - \frac{G_A}{2} + \frac{G_A^s}{2} \right]
\end{equation}
while the NC vector form factors can be written as:
\begin{equation}
F_{1,2} = \left[ \left(\frac{1}{2} - \sin^2 \theta_W \right) \left( F_{1,2}^p - F_{1,2}^n \right)
- \sin^2 \theta_W \left( F_{1,2}^p + F_{1,2}^n \right) - \frac{1}{2} F_{1,2}^s \right]
\end{equation}
where $F_1^{p(n)}$ is the Dirac form factor of the proton (neutron), $F_2^{p(n)}$ is the
corresponding Pauli form factor, and $F_{1,2}^s$ are the strange vector form factors.
These latter are expected to be small from the PVES measurements summarized above.
In the limit $Q^2 \to 0$, the differential cross section is
proportional to the square of the axial-vector form factor
$d \sigma / d Q^2 \propto G_1^2$ and $G_A^s \to \Delta s$.
The value of $\Delta s$ can therefore be extracted experimentally by extrapolating
the NC differential cross section to $Q^2=0$.

Previous neutrino scattering experiments have been limited by the statistics
and by the systematic uncertainties on background subtraction.
The only information available
comes from the analysis of 951 NC $\nu p$ and 776 NC $\bar{\nu}p$ collected
by the experiment BNL E734~\cite{E734,E734-re1,E734-re2}.
The LBNE neutrino beam will be sufficiently intense that
a measurement of NC elastic scattering on proton in the fine-grained water ND
can provide a definitive statement on the contribution of the strange sea to
either the axial or vector form factor.

\begin{figure}[htb]
\centering\includegraphics[width=.5\textwidth]{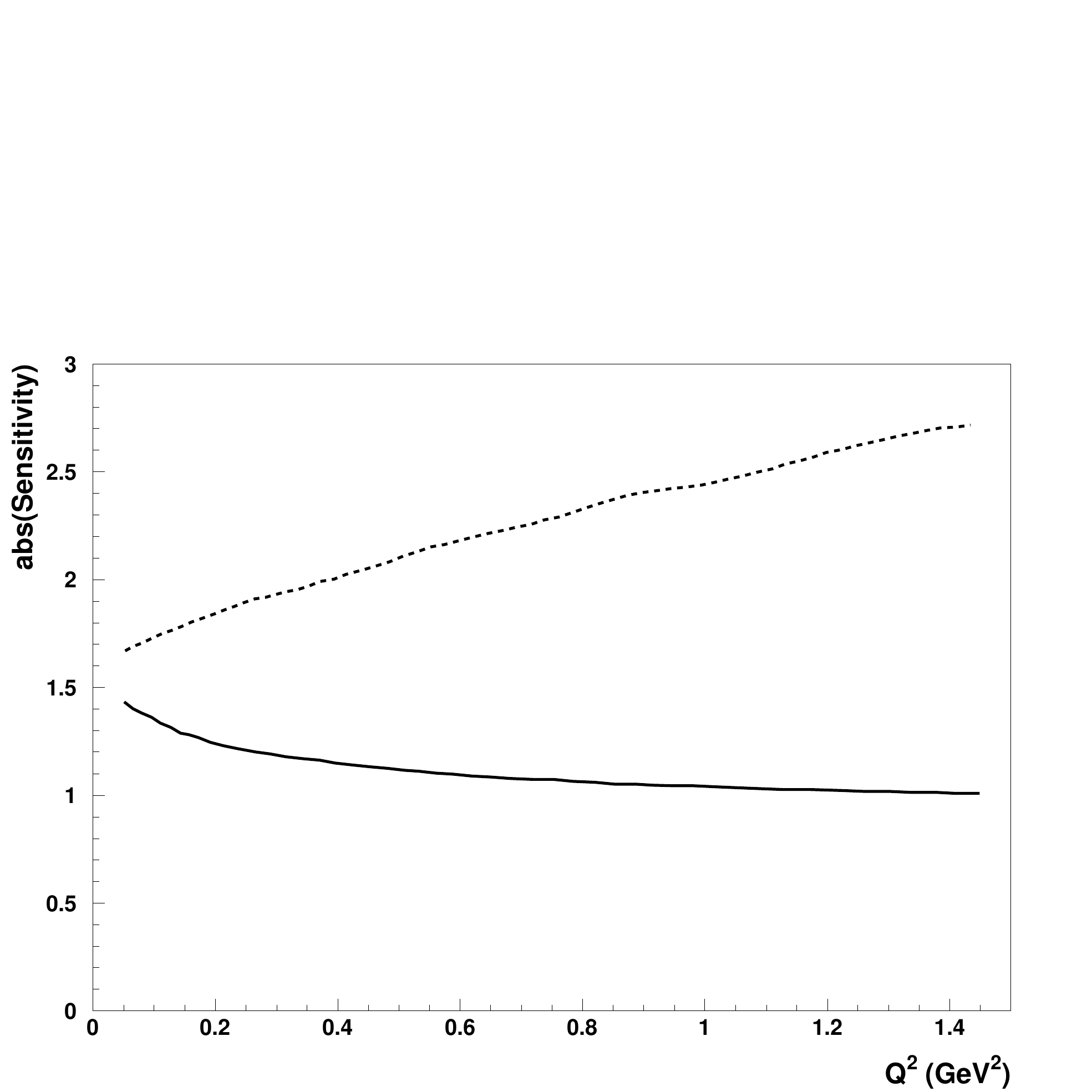}
\caption{Absolute sensitivity of the ratios ${\cal R}_{\nu p}$ (solid) and
${\cal R}_{\bar{\nu} p}$ (dashed) to the strange contribution to the spin of the nucleon, $\Delta s$,
as a function of $Q^2$.
}\label{fig:deltas}
\end{figure}

Systematic uncertainties can be reduced by measuring the NC/CC ratios for both neutrinos and
anti-neutrinos:
\begin{equation}
{\cal R}_{\nu p} (Q^2) \equiv \frac{\sigma(\nu_\mu p \to \nu_\mu p)}{\sigma(\nu_\mu n \to \mu^- p)}(Q^2); \;\;\;\;\;
{\cal R}_{\bar{\nu} p} (Q^2) \equiv \frac{\sigma(\bar{\nu}_\mu p \to \bar{\nu}_\mu p)}{\sigma(\bar{\nu}_\mu p \to \mu^+ n)}(Q^2)
\end{equation}
as a function of $Q^2$. Figure~\ref{fig:deltas} shows the absolute sensitivity of both ratios to
$\Delta s$ for different values of $Q^2$. The sensitivity for $Q^2\sim0.25$~GeV$^2$ is about 1.2
for neutrinos and 1.9 for anti-neutrinos, which implies that a measurement of ${\cal R}_{\nu p}$ and
${\cal R}_{\bar{\nu} p}$ of 1\% precision would enable the extraction of $\Delta s$ with an
uncertainty of 0.8\% and 0.5\%, respectively.

The design of the fine-grained water ND involves a combination of
different nuclear targets. Therefore, most of the neutrino scattering is from nucleons embedded in a
nucleus, requiring nuclear effects to be taken into account. Fortunately, in the ratio of
NC/CC yields, the nuclear corrections are expected to largely cancel out.
The $\Delta s$ analysis requires a good proton reconstruction efficiency as well as high
resolution on both the proton angle and energy. To this end, the low density magnetized
tracker at LBNE can increase the range of the protons inside the ND, allowing the
reconstruction of proton tracks down to $Q^2\sim0.07$~GeV$^2$. This fact will reduce the
uncertainties in the extrapolation of the form factors to the limit $Q^2 \to 0$.

Table~\ref{tab:prange} summarizes the expected proton ranges for both the scintillator
and the low density ($\rho\sim0.1~g/cm^3$) ND options.
With the reference $700~kW$ beam we expect about $2.5 (1.5) \times 10^5$ $\nu p (\bar{\nu} p)$
events after the selection cuts in the low density tracker, yielding a statistical
precision of the order of 0.2\%.

\begin{table}[htb]
\centering
\begin{tabular}{ccccc}
  $T_p$  &  $Q^2$ & Range Sci & Range STT & $P_p$  \\
  MeV   &  $\gev^2/c^2$ & $cm$ & $cm$ &  $\gev/c$  \\ \hline\hline
20 & 0.038  & 0.42 & 4.2 & 0.195  \\
40 & 0.075  & 1.45 & 14.5 & 0.277  \\
60 & 0.113  & 3.03 & 30.3 & 0.341   \\
80 & 0.150  & 5.08 & 50.8 & 0.395  \\
100 & 0.188 & 7.57 & 75.7 & 0.445  \\ \hline
\end{tabular}
\caption{Expected proton range for the scintillator and low density
($\rho\sim0.1~g/cm^3$) tracker options. The first column gives the proton kinetic energy
and the last column the proton momentum. The $Q^2$ value producing $T_p$ is calculated
assuming the struck nucleon was initially at rest.}
\label{tab:prange}
\end{table}

We follow the analysis performed by the FINeSSE collaboration~\cite{Finesse} and in the SciBooNE
experiment for the determination of $\Delta s$. In particular, based upon the latter,
with the scintillator tracker we expect a purity
of about 50\%, with background contributions of 20\% from neutrons produced outside of the
detector, 10\% $\nu n$ events and 10\% NC pion backgrounds. The dominant systematic uncertainty
will be related to the background subtraction. The low energy beam spectrum at LBNE
provides the best sensitivity for this measurement since
the external background from neutron-induced
proton recoils will be reduced by the strongly suppressed high energy tail.
The low density magnetized tracker is expected to increase the purity by reducing
the neutron background and the NC pion background. Overall, it seems possible to
achieve a precision on $\Delta s$ of about $0.02-0.03$ in this case. The maximal
exposure available with Scenario B will further improve the sensitivity due to higher
statistics and better constraints on the background subtraction procedure
from data control samples.

\subsubsection{Search for New Physics}

The Standard Model (SM) of elementary particles is in conflict with
several experimental observations: non-zero neutrino masses and oscillations,
the excess of matter over antimatter in the Universe, and the presence of non-baryonic dark
matter. In addition, a number of fine-tuning problems (such as the gauge hierarchy
problem, and cosmological constant problem) may indicate the existence of new physics
between the electroweak and the Planck scales.

The energy scale of new physics is not known at present. If it exists
at energies above the Fermi scale (examples include supersymmetry,
large or warped extra dimensions, models with dynamical electroweak
symmetry breaking), the  search for new particles can be carried out
in direct experiments, such as ATLAS or CMS at LHC. In addition, new
hypothetical heavy particles inevitably appear as virtual states,
leading to different rare processes, absent in the SM. These effects
can be found at experiments such as LHCb, charm and beauty factories
and are competitive with the direct searches. If the new physics is
associated with existence of new relatively light particles  (an
example is given by the $\nu$MSM, see below) then the search for rare
processes is superior to high energy experiments. Moreover, it
provides a unique possibility for discovery of new physics, not
accessible by any of the LHC experiments. We will argue here that the
LBNE near detector can be used for these searches.

Main topics:
\begin{itemize}[parsep=-2pt]
\item Search for $\nu$MSM Neutral Leptons
\item High $\Delta m^2$ Neutrino Oscillations
\item Radiative Decay of Sterile Neutrinos
\item MiniBooNE Low Energy Anomaly
\item Decay of Heavy Weakly Interacting Particles
\item $\nu_\mu \to \nu_e$ Transitions
\end{itemize}

\paragraph{\bf Search for \boldmath{$\nu$}MSM Neutral Leptons}

The most economic way to handle in a unified way the problems
of neutrino masses, dark matter and baryon asymmetry of the Universe
may be to add to the SM three Majorana singlet fermions with masses
roughly of the order of masses of known quarks and leptons. The
appealing feature of this theory (called the $\nu$MSM for  ``Neutrino
Minimal SM'') is the fact that there every left-handed fermion has a
right-handed counter-part, leading thus an equal way of treating of
quarks and leptons. The lightest of the three new leptons is expected
to have a mass from 1~keV to 50~keV and play the role of the dark
matter particle. Two other neutral fermions are responsible for giving
masses to ordinary neutrinos via the see-saw mechanism at the {\em
electroweak scale} and to creation of the baryon asymmetry of the
Universe (for a review see~\cite{Boyarsky:2009ix}). The masses of
these particles and their coupling to ordinary leptons are constrained
by particle physics experiments and cosmology. They should be almost
degenerate, forming thus nearly Dirac fermion (this is coming from the
requirement of successful baryogenesis). Different considerations
indicate that their mass should be in  ${\cal O}(1)$~GeV region
~\cite{Shaposhnikov:2008pf}.

The $\nu$MSM is described by the most general renormalizable
Lagrangian containing all the particles of the SM and three singlet
fermions. For the purpose of the present discussion we take away from
it the lightest singlet fermion $N_1$ (the ``dark matter sterile
neutrino''), which is coupled extremely weakly to the ordinary
leptons. In addition, we take $N_2$ and $N_3$ degenerate in mass,
$M_2=M_3=M$. Then the convenient parametrization of the interaction of
$N's$ with the leptons of SM is:
\begin{equation}
L_{\rm singlet}=\left(\frac{\kappa M m_{atm}}{v^2}\right)^{\frac{1}{2}}
\left[\frac{1}{\sqrt{\epsilon e^{i\eta}}}\bar L_2 N_2 +
\sqrt{\epsilon e^{i\eta}}\bar L_3 N_3\right] \tilde{H}
- M \bar {N_2}^c N_3 + \rm{h.c.} \,,
\label{pmm}
\end{equation}
where $L_2$ and $L_3$ are the combinations of $L_e,~L_\mu$ and
$L_\tau$
\begin{equation}
L_{2}=\sum_\alpha x_{\alpha}L_\alpha~,~~~~
L_{3}=\sum_\alpha y_{\alpha}L_\alpha~.
\label{L23def}
\end{equation}
with $\sum_\alpha |x_{\alpha}|^2=\sum_\alpha |y_{\alpha}|^2=1$.

In eq.(\ref{pmm}) $v=246$~GeV is the vacuum expectation value of the Higgs
field $H$, $\tilde{H}_i=\epsilon_{if}H_j^*$, $m_{atm}\simeq 0.05$ eV
is the atmospheric neutrino mass difference, $\kappa=1~(2)$ for normal
(inverted) hierarchy of neutrino masses. The $x_{\alpha}$ and
$y_{\alpha}$ can be expressed through the parameters of the active
neutrino mixing matrix (explicit relations can be found in
~\cite{Shaposhnikov:2008pf}). The parameter $\epsilon$ (by definition,
$\epsilon <1$) and the CP-breaking  phase $\eta$  cannot be fixed from
consideration of neutrino masses and mixings.

\begin{figure}[!htb]
\centerline{
\includegraphics[width=0.45\textwidth]{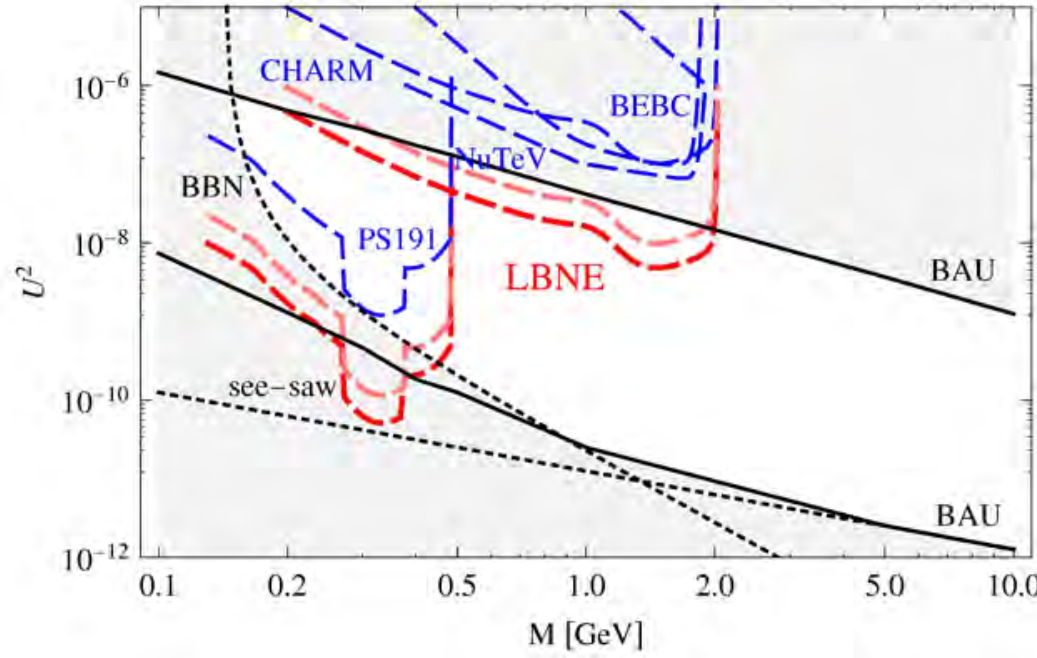}\hskip
0.08\textwidth
\includegraphics[width=0.45\textwidth]{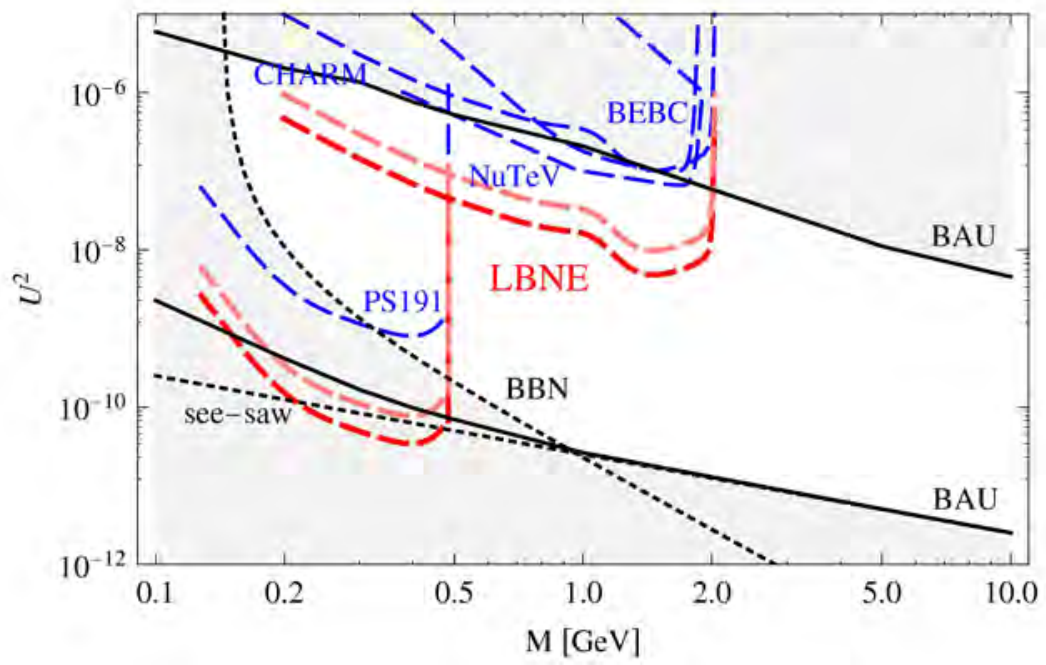}}
\caption{Constraints on $U^2$ coming from the baryon asymmetry of the
Universe (solid lines), from the see-saw formula (dotted line) and
from the big bang nucleosynthesis (dotted line). The regions corresponding to
different experimental searches
are outlined by blue dashed lines. Left panel:  normal hierarchy;
right panel: inverted hierarchy (adopted from
Ref.~\cite{Canetti:2010aw}).
Pink and red curves indicate the expected sensitivity of the LBNE near detector
with the Scenario B exposure of $10^{22}$ pot for detector lengths of 7~m and 30~m,
respectively (see text for details).}
\label{exp}
\end{figure}

If the mass of $N$ is fixed, smaller $\epsilon$ yields stronger
interactions of singlet fermions to the SM leptons. This leads to
equilibration of these particles in the early Universe above the
electroweak temperatures, and, therefore, to erasing of the baryon
asymmetry. In other words, the mixing angle $U^2$ between neutral
leptons and active neutrinos must be small, explaining why these new
particles has not been seen previously. For small $\epsilon$,
\be
U^2 = \frac{\kappa m_{atm}}{4 M \epsilon}~.
\label{U2}
\ee
The region, where baryogenesis is possible in $U^2-M$ plane is shown
in Fig.~\ref{exp}.
Also shown are the exclusion regions coming from
different experiments such as  BEBC~\cite{CooperSarkar:1985nh},  CHARM~\cite{Bergsma:1985is}, and NuTeV~\cite{Vaitaitis:1999wq}  and CERN
PS191 experiment~\cite{Bernardi:1985ny,Bernardi:1987ek} (see also
discussion of different experiments in~\cite{Atre:2009rg}).  For the
case of normal hierarchy, only  CERN PS191  have significantly entered
into the cosmologically interesting part of the parameter space of the
$\nu$MSM,  situated below the mass of the kaon.
If the hierarchy  is
inverted, there are some constraints even for higher $N$ masses.  The
lower constraint on $U^2$, coming from baryon asymmetry of the
Universe, is somewhat stronger than the ``see-saw'' constraint.

\begin{figure}[!htb]
\centerline{\hskip
0.05\textwidth
\includegraphics[width=0.2\textwidth]{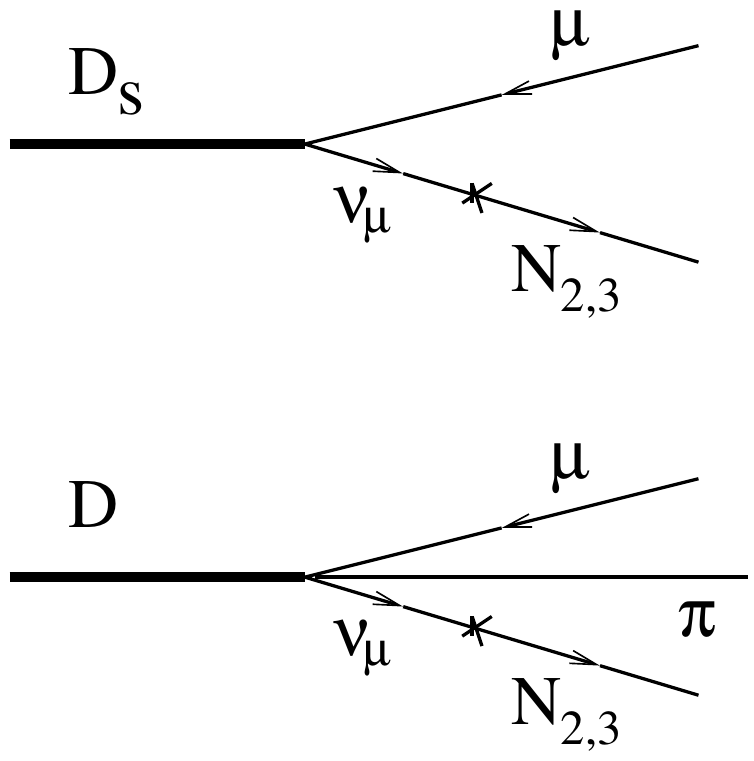}\hskip
0.1\textwidth
\includegraphics[width=0.22\textwidth]{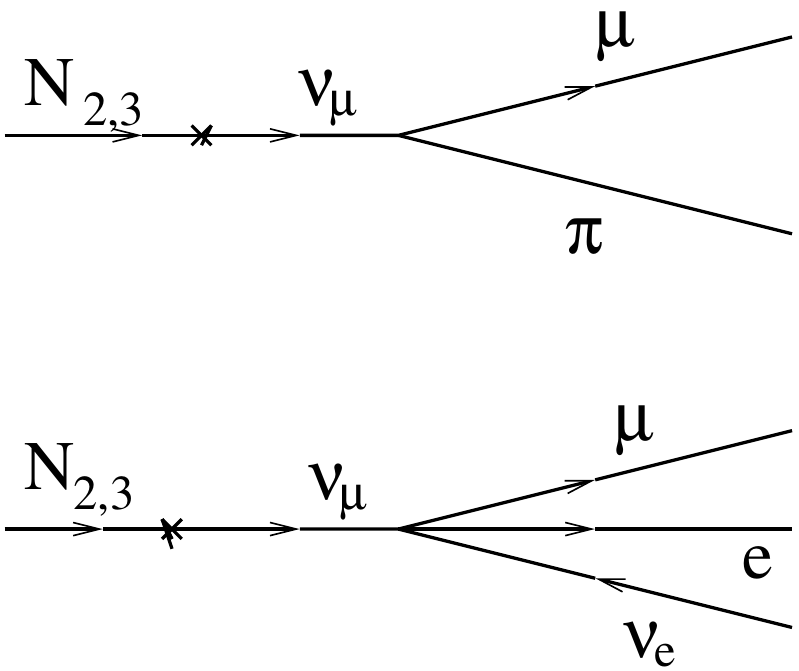}}
\caption{Left panel: Feynman  diagrams of meson decays producing
heavy sterile neutrinos. Right panel: Feynman diagrams of sterile
neutrino decays.}
\label{production-and-decays}
\end{figure}

The most efficient mechanism of sterile neutrino production is through
weak decays of heavy mesons and baryons, as can be seen from the left panel of
Fig.~\ref{production-and-decays}, showing some examples of relevant two- and
three-body decays.
Heavy mesons can be produced by energetic protons scattering off the
target material.

In case of the LBNE experiment the relevant heavy mesons are
charmed ones. With a typical lifetime (in the rest frame) of about
$10^{-10}$~s these mesons mostly decay before further interaction,
yielding the sterile neutrino flux.
Since these sterile neutrinos are very weakly interacting (see
Fig.~\ref{exp}) they can cover quite a large distance before decay,
significantly exceeding the distance of 670~m
from  the target to the near LBNE near detector.
Neutrino decays into SM particles due to mixing with active neutrinos can be searched
for in the ND, provided a sufficiently long instrumented decay region is available.
Two examples of the interesting decay modes
are presented on the right panel of Fig.~\ref{production-and-decays}. More
examples can be found in Ref.~\cite{Gorbunov:2007ak}.

We can obtain an estimate of sterile neutrino events to be
observed in the LBNE near detector, $N^{LBNE}_{signal}$,
by comparing the relevant parameters of the LBNE and CHARM experiments,
which are summarized in Table~\ref{LBNE-and-CHARM}.
\begin{table}[!htb]
{\renewcommand{\arraystretch}{1.3}%
\begin{center}
\begin{tabular}{|c|c|c|c|c|c|c|}
\hline
 & PoT & detector & distance  & beam  & detector & charm
\\
& & length & to target & energy & area & production \\
\hline
\hline
LBNE & { $1.0\times 10^{22}$ } & { 7~m } & { 670~m }
& { 120~GeV } &
{ $4\times4$~m$^2$ } & { $1.0\times 10^{-4}$ } \\
\hline
CHARM & { $2.5\times 10^{18}$ } & { 34~m } & { 480~m } & {
  400~GeV } & {
$3\times3$~m$^2$ } & { $4.5\times 10^{-4}$ } \\ \hline
\end{tabular}
\end{center}
}
\caption{Characteristics of LBNE and CHARM experiments.
\label{LBNE-and-CHARM}
}
\end{table}

The number of events linearly grows with the number of proton on
target, the number of produced charmed mesons,
the detector length (decay region) and the detector area.
In particular, this latter condition is valid if
the angular spread of the neutrino flux, which is of the order of $N_mM_D/E_{beam}$,
is larger than the angle at which the ND is seen from the target.
Here $N_m$ is the multiplicity of the produced hadrons, and the above
condition is valid for both LBNE and CHARM. The number of events
also decreases linearly when the energy increase, since it increases the
lifetime, reducing the decay probability within the detector.

Finally, the number of mesons
decreases quadratically with the distance between the target and the detector.
These considerations imply a search for $\nu$MSM sterile neutrinos can be only competitive
with the Scenario B exposure of $10^{22}$ pot and with a proton energy of 120~GeV.

The analysis can be performed with all detector options, although a low density
long decay region reduces backgrounds and allows the detection of both leptonic and
hadronic decay modes.
Assuming the length of the magnetized tracker (7~m) as decay region, we can then
estimate the ratio between the signal event to be observed in the LBNE ND
and the ones observed by the CHARM experiment to be about 50.

The CHARM experiment found no sterile neutrino events with zero background and very
high (about 65\%) efficiency.
Since both production and decay rates are proportional
to the neutrino mixing angles squared, LBNE will be able to achieve an improvement
by about a factor of seven in the neutrino mixing angle squared $U^2$ with respect
to the CHARM experiment.
Somewhat smaller numbers can be obtained for kaons by comparing the LBNE
and PS191 experiments. The expected sensitivity is indicated with red
dashed curves in Fig.~\ref{exp}.

The presence of a long LAr detector in front of the fine-grained tracker
can enhance the sensitivity to sterile neutrinos by providing an additional,
fully instrumented, decay region. In addition, a combined analysis of
a massive LAr detector with the fine-grained magnetized tracker would
allow the search for new weakly interacting particles produced in NC
neutrino interactions within LAr and decaying inside the fiducial
volume of the fine-grained tracker.

A better sensitivity to $\nu$MSM can be achieved by instrumenting the upstream
region of the ND hall (e.g. with the LAr detector and some minimal tracking
device upstream). The fiducial volume of the
new detector has to be empty (material-free) or fully sensitive in order to
suppress background events. The geometry of the ND hall would allow a maximal
decay length of about 30~m. The sensitivity of this configuration can be estimated
by rescaling the expected limits on mixing $U^2$. The expected number of signal
events with a total decay length of $\sim30$~m exceeds by about 200 times the number
of events in CHARM, which
implies an improvement by a factor of 15 in sensitivity to $U^2$ with
respect to the CHARM experiment.

It must be noted that if the magnetic moment of the sterile neutrinos is sizeable, the
dominant decay channel would be a radiative electromagnetic decay into $\gamma \nu$, which
has also been proposed as a possible explanation for the observed MiniBooNE low energy
excess~\cite{miniboone-lowE}. This fact, in turn, requires a detector capable of
identifying and reconstructing single photon events.

\paragraph{\bf High $\Delta m^2$ Neutrino Oscillations}
\label{SBL:sec:highMosc}

The evidence for neutrino oscillations obtained from atmospheric,
long-baseline accelerator, solar, and long-baseline reactor data from different
experiments consistently indicates two different scales with $\Delta_{32} m^2\sim2.4 \times 10^{-3} \evsq$ defining the atmospheric oscillations and
$\Delta_{21} m^2\sim7.9 \time 10^{-5} \evsq$ defining the solar oscillations.
The only way to accommodate oscillations with relatively high $\Delta m^2$
at the eV$^2$ scale is therefore to add one or more sterile neutrinos to the
conventional three light neutrinos.

The MiniBooNE experiment reported their anti-neutrino data might be consistent with the LSND
$\bar{\nu}_\mu \to \bar{\nu}_e$ oscillation with $\Delta m^2\sim\evsq$.
Contrary to the anti-neutrino data, the MiniBooNE neutrino data seem to
exclude high $\Delta m^2$ oscillations, possibly indicating a
a different behavior between neutrinos and anti-neutrinos. This difference,
which would require CP or CPT violation, could also be consistent with
the MINOS disappearance analysis.

Models with five (3+2) or six (3+3) neutrinos can potentially explain
the MiniBooNE results~\cite{Maltoni07}. In addition to the cluster of the three neutrino
mass states accounting for "solar" and "atmospheric" mass splitting two
(or three) states at the eV scale are added, with a small
admixture of $\nu_e$ and $\nu_\mu$ to account for the LSND signal.
One distinct prediction from such models is a significant probability
for $\bar{\nu}_\mu$ disappearance into sterile neutrinos, of the order
of 10\%, in addition to the small probability for $\bar{\nu}_e$ appearance.

The Near Detector at LBNE is located at a baseline of 670~m and with the reference
``Low Energy (LE)'' beam spectrum so it can reach the same value $L/E_\nu\sim1$ of MiniBooNE and
LSND. The large fluxes and the availability of fine-grained detectors make
the LBNE program well suited to search for oscillations at the $\evsq$ scale.
Due to the potential differences between neutrinos and anti-neutrinos
four possibilities have to be considered in the analysis: $\nu_\mu$
disappearance, $\bar{\nu}_\mu$ disappearance, $\nu_e$ appearance and
$\bar{\nu}_e$ appearance. As discussed in Section~\ref{SBL:sec:fluxosc},
the search for high $\Delta m^2$ oscillations has to be performed
simultaneously with the in situ determination of the fluxes.

To this end, we need to obtain an independent prediction of the $\nu_e$ and
$\bar{\nu}_e$ fluxes starting from the measured $\nu_\mu$ and $\bar{\nu}_\mu$
CC distributions since the $\nu_e$ and $\bar{\nu}_e$ CC distributions could
be distorted by the appearance signal. The low-$\nu_0$ method can provide
such predictions if external measurements for the $K_L^0$ component
are available from hadro-production experiments (Section~\ref{SBL:sec:flavorbeam}).

We will the follow an iterative procedure:
\begin{itemize}[parsep=-1pt]
\item Extract the fluxes from $\nu_\mu$ and $\bar{\nu}_\mu$ CC distributions assuming
no oscillations are present;
\item Comparison with data and determination of oscillation parameters (in any);
\item New flux extraction after subtraction of the oscillation effect;
\item Iterate until convergence.
\end{itemize}
The analysis has to be performed separately for neutrinos and anti-neutrinos due to
potential CP or CPT violation according to MiniBooNE/LSND data.

We measure the ratio of electron to muon CC events:
\begin{equation}
{\mathcal{R}}_{e \mu} (L/E)  \equiv  \frac{\#~of~\nu_e N \to e^- X}{\#~of~\nu_\mu N \to \mu^- X }(L/E); \;\;\;\;\;\;\;  \bar{\mathcal{R}}_{e \mu} (L/E) \equiv \frac{\#~of~\bar{\nu}_e N \to e^+ X}{\#~of~\bar{\nu}_\mu N \to \mu^+ X }(L/E)
\end{equation}
which is then compared with the predictions obtained from the low-$\nu_0$ method.
Deviations of ${\mathcal{R}}_{e \mu}$ or $\bar{\mathcal{R}}_{e \mu}$ from the expectations
as a function of $L/E$ would provide evidence for oscillations.
It must be noted that this procedure only provides a relative measurement of $\nu_e (\bar{\nu}_e)$
vs. $\nu_\mu (\bar{\nu}_\mu)$. Actually, since the fluxes
are extracted from the observed $\nu_\mu$ and $\bar{\nu}_\mu$ CC distributions an analysis
of the ${\mathcal{R}}_{e \mu} (\bar{\mathcal{R}}_{e \mu})$ ratio cannot distinguish
between $\nu_\mu (\bar{\nu}_\mu)$ disappearance and $\nu_e (\bar{\nu}_e)$ appearance.

The process of NC elastic scattering off protons (Section~\ref{sec:deltas})
can provide the complementary measurement
needed to disentangle the two hypotheses of $\nu_\mu (\bar{\nu}_\mu)$ disappearance into
sterile neutrinos and $\nu_e (\bar{\nu}_e)$ appearance. In order to cancel systematic
uncertainties, we will measure the NC/CC ratio with respect to quasi-elastic scattering:
\begin{equation}
{\mathcal{R}}_{NC} (L/E)  \equiv  \frac{\#~of~\nu p \to \nu p}{\#~of~\nu_\mu n \to \mu^- p }(L/E); \;\;\;\;\;\;\; \bar{\mathcal{R}}_{NC} (L/E) \equiv \frac{\#~of~\bar{\nu} p \to \bar{\nu} p}{\#~of~\bar{\nu}_\mu p \to \mu^+ n }(L/E)
\end{equation}

We can reconstruct the neutrino energy from the proton angle and momentum under the
assumption of neglecting the nuclear smearing (the same for the neutrino CC sample).
In the oscillation analysis we are only interested in relative distortions of the
ratio ${\mathcal{R}}_{NC} (\bar{\mathcal{R}}_{NC})$ as a function of $L/E$ and not
in the absolute values of the ratios. For $Q^2>0.2~\gev^2$ the relative shape of the
total cross sections is not very sensitive to the details of the form factors.
To improve the energy resolution we can use events originating from the
deuterium inside the D$_2$O target embedded into the fine-grained tracker.

An improved oscillation analysis is based on a simultaneous fit to both
${\mathcal{R}}_{e \mu} (\bar{\mathcal{R}}_{e \mu})$ and
${\mathcal{R}}_{NC} (\bar{\mathcal{R}}_{NC})$. The first ratio provides a measurement
of the oscillation parameters while the latter constrains the $\nu_e(\bar{\nu}_e)$
appearance vs. the $\nu_\mu(\bar{\nu}_\mu)$ disappearance. This analysis results in
two main requirements for the ND:
\begin{itemize}[parsep=-1pt]
\item $e^+/e^-$ separation to provide an unambiguous check of the different
behavior between neutrinos and anti-neutrinos suggested by MiniBooNE;
\item Accurate reconstruction of proton momentum and angle.
\end{itemize}

In order to validate the unfolding of the high $\Delta m^2$ oscillations from
the in situ extraction of the (anti)-neutrino flux, we would also need to change
the beam conditions, since the ND cannot be easily moved. To this end,
it will be important to have the possibility of a short run with a high energy beam
and to change/switch off the beam focusing system.

\subsubsection{Isospin Relations}
\label{SBL:sec:isospin}
Main topics:
\begin{itemize} [parsep=-2pt]
\item Adler Sum Rule
\item Tests of Isospin (Charge) Symmetry in Nucleons and Nuclei
\end{itemize}
The Adler sum rule relates the integrated difference of the
anti-neutrino and neutrino $F_2$ to the isospin of the target:
\begin{equation}
\label{ASR}
{\cal S}_A (Q^2) =\int_0^1 \; dx \; \left[ F_2^{\bar\nu} (x,Q^2) - F_2^{\nu}(x,Q^2) \right]/(2x)
= 2\,I_z,
\end{equation}
where the integration is performed over the entire kinematical range of the
Bjorken variable $x$ and $I_z$ is the projection of
the target isospin vector on the quantization axis ($z$ axis).
For the proton ${\cal S}_A^{p}=1$ and for the neutron ${\cal S}_A^{n}=-1$.

In the quark parton model the Adler sum is the difference between the
number of valence $u$ and $d$ quarks of the target. The Adler sum rule
survives the strong interaction effects because of CVC and provides an
exact relation to test the local current commutator algebra of the weak
hadronic current. We note in the derivation of the Adler sum rule the effects of
non-conservation of the axial current as well as heavy quark production are
neglected (see, e.g. \cite{IoKhLi84}).

Experimental tests of the Adler sum rule require the use of a hydrogen target
to avoid nuclear corrections to the bound nucleons inside nuclei.
The structure functions $F_2^{\bar{\nu}}$ and $F_2^\nu$ have to be determined
from the corresponding differential cross sections and must be extrapolated
to small $x$ values in order to evaluate the integral~(\ref{ASR}).
The only test available is limited by the modest statistics and
was performed in bubble chambers by the BEBC
collaboration using about 9,000 $\bar{\nu}$ and 5,000 $\nu$ events
collected on hydrogen~\cite{Allasia85}.

The LBNE program can provide the first precision test of the Adler sum rule.
To this end, the use of the high
energy beam configuration would significantly increase the sensitivity
allowing to reach higher $Q^2$ values. Since the use of a liquid H$_2$
bubble chamber is excluded in the ND hall, the (anti)-neutrino interactions
off a hydrogen target can only be extracted with a subtraction method from
the composite materials of the ND targets. The position resolution in the location
of the primary vertex is crucial with this technique to reduce systematic
uncertainties. For this reason a precision test of the Adler sum rule
can be only performed with the low density magnetized ND.

Two different
targets are used resulting in a fiducial hydrogen mass of about 1 ton:
the polypropylene $(C_3H_6)_n$ foils placed in front
of the tracking modules and pure carbon foils. The statistical subtraction
increases the statistical uncertainty by a factor of four.
With the LBNE fluxes of the minimal exposure (Scenario A) we would collect
about $1.4(1.0) \times 10^6$ inclusive $\nu(\bar{\nu})$ CC events on
the hydrogen target. The number of events will increase to $6.4(4.5) \times 10^6$
for $\nu(\bar{\nu})$ CC with the Scenario B exposure of $10^{22}$ pot.
This level of precision will offer the possibility to make new discoveries
in the quark and hadron structure of the proton.

\subsubsection{Exclusive and Semi-exclusive Processes in NC and CC}
\label{sec:xsecs}
Main topics:
\begin{itemize}[parsep=-2pt]
\item Quasi-elastic Interactions
\item Resonance Production
\item Coherent Meson Production
\item Diffractive Production
\item Deep Inelastic Scattering
\end{itemize}

At the time of LBNE running it is expected that our knowledge of the (anti)-neutrino
cross sections will be much improved over present levels. The MINER$\nu$A experiment is
now running and is expected to collect a total number of inclusive CC
interactions of $9 \times 10^6$ on a CH target, and $2.7 \times 10^6$ on both Fe and
Pb targets. These raw (excluding efficiencies) rates are assuming a 3-ton fiducial mass,
a LE beam run with an accumulated POT of 4.0$\times 10^{20}$ and 12.0$\times 10^{20}$ with
the ME NuMI beam configurations in neutrino running mode.
The statistical errors on most of the relevant processes will be of the order of 1\%.

The goal of the MINER$\nu$A experiment is a 7\% precision on the
relative flux shape and 10\% for the absolute normalization.
Therefore, the uncertainties on the cross section measurements will be dominated by
the systematic uncertainties associated with the flux determination.
This level of precision on the flux is expected to be achieved using a series
of special proton on target runs to probe different pion production kinematic
regions and the neutrino data from these runs
will be used to generate the hadron production off the target.
The flux shape will be determined using the quasi-elastic sample and the absolute
normalization using the high energy tail fixed to the CCFR/CHARM
cross section measurements.

The analysis of data from the NOMAD experiment will provide complementary measurements
of neutrino cross sections at energies which are much higher than the ones in MINER$\nu$A.
The uncertainties on the neutrino fluxes were constrained to $2.5-8\%$ in NOMAD by detailed
simulations of the beam transport, by external hadron-production measurements (SPY) and
by the low-$\nu_0$ technique. The reduced flux uncertainties coupled with the good
detector resolution allowed a measurement of the NC coherent $\pi^0$ cross section
to about 14\% and a measurement of the neutrino QE cross section to about 6\%.
The main limitation of the NOMAD cross section measurements is the limited
statistics collected, especially at low neutrino energies.

Due to the large statistics expected in the ND and the increased precision in the
determination of the (anti)-neutrino fluxes described in Section~\ref{SBL:sec:fluxes},
LBNE will overcome the two main limitations of existing and past experiments.
Even with the minimal exposure given by Scenario A, it would be possible to measure exclusive
cross sections to a precision comparable to the one of the flux predictions,
which will be in the range $1-3\%$.

\subsubsection{Structure of the Nucleon}
Main topics:
\begin{itemize}[parsep=-2pt]
\item Measurement of Form Factors and Structure Functions
\item QCD Analysis of Parton Distribution Functions
\item $d/u$ Parton Distribution Functions at Large $x$
\item GLS Sum Rule and $\alpha_s$
\item Non-perturbative Contributions and High Twists
\item Quark-hadron Duality
\item Generalized Parton Distributions
\end{itemize}

\label{SBL:sec:PDFs}

A QCD analysis of the ND data in the framework of global fits to extract
parton distribution functions is a crucial step to constrain systematic uncertainties
on the electroweak measurements (Section~\ref{SBL:sec:sin2thetaW}). In addition,
precision measurements of (anti)-neutrino structure functions and differential cross sections
would directly affect the LBL oscillation searches, providing an estimate of all background
processes which are dependent upon the angular distribution of the outgoing particles
in the FD.

For quantitative studies of inclusive deep-inelastic lepton-nucleon
scattering, it is vital to have precise $F_3$ structure functions,
which can only be measured with neutrino and antineutrino beams, as
input into global PDF fits.  Because it depends on weak axial quark
charges, the $F_3$ structure function is unique in its ability to
differentiate between the quark and antiquark content of the nucleon.
On a proton target, for instance, the neutrino and antineutrino $F_3$
structure functions (at leading order in $\alpha_s$) are given by
\begin{eqnarray}
xF_3^{\nu p}(x)
&=& 2 x \left( d(x) - \bar u(x) + \bar s(x) + \cdots \right)\, , \\
xF_3^{\bar\nu p}(x)
&=& 2 x \left( u(x) - \bar d(x) - \bar s(x) + \cdots \right)\, .
\end{eqnarray}
In contrast, electromagnetic probes are sensitive only to a sum of
quark and antiquark PDFs.  Unfortunately, the neutrino scattering
cross sections have considerably larger uncertainties than the
electromagnetic inclusive cross sections at present.
The high statistics of  Scenario B offer the promise to reduce the gap between the uncertainties on the weak and electromagnetic structure functions, and would have a major impact on global PDF analyses.

Recent experiments at JLab have collected high-precision data on the
individual $F_1$ and $F_2$ (or $F_T$ and $F_L$) structure functions at
large $x$ from Rosenbluth-separated cross sections.
This avoids the need for model-dependent assumptions about the ratio
$R = \sigma_L/\sigma_T$ of the longitudinal to transverse cross
sections in the extraction of the structure functions from the
measured cross sections.
Similar quality data on the individual $F_T$ and $F_L$ structure
functions from neutrino scattering would be available from the ND at LBNE,
to maximally complement and
facilitate the flavor decomposition of these functions.

In addition to data in the DIS region, there is considerable interest in
obtaining data at low $Q^2$ (down to $Q^2\sim1$~GeV$^2$) and low $W$
($W < 2$~GeV), to complement those from JLab.
Unpolarized structure functions can be expressed in terms of
powers of $1/Q^2$ (power corrections):
\begin{equation}
F_{2,T,3}(x,Q^2) = F_{2,T,3}^{\tau = 2}(x,Q^2)
+ {H_{2,T,3}^{\tau = 4}(x) \over Q^2}
+ {H_{2,T,3}^{\tau = 6}(x) \over Q^4} + .....
\label{eqn:ht}
\end{equation}
where the first term ($\tau=2$), expressed in terms of PDFs, represents the
Leading Twist (LT) describing the scattering off a free quark and
is responsible for the scaling of SF via perturbative QCD
$\alpha_s(Q^2)$ corrections. The Higher Twist (HT) terms ($\tau = 4,6$)
reflect instead the strength of multi-parton correlations (qq and qg).
The ND data at LBNE would allow a good separation of target mass and higher twist
corrections, both of which are $1/Q^2$ suppressed at high $Q^2$, from leading twist
contributions~\cite{Alekhin:2007fh,CTEQX}.

Global PDF fits show that at large values of $x$ ($x > 0.5-0.6$) the $d$
quark distribution (or the $d/u$ ratio) is very poorly determined.
The main reason for this is the
absence of free neutron targets.  Because of the larger electric charge
on the $u$ quark than on the $d$, the electromagnetic proton $F_2$
structure function data provide strong constraints on the $u$ quark
distribution, but are relatively insensitive to the $d$ quark distribution.

To constrain the $d$ quark distribution a precise knowledge
of the corresponding neutron $F_2^n$ structure functions is required, which in
practice is extracted from inclusive deuterium $F_2$ data.
At large values of $x$ the nuclear corrections in deuterium become
large and, more importantly, strongly model dependent, leading to
large uncertainties on the resulting $d$ quark distribution.

Several planned experiments at JLab with the energy upgraded 12~GeV
beam will measure the $d/u$ ratio up to $x\sim0.85$ using several
different method to minimize the nuclear corrections.  One method will
use semi-inclusive DIS from deuterium with a low-momentum
($|\vec{p}| < 100$~MeV) spectator proton detected in the backward
center-of-mass hemisphere, to ensure scattering on an almost free
neutron (so-called ``BONUS'' experiment~\cite{BONUS12}). Preliminary
results have confirmed the feasibility of this method at the current
6~GeV energies, and a proposal for the extension at 12~GeV has been
approved.

Perhaps the cleanest and most direct method to determine the $d/u$
ratio at large $x$ is from neutrino and antineutrino DIS on hydrogen.
Existing neutrino data on hydrogen have relatively large errors and
do not extend beyond $x\sim0.5$~\cite{NU-H}. A new measurement of
neutrino and antineutrino DIS from {\it hydrogen} at LBNE with significantly
improved uncertainties would therefore make an important discovery
about the $d/u$ behavior as $x \to 1$. This measurement is possible with the Scenario B
exposure and a statistical subtraction of the hydrogen target
from the composite materials of the fine-grained ND(Section~\ref{SBL:sec:isospin}).
To be competitive with the proposed JLab 12~GeV experiments, the kinematical reach would need
to be up to $x\sim0.85$ and with as large a $Q^2$ range as possible
to control for higher twist and other sub-leading effects in $1/Q^2$.

\subsubsection{Nuclear Effects}
\label{sec:nucleff}
Main topics:
\begin{itemize}[parsep=-2pt]
\item Nuclear Modifications of Form Factors
\item Nuclear Modifications of Structure Functions
\item Mechanisms for Nuclear Effects in Coherent and Incoherent Regimes
\item A Dependence of Exclusive and Semi-exclusive Processes
\item Effect of Final State Interactions
\item Effect of Short Range Correlations
\item Two Body Currents
\end{itemize}

The study of nuclear effects in (anti)-neutrino interactions off nuclei
is directly relevant for the LBL oscillation search. The use of
water or argon in the FD requires a measurement of nuclear
cross sections on the same targets in the ND.
In addition to the different $p/n$ ratio
in water and argon, nuclear modifications of cross sections can differ from 5\% to 15\%
between oxygen and argon, while final state interactions are expected to be about
a factor of two larger in argon~\cite{docdb740}.

Furthermore, nuclear modifications can introduce a substantial smearing of the
kinematic variables reconstructed from the observed final state particles.
Detailed measurements of the $A$ dependence of different processes
are then required in order to understand the absolute energy scale
of neutrino events and to reduce the corresponding systematic
uncertainties on the oscillation parameters.

In addition to understanding the structure of the free nucleon, an
important question in nuclear physics is how that structure is modified
when the nucleon is inside a nuclear medium.  Studies of ratios of
structure functions of nuclei to those of free nucleons (or in practice
the deuteron) reveal nontrivial deviations from unity as a function of
$x$ and $Q^2$.  These have been explored thoroughly in charged lepton
scattering experiments, but little
empirical information exists from neutrino scattering.

Another reason to investigate the medium modifications
of neutrino structure functions is that most neutrino scattering
experiments are performed on nuclear targets, from which information
on the free nucleon is inferred by performing a correction for the
nuclear effects.  In practice this often means applying the same
nuclear correction as for the electromagnetic structure functions,
which introduces an inherent model dependence in the result.

In particular, significant differences between photon-induced and weak
boson-induced nuclear structure functions are predicted, especially
at low $Q^2$ and low $x$, which have not been tested.  A striking
example is offered by the ratio $R$ of the longitudinal to transverse
structure functions~\cite{Kulagin:2007ju}.
While the electromagnetic ratio tends to zero in the photoproduction
limit, $Q^2 \to 0$, by current conservation, the ratio for neutrino
structure functions is predicted to be {\it finite} in this limit.
Thus significant discovery potential exists in the study of neutrino
scattering from nuclei.

Finally, the extraction of (anti)-neutrino interactions on deuterium from the
statistical subtraction of H$_2$O from D$_2$O, which is required to
measure the fluxes (Section~\ref{SBL:sec:absflux}), would allow the
first direct measurement of nuclear effects in deuterium at LBNE.
This measurement can be achieved since the structure function of free
isoscalar nucleon is given by the average of neutrino
and anti-neutrino structure functions on hydrogen ($F_2^{\nu n}=F_2^{\bar{\nu} p}$).
A precise determination of nuclear modifications of structure functions
in deuterium would play a crucial role in reducing systematic uncertainties
from global PDF fits.

%


\subsection{Requirements for the Near Detector Complex}
\label{SBL:sec:requirements}

\subsubsection{Requirements for the LBL Oscillation Analysis}

The detector requirements to support the LBL oscillation analysis are driven by the
determination of the fluxes. Three main measurements are needed to obtain accurate
predictions of the $\nu_\mu, \bar{\nu}_\mu, \nu_e, \bar{\nu}_e$ content of the beam:
a) NC elastic scattering off electrons; b) low-$\nu_0$ technique;
and c) QE CC scattering off deuterium in the limit $Q^2=0$.
The most important background measurement for the $\nu_e (\bar{\nu}_e)$ appearance search
is the determination of the $\pi^0/\gamma$ content in NC and high-$y_{Bj}$ CC
interactions.
Similarly, the $\nu_\mu (\bar{\nu}_\mu)$
disappearance search requires a precise measurement of the yields of
$\pi^\pm$ and $K^\pm$ and of their $\mu^\pm$ decays of as a function of energy and angle.
In addition, QE, single $\pi$ and DIS
processes and their nuclear dependence must be measured in CC interactions.
The above measurements define the key requirements for the ND complex:
\begin{itemize}
\item $\mu^-/\mu^+$ separation for the measurement of the $\nu_\mu$ and $\bar{\nu}_\mu$
beam content, the low-$\nu_0$ technique and the $\mu^\pm$ decays of $\pi^\pm/K^\pm$;
\item $e^-/e^+$ separation for the measurement of the $\nu_e$ and $\bar{\nu}_e$ beam content,
NC elastic scattering off electrons, and to unfold the fluxes in the presence of high $\Delta m^2$ oscillations;
\item D$_2$O target embedded in the fine-grained tracker for the $\nu_\mu$ flux extraction;
\item Possibility to accommodate different nuclear targets to constrain nuclear effects
on the predicted event rates and to study the calibration of the neutrino energy scale;
\end{itemize}
This list implies that at least part of the near detector complex must be magnetized.
The $e^-/e^+$ separation and the measurement of $\mu^\pm$ decays of $\pi^\pm/K^\pm$
are possible only in a magnetized LAr detector or in the low density magnetized tracker.

The reference exposure of a 3+3 year run with the 700~kW beam seems adequate
to achieve a precision of $\sim1-3\%$ on fluxes and backgrounds, if these detector
requirements are satisfied. The reference exposure will also provide the relevant
information for the LBL oscillation analysis well before the completion of the
nominal LBL data taking (5+5 years).
An additional requirement would be the possibility
to change the beam focusing system and to run with a higher energy beam spectrum.
These options would be crucial to unfolding the fluxes if high $\Delta m^2$ oscillations are present.

\subsubsection{Additional Requirements for the Study of Neutrino Interactions}

By satisfying the requirements described in the previous Section, there are no significant additional detector requirements for the ND complex in order to perform precision studies of fundamental interactions along with  satisfying the requirements of LB oscillation studies.
The only upgrade of the ND complex that might be considered is a modest
instrumentation of the upstream end of the ND hall to increase the decay
length for the search for neutral leptons to $\sim30$~m. Indeed, in most cases the physics processes studied are the same as those used to constrain fluxes and backgrounds for LB measurements.

From the point of view of the beam design, our sensitivity studies show that the
Scenario B high statistics exposure would be a crucial step to achieve a breakthrough
in the precision tests of fundamental interactions and of the structure of matter.
For most measurements we considered, the Scenario A reference exposure with the 700~kW beam provides
only limited improvements over existing or future programs, whereas, Scenario B has the potential to bring the LBNE program to a level of precision similar to $e^+e^-$ colliders. To this end, it is also worth noting
a run with the high energy beam option would substantially enhance the physics potential of LBNE.

\subsubsection{Conclusions}

The short baseline measurements are an integral part of the LBL program since
the LBL oscillation analysis requires precise predictions of fluxes and backgrounds.
The precision and resolution achieved by a fine-grained ND complex can therefore directly
affect the design of the FD, especially for the detector options characterized by
larger background levels and lower resolution (e.g. water Cherenkov).

The choice between water Cherenkov and LAr for the technology of the FD has,
in turn, some implications in the design of the ND complex.
Since sizeable nuclear modifications are expected on cross sections, the minimal requirement
is that the nuclear targets selected for the FD shall also be present in the ND.
This condition can be achieved either by using
a combination of different detectors at the near site, or by using different
dedicated nuclear targets within the same detector technology.

A hybrid FD with water Cherenkov and LAr technologies clearly implies the need for an ND complex
with both a LAr TPC and a fine-grained tracker with a water target embedded.
Our sensitivity studies showed that, with a magnetized tracker capable of $e^+/e^-$ separation,
such a combination in the ND has an advantage also in terms of physics potential over
a single detector, regardless of the specific choice of the FD technology.
A combined analysis can provide the absolute normalization of the fluxes
to $\sim1\%$ from NC elastic scattering off electrons in LAr and the relative
fluxes as a function of energy to $\leq 2\%$ from the low-$\nu_0$ method in the
magnetized tracker.
{\em We conclude that the inclusion of an LAr TPC with a magnetized tracker in the ND complex is preferred regardless of the FD configuration.}
The LAr TPC in the ND does not need necessarily to be
magnetized when coupled with the tracker since this latter will also serve
as a spectrometer for the LAr detector. The crucial parameter for the LAr
TPC in light of a combined performance is a fiducial mass of the
order of 100 tons.

With at least one water Cherenkov detector
in the far site the most effective solution would be to have
a large LAr TPC without magnetic field with a magnetized tracker.
If LAr alone is the choice of the technology for the FD, the possibility
of a single magnetized LAr detector might be considered, even if
this solution would still have a reduced physics potential with
respect to a two-detector complex with a LAr TPC and a magnetized tracker.
The magnetic field should be sufficiently high to allow $e^+/e^-$ separation
and the LAr active volume should provide a good containment of the neutrino events.
Detailed simulations of (anti)-neutrino interactions in the LAr and a
complete flux analysis are required in order to evaluate the figure of merit
of a single LAr magnetized detector.

We note that an important issue to be addressed by the LAr detector design is the
capability to handle the high event rates expected at LBNE given the
typical drift time of about 600~$\mu$s for a 1~m drift length, this issue would be more prominent if the beam power were increased above the nominal 700~kW.

Table~\ref{tab:NDperf} summarizes the scorecard for the different ND configurations.
In addition to the individual detector options, we list the two best configurations
from the point of view of a combined analysis including a LAr detector and a
fine-grained tracker. While a magnetized LAr is not needed with the low density
magnetized tracker, it would be required to extend the physics potential of the scintillator
tracker. The main limitation of the magnetized LAr detector currently
considered is the relatively small fiducial volume ($\sim$20 tons),
which would not allow a complete containment of the events so reducing the
usable statistics.

\begin{table}[ht]
\begin{tabular}{l|c|c|c|c|c|c}
Measurement & STT & Sci+$\mu$Det & LAr & LArB & LArB+Sci+$\mu$Det & LAr+STT \\
\hline\hline
 \multicolumn{7}{c}{~~~~~~~In Situ Flux Measurements for LBL:~~~~~~~ } \\
 $\nu e^- \to \nu e^-$  & Yes & No & Yes & No & No & Yes  \\
 $\nu_\mu e^- \to \mu^- \nu_e$  & Yes & Yes & No & Yes & Yes & Yes  \\
 $\nu_\mu n \to \mu^- p$ at $Q^2=0$  & Yes & Yes & No & No & Yes & Yes  \\
 Low-$\nu_0$ method & Yes & Yes & No & Yes & Yes & Yes  \\
 $\nu_e$ and $\bar{\nu}_e$ CC & Yes & No & No & Yes & Yes & Yes  \\ \hline
 \multicolumn{7}{c}{~~~~~~~Background Measurements for LBL:~~~~~~~ } \\
 NC cross sections & Yes & Yes & No & Yes & Yes & Yes  \\
 $\pi^0/\gamma$ in NC and CC & Yes & Yes & Yes & Yes & Yes & Yes  \\
 $\mu$ decays of $\pi^\pm, K^\pm$ & Yes & No & No & Yes & Yes & Yes  \\
 (Semi)-Exclusive processes & Yes & Yes & Yes & Yes & Yes & Yes  \\ \hline
 \multicolumn{7}{c}{~~~~~~~ Precision Measurements of Neutrino Interactions:~~~~~~~ } \\
 $\sin^2 \theta_W$ $\nu$ N DIS & Yes & No & No & No & No & Yes  \\
 $\sin^2 \theta_W$ $\nu e$ & Yes & No & Yes & No & No & Yes  \\
 $\Delta s$ & Yes & Yes & Yes & Yes & Yes & Yes  \\
 $\nu$MSM neutral leptons & Yes & Yes & Yes & Yes & Yes & Yes  \\
 High $\Delta m^2$ oscillations & Yes & No & No & Yes & Yes & Yes  \\
 Adler sum rule  & Yes & No & No & No & No & Yes \\
 $D/(p+n)$ & Yes & No & No & No & No & Yes \\
 Nucleon structure & Yes & Yes & Yes & Yes & Yes & Yes \\
 Nuclear effects & Yes & Yes & Yes & Yes & Yes & Yes \\  \hline
\end{tabular}
\caption{Summary of measurements that can be performed by different ND reference configurations.
}
\label{tab:NDperf}
\end{table}


\vfill\eject
\clearpage
%
\section{Summary}
\label{sec:summary}

In this report we have summarized the interim findings of a 2010 study by the LBNE Science Collaboration Physics Working Group. It provides a snapshot of the projected science capability of various potential LBNE far and near detector configurations based on a particular set of assumptions, particularly with respect to the far detector performance parameters. Table~\ref{tab:crit_assumptions} summarizes some of the assumptions that may significantly affect the conclusions if they are incorrect. Work to mitigate the uncertainties in the detector and physics models continues and will be reported in future documents.

\begin{table}[ht]
\begin{tabular}{cp{17cm}}
\hline
\\
A. &The muon veto for the 800-foot LAr option is $>$80\% effective in conserving fiducial volume in liquid argon.\\
B. &Gadolinium-loaded water reduces SRN background by a factor of five.\\
C. &For LAr, background during a supernova burst is not significant for any depth, and that an appropriate triggering scheme will be implemented for high signal efficiency during a burst.\\
D. &SK-1(2) detector performance parameters are used for the LBNE WC 30\%(15\%) pmt coverage studies.\\
E. &For LAr, 80\% efficiency and $<$1\% background are used for Long-Baseline studies.\\
F. &In evaluating impact for proton decay studies, 10 years LBNE running is compared to continued running of SK.\\
G. &For WC, current Super-K background rates are assumed for proton decay.\\
H. &{\em K select} trigger for proton decay mitigates the need for photon trigger in the 300-foot LAr option.\\
\\
\hline
\end{tabular}
\caption{Summary of some critical performance assumptions on which the section conclusions are based (in no special order). The full set of assumptions is detailed in the relevant sections.}
\label{tab:crit_assumptions}
\end{table}

\begin{acknowledgments}
We would like to thank the Institute for Nuclear Theory, University of Washington for hosting the LBNE Science Collaboration mini-Workshop 9-11 August 2010~\cite{int10-2b}, during which much of the content of this document was finalized. A preliminary version of this report was made available as an INT publication in spring 2011~\cite{INT-PUB-11-002}.
This work was supported in part by grants from the US Department of Energy and the National Science Foundation.
\end{acknowledgments}

\clearpage
\vfill\eject

\appendix
\section{Long-Baseline Oscillation Sensitivity Assumptions}\label{lbl_appendix}

\begin{figure}[htb]
 \centering\includegraphics[width=0.45\textwidth]{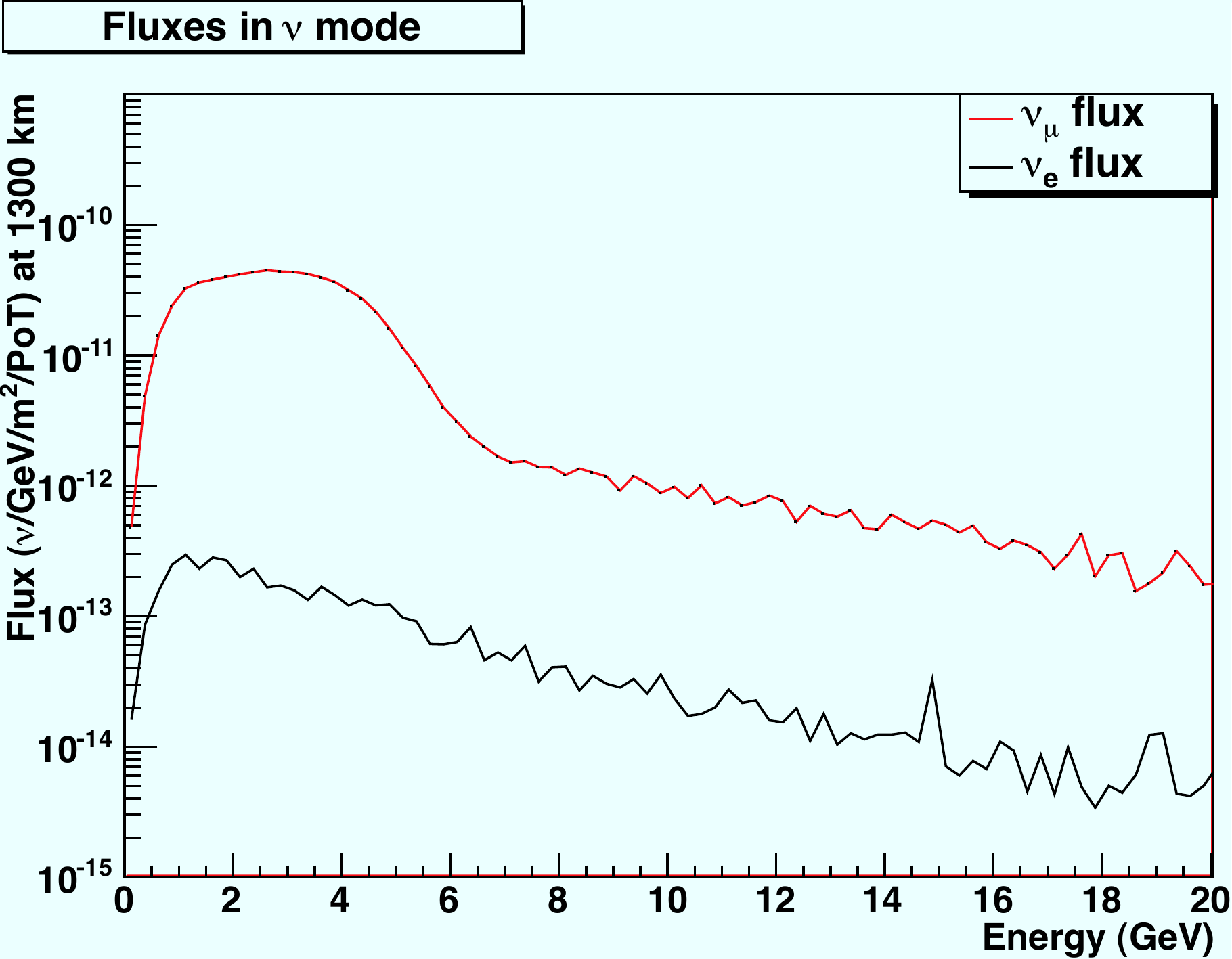}
 \centering\includegraphics[width=0.45\textwidth]{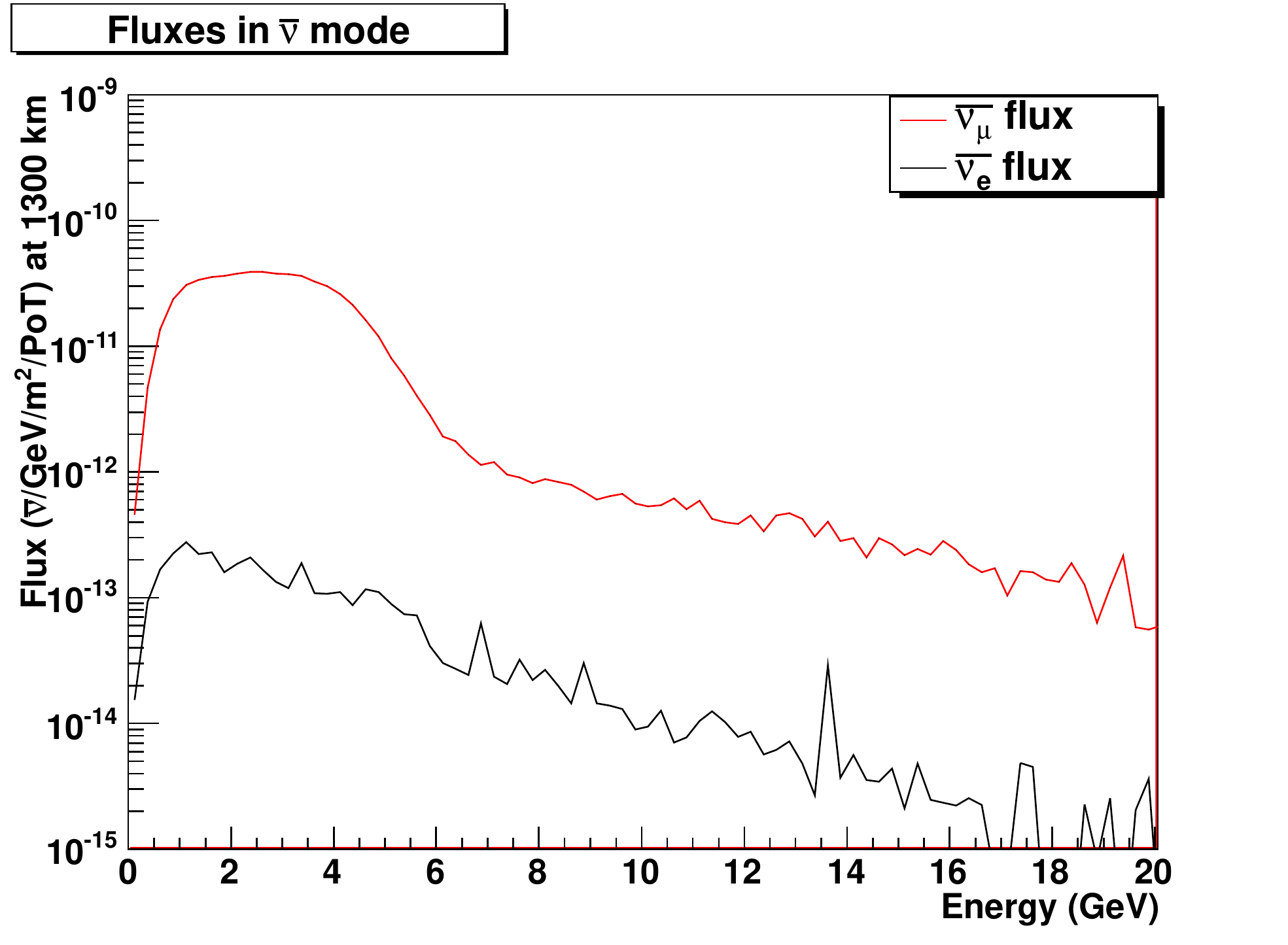}
  \caption{The ``August 2010'' $\nu$ (left) and $\nubar$ (right) mode fluxes
           (\texttt{dusel120e250(n)i002dr280dz-tgtz30-1300km-0kmoa-flux.txt}).
            This is a 120~GeV optimized beam tune using two
           parabolic NuMI horns with the target pulled out 30~cm
           and assuming a 250~kA horn current and 280~m long,
           2~m radius decay region (red curve in
           Fig.~\ref{fig:fig_beam_spectra1}). This same set of fluxes is
           consistently used for all $\nue$ appearance and $\numu$
           disappearance studies in this document.}
  \label{fig:lbl_globes_fluxes}
\end{figure}

\begin{figure}[htb]
 \centering\includegraphics[width=0.85\textwidth]{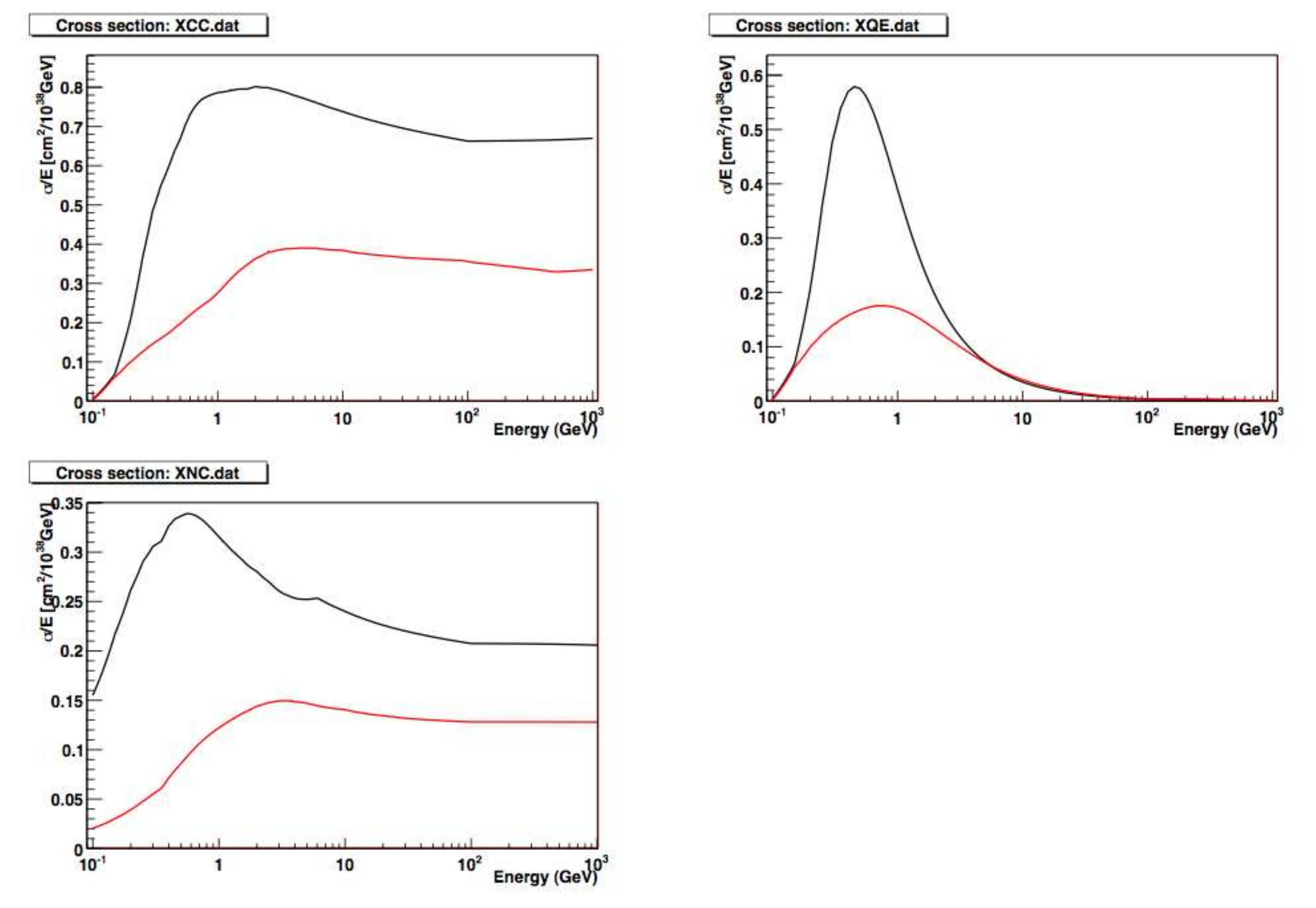}
  \caption{The QE, total CC, and total NC $\numu$ (black) and $\numubar$ (red)
           cross sections used in the long-baseline calculations.
           The same cross sections are currently used for both WC and LAr.
           These files were provided with GLoBES and have not been
           additionally modified~\cite{globes-xsec}.
           Using the same input GLoBES cross sections for WC and LAr
           is not a bad assumption for now given that all of the signal
           samples in LAr are selected to be inclusive CC channels where the
           cross sections for Ar and H$_2$O are similar~\cite{docdb740}.
           However, if a QE signal channel is used for LAr in the future, the
           $\sim20\%$ neutron excess in argon would have to be accounted for
           in GLoBES. Also, $\pi^0$ absorption rates are about a factor of two
           larger in Ar than in H$_2$O~\cite{docdb740}. This potentially
           leads to a factor of two fewer $\pi^0$'s that can decay and lead
           to $\nue$-like backgrounds in LAr. This difference can be accounted
           for in the $\pi^0$ efficiency factors for LAr rather than in the
           cross sections. }
  \label{fig:lbl_globes_xsec}
\end{figure}

This section details the assumptions used in the long-baseline oscillation
sensitivity projections for LBNE. All calculations (with the exception of
those that did not include detector response) were performed using the
GLoBES~\cite{globes} software package.
The same neutrino fluxes (Fig.~\ref{fig:lbl_globes_fluxes}) and
cross sections (Fig.~\ref{fig:lbl_globes_xsec})
are used in GLoBES for both WC and LAr detectors.
Inputs that differ between the two sets of detectors include assumptions about the
detector performance (i.e. energy resolutions and detection
efficiencies), specific signal processes, and backgrounds. These differences
are outlined in the following sections.
A summary of preliminary liquid argon TPC performance parameters for use with GLoBES
can be found in a presentation by Bonnie Fleming to the Long-Baseline
group~\cite{LAr_perf_fleming}. All estimates assume a 700~kW LBNE beam,
which implies a beam delivery of $7.3\times10^{20}$ POT/year.

\subsection{Inputs for $\nue$ Appearance}

For the estimate of $\theta_{13}$, $\delta_{\mathrm CP}$, and mass hierarchy
sensitivities, GLoBES requires information on $\nue$ efficiencies,
backgrounds, and energy resolutions for both WC and LAr detectors. These
inputs are summarized in Table~\ref{table:lbl_globes_inputs_nue} and
detailed below.

\begin{table} [htb]
\begin{center}
\begin{tabular}{c||c|c}
Input  & Water Cerenkov  & Liquid Argon TPC \\ \hline
signal channel           & $\nue$ QE & $\nue$ CC \\
signal efficiencies      & Fig.~\ref{fig:lbl_eff_nue} & $80\%$ \\
signal $E_\nu$ resolution, $\sigma(E)/E$  & Fig.~\ref{fig:lbl_eres_nue} & $0.15/E$  \\
signal normalization error  & $1\%$  & $1\%$ \\
\hline
background channels        & $\numu$ CC, $\numu$ NC, intrinsic $\nue$
                           & $\numu$ CC, $\numu$ NC, intrinsic $\nue$  \\
background efficiencies    & Fig.~\ref{fig:lbl_eff_nue}
                           & $1\%$, $1\%$, $80\%$ \\
background $E_\nu$ resolution, $\sigma(E)/E$  & Figs.~\ref{fig:lbl_eres_dis_bkg},~\ref{fig:lbl_eres_nc_bkg} &  $0.15/E$, Fig.~\ref{fig:lbl_eres_nc_bkg}, $0.15/E$\\
background normalization error   & $5\%$ & $5\%$
\end{tabular}
\caption{\label{table:lbl_globes_inputs_nue} Summary of GLoBES inputs for both
detector technologies for the LBNE $\nue$ appearance sensitivity estimates.
While not listed above, the $\nu$ contribution for each channel
is explicitly included in GLoBES in the case of $\nubar$ running.}
\end{center}
\end{table}

It should be noted that the treatment of systematic errors
is admittedly not very sophisticated at present. In all long-baseline
oscillation studies up to now, the total background normalization
is allowed to vary by $5\%$ as a source of systematic error. Background
contributions are assumed to have a perfectly known shape and are not
allowed to vary independently. Also, separate uncertainties
on neutrino vs. antineutrino interactions have not yet been assessed.
A more rigorous and complete systematic error handling in GLoBES is
currently being developed for LBNE.

\begin{figure}[htb]
 \centering\includegraphics[width=0.9\textwidth]{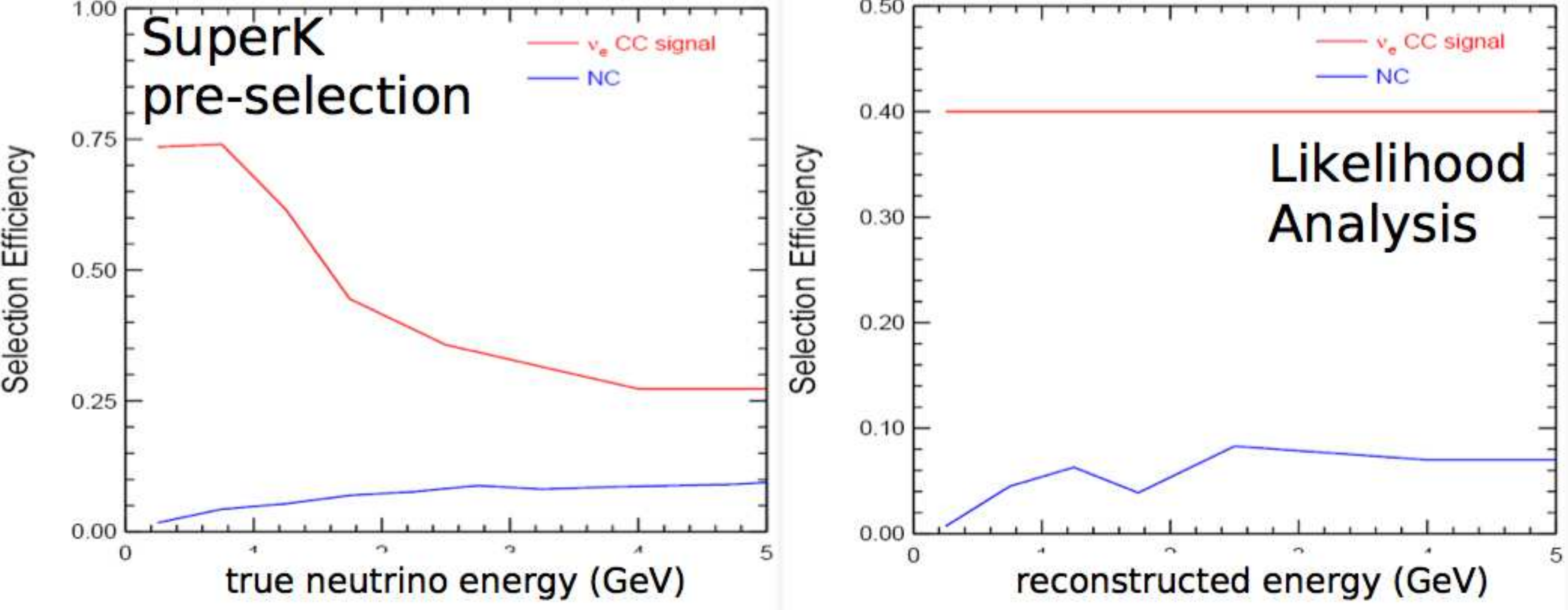}
  \caption{Efficiencies for both $\nue$ (red) and NC background (blue)
           events assumed for a WC detector in GLoBES. The
           efficiencies are based on improvements to the Super-K reconstruction
           using a specialized fitter (POLfit) and are based on $40\%$
           photocathode coverage~\cite{chiaki}.
           The efficiencies associated with the pre-cuts (left) are a function
           of true neutrino energy and reflect the Super-K requirement of a
           single-ring, electron-like event with no decay electron in the
           fiducial volume. The efficiencies  resulting from the likelihood
           analysis (right) are functions of reconstructed neutrino energy
           and were set to retain $40\%$ of $\nue$ signal events (which in turn
           specifies the level of NC background). The product of these two
           efficiencies yields the total efficiency. Combined, this results in
           an overall $16\%$ ($28\%$) $\nue$ efficiency at 2~GeV (0.8~GeV)
           for WC (see also Fig.~\ref{fig:lbl_nue_net}).}
  \label{fig:lbl_eff_nue}
\end{figure}

\begin{figure}[htb]
 \centering\includegraphics[width=.4\textwidth]{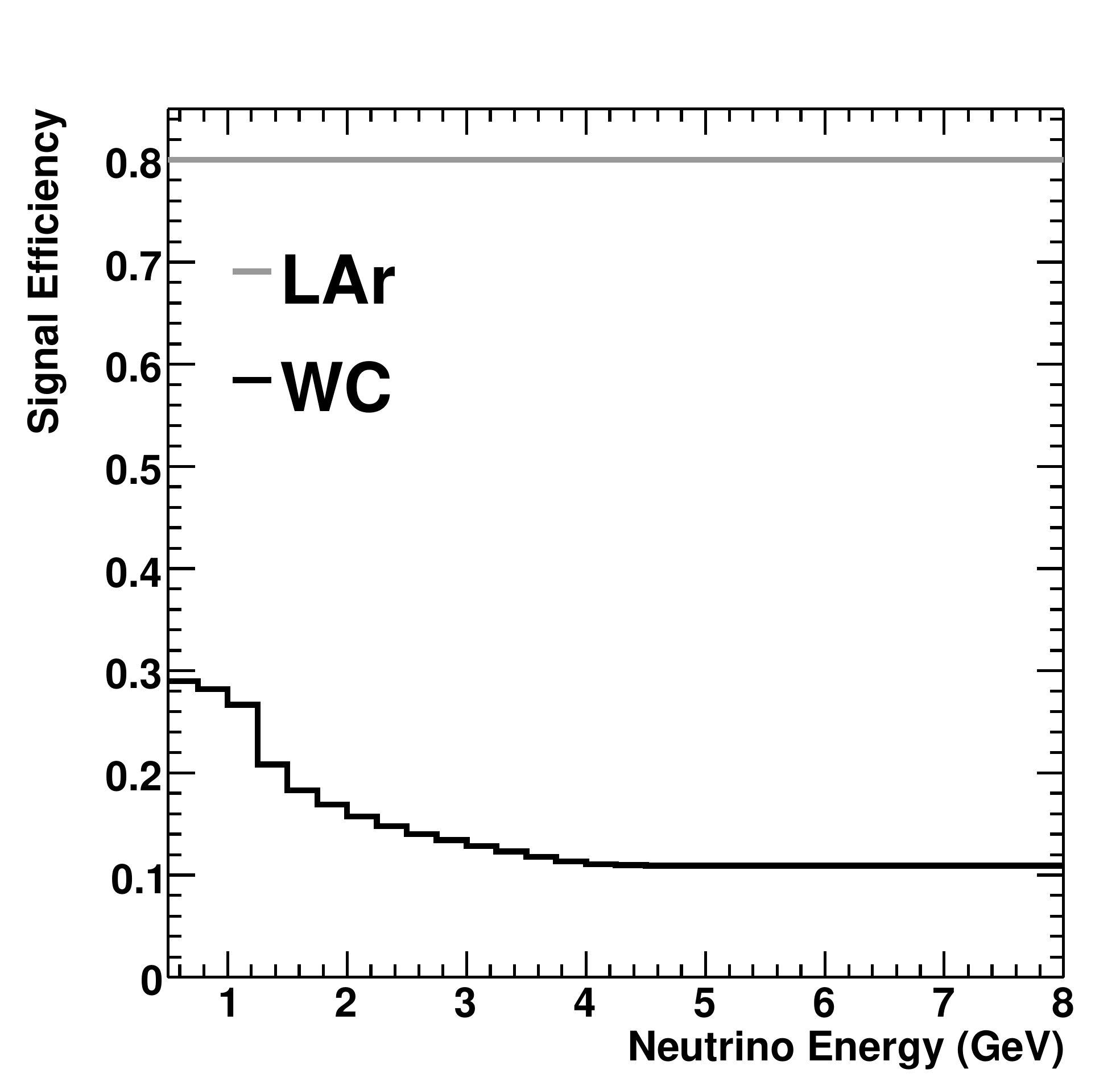}
 \centering\includegraphics[width=.4\textwidth]{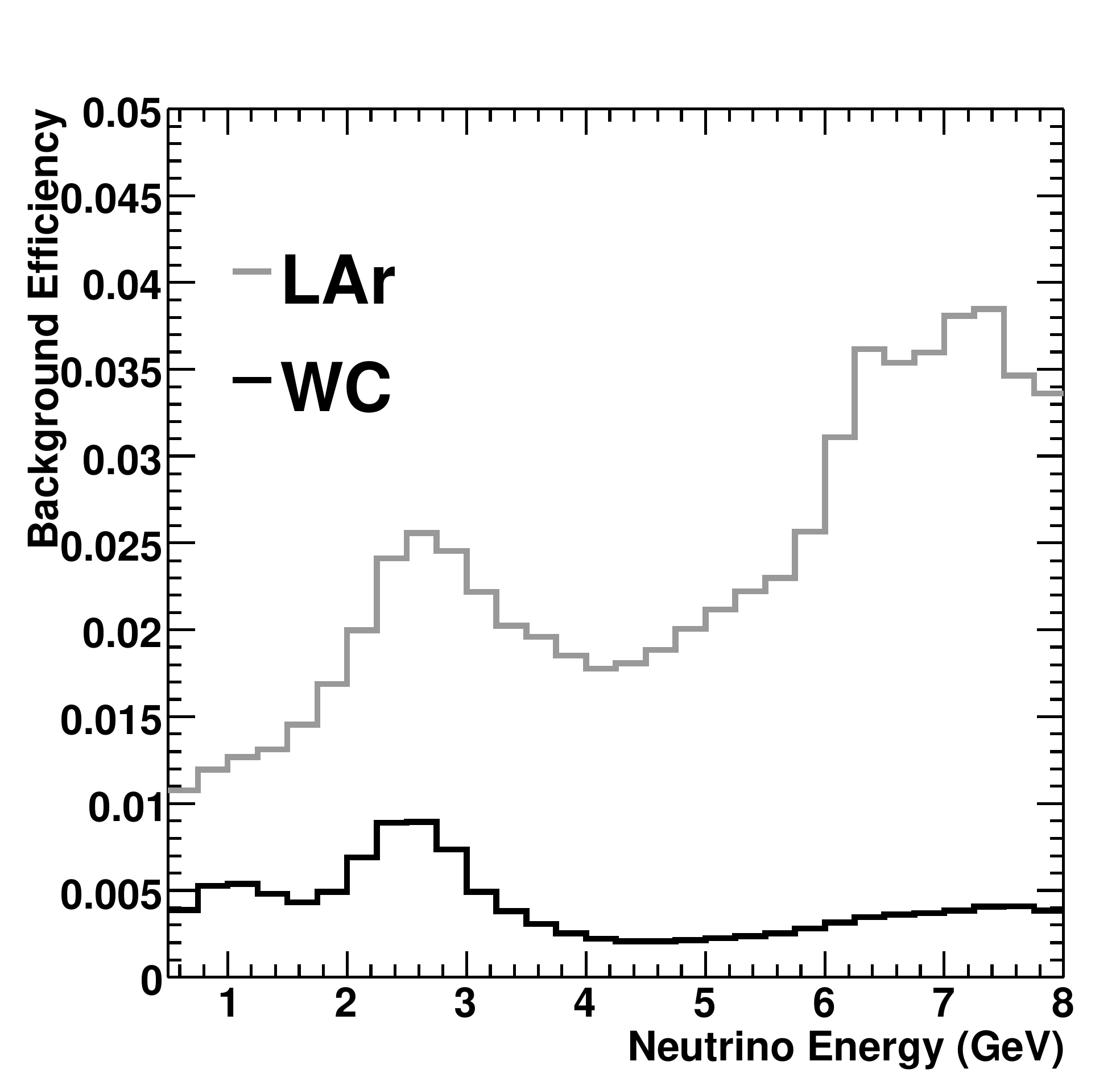}
  \caption{Assumed $\nue$ signal (left) and total background (right)
   efficiencies plotted as a function of neutrino energy for WC (black)
   and LAr (gray). This reflects the fraction of signal and background
   events that enter the $\nue$ appearance sample. The individual factors
   that make up the total signal $\nue$ efficiency in the case of WC is
   provided in Fig.~\ref{fig:lbl_eff_nue}. Note: while not shown, the
   efficiency for $\nuebar$'s is slightly higher than that for $\nue$'s in
   WC to account for the fact that the QE fraction is larger for
   antineutrinos as one moves up in energy. The efficiency for the total
   background is the weighted average of the individual efficiencies, i.e.
   the efficiency has been weighted by the amount of background in each
   energy bin. This is why the total background efficiencies are not
   constant in energy. A breakdown of the efficiencies for each of the
   individual background sources is provided in
   Fig.~\ref{fig:lbl_nue_eff_bkg}.}
  \label{fig:lbl_nue_net}
\end{figure}

\begin{figure}[htb]
 \centering\includegraphics[width=.4\textwidth]{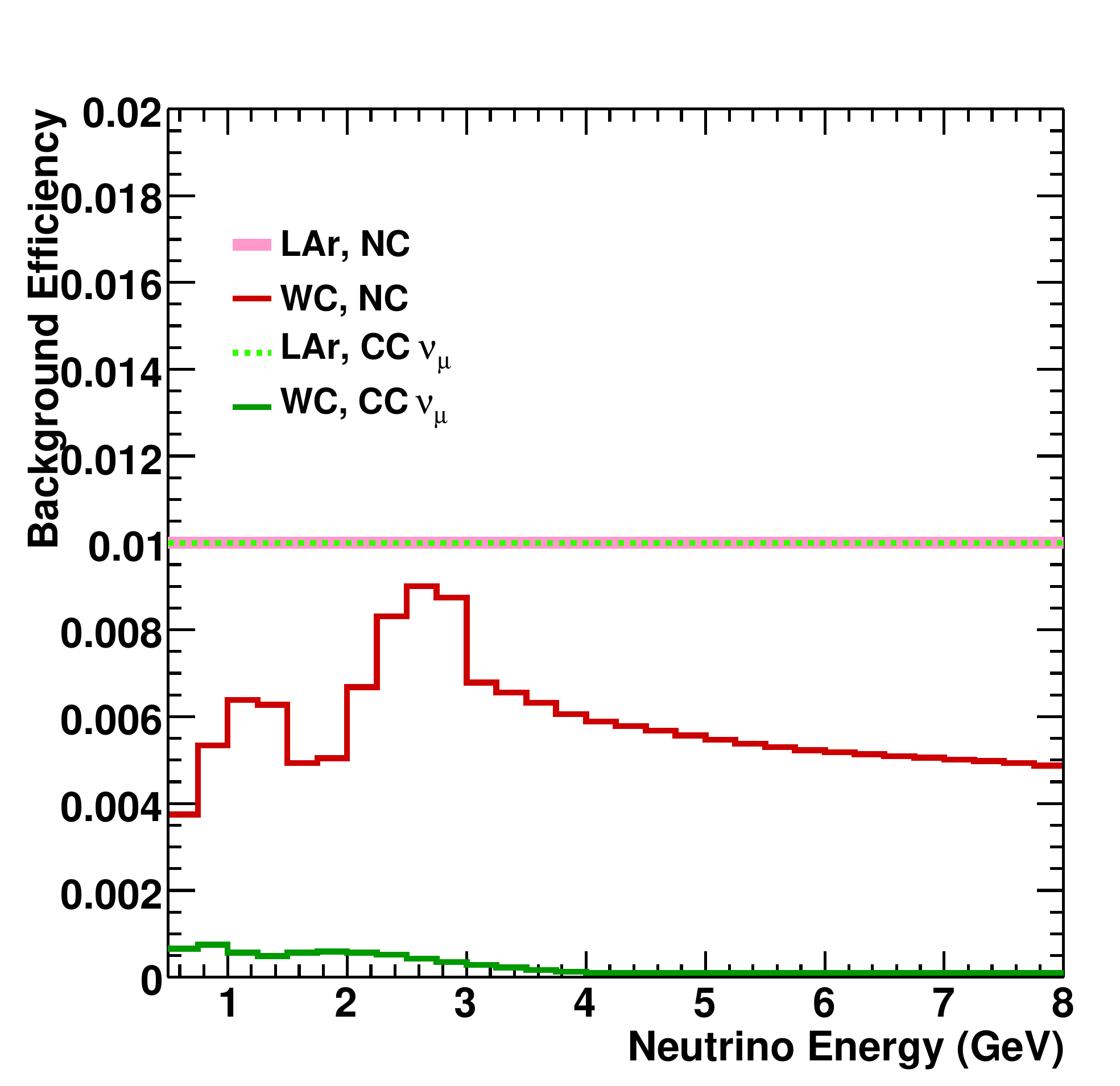}
  \caption{Assumed efficiencies plotted as a function of neutrino energy for
   each background source in WC and LAr. This reflects the fraction of the
   background that enters the $\nue$ sample. A $1\%$ energy-independent
   efficiency is assumed for each of the NC and $\numu$ CC backgrounds in LAr
   (pink and dashed green lines). The efficiencies for NC and $\numu$ CC
   are energy-dependent in the case of WC (red and solid green
   lines). Not shown are the intrinsic $\nue$ backgrounds which are assumed
   to have the same efficiency as $\nue$ signal events
   (left panel in Fig.~\ref{fig:lbl_nue_net}).}
  \label{fig:lbl_nue_eff_bkg}
\end{figure}

Fig.~\ref{fig:lbl_nue_net} shows the net signal and background
efficiencies plotted as a function of neutrino energy for each detector type.
Fig.~\ref{fig:lbl_nue_eff_bkg} further breaks down the background
efficiencies by source.  \\

There are several things to note. In the case of WC, the efficiencies are
based on modified Super-K simulations set to retain $40\%$ of all $\nue$
signal events and assume $40\%$ PMT coverage~\cite{chiaki}. The energy
resolutions for each process were obtained from Nuance
simulations~\cite{nuance} and checked against Super-K Monte
Carlo~\cite{diwan}. In the case of LAr, the background efficiency factors
imply that $1\%$ of $\numu$ CC events, $1\%$ of NC events, and $80\%$ of
intrinsic $\nue$'s pose backgrounds to the $\nue$ appearance search and not
that $1\%$ of the total sample is $\numu$ CC, $1\%$ is NC, etc. Also, in
GLoBES, the background sources are varied by a $\pm 5\%$ uncertainty in a
process and energy independent way (i.e., the overall normalizations of the
$\numu$ CC, NC, and $\nue$ backgrounds are varied together). This is true
for both WC and LAr and will be upgraded to more realistic spectral
uncertainties in the near future. Overall, the inputs were chosen to
conservatively reflect the performance that we think we can realistically
achieve with each detector technology and are subject to further change.

\begin{figure}[htb]
 \centering\includegraphics[width=.4\textwidth]{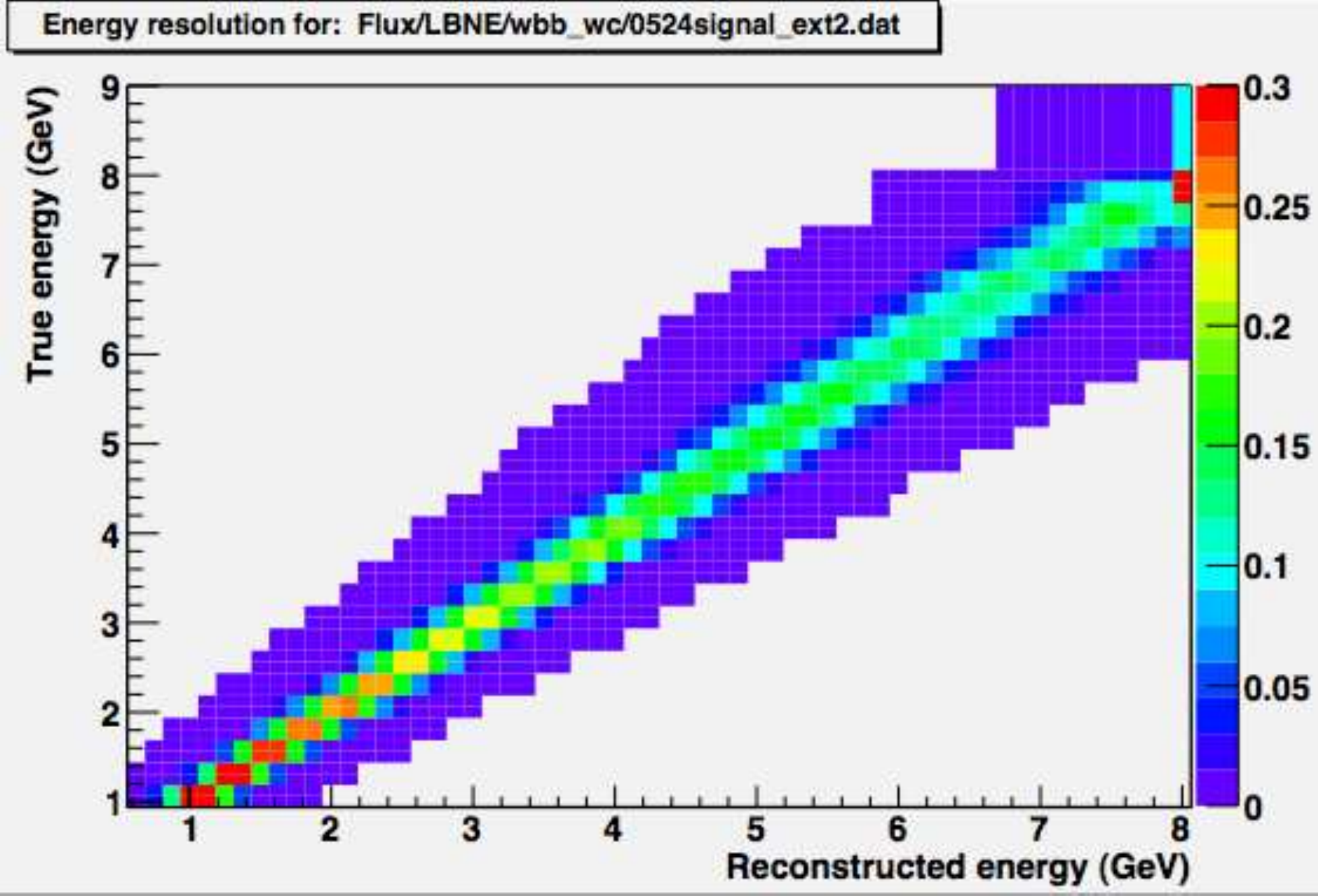}
 \centering\includegraphics[width=.4\textwidth]{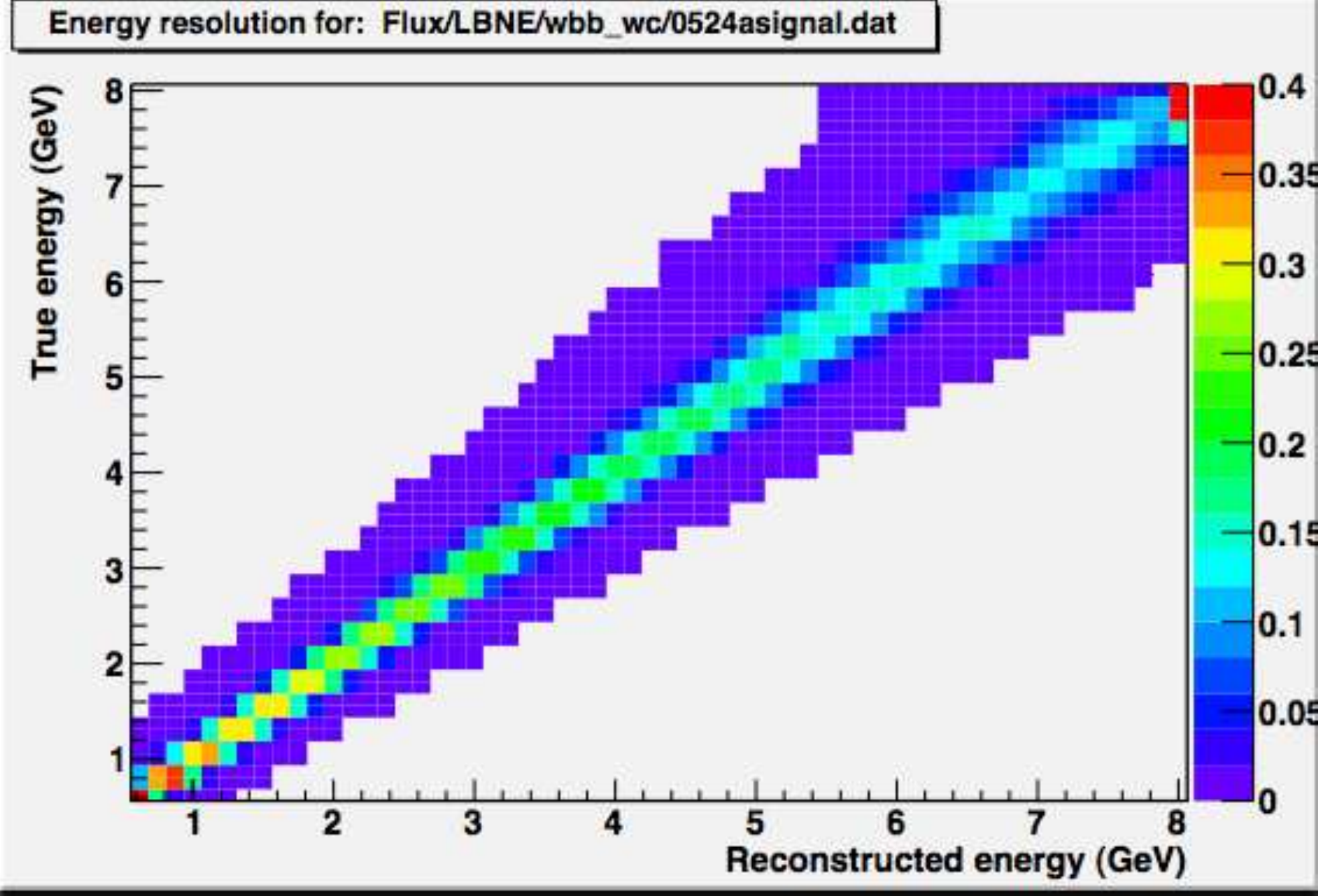}
  \caption{Energy resolutions for $\nue$ (left) and $\nuebar$
          (right) events for water Cerenkov
          (\texttt{0524signal-ext2.dat} and \texttt{0525asignal-ext2.dat}).
          This smearing is used to simulate the energy resolution of the
          detector for both $\nue$ signal and intrinsic $\nue$ backgrounds.}
  \label{fig:lbl_eres_nue}
\end{figure}

\begin{figure}[htb]
 \centering\includegraphics[width=.4\textwidth]{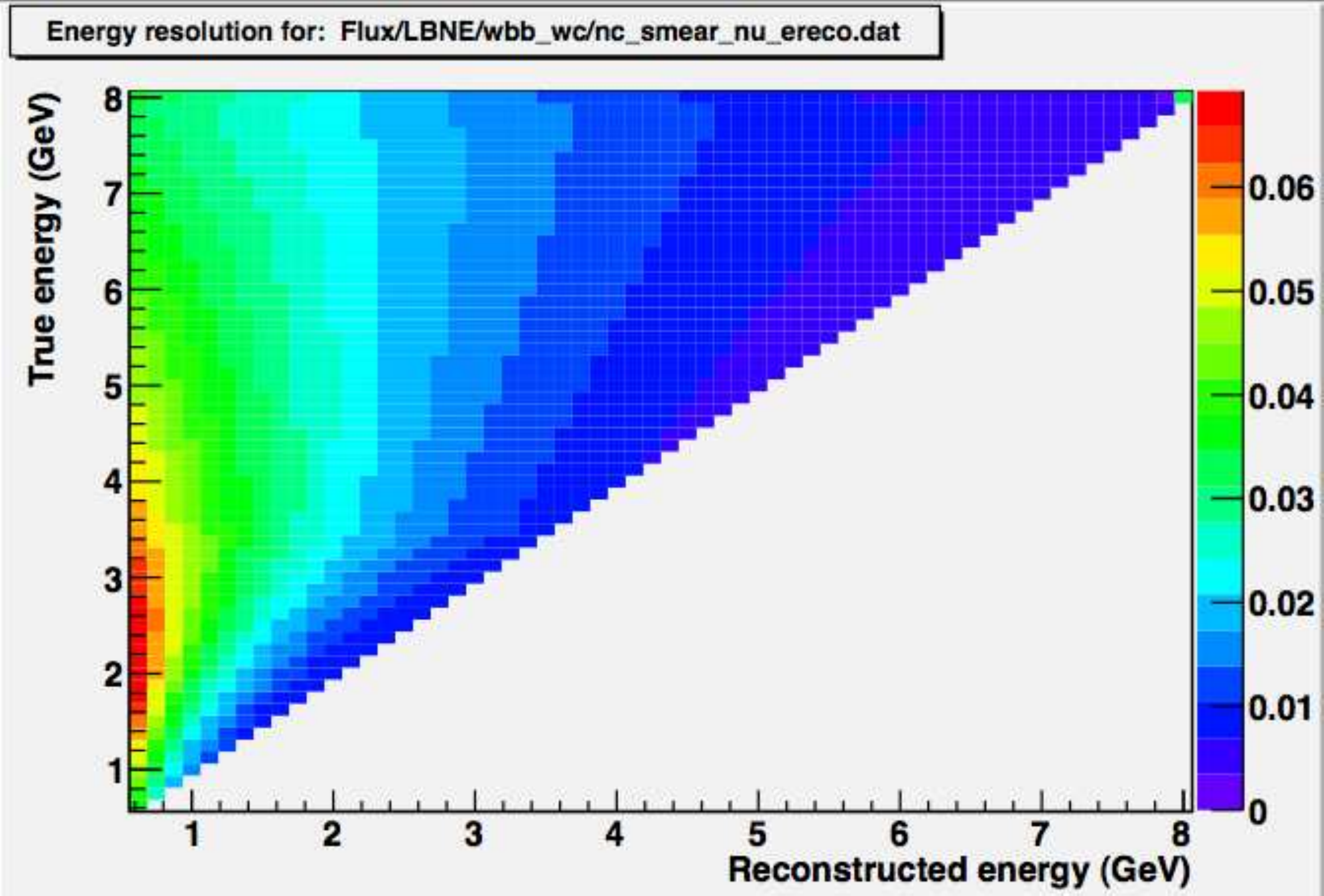}
 \centering\includegraphics[width=.4\textwidth]{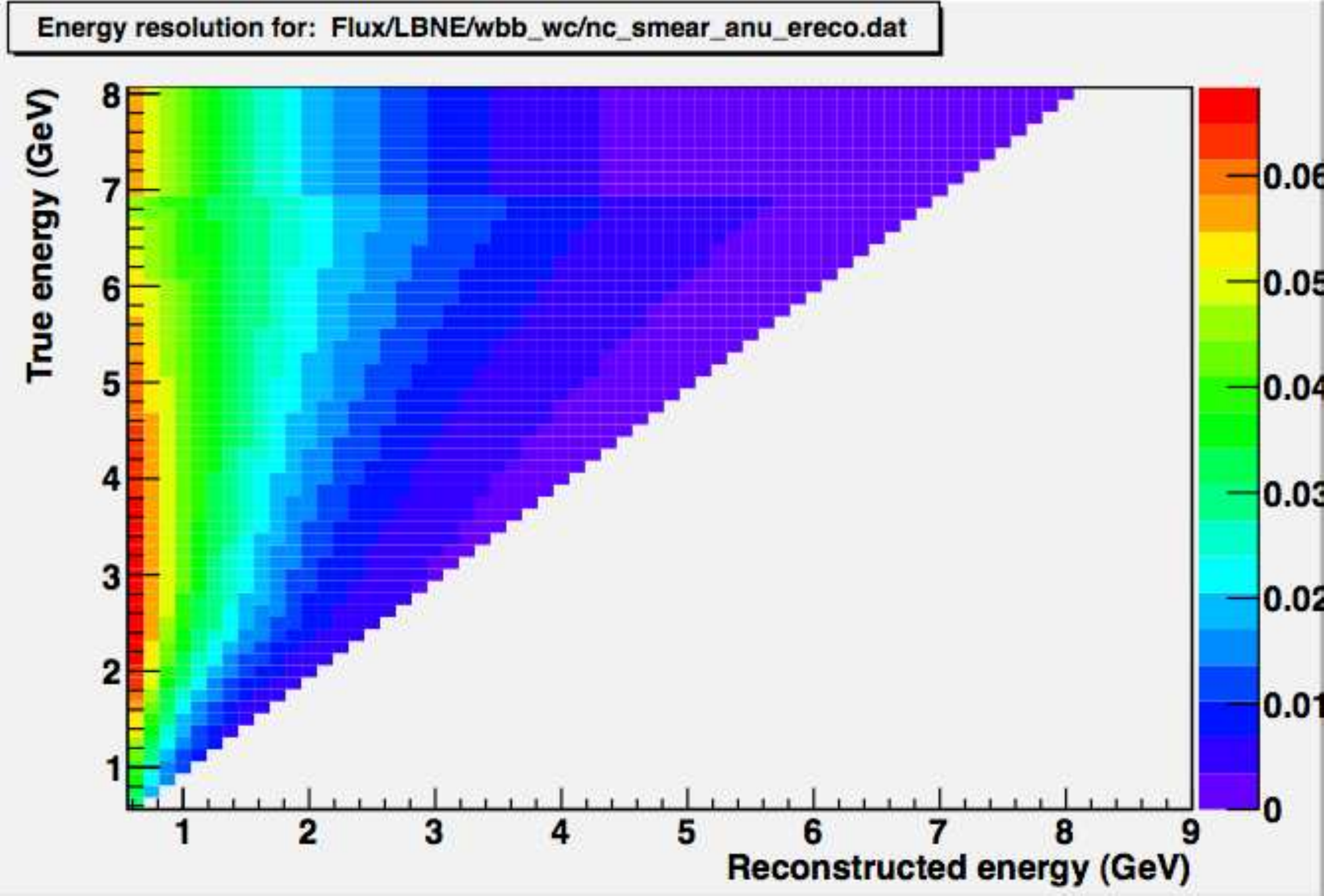}
  \caption{Energy resolutions for NC backgrounds in neutrino (left)
          and antineutrino (right) modes for water Cerenkov
          (\texttt{nc-smear-nu-ereco.dat} and
          \texttt{nc-smear-anu-reco.dat}). This smearing is used to simulate
          the energy resolution of the detector for $\numu$ NC events.}
  \label{fig:lbl_eres_nc_bkg}
\end{figure}

\begin{figure}[htb]
 \centering\includegraphics[width=.4\textwidth]{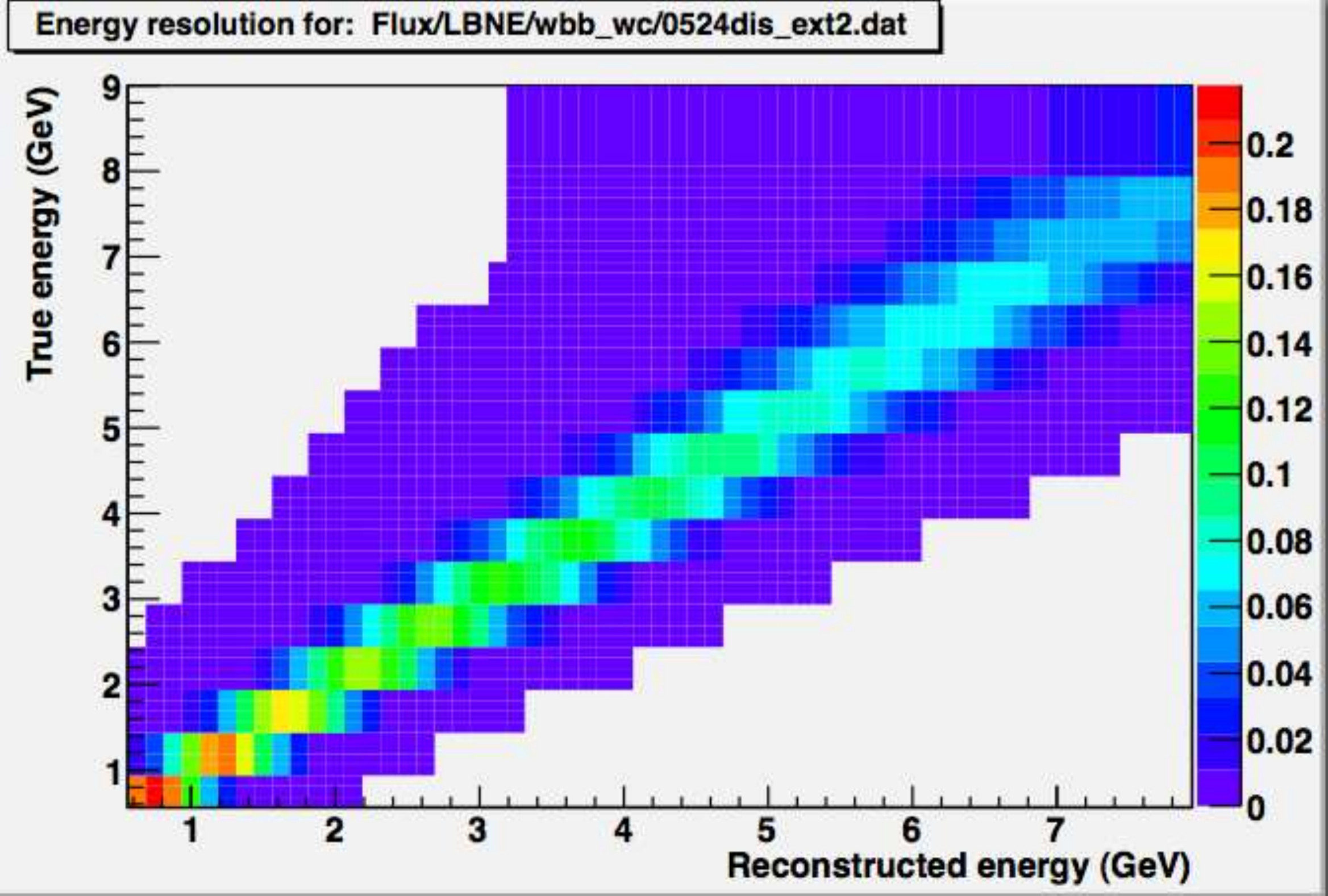}
 \centering\includegraphics[width=.4\textwidth]{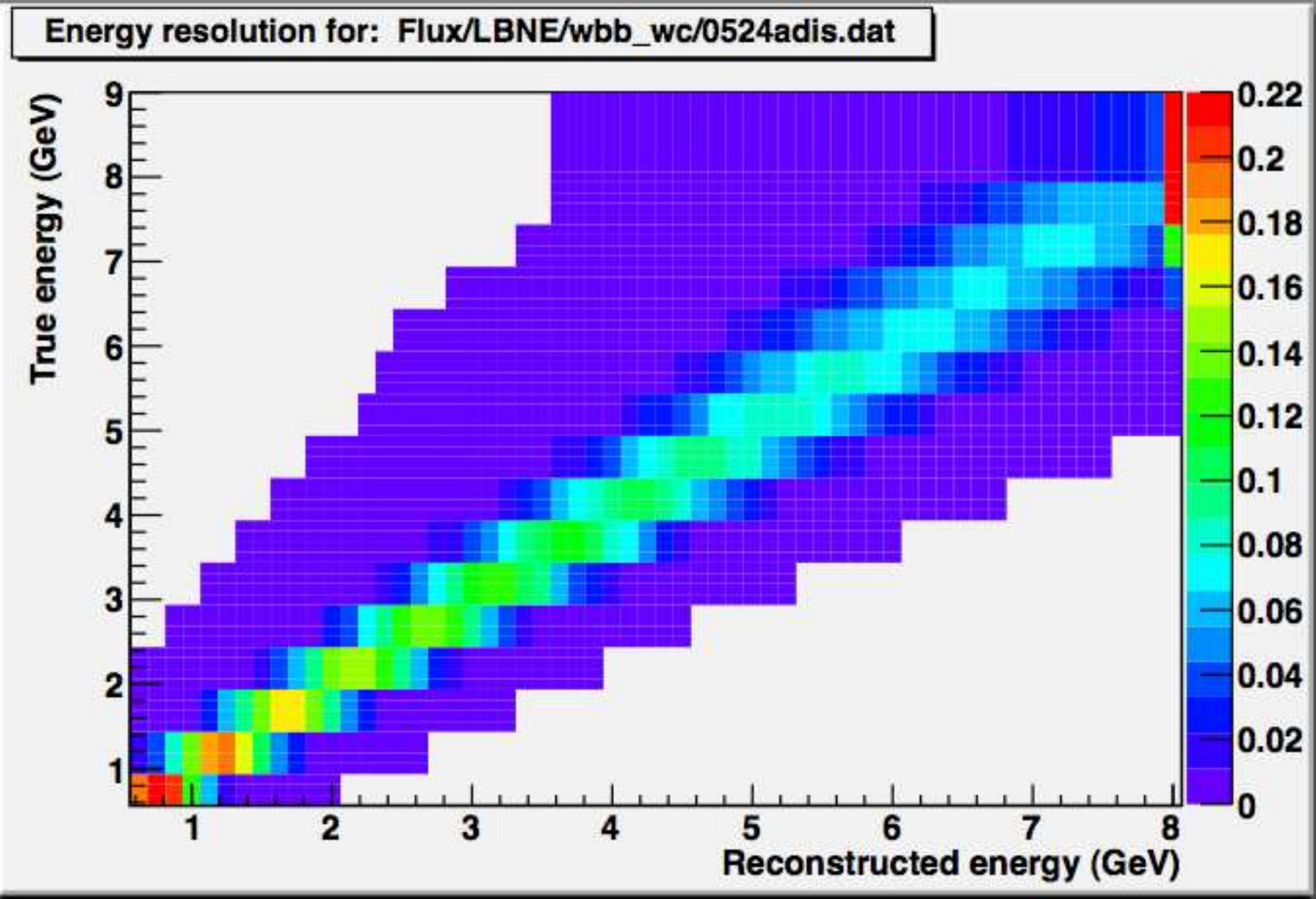}
  \caption{Energy resolutions for $\numu$ mis-ID backgrounds in neutrino
         (left) and antineutrino (right) modes for water Cerenkov
         (\texttt{0524dis-ext2.dat} and \texttt{0524adis.dat}).
          This smearing is used to simulate the energy resolution of the
         detector for $\numu$ CC events.}
  \label{fig:lbl_eres_dis_bkg}
\end{figure}

\clearpage
\subsection{Inputs for $\nu_\mu$ Disappearance}

To estimate how accurately LBNE can measure $\sin^22\theta_{23}$ and
$\Delta m^2_{32}$, GLoBES requires information on $\numu$ efficiencies,
backgrounds, and energy resolutions for both WC and LAr detectors. These
inputs are summarized in Table~\ref{table:lbl_globes_inputs_numu} and
detailed below.

\begin{table} [H]
\begin{center}
\begin{tabular}{|c||c|c|} \hline
Input  & Water Cerenkov  & Liquid Argon TPC \\ \hline\hline
signal channel                & $\numu$ QE  & $\numu$ CC \\ \hline
signal efficiencies           & $97\%$      & $85\%$     \\ \hline
signal $E_\nu$ resolution, $\sigma(E)/E$    & Fig.~\ref{fig:lbl_eres_dis_bkg} & $0.20/\sqrt{E}$ \\ \hline
signal normalization error    & $5\%$  & $5\%$ \\ \hline
signal $E_\nu$ scale error    & $3\%$  & $2\%$ \\ \hline
\hline
background channels            & $\numu$ CC $\pi^+$ & $\numu$ NC \\ \hline
background efficiencies        & $97\%^{(*)}$       & $0.5\%$ \\ \hline
background $E_\nu$ resolution, $\sigma(E)/E$  & Fig.~\ref{fig:lbl_eres_ccpip_bkg} & $0.20/\sqrt{E}$\\ \hline
background normalization error  & $10\%$  & $10\%$ \\ \hline
background $E_\nu$ scale error  & $3\%$   & $2\%$ \\ \hline
\end{tabular}
\caption{\label{table:lbl_globes_inputs_numu} Summary of GLoBES inputs
for both detector technologies for the LBNE $\numu$ disappearance sensitivity
estimates. While not listed above, the neutrino contribution for each channel
is explicitly included in GLoBES in the case of antineutrino running.
(*) The CC $\pi^+$ event rate is calculated from the QE cross section
rescaled by a factor of 0.21. This value of 0.21 appears as an additional
efficiency factor for this process in the GLoBES input file. Even though
the QE cross section is used, the energy smearing takes into account
the fact that this background is largely due to CC $\pi^+$ events
(Fig.~\ref{fig:lbl_eres_ccpip_bkg}). This was done for convenience.}
\end{center}
\end{table}

\begin{figure}[htb]
 \centering\includegraphics[width=.45\textwidth]{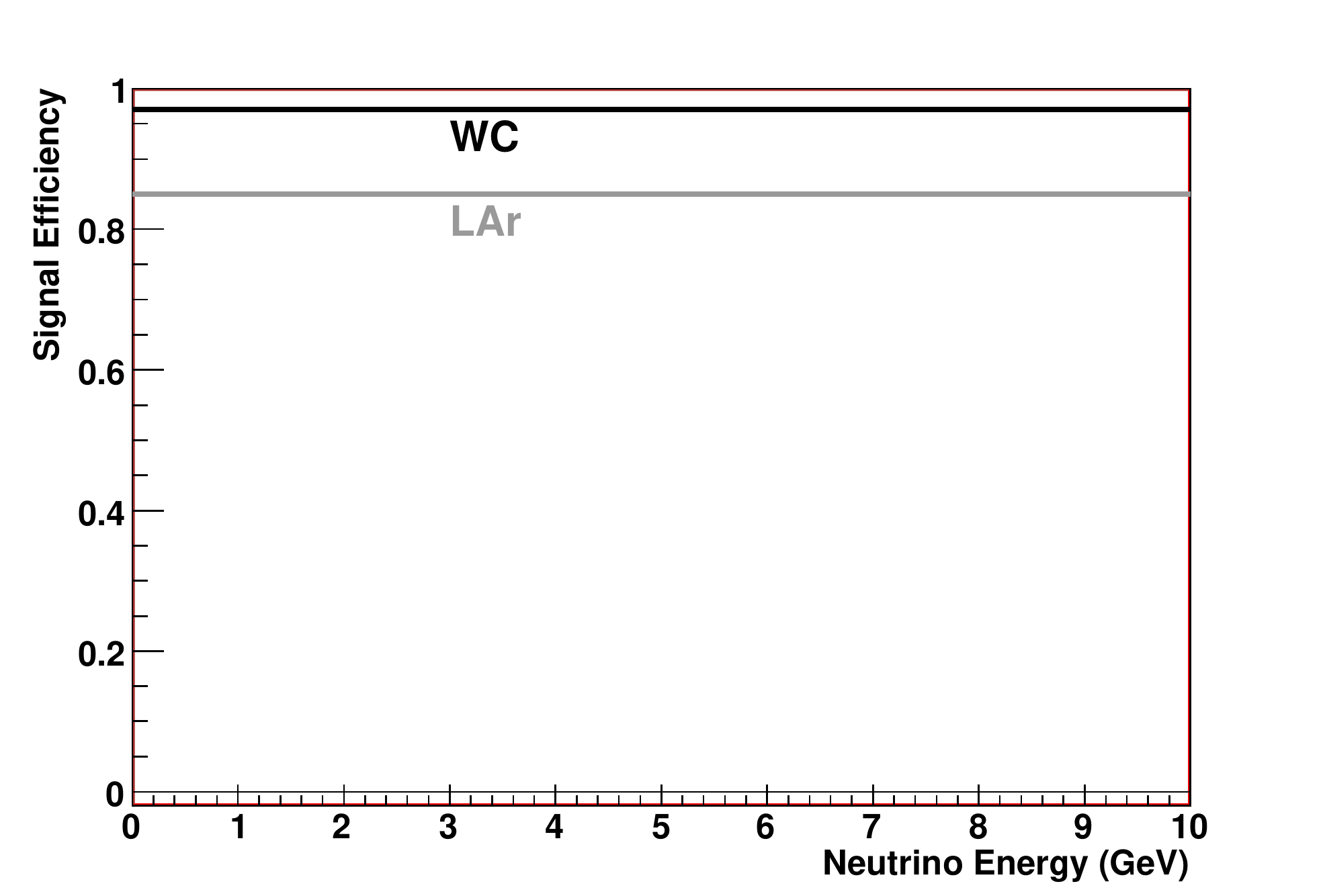}
 \centering\includegraphics[width=.45\textwidth]{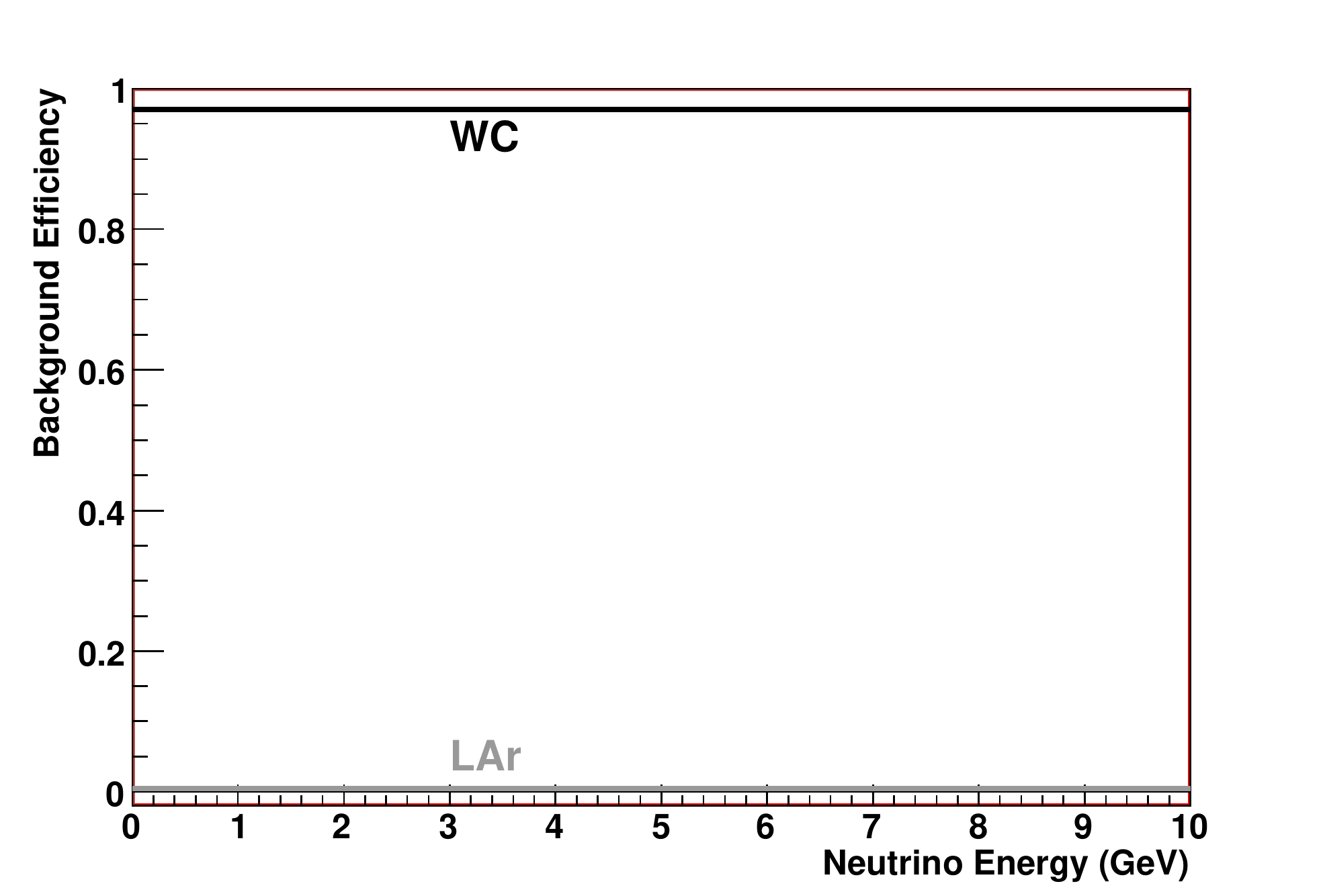}
  \caption{Assumed $\numu$ signal (left) and background (right)
   efficiencies plotted as a function of neutrino energy for WC (black)
   and LAr (gray). This reflects the fraction of signal and background
   events that enter the $\numu$ disappearance sample. For WC, a $97\%$
   efficiency is assumed for both $\numu$ QE signal and non-QE
   backgrounds. For LAr, a signal efficiency of $85\%$ and a NC background
   efficiency of $0.5\%$ is assumed.}
  \label{fig:lbl_numu_net}
\end{figure}

Fig.~\ref{fig:lbl_numu_net} shows the net signal and background
efficiencies plotted as a function of neutrino energy for each detector type.
In both cases, efficiencies that are flat in energy are used.

There are several things to note. The WC and LAr disappearance fits differ
in that a $\numu$ QE sample is selected as the signal in one case (WC) and a
$\numu$ CC sample in the other (LAr). This does not mean than a QE sample could
not be used in LAr or a CC sample in WC. This was simply the first choice for
these initial LBNE sensitivity estimates. Optimization of the signal choice
will be the subject of further study. Also, as has been pointed out,
a $97\%$ $\numu$ QE efficiency in WC is probably too optimistic. This estimate
will be better quantified and revised in future sensitivity estimates. In
addition, there are several input values that have recently changed. Explicit
energy scale uncertainties have been added for both WC and LAr and assigned
conservative $2-3\%$ uncertainties based on the projected performance of
similar detectors. Also, the neutrino energy resolutions for LAr have been
increased to $20\%$ to reflect the ability of a LAr detector to reconstruct
$E_\nu$ from the energies of the outgoing muon and hadronic shower.
Specifically, the value of $20\%/\sqrt{E}$ is based on an estimate of the
neutrino energy resolution in ICARUS for events with $E_\nu<1.25$
GeV~\cite{fleming, LAr_perf_fleming} and will be re-evaluated for LBNE energies using Monte
Carlo. In addition, the uncertainty on the normalization of signal events
for both WC and LAr has been increased to $5\%$. Uncertainties on the
normalization of the non-QE and non-CC backgrounds are conservatively set
at $10\%$.

\begin{figure}[htb]
 \centering\includegraphics[width=.4\textwidth]{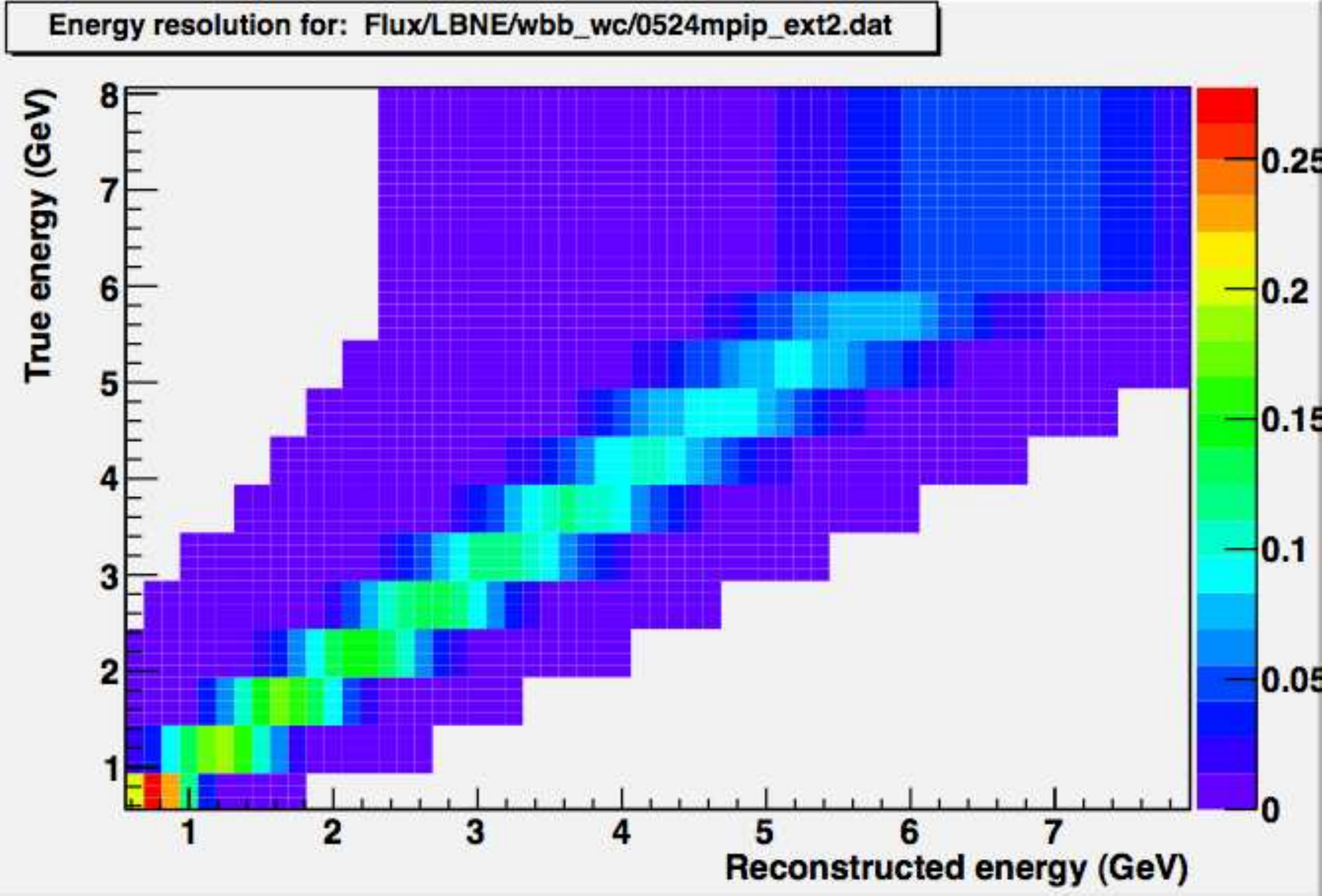}
 \centering\includegraphics[width=.4\textwidth]{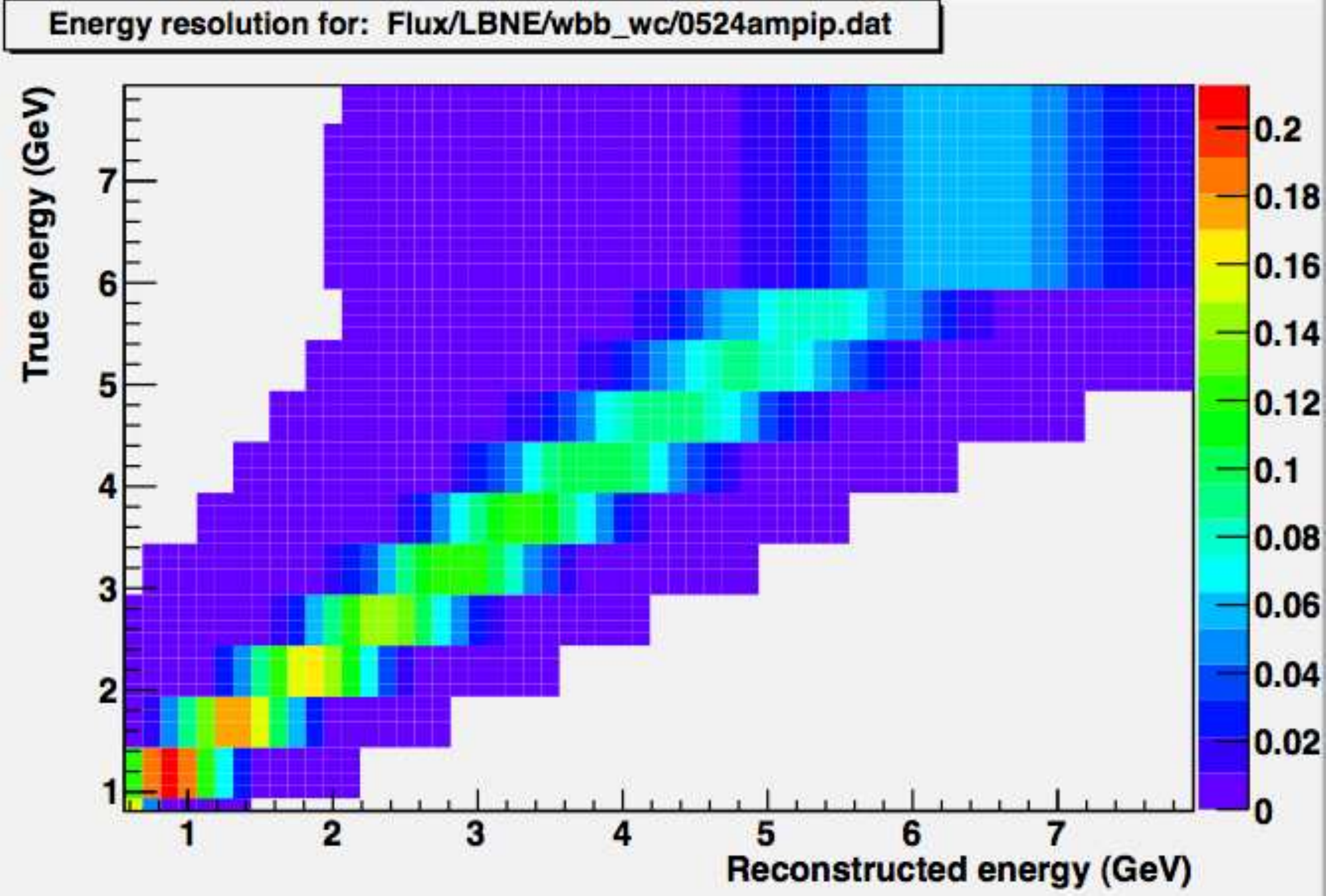}
  \caption{Energy resolutions for $\numu$ CC $\pi^+$ backgrounds in neutrino
         (left) and antineutrino (right) modes for water Cerenkov
         (\texttt{0524mpip-ext2.dat} and \texttt{0524ampip.dat}).
         These events will preferentially mis-reconstruct at lower energies
         when using a QE assumption, however the current smearing doesn't accurately
         reproduce this effect. This is under investigation.}
  \label{fig:lbl_eres_ccpip_bkg}
\end{figure}

For the measurement of the atmospheric parameters ($\sin^22\theta_{23}$ and
$\Delta m^2_{32}$), assumptions must be made for $\theta_{13}$, $\delta_{\mathrm CP}$,
and the solar parameters. The values that were used are provided in
Table~\ref{table:lbl_osc_pars}.

\begin{table} [H]
\begin{center}
\begin{tabular}{ccc}
parameter  & value  & uncertainty \\ \hline
$\theta_{12}$ & 0.601 & $\pm 10\%$ \\
$\theta_{13}$ & 0     & $\pm 0\%$ \\
$\delta_{\mathrm CP}$ & 0     & $\pm 0\%$ \\
$\Delta m^2_{21}$ & $7.59\times10^{-5}$     & $\pm 4\%$ \\
$\Delta m^2_{31}$ & $2.41\times10^{-3}$     & $\pm 5\%$ \\
\end{tabular}
\caption{Oscillation parameters and uncertainties used in the LBNE $\numu$
disappearance sensitivity estimates.}
\label{table:lbl_osc_pars}
\end{center}
\end{table}

\subsection{Inputs for Resolving $\theta_{23}$ Octant Degeneracy}

The projections for LBNE's ability to resolve the $\theta_{23}$ octant
degeneracy were generated by Joachim Kopp assuming the ``August 2010'' beam
(\texttt{dusel120e250(n)i002dr280dz-tgtz30-1300km-0kmoa-flux}, red
curve in Fig.~\ref{fig:fig_beam_spectra1}). The same GLoBES inputs are used
as the aforementioned $\nue$ appearance and $\numu$ disappearance studies
with a few exceptions. The energy scale uncertainties (as listed in
Table~\ref{table:lbl_globes_inputs_numu}) are not yet included. Also, the
energy resolutions from Table~\ref{table:lbl_globes_inputs_nue} are used
for LAr in the disappearance case rather than those from
Table~\ref{table:lbl_globes_inputs_numu}.

\subsection{Inputs for Non-Standard Interactions (NSI)}

The projection for LBNE's sensitivity to new physics is based on work
done by Joachim Kopp and Patrick Huber (see also~\cite{kopp-tn} for an earlier
version of this work). All of the GLoBES inputs are the same as in the LBNE
$\nue$ appearance studies. The only input that differs is the neutrino flux.
Kopp's NSI plots and estimates were made assuming the 2008/2009 LBNE beam
design (\texttt{lbne120e300(n)i002dr250dz-1300m-0kmoa-flux.txt},
black curve in Fig.~\ref{fig:fig_beam_spectra1}). The NSI projections for
LBNE will likely be improved with the newer ``August 2010'' beam
(\texttt{dusel120e250(n)i002dr280dz-tgtz30-1300km-0kmoa-flux}, red
curve in Fig.~\ref{fig:fig_beam_spectra1}) as a result of the higher
flux in the 2-6~GeV region.

%

\section{Supernova Burst Physics Sensitivity Assumptions}\label{snb_assumptions}

\subsection{Assumptions for Event Rates in Water}

\begin{figure}[htb]
  \centering\includegraphics[width=.45\textwidth]{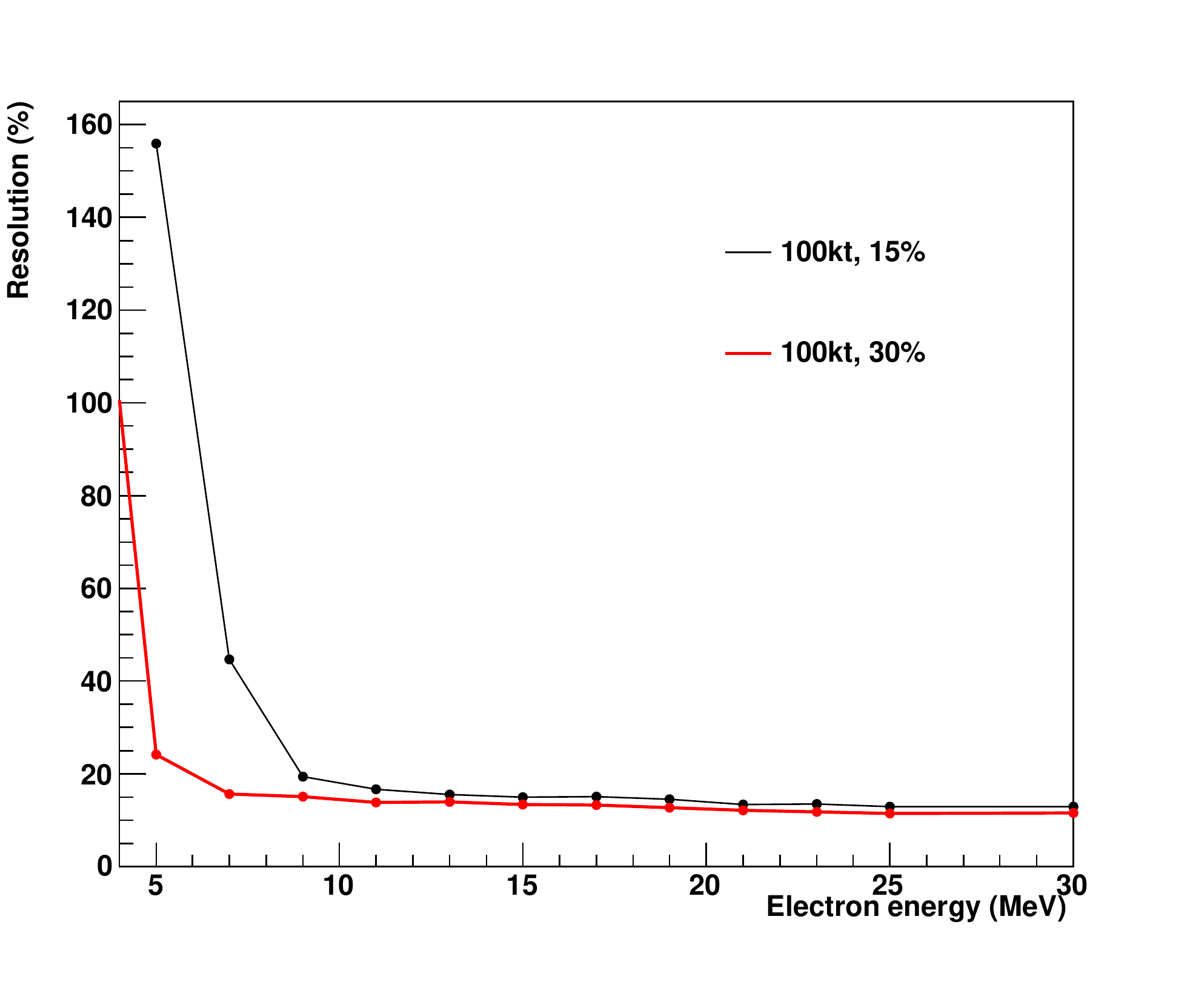}
  \centering\includegraphics[width=.45\textwidth]{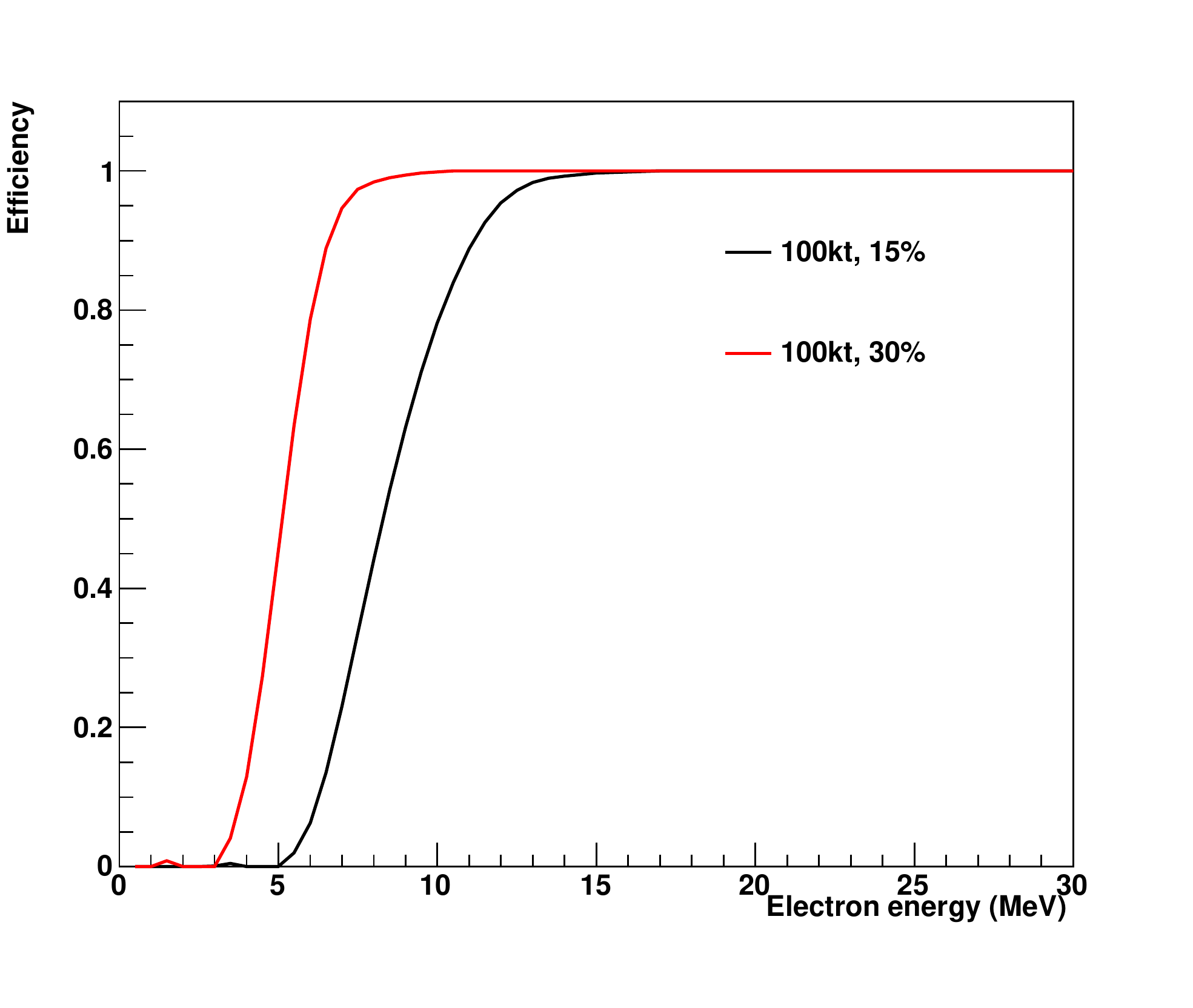}
   \caption{Energy resolution (left) and trigger efficiency (right) as a function of electron energy, for 30\% and 15\% PMT coverages.}
   \label{fig:res_trigger_water}
\end{figure}

We used WCsim~\cite{wcsim} to evaluate the detection response in water for the 15\% and 30\% PMT coverage configurations.  At the moment, we do not yet have a perfect match between WCsim output in Super-Kamiokande mode and published SK detector parameters in the few to few tens of MeV range~\cite{Hosaka:2005um,Cravens:2008zn}-- the agreement is at about the 10\% level.  The resolution used as a function of electron energy is shown in Fig.~\ref{fig:res_trigger_water}~(left); it was scaled by a factor of 0.66 in order
that WCsim SK-mode output match the resolution in Ref.~\cite{Hosaka:2005um} (in addition to minor simulation mismatch, we do not have all SK software tools for energy response
reconstruction at our disposal for WCsim output, which likely accounts for some of the discrepancy).

Trigger efficiency is shown in Fig.~\ref{fig:res_trigger_water}~(right) for the different configurations; this assumes a trigger requirement of 33 hits in 300~ns for the 15\% PMT coverage configuration and 39 hits in 300~ns for the 30\% configuration. (Note that we are assuming supernova neutrino events in water are self-triggered; one can imagine a configuration in which all digitized PMT hits are saved in the event of a high rate, which could improve efficiency.)  These efficiencies are slightly worse than SK~I~\cite{Hosaka:2005um} and SK~II~\cite{Cravens:2008zn} efficiencies.

\subsection{Assumptions for Event Rates in Argon}

For event rate estimates in liquid argon, we assume a detection threshold of 5~MeV.
We assume also that suitable triggering will be provided from photons, charge collection or from some external trigger.
The energy resolution used is from Ref.~\cite{Amoruso:2003sw},~$\frac{\sigma}{E}~=~\frac{11\%}{\sqrt{E}}~+~2\%$.

A summary of preliminary liquid argon TPC performance parameters can be found in a presentation by Bonnie Fleming to the Long-Baseline group~\cite{LAr_perf_fleming}.

\clearpage
\vfil\eject


%

\vfill\eject

\end{document}